\documentclass[
nofootinbib,amsmath,amssymb,
aps,
floatfix
]{revtex4-1}
\pdfoutput = 1

\usepackage{graphicx}          
\usepackage{dcolumn}          
\usepackage{multirow}
\usepackage{bm}                
\usepackage{color}

\usepackage[table]{xcolor}
\usepackage{subfig}
\usepackage[utf8]{inputenc}

\definecolor{darkblue}{rgb}{0.0,0,0.5}
\definecolor{darkred}{rgb}{0.7,0,0.0}
\definecolor{medmagenta}{rgb}{0.3,0,0.7}
\definecolor{darkmagenta}{rgb}{0.4,0,0.4}
\definecolor{lightgray}{gray}{0.87}
\usepackage[unicode=true, bookmarks=false, linkcolor = darkblue, citecolor = medmagenta, urlcolor=darkmagenta,breaklinks=false, colorlinks=true, hyperfootnotes=true]{hyperref}
\DeclareUnicodeCharacter{FF0C}{,}

\usepackage{epsfig,amssymb,amsmath,psfrag,epstopdf,color}
\usepackage{ulem}
\usepackage{amsthm}

\newcommand{\be}{\begin{equation}}
\newcommand{\ee}{\end{equation}}
\newcommand{\bea}{\begin{eqnarray}}
\newcommand{\eea}{\end{eqnarray}}

\newcommand{\beaa}{\begin{equation}\begin{aligned}}
\newcommand{\eeaa}{\end{aligned}\end{equation}}

\newcommand{\Q}{Q}

\allowdisplaybreaks
\DeclareGraphicsExtensions{.eps,.pdf,.png,.jpg,.jpeg}

\newcommand*{\LEinline}[1]%
{\todo[inline, size=\footnotesize]{#1}}

\newcommand{\CTHERAII}{CT14$_{\textrm{HERAII}}$}
\begin{document}

\preprint{MSUHEP-19-025, PITT-PACC-1911, SMU-HEP-19-03}

\title{New CTEQ global analysis of quantum chromodynamics with high-precision data from the LHC}

\author{Tie-Jiun Hou$^{(a,1)}$, Jun Gao$^{(b)}$,  T.~J.~Hobbs$^{(c,d)}$,
Keping Xie$^{(c,e)}$,
Sayipjamal Dulat$^{(f,2)}$, Marco Guzzi$^{(g)}$, Joey Huston$^{(h)}$, Pavel Nadolsky$^{(c,3)}$, Jon Pumplin$^{(h,\dagger)}$, Carl Schmidt$^{(h)}$, Ibrahim Sitiwaldi$^{(f)}$, Daniel Stump$^{(h)}$,  C.-P. Yuan$^{(h,4)}$}

\affiliation{$^{(a)}$Department of Physics, College of Sciences, Northeastern University, Shenyang 110819, China \\
$^{(b)}$INPAC, Shanghai Key Laboratory for Particle Physics and Cosmology \\ \& School of Physics and Astronomy, Shanghai Jiao Tong University, Shanghai 200240, China \\
Center for High Energy Physics, Peking University, Beijing 100871, China \\
$^{(c)}$Department of Physics, Southern Methodist University, Dallas, TX 75275-0181, U.S.A. \\
$^{(d)}$Jefferson Lab, EIC Center, Newport News, VA 23606, U.S.A. \\
$^{(e)}$PITT PACC, Department of Physics and Astronomy, University of Pittsburgh, Pittsburgh, PA 15260, U.S.A. \\
$^{(f)}$School of Physics Science and Technology \\ 
Xinjiang University, Urumqi, Xinjiang 830046 China \\
$^{(g)}$Department of Physics, Kennesaw State University, 370 Paulding Ave.,  30144 Kennesaw, GA, U.S.A. \\
$^{(h)}$Department of Physics and Astronomy, Michigan State University, East Lansing, MI 48824 U.S.A. \\
}%
\email{$^{1}$tjhou@msu.edu, 
$^{2}$sdulat@hotmail.com, $^{3}$nadolsky@smu.edu, $^{4}$yuan@pa.msu.edu}
\thanks{\\ $^{\dagger}$Deceased}

 \date{\today}

\newpage
\begin{abstract}
We present the new parton distribution functions (PDFs) from the CTEQ-TEA collaboration, obtained using a wide variety of high-precision Large Hadron Collider (LHC) data, in addition to the combined HERA I+II deep-inelastic scattering data set, along with the data sets present in the CT14 global QCD analysis. 
New LHC measurements in single-inclusive jet production with the full rapidity coverage, as well as production of Drell-Yan pairs, top-quark pairs, and high-$p_T$ $Z$ bosons, are included to achieve the greatest sensitivity to the PDFs. The parton distributions are determined at NLO and NNLO, with each of these PDFs accompanied
by error sets determined using the Hessian method. Fast PDF survey techniques, based on the Hessian representation and the Lagrange Multiplier method, are used to quantify the preference of each data set to quantities such as $\alpha_s(m_Z)$, and the gluon and strange quark distributions. We designate the main resulting PDF set as CT18. The ATLAS 7 TeV precision $W/Z$ data are not included in CT18, due to their tension with other data sets in the global fit.  Alternate PDF sets are generated including the ATLAS precision 7 TeV $W/Z$ data (CT18A), a new scale choice for low-$x$ DIS data (CT18X), or all of the above with a slightly higher choice for the charm mass (CT18Z). Theoretical calculations of standard candle cross sections at the LHC (such as the $gg$ fusion Higgs boson cross section) are presented. 
\end{abstract}


\maketitle

\tableofcontents\newpage

\renewcommand*{\thefootnote}{\arabic{footnote}}
\setcounter{footnote}{0}

\section{\label{sec:Introduction} Introduction}

With an accumulated data sample of over 140 fb$^{-1}$ at the 13 TeV run for both ATLAS and CMS collaborations, the Large Hadron Collider (LHC) has entered an era of precision physics.
The experimental precision has been matched by improvements to the theoretical predictions, with a number of collider processes now available at the next-to-next-to-leading order
(NNLO) in the QCD coupling strength. Such precision is necessary for rigorous tests of the Standard Model (SM) and in searches for signs of physics beyond the Standard Model
(BSM), as there have been no `smoking-gun' signs of BSM physics to date. Precise predictions in QCD theory require correspondingly precise parton distribution
functions (PDFs)~\cite{Dulat:2015mca,Harland-Lang:2014zoa,Ball:2017nwa,Alekhin:2017kpj,Accardi:2016qay,Harland-Lang:2019pla,Bertone:2017bme,Manohar:2017eqh},
which in turn warrant advances in interpreting LHC experiments to extract important information about the SM, and possibly, BSM physics.

To this end, we present a new family of CTEQ-TEA parton distribution functions, designated as CT18. These PDFs are produced at both next-to-leading order (NLO) and NNLO in the QCD coupling constant, $\alpha_s$. The CT18 PDFs update those of the CT14 family presented in Ref.~\cite{Dulat:2015mca}.
In the new analysis, we include a variety of new LHC data, at the center-of-mass energies of 7 and 8 TeV, on production of single-inclusive jets, $W/Z$ bosons, and top-antitop
quark pairs, obtained by the ATLAS, CMS and LHCb collaborations. At the same time, the update retains crucial ``legacy'' data from the previous CT global QCD analyses, such as the HERA I+II combined data on deep-inelastic scattering (DIS) and measurements in fixed-target experiments and at the Fermilab Tevatron $p\bar p$ collider. Measurements of processes in similar kinematic regions, by both ATLAS and CMS, allow crucial cross-checks of the data. Measurements by LHCb often allow extrapolations into new kinematic regions not covered by the other experiments. Some processes, such as $t\bar{t}$ production, allow for the measurement of multiple observables that provide similar information for the determination of PDFs. 
In addition to the PDFs themselves, we also present relevant PDF luminosities, and predictions with uncertainties for standard candle cross sections at the LHC.

The goal of the CT18 analysis is to include as wide a kinematic range for each measurement as possible, while still achieving reasonable agreement between data and theory. For the ATLAS
7 TeV jet data~\cite{Aad:2014vwa}, for example, all rapidity intervals cannot be simultaneously used without the introduction of systematic error decorrelations provided by the ATLAS collaboration~\cite{Aaboud:2017dvo}. Even with that decorrelation, the resultant $\chi^2$ for the new jet experiments is not optimal, resulting in less effective PDF constraints. Inclusive cross section measurements for jet production have been carried out for two different jet-radius values, $R$, by both ATLAS and CMS. For both experiments, we have chosen the data with the larger $R$-value, for which the NNLO (fixed order) prediction should have a higher accuracy. We evaluate the jet cross section predictions using a QCD scale of inclusive jet transverse momentum $\Q\! =\! p_T^{jet}$, consistent with past usage at NLO. The  result is largely consistent with similar evaluations using
$\Q\! =\! H_T$~\cite{Currie:2016bfm,Currie:2017ctp,Currie:2018xkj}.

Theoretical predictions for comparison to the data used in the global fit have been carried out at NNLO, either indirectly through the use of fast interpolation tables such
as \texttt{fastNLO}~\cite{Britzger:2012bs,Wobisch:2011ij} and  \texttt{ApplGrid}~\cite{Carli:2010rw}, together with NNLO/NLO $K$-factors, or directly (for top-quark related observables) through
the use of \texttt{fastNNLO} grids~\cite{Czakon:2017dip,fastnnlo:grids}. 

In an ideal world all such data sets would perfectly be compatible with each other, but differences are observed that do result in some tension between data sets and pulls in opposite directions.
One of the crucial aspects of carrying out a global PDF analysis is dealing with data sets that add some tension to the fits, while preserving the ability of the combined data set to improve on the existing constraints on the PDFs. In some cases, a data set may be in such tension as to require its removal from the global analysis, or its inclusion only in a separate iteration of the new PDF set. 

In this paper, we will describe how the high-precision ATLAS 7 TeV $W/Z$ rapidity distributions, which, as we find, favor an increase of the strange quark distribution at low $x$, require such special treatment. In particular, while other PDF-analysis groups ({\it e.g.}, MMHT, see Ref.~\cite{Thorne:2019mpt}) have noted that these ATLAS $W/Z$ data can be fitted with  $\chi^2/\mathrm{d.o.f.}$ that is comparable to the CT18 one, we find that such $\chi^2$ reflects systematic tensions with many of the other data in our global analysis.
Furthermore, the standard Hessian profiling technique used by the experimental collaborations significantly underestimates the minimal $\chi^2$ that can be reached for the ATLAS 7 TeV $W/Z$ data when they are included in the CT18 fit.
We therefore treat these measurements separately in an alternative fit, CT18A, introduced in Sec.~\ref{sec:alt}.
In another variant, CT18X, a special scale $\mu^2_{F,x} \equiv 0.8^2 \left(Q^2 + 0.3\mbox{ GeV}^2/x_B^{0.3}\right)$ is used for the calculation of low-$x$ DIS cross sections; the scale mimics the impact of low-$x$ resummation. Both modifications cause an increase in the low-$x$ quark and gluon distributions. Finally, these two variants of the CT18 fit are
amalgamated into a combined alternative fit, CT18Z. Since the CT18Z PDFs are most dissimilar from the CT18 ones, we show numerous results based on the CT18 and CT18Z PDFs throughout the article, while deferring the comparisons to CT18A and X to  Appendix~\ref{sec:AppendixCT18Z}, where additional in-depth comparisons to CT18Z are also provided. A recommendation on selecting one of the four PDF ensembles depending on the user's needs is given at the beginning of Sec.~\ref{sec:Conclusions}.

Our current global analyses are carried out in four stages. First, \texttt{PDFSense}~\cite{Wang:2018heo,Hobbs:2019gob}, a program for a rapid survey of QCD data using the Hessian
approach \cite{Pumplin:2001ct,Pumplin:2002vw}, is used to select the  data sets that are expected to have the greatest impact on the global PDF sets. This selection takes into account the sensitivity of the data to specific PDFs in a given $x$ range, 
which reflects both the correlation of these data with a given PDF, as well the size of the data set and magnitudes of its statistical
and correlated systematic errors. For example, both the collider inclusive jet data and the top-quark data have a strong correlation with the high-$x$ gluon,
but the inclusive jet data has a larger sensitivity due to a much larger number of data points. Next, \texttt{ePump}~\cite{Schmidt:2018hvu,Hou:2019gfw} is used to quickly examine the quantitative
impact of each selected data set, within the Hessian approximation. Third, the full global PDF fit is carried out using all such data sets. Recent enhancements to the CT global analysis code have
greatly improved the speed of the calculations. Lastly, the impact of key data sets on certain PDFs at specific kinematic points of interest, as well as on the value of $\alpha_s(M_Z)$,
is assessed using the Lagrange Multiplier (LM) method~\cite{Stump:2001gu}. 
In order to minimize any parametrization bias, we have tested different parametrizations for CT18: {\it e.g.}, using a more flexible parametrization for the strange quark PDF.
In some kinematic regions, there are fewer constraints from the data on certain PDFs. In particular, LM constraints have been applied to limit the strangeness PDF at $x < 10^{-5}$ to
physically reasonable values, as summarized in App.~\ref{sec:AppendixParam}.

Our paper is organized as follows.
Section~\ref{sec:Datasets} begins with an executive summary of the key stages and results of the CT18 global analysis. It continues with an overview of the chosen experimental data
and alternative fits (CT18Z, CT18A, and CT18X) in the CT18 release. This section concisely summarizes the key results that are of interest to most readers. The subsequent sections and appendices elaborate on specific aspects and outcomes. 

In Sec.~\ref{sec:Theory} we detail theoretical/computational updates to the CT
fitting methodology and details for specific process-dependent calculations.
Sec.~\ref{sec:OverviewCT18} presents the main results obtained in CT18 ---
the fitted PDFs as functions of $x$ and $\Q$, determinations of QCD parameters ($\alpha_s$, $m_c$), calculated parton luminosities, and various PDF moments and sum rules. These comparisons be of interest to a broad group of researchers who will use the PDFs for theoretical predictions at LHC experiments.

Sec.~\ref{sec:Quality} describes the ability of CT18 to provide a successful theoretical
description of the fitted data.
In addition to characterizing the fit of individual data sets, in Sec.~\ref{sec:StandardCandles}
we also compute the various standard candle quantities of relevance to LHC phenomenology, for
instance, Higgs boson production cross sections at $13$ and $14$ TeV, and various correlations among
electroweak boson and top-quark pair production cross sections.
In Sec.~\ref{sec:Conclusions}, we discuss the broader implications of this work and highlight
our main conclusions.

Several appendices present a number of important supporting details.
In Appendix~\ref{sec:AppendixCT18Z}, we review the CT18Z and other alternative fits,
including descriptions of various data sets admitted into these separate analyses.
A number of more formal details related to our likelihood functions and relations
among covariance matrices are summarized in Appendix~\ref{sec:chi2_app}.
Appendix~\ref{sec:AppendixParam} presents the analytical fitting form adopted in CT18
and best-fit values of the PDF parameters.
Appendix~\ref{sec:AppendixCodeDevelopment} presents a number of technical advances
in the CT fitting framework, including code parallelization, while Appendix~\ref{sec:ATLASjetdecorrel}
enumerates the decorrelation models utilized in fitting the newly included inclusive
jet data from the LHC.
Lastly, in Appendix~\ref{sec:Appendix4xFitter}, we present the results of a short study based on Hessian profiling
methods to assess the impact of the 7 TeV $W/Z$ production data taken by ATLAS.

\section{\label{sec:Datasets} Overview of the CT18 global QCD analysis
}

\subsection{\label{sec:summary} Executive Summary}

\subsubsection{Input experimental data and final PDF ensembles}

\begin{figure}[tb]
	\includegraphics[width=0.99\textwidth]{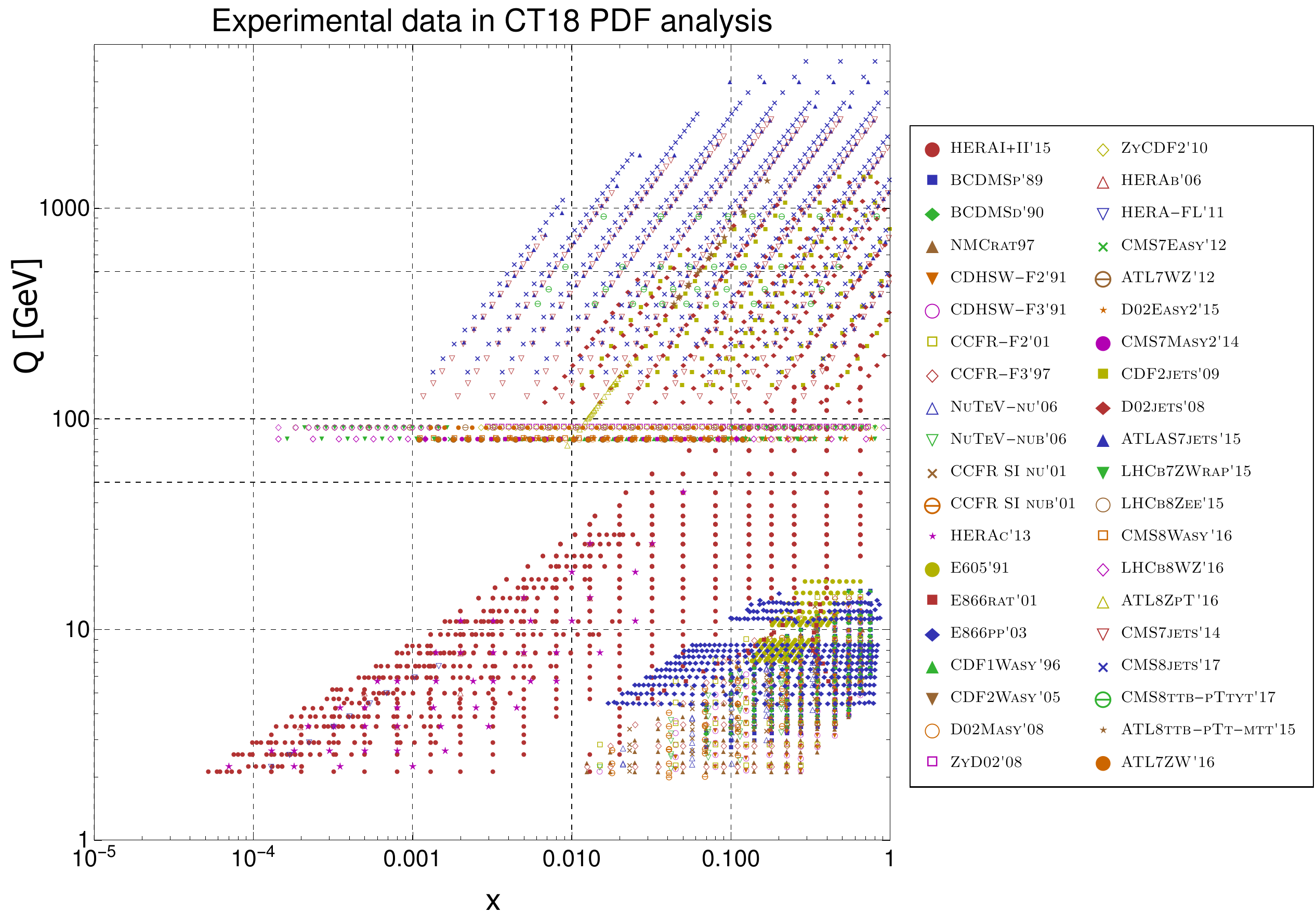}
	\caption{The CT18 data set, represented in a space of partonic $(x,\Q)$,
		based on Born-level kinematical matchings, $(x,\Q) = (x_B, Q)$, in DIS, {\it etc.}.
		The matching conventions used here are described in Ref.~\cite{Wang:2018heo}. Also
		shown are the ATLAS 7 TeV $W/Z$ production data (ID=248), labeled ATL7WZ'12, fitted
		in CT18Z.
		\label{fig:ct18data_xmu}}
\end{figure}

The CT18 analysis updates the widely used CT14 PDF sets~\cite{Dulat:2015mca}
by applying NNLO and NLO global fits to an expanded set of experimental
measurements that include high-statistics data from the $ep$ collider HERA and the LHC. The
CT18 experimental data set includes high-statistics measurements from ATLAS, CMS, and LHCb on 
production of inclusive jets, $W/Z$ bosons, and top-quark pairs, 
while it retains the crucial {\it legacy} data, such as 
the HERA Run I and Run II combined data and measurements
from the Tevatron. 
By 2018, the LHC collaborations published
about three dozen experimental data sets that can potentially constrain 
the CT PDFs. We selected the most promising experiments available by mid-2018 using the methods reviewed in Secs.~\ref{sec:ExptSelection} and \ref{sec:Advances}. We then extensively examined the impacts of the data sets within the full fitting framework. Sec.~\ref{sec:data_overview} contains an overview of these experiments. 
The kinematic distribution of the data points included in CT18 is shown in Fig.~\ref{fig:ct18data_xmu} as a function of the typical parton momentum fraction, 
$x$, and QCD factorization scale, denoted here as $\Q$.\footnote{ The typical momentum fractions and factorization scales are estimated as in Ref.~\cite{Wang:2018heo}.} As has been true for global PDF fits for some time, the data included cover a large kinematic range, both in $x$ and $\Q$. 

\begin{figure}[htbp]
\hspace*{-0.9cm}\includegraphics[width=0.52\textwidth]{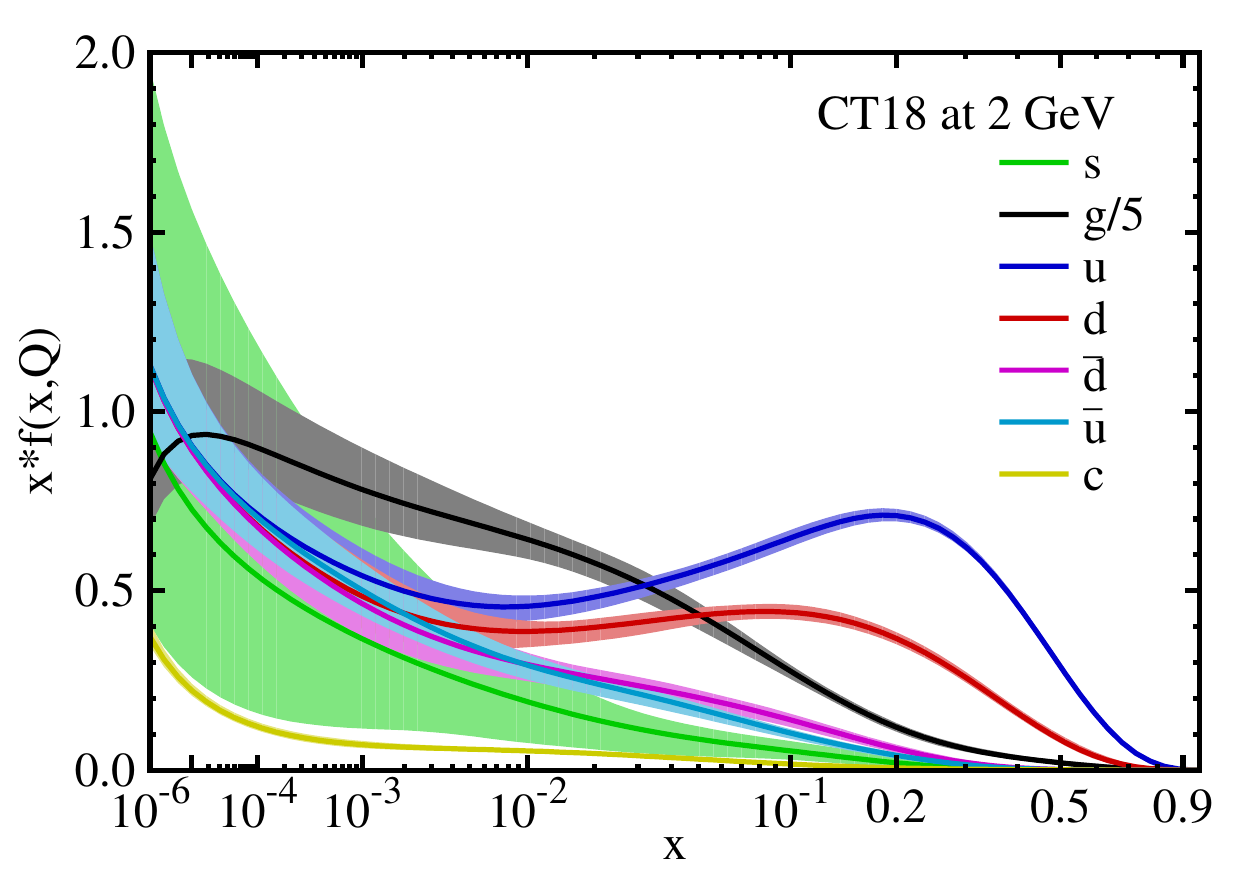}
\includegraphics[width=0.52\textwidth]{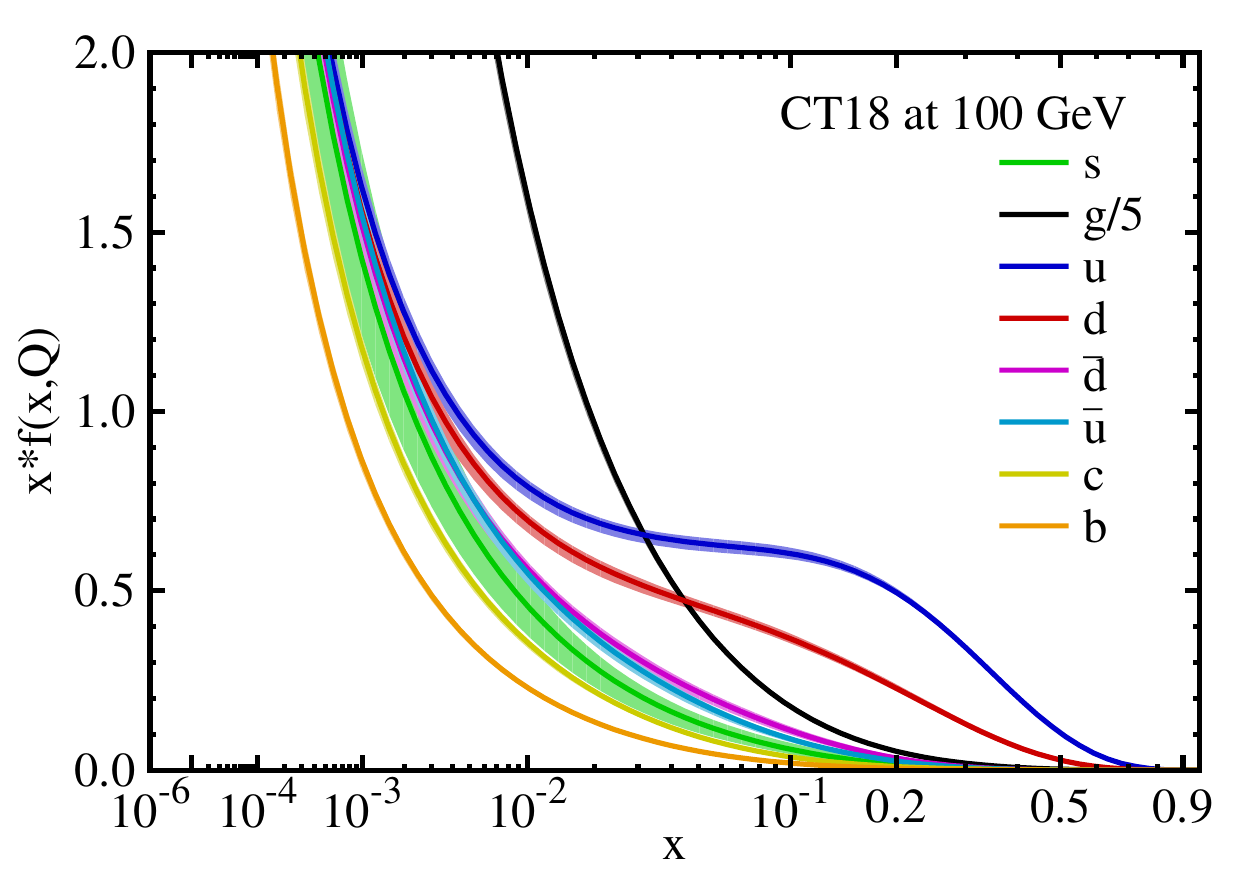}
\hspace*{-0.9cm}\includegraphics[width=0.52\textwidth]{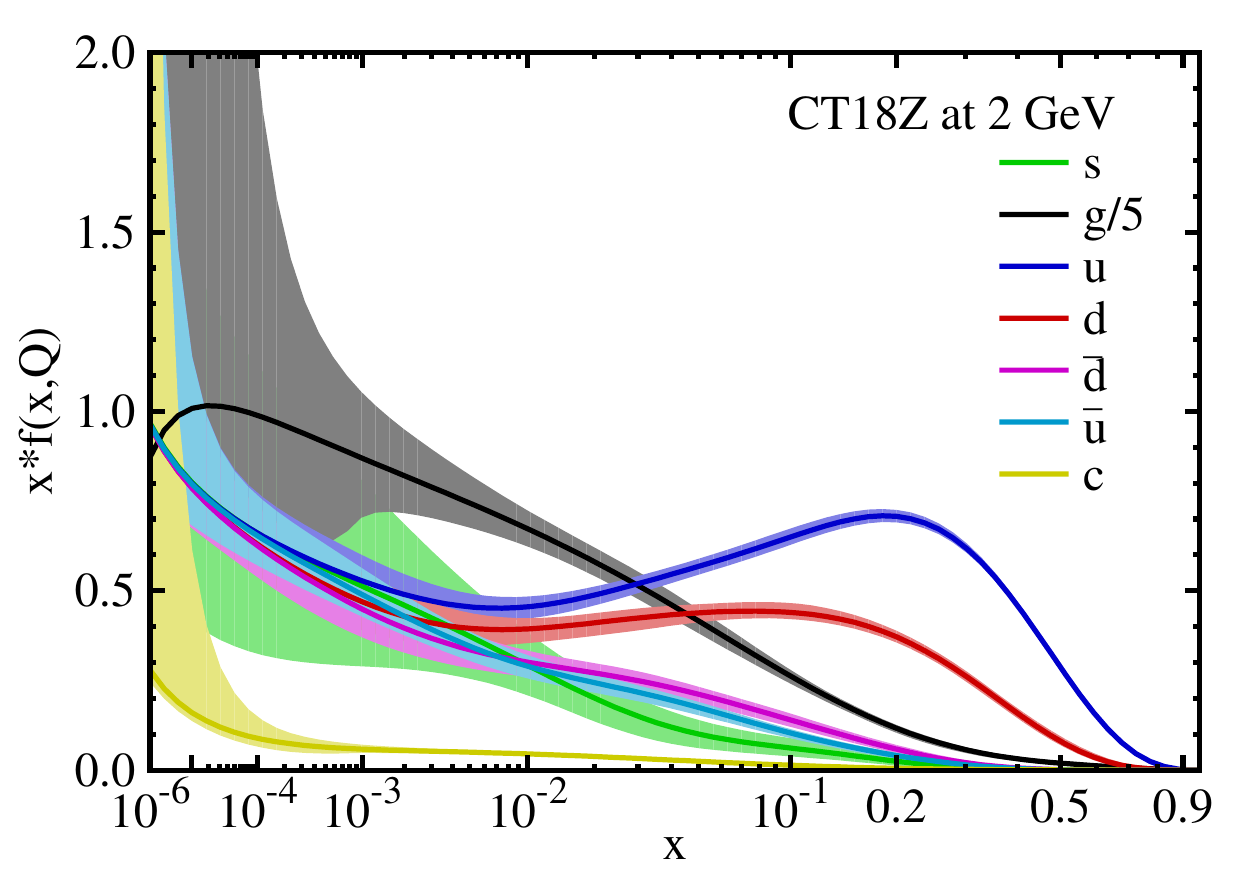}
\includegraphics[width=0.52\textwidth]{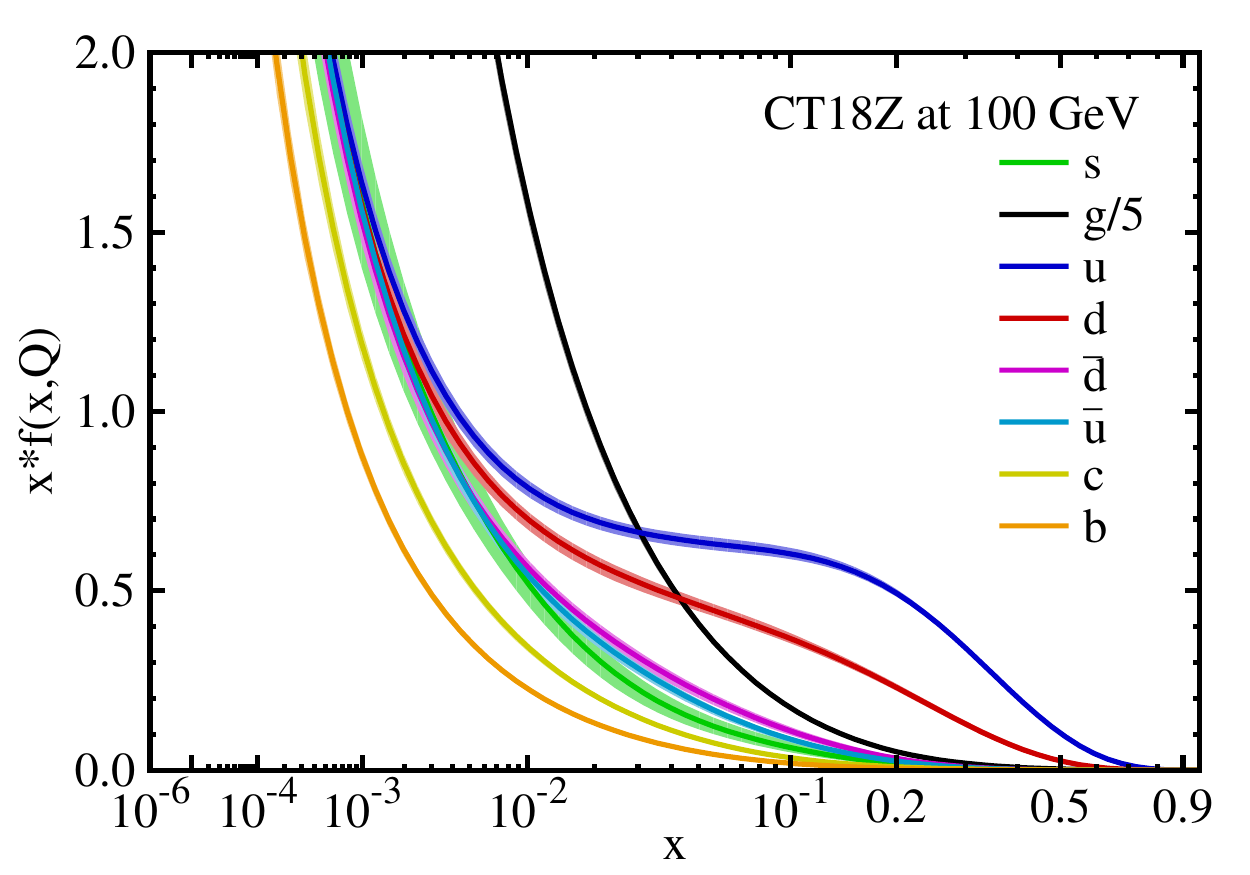}
\caption{Upper panels: The CT18 parton distribution functions at $\Q\! =\! 2$ GeV and $\Q\! =\! 100$ GeV for
	$u, \overline{u}, d, \overline{d}, s = \overline{s}$, and $g$. 
		Lower panels: The analogous curves, but obtained for CT18Z. In all
		instances, the gluon PDF has been scaled down as $g(x,\Q)/5$.
		The charm distribution, $c(x,\Q)$, which
		is perturbatively generated by evolving from $Q_0\! =\! 1.3$ and 1.4 GeV, respectively, in CT18 and CT18Z, is also shown.
		\label{fig:ct18pdf}}
\end{figure}

In light of the unprecedented precision reached in some 
measurements, the latest LHC data must be analyzed using
NNLO theoretical predictions in perturbative QCD.
The fitted PDFs we obtain in this analysis are plotted in Fig.~\ref{fig:ct18pdf},
which displays in the upper panels the CT18 PDFs at two widely-separated scales,
$\Q\!=\!2$ and $\,100$ GeV (on the left and right, respectively). In the lower panels,
we show the corresponding PDFs found in our amalgamated alternative analysis,
CT18Z.

The final CT18(Z) data ensemble contains a total of $N_\mathit{pt}\! =\!$ 3681 (3493) data points and results in $\chi^2/N_\mathit{pt}=1.17 (1.19)$ at NNLO.
The PDF uncertainties are constructed at the 90\% probability level based on two tiers of criteria as in the CT14 global analysis \cite{Dulat:2015mca}.  These PDFs are obtained by assuming a world-average QCD coupling constant, $\alpha_s(M_Z)=0.118$ \cite{Tanabashi:2018oca}. The combined PDF+$\alpha_s$ uncertainty can be computed using the special $\alpha_s$ series of the PDFs for each family by adding the PDF and $\alpha_s$ uncertainties in quadrature, as explained in Ref.~\cite{Lai:2010nw}.

Among the four ensembles (CT18, A, X, and Z) of PDFs,  the CT18 and CT18Z ensembles are the most dissimilar in terms of the shapes of PDFs, notably in the $x$ dependence of the fitted gluon and strangeness
distributions, $g(x,Q)$ and $s(x,\Q)$, as well as in some PDF uncertainties. For CT18, we obtain modest improvements in the precision
for the gluon density $g(x,\Q)$, as compared to CT14, following the inclusion of the LHC Run-1 data discussed below. For CT18Z, however, we obtain a somewhat enlarged uncertainty for the gluon and perturbatively-generated charm PDFs, especially at the lowest values
of $x\! <\! 10^{-3}$, due to the modified treatment of the DIS data described in
Sec.~\ref{sec:alt} and App.~\ref{sec:AppendixCT18Z}.
These final PDFs depend on numerous systematic factors in the experimental data. Scrupulous examination of the systematic effects was essential for trustworthy estimates of PDF uncertainties, 
and the scope of numerical computations also needed to be expanded. 

\subsubsection{\label{sec:summary-HERA2} Combined HERA I+II DIS data and the $x_B$-dependent factorization scale }
Even in the LHC era, DIS data from the $ep$ collider HERA provide the dominant
constraints on the CT18 PDFs. This dominance is revealed by independently applying the \texttt{ePump}, \texttt{PDFSense},  and Lagrange Multiplier methods.
CT18 implements the final (``combined'') data set from DIS at
HERA Run-I and Run-II \cite{Abramowicz:2015mha} that supersedes the HERA Run-I
only data set \cite{Aaron:2009aa} used in CT14 \cite{Dulat:2015mca}.
A transitional PDF set, \CTHERAII, was released based on fitting the
final HERA data \cite{Hou:2016nqm}. We found fair overall agreement of the HERA I+II
data with both CT14 and \CTHERAII~PDFs, and that both PDF ensembles
describe equally well the non-HERA data included in our global analysis.
At the same time, we observed some disagreement (``statistical tension'') 
between the $e^{+}p$ and $e^{-}p$ DIS cross sections of the HERA I+II data set. 
We determined that, at the moment, no plausible explanation could be provided to describe 
the full pattern of these tensions, as they are distributed across
the whole accessible range of Bjorken $x$ and lepton-proton momentum
transfer $Q$ at HERA.
Extending these studies using the CT18 fit, we have investigated the impact of the choice of
QCD scales on inclusive DIS data in the small-$x_B$ region, as will be explained later in
Sec.~\ref{sec:alt}.

\begin{figure}[tb]
	\includegraphics[width=0.59\textwidth]{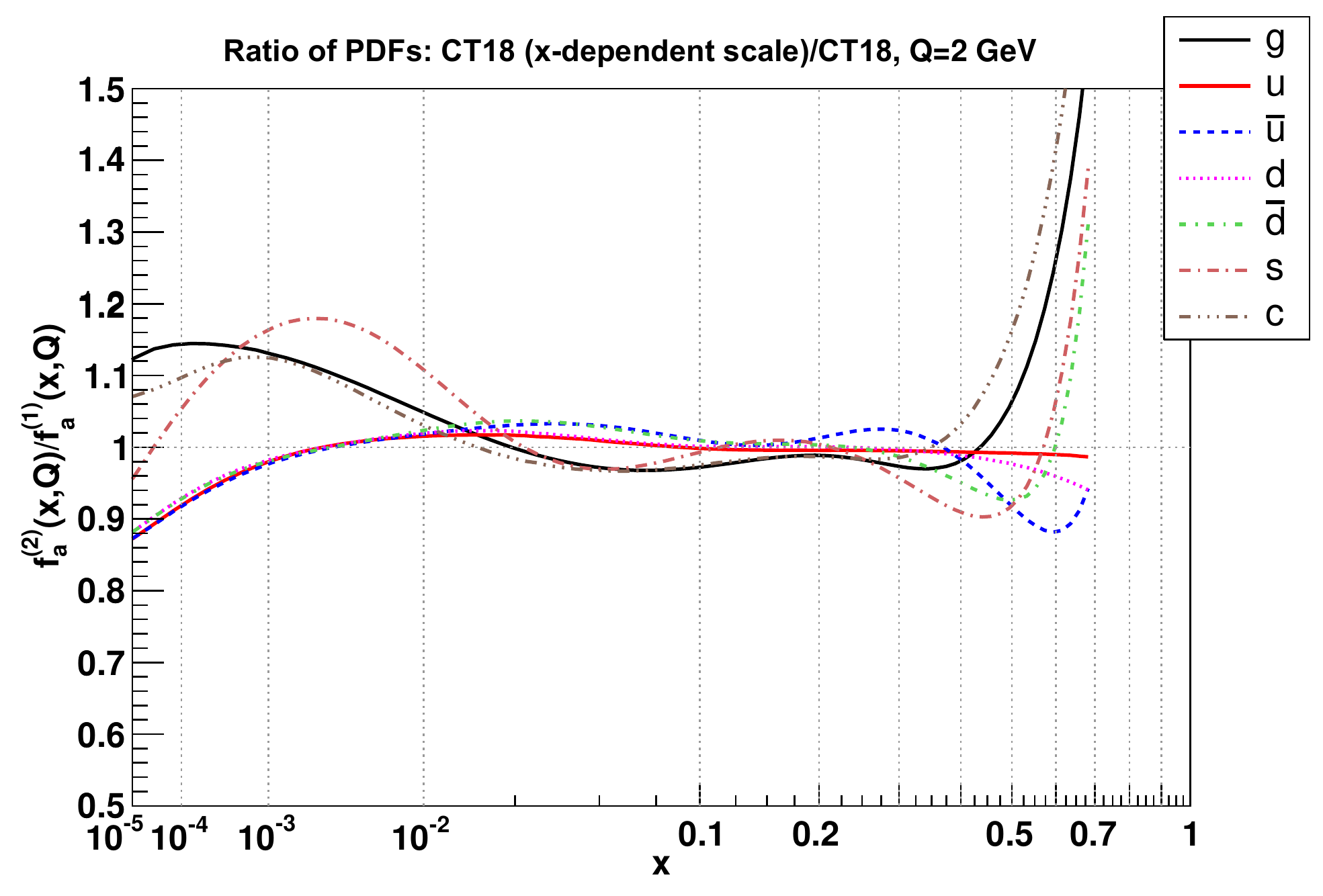}
	\includegraphics[width=0.4\textwidth]{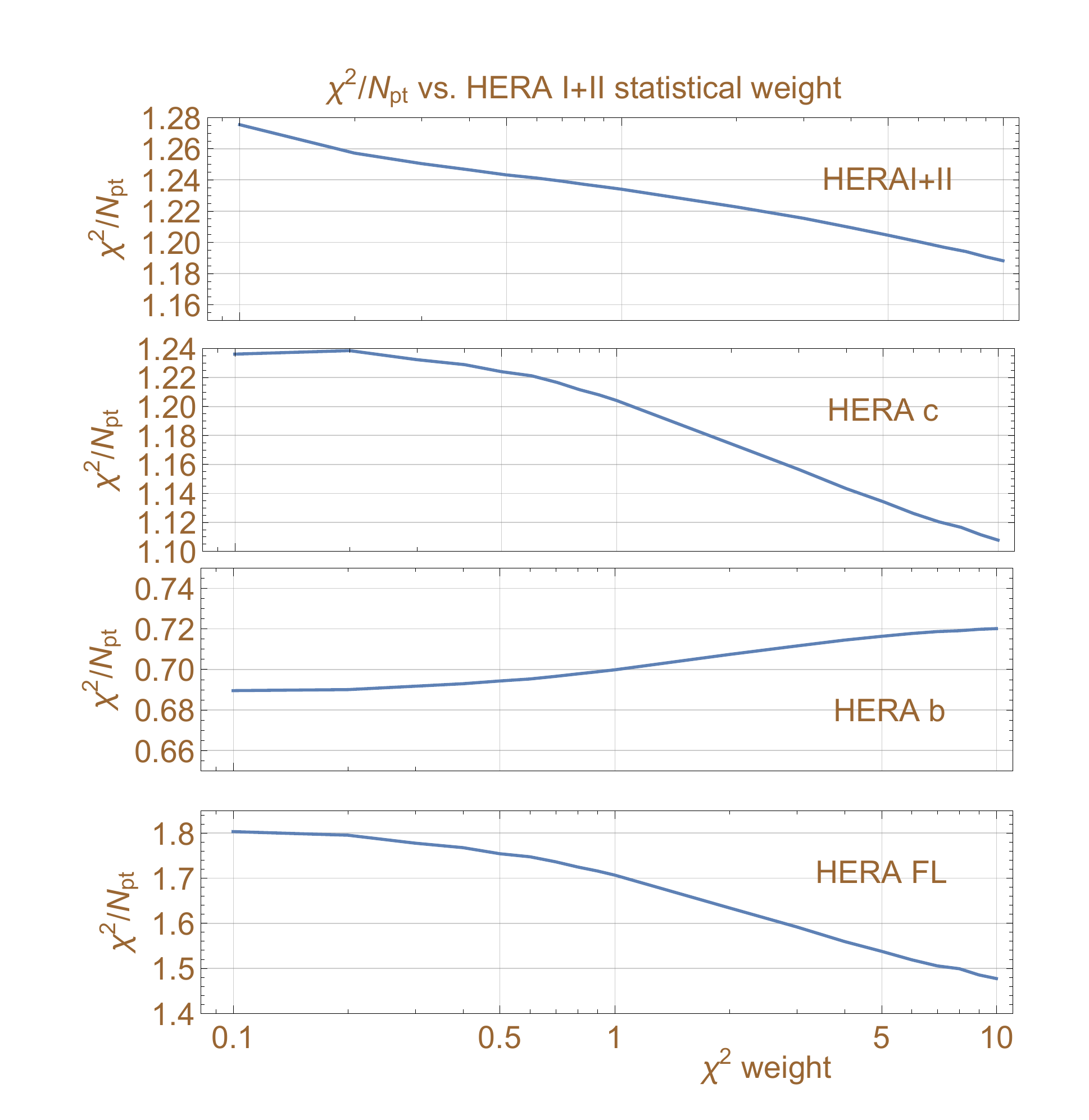}
\caption{Left: The ratios of the candidate CT18 NNLO PDFs obtained with the
  $x_B$-dependent and standard factorization scales in DIS data
  sets. Right: The $\chi^2/N_{pt}$ values for four HERA data sets in
  the CT18Z fit with the $x_B$-dependent DIS factorization scale and
  varied statistical weight of the HERA I+II inclusive DIS data set.}
\label{fig:saturation}
\end{figure}

We find that the quality of fit to HERA data is improved by about 50 units by evaluating the NNLO theoretical cross sections in DIS with a special factorization scale, $\mu_{F,x}$, that depends on Bjorken $x_B$ (not the momentum fraction $x$) and is introduced in Section~\ref{sec:alt}. Fig.~\ref{fig:saturation} (left) shows
the changes in the candidate CT18 PDFs obtained by fitting the DIS data sets with the factorization scale $\mu_{F,x}$, as compared to the CT18 PDFs
with the nominal scale $\mu_F=Q$. With the scale $\mu_{F,x}$, we observe
reduced $u$ and $d$ (anti-)quark PDFs and increased gluon and
strangeness PDFs at $x < 10^{-2}$, as compared to the nominal CT18 fit,
with some compensating changes occurring in the same PDFs in the
unconstrained region $x > 0.5$ in order to satisfy the valence and
momentum sum rules. 

The right panel of Fig.~\ref{fig:saturation} shows the
$\chi^2/N_{pt}$ values ($\chi^2$ divided by the number, $N_\mathit{pt}$, of experimental data
points) for four HERA data sets (inclusive neutral and charged current DIS
\cite{Abramowicz:2015mha},
reduced charm, bottom production cross sections, and H1 longitudinal function
$F_L(x_B,Q^2)$ \cite{Collaboration:2010ry}) in the fits as a function of the
statistical weight $w$ of the HERA I+II inclusive DIS data set \cite{Abramowicz:2015mha}. The
default CT18Z fit corresponds to $w=1$; with $w=10$, the CT18Z fit
increasingly behaves as a HERA-only fit. We see that, with the scale
$\mu^2_{F,x}$ and $w=10$, $\chi^2/N_{pt}$ for the inclusive DIS data set
improves almost to the levels observed in the ``resummed'' HERA-only fits
without intrinsic charm \cite{Ball:2017otu,Abdolmaleki:2018jln}. The
quality of the fit to the charm semi-inclusive DIS (SIDIS) cross section and H1 $F_L$ also improves.~\footnote{The use of the separate H1 $F_L$ data as well as the HERA-II combined data introduces some double counting. However, we have checked that this choice does not appreciably change the PDFs, while it does provide a useful indicator of the goodness of fit in the small-$x$ region.  In particular, it is telling that the total $\chi^2$ value of H1 $F_L$ data (ID=169) becomes smaller, not larger, in the CT18X and CT18Z fits, as compared to CT18.}

The new combined charm and bottom production measurements from the H1 and ZEUS collaborations published in Ref.~\cite{H1:2018flt} (2018) have been investigated and in their current version, when these measurements replace the previous ones in the CT18 global analysis, they cannot be fitted with a reasonable $\chi^2$.
Moreover, a mild tension is observed between these new combined data and several CT18 data sets such as the LHCb 7 and 8 TeV $W/Z$ production data, $Z$-rapidity data at CDF run II, CMS 8 TeV single inclusive jet production, and $t\bar{t}$ double differential $p_T$ and $y$ cross section. Therefore, we decided not to include these data in the CT18 global analysis as they require a dedicated investigation.
In the H1+ZEUS analysis of Ref.~\cite{H1:2018flt}, the $\chi^2$ for these measurements is also found not to be optimal. This is ascribed to a difference in the slope between data and theory in the intermediate/small $x$ region. In our attempt to fit the new combined charm and bottom production measurements, we have noticed a preference for a harder gluon at intermediate/small $x$.
We are currently investigating these data separately and, in particular, we are exploring the impact of the new correlated systematic uncertainties as they increased from 42 in the old version of the data, to 167 in the new version.
The results of this new study are going to be published in a separate forthcoming paper.

\begin{figure}[tb]
	\includegraphics[width=0.6\textwidth]{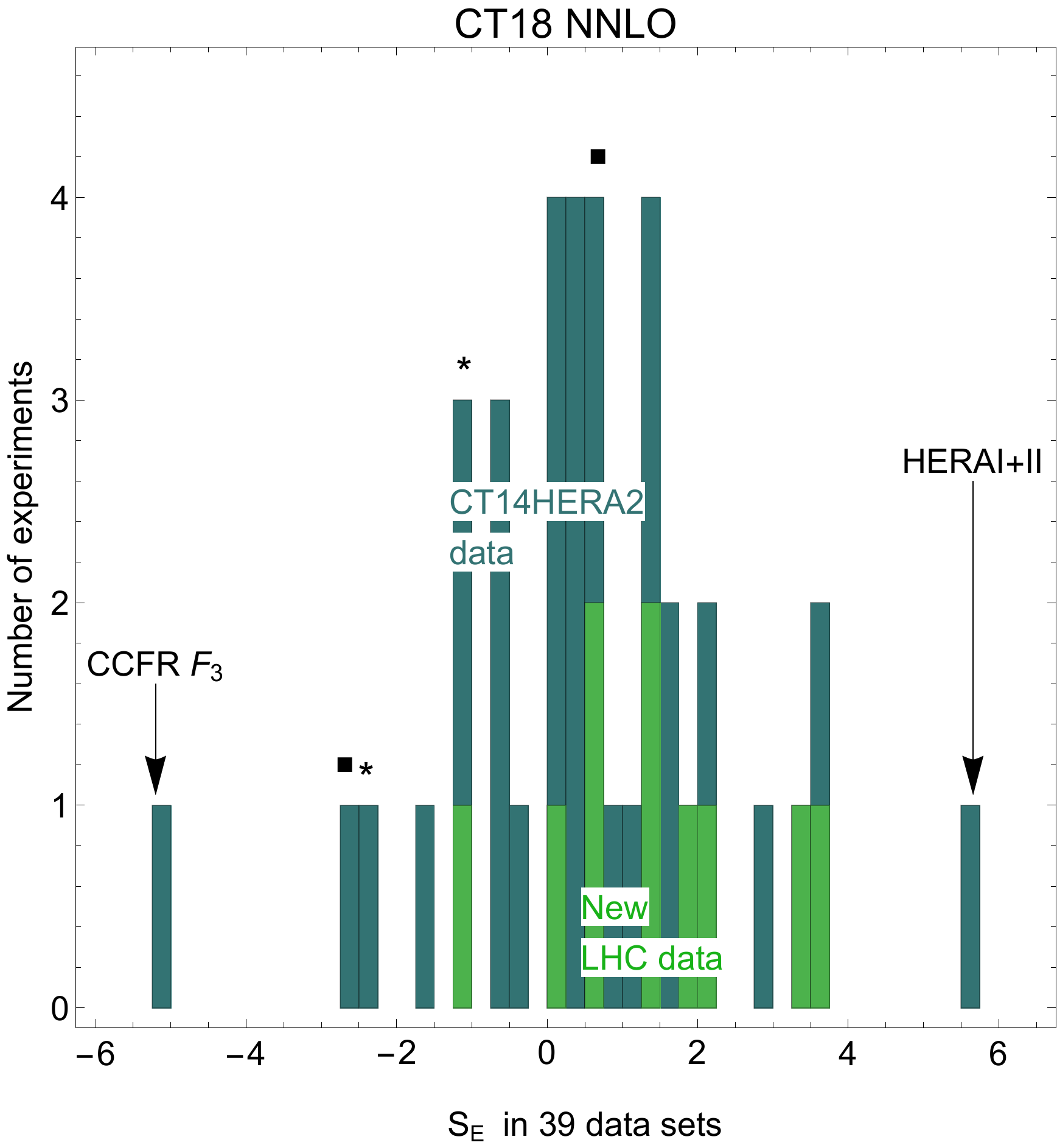}
	\caption{A histogram of the effective Gaussian variable ($S_E$) distributed over all CT18 data sets. 
	Two squares and two stars indicate the $S_E$ values for the NuTeV dimuon and CCFR dimuon data, respectively.
\label{fig:sn_ct18}}
\end{figure}

\subsubsection{Selection of new LHC experiments \label{sec:ExptSelection} }
When selecting the 
most promising LHC experiments for the CT18 fit, 
we had to address a recurrent challenge --- the presence of statistical
tensions among various (sub)sets of the latest experimental data from
HERA, LHC, and the Tevatron. The quickly improving precision of the collider
data reveals previously irrelevant anomalies either in the experiment
or theory. These anomalies are revealed by applying strong goodness-of-fit
tests~\cite{Kovarik:2019xvh}. Figure~\ref{fig:sn_ct18}
illustrates the degree of tensions using a representation based on the effective
Gaussian variables $S_E\equiv \sqrt{2\chi_E^2}-\sqrt{2 N_E-1}$~\cite{Lai:2010vv}
constructed from the $\chi^2$ values and numbers of data points $N_\mathit{E}$ for
individual data sets $E$. In an ideal fit in which the differences between theory and data are consistent with Gaussian random fluctuations, the probability
distribution for $S_E$ must be approximately a standard normal
distribution (with a unit half-width). In the global fits by CTEQ-TEA and  external groups, we rather observe wider $S_E$ distributions
as in Fig.~\ref{fig:sn_ct18}, with some of the more comprehensive
and precise data sets (notably, HERA I+II inclusive DIS
\cite{Abramowicz:2015mha} and ATLAS 7 TeV $W/Z$ production
\cite{Aaboud:2016btc}) having $S_E$ values as high as five units or more.  
The question, then, is how to select clean and accurate experiments for
the global analysis from an ever-growing list of measurements, while
maximally preserving the consistency of the selected experiments. 
For example, there are many LHC data sets~\cite{Rojo:2015acz}
that are potentially sensitive to the PDFs, including novel measurements
involving the production of high-$p_{T}$ $Z$ bosons, $t\bar{t}$ pairs,
isolated photon, and small-$x$ heavy flavor (charm or bottom) quarks.
Including all such candidate experiments
into the full global fit is impractical: CPU costs grow quickly with
the number of experimental data sets at NNLO. Poorly fitted experiments
would increase, not decrease, the final PDF uncertainty. The generation
of one error PDF set took several days of CPU time in the CT14 fit
to 33 experiments in single-thread mode. Adding 20-30 additional
experiments with this setup was thus impossible.
The CTEQ-TEA group resolved
these challenges through a multi-pronged effort that allowed us
to include eleven new LHC data sets at 7 and 8 TeV on $W^\pm$, $Z$, jet, and $t\bar{t}$ production.

\subsubsection{Advances in fitting methodology \label{sec:Advances} }
To identify the eligible experimental
data sets for the global fit, we developed two programs for fast preliminary analysis. The \texttt{PDFSense} program~\cite{Wang:2018heo} 
was developed at Southern Methodist University (SMU)
to predict quantitatively, and before doing the fit, which data sets
will have an impact on the global PDF fit. The \texttt{ePump} program~\cite{Schmidt:2018hvu} developed at Michigan State University (MSU)
applies Hessian profiling to quickly estimate the impact of data on the
PDFs prior to the global fit. These programs provide helpful guidelines
for the selection of the most valuable experiments based entirely on the previously published Hessian error PDFs.  Section~\ref{sec:Sensitivity} demonstrates an application of \texttt{PDFSense}.

 As we will discuss in Appendix~\ref{sec:Appendix4xFitter}, the out-of-the-box algorithm for Hessian profiling implemented in the commonly used version 2.0.0 of the \texttt{xFitter} program~\cite{Alekhin:2014irh} is inconsistent with the CTEQ-TEA definitions of PDF uncertainties and has predicted  too optimistic $\chi^2$ values and PDF uncertainties in a number 
 of studies for profiling the CTEQ-TEA PDFs. The \texttt{ePump} program does not have this caveat. 
 Its Hessian updating algorithm better reproduces the $\chi^2$ values of the data sets in the full CT14 and CT18 fits, 
 as well as the respective PDF uncertainties defined according to the two-tier definition of $\chi^2$ adopted in the CTEQ-TEA analyses since CT10 NLO~\cite{Gao:2013xoa}. 

The CTEQ fitting code was parallelized to allow a faster turnaround
time (one fit within a few hours instead of many days) on high-performance
computing clusters. For as much relevant LHC data as possible, we computed
the NLO cross sections with the \texttt{APPLGrid/fastNLO} tables~\cite{Kluge:2006xs,Carli:2010rw}  (to be multiplied by tabulated point-by-point NNLO/NLO $K$-factor corrections)
for various new LHC processes: production of $W/Z$ bosons, high-$p_{T}$
$Z$-bosons and inclusive jets; the NNLO cross section with the \texttt{fastNNLO} tables \cite{Czakon:2017dip,fastnnlo:grids} for the $t\bar{t}$ pair production at the LHC. The \texttt{APPLgrid} tables were cross validated
against similar tables from other groups (available in the public
domain) and optimized for speed and accuracy. 

\subsubsection{Estimates of theoretical and parametrization uncertainties}
Significant effort was spent on understanding the sources
of PDF uncertainties. Theoretical uncertainties associated with the
scale choice were investigated for the affected processes, such as
DIS as well as inclusive jet and high-$p_{T}$ $Z$ boson production.
Other considered theoretical uncertainties were due 
to the differences among the NNLO and resummation codes ({\it e.g.}, 
\texttt{DYNNLO}~\cite{Catani:2007vq,Catani:2009sm}, 
\texttt{MCFM}~\cite{Campbell:2010ff, Boughezal:2016wmq, MCFM8}, 
\texttt{FEWZ}~\cite{Gavin:2010az, Gavin:2012sy, Li:2012wna},
\texttt{NNLOJet}~\cite{Ridder:2015dxa,Gehrmann-DeRidder:2017mvr,Currie:2016bfm,Currie:2017ctp}, and
\texttt{ResBos}~\cite{Ladinsky:1993zn, Balazs:1997xd})
and Monte-Carlo (MC) integration error, see Sec.~\ref{sec:calcs}.
Specifically, we have included the MC errors in the CT18(Z) analysis for the inclusive jet and high-$p_T$ $Z$ boson production data. But, the PDF  uncertainties related to the choice of the QCD scales and the codes for theoretical calculations have not been systematically included in this analysis. 
The important PDF parametrization uncertainty was investigated by repeating
the fits for $\mathcal{O}(250)$ trial functional forms of the PDFs. [Our post-CT10
fits parametrize PDFs in terms of Bernstein polynomials, which simplify
trying a wide range of parametrization forms to quantify/eliminate
potential biases. Appendix~\ref{sec:AppendixParam} presents an example of such parametrization.] The final uncertainty on the nominal CT18 PDF set is determined so as to cover central solutions obtained with alternative parametrization forms and alternative fit settings or scale choices, see Sec.~\ref{sec:Paramstudies}.

\subsection{Experimental data sets fitted in CT18}
\label{sec:data_overview}

The CT18 global analysis starts with the data set baseline of \CTHERAII~\cite{Hou:2016nqm} and adds the LHC results published before mid-2018. The experiments in the \CTHERAII~baseline
are listed in Table \ref{tab:EXP_1}, while the new LHC data sets included in the CT18(Z) fit are shown in Table \ref{tab:EXP_2}. 
Tables~\ref{tab:EXP_1} and~\ref{tab:EXP_2}
also include information on the number of data points, $\chi^2$, and the effective Gaussian variable, $S_E$, for each individual data set appearing in the global fit.   
Most of the data sets are included in all four PDF ensembles; we  will identify differences between the specific selections as they arise.

\subsubsection{Charting sensitivity of new data sets to the PDFs \label{sec:Sensitivity} }

As discussed in Secs.~\ref{sec:ExptSelection} and \ref{sec:Advances}, we employed a new method based on the Hessian sensitivity variables \cite{Wang:2018heo,Hobbs:2019gob} ( informative descendants of the Hessian correlation between theoretical observables and PDFs~\cite{Pumplin:2001ct,Nadolsky:2001yg,Nadolsky:2008zw}) 
to determine quantitatively a hierarchy of impact of data on the global fit, and on specific cross sections. 

\begin{figure}[tb]
	\includegraphics[width=0.48\textwidth]{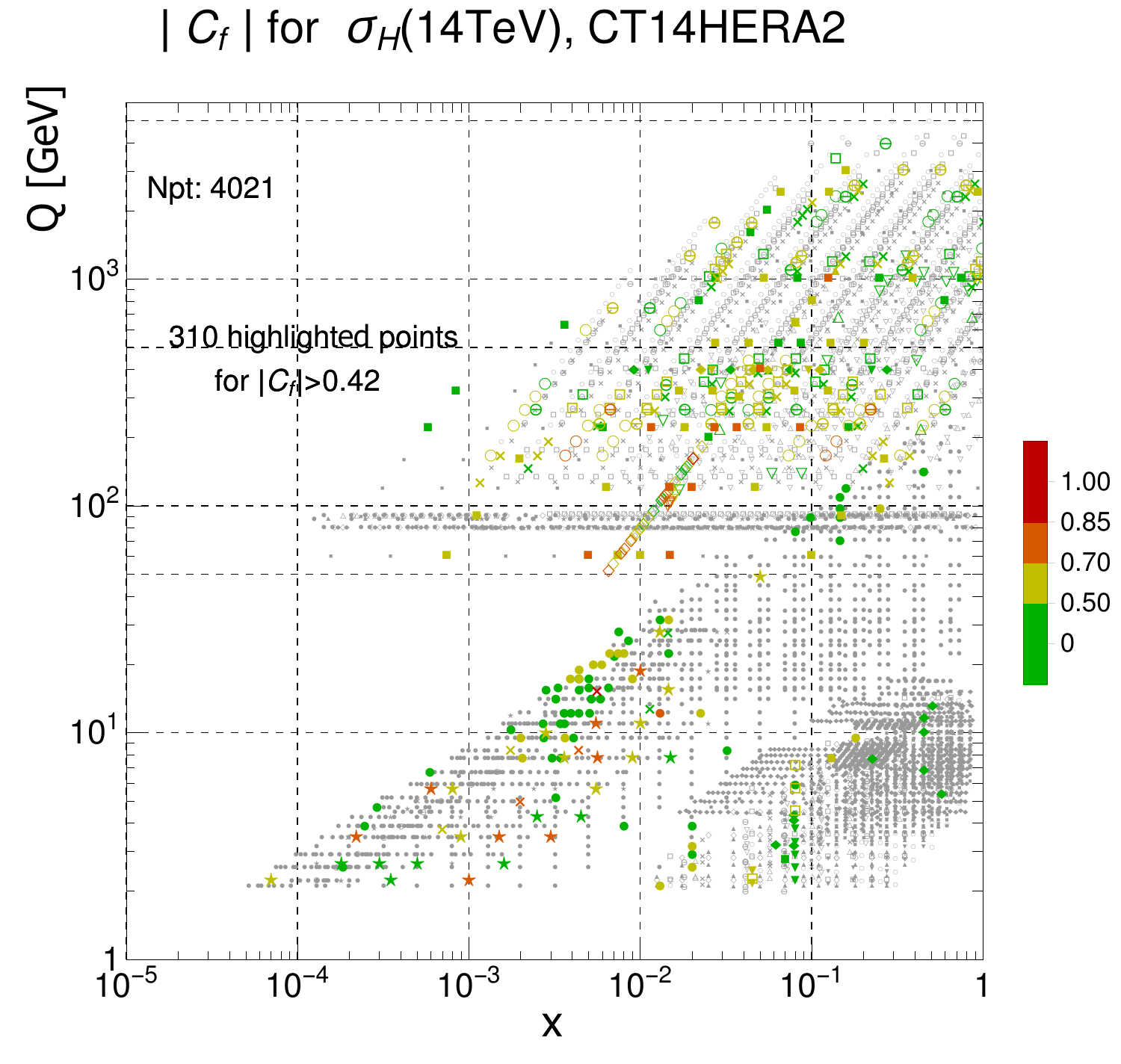}
	\includegraphics[width=0.48\textwidth]{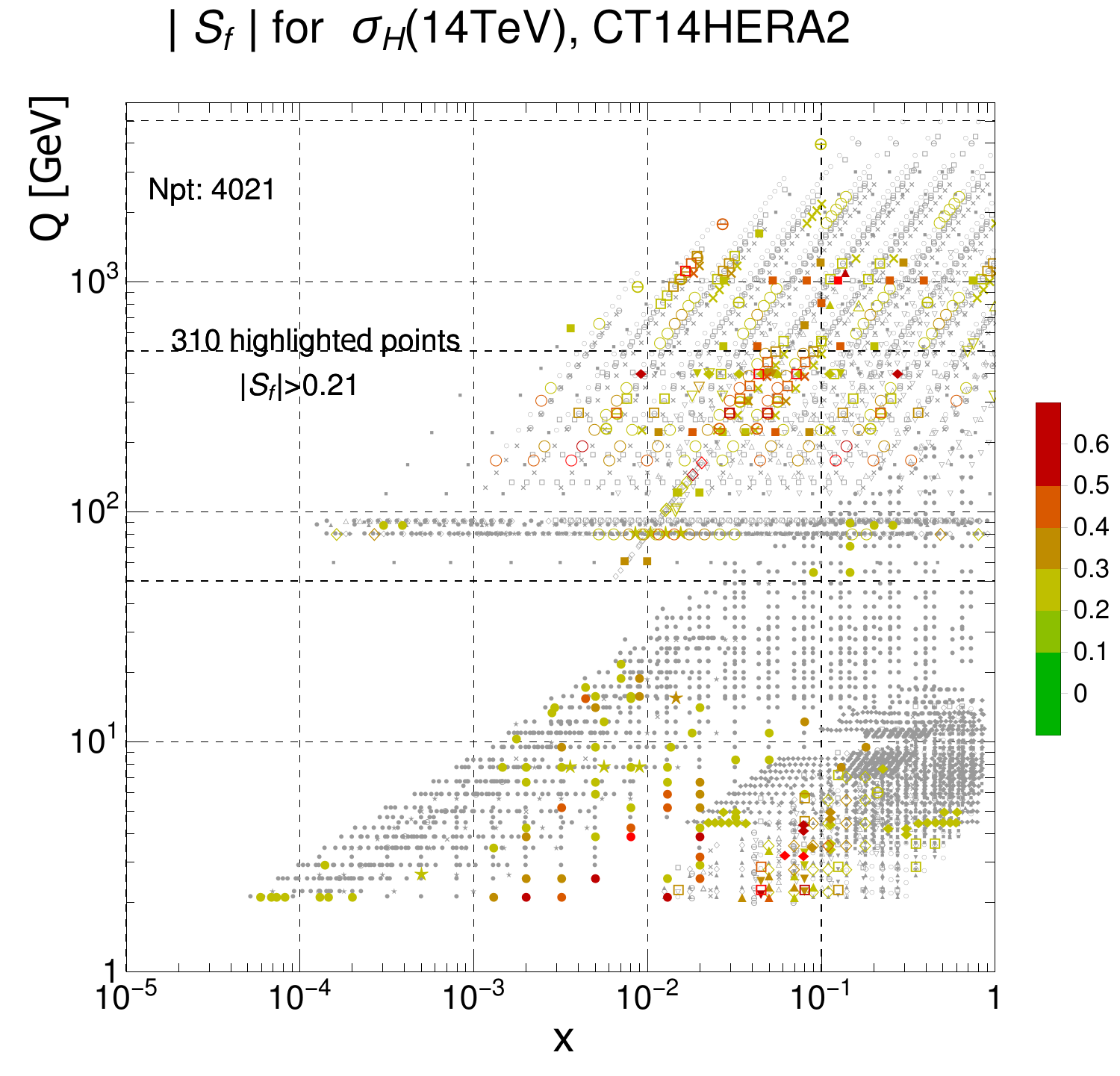}
\caption{
	Left: Candidate data considered for inclusion in CT18
	evaluated according to the magnitude of the Pearson
	correlation $C_f$ between the total Higgs cross section at 14 TeV,
	$\sigma_{H}(14\, \mathrm{TeV})$ and the residual of
	each point as determined within the \texttt{PDFSense}
	framework~\cite{Wang:2018heo}.
	Right: A similar assessment of the CT18 candidate data,
	but computed on the basis of the {\it sensitivity},
	$|S_f|(x_i,\Q_i)$. In both panels, a highlighting
	cut has been imposed to draw attention to the $\sim\!300$
	highest-impact points according to each metric.
  }
\label{fig:PDFSense}
\end{figure}

As a demonstration of this, the \texttt{PDFSense} framework can predict in advance which fitted data sets may have the most impact on one of the most crucial predictions at the LHC, such as the Higgs boson cross section ($\sigma_H$) through the $gg$ fusion process (at $\sqrt{s}=14$ TeV). Often, to get an indication which data sets will have the most impact on such cross sections, one examines the Pearson correlations \cite{Pumplin:2001ct,Nadolsky:2001yg,Nadolsky:2008zw} between the experimental data points and the gluon distribution in the kinematic region responsible for Higgs boson production.  The
left-hand side of Fig.~\ref{fig:PDFSense} shows the data points with the highest absolute correlations $|C_f|$ (defined by the statistical residuals as in Ref.~\cite{Wang:2018heo}) with the Higgs boson cross section at 14 TeV. 

By this measure, there may be a number of high-impact data sets, notably HERA neutral current DIS, LHC and Tevatron jet production, and HERA charm quark production.
The correlations, however, do not reflect the experimental uncertainties of the data points: an experimental cross section could be highly correlated with the gluon distribution in the $x$ range  responsible for Higgs boson production and still not provide much of a constraint
on the Higgs boson cross section if the experimental uncertainties are large. Conversely, an experimental cross section that might not have as large a correlation, but which has smaller (statistical and systematic) uncertainties,
may provide a stronger constraint. 

The level of constraint is thus better predicted by the sensitivity variable $S_f$, defined in Ref.~\cite{Wang:2018heo}. The experimental data points used in the CT18 global fit that have the
highest absolute sensitivity $|S_f|$ to the PDF dependence of $\sigma_H$ at 14 TeV are shown on the right-hand side of Fig.~\ref{fig:PDFSense}. More data points (from a larger number of experiments) have high sensitivity
than those identified by high correlation. In addition to the DIS data from HERA I+II, there are also contributions from the fixed-target DIS experiments, as well as measurements from the LHC. 

As will be shown in Sec.~\ref{sec:LMScans} using Lagrange Multiplier scans, the HERA I+II data set, with its abundant data points and small experimental errors, still dominates the constraints on the gluon distribution in the range sensitive to Higgs boson production at the LHC. 
Because of the continuing influence of the older data sets, we will find that the reduction of the PDF uncertainty for Higgs boson production is less significant in CT18 than in CT14. In addition, tension between some of the most sensitive data sets limits the reduction on the uncertainty
of the Higgs cross section. These effects are explored in detail in Sec.~\ref{sec:LMScans}.

We will now discuss the new data sets included in the CT18 analysis, and highlight the differences in the alternative fits.

\subsubsection{Baseline data sets \label{sec:Baseline}}

The CT18 global analysis inherits from \CTHERAII~a number of precision non-LHC experiments listed in Table 1. Among those, the HERA I+II DIS data set provides the most significant constraints, followed by a group of fixed-target neutral-current DIS experiments: BCDMS, NMC, and CCFR.
Similarly, a number of neutrino DIS measurements have previously been included
and provide valuable constraints on sea (anti-)quarks.
Among them, we find that the single-nucleon structure functions $F_2^p$ and $F_3^p$ extracted from CDHSW data on neutrino-iron deep inelastic scattering exhibit a preference for a harder gluon PDF at $x \gtrsim 0.1$, compared to CCFR and other experiments, cf. Fig.~\ref{fig:LMg18}. 
This well-known behavior reflects  larger logarithmic slopes of $F_2^p$ and $x F_3^p$ measured by CDHSW, as compared to the 
analogous CCFR measurements~\cite{Seligman:1997fe}, which in turn may reflect differences in the energy calibration and resolution smearing between the two experiments~\cite{Bodek:1996zu}. 
Thus, to help obtain a softer large-$x$ gluon behavior, as being favored by recent LHC data, we exclude the CDHSW $F_2$ and $x_B F_3$ data sets from the CT18Z analysis, while including these sets in the rest of the CT18 PDF ensembles. 

We continue to include a variety of lepton pair production measurements from the Tevatron and fixed-target experiments, as summarized in Table~\ref{tab:EXP_1}. The low-statistics data on $W/Z$ production at LHCb 7 TeV~\cite{Aaij:2012vn} and ATLAS, CMS 7 TeV jet production \cite{Aad:2011fc, Chatrchyan:2012bja} are replaced in the CT18 analyses by more recent measurements, as summarized in the next section. 

\begin{widetext}

\begingroup
\squeezetable
\begin{table}[htbp]
\begin{tabular}{|l|lr|c|c|c|c|}
\hline
\textbf{ID\# }  & \textbf{Experimental data set} &  & $N_{pt, E}$  & $\chi^2_E$ & $\chi^{2}_E/N_{pt, E}$  & $S_E$ \tabularnewline
\hline
\hline
 160 & HERAI+II 1 fb$^{-1}$, H1 and ZEUS NC and CC $e^\pm p$ reduced cross sec. comb.      & \cite{Abramowicz:2015mha}   & 1120  &  1408(1378) &   1.3( 1.2) &   5.7(  5.1)   \tabularnewline\hline
 101 & BCDMS $F_{2}^{p}$                                                                   & \cite{Benvenuti:1989rh}     &  337  &   374 ( 384) &   1.1( 1.1) &   1.4(  1.8)   \tabularnewline\hline
 102 & BCDMS $F_{2}^{d}$                                                                   & \cite{Benvenuti:1989fm}     &  250  &   280 ( 287) &   1.1( 1.1) &   1.3(  1.6)   \tabularnewline\hline
 104 & NMC $F_{2}^{d}/F_{2}^{p}$                                                           & \cite{Arneodo:1996qe}       &  123  &   126 ( 116) &   1.0( 0.9) &   0.2( -0.4)   \tabularnewline\hline
 108$^{\dagger}$ & CDHSW $F_{2}^{p}$                                                       & \cite{Berge:1989hr}         &   85  &    85.6 ( 86.8) &   1.0( 1.0) &   0.1(  0.2)   \tabularnewline\hline         
 109$^{\dagger}$ & CDHSW $x_B F_{3}^{p}$                                                       & \cite{Berge:1989hr}         &   96  &    86.5 (  85.6) &   0.9( 0.9) &  -0.7( -0.7)   \tabularnewline\hline
 110 & CCFR $F_{2}^{p}$                                                                    & \cite{Yang:2000ju}          &   69  &    78.8(  76.0) &   1.1( 1.1) &   0.9(  0.6)   \tabularnewline\hline
 111 & CCFR $x_B F_{3}^{p}$                                                                   & \cite{Seligman:1997mc}      &   86  &    33.8(  31.4) &   0.4( 0.4) &  -5.2( -5.6)   \tabularnewline\hline
 124 & NuTeV $\nu\mu\mu$ SIDIS                                                             & \cite{Mason:2006qa}         &   38  &    18.5(  30.3) &   0.5( 0.8) &  -2.7( -0.9)   \tabularnewline\hline
 125 & NuTeV $\bar\nu \mu\mu$ SIDIS                                                        & \cite{Mason:2006qa}         &   33  &    38.5(  56.7) &   1.2( 1.7) &   0.7(  2.5)   \tabularnewline\hline
 126 & CCFR $\nu\mu\mu$ SIDIS                                                              & \cite{Goncharov:2001qe}     &   40  &    29.9(  35.0) &   0.7( 0.9) &  -1.1( -0.5)   \tabularnewline\hline
 127 & CCFR  $\bar\nu \mu\mu$ SIDIS                                                        & \cite{Goncharov:2001qe}     &   38  &    19.8(  18.7) &   0.5( 0.5) &  -2.5( -2.7)   \tabularnewline\hline
 145 & H1 $\sigma_{r}^{b}$                                                                 & \cite{Aktas:2004az}         &   10  &     6.8(   7.0) &   0.7( 0.7) &  -0.6( -0.6)   \tabularnewline\hline
 147 & Combined HERA charm production                                                      & \cite{Abramowicz:1900rp}    &   47  &    58.3(  56.4) &   1.2( 1.2) &   1.1(  1.0)   \tabularnewline\hline
 169 & H1 $F_{L}$                                                                          & \cite{Collaboration:2010ry} &    9  &    17.0(  15.4) &   1.9( 1.7) &   1.7(  1.4)   \tabularnewline\hline
 201 & E605 Drell-Yan process                                                              & \cite{Moreno:1990sf}        &  119  &   103.4( 102.4) &   0.9( 0.9) &  -1.0( -1.1)   \tabularnewline\hline
 203 & E866 Drell-Yan process  $\sigma_{pd}/(2\sigma_{pp})$                                & \cite{Towell:2001nh}        &   15  &    16.1(  17.9) &   1.1( 1.2) &   0.3(  0.6)   \tabularnewline\hline
 204 & E866 Drell-Yan process  $Q^3 d^2\sigma_{pp}/(dQ dx_F)$                              & \cite{Webb:2003ps}          &  184  &   244 ( 240) &   1.3( 1.3) &   2.9(  2.7)   \tabularnewline\hline
 225 & CDF Run-1 lepton $A_{ch}$, $p_{T\ell}>25$ GeV                                    & \cite{Abe:1998rv}           &   11  &     9.0(   9.3) &   0.8( 0.8) &  -0.3( -0.2)   \tabularnewline\hline
 227 & CDF Run-2 electron $A_{ch}$, $p_{T\ell}>25$ GeV                                     & \cite{Acosta:2005ud}        &   11  &    13.5(  13.4) &   1.2( 1.2) &   0.6(  0.6)   \tabularnewline\hline
 234 & D\O~ Run-2 muon $A_{ch}$, $p_{T\ell}>20$ GeV                                        & \cite{Abazov:2007pm}        &    9  &     9.1(   9.0) &   1.0( 1.0) &   0.2(  0.1)   \tabularnewline\hline
 260 & D\O~ Run-2 $Z$ rapidity                       & 
 \cite{Abazov:2007jy}        &   28  &    16.9(  18.7) &   0.6( 0.7) &  -1.7( -1.3)   \tabularnewline\hline
 261 & CDF Run-2 $Z$ rapidity                                                              & \cite{Aaltonen:2010zza}     &   29  &    48.7(  61.1) &   1.7( 2.1) &   2.2(  3.3)   \tabularnewline\hline
 266 & CMS 7 TeV $4.7\mbox{ fb}^{-1}$, muon $A_{ch}$, $p_{T\ell}>35$ GeV                   & \cite{Chatrchyan:2013mza}   &   11  &     7.9(  12.2) &   0.7( 1.1) &  -0.6(  0.4)   \tabularnewline\hline
 267 & CMS 7 TeV $840\mbox{ pb}^{-1}$, electron $A_{ch}$, $p_{T\ell}>35$ GeV               & \cite{Chatrchyan:2012xt}    &   11  &    4.6(  5.5) &   0.4( 0.5) &   -1.6(  -1.3)   \tabularnewline\hline
 268$^{\ddagger\ddagger}$ & ATLAS 7 TeV $35\mbox{ pb}^{-1}$ $W/Z$ cross sec., $A_{ch}$                          & \cite{Aad:2011dm}           &   41  &    44.4   (50.6) &   1.1( 1.2) &   0.4(  1.1)   \tabularnewline\hline
 281 & D\O~ Run-2 $9.7 \mbox{ fb}^{-1}$ electron $A_{ch}$, $p_{T\ell}>25$ GeV              & \cite{D0:2014kma}           &   13  &    22.8(  20.5) &   1.8( 1.6) &   1.7(  1.4)   \tabularnewline\hline
 504 & CDF Run-2 inclusive jet production                                                  & \cite{Aaltonen:2008eq}      &   72  &   122 ( 117) &   1.7( 1.6) &   3.5(  3.2)   \tabularnewline\hline
 514 & D\O~ Run-2 inclusive jet production                                                 & \cite{Abazov:2008ae}        &  110  &   113.8 ( 115.2) &   1.0( 1.0) &   0.3(  0.4)   \tabularnewline
\hline
\hline
\end{tabular}
\caption{Data sets included in the CT18(Z) NNLO global analyses. Here we directly compare the quality-of-fit found for CT18 NNLO vs.~CT18Z NNLO on the basis of $\chi^2_E$,
	$\chi^2_E/N_{pt, E}$, and $S_E$, in which $N_{pt, E}$, $\chi^2_E$ are the number of points and value of $\chi^2$ for  experiment  $E$ at the global minimum. $S_E$ is
	the effective Gaussian parameter \cite{Lai:2010vv, Gao:2013xoa, Dulat:2013hea} quantifying agreement with each experiment. The ATLAS 7 TeV $35\mbox{ pb}^{-1}$ $W/Z$ data set, marked by $\ddagger\ddagger$, is replaced by the updated one (4.6 fb$^{-1}$) in the CT18A and CT18Z fits. The CDHSW data, labeled by $\dagger$, are not included in the CT18Z fit. The numbers in parentheses are for the CT18Z NNLO fit.
\label{tab:EXP_1} }
\end{table}
\endgroup
\end{widetext}

\begin{widetext}
\begingroup
\squeezetable
\begin{table}[tb]
\begin{tabular}{|l|lr|c|c|c|c|}
\hline
\textbf{ID\# }  & \textbf{Experimental data set} &  & $N_{pt, E}$  & $\chi^2_E$ & $\chi^{2}_E/N_{pt, E}$  & $S_E$ \tabularnewline
\hline
\hline
 245 & LHCb 7 TeV 1.0 fb$^{-1}$ $W/Z$ forward rapidity cross sec.                          & \cite{Aaij:2015gna}        &  33  &   53.8 (  39.9) &  1.6 (  1.2) &  2.2 ( 0.9) \tabularnewline\hline
 246 & LHCb 8 TeV  2.0 fb$^{-1}$ $Z\rightarrow e^{-} e^{+}$ forward rapidity cross sec.    & \cite{Aaij:2015vua}        &  17  &   17.7 (  18.0) &  1.0 (  1.1) &  0.2 ( 0.3) \tabularnewline\hline
 248$^{\ddagger}$ & ATLAS 7 TeV 4.6 fb$^{-1}$, $W/Z$ combined cross sec.                   & \cite{Aaboud:2016btc}      &  34  &  287.3 (  88.7) &  8.4 (  2.6) & 13.7 ( 4.8) \tabularnewline\hline
 249 & CMS 8 TeV 18.8 fb$^{-1}$ muon charge asymmetry $A_{ch}$                                & \cite{Khachatryan:2016pev} &  11  &   11.4 (  12.1) &  1.0 (  1.1) &  0.2 ( 0.4) \tabularnewline\hline
 250 & LHCb 8 TeV 2.0fb$^{-1}$ $W/Z$ cross sec.                                            & \cite{Aaij:2015zlq}        &  34  &   73.7 (  59.4) &  2.1 (  1.7) &  3.7 ( 2.6) \tabularnewline\hline 
 253 & ATLAS 8 TeV 20.3 fb$^{-1}$, $Z$ $p_T$ cross sec.                                    & \cite{Aad:2015auj}         &  27  &   30.2 (  28.3) &  1.1 (  1.0) &  0.5 ( 0.3) \tabularnewline\hline
 542 & CMS 7 TeV 5 fb$^{-1}$, single incl. jet cross sec., $R=0.7$ (extended in y)         & \cite{Chatrchyan:2014gia}  & 158  &  194.7 ( 188.6) &  1.2 (  1.2) &  2.0 ( 1.7) \tabularnewline\hline
 544 & ATLAS 7 TeV  4.5 fb$^{-1}$, single incl. jet cross sec., $R=0.6$                    & \cite{Aad:2014vwa}         & 140  &  202.7 ( 203.0) &  1.4 (  1.5) &  3.3 ( 3.4) \tabularnewline\hline
 545 & CMS 8 TeV 19.7 fb$^{-1}$, single incl. jet cross sec., $R=0.7$, (extended in y)     & \cite{Khachatryan:2016mlc} & 185  &  210.3 ( 207.6) &  1.1 (  1.1) &  1.3 ( 1.2) \tabularnewline\hline
 573 & CMS 8 TeV 19.7 fb$^{-1}$, $t\bar{t}$ norm. double-diff. top $p_T$ and $y$ cross sec. & \cite{Sirunyan:2017azo}    &  16  &   18.9 (  19.1) &  1.2 (  1.2) &  0.6 ( 0.6) \tabularnewline\hline
 580 & ATLAS 8 TeV 20.3 fb$^{-1}$, $t\bar{t}$ $p_{T}^{t}$ and $m_{t\bar{t}}$ abs. spectrum    & \cite{Aad:2015mbv}         &  15  &    9.4 (  10.7) &  0.6 (  0.7) & -1.1 (-0.8) \tabularnewline
 \hline
\end{tabular}
\caption{Like Table~\ref{tab:EXP_1}, for newly-included LHC measurements.
	The ATLAS 7 TeV $W/Z$ data (4.6 fb$^{-1}$), labeled by $\ddagger$, are included in the CT18A and CT18Z global fits, but not in CT18 and CT18X. 
\label{tab:EXP_2} }
\end{table}
\endgroup
\end{widetext}

\subsubsection{LHC precision data from $W/Z$ vector boson production
\label{sec:DataWZ}
}

The CT18(Z) global analysis uses $W/Z$ vector boson production data from LHC Run-I, including measurements from the ATLAS, CMS and LHCb collaborations. 

The measurements from ATLAS included in the fit are: 

\begin{itemize}
	
	\item The $\sqrt{s}=7$ TeV $W/Z$ combined cross section measurements \cite{Aaboud:2016btc} (ID=248) with 4.6 fb$^{-1}$ 
	of integrated luminosity. The ATLAS group has performed 7 measurements with a total of 61 data points: distributions in the pseudorapidity of charged lepton in $W^+$ (11 points) and $W^-$ (11 points) production; rapidity of lepton pairs for low-mass Drell-Yan (DY) process in the central region (6 points); $Z$-peak DY process in the central (12 points) and forward (9 points) regions; high-mass DY process in the central (6 points) and forward (6 points) regions. In the published fits, we include 3 measurements: $W^+$, $W^-$ and $Z$-peak central DY production (34 points in total).
	These data are used only to fit the CT18A and CT18Z PDFs, but not the CT18 and CT18X PDFs.
	Other data are ignored due to the sizable EW corrections and/or photon-induced contribution ($\gamma\gamma\to l^+l^-$), as discussed in Sec. \ref{sec:EW}.

	\item The $\sqrt{s}=8$ TeV distribution of transverse momentum $p_T$ of lepton pairs in the $Z/\gamma^*$ production (ID=253) \cite{Aad:2015auj} with 20.3 fb$^{-1}$ of integrated luminosity. The ATLAS collaboration measured the $p_{T,  \ell \bar \ell}$ distribution up to 900 GeV for the lepton pairs in the invariant mass range $12<M_{\ell \bar \ell}<150$ GeV. Meanwhile, the experimentalists presented both the normalized and absolute cross sections for the singly differential distribution $d\sigma/d p_{T,  \ell \bar \ell}$ and doubly differential distribution $d^2\sigma/(d p_{T,  \ell \bar \ell}d y_{\ell \bar \ell})$. To select the cleanest and most sensitive data for the CT18 fits, 
	we only include 3 invariant-mass bins around the $Z$-peak region: $M_{\ell \bar \ell}\in[46-66,\ 66-116,\ 116-150]$ GeV. We do not include the data at $M_{\ell \bar \ell} < 46$ GeV, for which the kinematic cut $p_T^l>20$ GeV restricts the cross section to be coming predominantly from the region $p_{T,  \ell \bar \ell} \gtrsim M_{\ell \bar \ell}$, where the higher-order corrections beyond the current ${\cal O}(\alpha_s^3)$ calculation are significant. We fit neither the normalized $p_{T,  \ell \bar \ell}$ distributions, as they introduce artificial interdependence between the particle rates in the disparate $p_{T,  \ell \bar \ell}$ regions through their shared overall normalization, nor the doubly differential distributions, which are not very sensitive. 
	In total, we include 27 data points in the pair's transverse momentum region $45\! \leq\! p_{T,  \ell \bar \ell}\! \leq\! 150$ GeV, where the fixed-order NNLO cross section is most reliable. The data at lower $p_{T,  \ell \bar \ell}$ and higher $p_{T,  \ell \bar \ell}$ regions are excluded because of contributions from small-$p_T$ resummation and electroweak corrections, as discussed in Sec. \ref{sec:EW}.
\end{itemize}

For CMS, measurements of the charge asymmetry for inclusive ${W}^{\pm}$ production at ${\sqrt{s}} = 8$ TeV \cite{Khachatryan:2016pev} (ID=249) are included, with 18.8 fb$^{-1}$ of integrated luminosity.
These consist of 11 bins of muon pseudo-rapidity (over the range $0\leq |\eta^\mu|\leq 2.4$) with $p^\mu_T \geq 25$ GeV. The correlated systematic errors are implemented using a decomposition
of the covariance matrix to convert it to the
correlation matrix representation according to the procedure described in Appendix~\ref{sec:chi2_app}.
The same decomposition method will also be applied to the LHCb $W/Z$ experiments (Exp ID 245 and 250) to convert the published covariance matrix to a correlation matrix.
We have explicitly verified their equivalence using \texttt{ePump}~\cite{Hou:2019gfw}, capable of operating with both the covariance matrix and correlation matrix representations. 

In CT18, we include three experimental data sets published by LHCb: 
\begin{itemize}
\item The $\sqrt{s}=7$ TeV $W/Z$ forward rapidity cross section measurements \cite{Aaij:2015gna} (ID=245), with 1.0 fb$^{-1}$ of integrated luminosity, consist of 17 bins of $Z$-boson
rapidity ($2.0 \leq y_Z \leq 4.25$) for $Z$ boson production cross sections and eight bins of muon pseudo-rapidity ($2.0 \leq \eta^\mu \leq 4.5$) for $W^+$ or $W^-$ boson productions.
Similarly to the CMS charge asymmetry measurements discussed above, the systematic errors
are included by converting the published covariance matrix into
a correlation matrix.
The beam energy and luminosity
uncertainties are taken to be fully correlated between the cross-section measurements. This data set replaces previous LHCb measurements~\cite{Aaij:2012vn} of
inclusive vector boson production and lepton-charge asymmetry in the forward region, with 35 pb$^{-1}$ of integrated luminosity.

\item The $\sqrt{s}=8$ TeV $Z\rightarrow e^+e^-$ cross section measurements \cite{Aaij:2015vua} (ID=246) at forward rapidity, with 2.0 fb$^{-1}$ of integrated luminosity, consist of 17
bins of $Z$-boson rapidity ($2.0 \leq y_Z \leq 4.25$). 
The luminosity uncertainty is taken to be fully correlated, but the other correlated uncertainties are simply added in quadrature. We have used \texttt{ePump} to confirm that this approximation yields similar updated PDFs as those obtained from using the covariance matrix representation.

\item The $\sqrt{s}=8$ TeV $W/Z$ production cross section measurements \cite{Aaij:2015zlq} (ID=250), with 2.0 fb$^{-1}$ of integrated luminosity, consist of 18 bins of $Z$-boson rapidity
($2.0 \leq y_Z \leq 4.5$) and 8 bins of muon pseudo-rapidity $(2.0 \leq \eta^\mu \leq 4.5)$  for $W^+$ or $W^-$ boson productions. As in the $\sqrt{s}=7$ TeV case, the correlated systematic errors are included by converting the covariance matrix into the correlation matrix representation.
 The beam energy and luminosity uncertainties are taken to be fully correlated between the cross-section measurements.

\end{itemize}

\subsubsection{LHC inclusive jet production
\label{sec:DataJets}
}
For CMS, double-differential cross section measurements, $d^2\sigma /(dp_T dy)$, for jet production at both $\sqrt{s}=7$ TeV and 8 TeV, are used. 
We use the larger of the two jet radii ($R=0.7$)~\cite{Chatrchyan:2014gia}. 
The data sets consist of  5 fb$^{-1}$ of integrated luminosity at $\sqrt{s}=7$ TeV (ID=542), and 19.7 fb$^{-1}$ at $\sqrt{s}=8$ TeV~\cite{Khachatryan:2016mlc} (ID=545). The 7 TeV jet measurement contains 158 data points  in six bins of rapidity (with total rapidity coverage of $0\leq |y|\leq 3.0$),
covering the jet transverse momentum range $56\leq p_T\leq 1327$ GeV. At 8 TeV,  six rapidity bins of jet data are also used, covering the rapidity range $0\leq |y|\leq 3.0$.  There are a total of 185 data points, with a transverse momentum range of  $74\leq p_T\leq 2500$ GeV. 
These new CMS measurements replace the previous ones published in Ref.~\cite{Chatrchyan:2012bja} with 5 pb$^{-1}$ of integrated luminosity.  

In addition to the systematic error information provided in the \texttt{HEPData} files, jet energy corrections (JEC) in the CMS 7 TeV data have been decorrelated  according to the procedure in Ref.~\cite{Khachatryan:2014waa}.  In particular the JEC2 (``e05'') and an additional CMS-advocated decorrelation for $|y|>2.5$, have been implemented \cite{Voutilainen}. These decorrelations  improve the ability to fit the data. 

For ATLAS, we again use the larger of the two jet radii ($R=0.6$). Inclusive jet cross section measurements at $\sqrt{s}=7$ TeV with $R=0.6$ and 4.5 fb$^{-1}$ of integrated luminosity~\cite{Aad:2014vwa}  (Exp.~ID=544) are included in the global fit. 
This data set contains six bins covering the rapidity range $0.0 \leq y \leq 3.0$, with a total of 140 data points in the $74\leq p_T\leq 1992$ GeV range.
This data set replaces the previous data set~\cite{Aad:2011fc} which contains 37 pb$^{-1}$ data. 
 
Following the prescription given in Ref.~\cite{Aaboud:2017dvo}, two jet energy scale (JES) uncertainties have been decorrelated in the ATLAS 7 TeV jet data, namely, MJB (fragmentation) (``jes16''), and flavor response (``jes62'').
The decorrelation procedure reduces the $\chi^2$ value by approximately 92 units. The total contribution to the $\chi^2$ from the systematic error shifts is 28 (for 74 correlated systematic errors).
Only one of the systematic error sources requires a shift greater than 2 standard deviations. A further improvement of 52 units is obtained by including a 0.5\% theoretical error to account for
statistical noise associated with the Monte Carlo calculations of the needed NNLO/NLO $K$-factors \cite{Currie:2016bfm,Currie:2017ctp,Currie:2018xkj} in the NNLO fit, as detailed in Sec.~\ref{sec:TheorySettings}.
More details on the treatment of the ATLAS inclusive jet data are provided in Appendix~\ref{sec:ATLASjetdecorrel}.

\subsubsection{LHC top-quark pair production}
\label{sec:DataTop}

ATLAS and CMS have measured top-quark pair production differential cross sections as a function of the top-(anti)quark transverse momentum $p_{T,t}$, invariant mass $m_{t\bar{t}}$, rapidity of the pair $y_{t\bar{t}}$, transverse momentum of the top-quark pair $p_{T, t\bar{t}}$, and top-quark
rapidity $y_t$, individually for ATLAS, and in pairs for CMS. The individual impacts of the single differential $t\bar{t}$ cross section measurements have been analyzed, first by
using the \texttt{PDFSense} sensitivity framework of Ref.~\cite{Wang:2018heo}, and second in separate fits via \texttt{ePump} in Refs.~\cite{Hou:2019gfw,Hou:2019jxd}. There is some tension between the $t\bar t$ observables that leads to different pulls on the gluon distribution that each prefers. 
Difficulties in fitting simultaneously $p_{T,t}$, $m_{t\bar{t}}$, $y_{t}$, and $y_{t\bar{t}}$ distributions at 8 TeV were also found in~\cite{Bailey:2019yze,Hou:2019gfw,Hou:2019jxd}.

For the CT18 analysis, we thus decided to select a few top-quark production measurements with the best compatibility within the fit.  In the case of ATLAS, more than one $t\bar t$ observable can be included by making use of their published statistical correlations. We have chosen the  absolute differential cross sections $d\sigma/dp_{T,t}$ for the top-$p_T$ and $d\sigma/dm_{t\bar{t}}$ for the invariant mass, at $\sqrt{s}=8$ TeV with 20.3 fb$^{-1}$ of integrated luminosity (Exp.~ID=580)~\cite{Aad:2015mbv}, based on the recommendation from ATLAS.~\footnote{A. Cooper-Sarkar, private communication, and ATLAS-PHYS-PUB-2018-017.} 

The two ATLAS measurements are combined into one single data set which includes the full phase-space absolute differential cross-sections after the combination of the $e$+jets and
$\mu$+jets channels for the $p_{T,t}$ and $m_{t\bar{t}}$ distributions with statistical correlations. Both of these distributions are fitted together by decorrelating one of the
systematic uncertainties relative to the parton shower (PS)~\cite{Bogdan}. The QCD theoretical predictions at NNLO for these observables are obtained by
using \texttt{fastNNLO} tables provided in Refs.~\cite{Czakon:2017dip,fastnnlo:grids}.

 In an upcoming study, we find that the ATLAS rapidity distributions of a single quark and top pair, $y_t$ and $y_{t\bar t}$, can be fitted in the CT18 setup with $\chi^2_E/N_{pt,E}>2.3$ -- too high for the fit to be acceptable, which is consistent with the findings in Ref.~\cite{Kadir:2020yml}. These distributions show tensions with some other data sets, their inclusion, either in the single-differential or double-differential form, would not lead to the reduction of the PDF uncertainty. 

For CMS, we
have chosen the normalized double differential cross section $d^2 \sigma/dp_{T,t}dy_t$  at $\sqrt{s}=8$ TeV, with 19.7 fb$^{-1}$ (Exp.~ID=573)~\cite{Sirunyan:2017azo}. 

The observed effect of the $t\bar t$ data sets on the CT18 PDFs is modest, when they are included together with the Tevatron and LHC jet production. 
Their impact on the gluon PDF is compatible with the jet data, but the jet data provide stronger constraints due to their larger numbers of data points, wider kinematic range, and relatively small statistical and systematic errors.

In the course of the CT18 analysis, CMS measurements of top-quark pair production differential cross sections at 13 TeV were published~\cite{Sirunyan:2018ucr}, 
and bin-by-bin data correlations were made available on the \texttt{HEPData} repository.
While these measurements are not currently included in the CT18 global fit, their description is discussed later in Sec.~\ref{sec:StandardCandles}.
 
\subsubsection{Other LHC measurements not included in the CT18 fits
\label{sec:DataOther}
}

Besides the CT18(Z) data ensemble, we have carefully investigated several other high-luminosity measurements from LHC Run-I. In certain cases we observed either no significant impact or substantial tensions with the CT18(Z) baseline. The following vector boson production data were examined using {\tt PDFSense}, {\tt ePump}, or full fits, but not included in the final CT18(Z) global analysis:
\begin{itemize}
\item Difficulties were encountered in obtaining a good agreement between theory and the ATLAS $\sqrt{s}=7$ TeV $Z$-boson transverse momentum distribution ($p_{T,  \ell \bar \ell}$) data with 4.7 fb$^{-1}$ of integrated luminosity \cite{Aad:2014xaa}. The subset of these data with $p_{T,  \ell \bar \ell} \sim M_{\ell \bar \ell}$ (in the kinematic region most amenable to a fixed-order calculation) has rendered unacceptably high $\chi^2$ values for various combinations of the renormalization and factorization scales that we have tried. 
No significant constraints could be ascribed to the CMS $\sqrt{s}=8$ TeV $p_{T,  \ell \bar \ell}$ and  $y_{\ell \bar \ell}$ distributions with 19.7 fb$^{-1}$ of integrated luminosity~\cite{Khachatryan:2015oaa} in the
$Z$ peak kinematic region, and to the CMS $\sqrt{s}=8$ TeV normalized $W$ $p_T$ and $Z$ $p_T$ spectra with 18.4 pb$^{-1}$ \cite{Khachatryan:2016nbe}. 
When comparing to the CMS double-differential distributions in ($p_{T,  \ell \bar \ell},y_{\ell \bar \ell}$), we observed a large discrepancy between theory and data in the last rapidity bin.
For the {\it normalized} $p_{T,  \ell \bar \ell}$ distributions of lepton pairs presented by
both ATLAS and  CMS groups, it was not clear how to consistently compare to data in a limited range $p_{T,  \ell \bar \ell} \sim M_{\ell \bar \ell}$ when the normalization of data points was dependent on the cross section outside of the fitted range. 

\item  No substantial changes in the candidate PDFs were observed after including either
the single- or double-differential distributions, $d \sigma/dQ$ or $d^2 \sigma/(dQ\ dy)$, of the ATLAS $\sqrt{s}=8$ TeV Drell-Yan cross section measurements at $116 \leq Q\leq 1500$ GeV and $0 \leq y_Z\leq2.5$ with 20.3 fb$^{-1}$ of integrated luminosity \cite{Aad:2016zzw}. These high-mass data are impacted by non-negligible EW corrections and photon-induced (PI) dilepton production, the point that is further addressed in Sec. \ref{sec:EW}. For the same reason, we do not include the data of ATLAS 7 TeV high-mass Drell-Yan production with 4.7 fb$^{-1}$ of integrated luminosity \cite{Aad:2013iua}.

\item The low-mass Drell-Yan data by the ATLAS collaboration at 7 TeV \cite{Aad:2014qja} were also explored and found to have no significant impact on the PDFs. 

\item We have explored the impact of the data of $W$-boson associated with charm-jet production from ATLAS \cite{Aad:2014xca} and CMS \cite{Chatrchyan:2013uja} measurements at 7 TeV. As the NNLO calculations for $W+\mbox{charm jet}$ are not available, we use these data only to compare against the NLO theoretical predictions in Sec. \ref{sec:Wcharm}. 

\item No significant impact is found by including the single- or double-differential cross sections, $d\sigma/dQ$ or $d^2\sigma/(dQdy)$, of the CMS DY data taken at 7 \cite{Chatrchyan:2013tia} and 8 \cite{CMS:2014jea} TeV. These data are not included in the CT18 fits for the following reasons. First, the EW corrections and photon-induced contributions to these data are non-negligible in the high-mass region. Second, these data are presented as cross sections over the full phase space, a fact which introduces additional uncertainties from the unfolding procedure.
The 8 TeV data set has $\chi^2_E/N_{pt, E} \approx 2$ for CT18(Z) PDFs and does not modify the PDFs when examined using \texttt{ePump} and \texttt{PDFSense}.
\end{itemize}

\subsection{Alternative PDF fits: CT18A, CT18X, CT18Z}
\label{sec:alt} 
We are now ready to review the three additional fits that were explored in parallel with CT18 by making alternative choices for data selection and theoretical calculations.
The key differences among these fits are listed in Table~\ref{tab:AXZ}. Their predictions will be compared in Appendix~\ref{sec:AppendixCT18Z}.

\begin{table}
  \begin{tabular}{ccccc}
\hline 
\textbf{PDF} & \textbf{Factorization scale } & \textbf{ATLAS 7 TeV $W/Z$} & \textbf{CDHSW $F_{2}^{p,d}$ } & \textbf{Pole charm }\tabularnewline
\textbf{ensemble}
 & \textbf{in DIS} & \textbf{data included?\quad} & \textbf{data included?\quad} & \textbf{mass, GeV}\tabularnewline
\hline 
\hline 
\noalign{\vskip6pt}
CT18 & $\mu_{F,DIS}^{2}=Q^{2}$ & No & Yes & 1.3\tabularnewline[6pt]
\hline 
\noalign{\vskip6pt}
CT18A & $\mu_{F,DIS}^{2}=Q^{2}$ & Yes & Yes & 1.3\tabularnewline[6pt]
\hline 
\noalign{\vskip6pt}
CT18X & $\mu_{F,DIS}^{2}=0.8^{2}\left(Q^{2}+\frac{0.3\mbox{ GeV}^{2}}{x_B^{0.3}}\right)$ & No & Yes & 1.3\tabularnewline[6pt]
\hline 
\noalign{\vskip6pt}
CT18Z  & $\mu_{F,DIS}^{2}=0.8^{2}\left(Q^{2}+\frac{0.3\mbox{ GeV}^{2}}{x_B^{0.3}}\right)$ & Yes & No & 1.4\tabularnewline[6pt]
\hline 
\end{tabular}
        \caption{
                A summary of theoretical settings and data set choices in CT18 and
		each of the three alternative fits: CT18A, CT18X and CT18Z. The
		lattermost of these is compared with CT18 throughout the main text
		of this article, whereas more detail regarding each of the alternative
		fits is presented in App.~\ref{sec:AppendixCT18Z}.
        }
\label{tab:AXZ}
\end{table}

({\it i}) {\bf CT18X} differs from CT18 in adopting an alternate
scale choice for the DIS data sets.
It is most common to compute the inclusive DIS cross sections using the photon's virtuality as the factorization scale,   $\mu^2_{F,\mathit{DIS}} = Q^2$.
It has been argued, however, that resummation of logarithms $\ln^p(1/x)$ at
$x\ll 1$ improves agreement with HERA Run I+II data by several tens
of units of $\chi^2$ \cite{Ball:2017otu,Abdolmaleki:2018jln}. In our
analysis, we observe that, by evaluating the DIS cross sections in a {\it fixed-order} calculation at NNLO accuracy, with a tuned factorization scale
$\mu^2_{F,x} \equiv 0.8^2 \left(Q^2 + 0.3\mbox{ GeV}^2/x_B^{0.3}\right)$, 
instead of the conventional $\mu^2_F =Q^2$, we achieve nearly the same quality of improvement in the description of the HERA DIS
data set as in the analyses with low-$x$ resummation \cite{Ball:2017otu,Abdolmaleki:2018jln}. The fit done with these
modified settings is designated as CT18X. For this fit, the
$\chi^2$ of $\mbox{HERA I+II}$ reduces by more than $50$ units in the kinematical region with 
$Q > 2\mbox{ GeV}$ and $x\! >\! 10^{-5}$, assessed in the CT18
global fit.  The CT18X prediction
for H1 $F_L$ is moderately higher than that for CT18, which improves $\chi^2_E$ for H1 $F_L$ by a few units. See an illustration in  Fig.~\ref{fig:saturation} and its discussion in the Executive Summary~\ref{sec:summary-HERA2}.

The parametric form of the $x_B$-dependent scale
$\mu^2_{F,x}$ is inspired by saturation arguments (see, e.g.,
\cite{GolecBiernat:1998js,Caola:2009iy}). The numerical coefficients in $\mu^2_{F,x}$
are chosen to  minimize $\chi^2$ for the HERA DIS data.
At $x\gtrsim 0.01$, $\mu_{F,x}^2\approx 0.8\ Q^2$ results in larger NNLO DIS cross sections than with $\mu_{F}^2=Q^2$, as it might happen due to contributions from next-to-NNLO (N3LO) and beyond. At $x \lesssim 0.01$, $\mu_{F,x}^2$ numerically reduces the 
$Q^2$-derivative of NNLO DIS cross sections. In turn, these changes result in the enhanced gluon PDF at small $x$ and
reduced gluon at $2.5\cdot 10^{-2} \lesssim  x\lesssim 0.2$.

({\it ii}) Unlike CT18, the {\bf CT18A} analysis includes
high-luminosity ATLAS 7 TeV $W/Z$ rapidity distributions~\cite{Aaboud:2016btc}
that show some tension with DIS experiments and prefer a larger strangeness PDF than the
DIS experiments in the small $x_B$ region. Inclusion of the ATLAS 7 TeV $W/Z$ data leads to
significant deterioration in the $\chi^2_E$ values ({\it i.e.}, larger $S_E$ values)
for the dimuon SIDIS production data (NuTeV, CCFR), which are strongly
sensitive to the strangeness PDF. One way to see this is to compare
the $S_E$ distributions for the CT18 fit in Fig.~\ref{fig:sn_ct18} and the counterpart figure for CT18Z in Fig.~\ref{fig:sn_ct18z} of App.~\ref{sec:AppendixCT18Z}.
The comparison shows that the $S_E$ values for CCFR and NuTeV dimuon data sets are elevated in the CT18Z fit, as compared to the CT18 fit, as a consequence of inclusion of the
ATLAS $W/Z$ data in the CT18Z fit.
Another way to see this was carried out in Ref.~\cite{Hou:2019gfw}, using the~\texttt{ePump} program.   

({\it iii}) {\bf CT18Z} represents the accumulation of these settings introduced to obtain a PDF set that is maximally different from CT18,  despite achieving about the same global $\chi^2/N_{pt}$ as CT18. The CT18Z fit
includes the 7 TeV $W/Z$ production data of ATLAS like CT18A, but
also includes the modified DIS scale choice, $\mu_{F,x}$,
as done for CT18X. In addition to these modifications, CT18Z
excludes the CDHSW extractions of the $F_2$ and $x_B F_3$ structure functions
from $\nu\mathrm{Fe}$ scattering, which otherwise would oppose the trend of CT18Z to have a softer gluon at $x>0.1$, cf. Sec.~\ref{sec:Baseline}. Finally, CT18Z is done by assuming a slightly higher value of
the charm quark mass (1.4 GeV compared to 1.3 GeV) in order to modestly improve the fit to the vector boson production data.

The combination of these choices in the
CT18Z analysis results in a Higgs boson production cross section via gluon
fusion that is reduced by about 1\% compared to the corresponding
CT14 and CT18 predictions. Thus, the various choices made during the
generation of four CT18(A,X,Z) fits allow us to more faithfully
explore the full range of the PDF behavior at NNLO that is consistent
with the available hadronic data, with implications for electroweak precision physics.

\section{Theoretical inputs to CT18
\label{sec:Theory}}

Modern global fits  determine the PDFs from 
a large number of data points ($N_\mathit{pt}\! >\! 3600$ for CT18), provided by a wide variety of experimental measurements
(39 data sets for CT18), and involving thousands of
iterations of multivariate fits, with the theoretical cross sections evaluated at NNLO.
In the CT18 fits, the $x$ dependence of the input PDFs, at the initial scale $Q_0$ equal to the pole mass of the charm quark, is parametrized by
Bernstein polynomials, multiplied by the standard $x^a$ and $(1-x)^b$ factors that determine the small-$x$ and large-$x$ asymptotics.  In these functions, there are 5-8 independent
fitting parameters for each parton flavor except strangeness; additional parameters may be determined by momentum and 
flavor sum rules or (if poorly constrained) fixed at physically reasonable values.

In the present section, we review the essential components of our theoretical setup: the goodness-of-fit function in  Sec.~\ref{sec:chi2}, 
computer programs for (N)NLO computations for various processes  in Sec.~\ref{sec:calcs}, and 
input parametric forms for the PDFs in Sec.~\ref{sec:Paramstudies}. The explicit parametric forms for the best-fit CT18 PDFs are presented in 
Appendix~\ref{sec:AppendixParam}. 

\subsection{Goodness of fit function and the covariance matrix}
\label{sec:chi2}
The CTEQ-TEA analyses quantify the goodness-of-fit to an experimental data set
$E$ with $N_{pt}$
data values by means of the log-likelihood function~\cite{Pumplin:2002vw},
\begin{equation}
\chi_{E}^{2}(a,\lambda)=\sum_{k=1}^{N_{pt}}\frac{1}{s_{k}^{2}}\left(D_{k}-T_{k}(a)-\sum_{\alpha=1}^{N_{\lambda}}\lambda_{\alpha}\beta_{k\alpha}\right)^{2}+\sum_{\alpha=1}^{N_{\lambda}}\lambda_{\alpha}^{2}.\label{Chi2sys}
\end{equation}
A $k$-th datum is typically provided as a central value $D_{k},$ an uncorrelated
statistical error $s_{k,\mbox{\scriptsize stat}}$, and possibly an
uncorrelated systematic error
$s_{k,\mbox{\scriptsize uncor sys}}$.
Then, $s_{k}\equiv \sqrt{s_{k,\mbox{\scriptsize stat}}^{2}+s_{k,\mbox{\scriptsize uncor sys}}^{2}}$
is the total uncorrelated error on the measurement $D_{k}$.

$T_{k}$ is the corresponding theory value that depends on the PDF
parameters $\left\{a_1,a_2,...\right\}\equiv a$. In addition, the $k$-th datum may depend on
$N_{\lambda}$ correlated systematic uncertainties, and those may
be fully correlated over all data points. To estimate such errors,
it is common to associate each source of the correlated error with
an independent random nuisance parameter $\lambda_{\alpha}$ that is assumed to
be sampled from a standard normal distribution, unless known
otherwise. The experiment does not tell us the values of
$\lambda_{\alpha}$
but it may provide
the change $\beta_{k\alpha}\lambda_\alpha$ of $D_k$ under a variation of $\lambda_{\alpha}$. Knowing $\beta_{k\alpha}$,
one can estimate the likely values of $\lambda_{\alpha}$, as well
as the uncertainty in the PDF parameters for a plausible range of
$\lambda_{\alpha}$.

For those experiments $E$ that provide
$\beta_{k\alpha}$, we find that, at the global minimum $a_0$,
the best-fit $\chi^2$ value is given as
\begin{equation}
\chi^{2}_E (a_{0},\overline \lambda(a_0))=
\sum_{i=1}^{N_{pt}} r_i^2(a_0)
+\sum_{\alpha=1}^{N_\lambda}\overline\lambda^2_\alpha(a_0) \label{Chi2a0l0}
\end{equation}
in terms of the best-fit {\it shifted residuals},
\begin{equation}
  r_{i}(a_0)\  =\ s_{i}\sum_{j=1}^{N_{\mathit{pt}}}(\mathrm{cov}^{-1})_{ij}\,\left(D_{j}-T_{j}(a_0)\right),\label{eq:res-cov}
\end{equation}
and best-fit nuisance parameters,
\begin{equation}
\overline{\lambda}_{\alpha}(a_0) =\sum_{i,j=1}^{N_{\mathit{pt}}}(\mathrm{cov}^{-1})_{ij}\frac{\beta_{i\alpha}}{s_{i}}\frac{\left(D_{j}-T_{j}(a_0)\right)}{s_{j}},
\label{eq:lam-cov}
\end{equation}
where
\begin{equation}
(\mathrm{cov}^{-1})_{ij}\ =\ \left[\frac{\delta_{ij}}{s_{i}^{2}}\,-\,\sum_{\alpha,\beta=1}^{N_{\lambda}}\frac{\beta_{i\alpha}}{s_{i}^{2}}A_{\alpha\beta}^{-1}\frac{\beta_{j\beta}}{s_{j}^{2}}\right]\ ,\label{eq:covmat}
\end{equation}
and
\begin{equation}
A_{\alpha\beta}\ =\ \delta_{\alpha\beta}\,+\,\sum_{k=1}^{N_{\mathit{pt}}}\frac{\beta_{k\alpha}\beta_{k\beta}}{s_{k}^{2}}\ .
\end{equation}
These relations are derived in Appendix~\ref{sec:chi2_app}.

Another instructive form expresses $r_i(a_0)$ in terms of the
shifted data values, $D_i^{sh}\equiv D_i -
\sum_{\alpha=1}^{N_{\lambda}}\overline \lambda_{\alpha}(a_0) \beta_{k\alpha}$:
\begin{equation}
r_{i}(a_0) = \frac{D_{i}^{sh}(a_0)-T_{i}(a_0)}{s_{i}}. \label{riShifted}
\end{equation}

Sometimes, we take extra steps to
convert the published table of correlated
uncertainties into the $\beta_{k\alpha}$ matrix formatted in accord
with Eq.~(\ref{Chi2sys}).
For example, when an experiment distinguishes between positive
and negative systematic variations, we average these for each data
point for consistency with the normally distributed $\lambda_{\alpha}$.
{[}We have verified that the choice of the averaging procedure
does not significantly affect the
outcomes, {\it e.g.}, if a central value is shifted to be in the middle of an originally
asymmetric interval, {\it etc.}{]}

In a small number of  experimental publications, only
a form based on the covariance
matrix $\left(\mbox{cov}\right)_{ij}$ is used in place of Eq.~(\ref{Chi2sys}):
\begin{equation}
\chi_{E}^{2}(a)=\sum_{i,j=1}^{N_{pt}}\left(\mbox{\mbox{cov}}^{-1}\right)_{ij}\left(D_{i}-T_{i}(a)\right)\left(D_{j}-T_{j}(a)\right).\label{Chi2CovMat1}
\end{equation}
While we can compute $\chi^2$ directly using
Eq.~(\ref{Chi2CovMat1}), when deriving the PDFs, we find it convenient
to go back to the form consisting of the uncorrelated errors $s_i$ and
the correlated contributions provided by $\beta_{k\alpha}$:
\begin{equation}
  (\mbox{cov})_{ij} \approx s_i^2 \delta_{ij}
  +\sum_{\alpha=1}^{N_\lambda} \beta_{i\alpha}\beta_{j\alpha}.\label{covapprox}
\end{equation}
An algorithm to construct such a representation with sufficient
accuracy is presented at the end of Appendix~\ref{sec:chi2_app}. In
all relevant cases, we have checked that both the input covariance matrix
$(\mbox{cov})_{ij}$ and its decomposed version (\ref{covapprox})
produce close values of $\chi^2$. With the latter representation, we
are also able to examine the shifted data values and shifted
residuals, Eq.~(\ref{riShifted}), to explore agreement with the
individual data points. 

In this article, we generally follow the CTEQ methodology and obtain
$r_{i}(a_0)$ directly from the CTEQ-TEA fitting program, together
with the optimal nuisance parameters $\overline{\lambda}_{\alpha}(a_0)$
and shifted central data values $D_{i}^{sh}(a).$

\subsection{Theoretical computations and programs \label{sec:calcs}}
\subsubsection{Overview
\label{sec:TheorySettings}}

For deep-inelastic scattering observables, we perform computations using an NNLO realization \cite{Guzzi:2011ew} of the SACOT-$\chi$ heavy-quark scheme \cite{Aivazis:1993pi,Aivazis:1993kh,Kramer:2000hn,Tung:2001mv} adopted since CT10 NNLO \cite{Gao:2013xoa}. These can be done using either the pole or $\overline{\mathrm{MS}}$ quark masses as the input \cite{Gao:2013wwa}, with the default choices of quark masses set to be $m_c^{pole}=1.3$ GeV in CT18, A, and X ($m_c^{pole}=1.4$ GeV in CT18Z), and $m_b^{pole}=4.75$ GeV.  The neutral-current DIS cross sections are evaluated at NNLO directly in the fitting code. For charged-current DIS cross sections,  the NNLO cross sections from heavy quarks can be obtained by fast interpolations with pre-generated grids based on the calculation presented in Ref.~\cite{Berger:2016inr}. The impact of the NNLO contribution on the description of the charged-current dimuon DIS data is further discussed in Sec.~\ref{sec:Qualitydimuon}.

The computational complexity of NNLO matrix elements precludes their direct evaluation for each fit iteration, particularly given the expansive size of the data sets fitted in CT18. Instead, for the
newly included high-precision data from the LHC, \texttt{ApplGrid}~\cite{Carli:2010rw} and \texttt{fastNLO}~\cite{Wobisch:2011ij} 
fast tables have been generated using programs such as \texttt{MCFM}~\cite{Campbell:2010ff}, \texttt{NLOJet++}~\cite{Nagy:2003tz} and 
\texttt{aMCfast}~\cite{Bertone:2014zva}, to allow fast evaluation of the matrix elements as the PDF parameters are varied. 
NNLO cross sections are then evaluated using NNLO/NLO point-by-point $K$-factors determined using the fast tables and NNLO programs such 
as \texttt{NNLOJET}~\cite{Buza:1997mg,Ridder:2015dxa,Gehrmann-DeRidder:2017mvr,Currie:2016bfm,Currie:2017ctp}, \texttt{FEWZ}~\cite{Gavin:2010az,Gavin:2012sy,Li:2012wna}, 
\texttt{MCFM}~\cite{Campbell:2010ff,Boughezal:2016wmq,MCFM8} and \texttt{DYNNLO}~\cite{Catani:2007vq,Catani:2009sm}. 
One exception is the top-quark data from ATLAS and CMS, for which \texttt{fastNNLO} tables have been 
provided by the authors for the NNLO cross sections~\cite{Czakon:2016dgf,Czakon:2018nun}. The programs used for the calculation of the cross sections for each data set are summarized 
in Table~\ref{Theory-Calc-II}. We have explored the impact of the choices of scales and NNLO programs for some data sets, but the variation is not included in the PDF uncertainties for various reasons. More details can be found in the rest of this paper.

\begin{table}[b]
\begin{tabular}{|c|c|c|c|c|c|c| }
\hline
Expt. ID\# & Process   &   Expt.   & fast table  & NLO code &  NNLO $K$-factors  & $\mu_{R,F}$\\
\hline
245  & \multirow{5}{*}{$W/Z$}  & LHCb 7 TeV    & \multirow{5}{*}{\texttt{APPLgrid}}   & \multirow{5}{*}{\texttt{MCFM/aMCfast}}   & \multirow{2}{*}{\texttt{FEWZ/MCFM}}   &  \multirow{5}{*}{$M_{W},M_{\ell \bar \ell}$}  \\
246  &  & LHCb 8 TeV $Z\rightarrow e^+e^-$    &  &  &  &  \\
248  &  & ATLAS 7 TeV  &  &  & \texttt{FEWZ/MCFM/DYNNLO}  &  \\
249  &  & CMS 8 TeV $A(\mu)$   &  &  & \multirow{2}{*}{\texttt{FEWZ/MCFM}} &  \\
250  &  & LHCb 8 TeV   &  &  &  &  \\
\hline
253 & high-$p_T$ $Z$  & ATLAS 8 TeV   & \texttt{APPLgrid} &  \texttt{MCFM} & \texttt{NNLOJET}  & $\sqrt{(p_{T,  \ell \bar \ell})^{2}+M_{\ell \bar \ell}^{2}}$ \\
\hline
542 &\multirow{3}{*}{Incl. jet} & CMS 7 TeV   & \texttt{fastNLO}  & \multirow{3}{*}{  \texttt{NLOJet++} }  & \multirow{3}{*}{ \texttt{NNLOJET}}   &  \multirow{ 3}{*}{$p_{T}$}    \\
544 &  &  ATLAS 7 TeV & \texttt{APPLgrid} & & &  \\
545 &  & CMS 8 TeV   & \texttt{fastNLO}  &  &   &      \\  
\hline
573 & \multirow{2}{*}{$t\bar{t}$} &  CMS 8 TeV  &
\multicolumn{3}{|c|}{\multirow{2}{*}{\texttt{fastNNLO} }  } &
\multirow{2}{*}{ $\frac{H_{T}}{4}$, $\frac{m_{T}}{2}$}\\
580 &   &   ATLAS 8 TeV & \multicolumn{3}{|c|}{} & \\
\hline
\end{tabular}
\caption{Theory calculations for the high-precision data from the LHC which are newly included in the CT18(Z) global fit. 
The $K$-factors of ATL7WZ (ID 248) extracted from \texttt{xFitter} are calculated with \texttt{DYNNLO} 
and compared with \texttt{FEWZ} and \texttt{MCFM} in App. \ref{sec:Appendix4xFitter}. 
\label{Theory-Calc-II}}
\end{table}

In the newest NNLO calculations for the high-$p_T$ $Z$ and inclusive jet production available to the CT18 analysis, the NNLO corrections were not perfectly smooth among the experimental bins because of the statistical uncertainty introduced by Monte-Carlo (MC) integration. The resulting artificial fluctuations (of the magnitude of less than a fraction of percent of the central cross section values) have elevated the values of $\chi^2$ in these precise measurements. Through examination of the kinematic dependence of the NNLO/NLO $K$ factors, we identified all such cases and approximated the $K$-factors by smooth functions during the PDF fit. To account for the uncertainty introduced by the smoothing of the $K$-factors, we included  uncorrelated {\it MC errors} equal to 0.5\% of the central data values in the affected processes in Secs.~\ref{sec:TheoryJets} and \ref{sec:LHCEWbosons}. The MC errors lowered the $\chi^2$ values for these processes without changing the central PDF fits. The MC errors were estimated from the maximal deviations of the individual $K$-factor values from the respective smooth functions, with 0.5\% being the conservative upper bound reached for a fraction of the fitted data points.

Error propagation must account for numerical theoretical errors of this kind. The non-negligible MC errors in some NNLO predictions were also noticed by other PDF fitting groups. The NNPDF group, for example, takes a similar approach in their analysis \cite{Ball:2017nwa}. For the inclusive jet data, they use NLO calculations as the theoretical predictions, together with an additional correlated uncertainty estimated from the renormalization and factorization scale variations. For the high-$p_T$ $Z$ boson data, an NNPDF-based analysis also adds an extra 1\% uncorrelated uncertainty to account for the Monte-Carlo fluctuations of the NNLO/NLO $K$-factor values \cite{Boughezal:2016wmq}.

For the legacy data on electroweak boson production, already included in the CT14 and \CTHERAII, 
we inherit the original CTEQ calculations summarized in Table~\ref{Theory-Calc-VB}. 
The NLO calculation is directly performed by the CT fitting code, while the point-by-point $K$-factors are calculated with 
\texttt{Vrap}~\cite{Anastasiou:2003ds,Anastasiou:2003yy}, \texttt{ResBos}~\cite{Ladinsky:1993zn,Konychev:2005iy} and
\texttt{FEWZ}~\cite{Gavin:2010az,Gavin:2012sy,Li:2012wna}.

\begin{table}[tb]
\begin{tabular}{|c|c|c|c|c| }
\hline
Expt. ID\# & Experiment   &  NLO code &  NNLO $K$-factors  & $\mu_{R,F}$\\
\hline
201 &  E605 DY  &  \multirow{3}{*}{ \texttt{CTEQ}}  & \multirow{3}{*}{ \texttt{FEWZ}  }   & \multirow{3}{*}{$M_{\ell \bar \ell}$} \\
203  & E866 DY $\sigma_{pd}/\sigma_{pp}$  &  &    & \\
204  & E866 DY $\sigma_{pp}$  &  &    & \\
\hline
225 & CDF Run-1 $A(e)$  &  \multirow{4}{*}{ \texttt{CTEQ}}  &  \multirow{4}{*}{ \texttt{ResBos}}  &  $M_{\ell \bar \ell}$   \\
227 & CDF Run-2 $A(e)$ &     &  &  \multirow{3}{*}{$M_{W}$} \\
234 &  D$\varnothing$ Run-2 $A(\mu)$  &    && \\
281 & D$\varnothing$ Run-2 $A(e)$  &   &&  \\
\hline
260 &   D$\varnothing$ Run-2 $y_{Z}$ &   \multirow{2}{*}{ \texttt{CTEQ}}  &  \multirow{2}{*}{\texttt{Vrap}}  &
\multirow{2}{*}{ $M_{\ell \bar \ell}$ } \\
261 &  CDF Run-2 $y_{Z}$ &        &  &\\
\hline 
266 & CMS 7 TeV $A(\mu)$   &   \multirow{3}{*}{\texttt{CTEQ} }  &\multirow{3}{*}{ \texttt{ResBos}} & \multirow{2}{*}{ $M_{W}$}   \\
267  & CMS 7 TeV $A(e)$  &        &  &\\
268 &  ATLAS 7 TeV 2011 $W/Z$  &   &   & $M_{W},M_{\ell \bar \ell}$ \\
\hline
\end{tabular}
\caption{Theory calculations for the CT14 and \CTHERAII's legacy data of electroweak vector boson production.
\label{Theory-Calc-VB}}
\end{table}

Even with the use of stored grids for fast evaluation of the matrix
elements, significant improvements on speed are needed. The CT fitting
code has been upgraded to a multi-threaded version with a two-layer
parallelization, through a rearrangement of the minimization
algorithm and via a redistribution of the data sets. As a result, the speed of calculations increased by up to a factor of 10. Details are provided in
Appendix~\ref{sec:AppendixCodeDevelopment}.  

We will now describe the theoretical calculations for each new LHC process included in the CT18(Z) fits.

\subsubsection{LHC inclusive jet data
\label{sec:TheoryJets}
}

LHC inclusive jet data are available with different jet radii. We have chosen the larger of the two nominal jet radii, 
0.6 for ATLAS and 0.7 for CMS, to reduce dependence on resummation/showering and hadronization effects~\cite{Bellm:2019yyh}. 
There is a non-negligible difference at low jet transverse momentum between theory predictions at NNLO 
using as the momentum-scale choice of either the inclusive jet $p_T$ or the leading jet $p_T$ ($p_{T1}$)~\cite{Currie:2018xkj} .
The nominal choice adopted by the CTEQ-TEA group is to use the inclusive jet $p_T$. 
We have observed that the fitted gluon PDF is not very sensitive to this choice even in the kinematic regions 
where the difference in NNLO predictions between these two scale choices is important.

Electroweak corrections from Ref.~\cite{Dittmaier:2012kx} were applied
to jet cross sections and can be as large as 10\% for the highest transverse momentum bin in the central
rapidity region, but decrease quickly with increasing rapidity and with decreasing jet transverse momentum.
Furthermore, in accord with the previous subsubsection, the QCD NNLO/NLO $K$-factors were fitted with smooth curves, and a 0.5\% theoretical error
assessed with respect to the data has been added to each data value to take into account the fluctuations
in integration of NNLO cross sections provided by \texttt{NNLOJET}.

\subsubsection{LHC electroweak gauge boson hadroproduction
\label{sec:LHCEWbosons}}
The Drell-Yan theory calculations at NNLO in the CT18(Z) global analysis consist of the following:

\begin{itemize}
\item ATLAS 7 TeV 4.6 fb$^{-1}$ measurements of $W^{\pm}$ and $Z/\gamma^{*}$ production cross 
sections in the $e$ and $\mu$ decay channels~\cite{Aaboud:2016btc}: 
the theory predictions at NLO are obtained by using \texttt{APPLgrid}~\cite{Carli:2010rw} fast tables 
generated with \texttt{MCFM}~\cite{Campbell:2010ff} and validated against \texttt{aMCfast}~\cite{Bertone:2014zva} 
interfaced with \texttt{MadGraph5\_aMC@NLO}~\cite{Alwall:2014hca}. The NNLO corrections are imported from the 
\texttt{xFitter} analysis published in Ref.~\cite{Aaboud:2016btc}. These corrections are 
obtained using the \texttt{DYNNLO-1.5} code  \cite{Catani:2007vq,Catani:2009sm}, and checked against  \texttt{FEWZ-3.1.b2}~\cite{Gavin:2010az,Gavin:2012sy,Li:2012wna} and \texttt{MCFM-8.0} \cite{Boughezal:2016wmq} codes. Some discrepancy among these codes (up to $\sim\!1\%$) were found. However,  
these discrepancies do not induce significant differences
in the calculated results like $\chi^2$. More details can be found in Appendix~\ref{sec:Appendix4xFitter}.

\item CMS 8 TeV 
$18.8 \mbox{ fb}^{-1}$ measurements of muon charge asymmetry~\cite{Khachatryan:2016pev}: 
The theory predictions at NLO are from \texttt{APPLgrid} generated with \texttt{MCFM}, while for the NNLO 
corrections, we use $K$-factors calculated with \texttt{FEWZ-3.1}. These predictions have also been validated with \texttt{MCFM-8.0}. 

\item LHCb 7 TeV $W/Z$ cross sections, $W$ charge asymmetry measurements with 1 fb$^{-1}$ of integrated luminosity~\cite{Aaij:2015gna}, 
and LHCb 8 TeV measurements including both the electron \cite{Aaij:2015vua} and muon \cite{Aaij:2015zlq} channels: 
the NLO theory calculation is obtained by using \texttt{APPLgrid} fast tables generated with \texttt{MCFM}. 
These have been validated against \texttt{MadGraph5\_aMC@NLO} + \texttt{aMCfast}. The NNLO corrections are calculated with  \texttt{FEWZ}, and validated by \texttt{MCFM}.

\item ATLAS~\cite{Aad:2014xaa,Aad:2015auj} and CMS~\cite{Khachatryan:2015oaa} measurements of transverse momentum of Drell-Yan lepton pairs at 7 TeV and/or 8 TeV. 
The CT18(Z) fit includes only the ATLAS 8 TeV absolute differential cross section measurements. The NLO theoretical calculation is performed with \texttt{APPLgrid} generated with \texttt{MCFM}. The NNLO corrections are provided by the \texttt{NNLOJET} group \cite{Ridder:2015dxa,Gehrmann-DeRidder:2017mvr}.
We have fitted the NNLO/NLO $K$-factors with smooth curves and include a 0.5\% MC error to account for the fluctuations in the NNLO calculations.
In addition, we have imposed the kinematic cut $45\!<\!p_{T,  \ell \bar \ell}\!<\!150$ GeV  to ensure reliability of the fixed-order calculation. 
The low-$p_{T,  \ell \bar \ell}$ region is dropped due to the non-negligible contribution from QCD soft-gluon resummation, and the high $p_{T,  \ell \bar \ell}$ region is dropped because the EW corrections there are expected to grow~\cite{Hollik:2015pja,Kallweit:2015fta}.

\end{itemize}

\begin{figure}[p]
        \center
	\hspace*{-0.5cm}\includegraphics[width=0.51\textwidth]{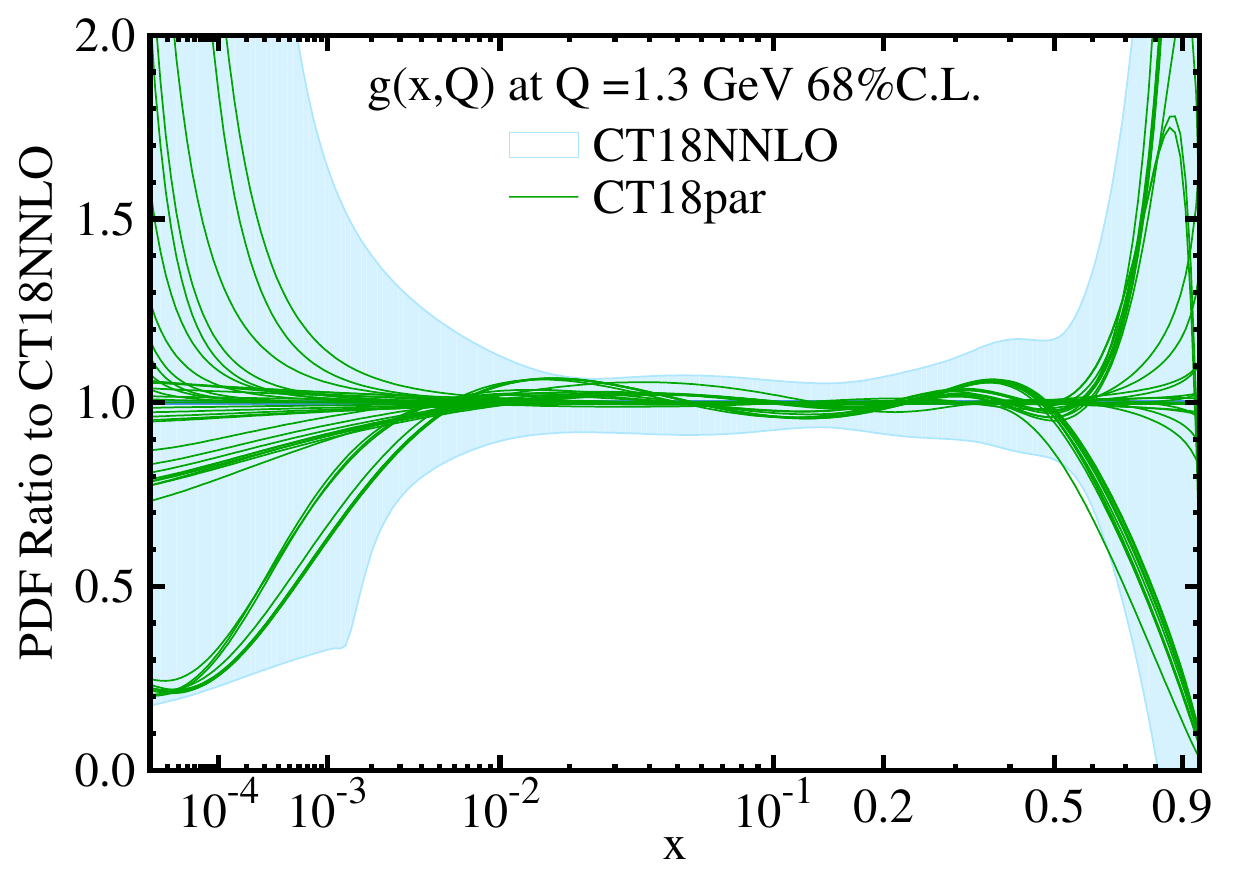}
        \includegraphics[width=0.51\textwidth]{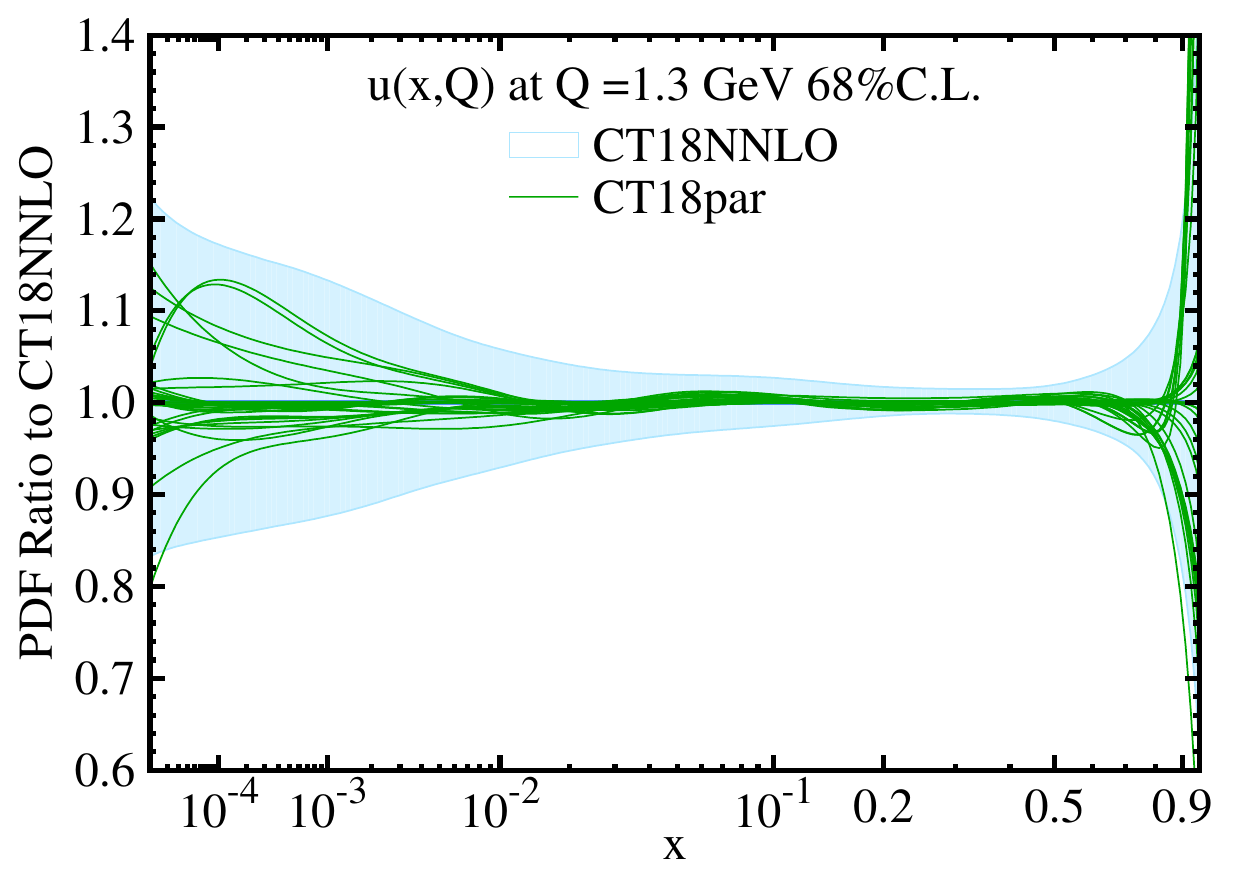}\\
        \hspace*{-0.5cm}\includegraphics[width=0.51\textwidth]{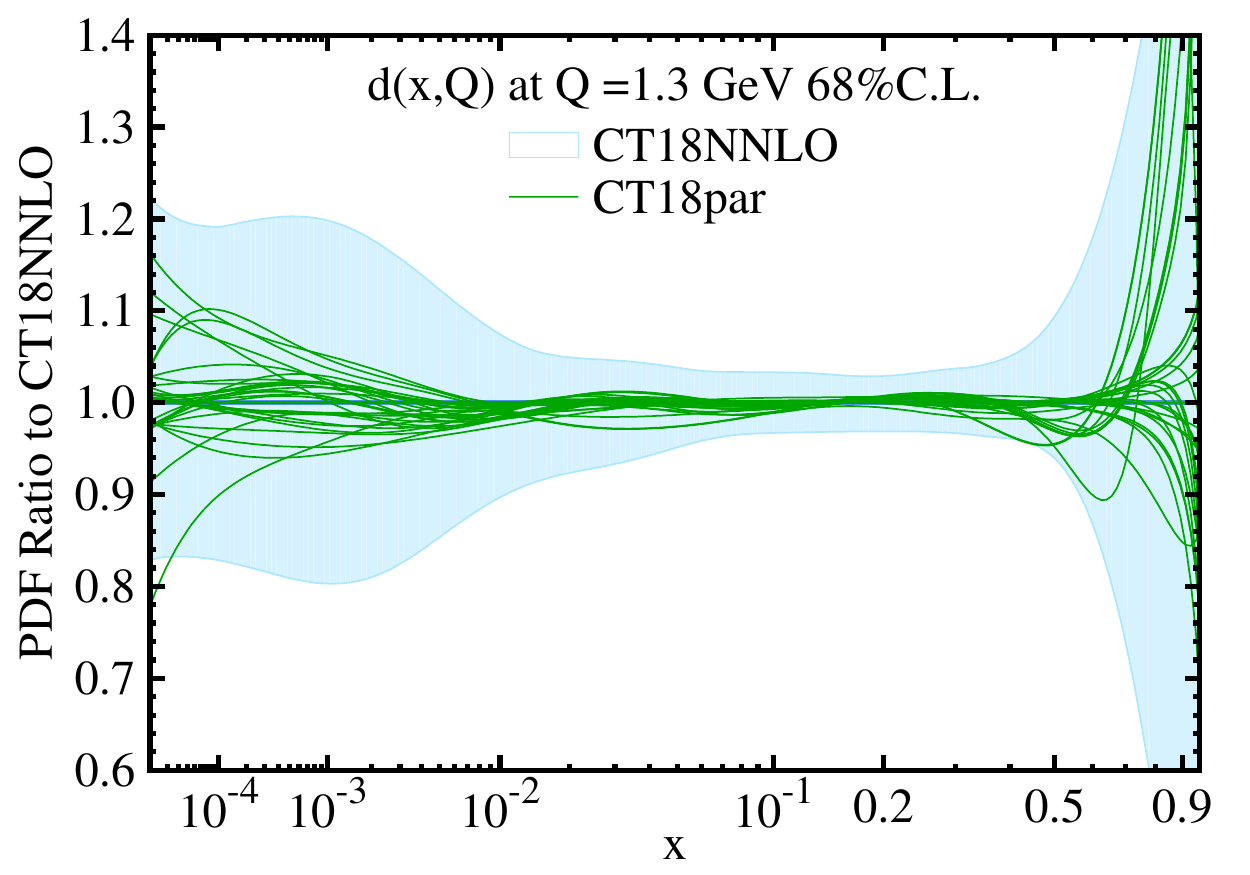}
        \includegraphics[width=0.51\textwidth]{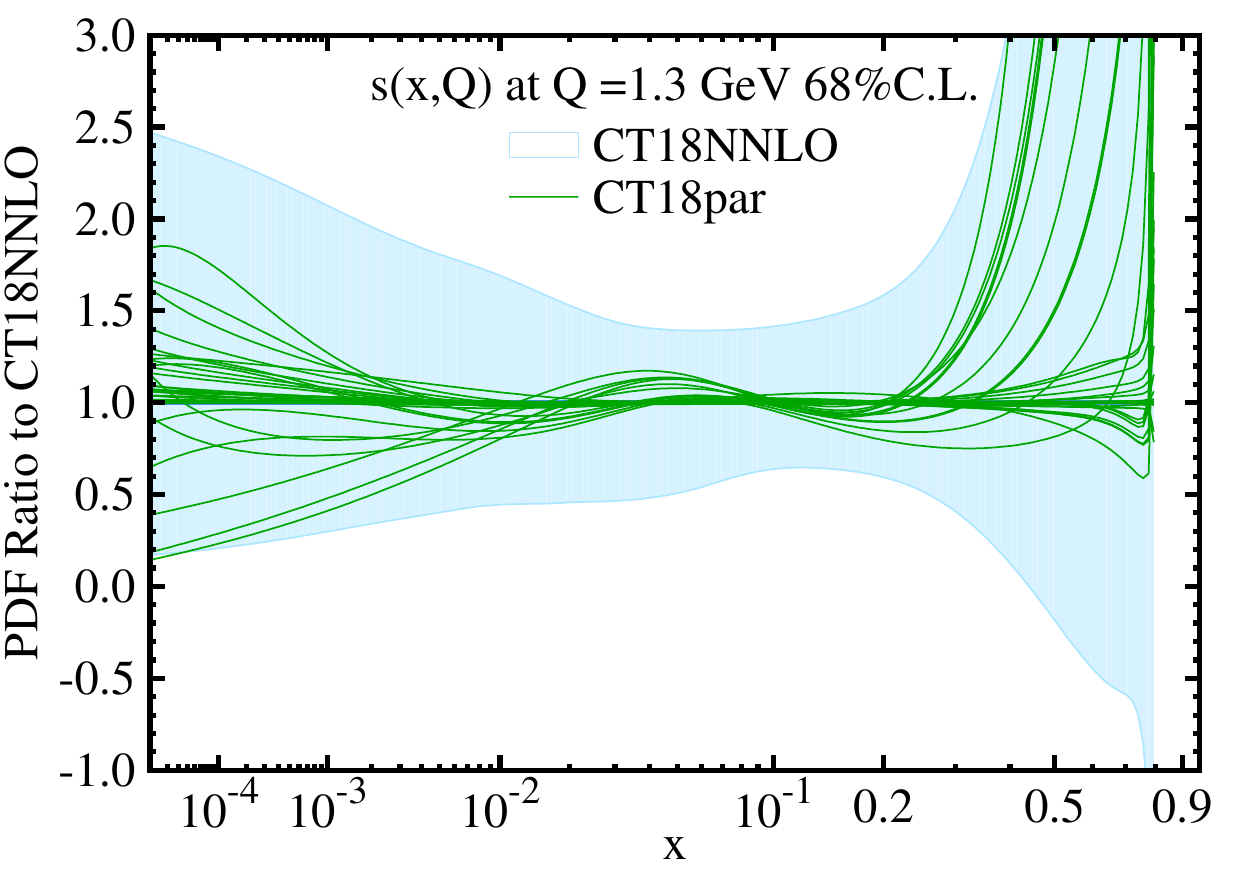}\\
        \hspace*{-0.5cm}\includegraphics[width=0.51\textwidth]{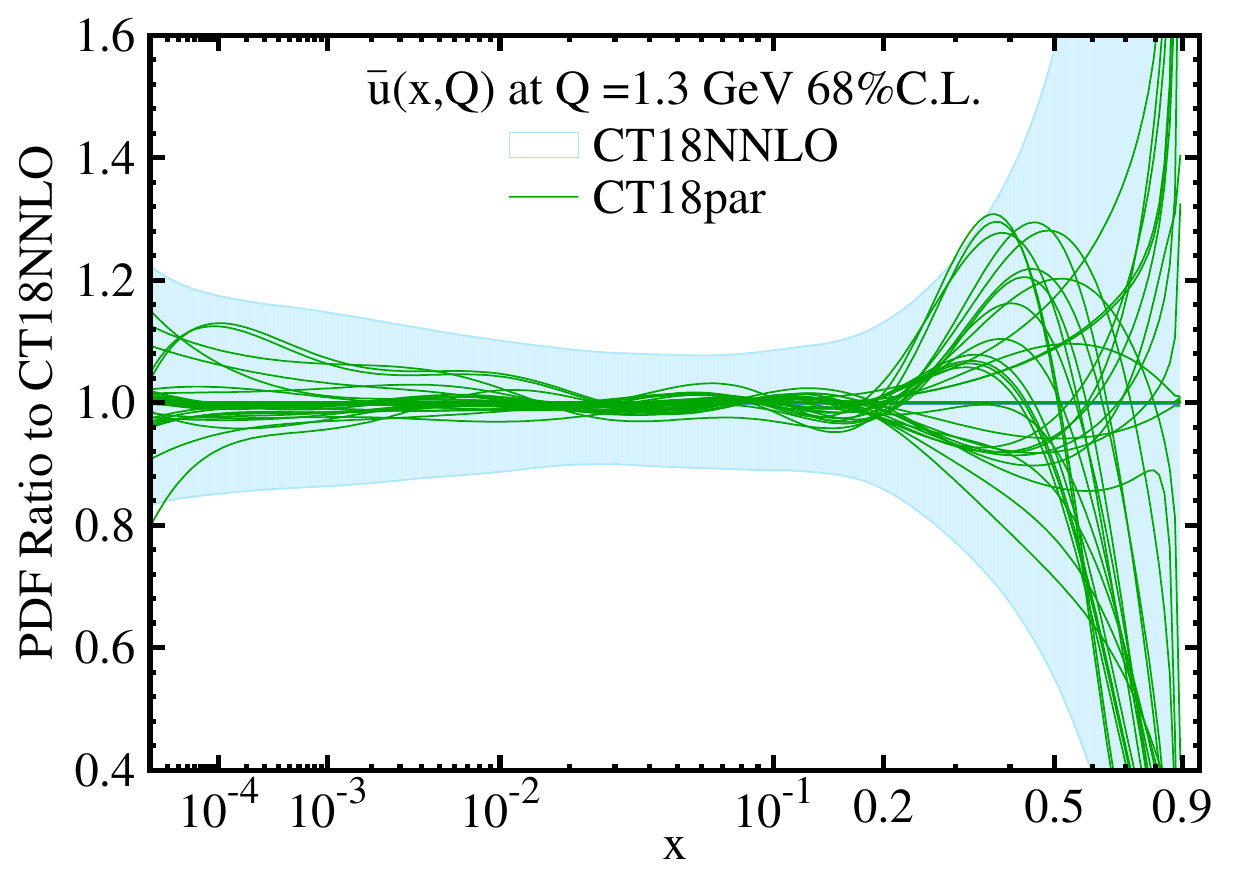} 
        \includegraphics[width=0.51\textwidth]{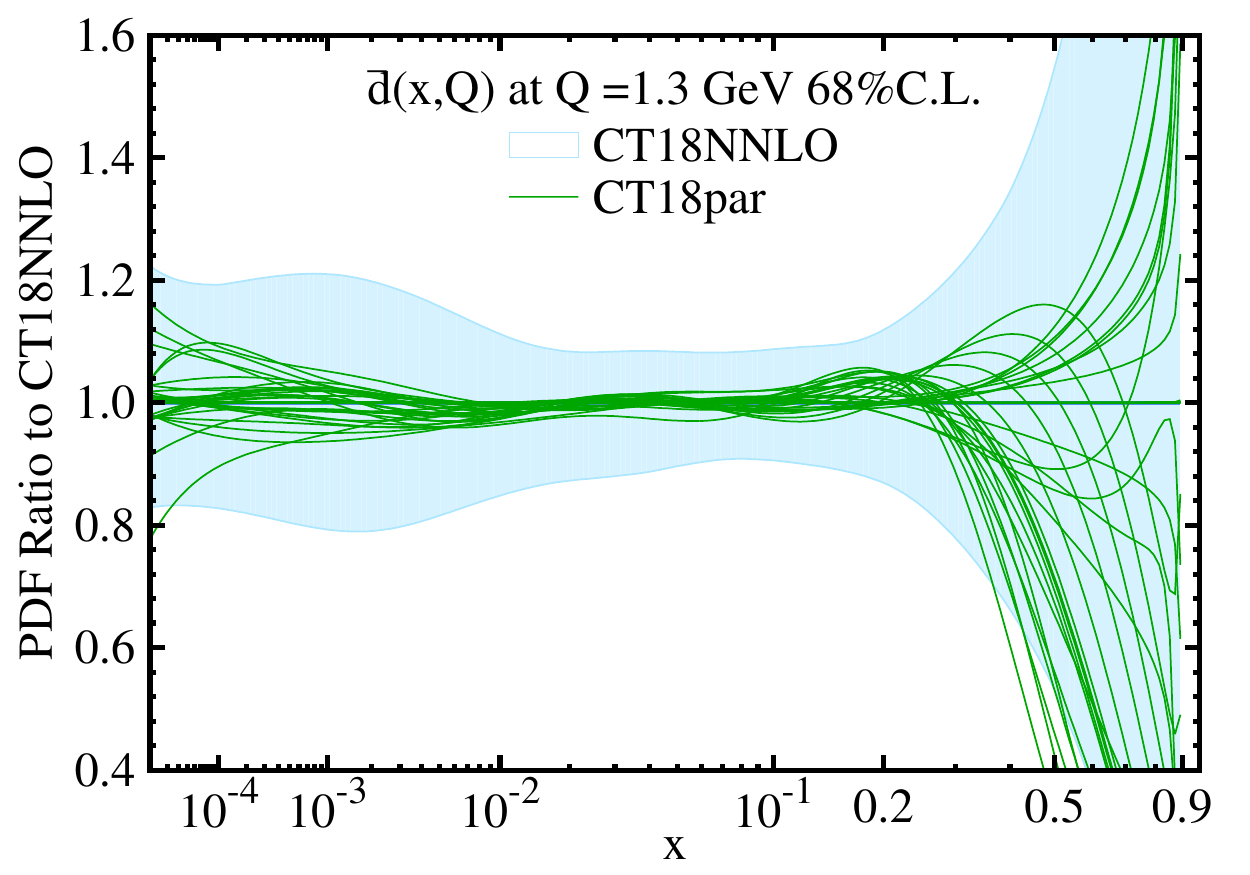}
        \caption{
	To understand the parametrization dependence in the CT18 fit, we performed $\mathcal{O}(250)$ candidate
	PDF analyses using a wide range of alternative functional forms for $f_a(x,Q_0)$. The green curves
	in the panels above illustrate the spread of central fits achieved with the various candidate fits,
	evaluated as ratios with respect to the central CT18 fit.
		}
\label{fig:params}
\end{figure}

\subsubsection{Top-quark pair production
\label{sec:TheoryTop}
}
Theory predictions for top-quark pair production differential distributions at the LHC 8 TeV are implemented at NNLO in QCD using
\texttt{fastNNLO} tables~\cite{Czakon:2017dip,fastnnlo:grids}. 
In the CT18 global fit, the top-quark mass has been set to $m_t^{\rm pole}=173.3$ GeV. Motivated by~\cite{Czakon:2016dgf}, we chose the default central scale
$\mu_{F,R}\equiv \mu =1/2\sqrt{m_t^2 + p_{T,t}^2}$ for the top-quark $p_T$ spectrum, while the rest of the distributions are obtained with
$\mu=1/4\left(\sqrt{m_t^2 + p_{T,t}^2} + \sqrt{m_t^2 + p_{T,\bar{t}}^2}\right)$.
The impact of the EW corrections on the theory predictions for $t \bar{t}$ differential distributions has been studied in~\cite{Czakon:2017wor} 
where the difference between the additive and
multiplicative approaches for combining QCD and EW corrections is also investigated. EW $K$-factors 
from an analytic fit for the QCD $\times$ EW/QCD contributions are  available~\cite{EW:kfac}. 
The CT18 global analysis does not include EW corrections in $\bar{t} t$ production. Their impact on the fitted PDFs is expected to be small in the kinematic range of the differential distributions currently considered.   

The impact of the EW corrections on the CT18 theory predictions at CMS 13 TeV is illustrated in Sec.~\ref{sec:StandardCandles}. In this case, the CT18 theory predictions include EW corrections evaluated using the multiplicative approach of~\cite{Czakon:2017wor}, and the recommended value of $m_t^{\rm pole}=172.5$ GeV has been used to compare theory and the CMS data (without fitting the data).

\subsection{Parametrization forms, systematic errors, and final PDF uncertainty} 
\label{sec:Paramstudies}

\subsubsection{Nonperturbative parametrization forms\label{sec:ParamForms}} 
An important source of the uncertainty in the CTEQ PDF analysis is associated with the choice of the parametric form for the fitted distributions at the lower boundary of QCD evolution, $f_a(x,\Q\! =\! Q_0)$. There is limited guidance
from theory as to the most appropriate PDF parametrizations, and it is
favorable to guarantee a maximal level of parametric flexibility without
overfitting experimental data~\cite{Kovarik:2019xvh}. In App.~\ref{sec:AppendixParam}, we present the explicit parametrization forms used in CT18. As usual, the PDFs at higher scales $Q>Q_0$ are computed
using the Dokshitser-Gribov-Lipatov-Altarelli-Parisi (DGLAP) equations at NNLO, with splitting kernels available from Refs.~\cite{Moch:2004pa,Vogt:2004mw}.\footnote{Independent recent computations of these kernels are available in \cite{Ablinger:2014nga,Ablinger:2017tan}.}

\subsubsection{Treatment of experimental systematic errors\label{sec:SystErrors}} The experimental systematic errors are commonly published in the form of percentage tables and belong to one of two types: additive or multiplicative. An additive error is the one whose absolute value is known, for example the uncertainty of the pileup energy, or the underlying event energy. Most errors, though, are multiplicative, meaning that the error is determined as a fraction of the experimental cross section for that bin. An example is the jet energy scale uncertainty. There are a number of options as to how to evaluate both types of systematic errors. This topic was explored in depth in previous CT papers~\cite{Stump:2003yu,Nadolsky:2008zw,Lai:2010vv,Ball:2012wy,Gao:2013xoa,Dulat:2015mca,Hou:2016nqm}. 

The most natural choice may seem to simply multiply the fractional uncertainty corresponding to a particular systematic error by the experimental cross section in that bin. However, due to fluctuations, this choice can result in a bias in favor of experimental data points with lower central values, the so-called D'Agostini bias~\cite{DAgostini:1993arp,DAgostini:1999gfj}. Instead, for CT18,  as for CT14 and CT10, we use what we have termed the 'extended-$T$' option, where the systematic error for each multiplicative term is determined by multiplying the fractional uncertainty times the theoretical prediction for that bin, a quantity which is not subject to the same fluctuations. The theory, and thus the multiplicative error, is recalculated for every iteration of the global PDF fitting. In the case of inclusive jet production, we observe that the additive treatment of experimental systematic errors produces the gluon PDF that is substantially softer at $x>0.1$, the pattern that was already observed in the CT10 NNLO analysis (cf. Figs. 18 and 19 in Section VI.D of \cite{Gao:2013xoa}).

\subsubsection{The final PDF uncertainty\label{sec:FinalPDFuncertainty}} 
To estimate the parametrization dependence, we repeated the CT fits multiple times using a large number (more than 250) of initial parametrization forms which have comparable numbers of fitting parameters. Some candidate fits are based on the functional forms like the ones shown in App.~\ref{sec:AppendixParam}, but with alternative choices for the orders of Bernstein polynomials, relations between the $x$ and $y$ variables, and relations between the Bernstein coefficients $a_i$. 
In many of these 250 fits, we increased the number of free parameters in Bernstein polynomials for some flavors up to 6 or 7, or we used a different form of the variable $y$ defined after Eq.~(\ref{eq:fiQ0})  and before Eq.~(\ref{eq:s_param}), or we did not require $a_2$ to be the same for $u_v$ and $d_v$, and similarly sometimes we relaxed the equality relations on $a_1$ for $\bar u$, $\bar d$, $\bar s$.

In addition, we repeated some fits by randomly changing the treatment of some experimental systematic errors from multiplicative to additive. Yet another class of candidate fits is obtained by choosing alternative QCD scales in sensitive experiments such as high-$p_T$ $Z$ boson production, or alternative codes to compute the NNLO $K$ factors, cf. Appendix~\ref{sec:Appendix4xFitter}. 
The final PDFs are obtained using a fixed parametrization form and systematic parameter settings, but the uncertainty is computed according to the two-tier convention adopted in Refs.~\cite{Lai:2010vv,Gao:2013xoa} so as to cover the bulk of the solutions obtained with the alternative choices.  The results of this study are illustrated in Fig.~\ref{fig:params}, showing a selection of central fits (green solid curves) for a range of alternative fitting forms and multiplicative/additive choices for systematic errors in the LHC and Tevatron jet production, superposed within
the uncertainty band (at the 68\% confidence level) for the published version of CT18. 

As we increased the number of free PDF parameters, a mild improvement (up to several tens of units) in the global $\chi^2$ or individual $S_E$ values was typically found, so long as $\lesssim\! 30$ free parameters were fitted. With more than about 30 parameters, the fits tend to destabilize, as expanded parametrizations attempt to describe statistical noise. The final PDFs are based on the parametrizations with a total of 29 free parameters. For each of the four fits, we provide twice as many Hessian error PDFs to evaluate the PDF uncertainties according to the CTEQ6 master formulas \cite{Pumplin:2002vw}.

\section{The CT18 output: PDFs, QCD parameters, parton luminosities, moments}
\label{sec:OverviewCT18} 
In this section, we review the behavior of CT18 PDFs and corresponding parton luminosities, Mellin moments, and parameters of the QCD Lagrangian. Given the large number of figures, for CT18Z fits, this section shows only the most critical comparisons. The rest of counterpart illustrations for CT18Z PDFs are presented in Appendices.~\ref{sec:ATL7ZWchi2} and \ref{sec:LMCT18Z}.

\subsection{Parton distributions as functions of $x$ and $Q$}
\subsubsection{PDFs for individual flavors}

Figure \ref{fig:ct18pdf} shows an overview of the CT18 parton
distribution functions, for $Q = 2$ and $100$ GeV.
The function $x f(x,Q)$ is plotted versus $x$, for flavors $u,
\overline{u}, d, \overline{d}, s = \overline{s}$, and $g$.
We assume $s(x,Q_0)=\bar s(x,Q_0)$, since their difference is
consistent with zero and has large uncertainty \cite{Lai:2007dq}.
The plots show the central fit to the global data listed
in Tables~\ref{tab:EXP_1} and \ref{tab:EXP_2}, corresponding to the
lowest total $\chi^2$ for our choice of PDF parametrizations. These
are displayed with error bands representing the PDF uncertainty at
the 90\% confidence level (C.L.).
\begin{figure}[p]
	\center
	\includegraphics[width=0.49\textwidth]{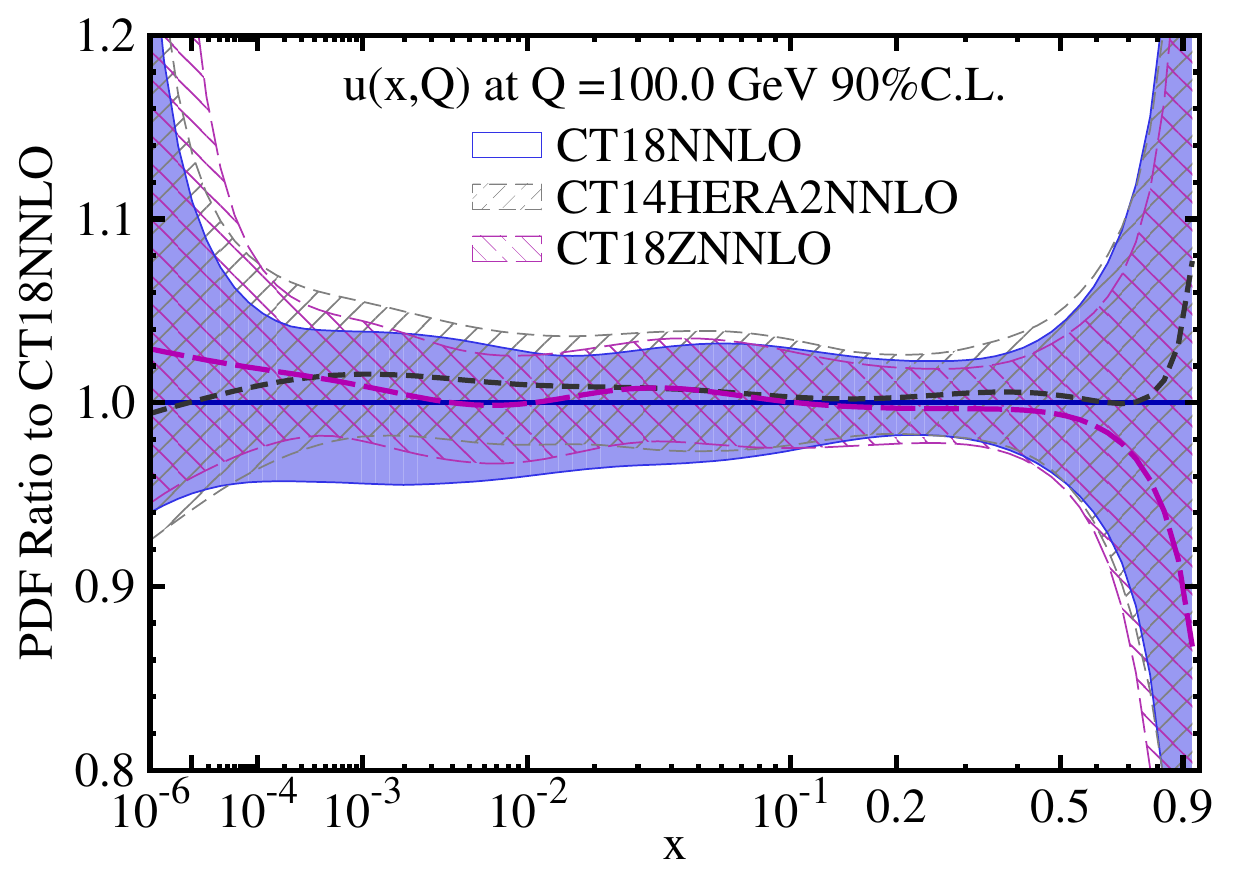}
	\includegraphics[width=0.49\textwidth]{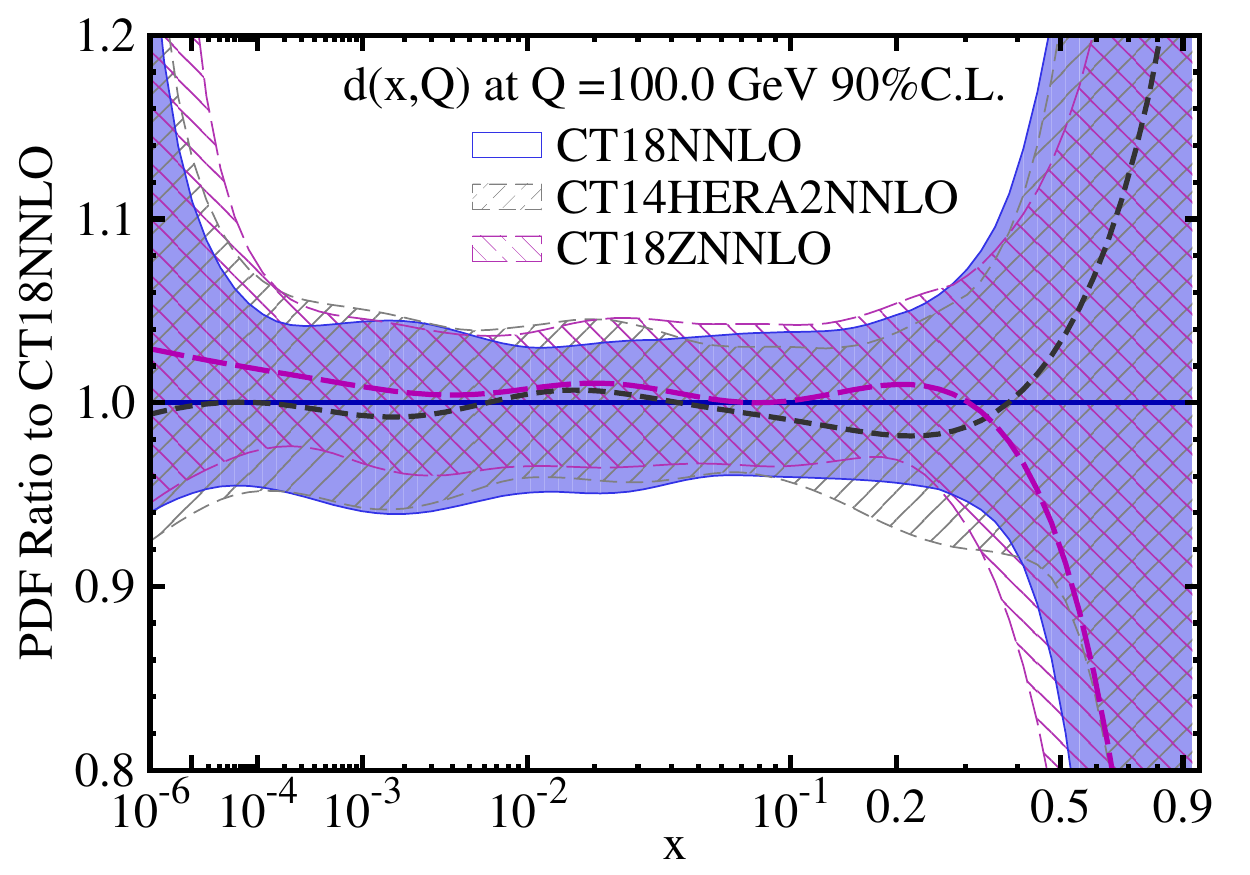}
	\includegraphics[width=0.49\textwidth]{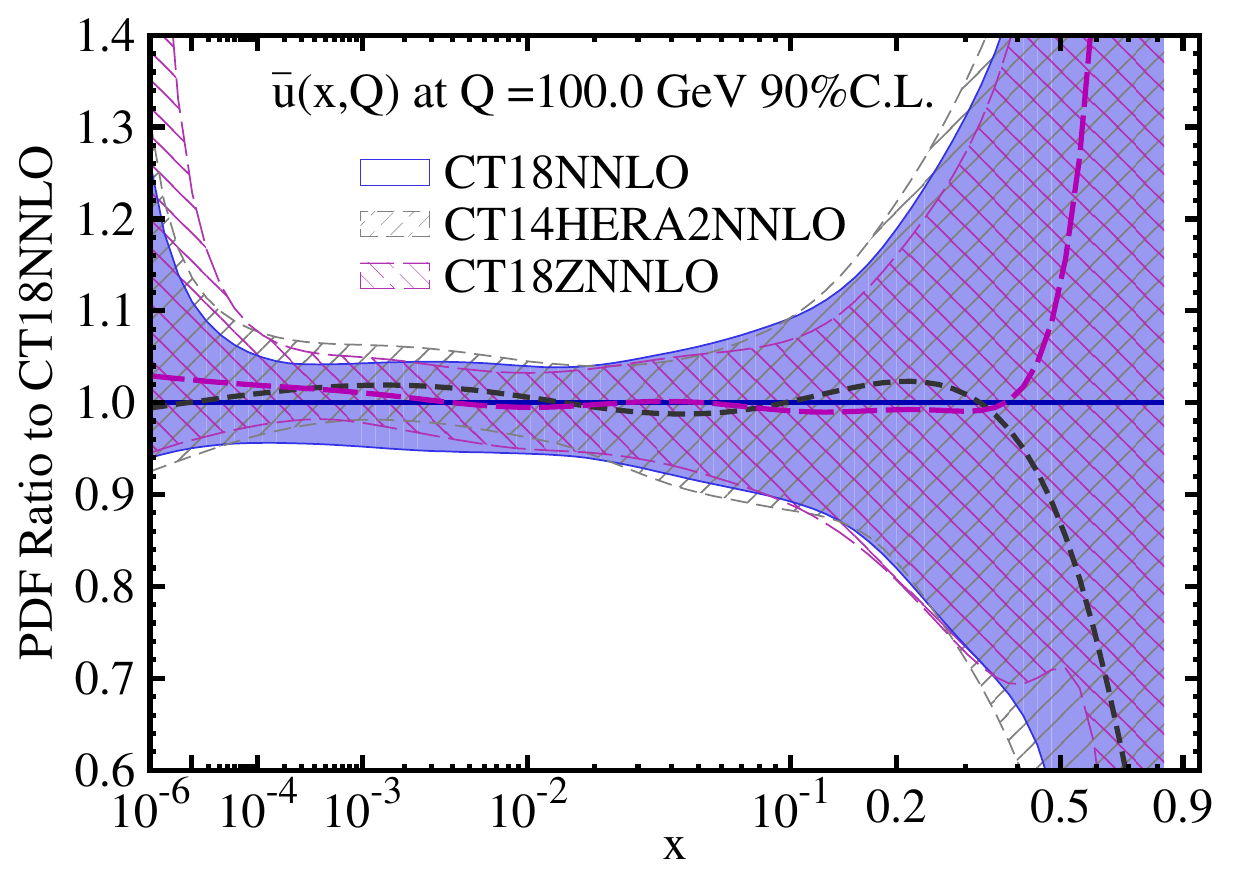}
	\includegraphics[width=0.49\textwidth]{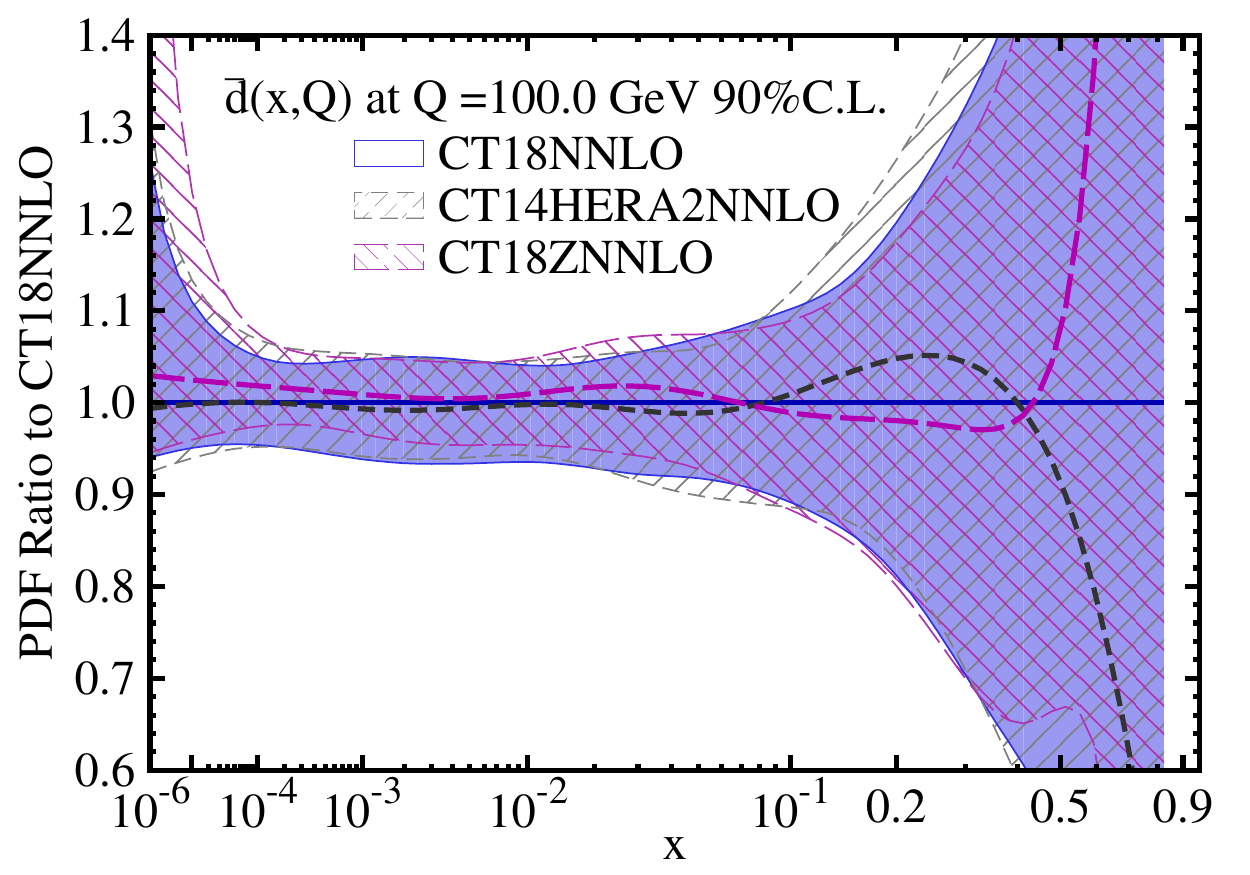}
	\includegraphics[width=0.49\textwidth]{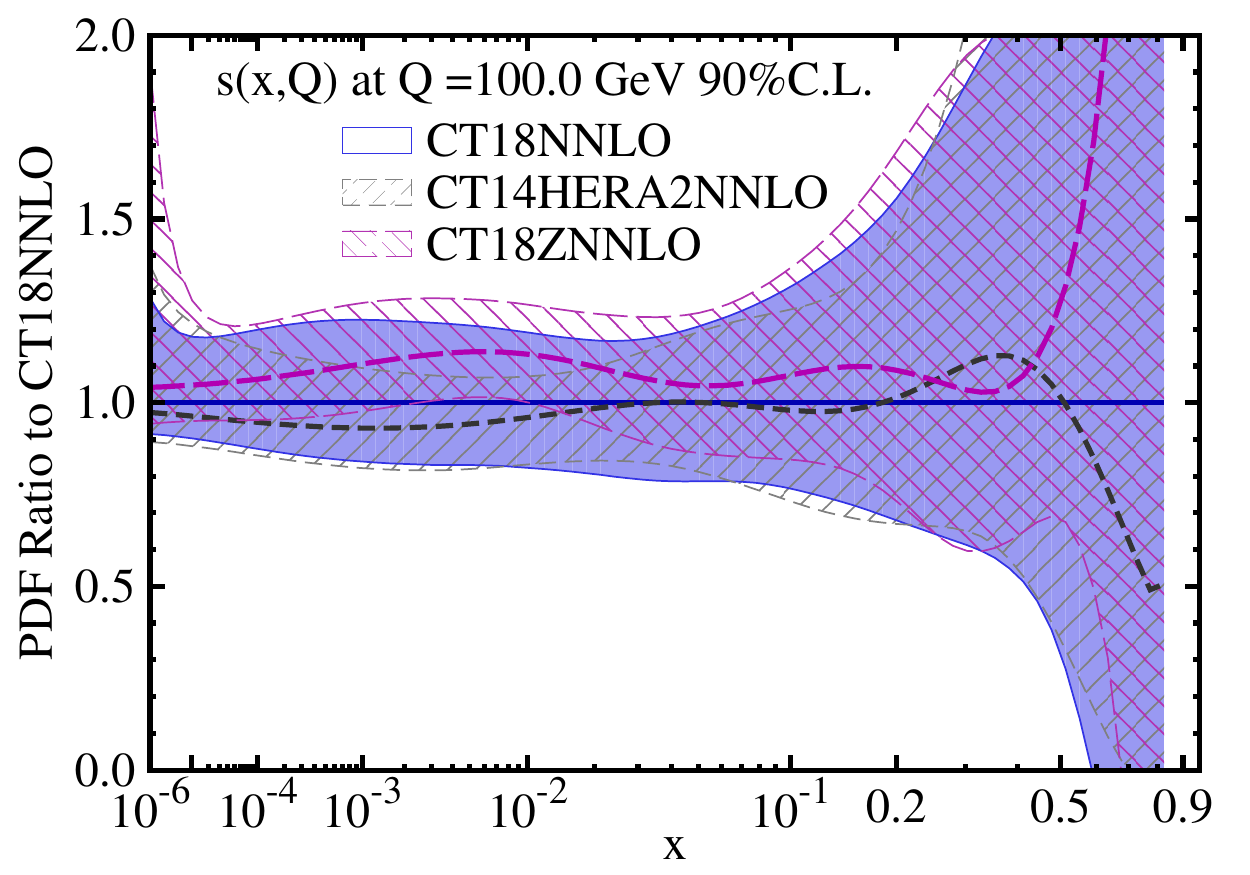}
	\includegraphics[width=0.49\textwidth]{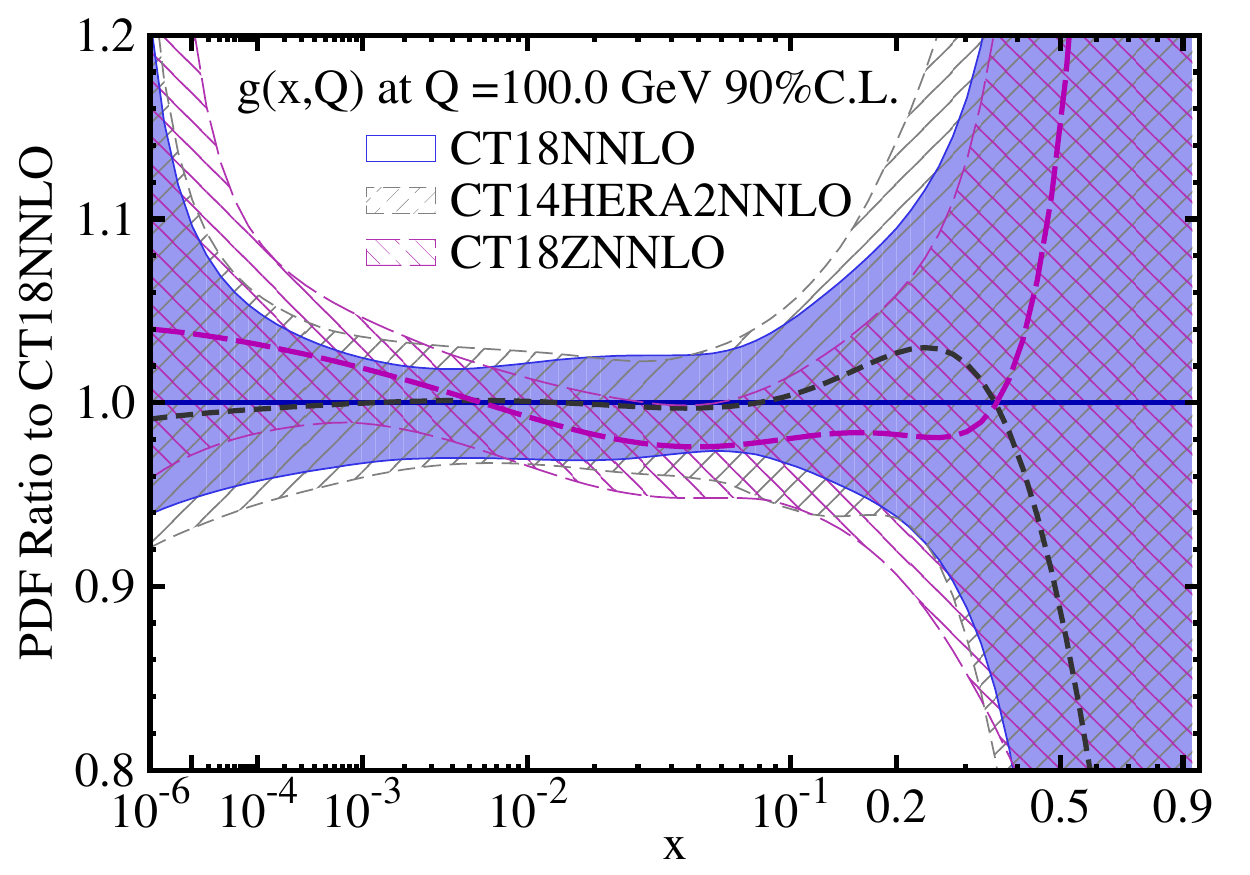}
	\caption{A comparison of 90\% C.L. PDF uncertainties from CT18 (violet solid), \CTHERAII (gray short-dashed), and CT18Z (magenta long-dashed) NNLO ensembles at $Q=100$ GeV. The uncertainty bands are normalized to the central CT18 NNLO PDFs.
		\label{fig:PDFbands1}}
\end{figure}

\begin{figure}[p]
	\center
	\includegraphics[width=0.49\textwidth]{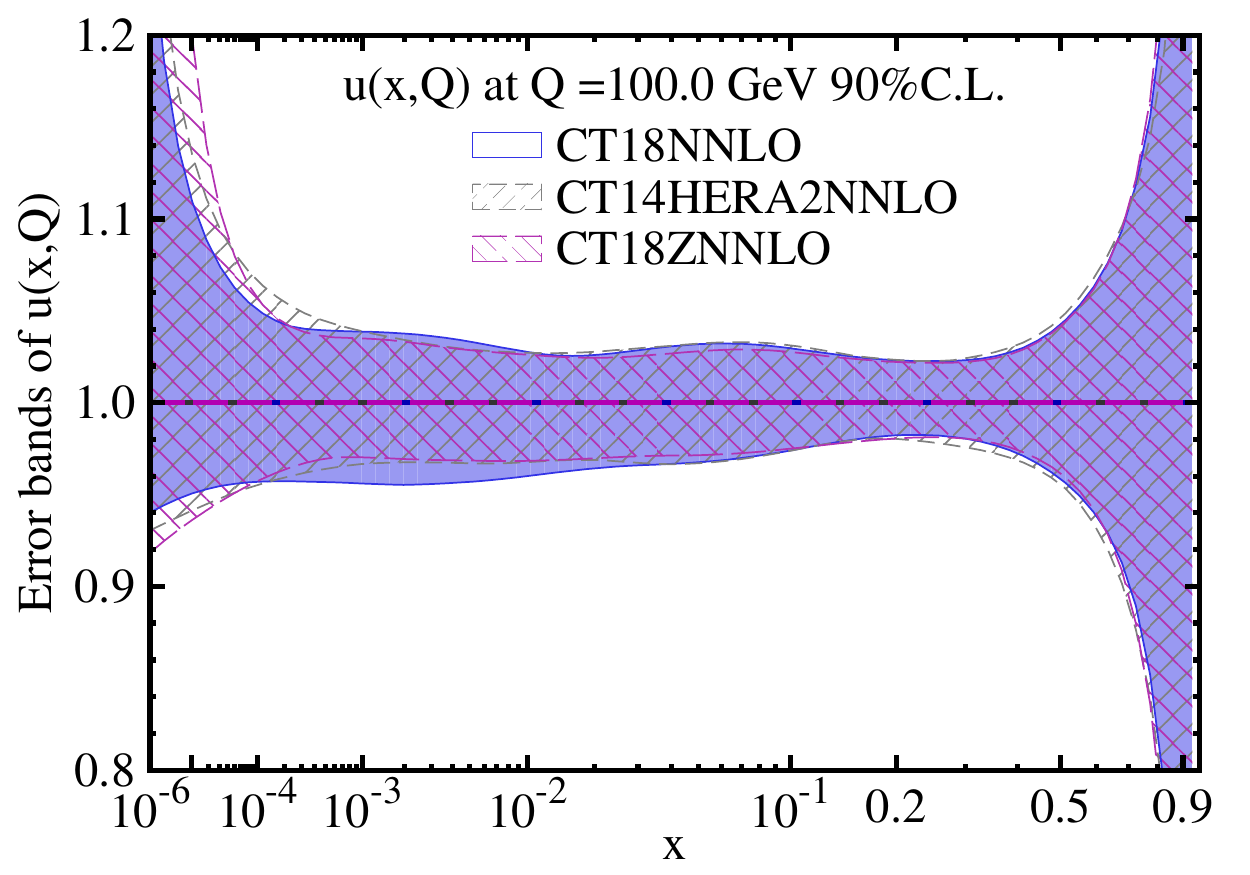}
	\includegraphics[width=0.49\textwidth]{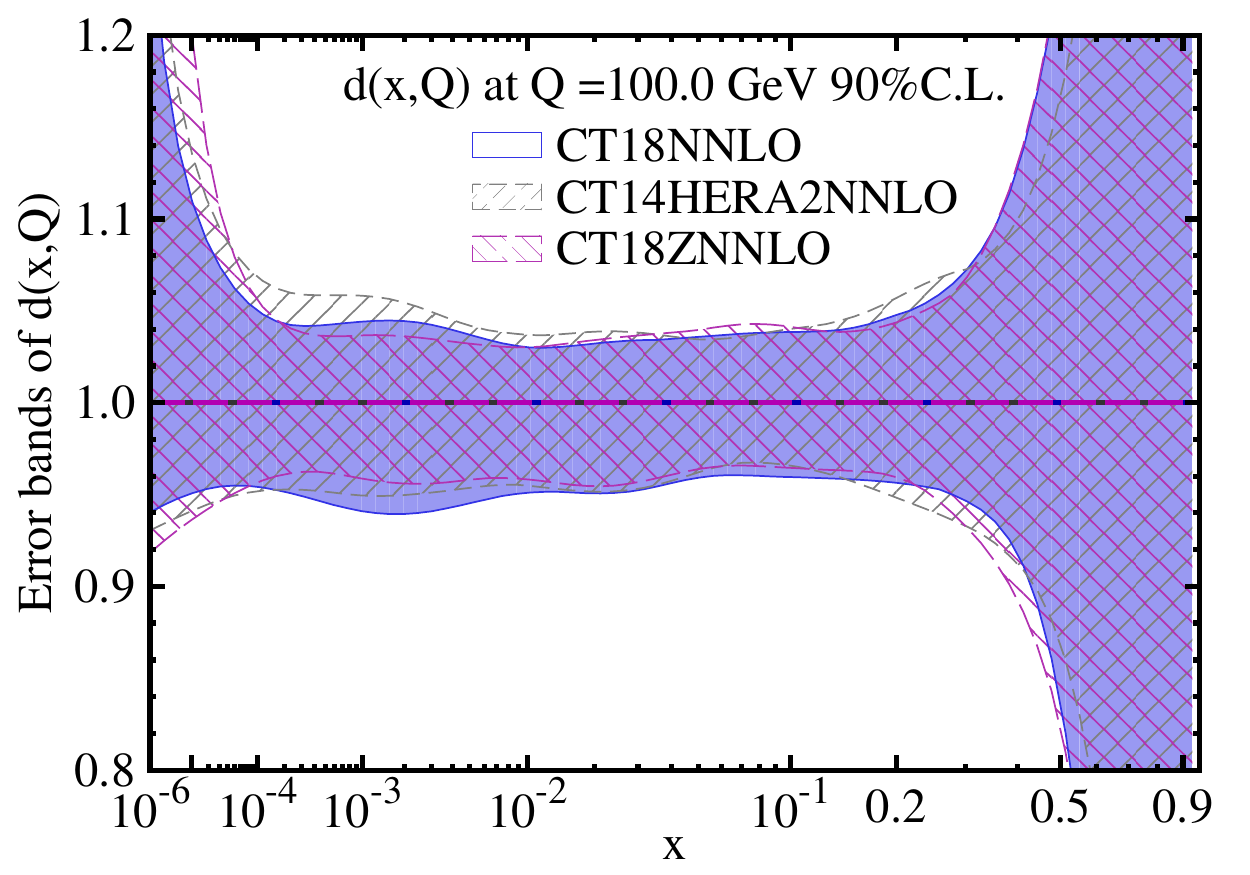}
	\includegraphics[width=0.49\textwidth]{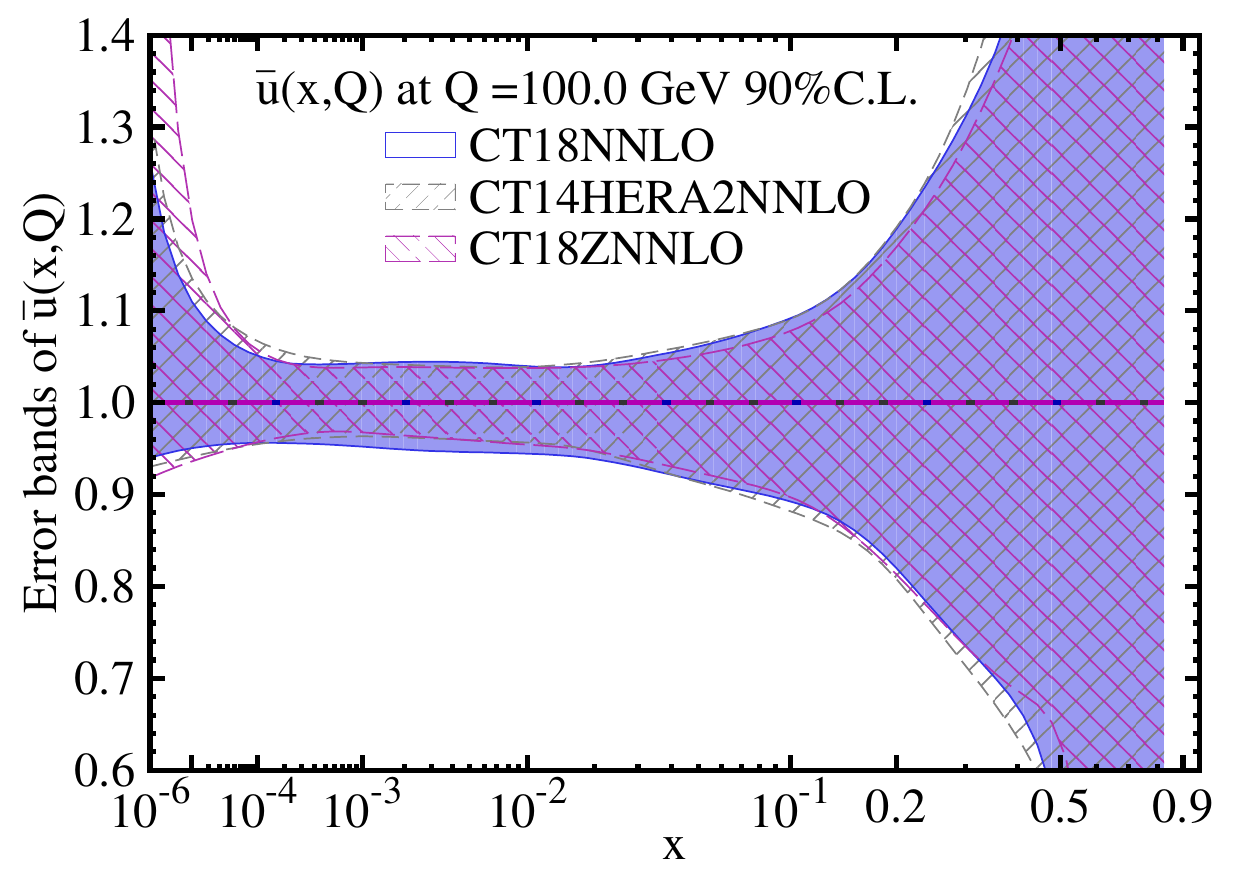}
	\includegraphics[width=0.49\textwidth]{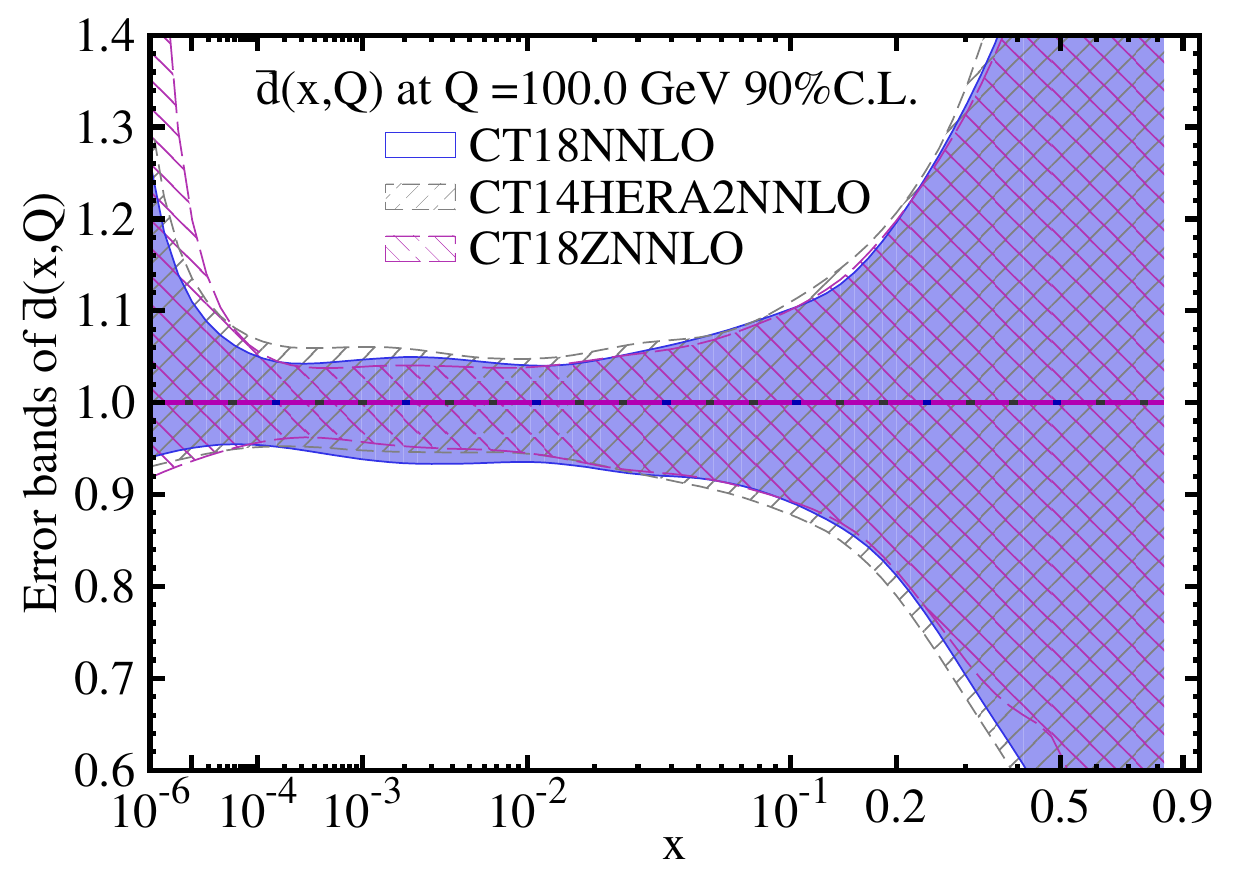}
	\includegraphics[width=0.49\textwidth]{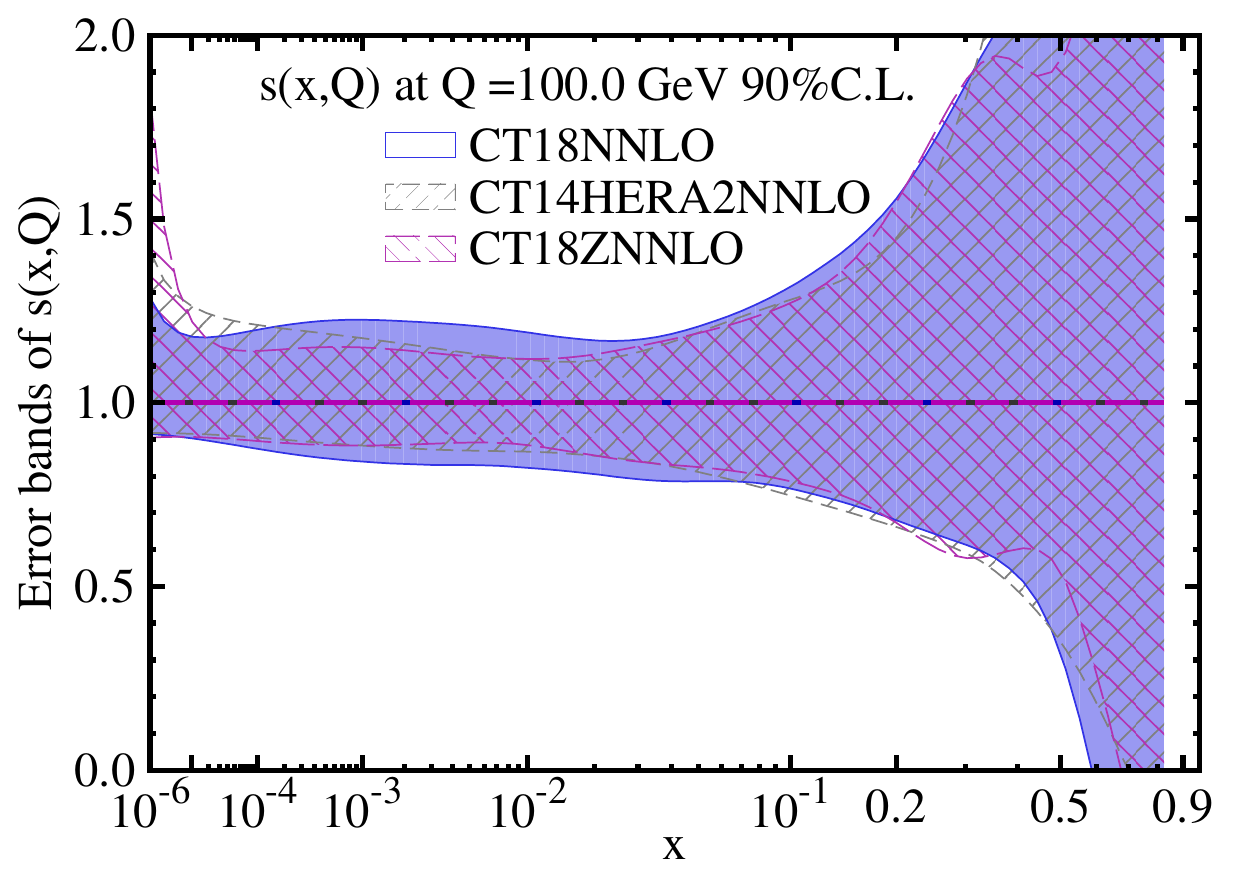}
	\includegraphics[width=0.49\textwidth]{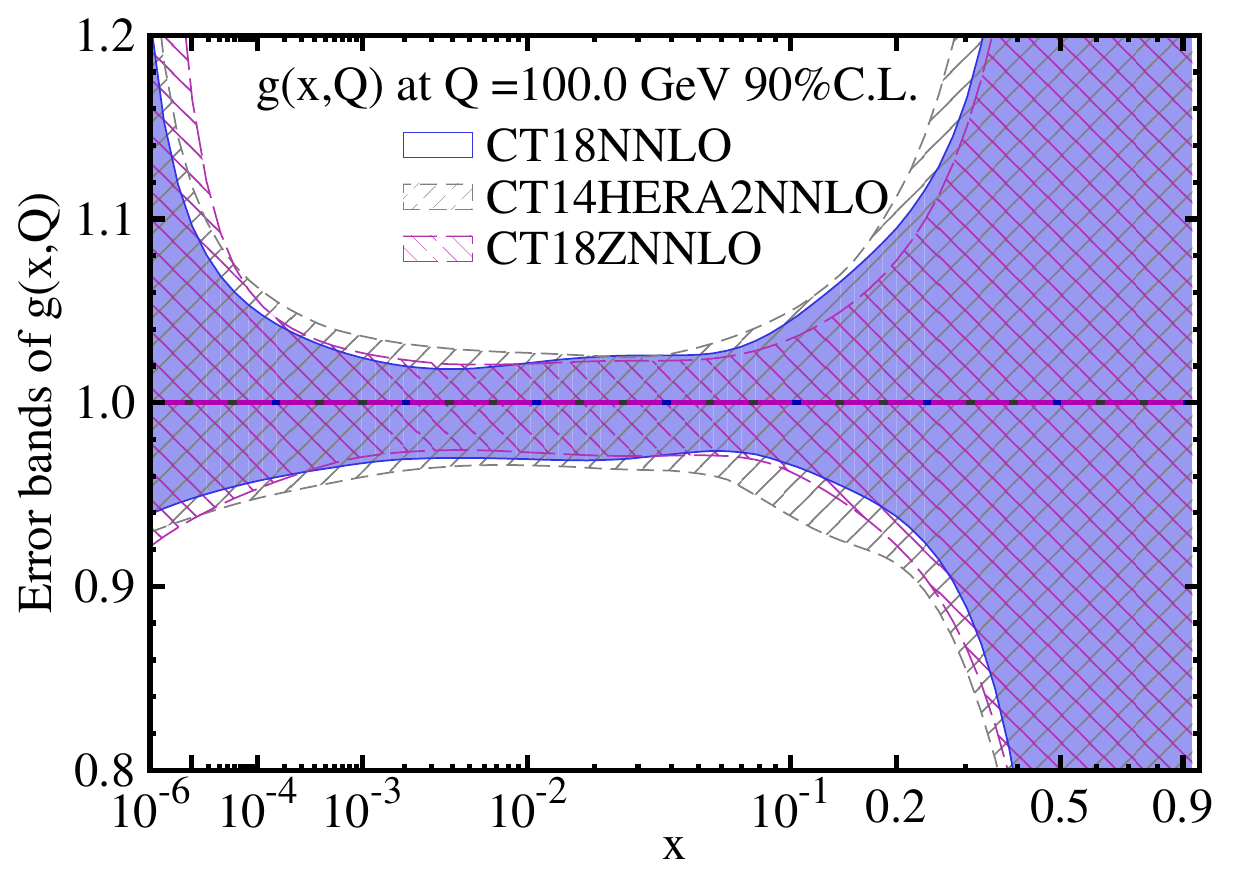}
	\caption{Like Fig.~\ref{fig:PDFbands1}, but with each error band normalized to its respective central PDF for the purpose of directly comparing the PDF uncertainties in CT18 NNLO with CT18Z and CT14$_\mathrm{HERAII}$ NNLO.
		\label{fig:PDFbands2}}
\end{figure}

The relative changes from CT14$_\mathrm{HERAII}$ NNLO to CT18 NNLO PDFs are
best visualized by comparing their associated PDF uncertainties. Fig.~\ref{fig:PDFbands1}
compares the PDF error bands at 90\% C.L. for the key flavors, with each band
normalized to the corresponding best-fit CT18 NNLO PDF, represented by the
solid violet line/bands. The long-dashed magenta and short-dashed gray curves/bands correspond
to the CT18Z and CT14$_\mathrm{HERAII}$ NNLO PDFs at $Q=100$ GeV, respectively. Figure~\ref{fig:PDFbands2} shows the same error bands normalized to their respective central fits to facilitate comparison of their PDF uncertainties. 

We make a number of observations for the NNLO PDFs. The CT18 $u$ PDF becomes slightly
smaller, compared to \CTHERAII, at almost all $x$ values, with the largest decrease at $x \sim 10^{-3}$.
The $d$ PDF has increased at $x \sim 10^{-3}$ and $x \sim 0.2$, while it slightly decreased at $ x\sim 0.01$. 
The $\bar u$ and $\bar d$ distributions are both smaller at $x \sim 0.3$ and larger at $x \sim 0.05$, 
though the decrease in $\bar d$ is larger. 
Furthermore, except for the $d$ PDF at $ x\sim 0.2$, the error bands of $u$, $d$, $\bar u$ and $\bar d$ 
are about the same as \CTHERAII.
The central strangeness ($s$) PDF has increased for $ x < 0.01$ and decreased for $0.2< x < 0.5$, where  
the strange quark PDF is essentially unconstrained in CT18, just as in \CTHERAII~NNLO.
Also, its uncertainty band is slightly larger than \CTHERAII~for $x > 10^{-4}$, 
as a consequence of the more flexible parametrization and the inclusion of the LHC data.
We have checked that the most important data sets that drive
the abovementioned changes 
in the quark and antiquark PDFs are the LHCb $W$ and $Z$ boson data, as listed in 
Table~\ref{tab:EXP_2} 
with Exp.~IDs=250, 245 and 246, with importance in that order. 
After including the LHCb $W$ and $Z$ boson data, the addition of
CMS 8 TeV $W$ charge-asymmetry data (Exp.~ID=249) leads only to very mild changes in the CT18 PDFs. The central gluon PDF has
decreased in CT18 at $x\approx 0.3$, with a smaller error band at $x \sim 0.1$ and below. 
The decrease of $g$ PDF for $0.1 < x < 0.4$ is caused by the inclusion of CMS and ATLAS jet data (with
Exp.~IDs=545, 543 and 544, in that order) and ATLAS 8 TeV $Z$ boson transverse momentum ($p_T$) data (Exp.~ID=253). 
With the LHC jet data sets already included, adding the ATLAS and CMS top-quark pair data (Exp.~IDs=580 and 573) into the fit does not change the PDFs by a statistically significant amount. 

\subsubsection{Ratios of PDFs \label{sec:PDFratios}}
\begin{figure}[t]
	\center
	\includegraphics[width=0.49\textwidth]{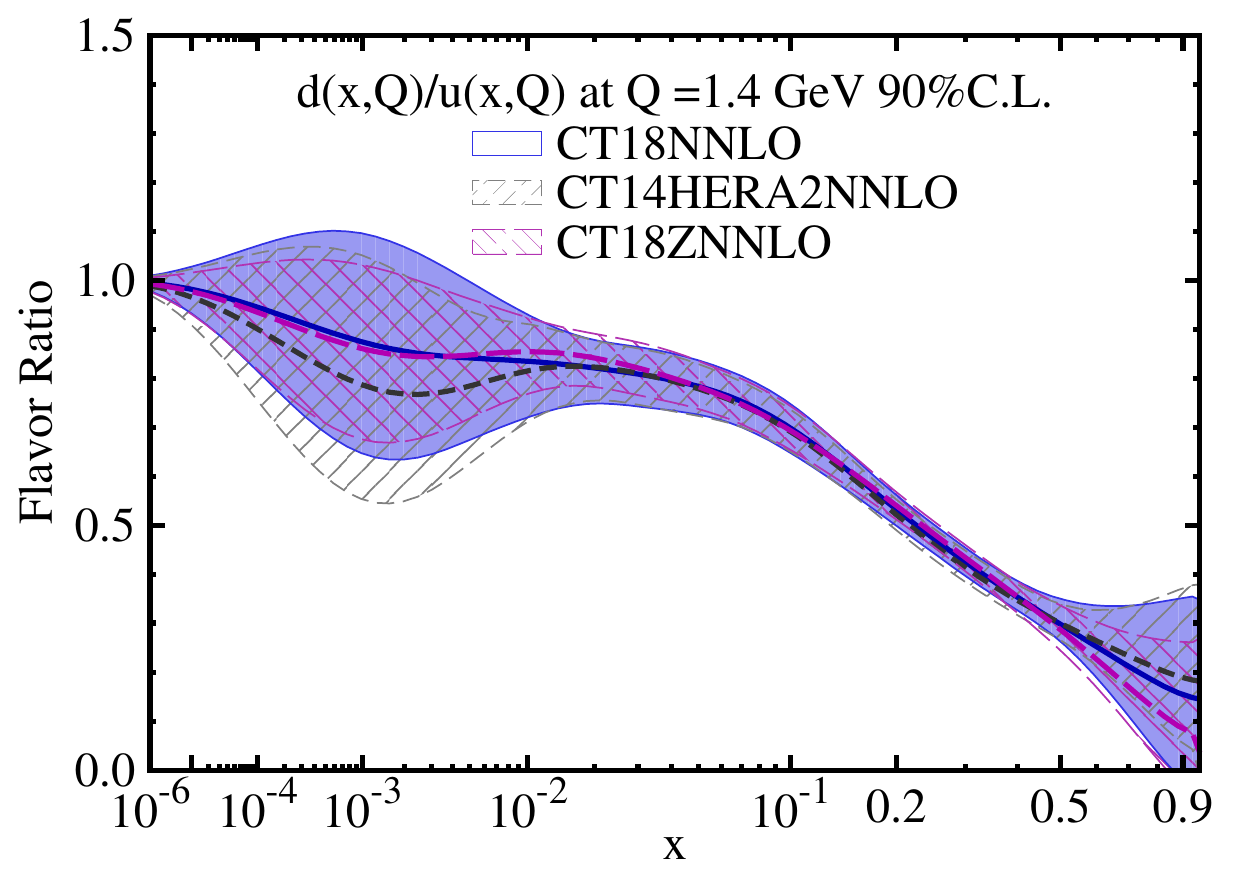}
	\includegraphics[width=0.49\textwidth]{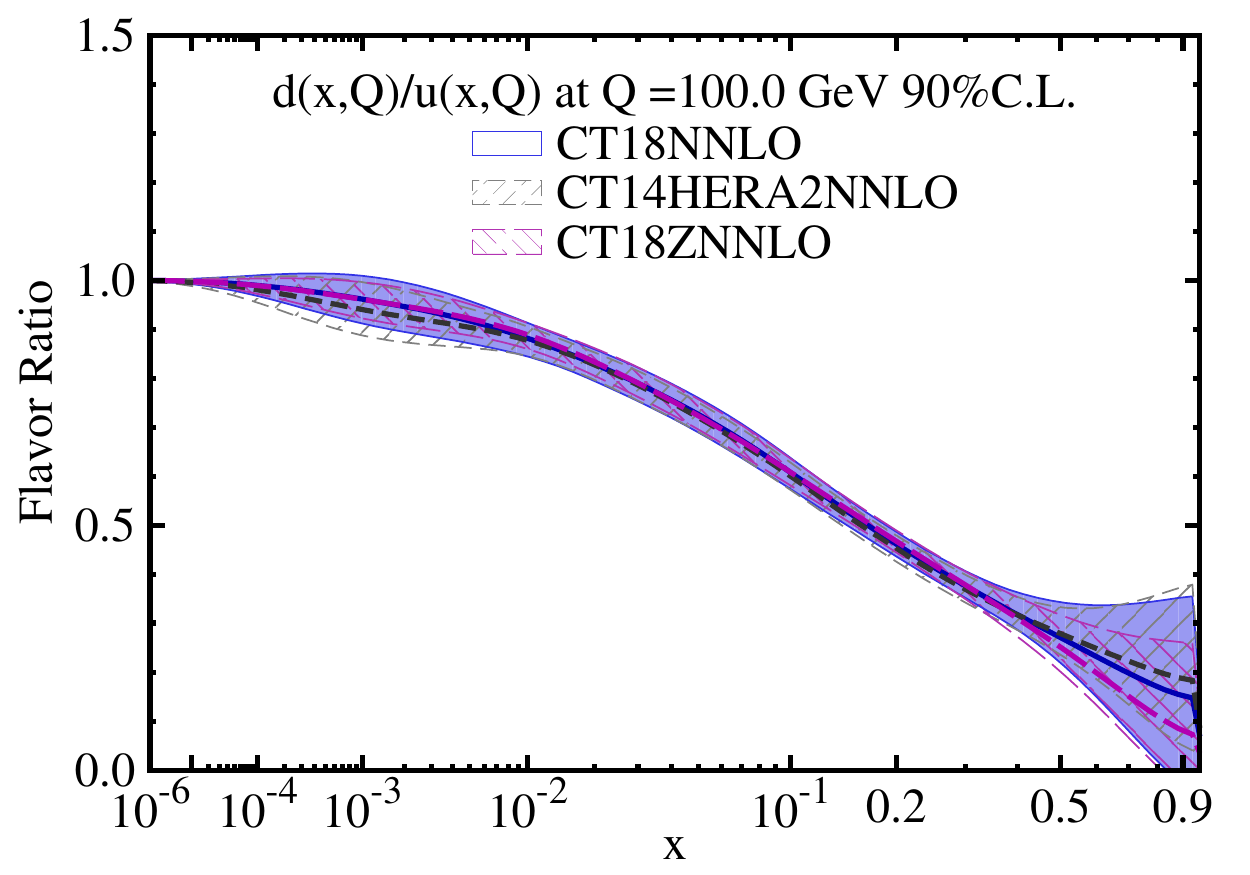}
	\includegraphics[width=0.49\textwidth]{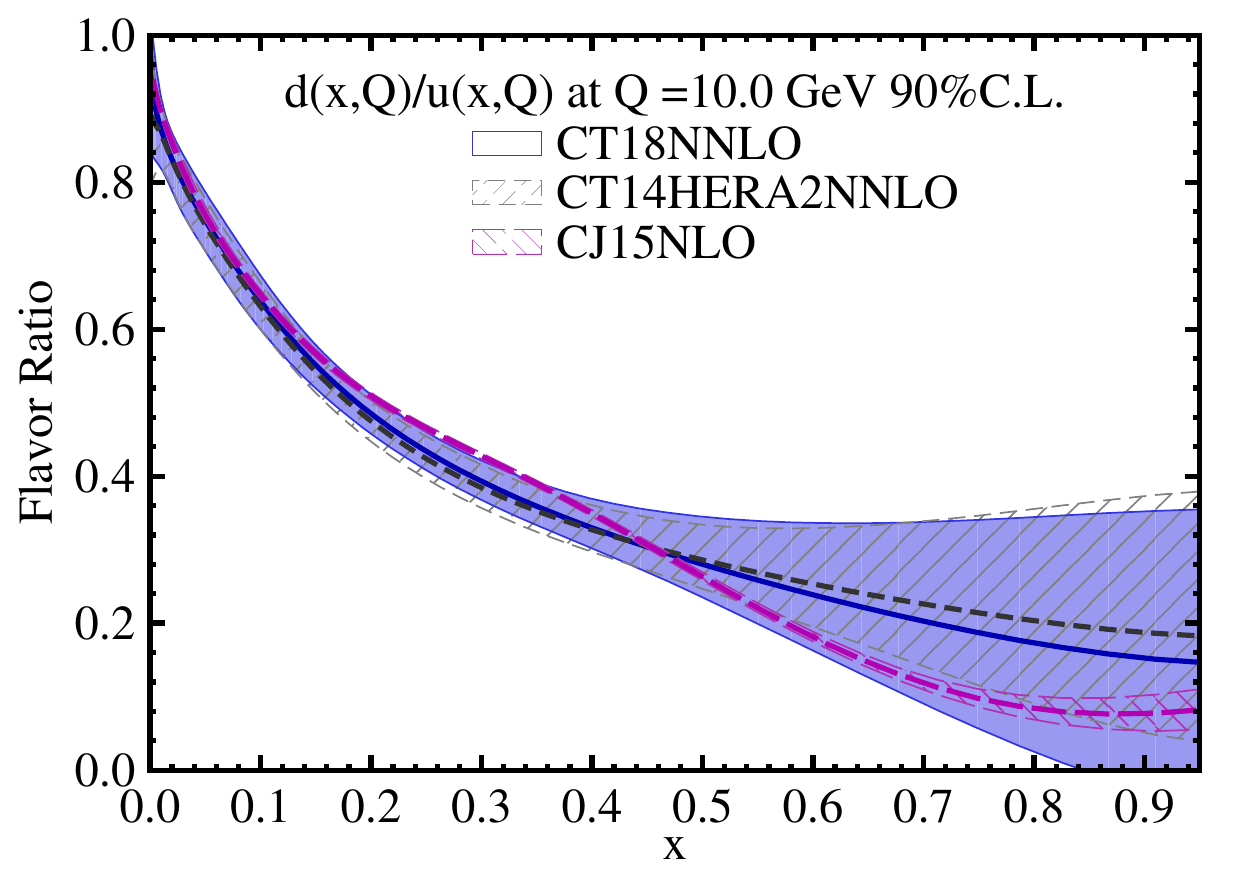}
	\includegraphics[width=0.49\textwidth]{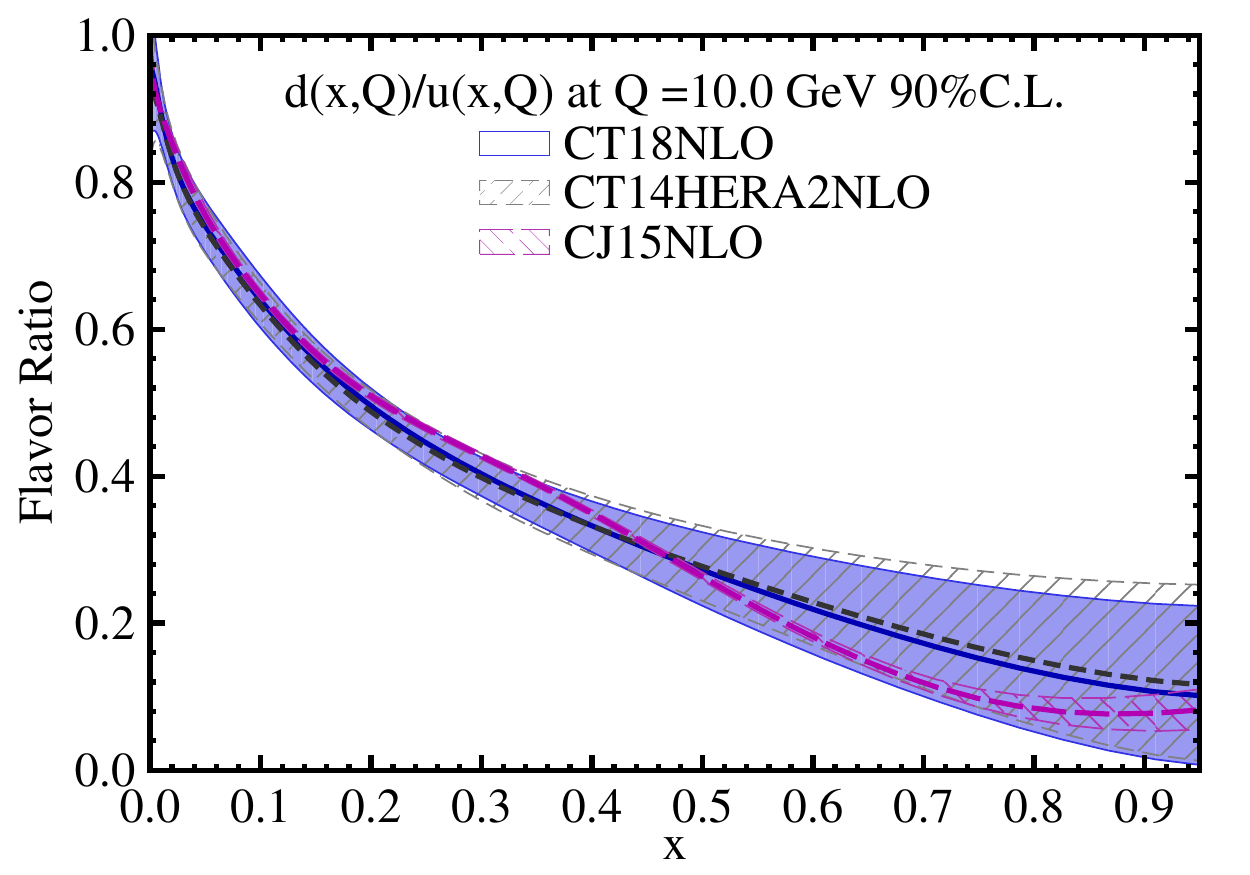}

	\caption{Top: 90\% C.L. uncertainties on the ratio
		$d(x,Q)/u(x,Q)$ for CT18, \CTHERAII, and CT18Z NNLO ensembles at $Q=1.4$ and $100$ GeV. Bottom: Same, comparing CT18 and \CTHERAII  NNLO ratios (bottom-left) and respective NLO ratios (bottom-right) to the CJ15 NLO ensemble
		at $Q=10$ GeV.
		\label{fig:DOUband}}
\end{figure}

\begin{figure}[t]
	\center
	\includegraphics[width=0.49\textwidth]{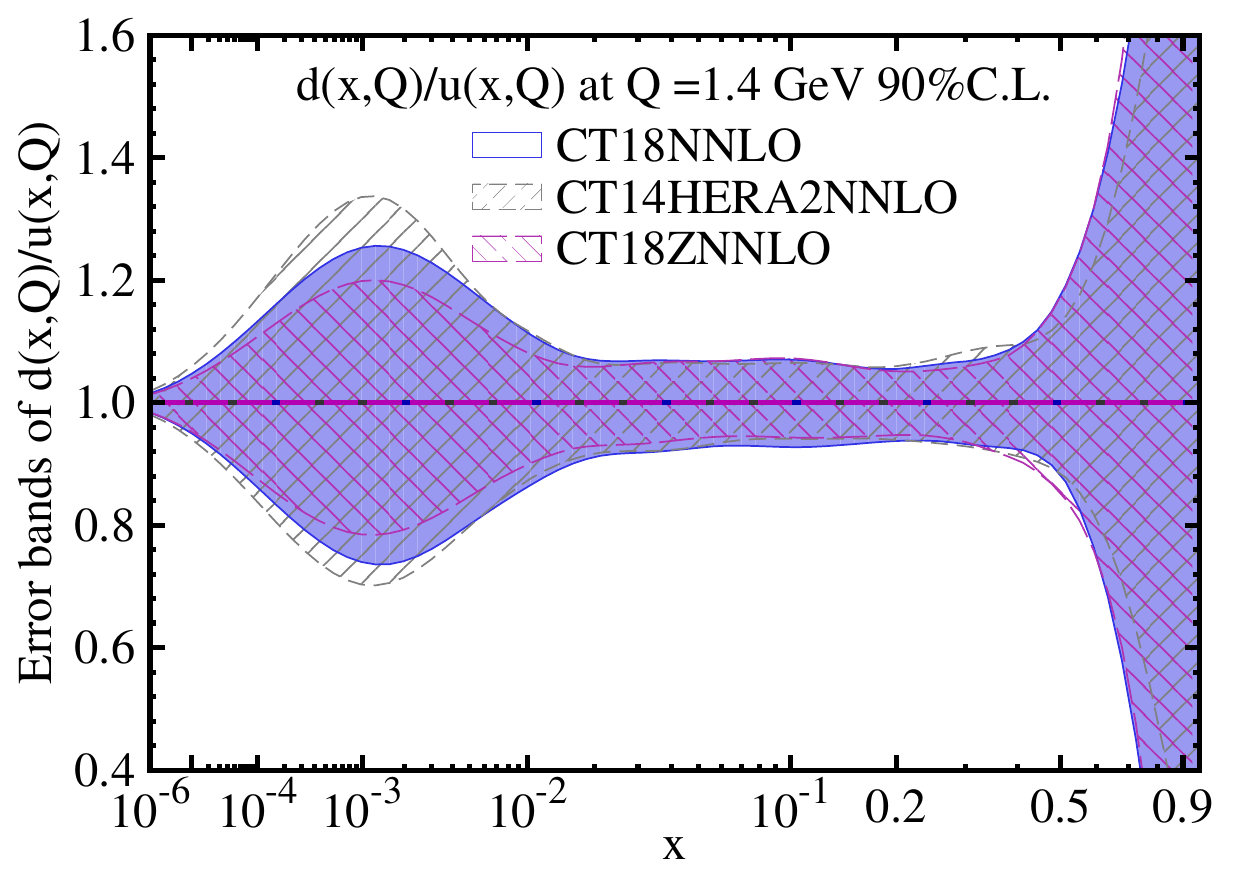}
	\includegraphics[width=0.49\textwidth]{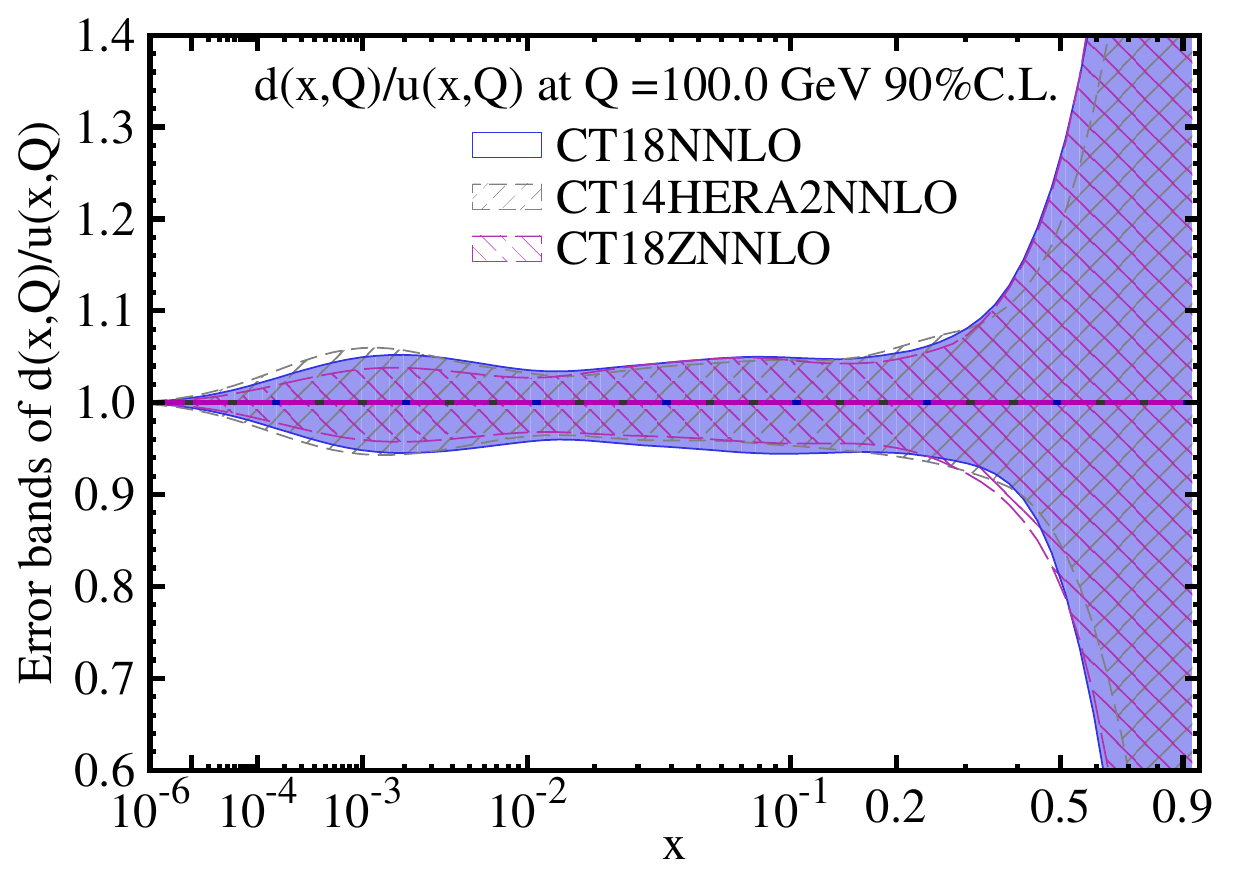}
	\caption{Like the upper panels of Fig.~\ref{fig:DOUband} for $d/u$, but normalizing each
	fit to its respective central value to compare PDF uncertainties.
		\label{fig:DOUband2}}
\end{figure}

Let us now review the ratios of various
PDFs, starting with the ratio $d/u$ shown in
Figs.~\ref{fig:DOUband} and \ref{fig:DOUband2}. 
The changes in $d/u$ from CT18, as compared
to \CTHERAII, can be summarized as a reduction (increase) of the central ratio at $x > 0.5$ ($x<10^{-2}$) and a decreased uncertainty at $x < 10^{-2}$.
Beyond $x=0.5$, the error band of $d/u$ ratio grows,  and
the parametrization form adopted since CT14 NNLO \cite{Dulat:2015mca}
guarantees that $d/u$ approaches a constant value as $x\rightarrow 1$,
as predicted by a wide array of theoretical models of nucleon
structure. This is realized by equating the $(1-x)^{a_2}$ exponents of the
$u_v$ and $d_v$ PDFs, {\it i.e.}, $a^{u_v}_2 = a^{d_v}_2$ (see App.~\ref{sec:AppendixParam}). This choice affects only the extrapolation to very high-$x$ values, $x\gtrsim 0.9$, beyond the range covered by the fitted data. At $x<0.9$, our parametrizations are flexible enough to cover the solutions and reproduce the uncertainty bands of the fits without this constraint. For example, the uncertainty band in Fig.~\ref{fig:DOUband} extends down to $d/u=0$ at $x=0.9$. Within the accessible $x$ reach, it also covers our candidate best fits with independent $a^{u_v}_2$ and $a^{d_v}_2$. Without this choice, the
PDF {\it ratio} for an individual fitted PDF Hessian set
would not have the parametric freedom to extrapolate to a
finite constant at $x = 1$. Instead, even minor differences in the fitted parameters, $a^{u_v}_2 \neq a^{d_v}_2$, will cause it either to diverge or go to $0$ at the highest-$x$ points, producing an infinite uncertainty on $d/u$ that is not compatible with the empirical electron-hadron data in that region or with common models of hadron structure. In a fit that does not constrain $a^{u_v}_2$ and $a^{d_v}_2$ to be the same, they may come up to be equal within the numerical precision of input parameters, however, in our extensive experience such coincidence hardly ever happens.

Similar logic applies to other PDF ratios,
including the low-$x$ forms of the sea-quark distributions described below.
As noted earlier, the parametrization form of $u$, $d$, $\bar u$ and $\bar d$ quarks in CT18 are the same as those in \CTHERAII. 

At such high $x$, the CTEQ-JLab analysis (CJ15) \cite{Accardi:2016qay}
has independently determined the ratio $d/u$ at NLO, by
including the fixed-target DIS data at lower $W$ and higher $x$
that are excluded by the selection cut $W > 3.5 \mbox{ GeV}$ in
CT18,  and by considering higher-twist and nuclear effects
important in that kinematic region. 
Fig.~\ref{fig:DOUband} shows that the central prediction of CT18 
differs from CJ15 at $x > 0.1$. The CT and CJ uncertainty bands are in mutual agreement, even though the error band of CJ15 is much smaller than 
CT18, a fact partly attributable
to the $\Delta \chi^2 = 1$ criterion used in CJ15.
Since the CJ15 PDFs are available only at NLO in $\alpha_s$,
we compare the CJ15 NLO $d/u$ ratios to the respective CT18 NNLO (NLO) ratios in the bottom-left (bottom-right) frame of Fig.~\ref{fig:DOUband}.

\begin{figure}[tb]
	\center
	\includegraphics[width=0.49\textwidth]{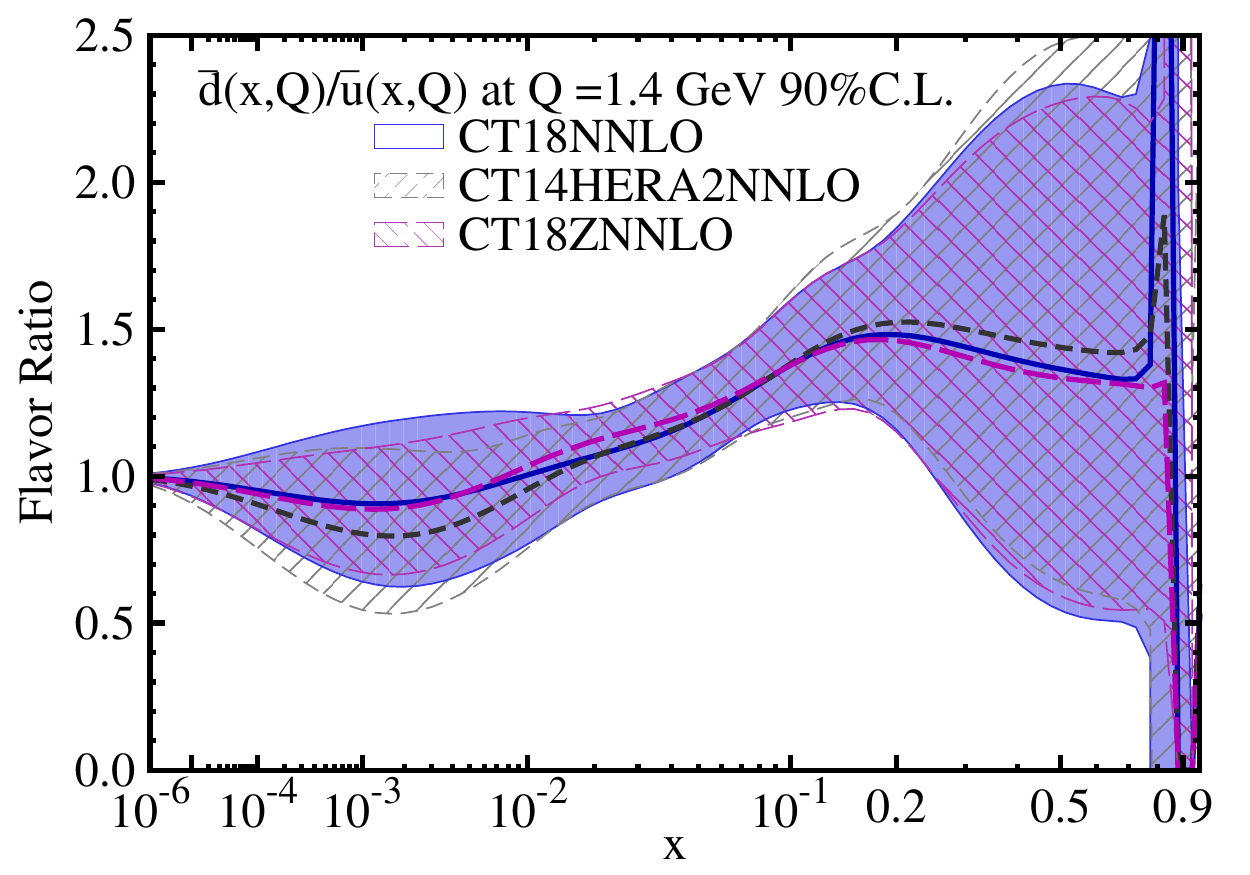}
	\includegraphics[width=0.49\textwidth]{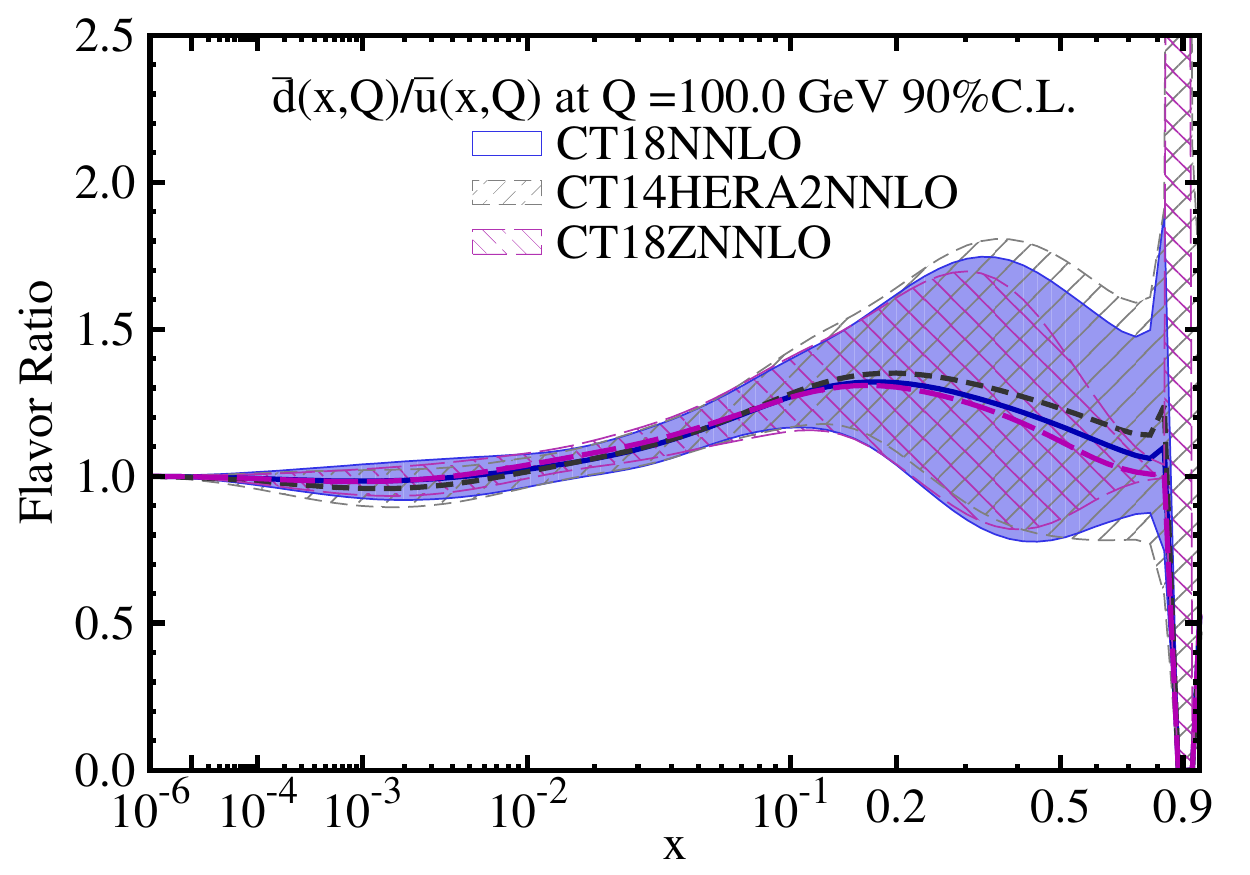}
	\includegraphics[width=0.49\textwidth]{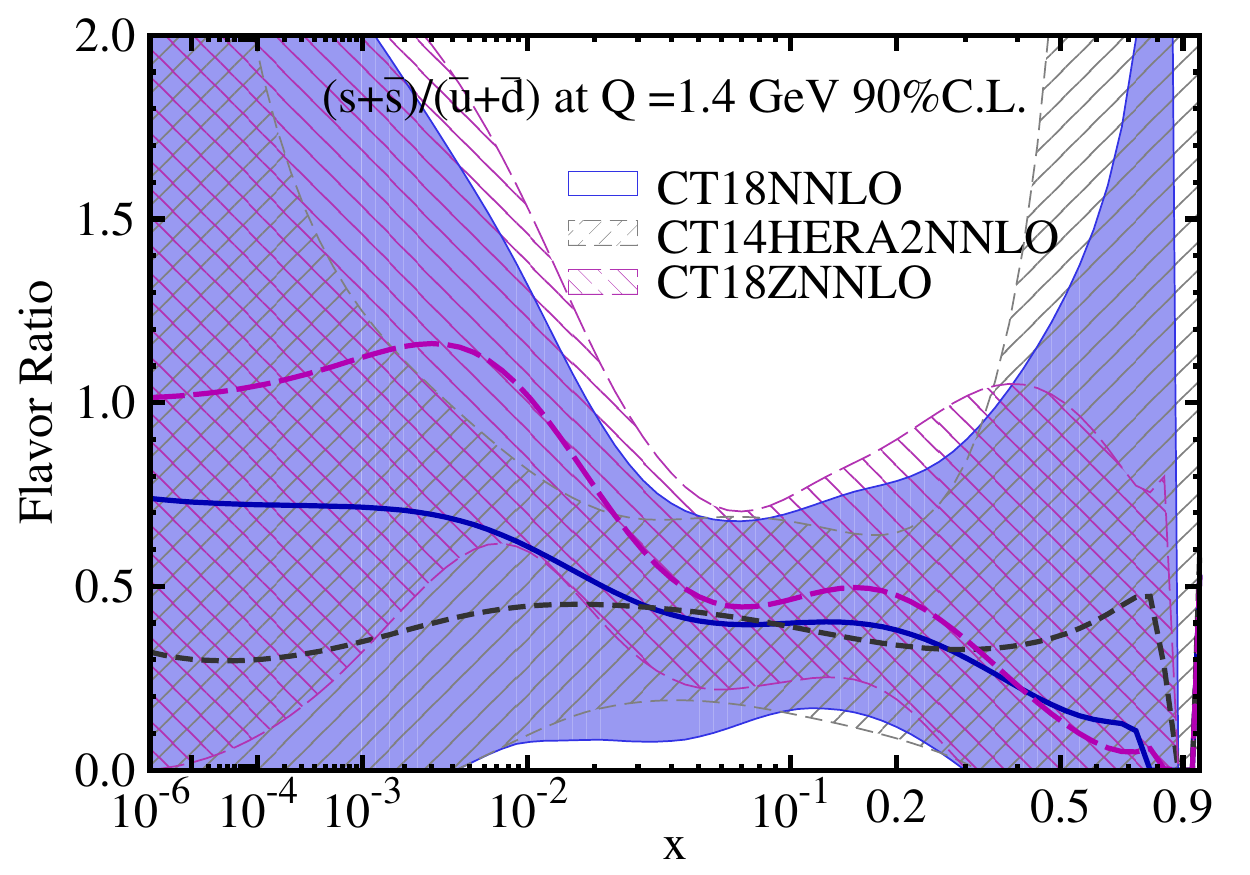}
	\includegraphics[width=0.49\textwidth]{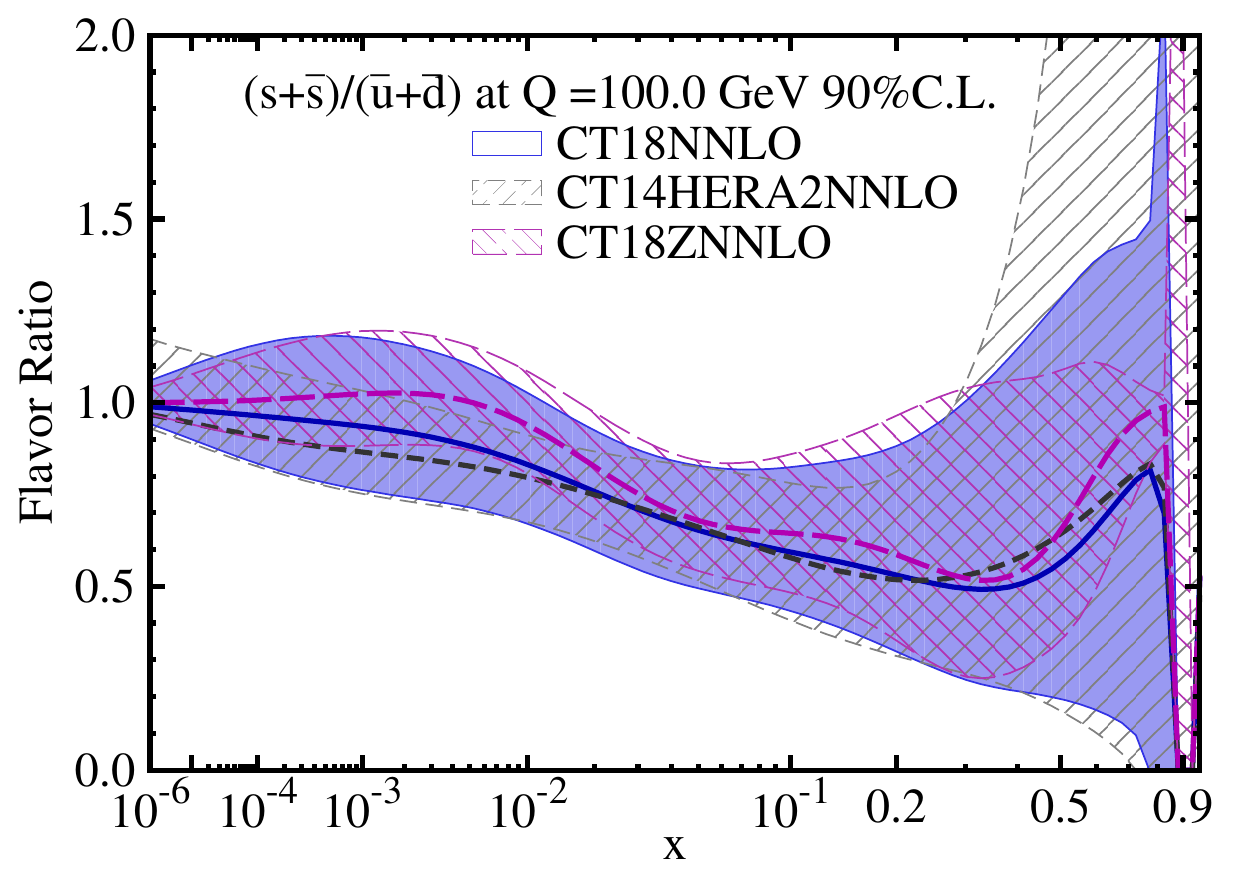}
	\caption{A comparison of 90\% C.L. uncertainties on the ratios
		$\bar d(x,Q)/\bar u(x,Q)$ and $\left(s(x,Q)+\bar
		s(x,Q)\right)/\left(\bar u(x,Q) +\bar d(x,Q)\right)$,
	for CT18 (solid
		blue), CT18Z (magenta long-dashed), and \CTHERAII~NNLO (gray short-dashed) ensembles
		at $Q=1.4$ or $100$ GeV.
		\label{fig:DBandSBbands}}
\end{figure}

\begin{figure}[tb]
	\center
	\includegraphics[width=0.49\textwidth]{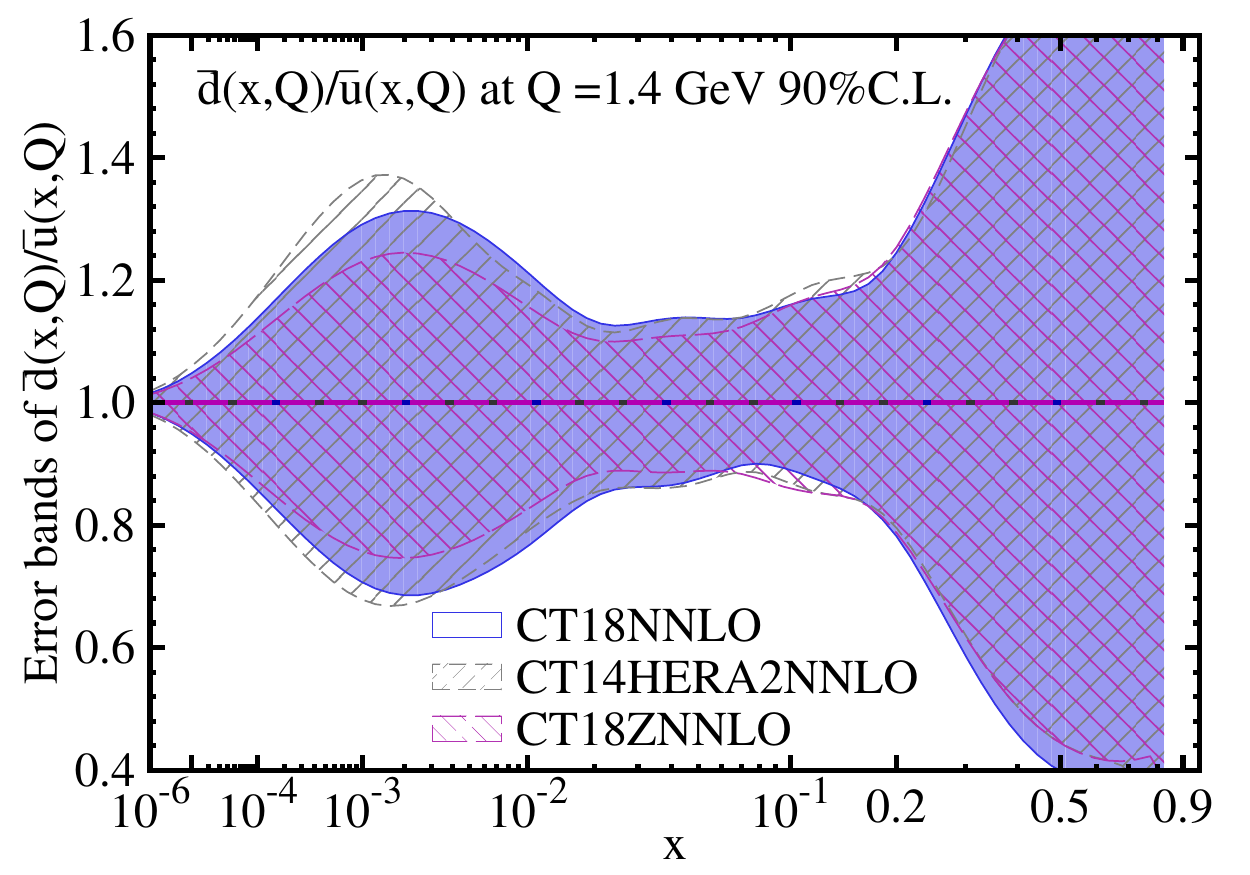}
	\includegraphics[width=0.49\textwidth]{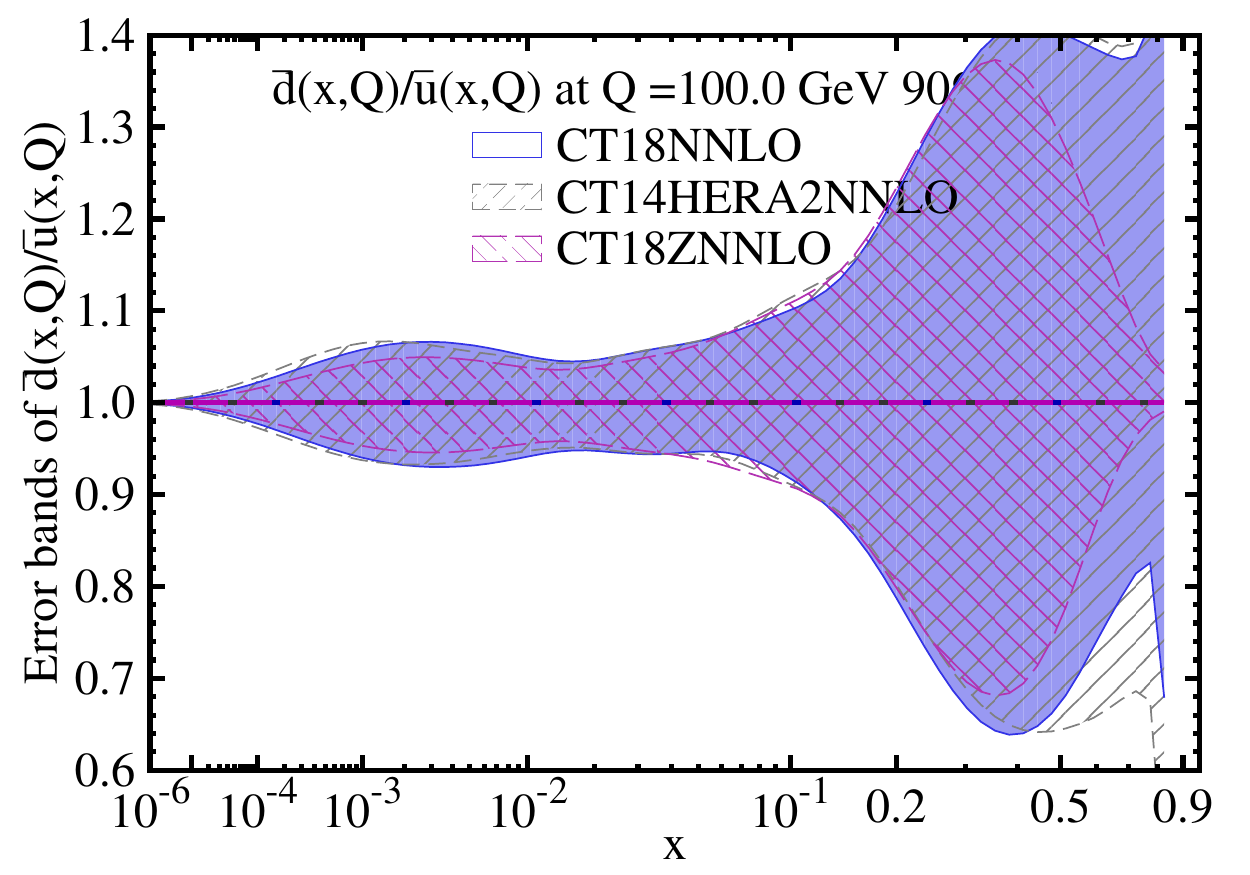}
	\includegraphics[width=0.49\textwidth]{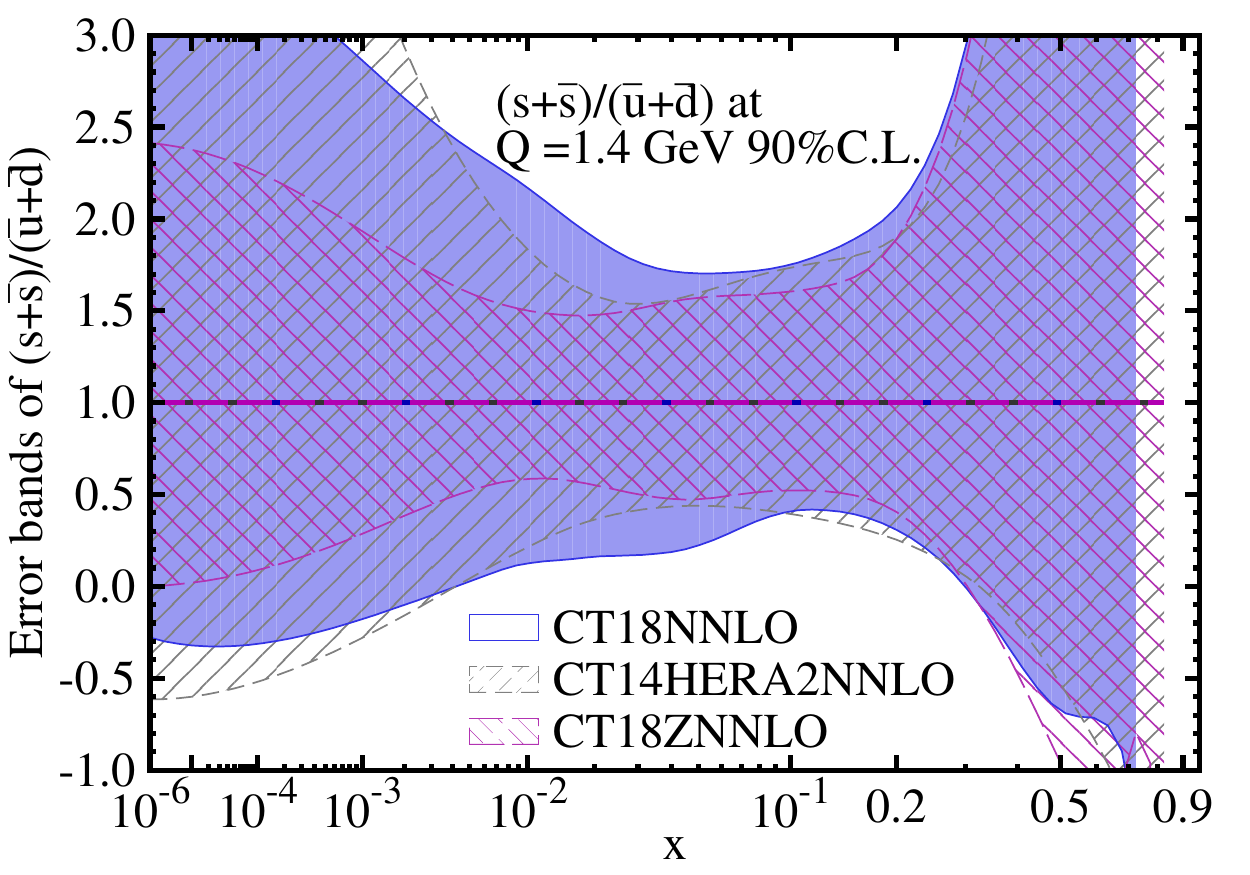} 
	\includegraphics[width=0.49\textwidth]{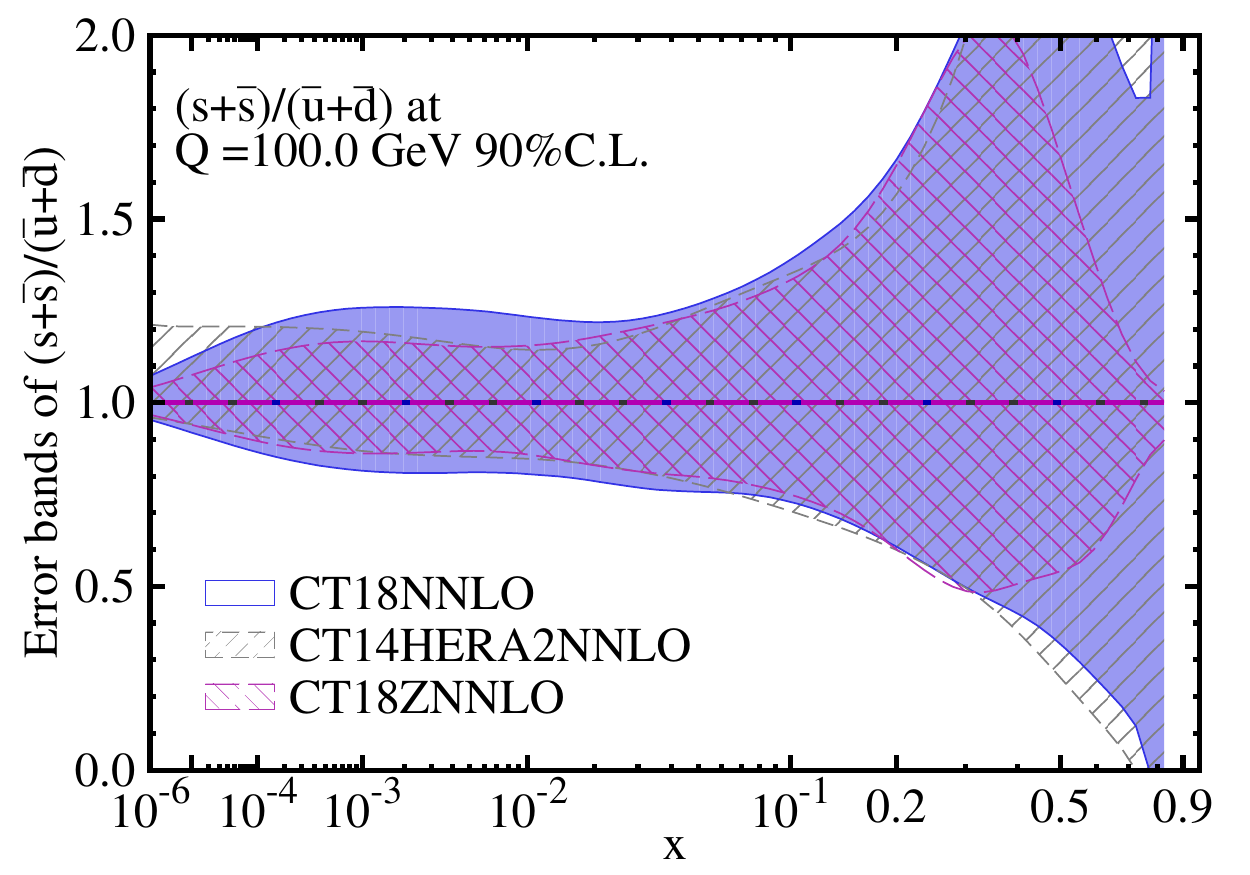} 
	\caption{A comparison of 90\% C.L. uncertainties on the ratios
		$\bar d(x,Q)/\bar u(x,Q)$ and $\left(s(x,Q)+\bar
		s(x,Q)\right)/\left(\bar u(x,Q) +\bar d(x,Q)\right)$,
	for CT18 (solid
		blue), CT18Z (magenta long-dashed), and \CTHERAII~NNLO (gray short-dashed) ensembles
		at $Q=1.4$ or $100$ GeV, relative to 
		their own central fit.  
		\label{fig:DBandSBbands2}}
\end{figure}

Turning now to the ratios of sea quark PDFs in
Fig.~\ref{fig:DBandSBbands}, we observe that the uncertainty on $\bar
d(x,Q)/\bar u(x,Q)$ in the left inset has decreased at small $x$
in CT18. For $x\! >\! 0.1$, the CT18 nonperturbative parametrization forms for $\bar u$ and $\bar d$ ensure that the ratio
$\bar d(x,Q_0)/\bar u(x,Q_0)$ can approach a constant value, which
turns out to be close to 1 in the central fit. The uncertainty on
$\bar d/\bar u$ has also decreased, most notably for $x\! \gtrsim\! 10^{-3}$, primarily due to the inclusion
of the LHCb data sets (Exp.~IDs=250, 245 and 246), cf.~the upper-left panel of Fig.~\ref{fig:DBandSBbands2} at $Q\!=\!1.4\,\mathrm{GeV}$. 

At high $Q$ values, such as in the right panels for $Q=100 $ GeV in Fig.~\ref{fig:DBandSBbands}, the ratios depend as much on the large-$x$ gluon behavior at $Q_0$ as on the quark PDFs themselves. As a result, for CT18Z that has an enhanced gluon PDF and suppressed sea quark PDFs at very large $x$ and $Q_0$, the uncertainties on the ratios $\bar d/\bar u$ and $R_s$ are reduced at $x\gtrsim 0.8$ and large $Q$, reflecting the flavor symmetry of $g\rightarrow q\bar q$ splittings that primarily drive the sea quark PDFs in this $\{x,Q\}$ region. 

The overall increase in the strangeness PDF at $ x < 0.03$ and decrease of $\bar u$ and $\bar d$ PDFs at $ x  < 10 ^{-3}$, cf.~Fig.~\ref{fig:PDFbands1}, lead to a
larger ratio of the strange-to-nonstrange sea quark PDFs,
\begin{equation}
	R_s(x,Q) \equiv \frac{s(x,Q) + \bar{s}(x,Q)}{\bar{u}(x,Q) + \bar{d}(x,Q)}\ ,
\label{eq:Rs}
\end{equation}
presented in
Fig.~\ref{fig:DBandSBbands}. 
$R_s(x,Q)$ measures the $x$ and $\Q$ dependence of the breaking of flavor-$\mathrm{SU}(3)$ symmetry,
with older analyses typically fixing $R_s = 0.5$. More recently, a number of previous CTEQ
studies \cite{Lai:2007dq,Olness:2003wz} examined contemporary constraints on $R_s$,
particularly driven by the neutrino-induced SIDIS dimuon production measurements by the CCFR and NuTeV Collaborations, but also by precise inclusive HERA measurements. 
These works found significant evidence of an independent $x$ dependence for $s^+(x)\! \equiv\! s(x)\! +\! \bar{s}(x)$, distinct from $\bar{u}\!+\!\bar{d}$, 
but were unable to exclude a vanishing strangeness momentum fraction  asymmetry,
$\langle x \rangle_{s^-}\! =\! \int^1_0 dx\ x [s-\bar{s}](x,\Q=m_c)\! =\! 0$.

In the present work, we continue to assume $s^-(x,Q)=0$ and focus on $s^+(x,Q)$
and the related $R_s(x,Q)$, the quantities that
both reflect the interplay of the older charged-current DIS data and new LHC measurements that are detailed later in Sec.~\ref{sec:Quality} and App.~\ref{sec:AppendixCT18Z}.
Here let us mention that, at $ x  \ll 10 ^{-3}$, the $R_s$
ratio is determined entirely by the parametrization form and was found
in CT10 to be consistent with the exact $\mathrm{SU}(3)$ symmetry of PDF
flavors, $R_s(x,Q) \rightarrow 1$
at $x\rightarrow 0$, albeit with a large uncertainty. 
The $\mathrm{SU}(3)$-symmetric asymptotic solution at $x\rightarrow 0$
was not enforced in CT14 or \CTHERAII, so that their $R_s$ ratio  
was around $0.3$ to $0.5$ at $x\approx 10^{-5}$ and $Q=1.4$ GeV.
In CT18, we have assumed a different $s$-PDF nonperturbative
parametrization form (with one more parameter added), but the one that
still ensures a stable behavior of $R_s$ for $x\! \to\! 0$, so that
$R_s (x \to 0)$ is about 0.7 and 1, respectively, in CT18 and CT18Z
fits. 

\subsubsection{Changes in the $x$ dependence of PDFs, summary}
We may summarize the pulls of specific processes on the central CT18 fit as follows. 

\begin{itemize}

\item The most noticeable overall impact of the LHC inclusive jet
  production on the central gluon PDF $g(x,Q)$ is to mildly reduce it
  at $x > 0.2$ within the original PDF uncertainty band. The pulls
  from the jet data sets change little after the decorrelation of some
  systematic errors, cf. Sec.~\ref{sec:DataJets},
  and when the $0.5\%$ MC uncertainty on theory values is added.
The pulls from various jet data sets on $g(x,Q)$ neither follow a
uniform trend across the whole $x$ range nor are consistent among
various measurements, as is demonstrated, e.g., by the $L_2$
sensitivity in Fig.~\ref{fig:L2glu} and LM scans in
Sec.~\ref{sec:QualityOverview}. 

\item The LHCb data, combined over all processes, have some impact on
  the $u$, $d$ and $s$ quarks, and pull the $s (x,Q)$ up at small $x$.   

\item The ATLAS 8 TeV $Z$ $p_T$ data (Exp.~ID=253), for the nominal
  QCD scales assumed in the CT18 NNLO fits, weakly pull the gluon PDF
  at $x>0.05$ downward, in the direction similar to the average pull
  of the LHC inclusive jet data. The relative magnitude of the pull
  from these data, as compared to those from the jet experiments,
  can be estimated from the $L_2$ sensitivity plot
  for $g(x,Q)$ in Fig.~\ref{fig:L2glu}.

\item The ATLAS 7 TeV data on $W$ and $Z$ rapidity distributions
  (Exp.~ID=248), included only in CT18A and Z, have the largest
  influence on the PDFs, as discussed in
  App.~\ref{sec:AppendixCT18Z}. The directions of their pulls are
  similar to LHCb. 

\item The LHC data on $t \bar t$ double differential cross sections
  also appears to favor a softer gluon at large $x$, but the pull is
  not statistically significant, {\it i.e.}, much weaker than that of
  the inclusive jet data with its much larger number of data points.  

\end{itemize}

These constraints are further explored in depth in
Sec.~\ref{sec:QualityOverview} using a combination of statistical techniques.

\subsection{The global fits for $\alpha_s$ and $m_c$}
\label{sec:AlphasDependence}

\begin{figure}[tbp]
\centering
\includegraphics[width=0.49\textwidth]{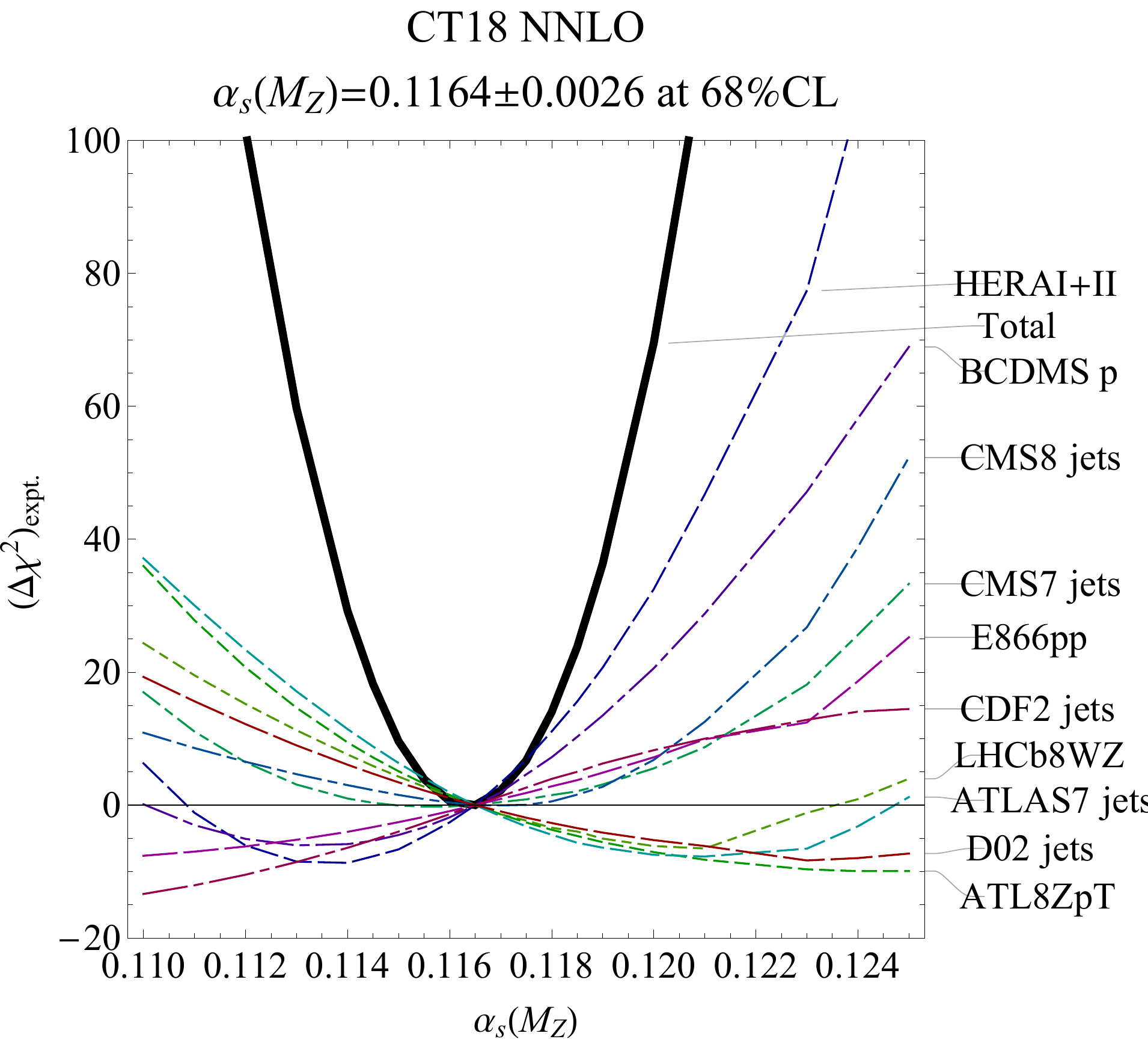} \ \
\includegraphics[width=0.46\textwidth]{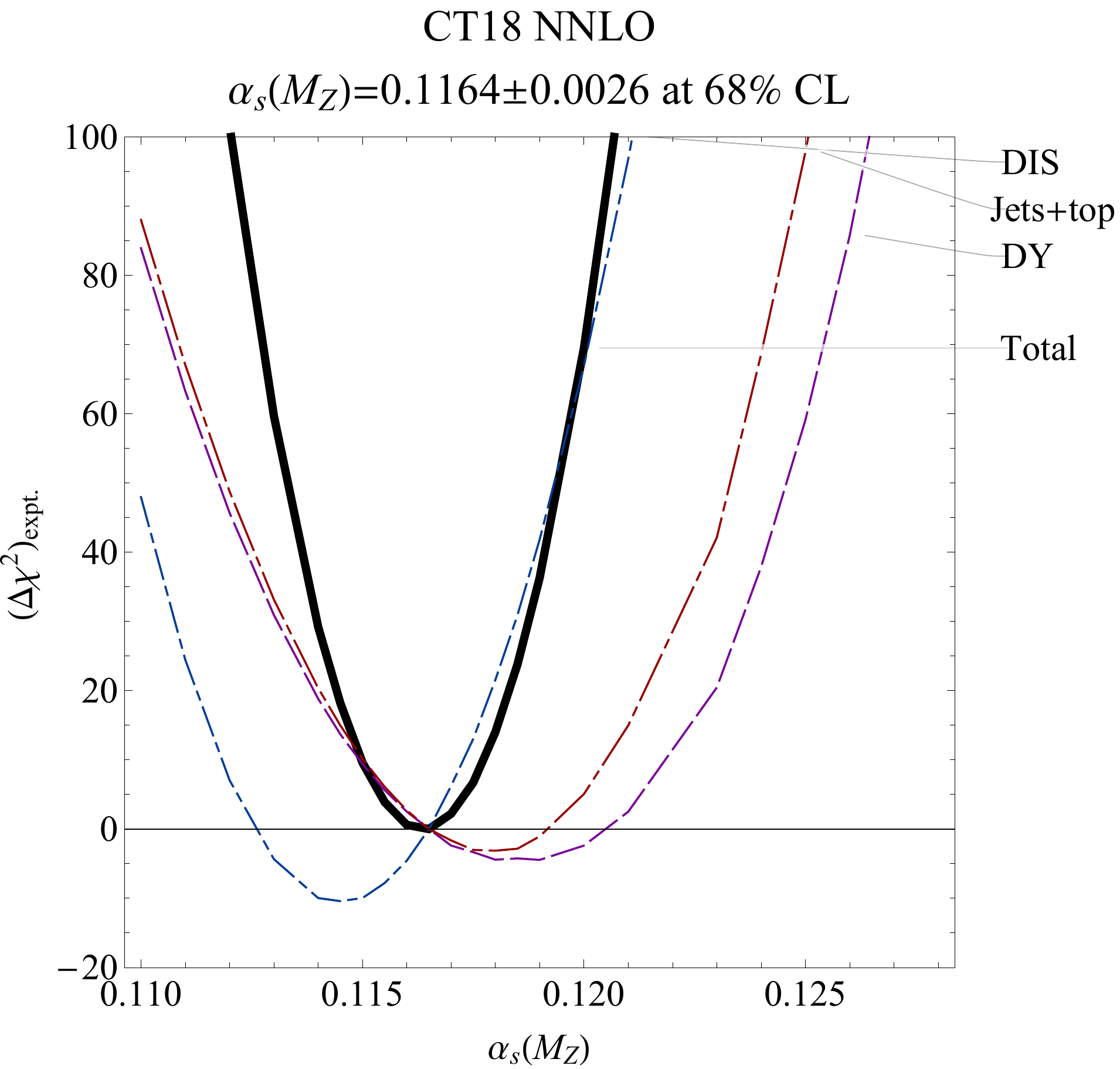}
\caption{The scan of the strong coupling constant at the scale of $M_Z$ for CT18 at NNLO. Left: changes of $\chi^2$ of all the data sets together
    (heavy black line) and of several individual experiments with especially strong pull on $\alpha_s(M_Z)$. Right: Values for the change in $\chi^2$ for all experiments
    fitted in CT18, but separately collected into combined DIS, DY and top/jets data sets. The growth in $\Delta \chi^2$ for the ``Jets+top'' curve in the right panel
    is mainly driven by the constraints from jet production. While $t\bar{t}$ production has important sensitivity to $\alpha_s$, the comparatively
    small number of top data points leads to a more intermediate impact in the full fit.
	}
\label{fig:lm_alphas}
\end{figure}

{\bf Determination of the QCD coupling}. Following the long-established practice \cite{Lai:2010nw}, in the canonical PDF sets such as CT18, the value of $\alpha_s(M_Z)$ is set to the world average of $\alpha_s(M_Z)\! =\! 0.118$ \cite{Tanabashi:2018oca};  alternate PDFs are produced for a range of fixed $\alpha_s(M_Z)$ above and below that central value ({\it i.e.}, an ``$\alpha_s$ series'') to evaluate the combined PDF+$\alpha_s$ uncertainty. 
In Ref.~\cite{Lai:2010nw}, we show how to evaluate the combined $\mathrm{PDF} \, + \, \alpha_s$ uncertainty in the global fit.
As shown, variations in $\alpha_s$ generally induce compensating adjustments in the preferred PDF parameters (correlation) 
to preserve agreement with those experimental data sets that simultaneously constrain $\alpha_s$
and the PDFs. At the same time, it is possible to define an ``$\alpha_s$ uncertainty'' that quantifies all correlation effects.
As the global QCD data set grows in size, more experiments introduce sensitivity to $\alpha_s(M_Z)$ either through radiative contributions to hard cross sections or through scaling violations, especially over a broad range of physical scales, $\Q$.  

\begin{figure}[tbp]
\centering
\includegraphics[width=0.8\textwidth]{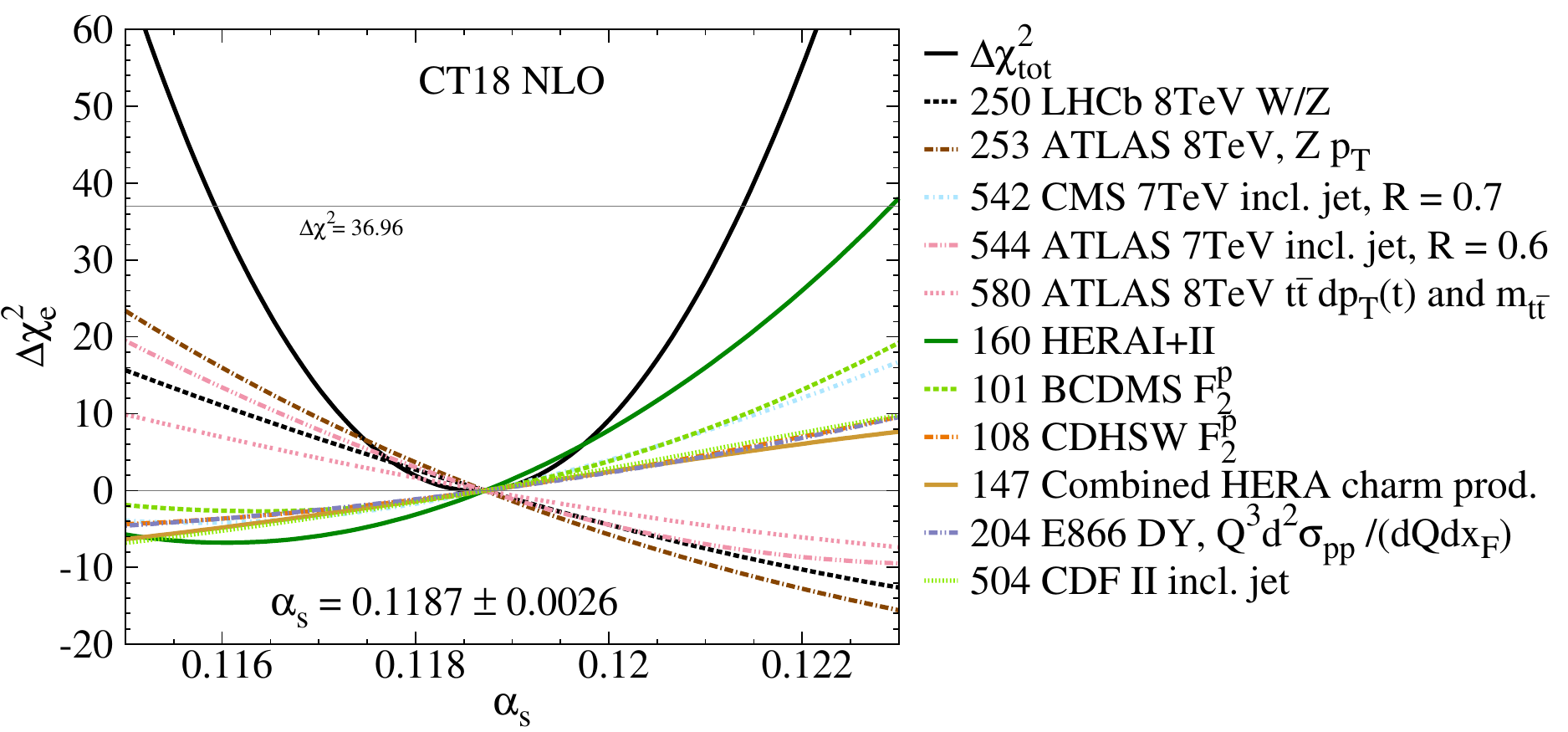}
\caption{
	Like Fig.~\ref{fig:lm_alphas}, but now showing the scan of $\alpha_s(M_Z)$ at NLO precision in $\alpha_s$.
        }
\label{fig:lm_alphas_nlo}
\end{figure}

Perhaps the best way to examine the sensitivity of each experiment,
and of the global ensemble of experiments, is to examine the variations
of their $\chi^2$ as the value of $\alpha_s(M_Z)$ is varied. Such
scans over $\alpha_s(M_Z)$ for CT18 NNLO and CT18 NLO are shown in
Figs.~\ref{fig:lm_alphas} and \ref{fig:lm_alphas_nlo}, respectively. In
all figures illustrating the scans in this and the next section,
we plot a series of curves for
\begin{equation}
  \Delta \chi_E^2(a) \equiv \chi_E^2(a)-\chi_E^2(a_0),
  \label{DelChi2Scan}
\end{equation}
as a function of some parameter $a$. The
variation $\Delta \chi^2_E(a)$ is the difference between the $\chi^2$
values for experiment $E$ at the fixed value of $a$ shown on the
horizontal axis (with $\chi^2_E(a)$
marginalized with respect to the rest of free
parameters), and when $a$ is determined at the global $\chi^2$ minimum
for the full CT18 data set, where $a=a_0$. The $\Delta \chi^2$ curves are
shown for all experiments (indicated as ``Total'' or
``$\chi^2_{\rm tot}$'') and for the top few experiments with the
largest variations $\Delta \chi^2_E$ in the shown range of $a$.
Thus, by definition $\Delta \chi^2_{\rm tot}(a_0) =0$. 

We note that we have varied $\alpha_s$ in the present scan in all exact radiative contributions, but kept  $\alpha_s$ fixed in the tabulated $K$ factors.  This approximation greatly simplifies the computations, and we have verified that it changes $\chi^2$ by only a small fraction of the higher-order uncertainty within the fitted $\alpha_s$ range. 

From Fig.~\ref{fig:lm_alphas}, we see that the various data sets
have different sensitivities to both
the central  value of $\alpha_s(M_Z)$ and its uncertainty.\
According to the scans, the greatest sensitivity to $\alpha_s(M_Z)$ is provided by the HERA I+II data set, followed by the BCDMS proton data. Relatively to the full CT18 data set, both experiments prefer a lower value of
$\alpha_s(M_Z)$, on the order of $0.114-0.116$, but with wider uncertainties. The dependence of those two DIS data sets on $\alpha_s(M_Z)$ is primarily through the effect of scaling violation, but the sheer number of data points,
and the experimental and theoretical precision, lead to their large sensitivities. 

The LHC inclusive jet production, especially the CMS 7 and 8 TeV data, generally prefer a large value of $\alpha_s(M_Z)$, as
does the ATLAS 8 TeV Z $p_T$ data. The full CT18 data set prefers a value of $\alpha_s(M_Z, \mbox{NNLO})=0.1164 \pm 0.0026$, at 68\% C.L., defined using the ``global tolerance" prescription to correspond to a $\Delta \chi^2\! =\! 37$ interval
(the corresponding 90\% interval is defined by $\Delta \chi^2\! =\! 100$). The extracted value of $\alpha_s(M_Z)$ obtained with CT18Z is very similar, $0.1169\pm0.0026$, cf. Fig.~\ref{fig:lm_alphasz}. These values are to be
compared with $\alpha_s(M_Z)\!=\!0.1150^{+0.0036}_{-0.0024}$ as obtained by CT14 with a smaller HERA+LHC data set. 

The $\Delta\chi^2$ distribution for the full data set is very
parabolic, less so for the individual data sets. The $\Delta\chi^2$
curves for collections of data sets, for example, all DIS data, all DY
data, and all jets and top data, also appear parabolic, as expected
from the central limit theorem. From the right panel of
Fig.~\ref{fig:lm_alphas}, it is clear that the totality of DIS data
prefer a smaller value of $\alpha_s(M_Z)$  than the DY pair, jet and
top-quark production.  
The exact size of the $\alpha_s$ uncertainty thus is not well
determined and depends on the convention, as the pulls from various
(types of) experiments are not consistent at the level of few tens of
units of $\chi^2$.  

The scan exercise can also be carried out at NLO in $\alpha_s$, as we show in Fig.~\ref{fig:lm_alphas_nlo}. In fact, any difference between the NLO and NNLO results can serve as a partial estimate of the
theoretical uncertainty of its determination. Although the uncertainty is similar to that obtained at NNLO, the central value is slightly higher: $\alpha_s(M_Z,\mbox{NLO})=0.1187\pm 0.0027$. We note that the qualitative interplay of the experiments with leading sensitivity to $\alpha_s(M_Z)$ is much the same at NLO as
found at NNLO, with the combined HERA (Exp.~ID=160) and BCDMS $F^p_2$ data (Exp.~ID=101) again preferring lower values, while the ATLAS 7 TeV jet data (Exp.~ID=544) and 8 TeV
$Z$ $p_T$ data (Exp.~ID=253) pulling in the opposing direction, but more strongly at NLO than at NNLO. The preference of a higher $\alpha_s$ value at NLO by an amount of about 0.002 is consistent with findings of other PDF groups~\cite{Ball:2011us,Harland-Lang:2015nxa,Abramowicz:2015mha,Alekhin:2017kpj}.

To summarize, we find that the CT18 data set prefers a larger value of $\alpha_{s}(M_Z)$ and a marginally smaller nominal uncertainty than in CT14.

\begin{figure}[tbp]
\centering
\includegraphics[width=0.6\textwidth]{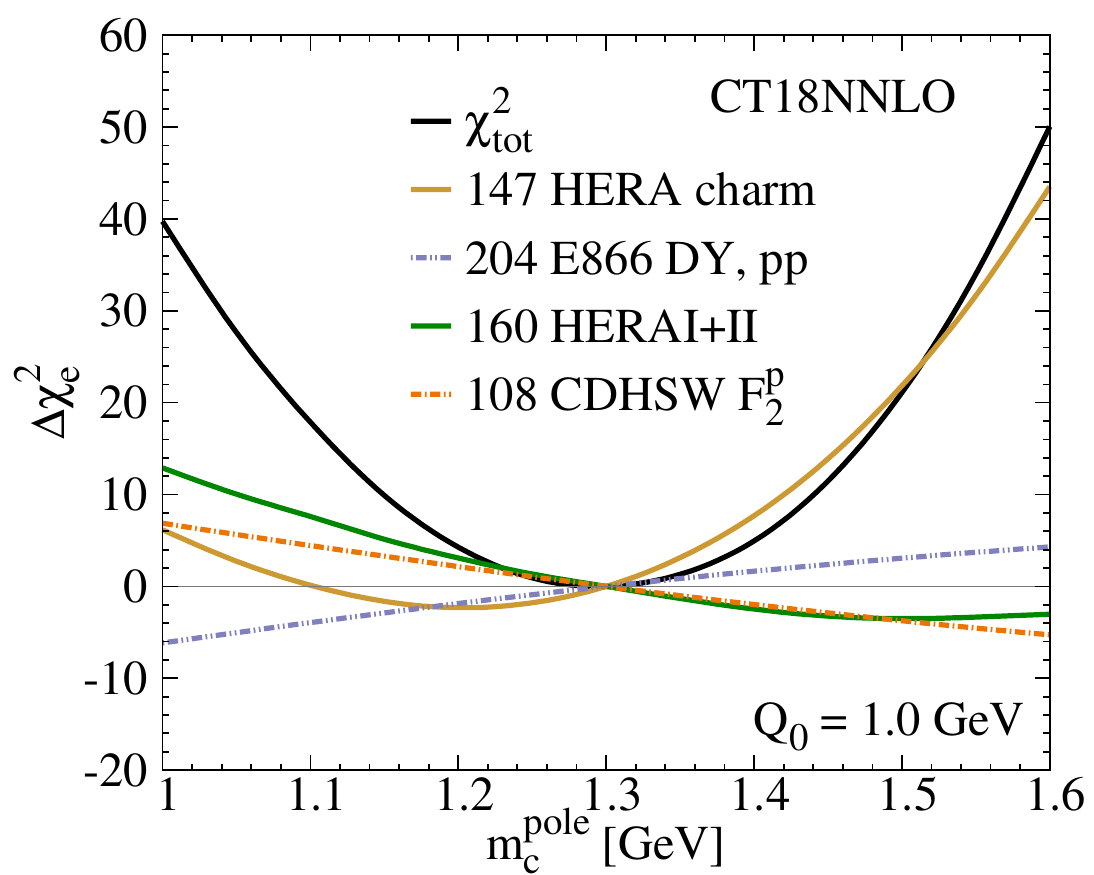}
\caption{
	A $\chi^2$ scan over values of the charm pole mass, $m_c$, at NNLO, using the CT18 data set. The settings of the fit are described in the text. The CT18Z counterpart to this $m_c$ scan is presented in Fig.~\ref{fig:lm_mcZ} in App.~\ref{sec:AppendixCT18Z}.
        }
\label{fig:lm_mc}
\end{figure}

{\bf Constraining the charm pole mass}.
Similar investigations can be carried out for other inputs of the perturbative theory, such as the pole mass of the charm
quark, $m_c$. A conclusive study on the charm mass dependence is beyond the scope of this article: the experimental preferences for $m_c$ may be affected by the initial scale $Q_0$, auxiliary settings in the heavy-quark scheme, and possibility of the nonperturbative charm~\cite{Gao:2013wwa,Hou:2017khm,Dulat:2013hea}. In the candidate fits we made, we observe that the traditional choice $m_c^{pole}=1.3$ GeV remains compatible with the CT18(Z) global data, however, the most recent HERA DIS and LHC vector boson production experiments in totality may mildly prefer the pole mass of 1.4 GeV or higher. 

An illustration of the observed trends can be viewed in
Fig.~\ref{fig:lm_mc}, where we show the $\chi^2$ variation for the
total data set and for the leading experiments in a NNLO fit to the
CT18 data set at different pole $m_c$. To separate the $m_c$ dependence from $Q_0$ dependence, we set $Q_0=1$ GeV and use a more flexible  (sign-indefinite at small $x$) gluon parametrization that better accommodates the full range of solutions at such low $Q$. 
The choice $Q_0=1$ GeV, also used in the CT10 study \cite{Gao:2013wwa} of $m_c$ dependence, allows us to widen the examined range of $m_c$, while the extra flexibility of the gluon at $Q=1 $ GeV is needed to accommodate the full range of the CT18 PDF uncertainty at $Q>1.3$ GeV.
As can be seen based on the minimum of the heavy black curve, the scan prefers a value of $m_c\! =\! 1.3$ GeV, with this mass
being somewhat larger than the preference of the combined charm production data from HERA (Exp.~ID=147) alone, which would otherwise suggest
$m_c\!\gtrsim\!1.2$ GeV. The combined HERA Run I and II inclusive data, on the other hand, essentially provide a lower bound to $m_c$, and prefer
a larger magnitude, $m_c\! > \!1.45$ GeV. 
However, these preferences are quite weak, yielding an overall change by ten units of $\chi^2$ over a large range of $m_c$.
The individual sensitivities of the other experiments presented in Fig.~\ref{fig:lm_mc} are even weaker.
It should be pointed out that the inclusion of the ATLAS 7 TeV $W$/$Z$ data (Exp.~ID=248) and other changes associated with CT18Z lead to a reconfiguration of the picture shown in Fig.~\ref{fig:lm_mc} and to an increase in the best-fit value of $m_c$, as we show in Fig.~\ref{fig:lm_mcZ} and discuss in App.~\ref{sec:AppendixCT18Z}.
In the same spirit, the patterns of the pulls change somewhat if we set $Q_0=m_c$ (another acceptable choice).

\subsection{Parton luminosities at the LHC
\label{sec:PDFLuminosities}
}

\begin{widetext}

\begin{figure}[!htbp]
	\begin{center}
		\hspace*{-0.4cm}\includegraphics[width=0.48\textwidth]{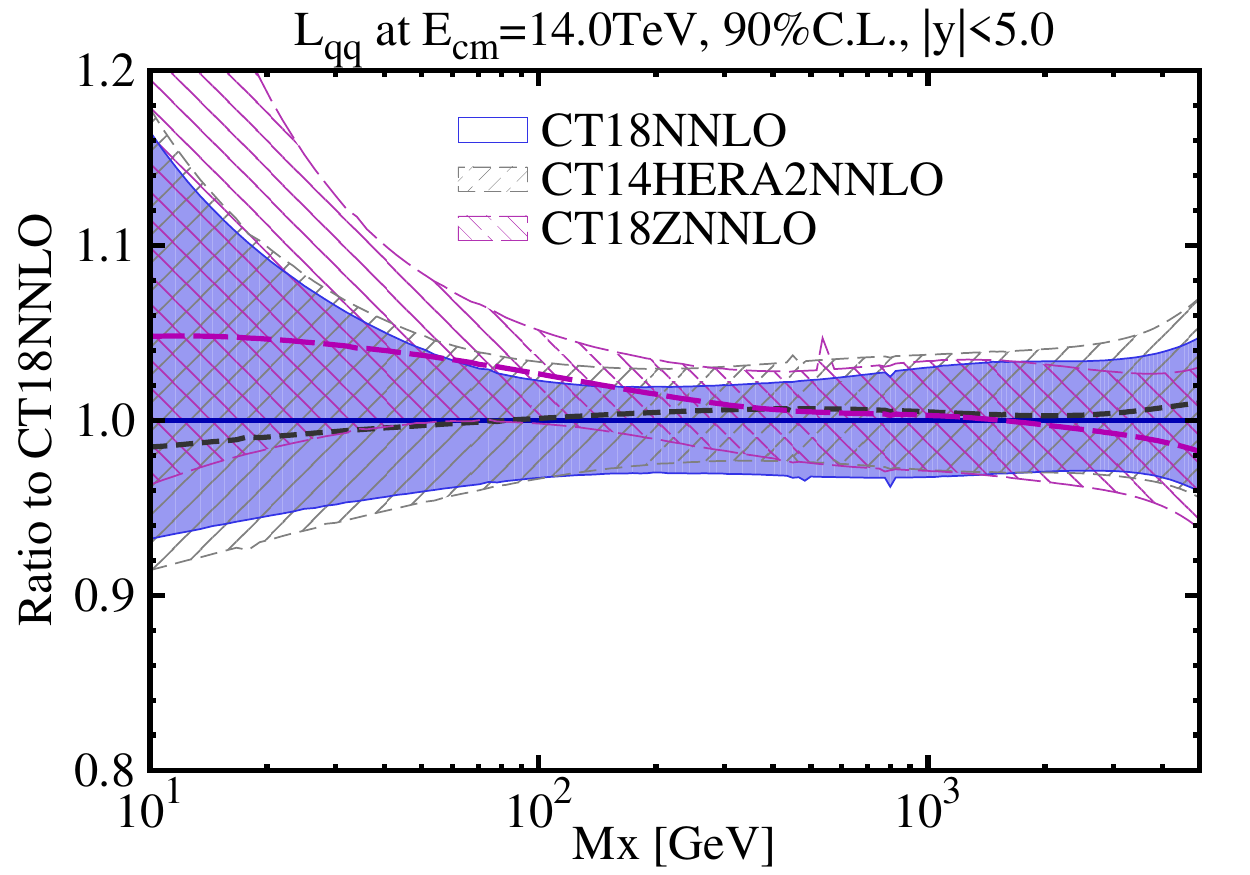}
		\includegraphics[width=0.48\textwidth]{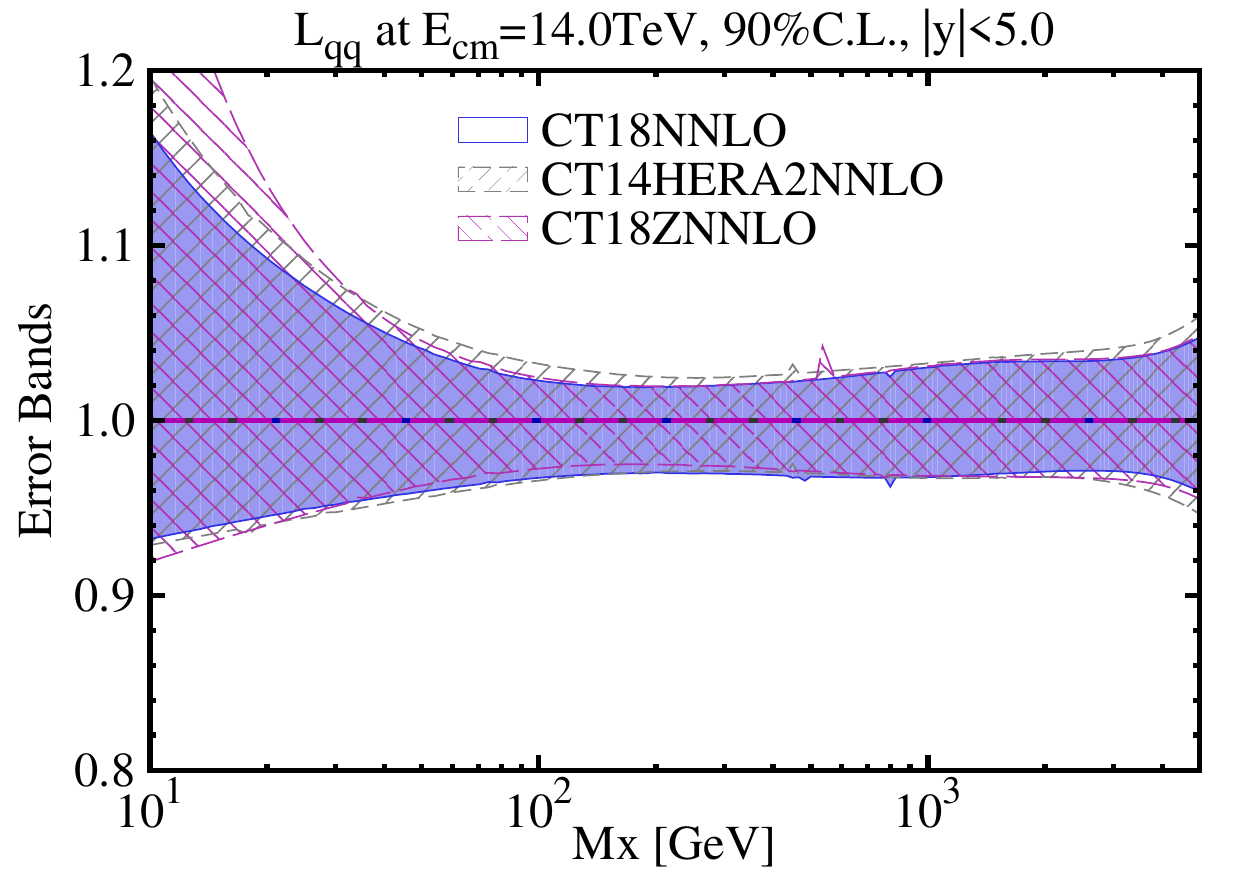} \\
		\includegraphics[width=0.48\textwidth]{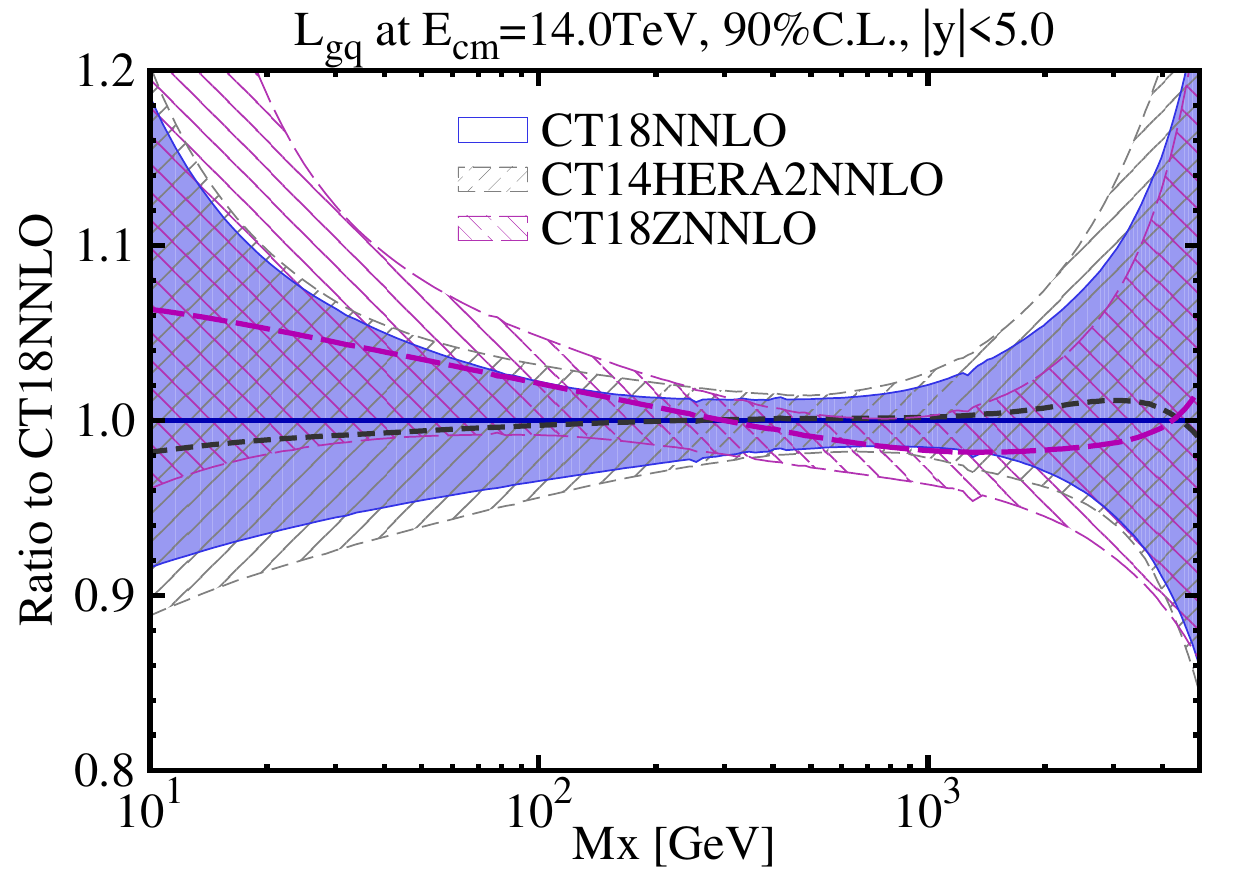}
		\includegraphics[width=0.48\textwidth]{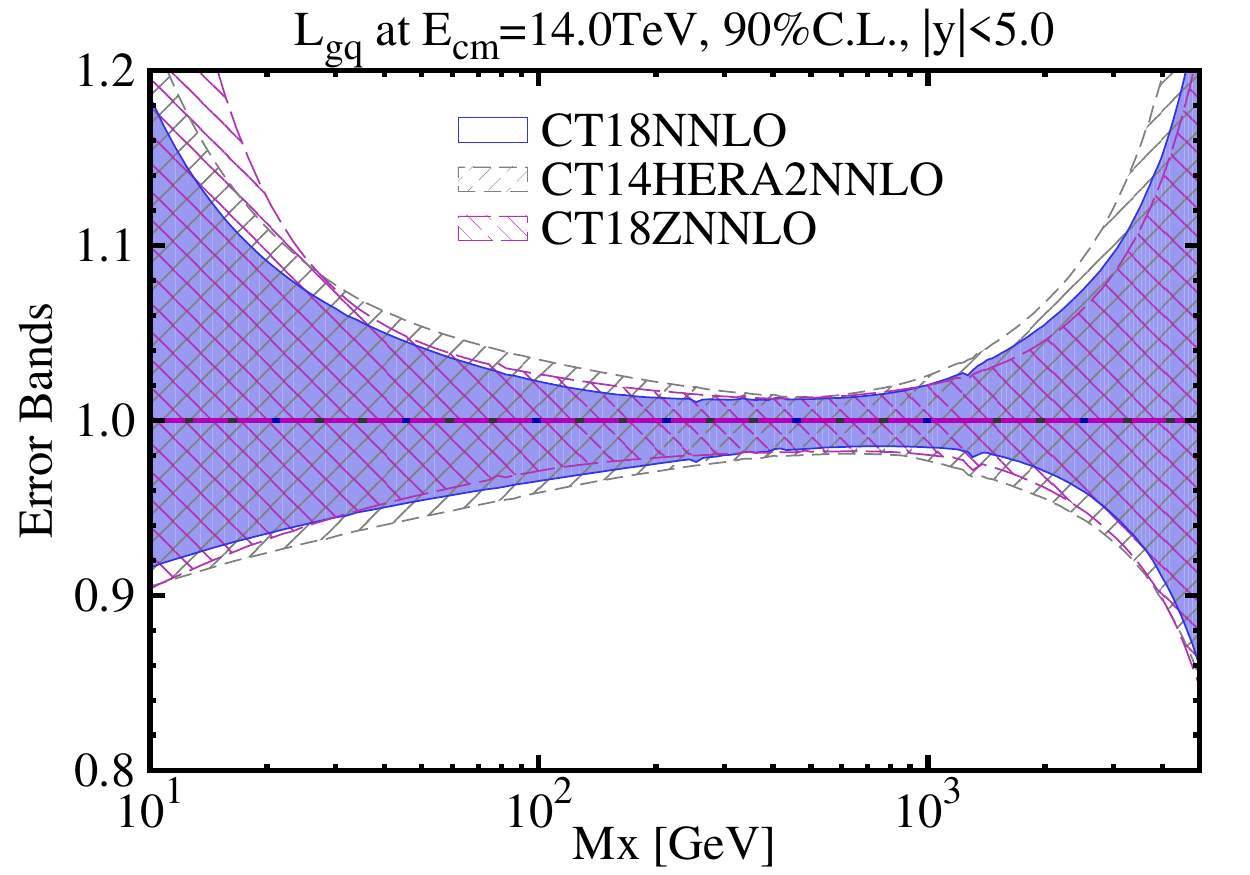} \\
		\includegraphics[width=0.48\textwidth]{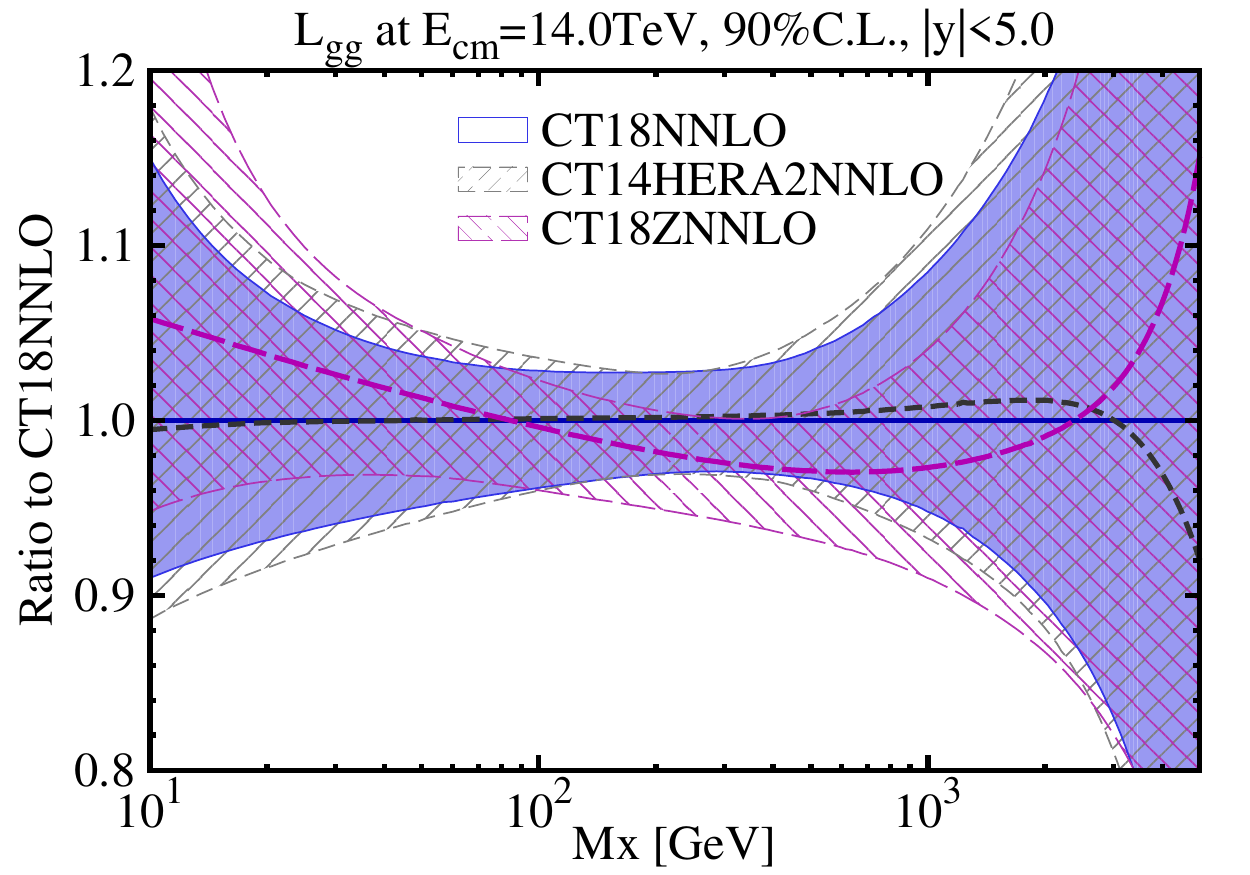}
		\includegraphics[width=0.48\textwidth]{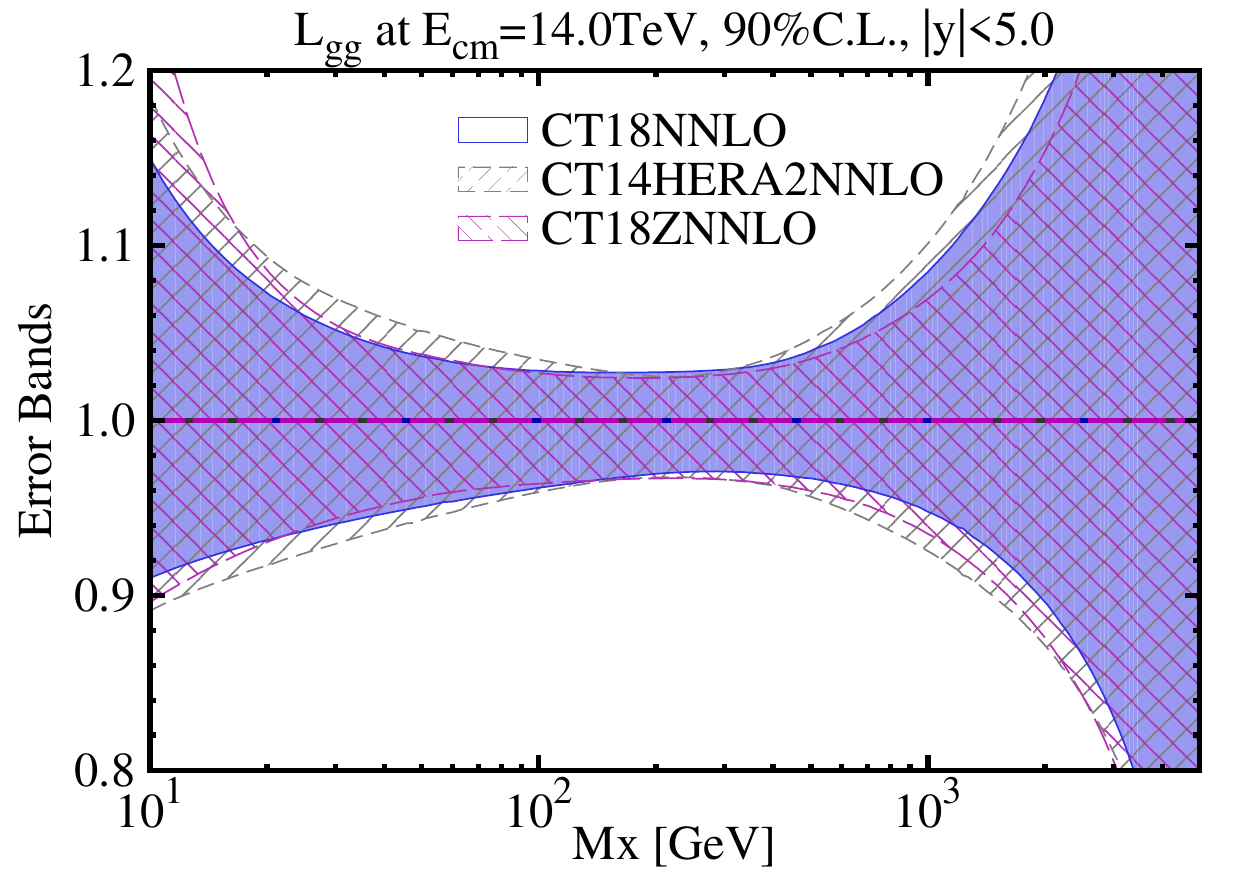}
	\end{center}
	\vspace{-2ex}
	\caption{
		Parton luminosities for processes
		at the LHC at $\sqrt{s} = 14$ TeV, in the central
                rapidity region $|y|<5$: $L_{qq}$ (upper panels), $L_{gq}$ (center panels),
		and $L_{gg}$ (lower panels); evaluated using CT18 (solid violet), CT18Z (short-dashed
		gray), and CT14$_\mathrm{HERAII}$ (long-dashed magenta) NNLO PDFs. The left panels give
		the luminosity ratios normalized to CT18, whereas the right panels show the error bands for
		each luminosity, normalized for each PDF ensemble to
                its own central prediction. 
	}
\label{fig:lumia}
\end{figure}

\end{widetext}

In Fig.~\ref{fig:lumia}, we show the parton luminosities
at the
LHC 14 TeV computed with the CT18 and CT18Z NNLO PDFs, contrasting them with
the previous \CTHERAII~release. To compare the luminosities only within the physically accessible regions, we compute the integrals of the luminosity with a restriction on the absolute rapidity of the final state to be
within 5 units, cf. Eq. (28) in \cite{Hou:2016sho}. In the comparisons for each
flavor combination, we again show results normalized either to a common reference 
(either CT18 NNLO or NLO) in the left-hand plots or to their respective central
predictions.

As in the case of individual PDFs, the CT18 central results for the parton
luminosities remain very close to \CTHERAII. On the other hand, the
PDF uncertainties of the luminosities for the individual PDF ensembles
are somewhat reduced, especially for those 
luminosities involving gluons. In the region of the Higgs 
boson mass, $M_X\! \sim\! 125$ GeV, the improvement on the $gg$ luminosity, $L_\mathit{gg}$, shown in the lowest panels of Fig.~\ref{fig:lumia}, is very small.
In the TeV-scale mass range, however, reductions in the PDF uncertainties of $L_\mathit{gg}$ are more sizable, closer to $\sim\!20\%$.
Parton luminosities computed using CT18Z NNLO behave distinctly from CT18 in several respects. For example, in the
$W/Z$ boson-mass region, the central predictions for the $qq$ luminosity are approximately $3\!-\!4\%$
higher in CT18Z relative to CT18. The other parton luminosities are similarly enhanced in CT18Z
in the low-mass region, $M_X\! \lesssim\! 100$ GeV, primarily because of the $x$-dependent DIS factorization scale used in CT18Z. This small-$x$ enhancement is about the same in CT18X and Z, in contrast to CT18, which more closely resembles \CTHERAII.
While the high-mass quark-quark luminosity, $L_{qq}$, is relatively unmodified in CT18Z, 
$L_{gq}$ and $L_{gg}$ are suppressed for $M_X\! \gtrsim\! 100\!-\!300$ GeV; for the
gluon-gluon luminosity, this suppression can be as large as $\sim\!4\%$
between 100 GeV and 1 TeV. 

\begin{figure}[!htbp]
	\begin{center}
		\hspace*{-0.4cm}
		\includegraphics[width=0.48\textwidth]{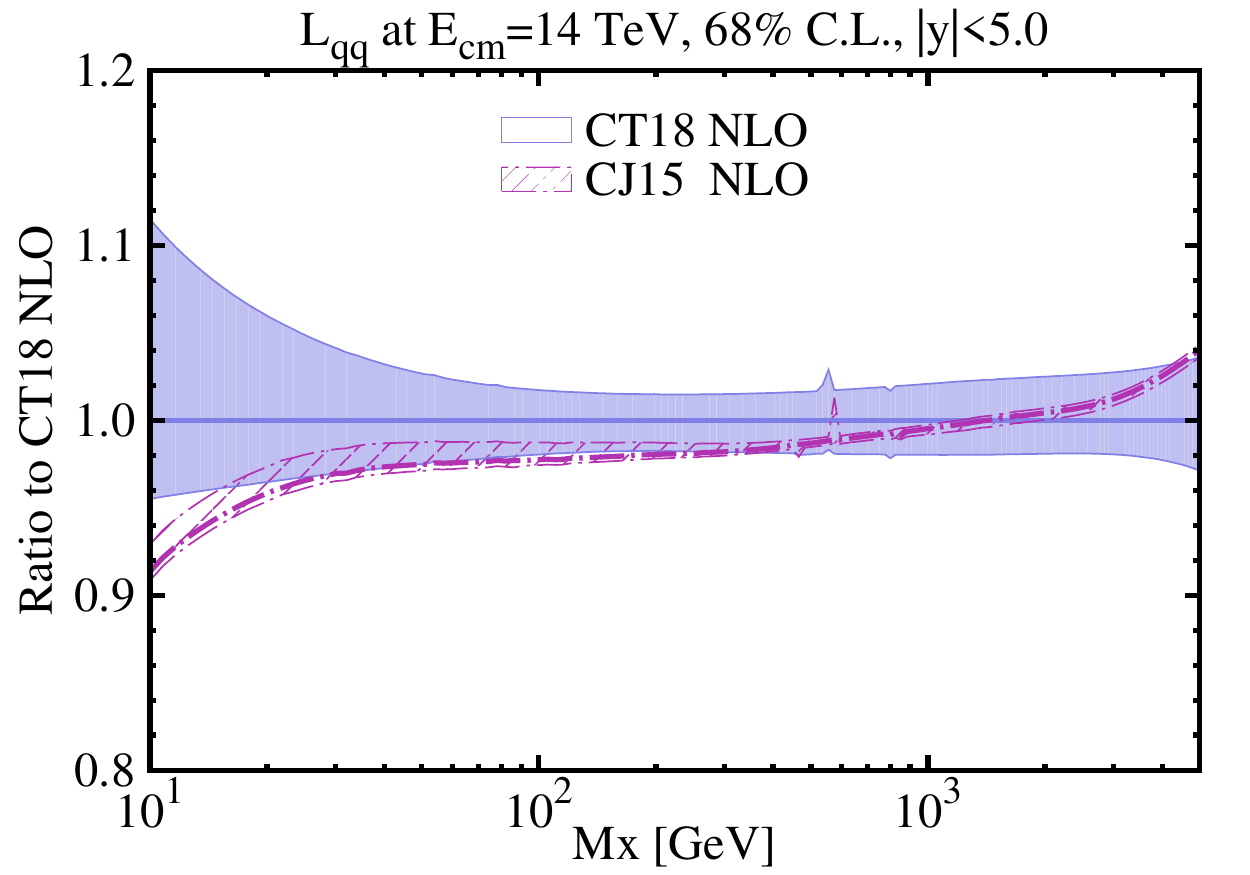} 
		\includegraphics[width=0.48\textwidth]{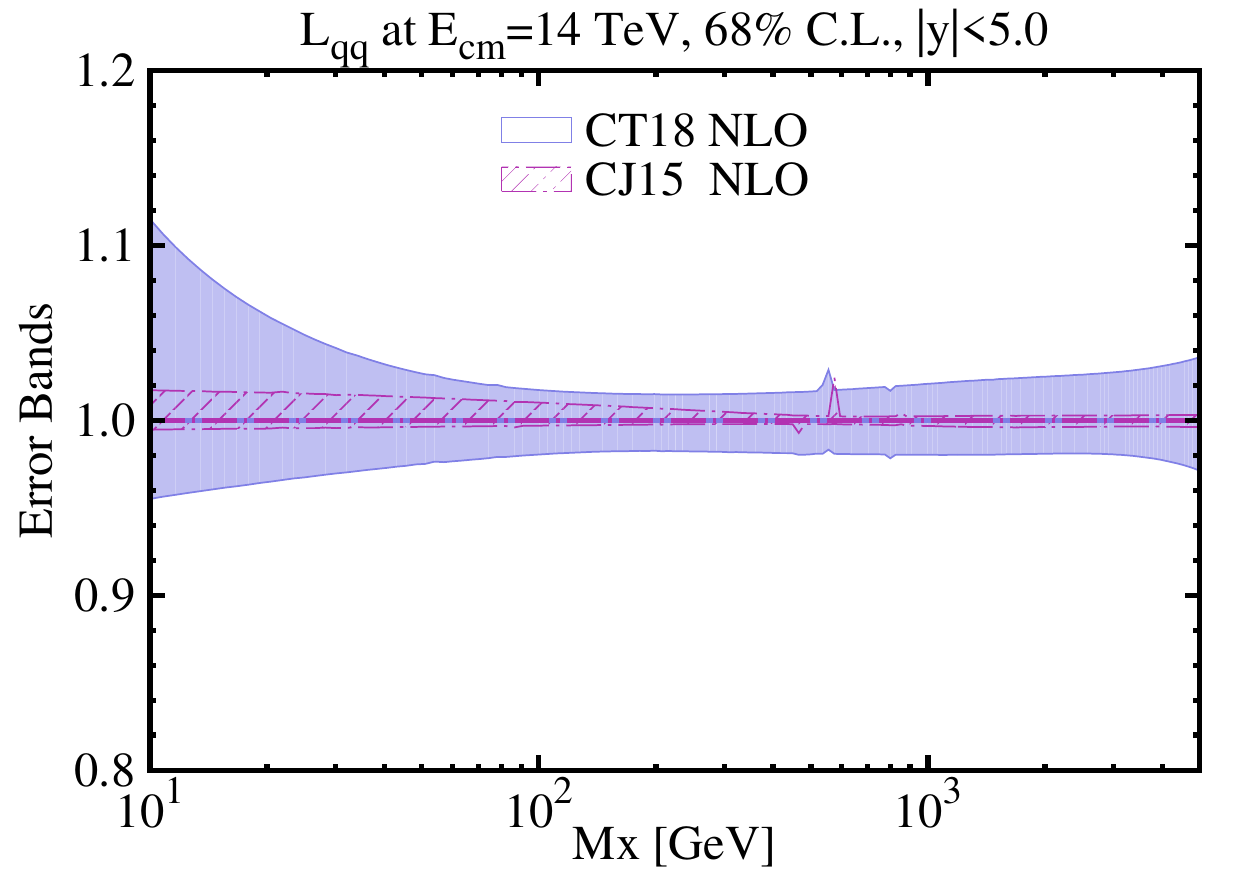} \\
		\includegraphics[width=0.48\textwidth]{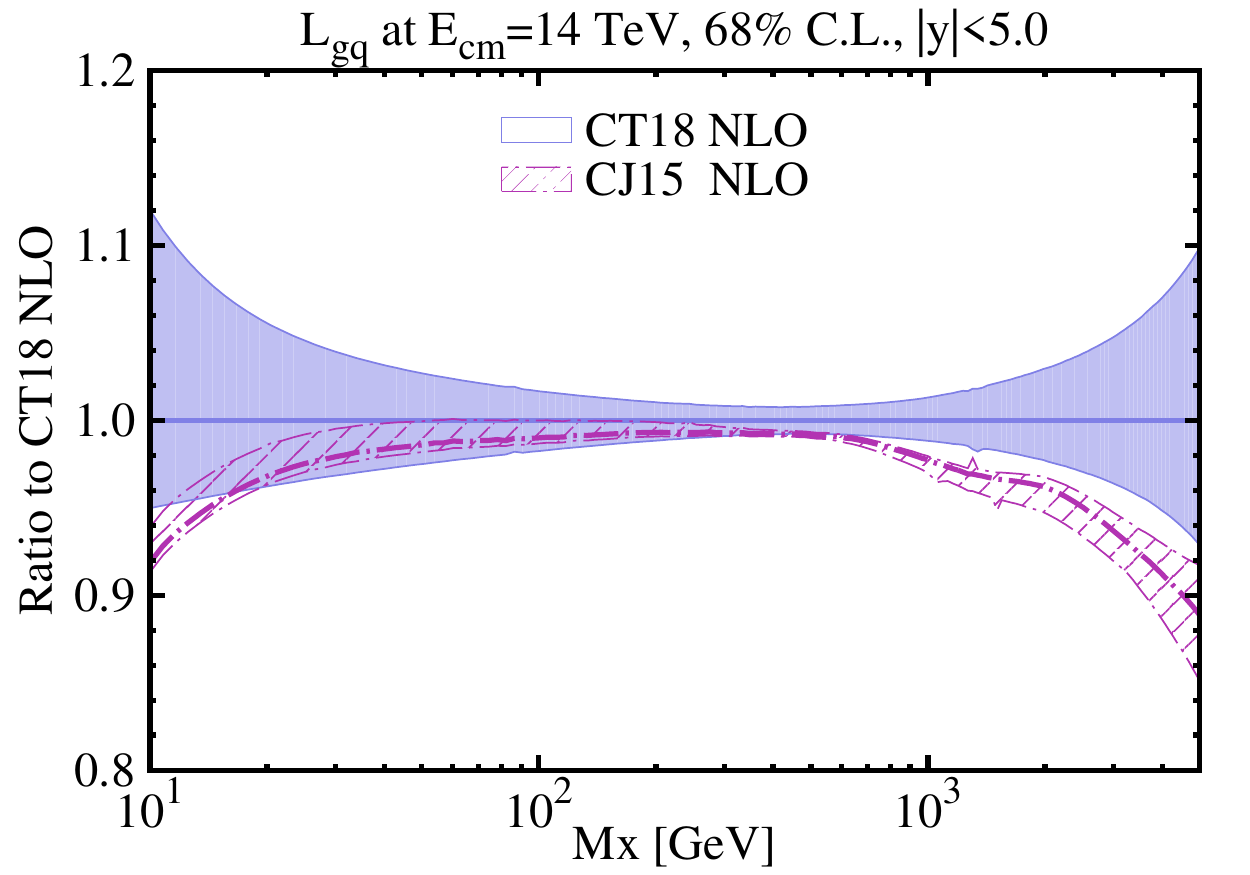} 
		\includegraphics[width=0.48\textwidth]{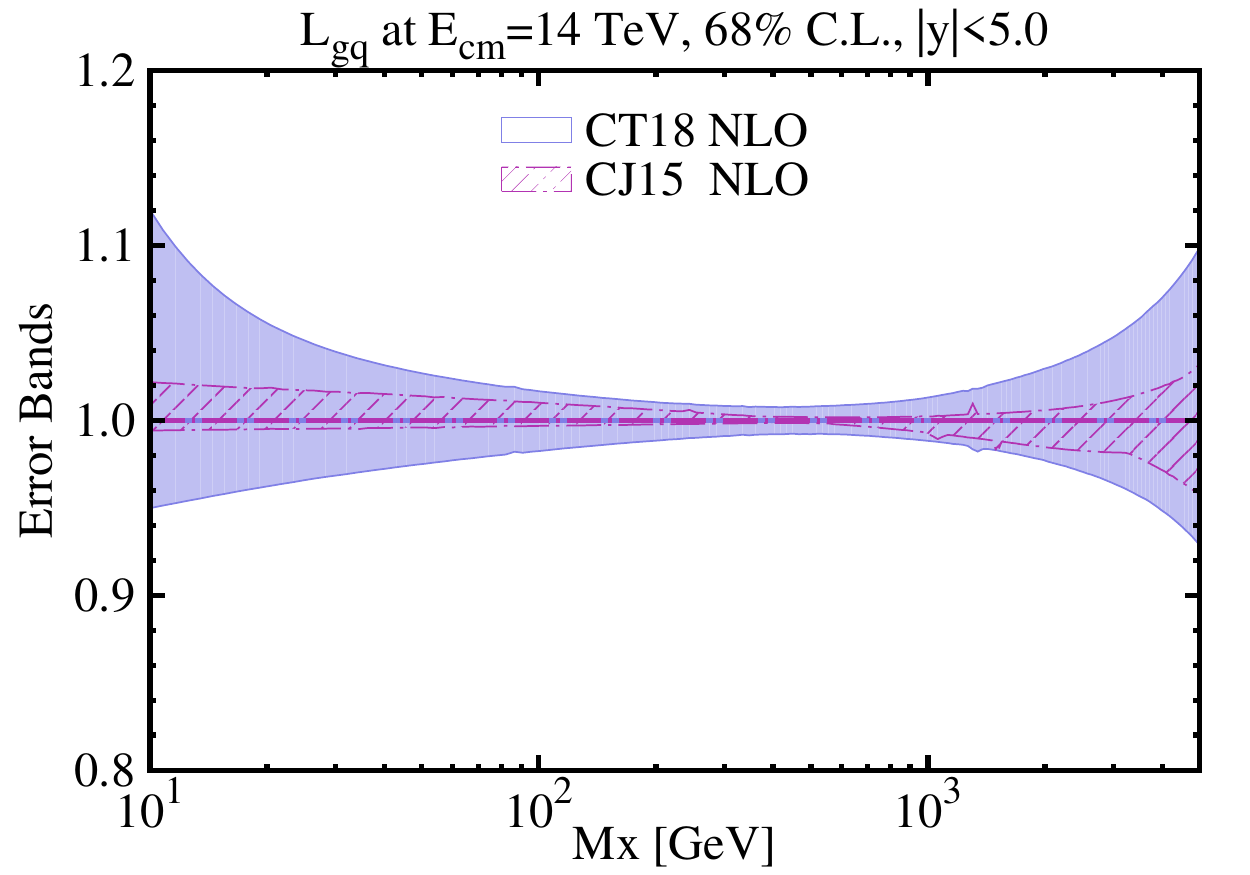} \\
		\includegraphics[width=0.48\textwidth]{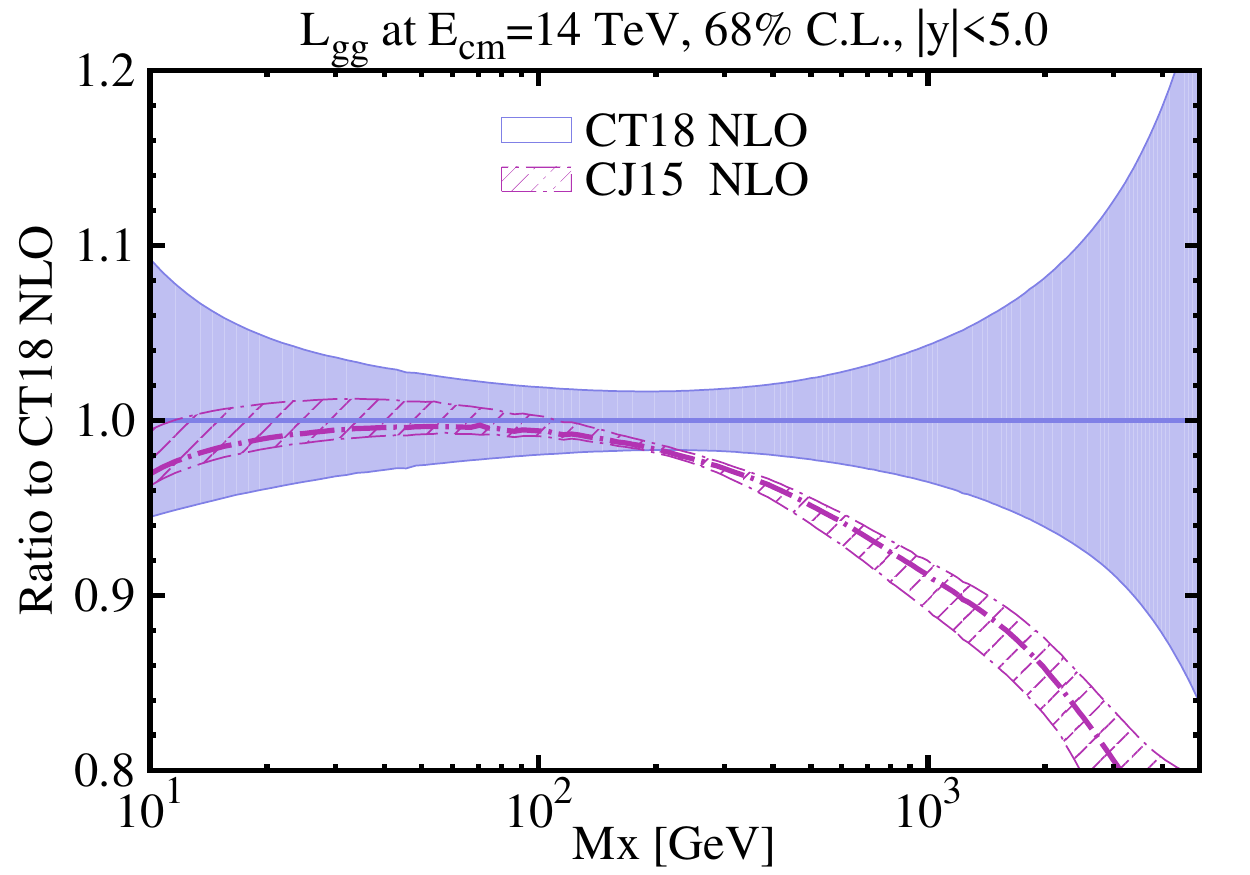}
		\includegraphics[width=0.48\textwidth]{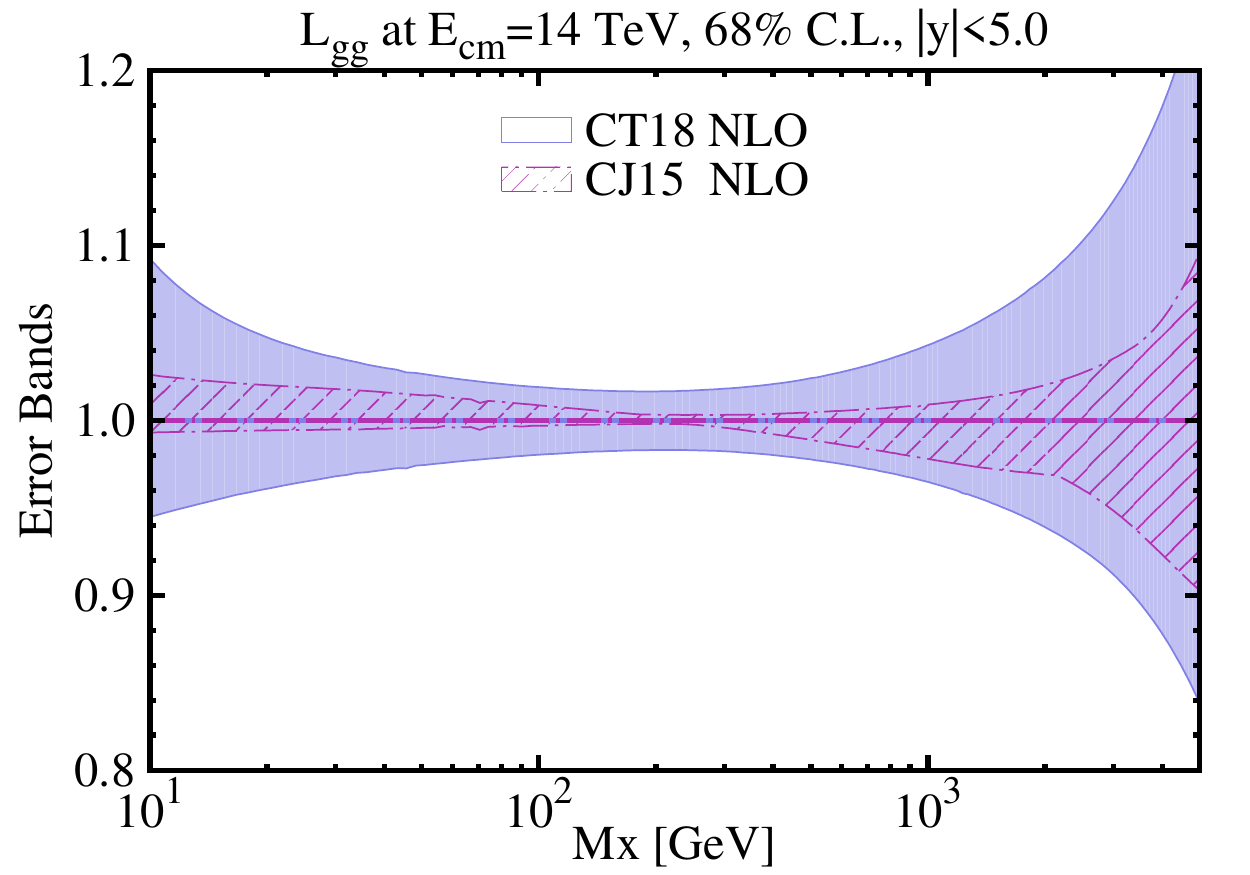}

	\end{center}
	\vspace{-2ex}
	\caption{Same as Fig.~\ref{fig:lumia}, comparing the CT18 NLO and CJ15 NLO parton luminosities.
	}
\label{fig:lumiCT18NLOvsothers}
\end{figure}

\begin{figure}[!htbp]
	\begin{center}
		\hspace*{-0.4cm}
		\includegraphics[width=0.48\textwidth]{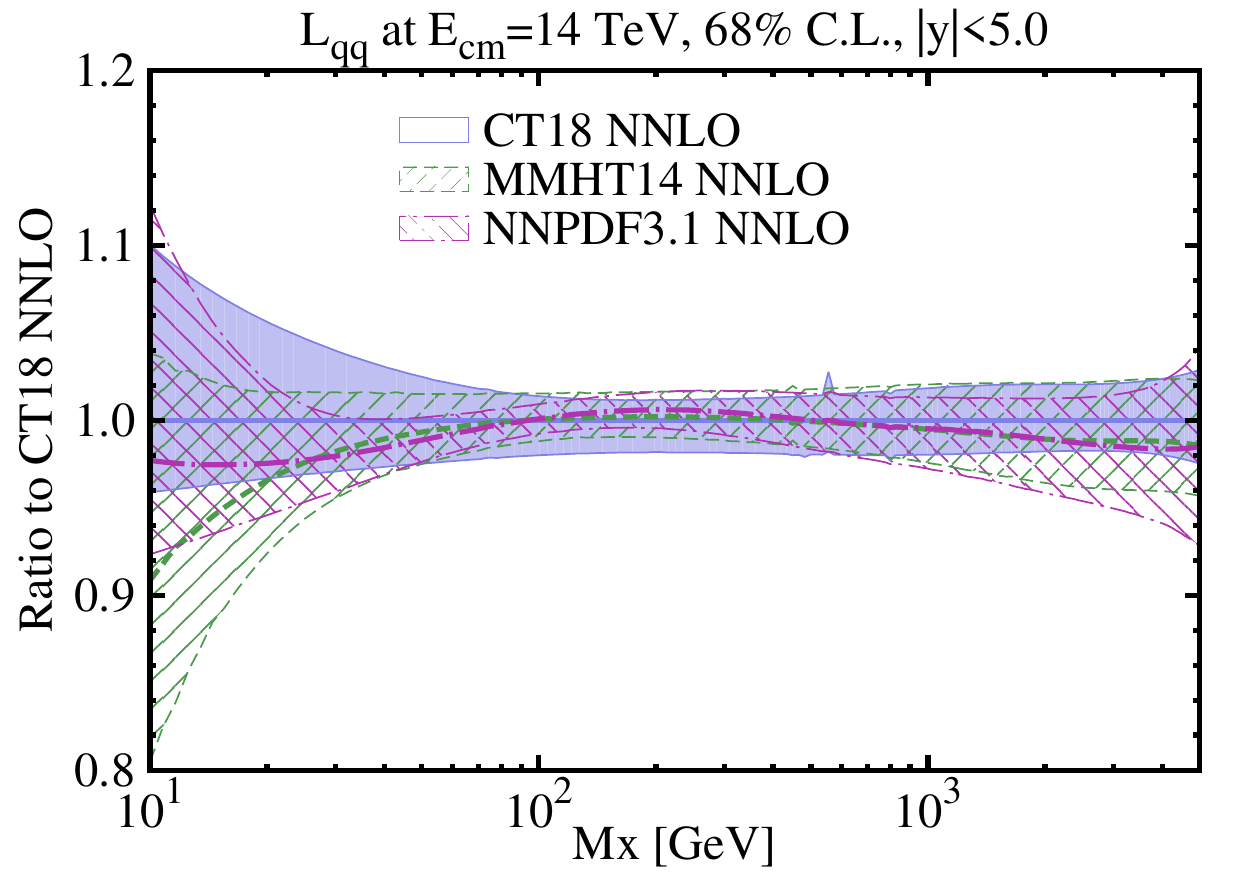} 
		\includegraphics[width=0.48\textwidth]{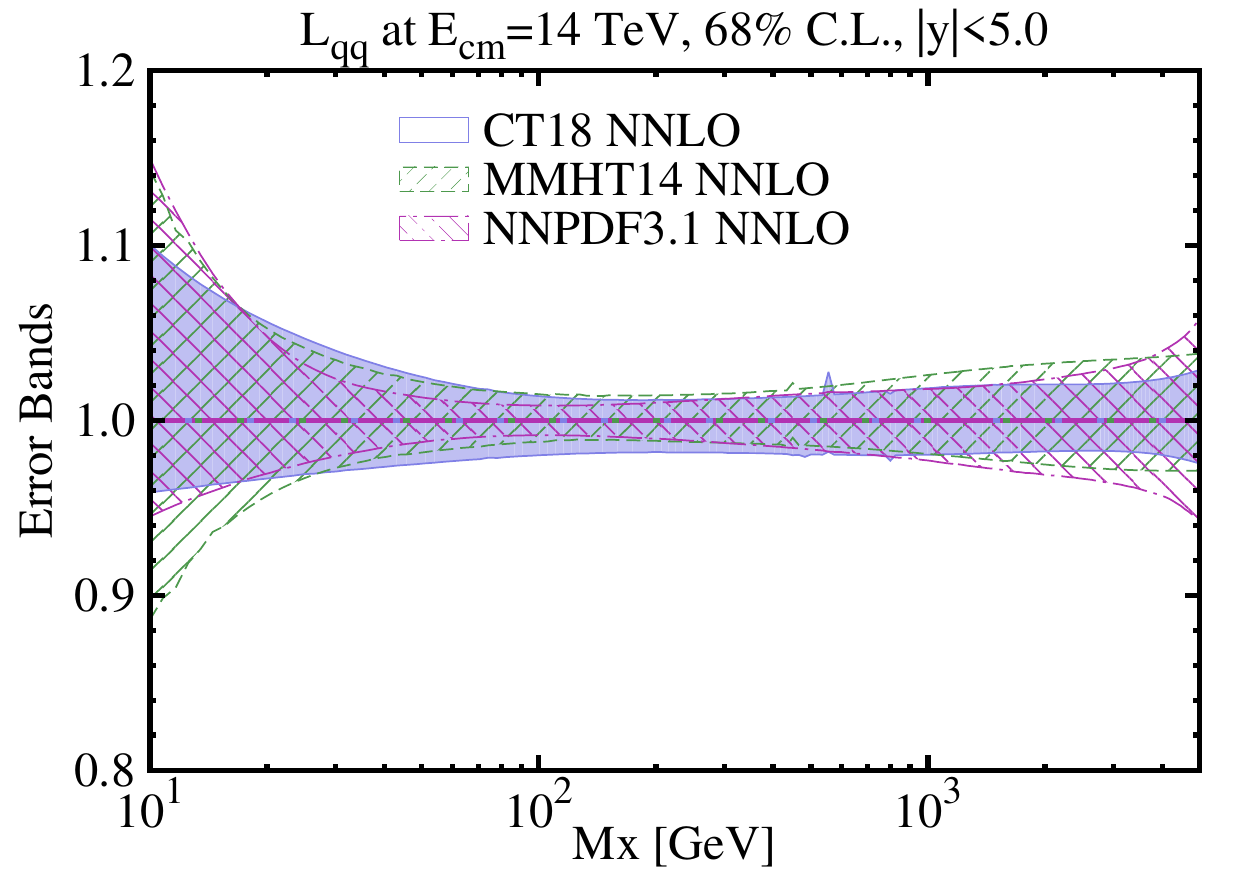} \\
		\includegraphics[width=0.48\textwidth]{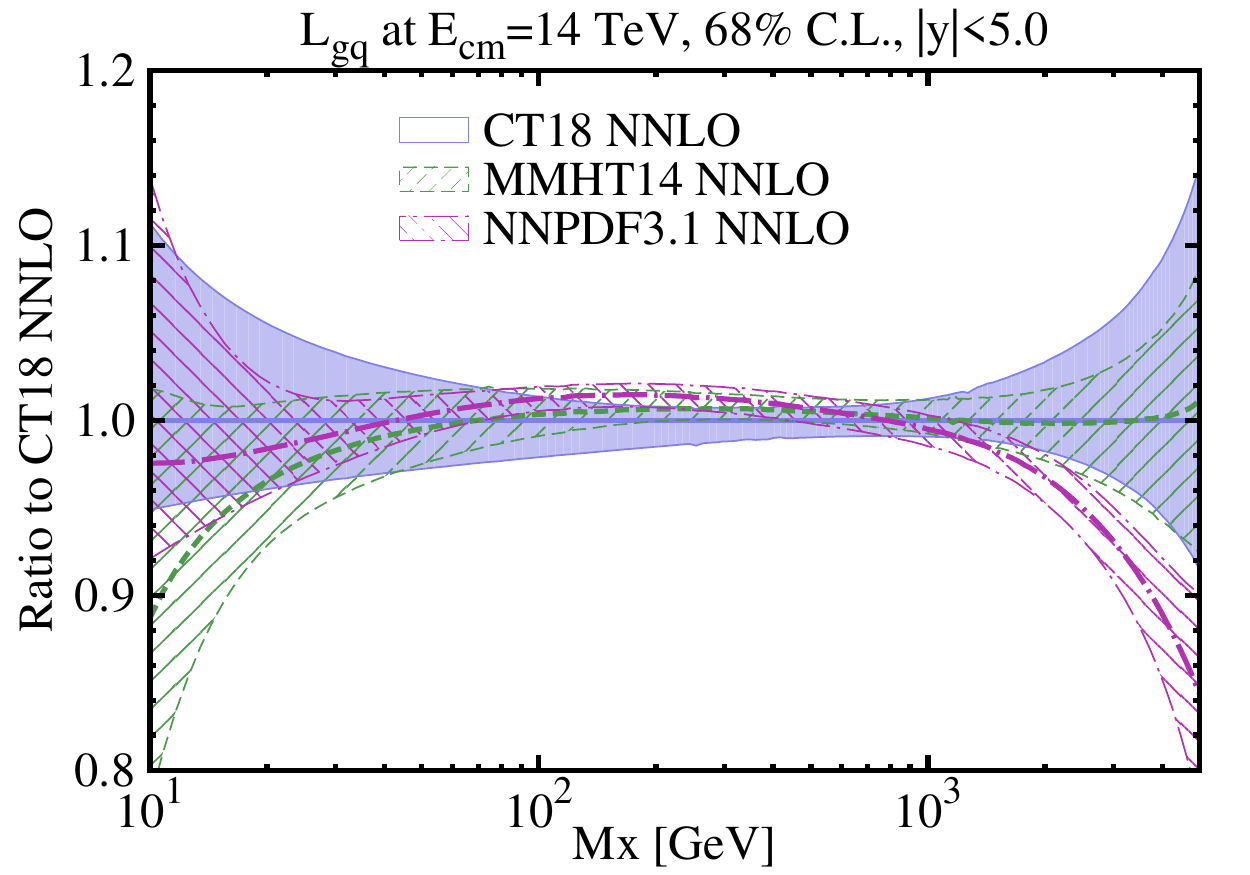} 
		\includegraphics[width=0.48\textwidth]{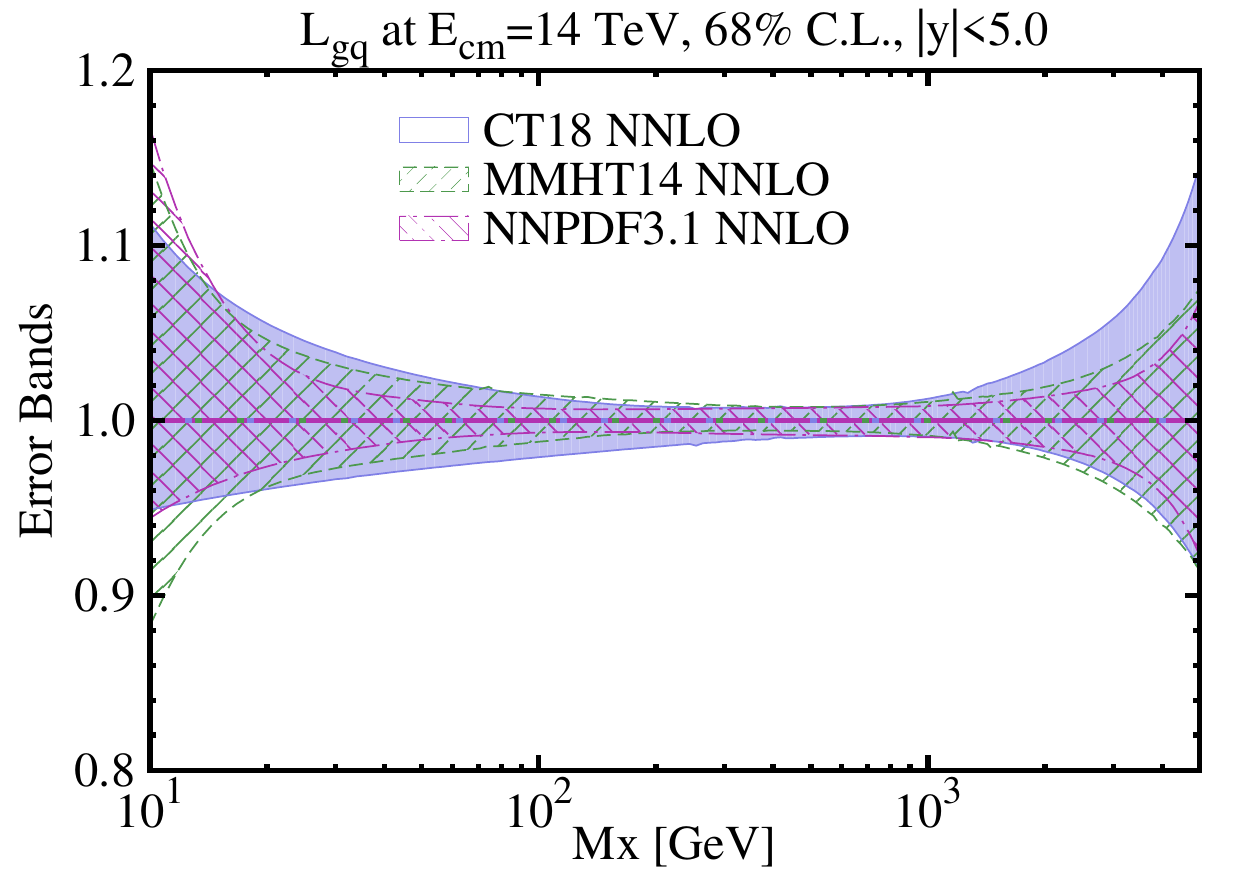} \\
		\includegraphics[width=0.48\textwidth]{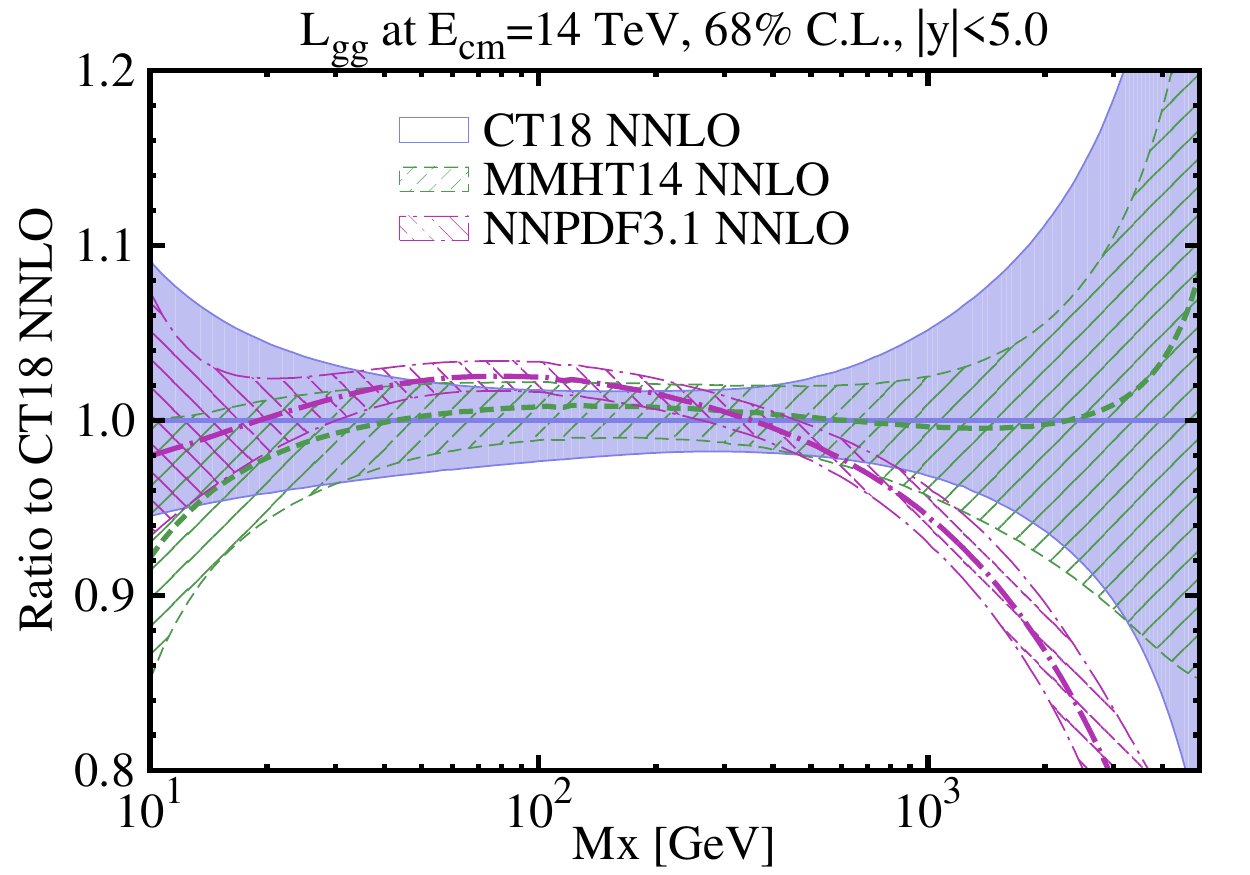}
		\includegraphics[width=0.48\textwidth]{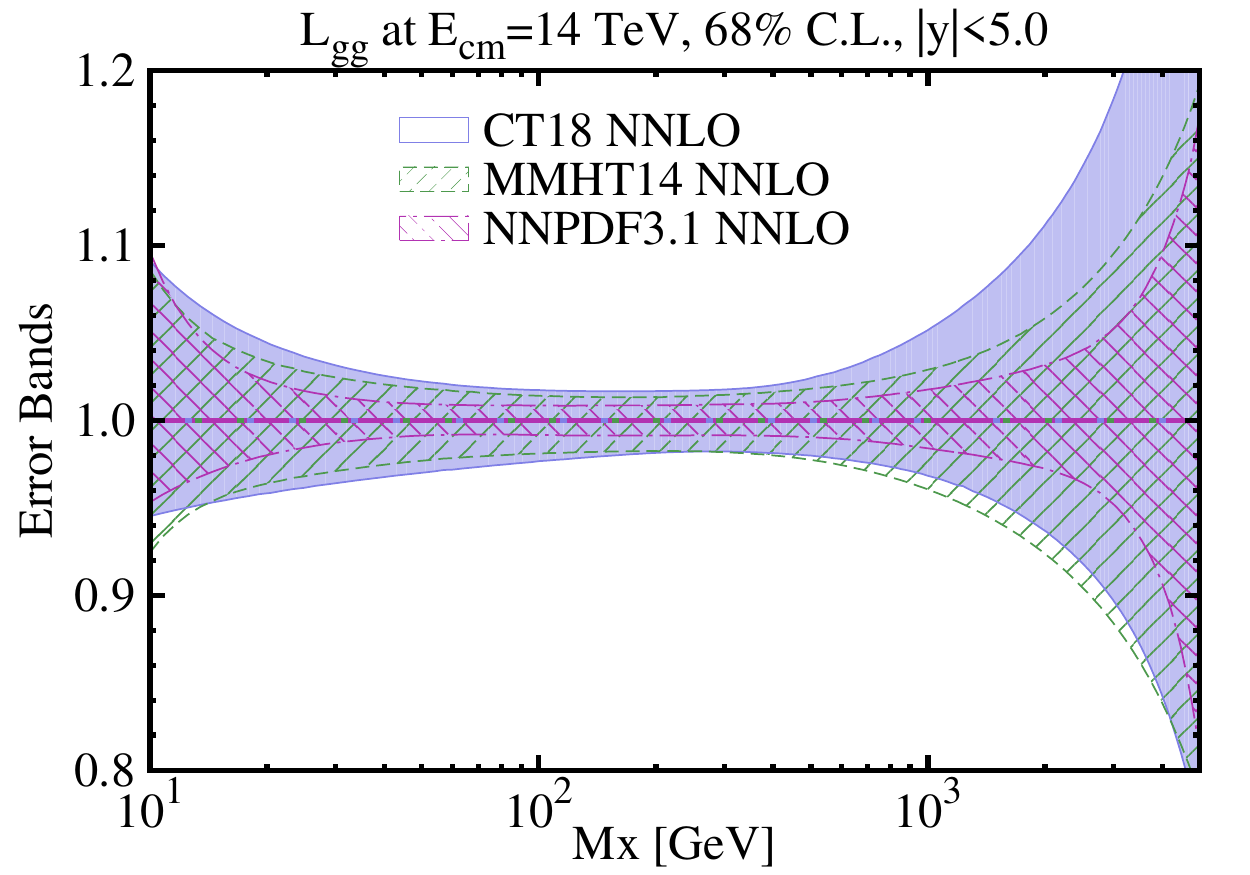}

	\end{center}
	\vspace{-2ex}
	\caption{Same as Fig.~\ref{fig:lumia}, comparing the CT18,
          MMHT'2014, and NNPDF3.1 NNLO parton luminosities with
          $\alpha_s(M_Z)=0.118$.  
	}
\label{fig:lumiCT18NNLOvsothers}
\end{figure}

In Figs.~\ref{fig:lumiCT18NLOvsothers}
and~\ref{fig:lumiCT18NNLOvsothers}, we compare these
parton luminosities against those from other groups:
CJ15~\cite{Accardi:2016qay}, MMHT14~\cite{Harland-Lang:2014zoa}, and
NNPDF3.1~\cite{Ball:2017nwa}. Here we adopt the rescaled 68\% C.L. for
the CT18 PDFs to match the convention of the other groups. 
The comparison in Fig.~\ref{fig:lumiCT18NLOvsothers} is done at NLO, 
because CJ15 PDFs are not available at NNLO in QCD. 
The PDF uncertainties on the CJ15 NLO luminosities
are smaller than those on the CT18 NLO (see right insets), in part 
due to a smaller tolerance criterion ($\Delta \chi^2 =1$) and less
flexible parametrization forms for the $\bar u$ and $\bar d$ PDFs
employed by CJ15.
The CT18 NLO $qq$ luminosity central value is approximately 8\% to 5\% higher than CJ15 for mass 
values $10\lesssim M_{X}\lesssim 2\cdot  10^{3}$ GeV and up to 8-10\% lower for higher values (see left insets). 
The CT18 PDF error band for $L_{qq}$ covers that of CJ15  over the mass range $20\lesssim M_{X}\lesssim 2\cdot  10^{3}$ GeV. 
The CT18 NLO $gq$ luminosity is approximately 2\% higher than CJ15 in
the mass region relevant for Higgs production,
$100\lesssim M_{X}\lesssim 300$ GeV. It is also higher everywhere
else, with major differences present at low masses, $M_X\lesssim 100$ GeV, where it is 8\% higher,
and at high masses, $M_X \gtrsim 650$ GeV, where differences are about 12\% at $M_X \approx 5$ TeV. 
For the $gg$ luminosity, CT18 at NLO is higher everywhere, in particular, differences are larger than 20\% at $M_X \approx 5$ TeV.

The luminosities obtained using CT18 NNLO PDFs are compared to those obtained with MMHT14 and NNPDF3.1 PDFs in Fig.~\ref{fig:lumiCT18NNLOvsothers}. 
The NNPDF3.1 PDFs set is selected with $\alpha_s(M_Z) = 0.118$. 
The  central values  of $qq$ and $gq$ luminosities for the three
groups agree within a few percent in the mass region $100\lesssim
M_{X}\lesssim 10^{3}$ GeV, where they also have comparable PDF
uncertainties.
Comparing the NNLO $gg$ luminosities in the mass range $20 < M_X < 300$ GeV,  we see that MMHT14 is within a percent or so of CT18, while NNPDF3.1 is $2-3\%$ higher. Furthermore, the uncertainty band for MMHT14 in this range is similar to that of CT18, while that of NNPDF3.1 is smaller.
At larger masses $M_X \gtrsim 300$ GeV, we observe a rapid drop of the
NNPDF3.1 luminosity. Moreover, the NNPDF3.1 uncertainty is smaller
over all the mass range.      

In 2012-2015, several detailed studies \cite{Ball:2012wy, Rojo:2015acz, Butterworth:2015oua} of the contemporary global PDF fits were carried out, including benchmark comparisons of their methodologies. The understanding gained from those studies has led to the 2015 recommendation on the usage of PDFs at the LHC \cite{Butterworth:2015oua}. The benchmarking also resulted in the improved agreement among the CT, MMHT, and NNPDF PDFs, which in turn allowed the PDF4LHC working group to combine these global PDFs as inputs  into the widely used PDF4LHC15 PDF ensembles. 

Since 2015 a great deal of LHC data has been added to the latest global fits. 
This has led in some cases to an increase in the differences among the central PDFs of the groups as compared to the corresponding 2015 PDF releases. This change may be attributable to various factors. In particular, small-$x$ resummation or a non-conventional choice of QCD scales in NNLO DIS cross sections modify the small-$x$ PDFs, as exemplified by CT18Z NNLO in Figs.~\ref{fig:PDFbands1} and \ref{fig:lumia}. The NNPDF3.1 parton luminosity show more pronounced differences vs. CT18 and MMHT2014 in some regions, cf. Fig.~\ref{fig:lumiCT18NNLOvsothers}. A followup study is currently underway to better understand the impact of the LHC data and methodological choices on each global PDF.

\clearpage

\clearpage

\subsection{PDF moments and sum rules}
\label{sec:moments}

Knowledge of the integrated PDF {\it Mellin} moments has long been of interest, both for their phenomenological
utility, and for their relevance to lattice QCD computations of hadronic structure \cite{Lin:2017snn,Hobbs:2019gob,Lin:2020rut}.
In the case of the former, PDF moments can serve as valuable benchmarks for the purpose of comparing various global
analyses and theoretical approaches, and can also be informative descriptors of the PDFs themselves. This follows
especially from the fact that numerical results obtained for PDFs of a given order are connected with the $x$
dependence of the underlying parton distribution, with, in general, higher-order Mellin moments mostly determined
by the PDFs' high-$x$ behavior.  In Ref.~\cite{Hobbs:2019gob}, an analysis of the sensitivities of HEP data to
lattice-calculable quantities --- specifically, the Mellin moments and parton quasi-distribution functions --- was
performed to further develop the still-emerging PDF-Lattice effort \cite{Lin:2017snn,Lin:2020rut}. 

Integrated moments can in general be evaluated for practically any phenomenological PDF from its underlying distribution,
provided the moment in question is convergent over the full range of support. However, in this analysis we concentrate
special attention on

\begin{equation}
\langle x^n\rangle_g(Q)=\int_{0}^{1}dx\ x^{n}\,  g(x,\Q) \, ,
\label{eq:gmom}
\end{equation}
with $n\!=\!1$ for the gluon, as well as
\begin{align}
	\langle x^n\rangle_{q^+}(Q) &= \int_{0}^{1}dx\ x^{n}\,  [q + \overline{q}](x,\Q) \ \ \mathrm{for}\ \ n = 1, 3, \dots \nonumber \\
	\langle x^n\rangle_{q^-}(Q) &= \int_{0}^{1}dx\ x^{n}\,  [q - \overline{q}](x,\Q) \ \ \mathrm{for}\ \ n = 2, 4, \dots
\label{eq:qmom}
\end{align}
for the quark distributions, where we denote the charge conjugation-even (odd) quark combinations as $q^\pm = q \pm \overline{q}$.	
We primarily consider these specific PDF moments of Eq.~(\ref{eq:gmom}) with $n\!=\!1$ and Eq.~(\ref{eq:qmom}) for compatibility with lattice QCD determinations,
which are only able to compute the odd ($n\! =\! 1, 3, \dots$) moments of $q\!+\!\bar{q}$-type distributions and even ($n\! =\! 2, 4, \dots$) moments
for $q\!-\!\bar{q}$-type distributions.
This follows from the fact that lattice calculations extract the integrated Mellin moments from hadronic matrix elements as,
\begin{equation}
	\frac{1}{2}\sum_{s}\langle p,s|\mathcal{O}_{\{\mu_{1},\cdots,\mu_{n+1}\}}^{q}|p,s\rangle = 2 \langle x^{n+1} \rangle_{q}\,[p_{\mu_{1}}\cdots p_{\mu_{n+1}}-{\rm traces}]\ .
\label{eq:OPE}
\end{equation}
In Eq.~(\ref{eq:OPE}), the lattice operators are
$\mathcal{O}_{\{\mu_{1},\cdots,\mu_{n+1}\}}\! \sim\! i^n \overline{q} \gamma_{\mu_1} \overleftrightarrow{D}_{\mu_2} \cdots \overleftrightarrow{D}_{\mu_{n+1}} q$, involving
covariant derivatives in such a way that successive derivative insertions increase the order of the extracted moment, but also alternate the evenness and oddness under charge conjugation.

\begin{table}
\begin{tabular*}{\textwidth}{c| @{\extracolsep{\fill}} cc|c|ccc}
\hline
PDF moment                             &   {\bf CT18}    &    {\bf CT18Z}   &    {\bf \CTHERAII}   &  {\bf MMHT14}   &   {\bf CJ15}      &     {\bf NNPDF3.1}    \tabularnewline
\hline                                                                                                                                                        
$\langle x \rangle_{u^+-d^+}$          &  $0.156(7)$     &    $0.156(6)$    &    $0.159(6)$     &   $0.151(4)$    &    $0.1518(13)$   &   $0.152(3)$          \tabularnewline
$\langle x^2 \rangle_{u^--d^-}$        &  $0.055(2)$     &    $0.055(2)$    &    $0.055(2)$     &   $0.053(2)$    &    $0.0548(2)$    &   $0.057(3)$          \tabularnewline
$\langle x^3 \rangle_{u^+-d^+}$        &  $0.022(1)$     &    $0.022(1)$    &    $0.022(1)$     &   $0.022(1)$    &    $0.0229(1)$    &   $0.022(1)$          \tabularnewline
\hline                                                                                                                                                        
$\langle x \rangle_{g}$                &  $0.414(8)$     &    $0.407(8)$    &    $0.415(8)$     &   $0.411(9)$    &    $0.4162(8)$    &   $0.410(4)$          \tabularnewline
\hline                                                                                                                                                        
$\langle x \rangle_{u^+}$              &  $0.350(5)$     &    $0.350(4)$    &    $0.351(5)$     &   $0.348(5)$    &    $0.3480(6)$    &   $0.348(4)$          \tabularnewline
$\langle x \rangle_{d^+}$              &  $0.193(5)$     &    $0.194(5)$    &    $0.193(6)$     &   $0.197(5)$    &    $0.1962(9)$    &   $0.196(4)$          \tabularnewline
$\langle x \rangle_{s^+}$              &  $0.033(9)$     &    $0.041(8)$    &    $0.031(8)$     &   $0.035(8)$    &    $0.0313(2)$    &   $0.039(4)$          \tabularnewline
\hline                                                                                                                                                        
$\langle x^2 \rangle_{u^-}$            &  $0.085(1)$     &    $0.084(1)$    &    $0.085(1)$     &   $0.083(1)$    &    $0.0853(2)$    &   $0.085(3)$          \tabularnewline
$\langle x^2 \rangle_{d^-}$            &  $0.030(1)$     &    $0.029(1)$    &    $0.030(1)$     &   $0.030(1)$    &    $0.0305(2)$    &   $0.028(3)$          \tabularnewline
$\langle x^2 \rangle_{s^-}$            &    ---          &      ---         &      ---          &   $0.001(1)$    &      ---          &   $0.001(4)$          \tabularnewline
\hline                                                                                                                                                        
$\langle 1 \rangle_{\bar{d}-\bar{u}}$  &  $-0.12(35)$    &    $-0.07(29)$   &    $-0.37(41)$    &   $0.084(15)$   &    $0.103(20)$    &     ---               \tabularnewline
$\kappa_s$                             &  $0.49(16)$     &    $0.61(14)$    &    $0.46(13)$     &   $0.51(14)$    &      ---          &   $0.563(82)$         \tabularnewline
\end{tabular*}\caption{
	We collect values of several PDF moments computed according to CT18, CT18Z, CT14$_\mathrm{HERAII}$, MMHT14, CJ15 NLO, and
	NNPDF3.1, all at the scale $Q=2$ GeV. The moments are chosen for their dual interest both as benchmarks for phenomenological calculations and relevance
	to lattice QCD calculations. In the descending order, we show the three lowest moments of the isovector ($u\!-\!d$) distribution, the first
	moment of the gluon, the first and second moments, respectively, for the flavor-separated $u, d, s$ distributions, and two measures
	of light quark flavor symmetry violation: the zeroth moment of the flavor $\mathrm{SU}(2)$ difference, $\langle 1 \rangle_{\bar{d}-\bar{u}}$,
	and the moment ratio related to the strangeness suppression, $\kappa_s$, defined in Eq.~(\ref{eq:str-supp}). We note that the $\bar{d}$ and
	$\bar{u}$ distributions are not constrained to coincide at $x\!\to\!0$ in NNPDF3.1, leaving $\langle 1 \rangle_{\bar{d}-\bar{u}}$ undefined,
	whereas the strange suppression factor was not fitted in CJ15.  Here, all computed moment uncertainties are either based on 68\% C.L., or have been rescaled accordingly for comparison.
}
\label{tab:moments}
\end{table}

\begin{figure*}
\includegraphics[scale=0.29]{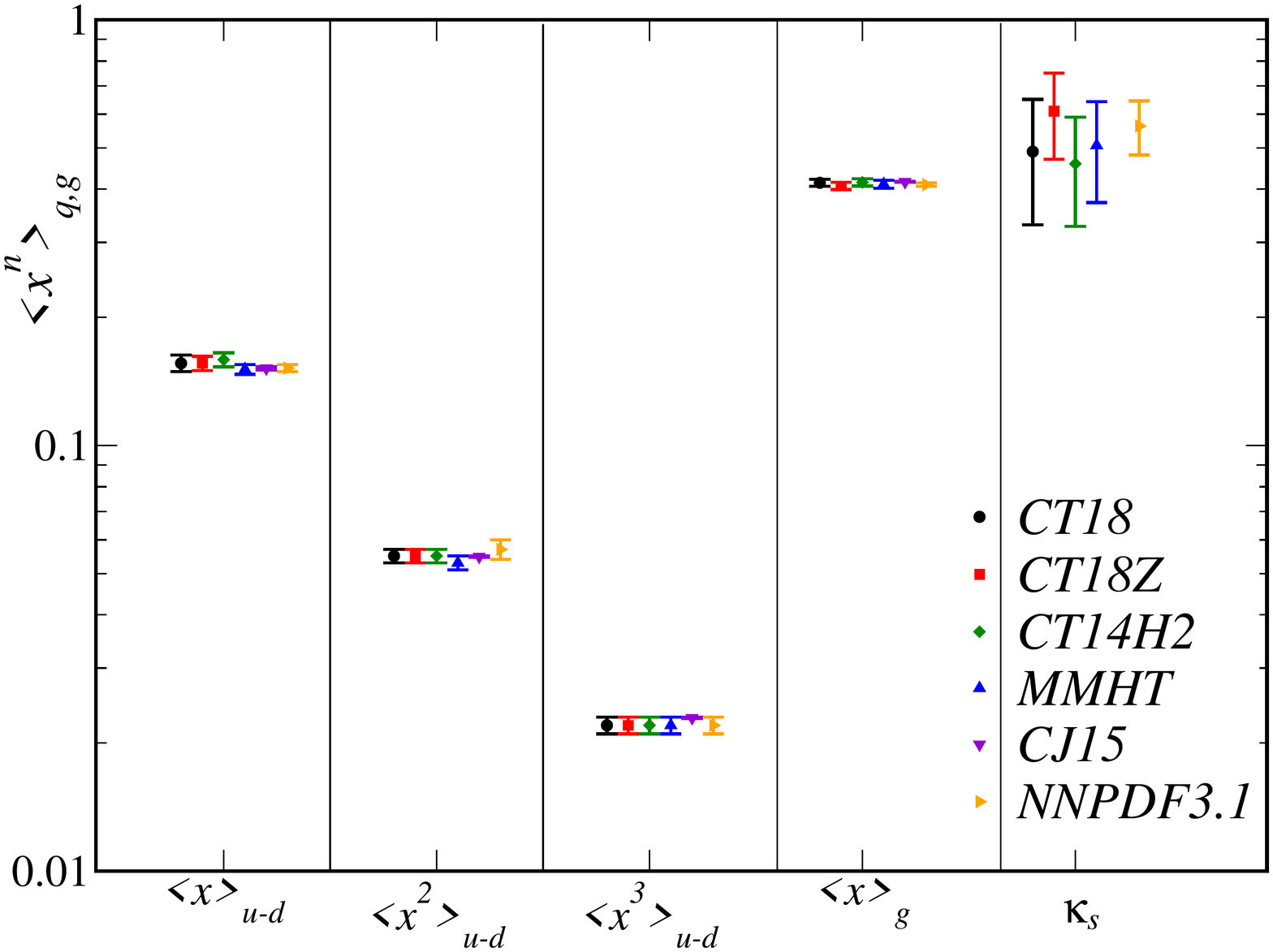}
\includegraphics[scale=0.29]{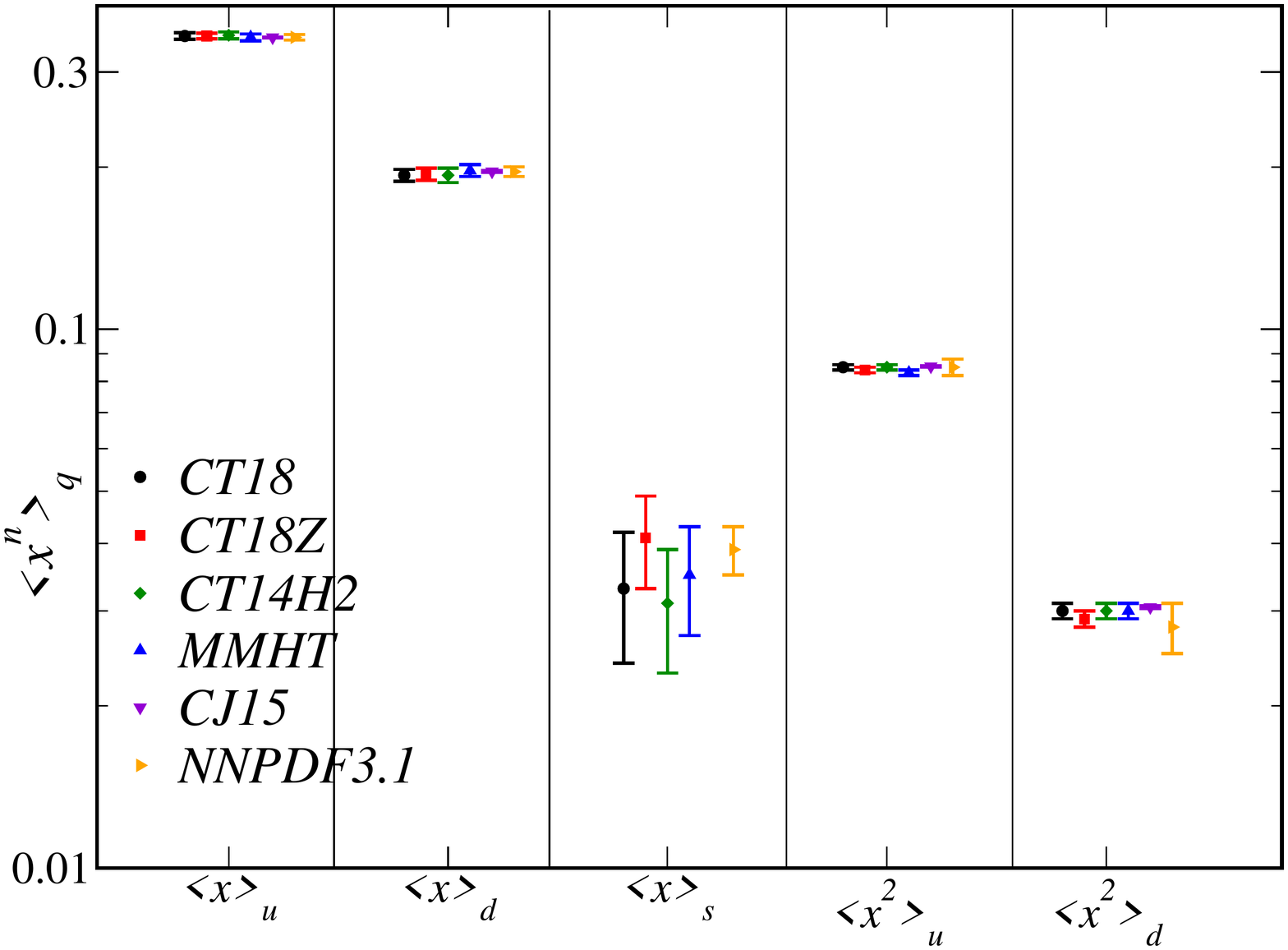}
\caption{
	A graphical comparison of the PDF moments summarized in Table~\ref{tab:moments}, with the exception of the
	results for the zeroth moment of $\bar{d}-\bar{u}$ combination, relevant for studies of the
	Gottfried Sum Rule; this latter quantity is given in
        Fig.~\ref{fig:moments2}. The CJ15 global fit does not
        determine $\kappa_s$ and $\langle x \rangle_{s}$ as
        independent entities from the data, their respective
        predictions are not shown. 
}
\label{fig:moments1}
\end{figure*}

\begin{figure*}
\includegraphics[scale=0.45]{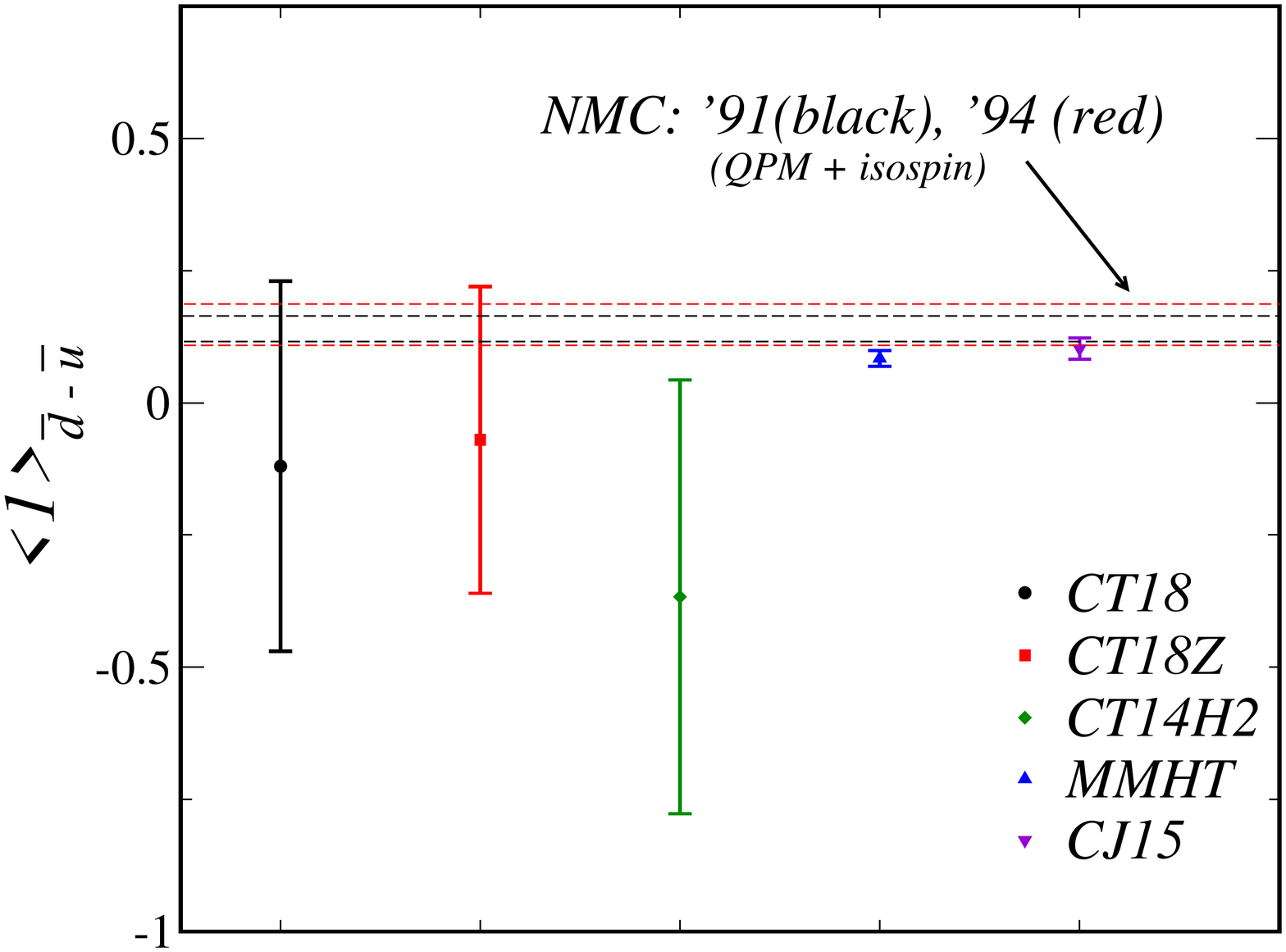}
	\vspace{-1.1cm}
\caption{
	A visual comparison of the results for $\int_0^1 dx [\bar{d}(x)-\bar{u}(x)]$. The horizontal
	nested black and red bands correspond to the values extracted from the original NMC analyses from 1991~\cite{Amaudruz:1991at} and 1994~\cite{Arneodo:1994sh}, respectively. These were based on direct
	quark-parton model extractions of the flavor asymmetry PDF from the deuteron-to-proton
	structure function ratio measured at $Q^2 = 4$ GeV$^2$ for a range of $x \lesssim 0.7$. While
	all but the highest $x$ bin in this data set is consistent with CTEQ kinematical cuts,
	the very low $Q$ is exactly at the boundary of the $Q$ cut,
        and likely subject to substantial higher-twist corrections, especially for the higher $x$ bins.
}
\label{fig:moments2}
\end{figure*}

We compute a number of the typical benchmark PDF Mellin moments using our updated CT18 and CT18Z NNLO fits and compare against
the older CT14$_\mathrm{HERAII}$ NNLO parametrization as well as the recent MMHT14 NNLO, CJ15 NLO, and NNPDF3.1 global analyses.
In all cases, moments are evaluated for an $\overline{\mathrm{MS}}$ factorization scale of $\Q = 2$ GeV, which is also the
standard matching scale computed in lattice QCD calculations. We summarize the numerical results of the PDF moment calculations
in the entries of Table~\ref{tab:moments} as well as in Figs.~\ref{fig:moments1}--\ref{fig:moments2}.
We point out that the comparatively small values of the CJ15 NLO uncertainties are primarily attributable to the use of
$\Delta \chi^2 = 1$ criterion, and, in some cases, a comparatively more restrictive parametrization.

{\it Observations.} In general, we observe concordance among the moments of the light distributions, including those of the isovector
({\it i.e.}, $u\!-\!d$) combination, $\langle x^{1,3} \rangle_{u^+-d^+}$ and $\langle x^2 \rangle_{u^--d^-}$. Notably, the CT results for the first isovector
moment, $\langle x \rangle_{u^+-d^+}\! \sim\! 0.156\!-\!0.159$, are marginally larger than those obtained under the other fits considered
here, which produce $\langle x \rangle_{u^+-d^+}\! \sim\! 0.151\!-\!0.152$, but are nevertheless in close agreement at the 1$\sigma$ level.
Similarly, we recover very robust agreement for the first moment of the gluon PDF, which can be understood to carry $\sim\! 41\%$
of the proton's longitudinal momentum at the scale $\Q = 2$ GeV. We find a slightly smaller total contribution
to the momentum sum rule from the gluon under CT18Z NNLO, which results in $\langle x \rangle_{g} = 0.407(8)$,  but is still easily in agreement within errors with the 
CT18 NNLO calculation, $\langle x \rangle_{g} = 0.414(8)$. This is consistent with the modest reduction in the central
gluon shown for CT18Z in the lower-right panel of Fig.~\ref{fig:PDFbands1}.

For the contributions of the individual flavor-separated quark densities to the proton's longitudinal momentum, we again find
in general strong convergence among our new global analysis and the results of previous and other fits. This is especially
true for the total $u^+$ and $d^+$ first moments, for which we find concordance at $\langle x \rangle_{u^+}\! \sim\! 0.35$
and $\langle x \rangle_{d^+}\! \sim\! 0.193\!-\!194$. The situation is similar for the total nucleon strangeness momentum,
but with a somewhat greater quantitative spread about $\langle x \rangle_{s^+}\! =\! 3-4\%$. For CT18 NNLO,
we obtain $\langle x \rangle_{s^+} = 3.3 \pm 0.9\%$ --- similar to CT14$_\mathrm{HERAII}$.
In shifting to CT18Z, a $24\%$ larger nucleon strange content is preferred, but with comparable error.

In addition to the first moments of the quark and gluon densities, $\langle x \rangle_{q^+,g}$, we also evaluate
the second moments of select $q-\bar{q}$ quark asymmetries according to Eq.~(\ref{eq:qmom}), finding for $\langle x^2 \rangle_{u^-,d^-}$
very close alignment among CT18(Z) and previous calculations. Recent CT fits and CJ15 do not independently parametrize $s$ {\it vs}.~$\overline{s}$,
and we therefore omit entries in Table~\ref{tab:moments} for $\langle x^2 \rangle_{s^-}$.

Results on the integrated PDF moments are also of interest to phenomenological sum rules --- for instance, the Gottfried Sum
Rule \cite{Gottfried:1967kk}, which relates $x^{-1}$-weighted moment of the $F^{p-n}_2 = F^{p}_2\! -\! F^{n}_2$ structure function difference,
\begin{equation}
\int_0^1\frac{dx}{x}F^{p-n}_2(x,\Q)\big|_\mathrm{QPM} =\frac{1}{3} - \frac{2}{3}\int_{0}^{1} dx\, [\bar{d}-\bar{u}](x,\Q)\ ,
\label{eq:Gott}
\end{equation}
to flavor-symmetry violation in the light quark sea via the breaking of the $\mathrm{SU}(2)$ relation
$\overline{d} = \overline{u}$.
For the zeroth moment related to the Gottfried Sum Rule, we obtain $\langle 1 \rangle_{\bar{d}-\bar{u}}\! =\! -0.12\!\pm\!0.35$ in CT18
($\langle 1 \rangle_{\bar{d}-\bar{u}}\! =\! -0.07\!\pm\!0.29$ under CT18Z), generally consistent with other PDF analyses. These other
analyses produce narrower uncertainties for $\langle 1 \rangle_{\bar{d}-\bar{u}}$, but this follows from comparatively
more restrictive parametrizations in the low-$x$ region, $x\! \le\! 0.001$. The zeroth moment is dominated by the low-$x$ behavior
of the $\bar{u},\,\bar{d}$ PDFs, for which high-energy data remain relatively sparse, as can be seen in Fig.~\ref{fig:ct18data_xmu}.
NNPDF, in contrast, imposes no restriction on the relative behavior of $\bar{u}$ and $\bar{d}$ for $x \to 0$, such that
$\langle 1 \rangle_{\bar{d}-\bar{u}}$ is not numerically defined; the corresponding NNPDF3.1~entry is therefore left blank in Table~\ref{tab:moments}
and Fig.~\ref{fig:moments2}.
CT uses a significantly more flexible parametrization for the light-quark sea (with 11 parameters for the combined $\bar{u}$ and $\bar{d}$-PDFs,
compared with 5 parameters in MMHT14 for $\bar{d}-\bar{u}$ \cite{Harland-Lang:2014zoa} and 5 parameters in CJ15 for $\bar{d}/\bar{u}$ \cite{Accardi:2016qay}),
with no constraint on the sign of $\bar{d}\!-\!\bar{u}$, as can be deduced from the $\bar{d}/\bar{u}(x,Q\!=\! 1.4\,\mathrm{GeV})$ ratio plot shown in the upper-left
panel of Fig.~\ref{fig:DBandSBbands}. We therefore find that, with this flexibility in the low-$x$ region important for
$\langle 1 \rangle_{\bar{d}-\bar{u}}$, modern high-energy data still allow a broad range for the zeroth moment.

The CT18(Z) values for $\langle 1 \rangle_{\bar{d}-\bar{u}}$ are in agreement with the moments calculated in the original 1991 and 1994 NMC analyses
\cite{Amaudruz:1991at,Arneodo:1994sh}, which we represent in Fig.~\ref{fig:moments2} as the inner-black and outer-red horizontal bands for the
1991 \cite{Amaudruz:1991at} and 1994 \cite{Arneodo:1994sh} extractions, respectively. Several aspects of the original NMC analysis can be expected
to underpredict the full experimental uncertainty on the Gottfried moment, but chief among these is the fact that NMC was sensitive only to the
region $0.004\! <\! x\! <\! 0.8$. In addition, directly matching the NMC structure function moment to $\langle 1 \rangle_{\bar{d}-\bar{u}}$, as in
Eq.~(\ref{eq:Gott}), entails a leading-order quark-parton calculation which necessarily induces corrections from missing higher orders and other QCD
effects not contained in the bands of Fig.~\ref{fig:moments2}. Moreover, in determining the isovector structure function $F^{p-n}_2$ from deuteron-to-proton cross section ratios,
NMC assumed a fairly restrictive parametrization to perform low-$x$ extrapolations as well as to represent the absolute deuteron structure function.
For the sake of comparison, it is instructive to consider the Gottfried Sum Rule in the region measured by NMC, for which we find reasonable agreement
between CT and NMC:
\begin{align}
	\frac{1}{3} - \frac{2}{3}\int_{0.004}^{0.8} dx\, [\bar{d}-\bar{u}](x,\Q=2\,\mathrm{GeV}) &= 0.227 \pm 0.016\ (\mathrm{NMC\, '91 }) \\
											       &= 0.221 \pm 0.021\ (\mathrm{NMC\, '94 }) \\
											       &= 0.260 \pm 0.053\ (\mathrm{CT18})\ .
\end{align}
We stress that the relatively narrow CT uncertainty about the NMC extractions obtained for the restricted integral over $0.004\! <\! x\! <\! 0.8$ underscores the
importance of low-$x$ PDF uncertainties on $\bar{u}, \bar{d}$ in the still lightly probed $x\! <\! 10^{-3}$ region. These must be brought under further control before
phenomenological analyses of high-energy data can make a definitive statement about the violation of $\mathrm{SU}(2)$ flavor symmetry at the moment-level of
Eq.~(\ref{eq:Gott}).

\begin{table}
\begin{tabular*}{\textwidth}{c| @{\extracolsep{\fill}} cc|c||c}
\hline
PDF moment                             &   {\bf CT18}   &    {\bf CT18Z}   &    {\bf \CTHERAII}   &   {\bf Lattice}     \tabularnewline
\hline                                                                                                              
                                       &                &                  &                   &   $0.153\!-\!0.194^{N_f=2+1+1} \star$             \tabularnewline
$\langle x \rangle_{u^+-d^+}$          & $0.156(7)$     &    $0.156(6)$    &    $0.159(6)$     &   $0.111\!-\!0.209^{N_f=2+1} \star$        \tabularnewline
                                       &                &                  &                   &   $0.166\!-\!0.212^{N_f=2} \star$         \tabularnewline
\hline
$\langle x^2 \rangle_{u^--d^-}$        & $0.055(2)$     &    $0.055(2)$    &    $0.055(2)$     &   $0.107(98) \star$        \tabularnewline
$\langle x^3 \rangle_{u^+-d^+}$        & $0.022(1)$     &    $0.022(1)$    &    $0.022(1)$     &     {\it N/A}             \tabularnewline
\hline                                                                                                             
                                       &                &                  &                   &   $0.427(92)$ \cite{Alexandrou:2020sml}        \tabularnewline
$\langle x \rangle_{g}$                & $0.414(8)$     &    $0.407(8)$    &    $0.415(8)$     &   $0.482(69)(48)$ \cite{Yang:2018nqn}        \tabularnewline
                                       &                &                  &                   &   $0.47(4)(11)$ \cite{Yang:2018bft}        \tabularnewline
\hline                                                                                                              
\raisebox{-0.25cm}{$\langle x \rangle_{u^+}$}              & \raisebox{-0.25cm}{$0.350(5)$}     &    \raisebox{-0.25cm}{$0.350(4)$}    &    \raisebox{-0.25cm}{$0.351(5)$}     &   \raisebox{-0.12cm}{$0.359(30)$ \cite{Alexandrou:2020sml}}       \tabularnewline
                                       &                &                  &                   &   \raisebox{0.12cm}{$0.307(30)(18)$ \cite{Yang:2018nqn}}        \tabularnewline
\hline                                       
\raisebox{-0.25cm}{$\langle x \rangle_{d^+}$}              & \raisebox{-0.25cm}{$0.193(5)$}     &    \raisebox{-0.25cm}{$0.194(5)$}    &    \raisebox{-0.25cm}{$0.193(6)$}     &   \raisebox{-0.12cm}{$0.188(19)$ \cite{Alexandrou:2020sml}}        \tabularnewline
                                       &                &                  &                   &   \raisebox{0.12cm}{$0.160(27)(40)$ \cite{Yang:2018nqn}}        \tabularnewline
\hline      
\raisebox{-0.25cm}{$\langle x \rangle_{s^+}$}              & \raisebox{-0.25cm}{$0.033(9)$}     &    \raisebox{-0.25cm}{$0.041(8)$}    &    \raisebox{-0.25cm}{$0.031(8)$}     &   \raisebox{-0.12cm}{$0.052(12)$ \cite{Alexandrou:2020sml}}        \tabularnewline
                                       &                &                  &                   &   \raisebox{0.12cm}{$0.051(26)(5)$ \cite{Yang:2018nqn}}        \tabularnewline
\hline                                                                                                             
$\langle x^2 \rangle_{u^-}$            & $0.085(1)$     &    $0.084(1)$    &    $0.085(1)$     &   $0.117(18)$ \cite{Deka:2008xr}        \tabularnewline
$\langle x^2 \rangle_{d^-}$            & $0.030(1)$     &    $0.029(1)$    &    $0.030(1)$     &   $0.052(9)$ \cite{Deka:2008xr}        \tabularnewline
$\langle x^2 \rangle_{s^-}$            &   ---          &      ---         &      ---          &      {\it N/A}            \tabularnewline
\hline                                                                                                             
$\langle 1 \rangle_{\bar{d}-\bar{u}}$  & $-0.12(35)$    &    $-0.07(29)$   &    $-0.37(41)$    &     ---             \tabularnewline
$\kappa_s$                             & $0.49(16)$     &    $0.61(14)$    &    $0.46(13)$     &   $0.795(95)$ \cite{Liang:2019xdx}       \tabularnewline

\end{tabular*}\caption{
	Like Table~\ref{tab:moments}, but now comparing the most recent results obtained under CT18(Z) and CT14$_\mathrm{HERAII}$
	with a representative selection of recent lattice QCD calculations in the rightmost column. For the latter, reported results
	are generally taken from the recent whitepapers in Refs.~\cite{Lin:2017snn,Lin:2020rut}. The information given in this table
	is not exhaustive, but summary, and we refer the interested reader to the detailed presentations in Refs.~\cite{Lin:2017snn,Lin:2020rut} for extensive
	surveys of modern lattice calculations. Those lattice entries corresponding to
	single calculations are given with the associated reference, whereas those which result from a combination of several
	lattice extractions are indicated with ``$\star$.'' In particular, for $\langle x \rangle_{u^+-d^+}$ we follow
	Ref.~\cite{Lin:2020rut} in supplying ranges obtained from various calculations, grouped according to the number
	of active flavors, $N_f$, in the lattice action used. Meanwhile, the corresponding result for $\langle x^2 \rangle_{u^--d^-}$
	shown above is an average over the result in Ref.~\cite{Gockeler:2004wp} and two separate calculations reported in
	Ref.~\cite{Dolgov:2002zm}.
}
\label{tab:moments2}
\end{table}

We may extend the analysis of the $\overline{d} \neq \overline{u}$ breaking to the $\mathrm{SU}(3)$ sector, by
analyzing the ratio of the first moments of the distributions appearing in Eq.~(\ref{eq:Rs}) leading to the
strange suppression factor moment ratio,
\begin{equation}
	\kappa_s(\Q) \equiv \frac{\langle x \rangle_{s^+}}{\langle x \rangle_{\bar{u}} + \langle x \rangle_{\bar{d}}}\ .
\label{eq:str-supp}
\end{equation}
as illustrated in Fig.~\ref{fig:moments1}. The final row of Table~\ref{tab:moments} lists the numerical results for this quantity 
for the PDF parametrizations considered above, with the exception of
CJ15, which sets $R_s(x,\Q)$ to a constant, making $s^+(x,Q)$
proportional to $\overline{u}(x,Q) + \overline{d}(x,Q)$.
Up to uncertainties, the moments we compute are generally consistent with the traditional strangeness suppression scenario,
$\kappa_s\! =\! 0.5$. In moving from CT14$_\mathrm{HERAII}$ to CT18, there is a modest enhancement, at $Q = 2\,\mathrm{GeV}$, in the preferred central
value and related growth of the associated uncertainty, which shifts from $\kappa_s({\mathrm{CT14}_\mathrm{HERAII}})\! =\! 0.46\! \pm\! 0.13$
  to $\kappa_s(\mathrm{CT18})\! =\! 0.49\! \pm\! 0.16$, in very close agreement with MMHT14, in particular. The inclusion of the ATLAS $W,\, Z$ production data, as well as other changes leading to CT18Z, noticeably increase the ratio
  to $\kappa_s(\mathrm{CT18Z})\! =\! 0.61\! \pm\! 0.14$ and marginally contract its uncertainty, as compared to CT18, making the value more similar to the one in NNPDF3.1.
Recently, a first lattice calculation of $\kappa_s$ was reported by the $\chi$QCD collaboration in
Ref.~\cite{Liang:2019xdx}, which found $\kappa_s(\Q = 2\,\mathrm{GeV}) = 0.795 \pm 0.079\, (\mathit{stat}) \pm 0.053\, (\mathit{sys})$.
Indeed, while this result lies just beyond the upper periphery of the values preferred by typical phenomenological fits,
$\kappa_s \sim 0.5$, it agrees at the $1\sigma$-level with the CT18Z result that follows from the inclusion of 7 TeV inclusive $W,Z$
production data taken by ATLAS.

This, as well as other entries for the PDF moments determined
on the QCD lattice as listed in the rightmost column of Table~\ref{tab:moments2}, have historically shown a general tendency to overestimate
the values extracted phenomenologically. More recent lattice calculations have in some cases begun to 
approach the phenomenological moments ---
{\it e.g.}, for the isovector $u\!-\!d$ moments, or for the total $u, d$-quark and gluon momenta, $\langle x \rangle_{u^+,d^+,g}$ --- for which the
lattice uncertainties are also sufficiently large as to allow agreement with global analyses.

Schematically, the PDF moments are extracted on the lattice from the ratio of 3-point to 2-point correlation functions
\cite{Gockeler:1995wg,Lin:2017snn}:
\begin{equation}
	R(t,\tau,\mathbf{p},\hat{O}) = \frac{\sum_{a,b} \Gamma_{b,a} \langle B_a(t,\mathbf{p}) | \hat{O}(\tau) |
	B_b(0,\mathbf{p}) \rangle}{\sum_{a,b} \Gamma_{b,a} \langle B_a(t,\mathbf{p}) | B_b(0,\mathbf{p}) \rangle}\ ,
\end{equation}
where the $B_{a,b}$ are baryon interpolating operators, $t$ the source-sink Euclidean time separation, and $\tau$ the Euclidean
time associated with the operator $\hat{O}$ insertion noted in Eq.~(\ref{eq:OPE}). For the lower moments of the nucleon
parton distributions, the lattice output is substantially governed by the interplay between excited-state contamination
of the correlation functions, which in general depend on Euclidean time as $\sim\! \exp{(-m_i t)}$, and the lattice signal-to-noise
ratio, which goes as $S/N\! \sim\! \exp{(-(E_N -[3/2]m_\pi) t)}$. As such, lattice calculations at physical pion mass (or chiral
extrapolations thereto) lead to more rapid deterioration of the signal-to-noise at precisely the larger lattice times at
which contributions from nucleon excited states are relatively suppressed. The subtle relationship between these lattice effects
(in addition to other systematic artifacts) complicate any straightforward interpretation of the presently large or small
lattice results for the PDF moments shown in Table~\ref{tab:moments2}.

\clearpage

\section{Description of individual data sets}
\label{sec:Quality}

The CT18 global analysis includes a wide range of data from Run-1 of the LHC, in addition to the extensive collection of data used in the previous CT14 analysis with the combined HERA measurements.
Sec.~\ref{sec:data_overview} and Tables~\ref{tab:EXP_1}--\ref{tab:EXP_2}
reviewed the CT18(Z) data sets and broadly summarized the overall
quality of the fits in terms of $\chi^2/N_\mathit{pt}$ and effective Gaussian
variables $S_E$ provided for each fitted experiment.
A successful fit of the global data, however, requires a far more fine-grained exploration of the degree to which individual experiments are well-described.
It is important to quantitatively evaluate the agreement between data and theory with a rigorous battery of statistical measures and tests \cite{Kovarik:2019xvh}, including a
comprehensive survey of potential tensions in fitting various
experiments. We survey the landscape of experimental constraints
in Sec.~\ref{sec:QualityOverview}, concentrating
primarily on the complementary techniques of Lagrange Multiplier (LM)
scans and sensitivity calculations to elucidate the level of agreement
within the fit and remaining sources of systematic
tension. Section~\ref{sec:Qualitydata} concentrates on the theoretical
description of specific fitted experiments, while Section~\ref{sec:EW}
examines the role of NLO electroweak corrections in describing the
fitted data. 

Procedurally, fitting in the CT approach is done as described in App.~\ref{sec:chi2_app}. Firstly, we minimize the difference between data and theory by computing the best-fit values of the nuisance parameters $\lambda$ associated with the correlated systematic errors of each experiments. Then, we minimize $\chi^{2}$ with respect to the parameters $a$ of the functional forms of the parton distribution functions. We arrive at the best-fit  $\chi^{2}$  given by Eq.~(\ref{Chi2a0l0}) as the sum of $(D^\mathit{sh}_{i}(a_0)-T_{i}(a_0))^{2}/s_{i}^{2}$ and squares of optimal individual nuisance parameters $\overline \lambda (a_0)$. Here $T_i$ is the $i$-th theory prediction, $D^\mathit{sh}_{i}$ denotes the respective data value shifted by the optimal systematic displacements of the nuisance parameters; $s_{i}$ is the published estimate for the total uncorrelated error.

In a high-quality fit, deviations of theory from data are consistent with random fluctuations associated with statistical and systematic uncertainties \cite{Kovarik:2019xvh}. To check that this is the case, we may
plot the shifted data points $D^\mathit{sh}_{i}$ and the theory values
$T_{i}$ for each fitted experiment. 
The error bars for the shifted data are the uncorrelated
errors $s_{i}$ only, because the correlated systematic errors are
already accounted for in the nuisance parameter values.  

There is
also a second comparison that needs to be considered: a histogram plotting
optimal nuisance parameter values $\bar \lambda_\alpha(a)$, associated
with the sources of systematic uncertainties.  The nuisance parameters, which are used to model the relation between the true and the experimentally-determined values of observables, are usually assumed to
be sampled from a normal distribution ${\cal N}(0,1)$ with the mean
equal to 0 and standard deviation equal to 1. Thus, if too many best-fit parameters $\bar \lambda_\alpha(a)$  are far from zero according to ${\cal N}(0,1)$, we should be concerned. On the other hand, the situation where many $\bar \lambda_\alpha(a)$ are close to zero, meaning that the empirical histogram is narrower than ${\cal N}(0,1)$, is common for several new data sets that have published large numbers of systematic uncertainties. This situation is generally less of a concern, as there may be benign reasons for having too many $\bar \lambda_\alpha(a)$ that are very small, see Sec. IV.E in \cite{Kovarik:2019xvh}.

\subsection{Overall agreement among experiments \label{sec:QualityOverview}}

\subsubsection{Revisiting effective Gaussian variables}
Let us first return to Fig.~\ref{fig:sn_ct18} illustrating the overall quality of individual description of experiments in the CT18 NNLO global fit based on the information collected in Tables~\ref{tab:EXP_1} and \ref{tab:EXP_2}. 
Instead of examining $\chi^2_E(N_{pt,E})/N_{pt,E}$ for individual experiments $E$, which have different probability distributions dependent on $N_{pt,E}$, we plot equivalent information in the form of a histogram of the effective Gaussian variables $S_E=\sqrt{2\chi^2_E}-\sqrt{2N_{pt,E}-1}$ listed in Tables \ref{tab:EXP_1} and \ref{tab:EXP_2} \cite{Lai:2010vv}.

If all deviations of theory from data are purely due to random
fluctuations, one would expect to recover an empirical distribution of
$S_E$ that is close to ${\cal N}(0,1)$ for any
$N_{pt,E}$. In practice, any recent global fit renders an $S_E$
distribution that is statistically incompatible with ${\cal N}(0,1)$
\cite{Kovarik:2019xvh}, indicating that too many experiments are
underfitted or overfitted compared to the textbook case.

For the CT18 NNLO fit, the observed $S_E$ distribution shown in Fig.~\ref{fig:sn_ct18} is most compatible with ${\cal N}(0.6,1.9)$. The probability that is compatible with ${\cal N}(0,1)$ is very small ($p=2.5\cdot 10^{-5}$ according to the Anderson-Darling test \cite{Kovarik:2019xvh}). In the figure, we labeled the
experiments with the largest deviations from $S_E\! =\! 0$. These are the combined HERAI+II data set on inclusive DIS \cite{Abramowicz:2015mha} with $S_E\approx 5.7$, which provides the dominant constraints on the PDFs and must be retained in the global analysis despite the quality-of-fit issues discussed in Sec.~\ref{sec:summary-HERA2}, and the CCFR measurement \cite{Seligman:1997mc} of the structure function $x_B F_3(x_B,Q)$ in charged-current DIS on iron, which has an unusually low $\chi^2/N_{pt}\approx 0.4$ for the central fit, but does constrain the PDF uncertainty for some flavors, as can be seen, {\it e.g.}, in the LM scans presented in the next section. 

We also note that the new LHC Run-1 data sets, indicated by the
light green color in Fig.~\ref{fig:sn_ct18}, have more positive than
negative $S_E$ values, indicating that their $\chi^2$ values are larger
than would be expected from random fluctuations consistent with the published experimental errors, as can be verified by consulting Table~\ref{tab:EXP_2}.

Two squares and two stars indicate the $S_E$ values for the NuTeV dimuon and CCFR dimuon data,
respectively, which we highlight for special attention given the importance of these data for probing the strangeness PDF. An analogous plot for the alternative CT18Z fit in Fig.~\ref{fig:sn_ct18z} shows increased $S_E$ values for the CCFR and NuTeV experiments, as compared to the CT18 fit, because of the conflicting pull of the ATLAS 7 TeV $W/Z$ production data. 

\begin{figure}[p]
  \hspace*{-0.6cm}\includegraphics[width=0.45\textwidth]{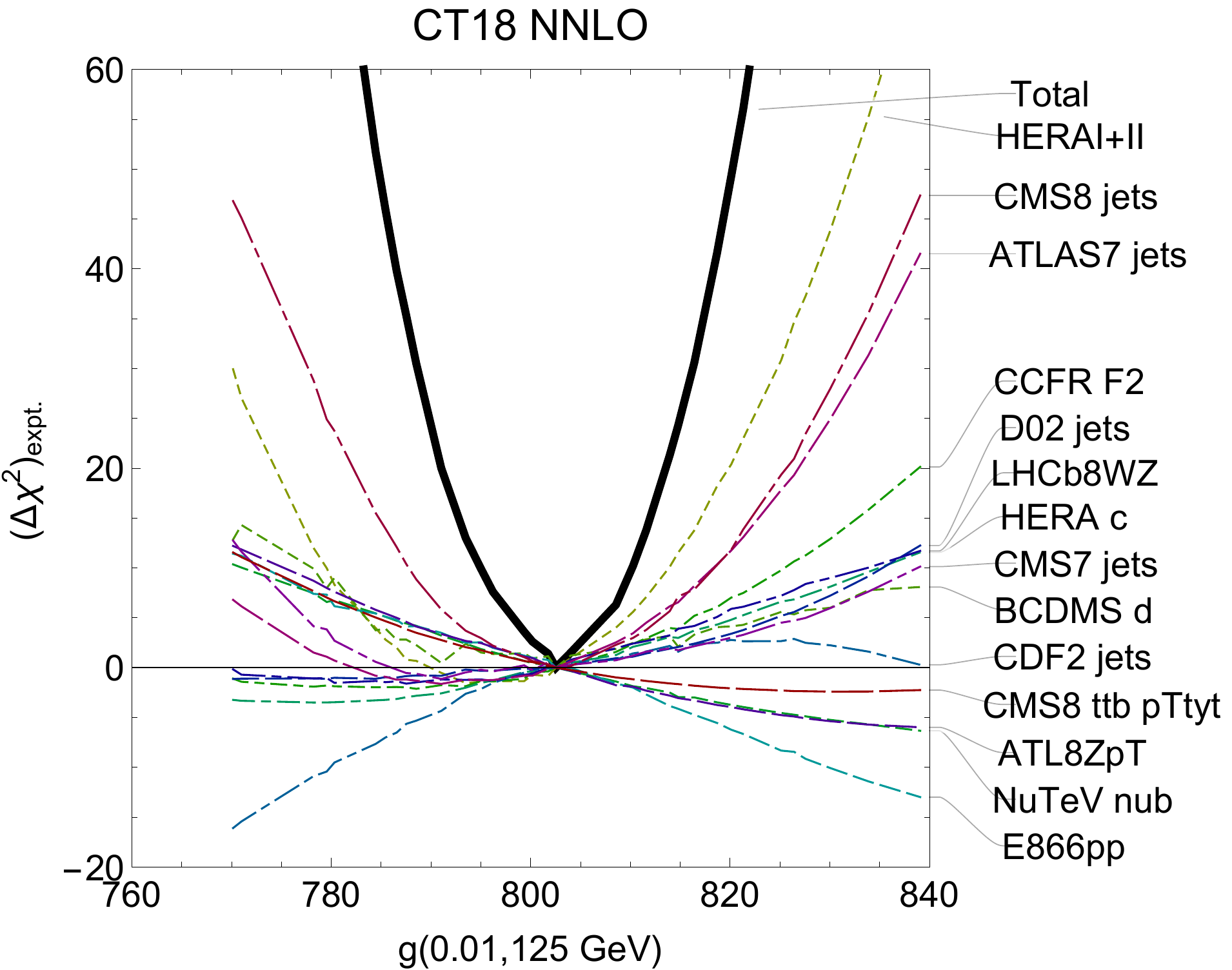}\quad
  \includegraphics[width=0.43\textwidth]{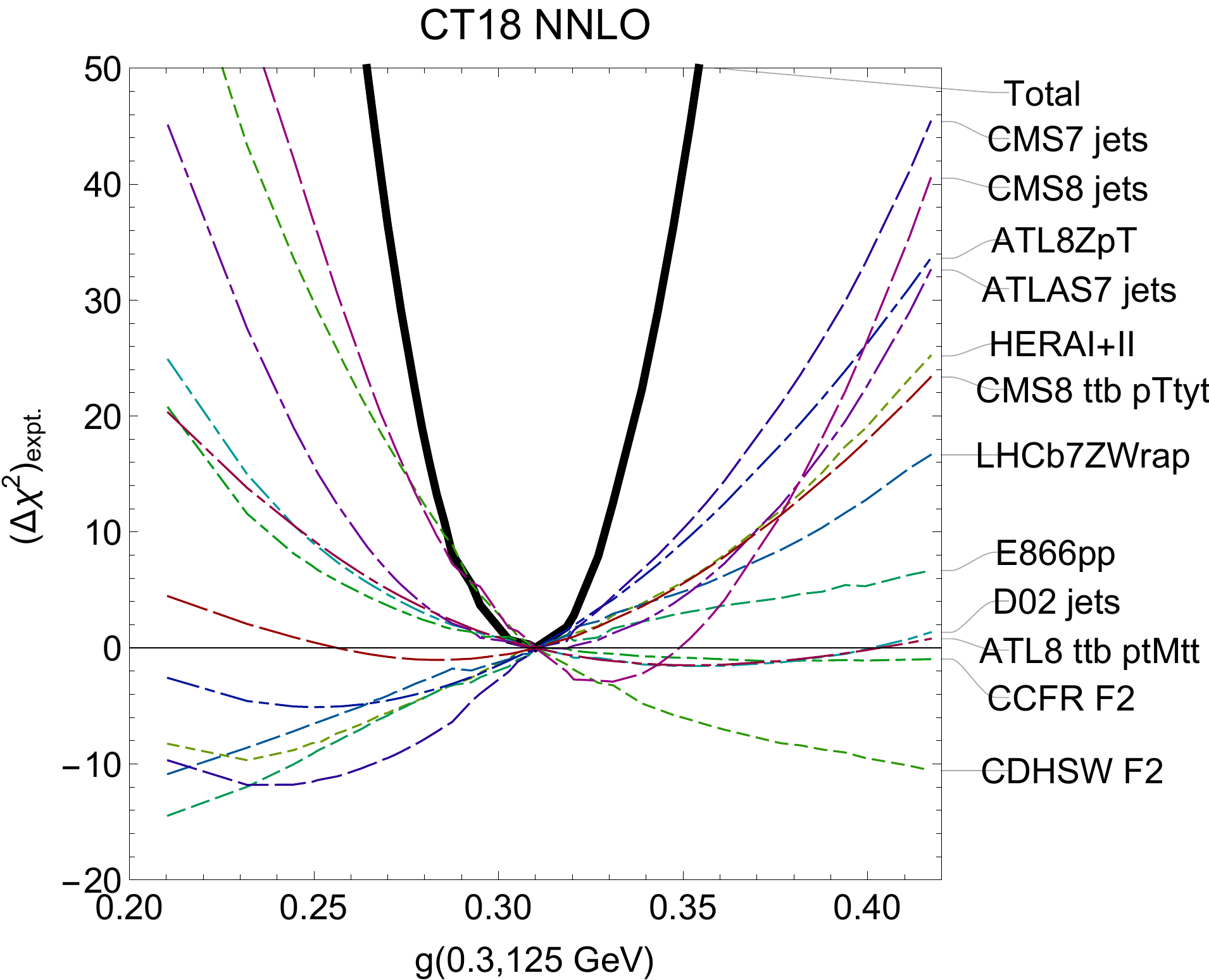}
	\caption{LM scans for the gluon PDF at $Q=125$ GeV and $x=0.01$ and $0.3$, based upon the CT18 NNLO fits.
		\label{fig:LMg18}}
\end{figure}

\begin{figure}[p]
	\includegraphics[width=0.49\textwidth]{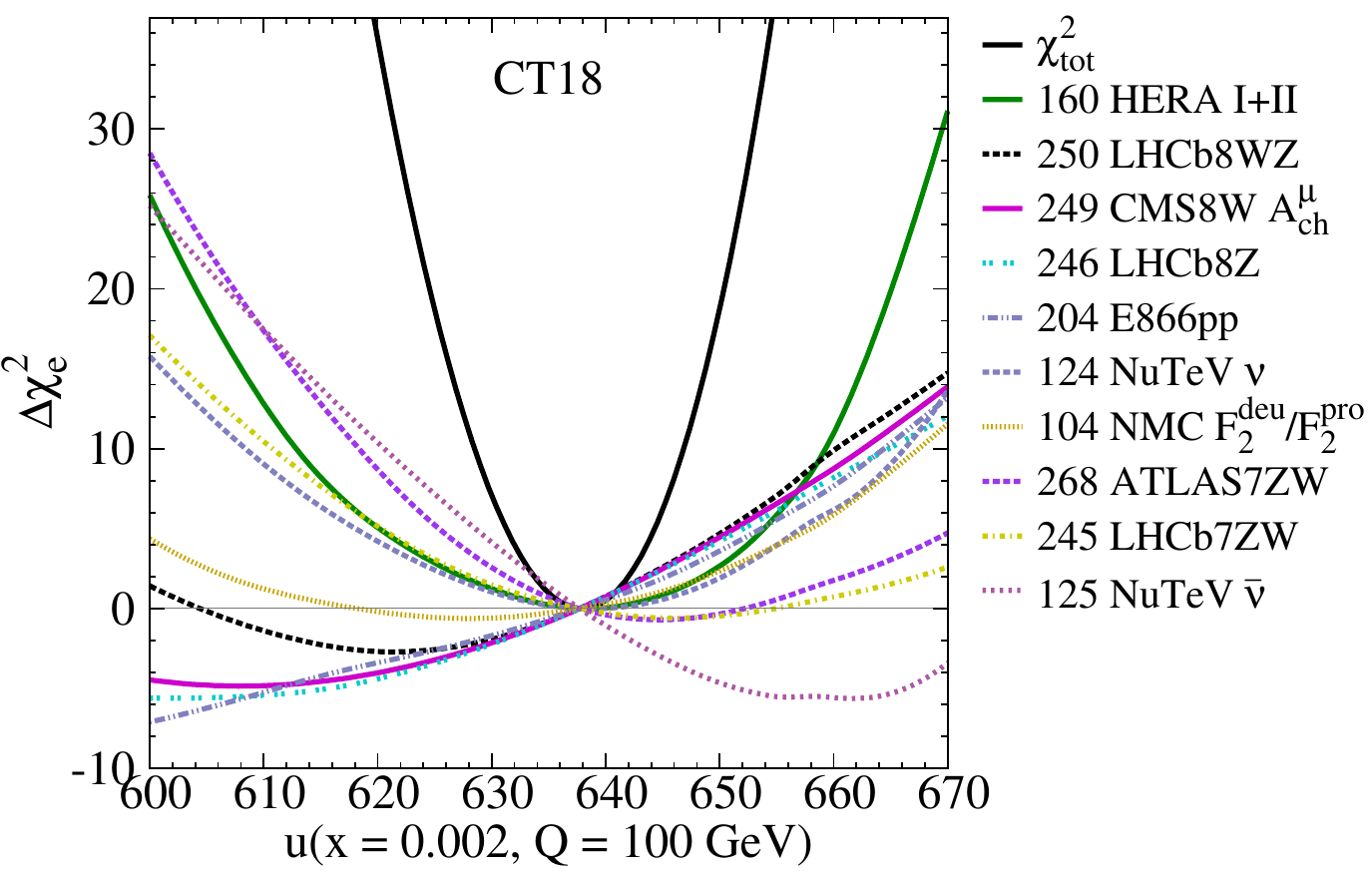} 
	\includegraphics[width=0.49\textwidth]{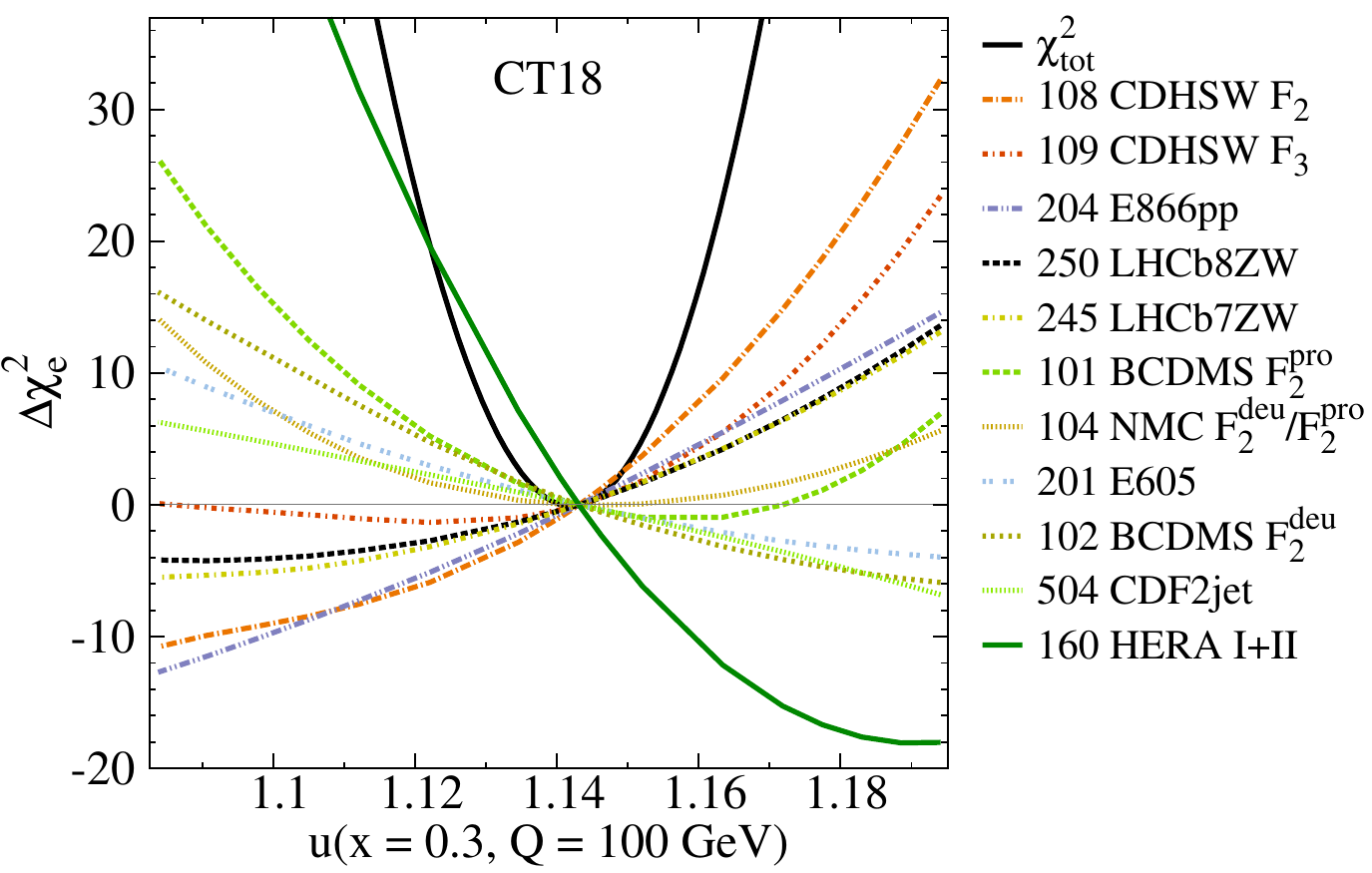}\\
	\includegraphics[width=0.49\textwidth]{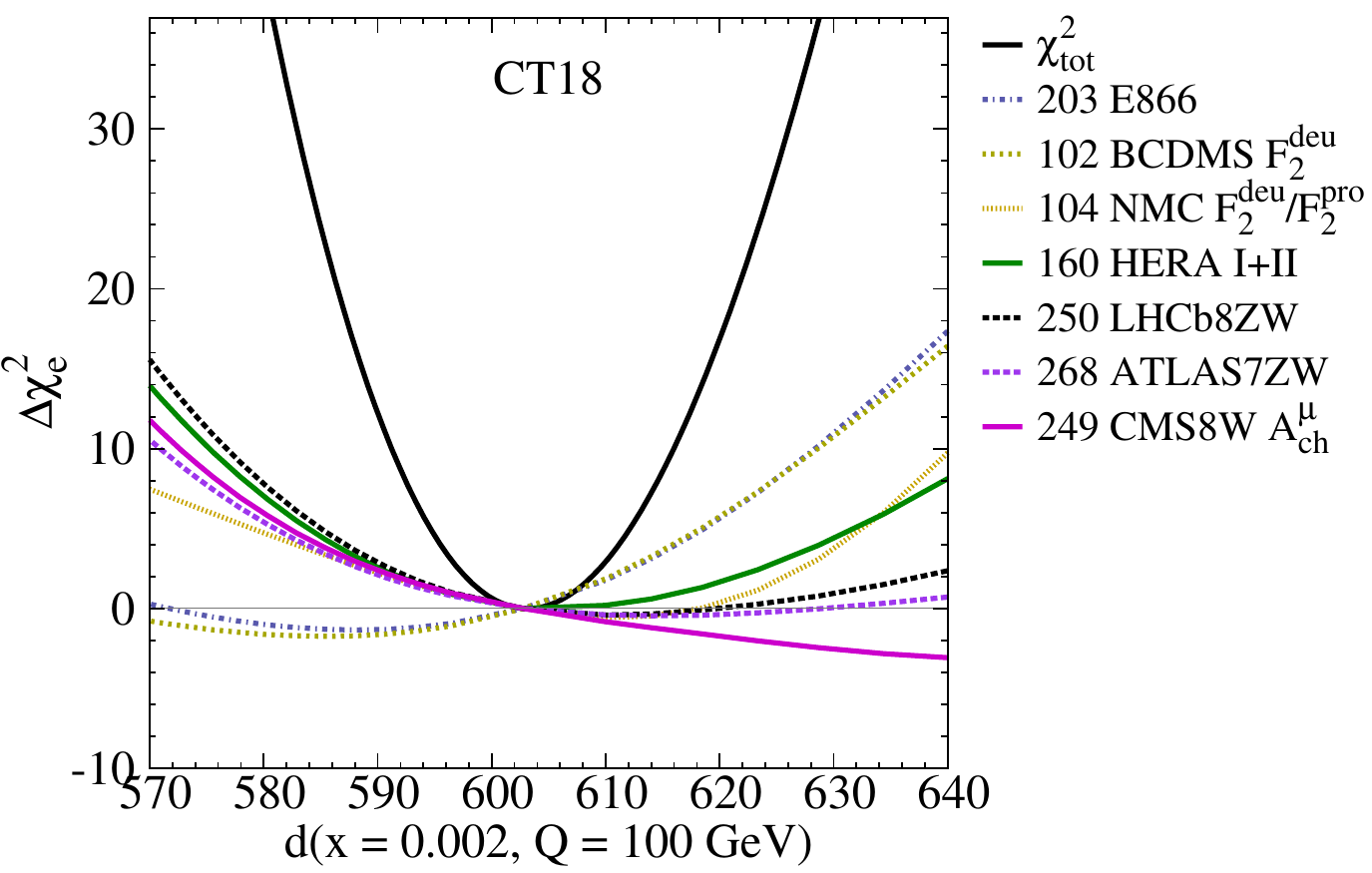}
	\includegraphics[width=0.49\textwidth]{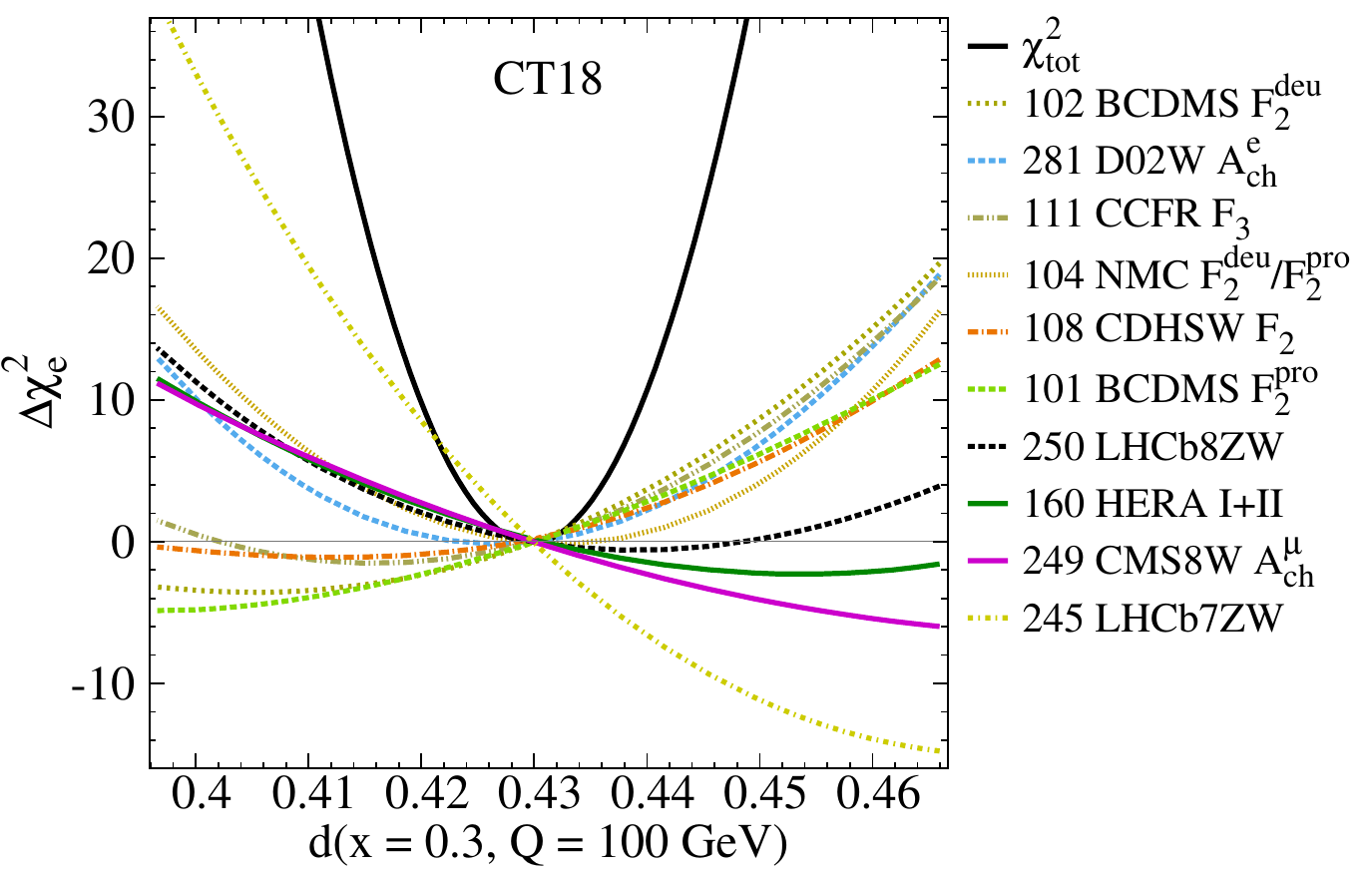}\\ 
        	\caption{LM scans for the up- and down-quark PDF at $Q=100$ GeV and $x=0.002$ and $0.3$, based upon the CT18 fits.
		\label{fig:LMud}}
\end{figure}

\begin{figure}[p]
	\center
	\includegraphics[width=0.49\textwidth]{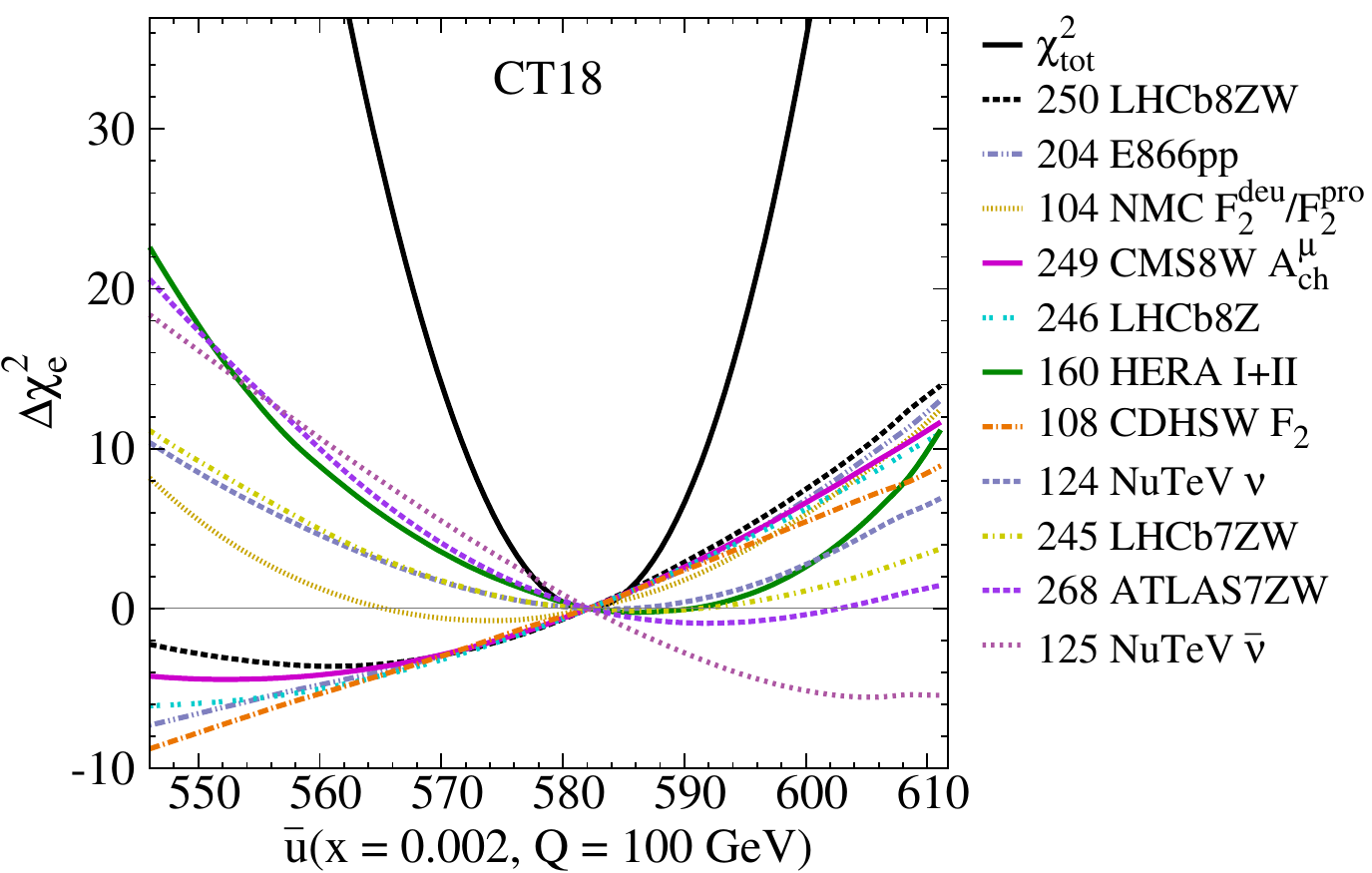}
	\includegraphics[width=0.49\textwidth]{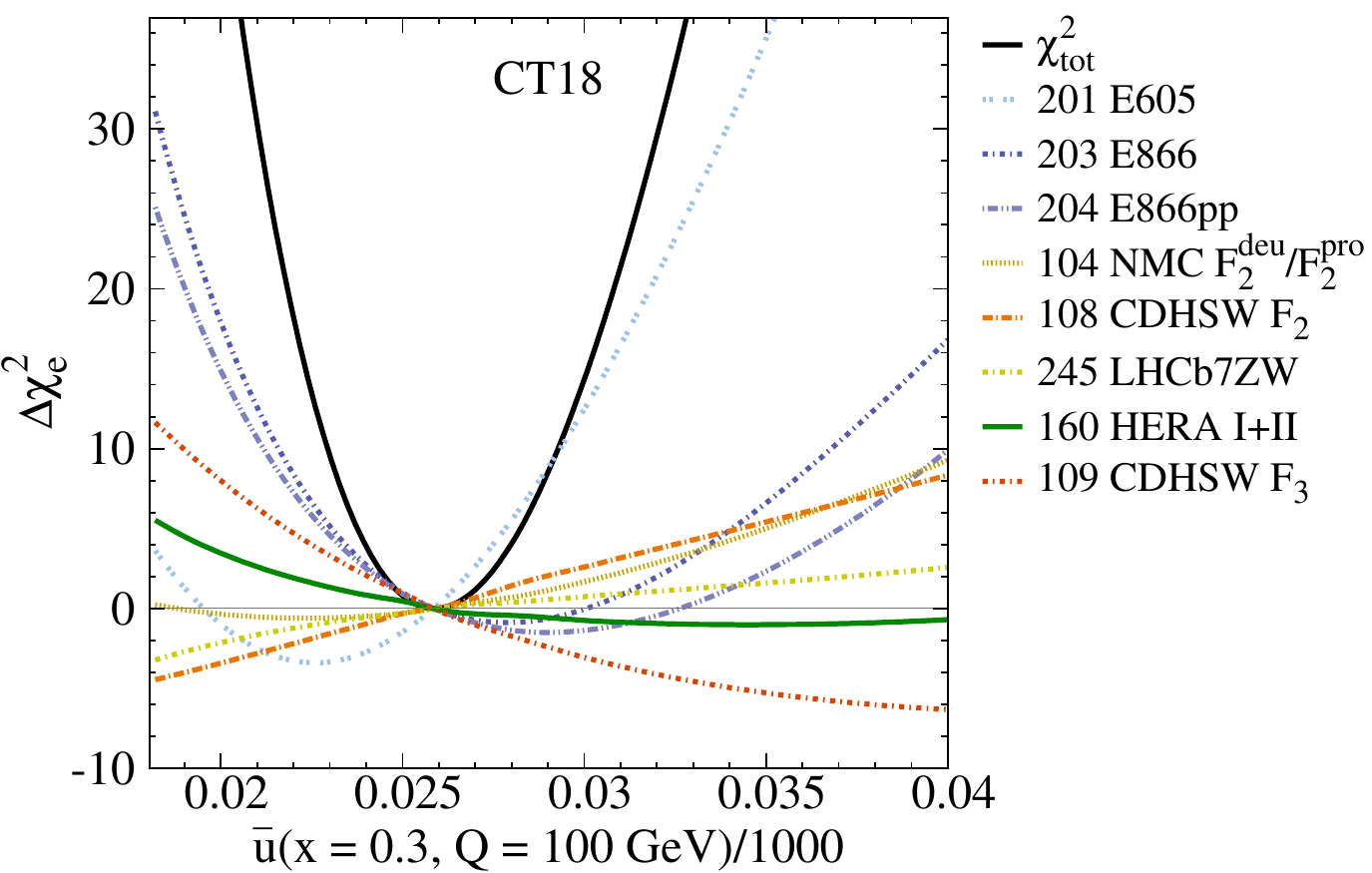}\\
	\includegraphics[width=0.49\textwidth]{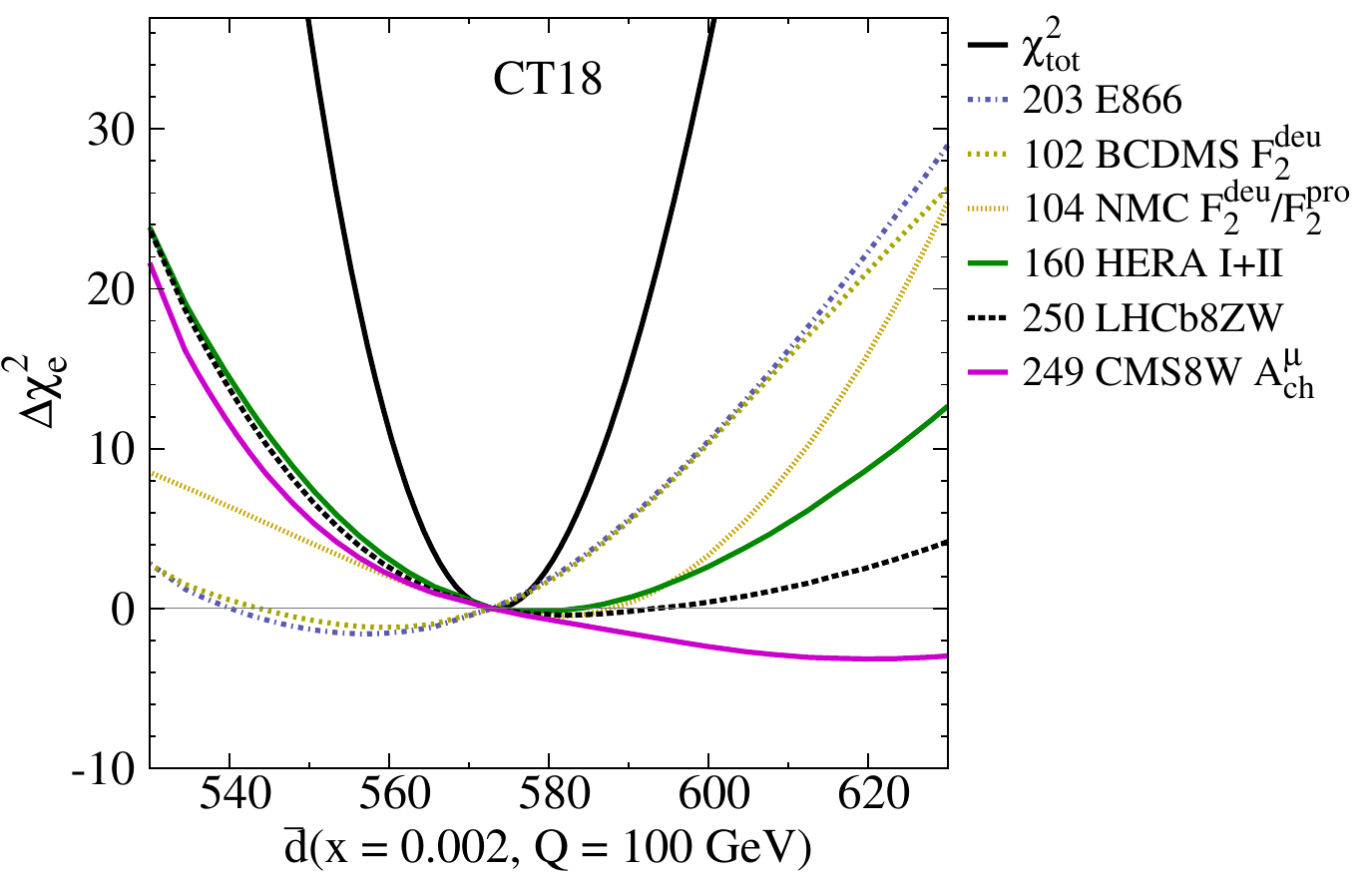}
	\includegraphics[width=0.49\textwidth]{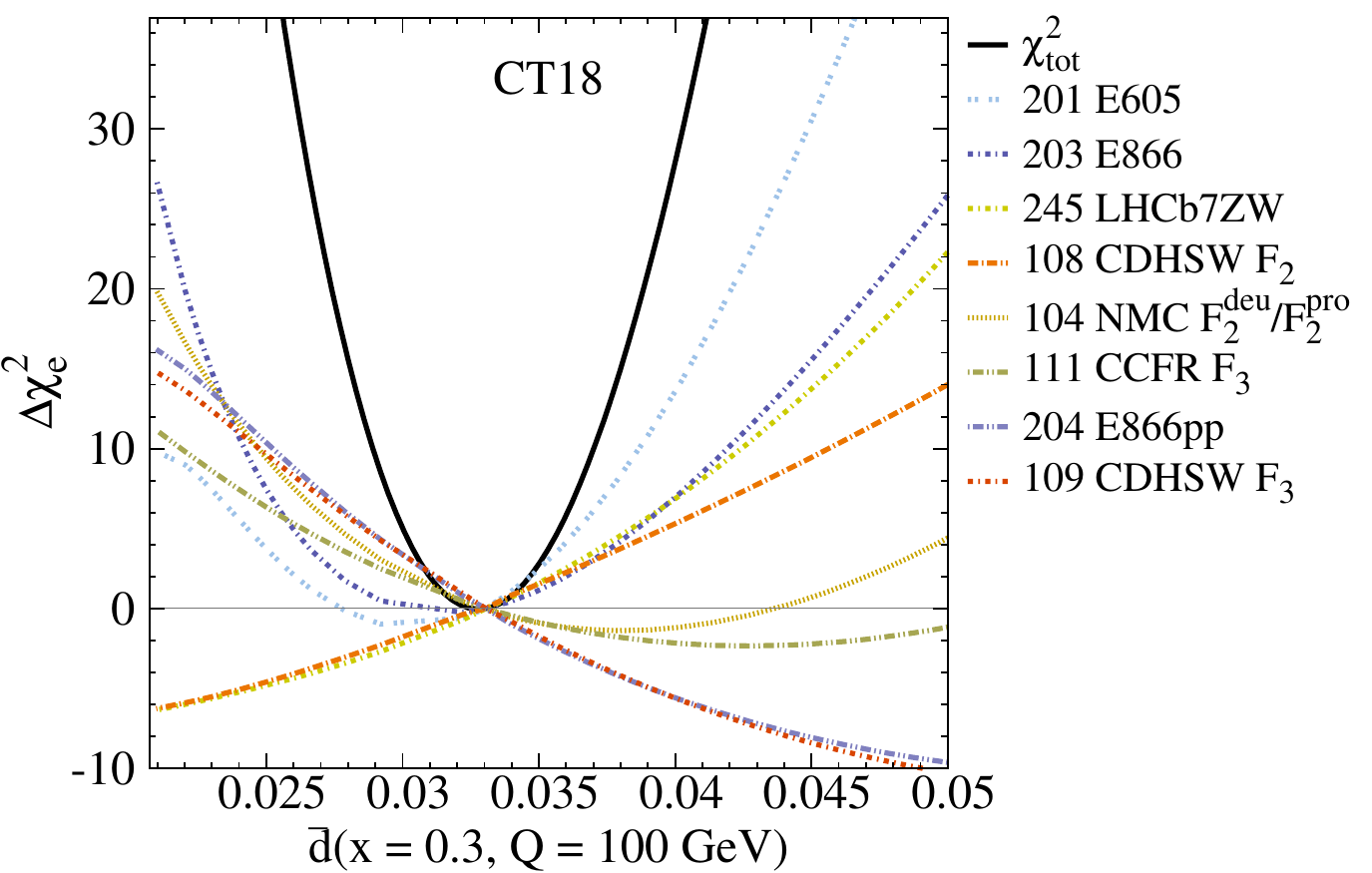}\\ 
         \includegraphics[width=0.49\textwidth]{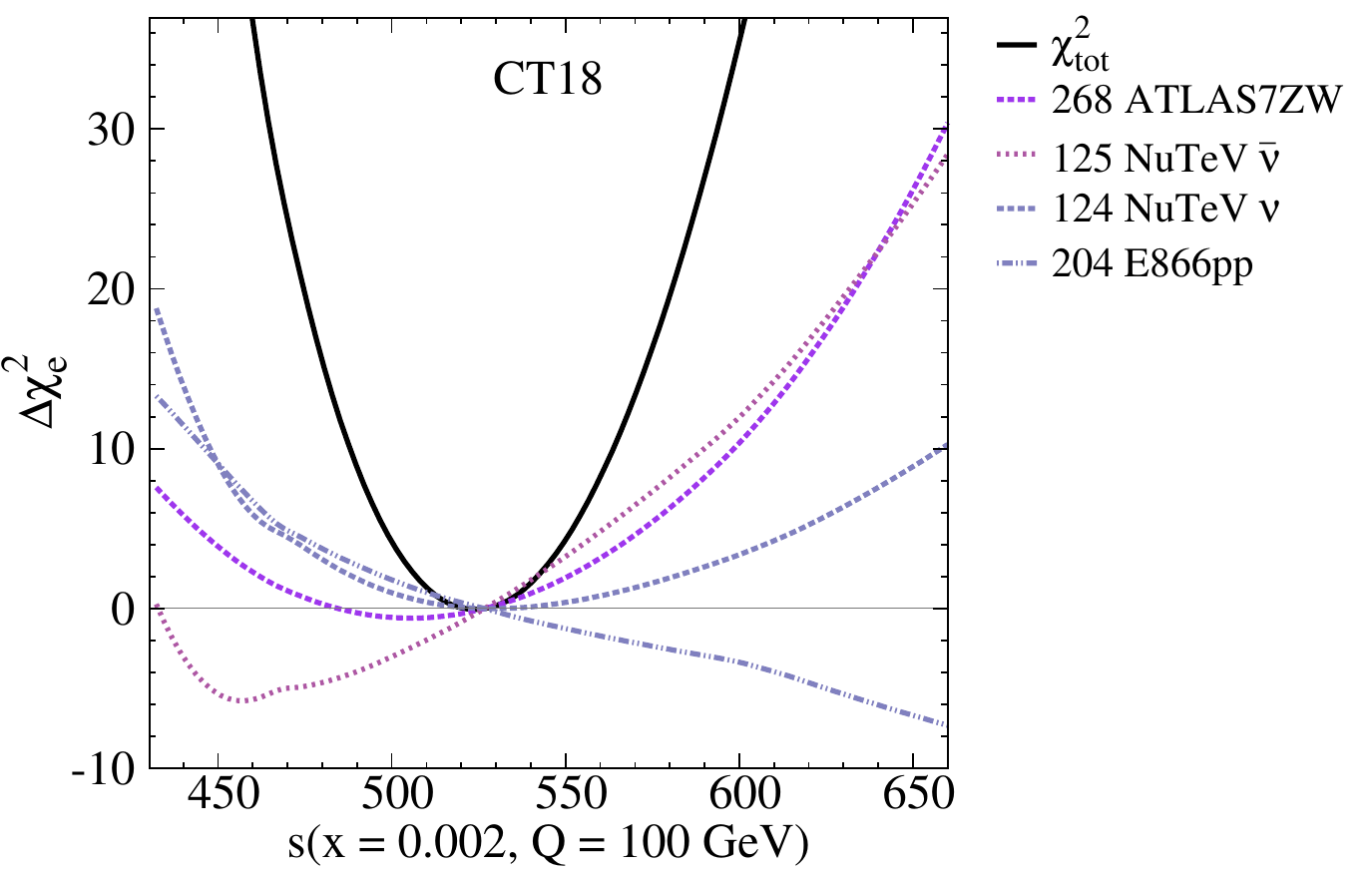} 
	\includegraphics[width=0.49\textwidth]{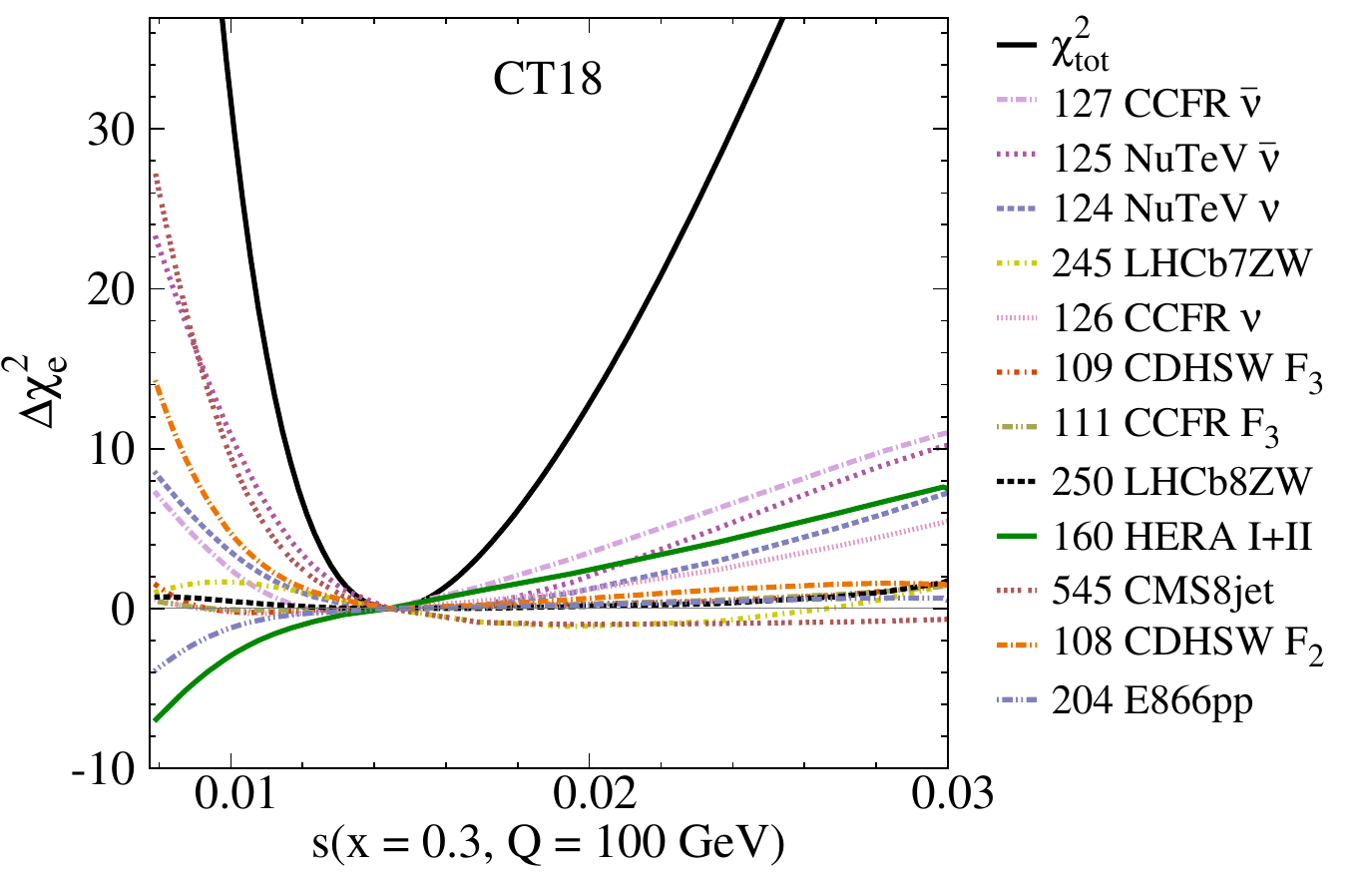} 
	\caption{
		Like Fig.~\ref{fig:LMud}, here giving LM scans for the
		$\bar{u}$-, $\bar{d}$-, and $s$-quark PDFs
		in CT18.
		}
\label{fig:LMubdbs}
\end{figure}

\subsubsection{Lagrange Multiplier scans
\label{sec:LMScans}
}

The Lagrange Multiplier (LM) scan technique,
which was introduced in Ref.~\cite{Stump:2001gu}, is among the most
robust methods of assessing the level of tension in a global fit. This method involves constraining a particular fitted distribution
to hold a chosen numerical value by means of Lagrange multipliers,
while refitting the rest of the PDF parameters with this constraint
in place. A PDF at a chosen $x$ and $\Q$ can then be systematically varied away from its value preferred in an unconstrained global fit. The profile of increases in $\chi^2$ (or $S_E$) as a result of this variation can be computed for each fitted experiment, revealing the extent to which numerical alteration
of the PDFs is connected to the ability to successfully describe specific data.

\begin{figure}[b]
\begin{center}
	\includegraphics[width=0.48\textwidth]{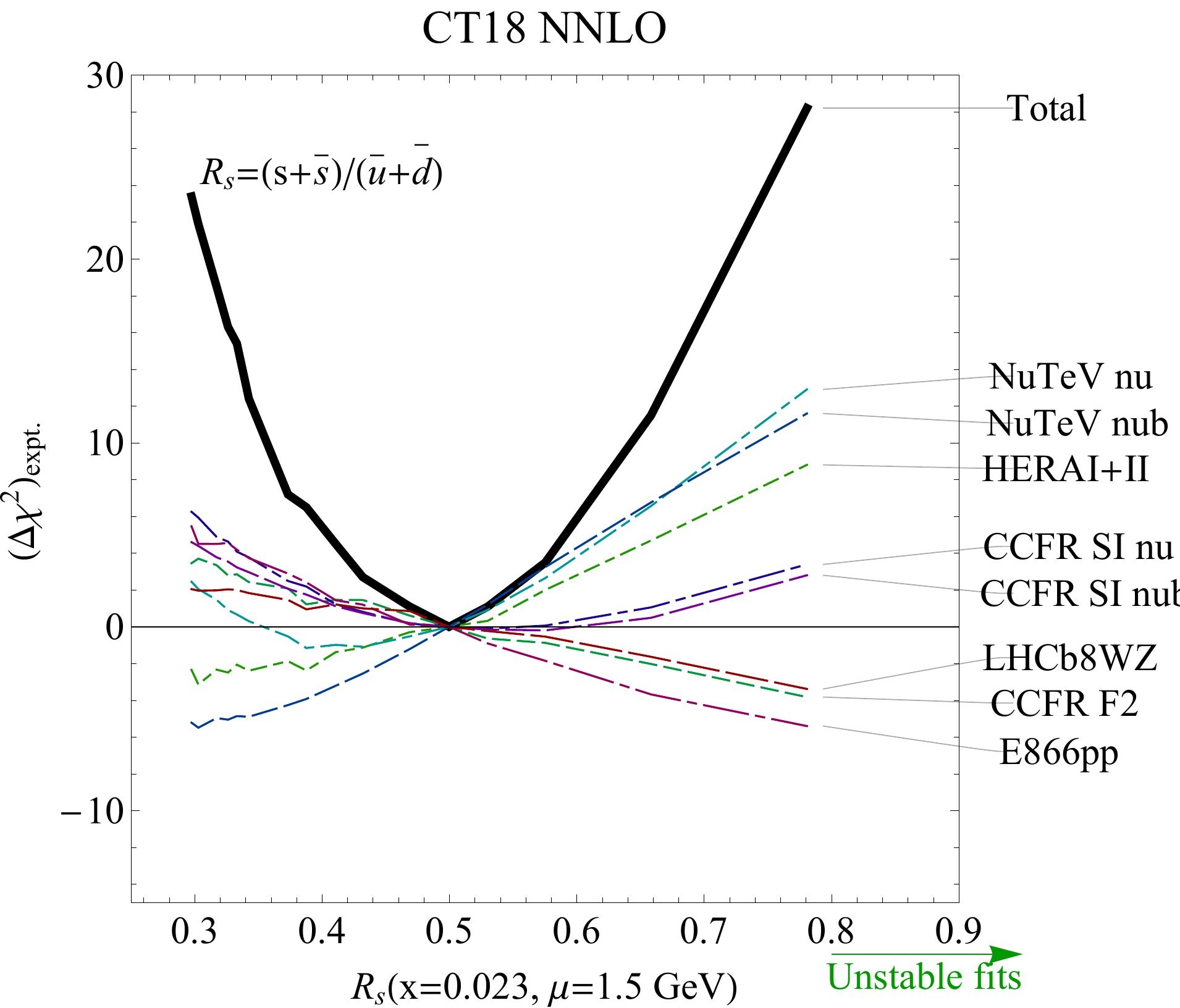}\quad
	\includegraphics[width=0.48\textwidth]{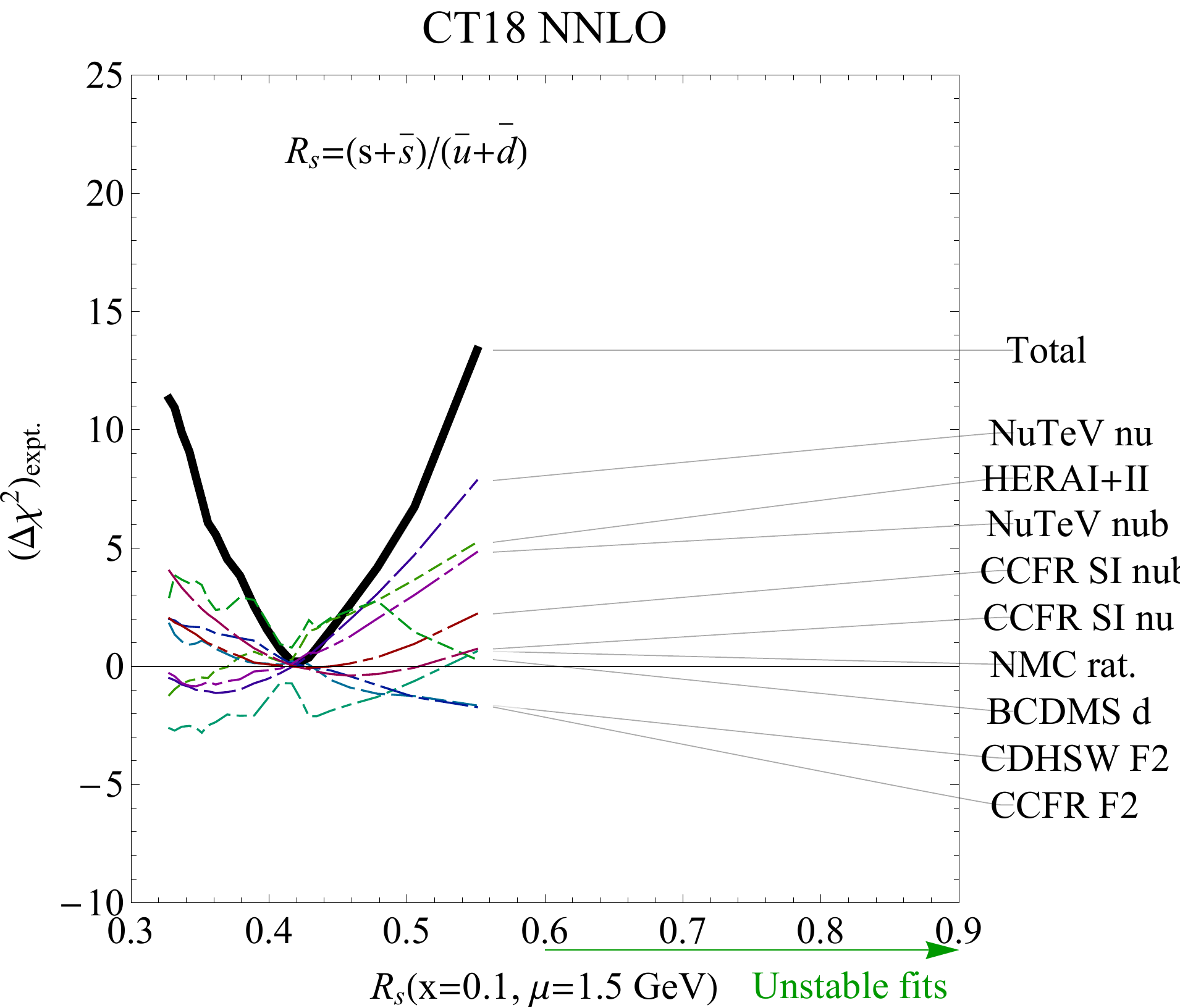}\\
(a)\hspace{2.6in}(b)\\
	\caption{The LM scan over $R_s$ at $Q=1.5$ GeV, with $x=0.023$ and $x=0.1$ respectively, for the CT18 NNLO fit.
\label{fig:LMRs}}
\end{center}
\end{figure}

\begin{figure}[tb]
	\center
	\includegraphics[width=0.49\textwidth]{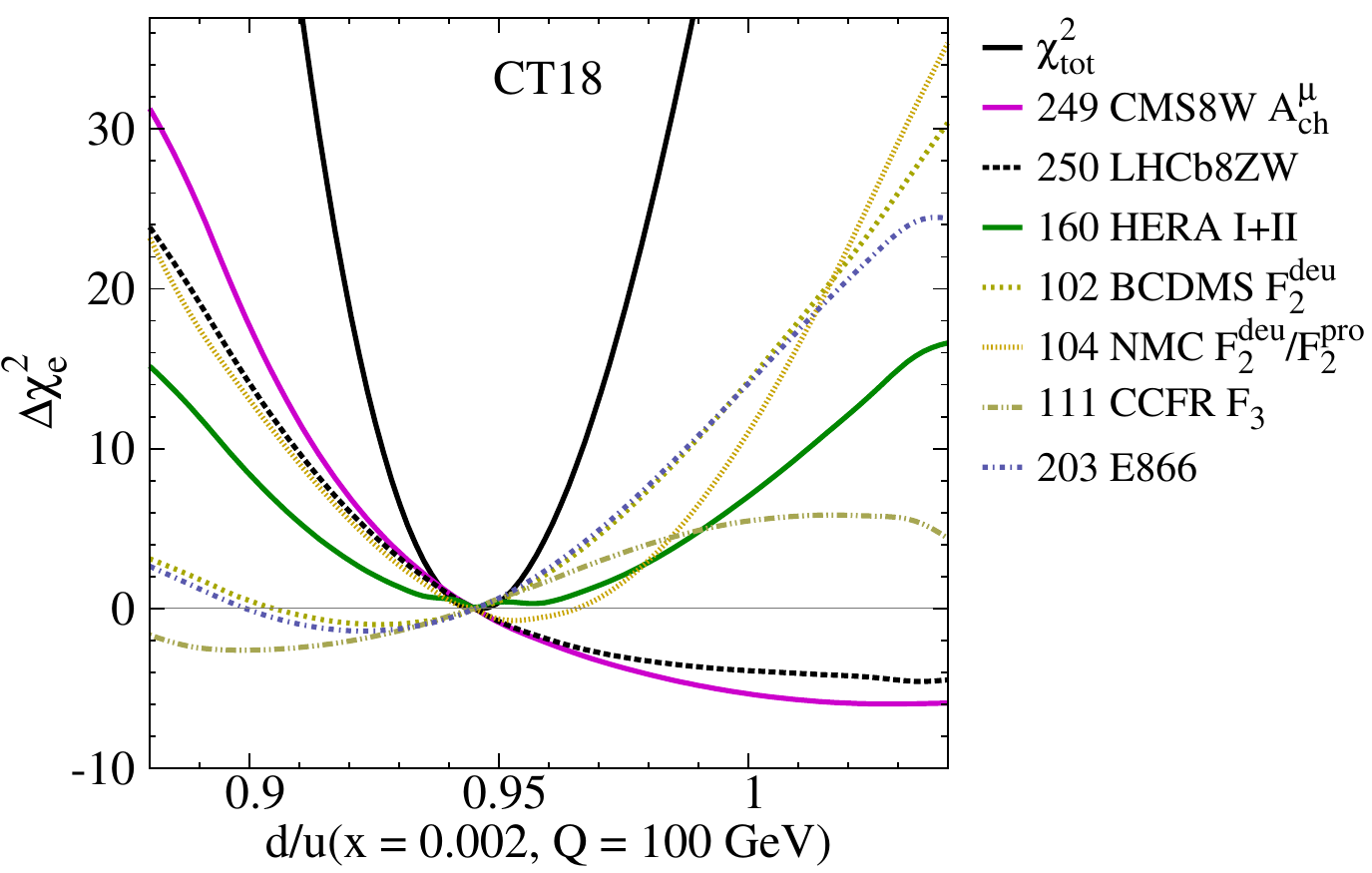} 
	\includegraphics[width=0.49\textwidth]{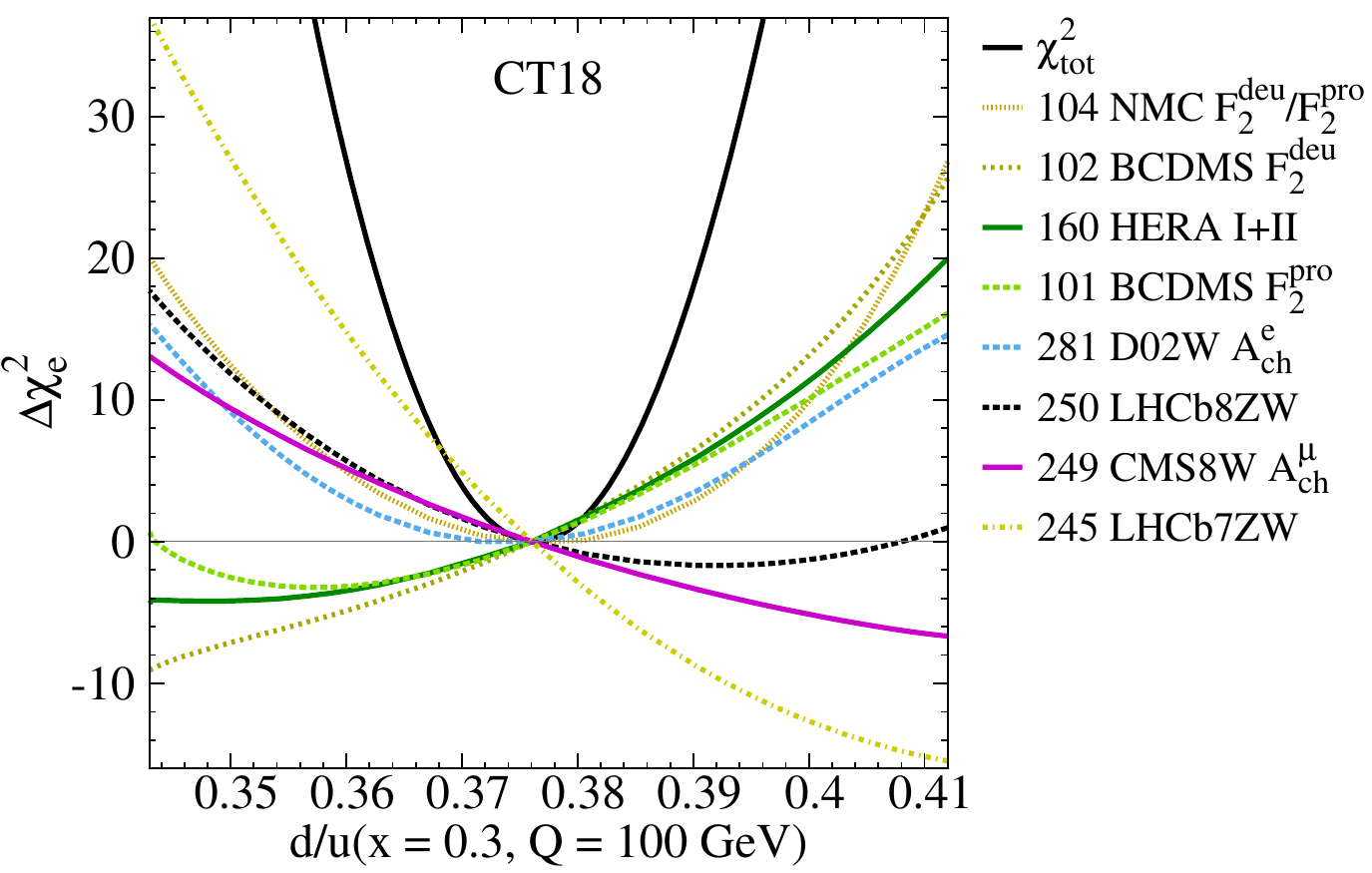}\\ 
	\includegraphics[width=0.49\textwidth]{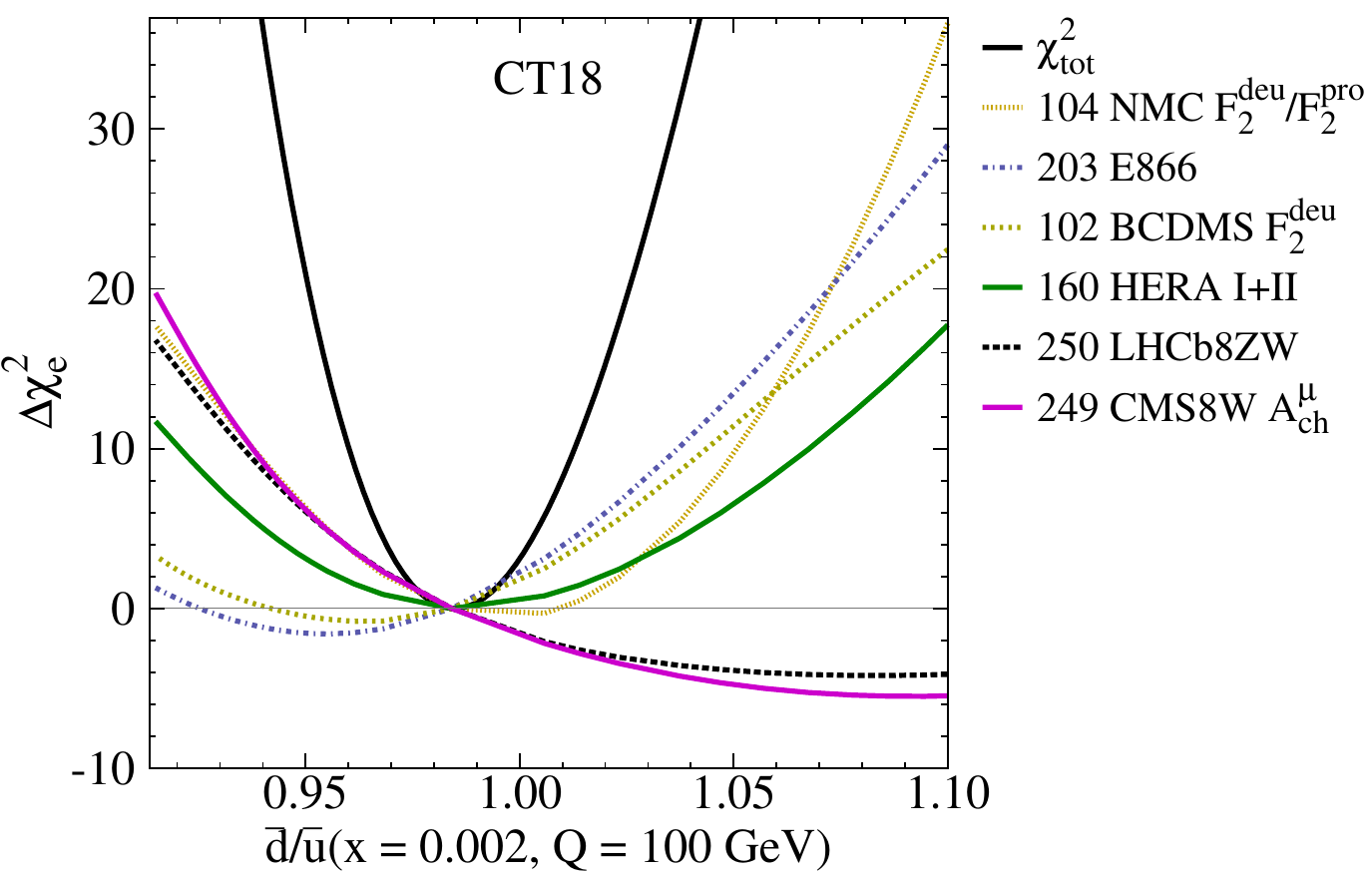}
	\includegraphics[width=0.5\textwidth]{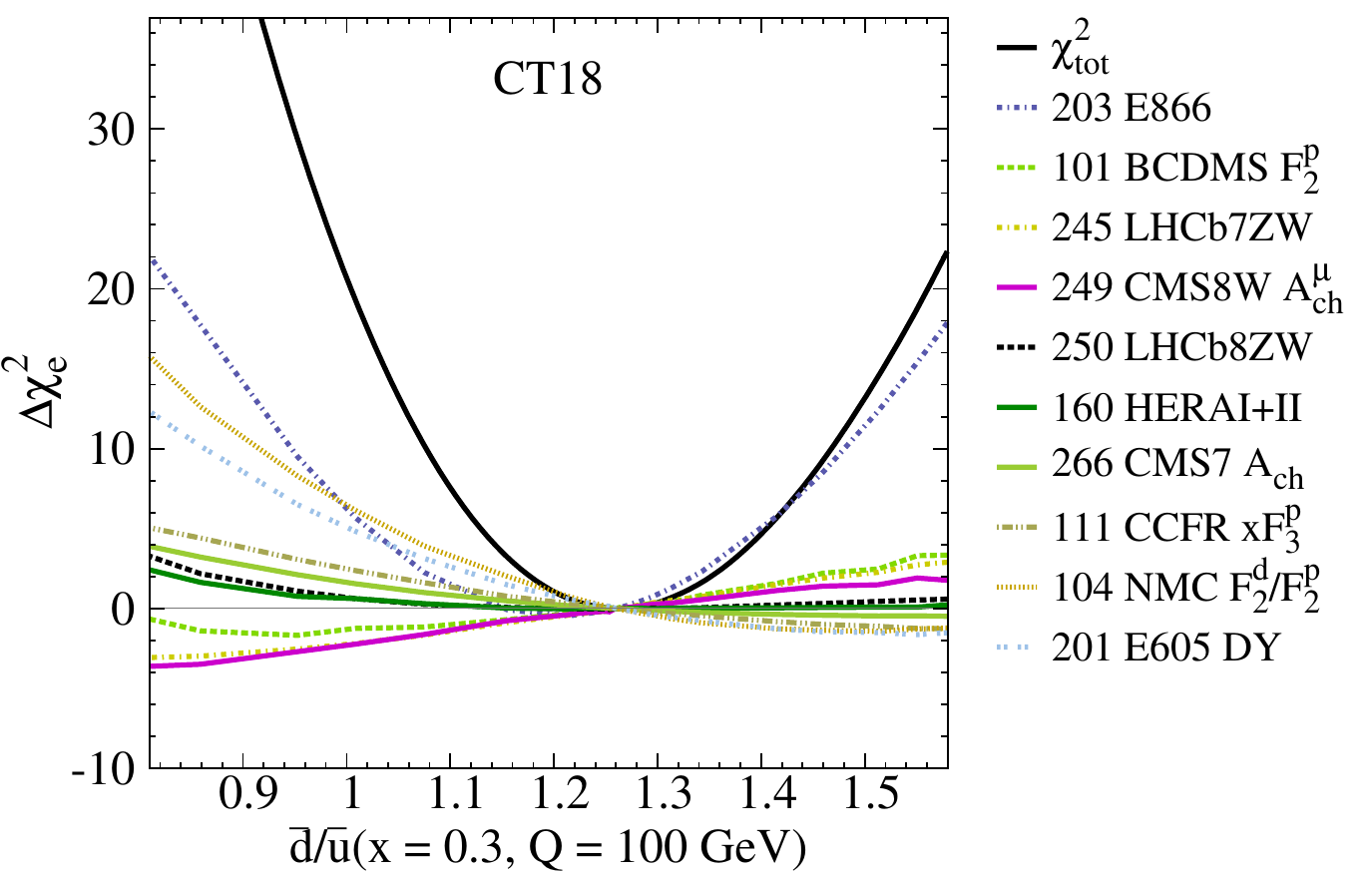}\\

	\caption{Like Fig.~\ref{fig:LMud}, for LM scans over the
          ratios $d/u$ and $\bar d/\bar u$. 
		}
\label{fig:LMduratios}
\end{figure}

 A collection of panels
in Figs.~\ref{fig:LMg18}--\ref{fig:LMduratios}
demonstrates $\chi^2$ profiles in LM scans
for a broad range of CT18 NNLO PDFs, typically at a high scale $\Q\!=\! 100$ GeV relevant for high-energy processes,
and for select parton fractions representative of the PDF
behavior at low $x$ ($x=0.002$ and $0.023$) and high $x$ ($x=0.1$ and
$0.3$). Among the generic features of the scans, we observe that,
while the global $\chi^2$ for all experiments is close to parabolic in
well-constrained $(x,Q)$ regions, some individual experiments may
prefer the PDF values that are quite different from the global
minimum. At the global minimum itself, the $\chi^2_E$ for such an
experiment may be elevated by up to tens of units.  

In Fig.~\ref{fig:LMg18}, for instance, we show two LM scans associated
with the gluon density, $g(x,\Q)$. In the left panel, the LM scan probes the pulls of the most sensitive measurements to the Higgs-region gluon PDF, which
contributes to Higgs boson production through the predominant $gg \to
H$ channel, especially in the neighborhood of $x = m_H /
(14\,\mathrm{TeV})\!\sim\! 0.01$ and for $\Q\!\sim\!m_H$. Evidently,
most constraints arise due to HERA inclusive DIS data as well as the
LHC jet data. 

In the right-hand plot for $x=0.3$, strong constraints spread over more data
sets, notably from high-$p_T$ $Z$ boson $p_T$ and top-quark
production. In particular, while the ATLAS 7 TeV inclusive jet data
prefer $g(0.3,125\mbox {GeV})\approx 0.3$, consistent with
the central value of the full fit, the CMS 7 TeV and 8 TeV jet
production prefer $g(0.3,125\mbox {GeV}) = 0.242^{+0.016}_{-0.020}$
and $0.327^{+0.015}_{-0.010}$ --- a $\approx 3\sigma$ difference
according to the $\Delta \chi^2=1$ criterion.

We notice that in some situations, when a significant tension
between the experiments is revealed, as in the right-hand plot of
Fig.~\ref{fig:LMg18}, a Hessian estimate based on the dynamic tolerance \cite{Martin:2009iq} may result in a much narrower PDF uncertainty
than the estimate based on the total $\chi^2$ in the LM scan, as a consequence of the trade-off between the opposite pulls on the PDF exerted by the conflicting experiments. We discuss this further in Appendix~\ref{sec:LMRsCT18Z}, with a specific example shown in Fig.~\ref{fig:rserrors}.

In Fig.~\ref{fig:LMud} we show LM scans for the $u$- and $d$-quark
PDFs at $x=0.002$ and $0.3$.
For the low-$x$ values, constraints from  LHC $W$ and $Z$ boson data 
(from the LHCb, CMS and ATLAS collaborations) stand out
as expected, in addition to constraints from HERA and NuTeV.
At $x=0.3$, several fixed target experiments, e.g., CDHSW, BCDMS, and
E866 make significant contributions.
The situations are similar for the $d$-quark density as well as for the $d/u$ ratio shown in Fig.~\ref{fig:LMduratios}.

For the $\bar u$ and $\bar d$ antiquarks in Fig.~\ref{fig:LMubdbs}, as
well as the  $\bar d/\bar u$ ratio in Fig.~\ref{fig:LMduratios},
the LHCb data and the CMS $W$ boson charge asymmetry data play an
important role at small-$x$, as can be seen from
Fig.~\ref{fig:LMubdbs}. 
On the other hand, at large-$x$, the flavor separation depends on the
E605, E866 and NMC deuteron data. 

The power of the LM method is most explicitly
demonstrated by the scans on the strange
quark PDF for CT18 in the third row of Fig.~\ref{fig:LMubdbs}, and the
strangeness ratio $R_s(x,Q)$ defined in Eq.~(\ref{eq:Rs}) and scanned at
$x=0.023$ and $x=0.3$ in Fig.~\ref{fig:LMRs}. We see from the lower
left inset of Fig.~\ref{fig:LMubdbs} at
$x=0.002$ that the CT18 data set provides no substantial direct
constraint on $s(x,Q)$ at $x < 0.01$. Rather, the behavior of $s(x,Q)$ is
weakly constrained by the low-luminosity ATLAS 7 TeV $W$ and $Z$ data
(Exp.~ID=268), as well as by the low-$x$ extrapolation of the constraints by
the NuTeV and CCFR dimuon data probing $x$ above $0.01$.

At $x=0.01\!-\!0.1$, the $R_s$ ratios in Fig.~\ref{fig:LMRs} indicate the
dominance of constraints from NuTeV and CCFR dimuon production,
together with HERA inclusive DIS, with weaker constraints from
LHCb $W/Z$ production
and the fixed-target experiments BCDMS, CDHSW, E866, and NMC. Here, the scans
reveal a salient feature, that the fits using the CT18
strangeness parametrization become unstable when $R_s(x,Q)$ is forced
to be close to 1 at $x > 0.01$. For such increased $R_s$ values, the
$\chi^2$ values fluctuate, or the fits fail to converge. Somewhat
larger values of $R_s$ are tolerated at $x < 0.01$. 

Finally, going back to $s(x,Q)$ at $x=0.3$ in the lower right inset of
Fig.~\ref{fig:LMubdbs}, the very large-$x$ behavior is again
determined by the extrapolation of the strangeness PDF from lower $x$,
where it is constrained by the combination of the experiments listed
in the figure. 

We see from this Section that the advantage of the Lagrange Multiplier approach lies in its systematic, robust nature, as well as its ability to reveal tensions or
instabilities that may be missed by the other techniques.
On the other hand, this calculation requires repeated refits
of the PDFs for many values of the LM parameter(s) ---
a limitation that makes the LM scans computationally expensive.

\subsubsection{The PDF sensitivity analysis
\label{sec:L2}
}
A technique complementary to the LM scans explored in Sec.~\ref{sec:LMScans}
is the calculation of the {\it $L_2$ sensitivity}. The $L_2$ sensitivity was
first introduced in Ref.~\cite{Hobbs:2019gob} for the purpose of analyzing
the interplay among the pulls of the CT18(Z) data upon the fitted PDFs.
Here we will review its essential definition. A closely related
implementation, based on the $L_1$ sensitivity detailed
in \cite{Wang:2018heo} and realized in the \texttt{PDFSense} program,
will be used at the end of this Section to rank the experiments of the
CT18 data set according to the sensitivity to various combinations of PDFs.
  
While the LM scans offer the most robust approach for exploring
possible tensions among fitted data sets in a given analysis, they are
very computationally costly to evaluate and 
done for specific choices of $x$ and $Q$.
As we explain here, the $L_2$ sensitivity can be rapidly
computed and provides a strong approximation to the $\Delta \chi^2$
trends in
a given global analysis. Moreover, the $L_2$ sensitivity can be
readily calculated across a wide range of $x$, allowing the $\Delta \chi^2$
variations shown in the LM scans to be visualized and interpreted
for multiple $x$ at once. We stress that the qualitative conclusions
revealed by consideration of the $L_2$ sensitivities, discussed and
presented below, are consistent with the picture based on the LM
scans themselves.
Although the $L_2$ sensitivities may not always provide the same
numerical ordering as the LM scans for the subdominant experiments,
they offer complementary information over broader reaches of $x$
that are not completely captured by the LM scans.

We work in the Hessian formalism \cite{Pumplin:2002vw,Nadolsky:2008zw,Pumplin:2001ct} and compute the $L_2$ sensitivity $S_{f, L2}(E)$ for each experiment, $E$, as

\begin{equation}
S_{f, L2}(E) = \vec{\nabla} \chi^2_E \cdot \frac{ \vec{\nabla} f } { |\vec{\nabla} f| }
             = \Delta \chi^2_E\, \cos \varphi (f, \chi^2_E)\ ,
\label{eq:L2}
\end{equation}
which yields the variation of the log-likelihood function $\chi^2_E$ due to a unit-length
displacement of the fitted PDF parameters away from the global minimum $\vec{a}_0$ of
$\chi^2(\vec{a})$  in the direction of $\vec{\nabla}f$. 
The PDF parameters $\vec a$ are normalized so that a unit displacement
from the best fit in any direction corresponds to the default
confidence level of the Hessian error set (90\% for CT18,
on average corresponding to slightly less than
$\Delta\chi^2_{\textrm{tot}}=100$ in a given direction.)

This displacement increases the
PDF $f(x,Q)$ by its Hessian PDF error $\Delta f$, and, to the extent its
PDF variation is correlated with that of $\chi^2_E$ through the correlation angle
\begin{equation}
	\varphi(f, \chi^2_E) = \cos^{-1} \left( \frac{\vec{\nabla} f}{|\vec{\nabla} f |} \cdot \frac{\vec{\nabla} \chi^2_E}{|\vec{\nabla} \chi^2_E |} \right)\ ,
\end{equation}
it changes $\chi^2_E$ by $\Delta \chi^2_E (\hat{a}_f) = \Delta \chi^2_E\, \cos \varphi (f, \chi^2_E) = S_{f, L2}(E)$.
The $L_2$ sensitivity, $S_{f, L2}(E)$, therefore quantifies the impact variations of PDFs
at fixed $x$ and $Q$ have upon the description of fitted data sets.  Plotting $S_{f, L2}(E)$
against $x$ yields useful information regarding the pulls of the CT18(Z) data sets 
upon PDFs (and PDF combinations) fitted in the global analysis. This also permits
the rapid visualization of possible tensions within the global fit, since
the PDF variation of some parton densities of given flavor are correlated with the
variation of $\chi^2_E$ ({\it i.e.}, $S_{f, L2}(E) > 0$), while others are
anti-correlated ($S_{f, L2}(E) < 0$), at the same values of $(x, Q)$.

The terms on the right-hand side of Eq.~(\ref{eq:L2}) for $S_{f,L2}$
are computed as
\begin{equation}
\Delta X=\left\vert \vec{\nabla}X\right\vert
=\frac{1}{2}\sqrt{\sum_{i=1}^{N_{\textrm{eig}}}\left(X_{i}^{(+)}-X_{i}^{(-)}\right)^{2}},\label{masterDX}
\end{equation}
and
\begin{equation}
\cos\varphi=\frac{\vec{\nabla}X\cdot\vec{\nabla}Y}{\Delta X\Delta
  Y}=\frac{1}{4\Delta X\,\Delta
  Y}\sum_{i=1}^{N_{\textrm{eig}}}\left(X_{i}^{(+)}-X_{i}^{(-)}\right)\left(Y_{i}^{(+)}-Y_{i}^{(-)}\right),\label{cosphi}
\end{equation}
from the values $X_{i}^{(+)}$ and $X_{i}^{(-)}$ that a quantity
$X$ takes for the parameter displacements
along the ($\pm$) direction of the $i$-th
eigenvector. With these symmetric master formulas, the sum of
$S_{f,L2}(E)$ over all experiments $E$ should be within a
few tens from zero, since the tolerance boundary for the total $\chi^2$ is close to being spherically
symmetric. The $S_{f,L2}(E)$ variables for individual experiments tend to
cancel among themselves to this accuracy; the order of magnitude 
of $S_{f,L2}(E)$ can be also interpreted as a measure  of tension of
$E$ against the rest of the experiments. 

The $L_2$ sensitivity can be computed for individual data point
residuals or optimal nuisance parameters, {i.e.}, for parts of
Eq.~(\ref{Chi2a0l0}). A related, similarly informative,
definition of sensitivity \cite{Wang:2018heo} is computed using the absolute values
of residuals, $|r_i|$, rather than their squares $r_i^2$ (using the
$L_1$ norm instead of the $L_2$ norm). 

\begin{figure}[!htbp]
	\begin{center}
		\includegraphics[width=0.85\textwidth]{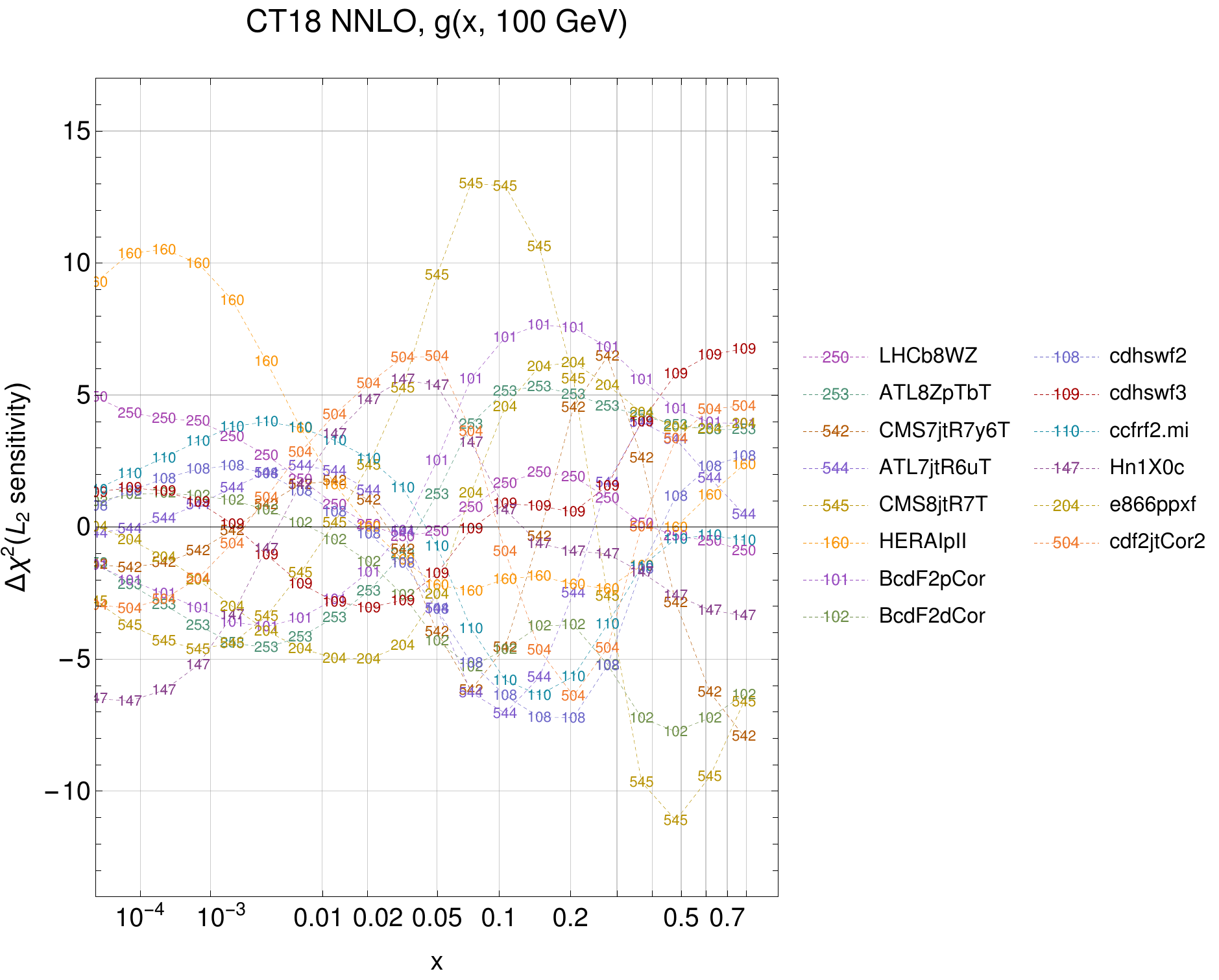}
	\end{center}
	\vspace{-2ex}
	\caption{
		The $x$-dependent $L_2$ sensitivity of the CT18 data sets with strongest
		pull upon the gluon PDF, $g(x,Q\!=\!100\,\mathrm{GeV})$. A number of tensions
		among the leading data sets are revealed by examining those regions of $x$
		where $S_{f, L2}(E)$ peaks for certain experiments in the `positive
		direction' while $S_{f, L2}(E)$ is sharply negative for others. For
		instance, at $x\! =\! 0.4$, the sensitivity curves indicate a strong competition of the CMS 8 TeV jet data (Exp.~ID=545) and the BCDMS $F^d_2$ data (Exp.~ID=102), both preferring a larger $g(0.4,100\mbox{ GeV}$), against the combined downward pull on the gluon by the BCDMS $F^p_2$ (101), CDHSW $F_3$ (109), E866 $pp$ Drell-Yan (204), and high-$p_T$ $Z$ boson production (253) data sets.
		At $x\! \approx \! 0.1$, the CMS 8 TeV
		jet data (Exp.~ID=545) strongly pulls against the ATLAS (544) and CMS 7 TeV (542) jet production, as well as measurements by CDHSW (108), CCFR (110), and BCDMS (102) of the DIS structure function $F_2(x,Q)$ on various targets.  
	}
\label{fig:L2glu}
\end{figure}

An extensive collection of the $L_2$
sensitivity plots for CT18(Z) PDFs and PDF ratios, reflecting the
interplay and competing pulls among the CT18 data sets, 
can be viewed at \cite{CT18L2Sensitivity}. Analogous calculations are shown for the
alternative CT18Z fit in Sec.~\ref{sec:LMCT18Z}.

In Fig.~\ref{fig:L2glu}, we show the $L_2$ sensitivity of the CT18 data
on the gluon PDF at fixed $Q\! =\! 100$ GeV, plotting curves for those
experiments that satisfy $|S_{f, L2}(E)| \ge 4$ for any value of $x$.
This criterion generally identifies the leading $5-10$ experiments with
strongest pulls on the PDF in the kinematical region under consideration.
By its proximity to the Higgs mass scale, $Q\!=\!100$ GeV,
Fig.~\ref{fig:L2glu} highlights the opposing pulls of a number of CT18
data sets relevant for the 14 TeV Higgs boson production cross section,
$\sigma_H (14\,\mathrm{TeV})$, and is the $L_2$-based counterpart
to Fig.~\ref{fig:LMg18} (left). In addition to several non-LHC experiments (Expt. ID~=101, 102, 108, 109, 160, 204) 
imposing significant pulls on the gluon PDF at various $x$, 
among the newly-fitted LHC Run-1 data, the 8 TeV $Z$ $p_T$ ATLAS data
(Exp.~ID=253) show the strongest overall pull in the immediate vicinity of $x=0.01$, $S_{g, L2}(E)\, \approx\, -(4\!-\!5)$,  approaching the pull of the E866 $pp$
absolute cross section data (Exp.~ID=204) in the same neighborhood.
Meanwhile, the corresponding pulls of the inclusive Tevatron (504) and CMS 8 TeV (545) jet-production data  are even larger
at slightly higher $x\! = \! 0.05-0.1$; at $x\approx 0.1$ the CMS 8 TeV jet data (545)
have a very strong pull of $S_{g, L2}(E)\! \approx\! +13$ against the opposing pulls of the ATLAS (544) and CMS 7 TeV (542) jet data data sets. At even higher $x$, the best-fit behavior of the gluon PDF reflects the tradeoffs among the pulls from multiple experiments, as explicated in the caption of Fig.~\ref{fig:L2glu}. 
At $x\! < \! 0.01$, we notice visible competition between the inclusive (160) and charm-production (147) data sets from HERA, with constraints from other experiments being less prominent in this region. The totality of observations based on Fig.~\ref{fig:L2glu} is consistent
with our findings based on the LM scans appearing in Sec.~\ref{sec:LMScans}, as typified by
Fig.~\ref{fig:LMg18} (left panel), wherein we identified the same experiments as imposing
the most stringent constraints upon $g(x\!=\!0.01, Q\!=\!m_H)$.

\begin{figure}[p]
\center
  \includegraphics[width=0.59 \textwidth]{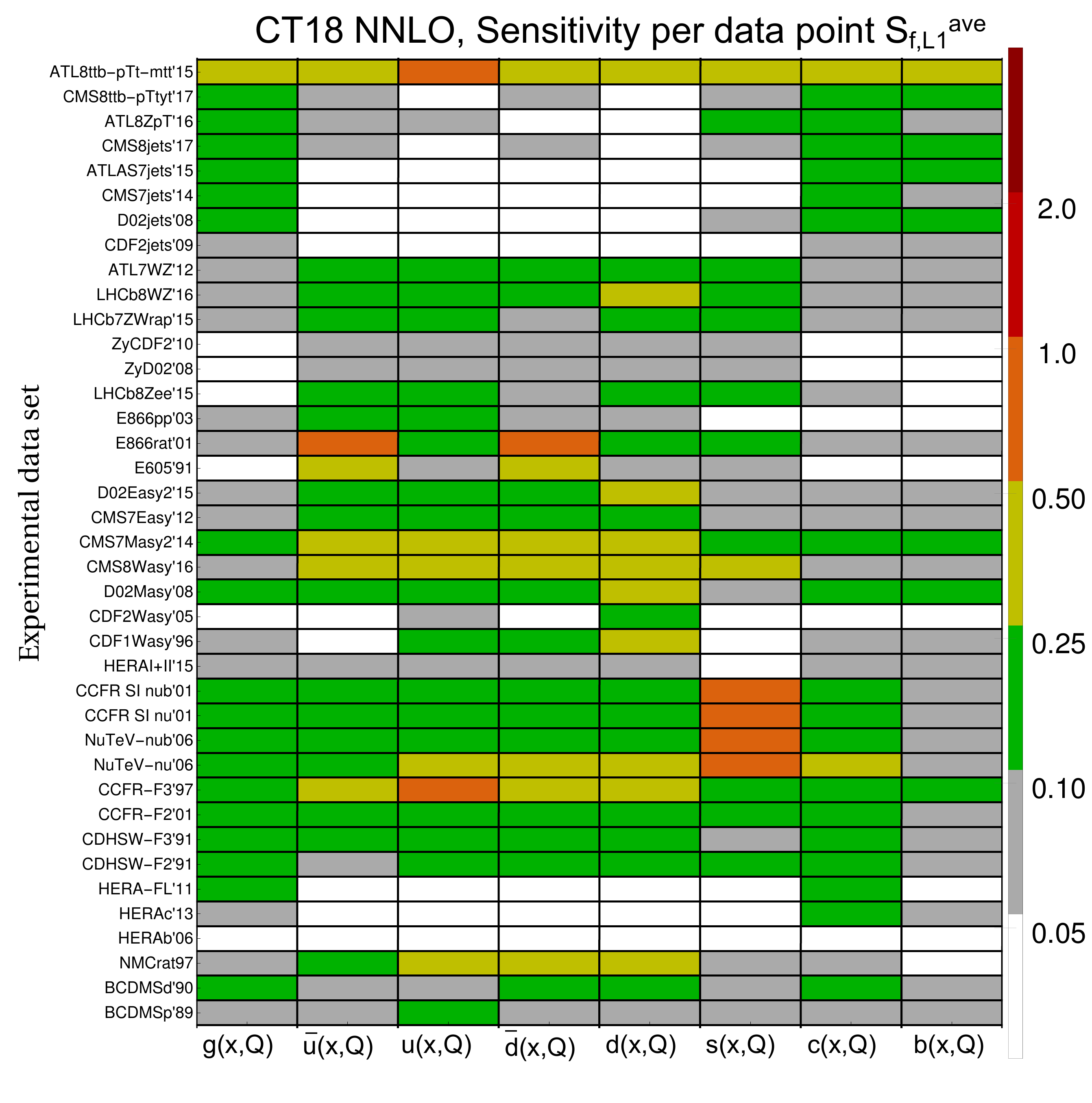}\\ 
  \includegraphics[width=0.59\textwidth]{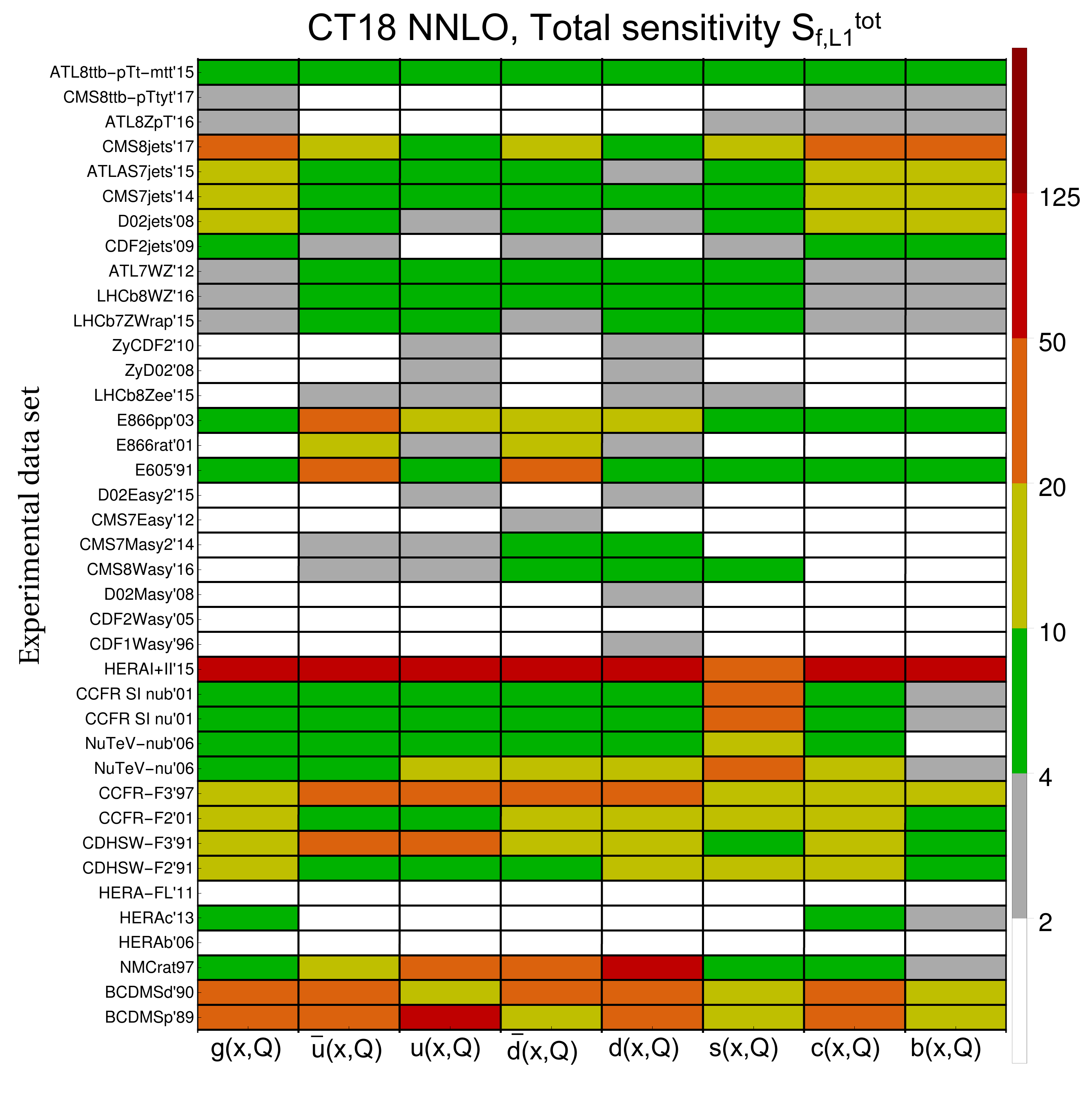}
        	\caption{$L_1$ sensitivities of experimental data
                  sets to PDF flavors in the CT18 NNLO analysis,
                  computed according to the methodology in
                  Ref.~\cite{Wang:2018heo}. The color of the cells
                  in the upper (lower) inset, 
                  chosen according to the palettes on the right,
                  indicates the point-average
                  (cumulative) sensitivity of the experimental set
                  on the vertical axis to the PDF flavor
                  on the horizontal axis. 
		\label{fig:CT18quilts}}
\end{figure}

We conclude this section by presenting 
Fig.~\ref{fig:CT18quilts} with the ranking plots of $L_1$
sensitivities computed by the \texttt{PDFSense} code
according to the approach in Ref.~\cite{Wang:2018heo}.
In that article, we presented tables that
rank the experiments in the \CTHERAII~NNLO analysis either according to
their total sensitivity to the PDFs, $f(x_i,Q_i)$, computed as
\begin{equation}
S_{f,L1}^{\textrm{tot}}(E)\equiv\sum_{i=1}^{N_{pt,E}} \left|S_{f,L1}(i)\right|,
\label{SfL1tot}
\end{equation}
or according to the average sensitivity per data point,
\begin{equation}
S_{f,L1}^{\textrm{ave}}(E)\equiv S_{f,L1}^{\textrm{tot}}(E)/N_{pt,E}.
\label{SfL1ave}
\end{equation}
These quantities respectively
estimate either the total sensitivity of the experiment $E$
to the PDF, $f(x_i,Q_i)$, at the typical $(x_i, Q_i)$ probed by data
points $i=1,..,N_{pt,E}$, and summed over all $N_{pt,E}$ points; or the
averaged sensitivity for a single data point in this experiment.
The two sensitivities allow
informative side-by-side comparison of the strengths of
constraints from individual experiments, once again estimated in the
Hessian approximation.  

In Fig.~\ref{fig:CT18quilts}, we present a graphical visualization of the ranking tables from
Ref.~\cite{Wang:2018heo}, now recomputed for the CT18 NNLO fit, and, for the most part, leading to similar conclusions as obtained for
\CTHERAII~NNLO.  The upper and lower panels of Fig.~\ref{fig:CT18quilts} correspond to the point-averaged and total
sensitivities, respectively, as discussed above. At right are placed palettes
relating the colors to the magnitudes of $S_{f,L1}(E)$.
The cells that vary from yellow to orange to red
indicate experiments (listed on the left)
with increasingly strong sensitivities to the PDFs,
$f(x,\mu)$, given at the bottom.
White or grey cells indicate experiments
with minimal sensitivity to $f(x,\mu)$.

We observe that, while the HERA I+II, BCDMS, and NMC data sets have
relatively low per-point sensitivity as seen in the upper panel,
when aggregated over their large number of points, the experiments have
very large total sensitivities to all PDF flavors
seen in the lower inset. The specialized fixed-target
measurements, such as CCFR, NuTeV, E605, and E866, are most sensitive
to certain flavors, such as $s$, $\bar u$, and $\bar d$, as expected. 

Several LHC experiments, on the other hand, have strong per-point sensitivities,
especially $t\bar t$ and high-$p_T$ $Z$ production, as
well as CMS $W$-charge asymmetries at 7 and 8 TeV (see the upper
inset). The total sensitivities of these experiments in the lower inset
are still quite low because of their small numbers of data points
($N_{pt,E}\approx 10-20$). On the other hand, the inclusive
jet production data sets
by ATLAS and CMS at 7 TeV, and especially by CMS at 8 TeV,
despite their modest sensitivities per data point, show the highest
total sensitivities among all LHC experiments because of their large
numbers of data points and extended kinematic coverage.

In aggregate, while the bulk of the sensitivity in the CT18 fit
still arises from HERA and fixed-target data, the LHC
experiments could already reduce some PDF uncertainties, given their
sizable per-point sensitivities. These uncertainty reductions have not
yet been fully realized in part due to the tensions among some LHC experiments
expounded upon earlier in the paper. 

\subsection{Description of data sets fitted in CT18}
\label{sec:Qualitydata}
In this subsection, we illustrate the ability of CT18 to describe the individual
experiments included in this analysis, with particular attention paid to the
newly included LHC Run-1 data. We organize this discussion according to the specific
physical process.

\begin{figure}[tb]
	\includegraphics[width=0.49\textwidth]{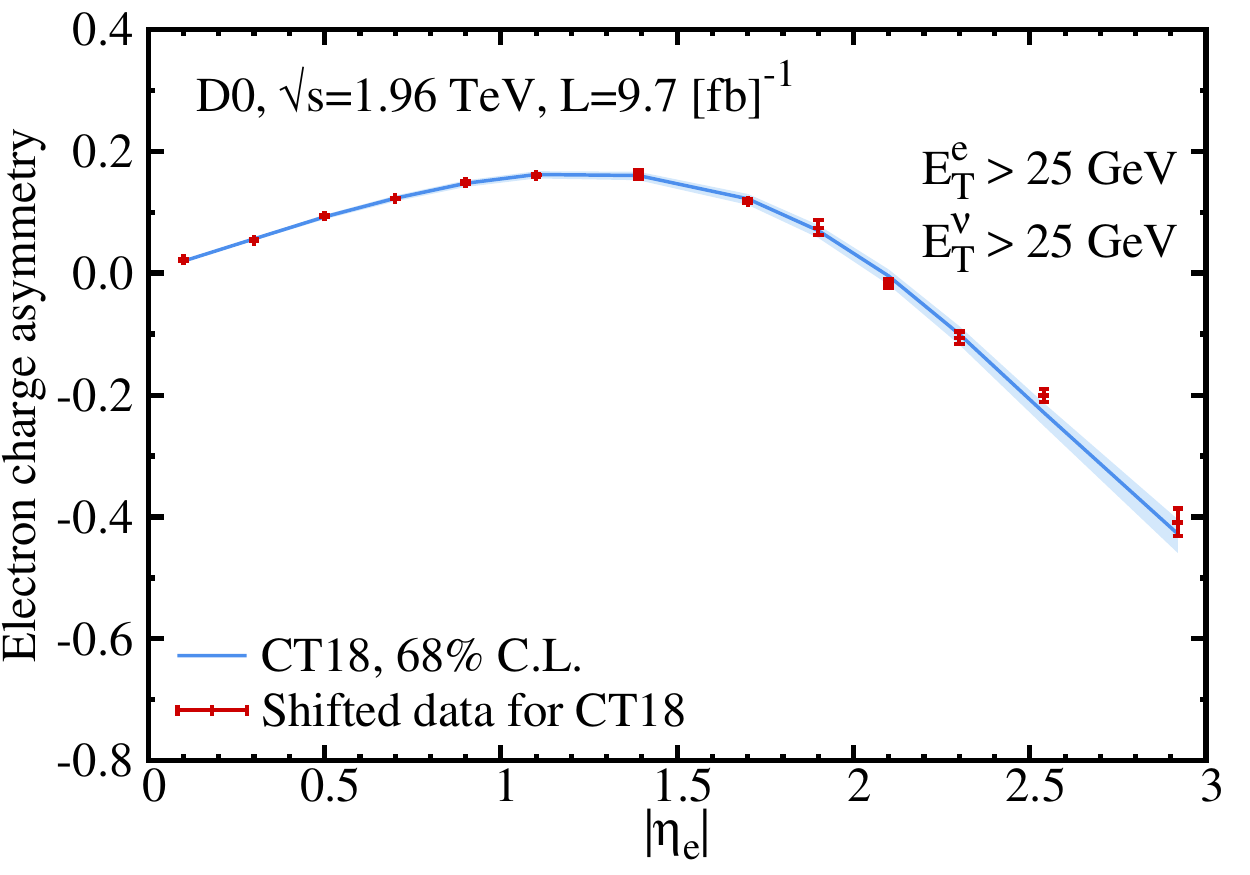}
	\includegraphics[width=0.49\textwidth]{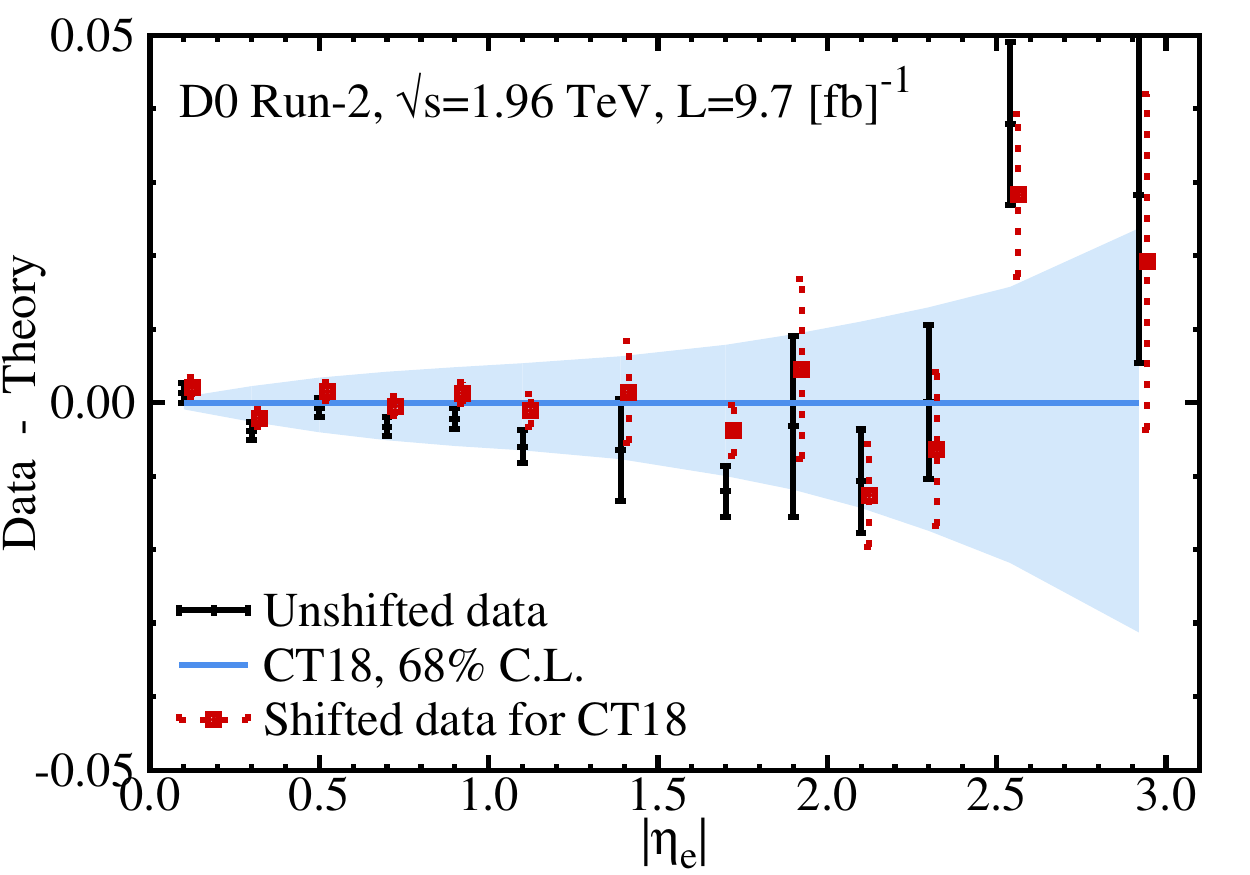}
	\caption{
	A comparison of the CT18 theory with the D0 Run II electron charge-asymmetry data (Exp.~ID=281). Since the asymmetry crosses zero in the shown range, the right panel shows the {\it difference}, $\mathrm{Data}\!-\!\mathrm{Theory}$, rather than the {\it ratio}, $\mathrm{Data}/\mathrm{Theory}$, as done elsewhere in this section.
	}
\label{fig:281}
\end{figure}

\subsubsection{Vector boson production data}
\label{sec:QualityDYdata}
{\bf Tevatron charge asymmetry.}
CT18 PDFs show a good overall agreement with the vector boson production data from fixed-target and Tevatron experiments. In particular, the high-luminosity charge asymmetry data set 281 from D0 Run-2 \cite{D0:2014kma}, used in our analysis since CT14 \cite{Dulat:2015mca} and sensitive to $d(x)/u(x)$ at $x>0.1$, is well described.\footnote{According to the $L_2$ sensitivity \cite{CT18L2Sensitivity}, the NMC DIS data 104 and the charge asymmetry data set 281 prefer to have a softer $d(x)/u(x)$ at large $x$ by about (15) 5 units of $\chi^2_E$, compared to the full data, in contrast to the LHCb 7 TeV W rapidity (245) and E866 $pp$ Drell-Yan (204) data sets that prefer a harder $d/u$ in the same $x$ region.}

Fig.~\ref{fig:281} shows a data versus theory comparison for the electron charge asymmetry as a function of the absolute value of the electron pseudorapidity.
Shifted data are represented by red points, while unshifted data are black. 
The absolute charge asymmetry is illustrated in the left inset of Fig.~\ref{fig:281}, while in the right one we show the 
$\mathrm{Data}\!-\!\mathrm{Theory}$ difference, where the error bars represent the total uncorrelated uncertainty (the quadrature sum of uncorrelated statistical and uncorrelated systematic errors) for both the shifted and unshifted data, as we show consistently
throughout this paper, unless specified otherwise. The differences of the shifted data from theory, relative to the error bars, exemplify the goodness-of-fit, while the movements between the shifted and unshifted data show the effect of the correlated nuisance parameters.
The theoretical predictions are computed using the code \texttt{ResBos} at approximate NNLO + NNLL in QCD. 
The blue band represents the CT18 PDF uncertainty evaluated using the Hessian symmetric errors at 
the 68\% C.L. We see that the data are described well by the CT18 predictions, with the 
exception of one high pseudorapidity bin ($|\eta_{e}|\sim2.6$), in which we observe a mild disagreement.

\begin{figure}[tbp]
	\includegraphics[width=0.49\textwidth]{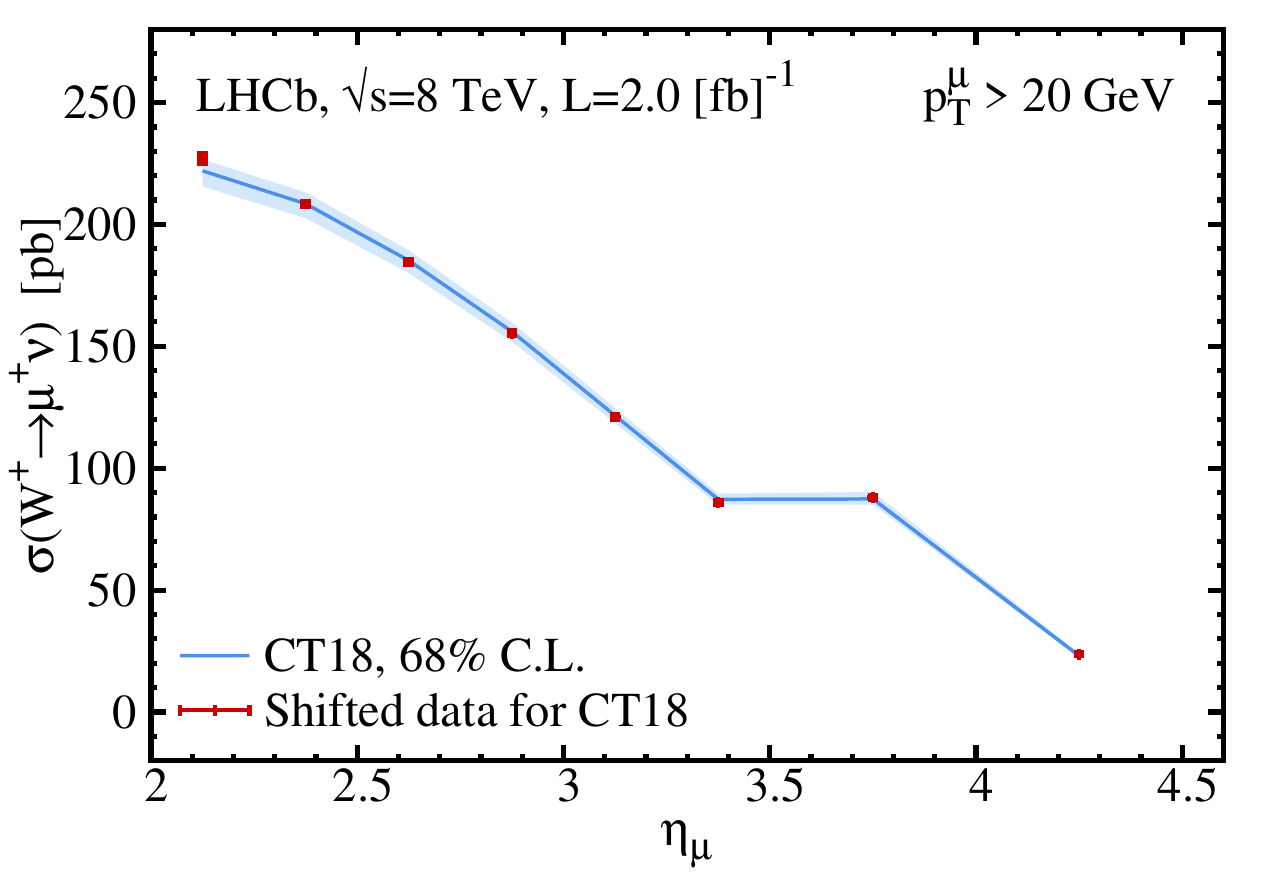}
	\includegraphics[width=0.49\textwidth]{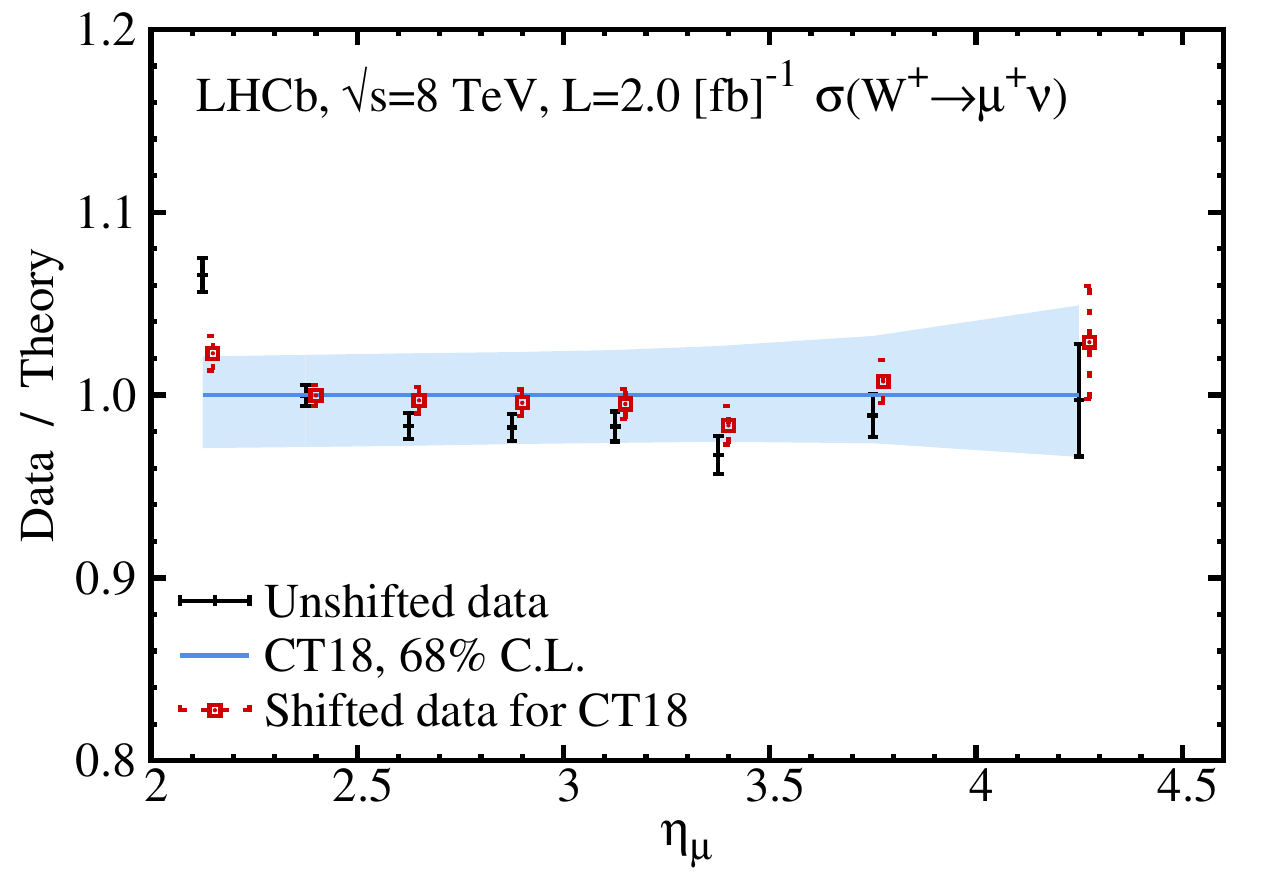}
	\includegraphics[width=0.49\textwidth]{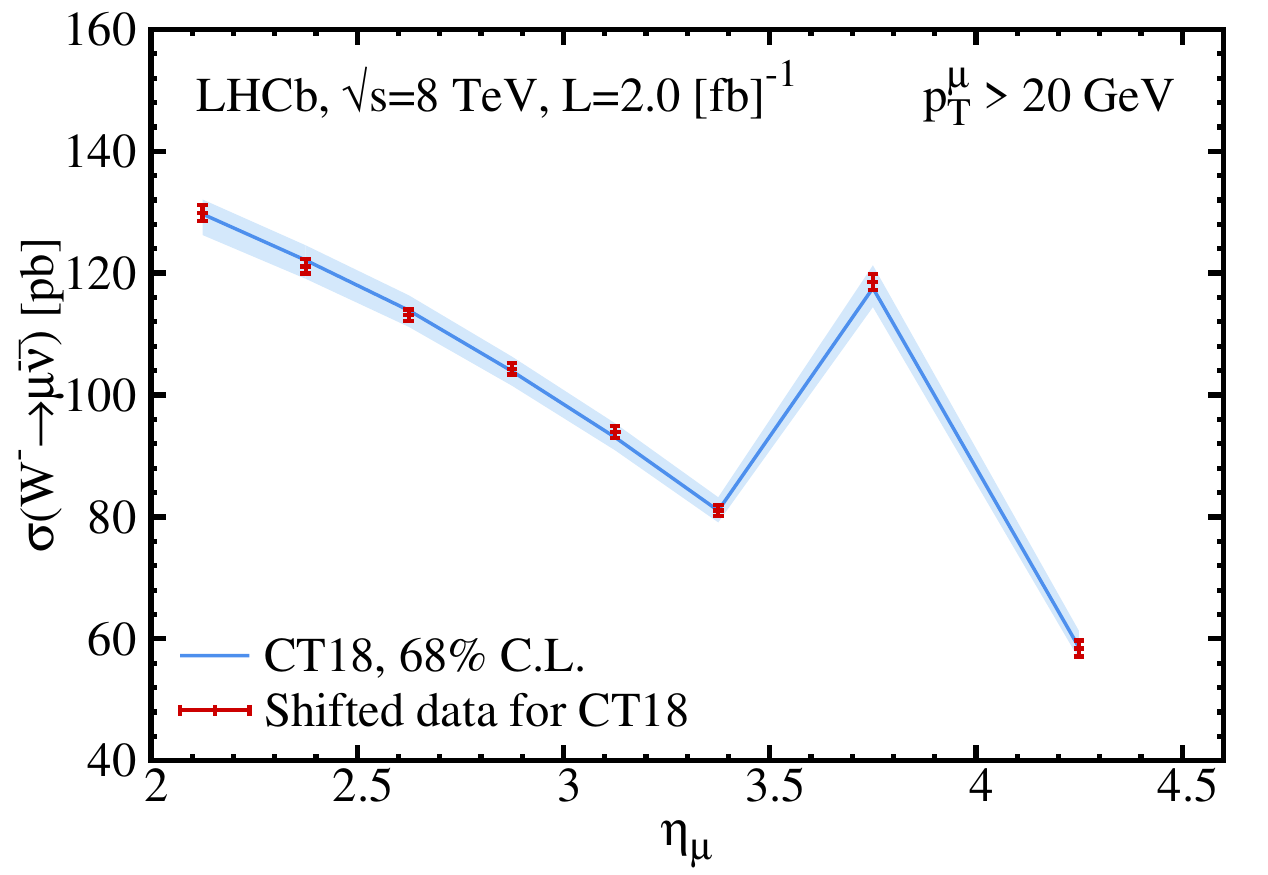}
	\includegraphics[width=0.49\textwidth]{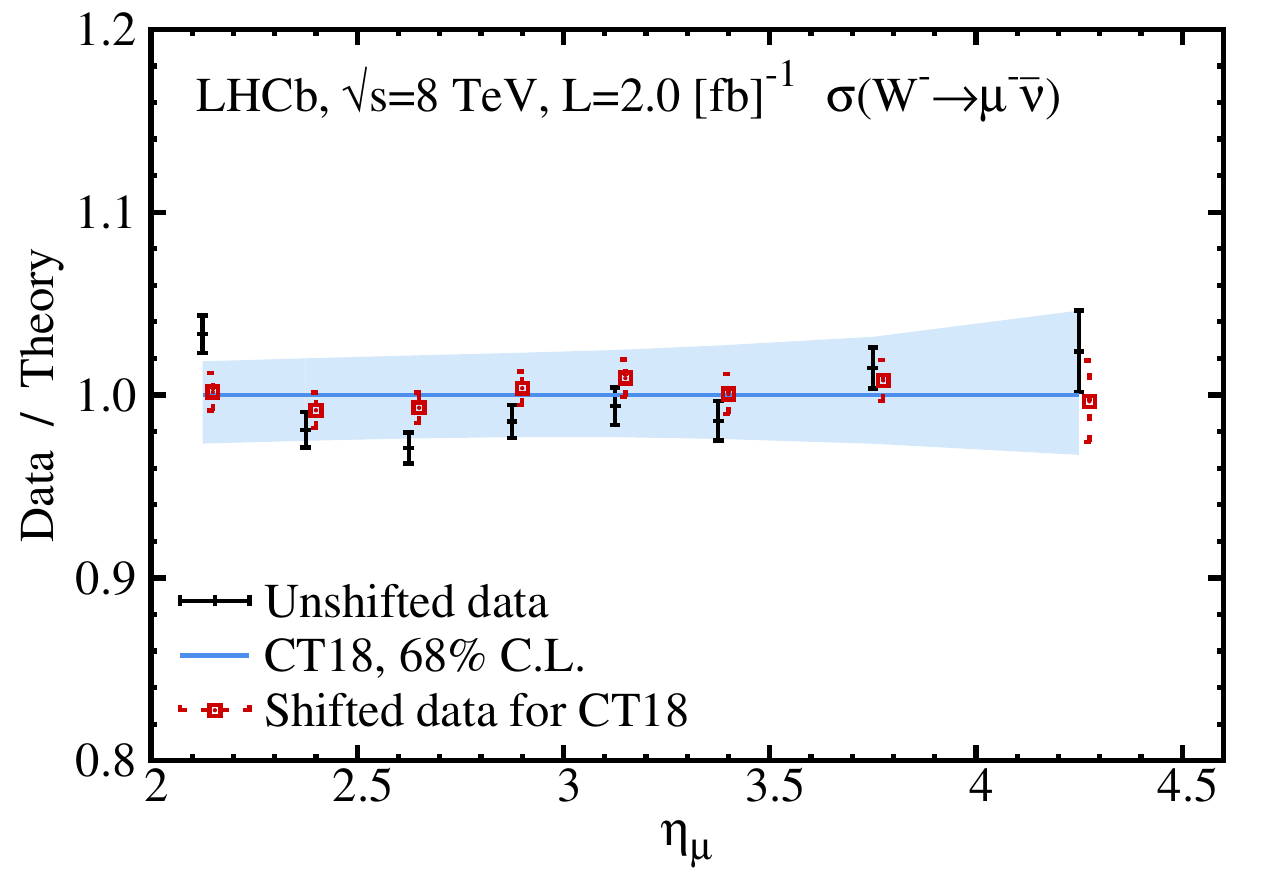}
	\includegraphics[width=0.49\textwidth]{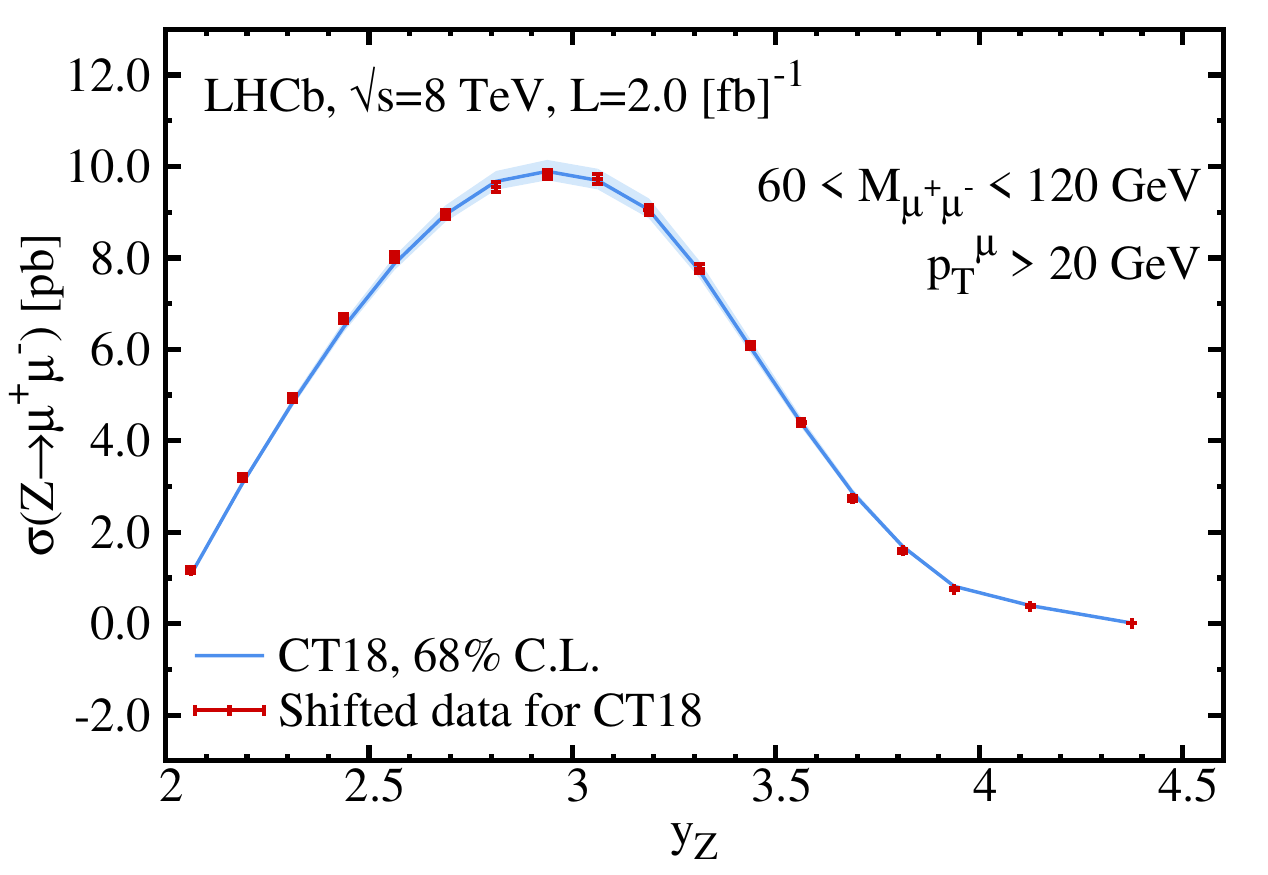}
	\includegraphics[width=0.49\textwidth]{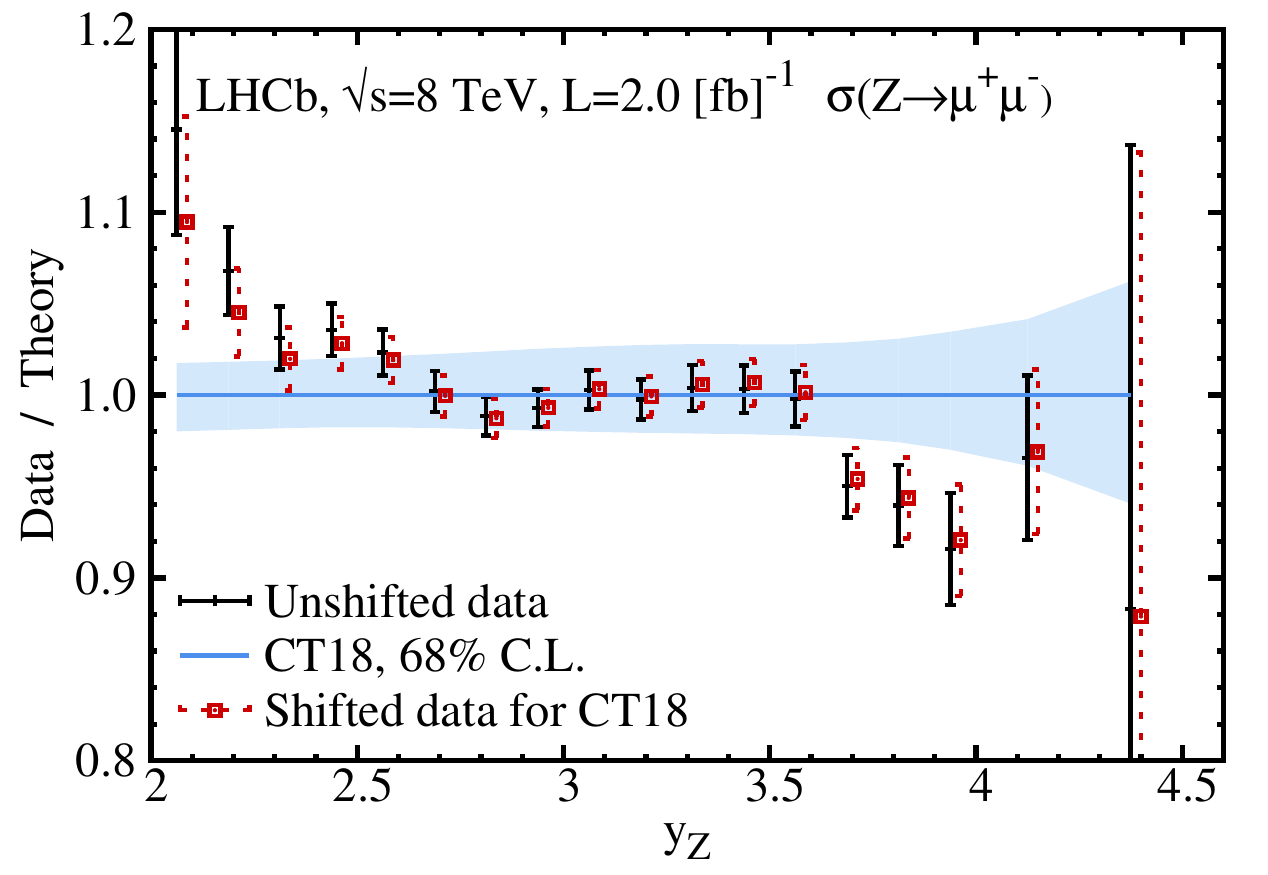}
	\caption{A comparison of the CT18 theoretical predictions to the $W^+$ (top), $W^-$ (middle), and $Z^0$ (bottom) cross section measurements by LHCb at 8 TeV in the muon decay channel (Exp.~ID=250). The data are presented as cross sections for each bin, $\sigma=\frac{d\sigma}{d\eta_\mu}\Delta\eta_\mu,\frac{d\sigma}{dy_Z}\Delta y_Z$, rather than as differential cross sections.
	The bump in the histogram bin $3.5<\eta_\mu<4.0$ of $W^-$ plot thus results from its larger bin width. A similar bump occurs in the plots for the LHCb 7 TeV $W/Z$ data (Exp. ID=245).
			\label{fig:id250}
}
\end{figure}

\begin{figure}[tb]
	\includegraphics[width=0.49\textwidth]{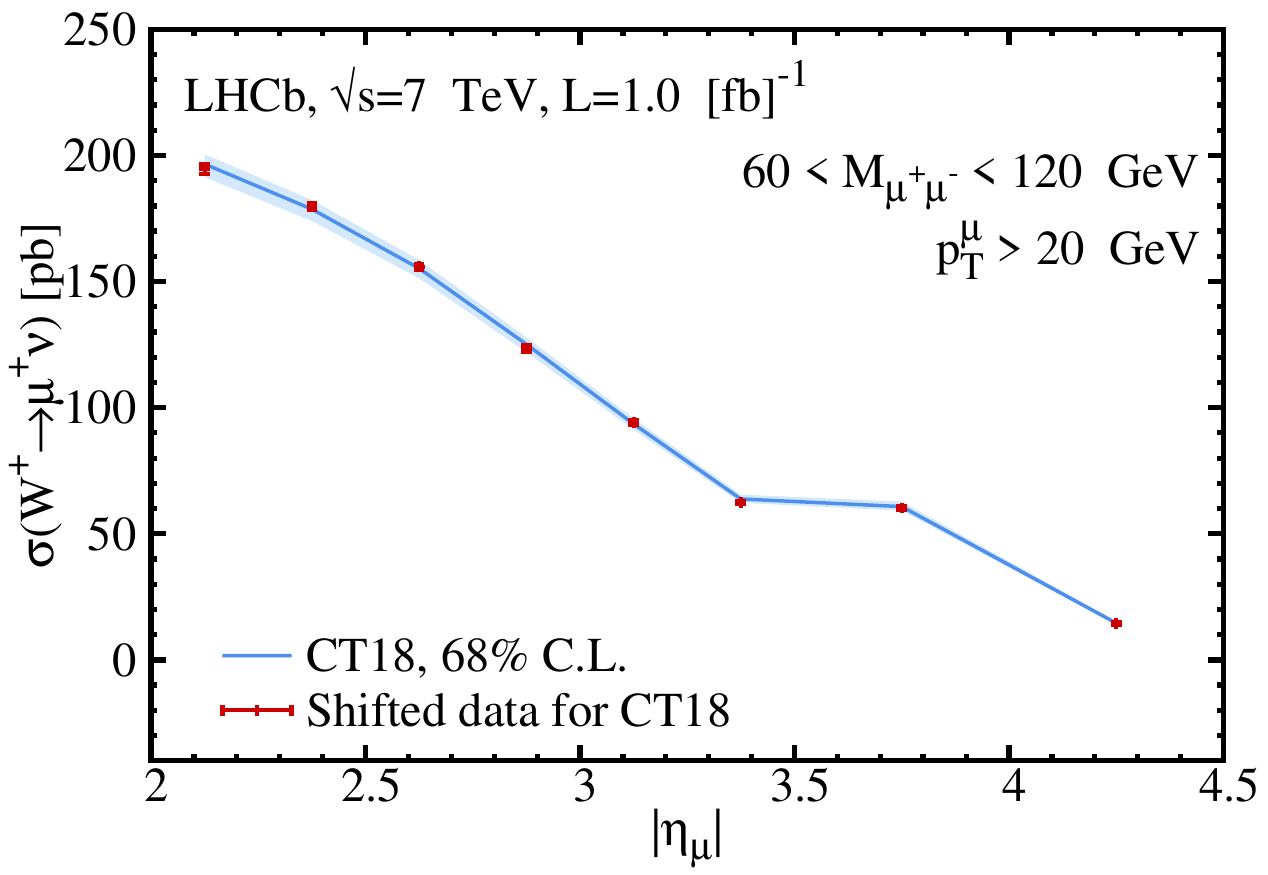}
	\includegraphics[width=0.49\textwidth]{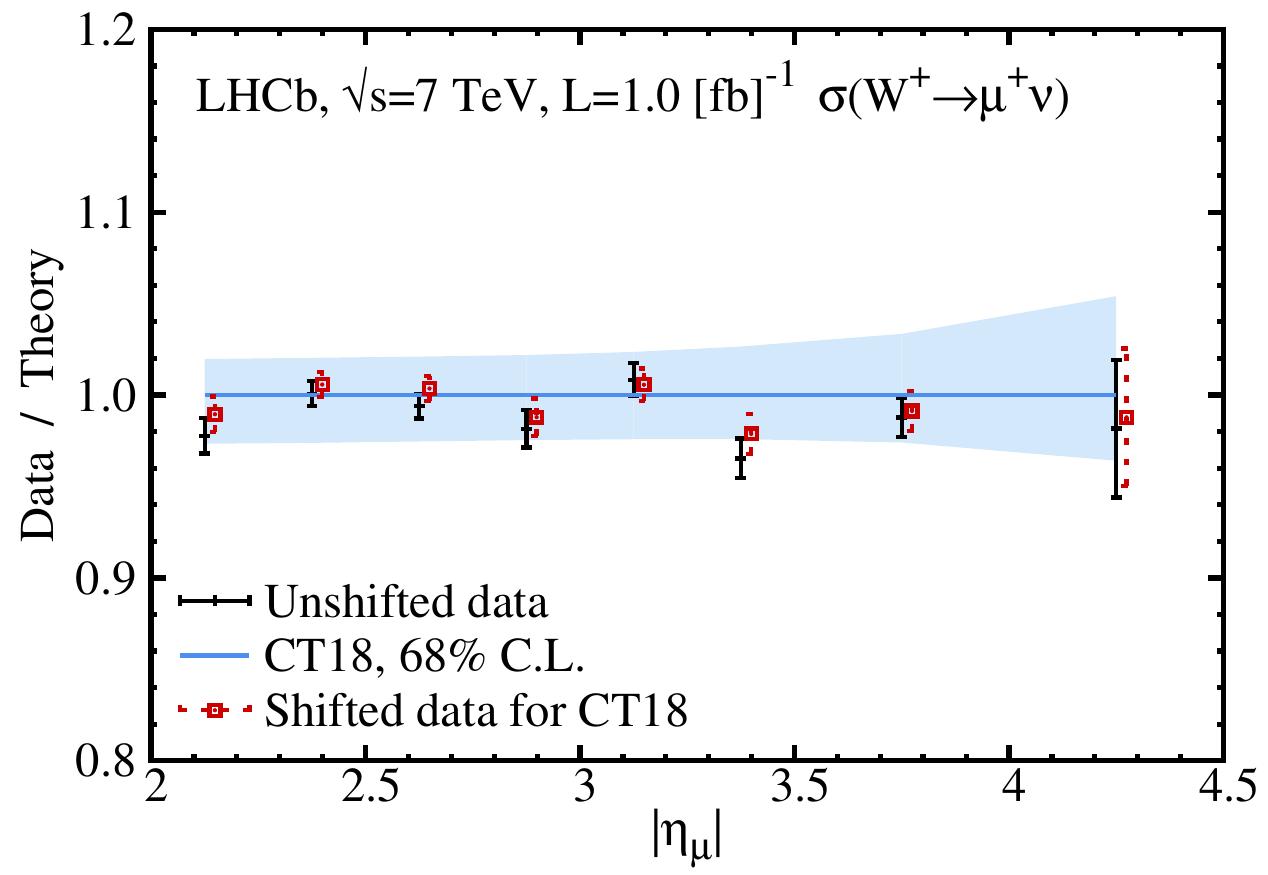}
	\includegraphics[width=0.49\textwidth]{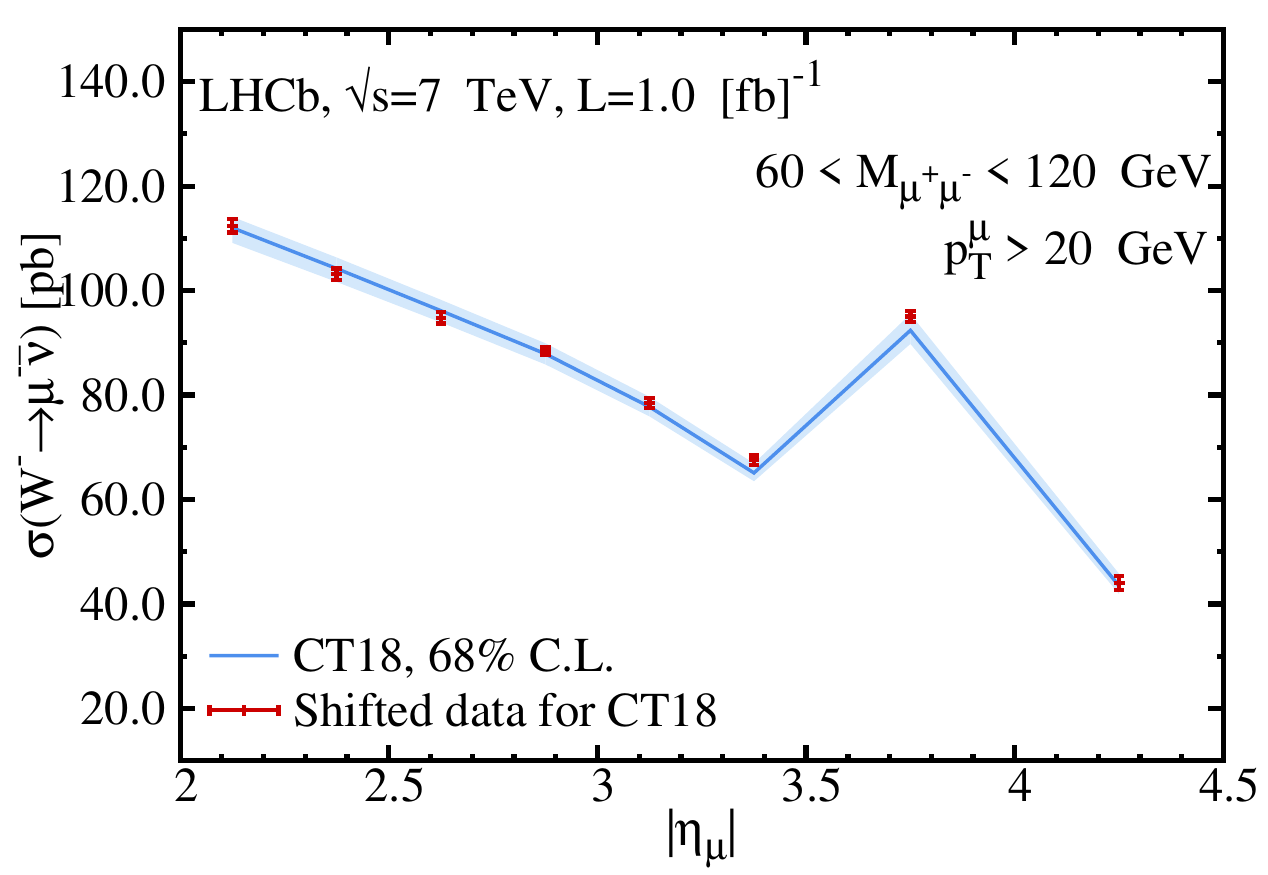}
	\includegraphics[width=0.49\textwidth]{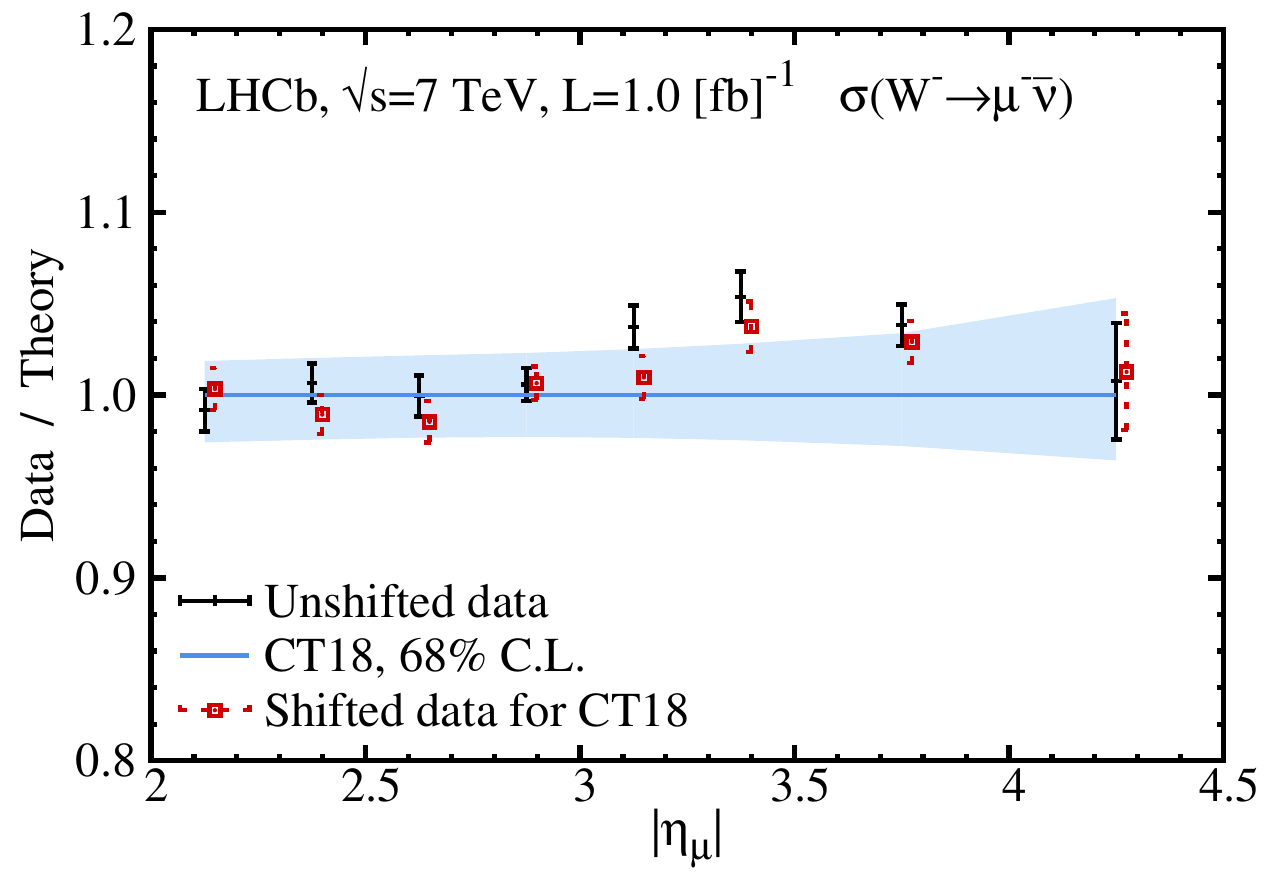}
	\includegraphics[width=0.49\textwidth]{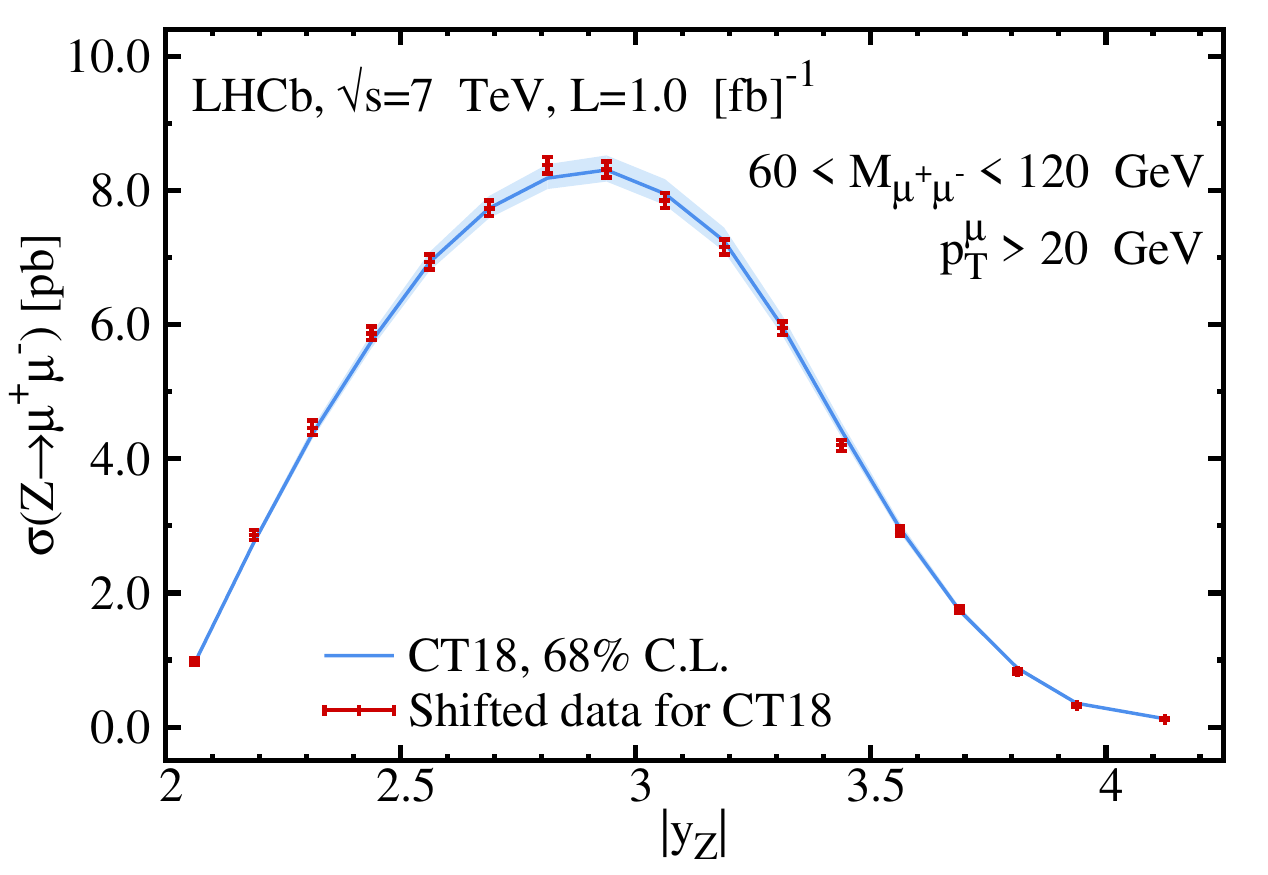}
	\includegraphics[width=0.49\textwidth]{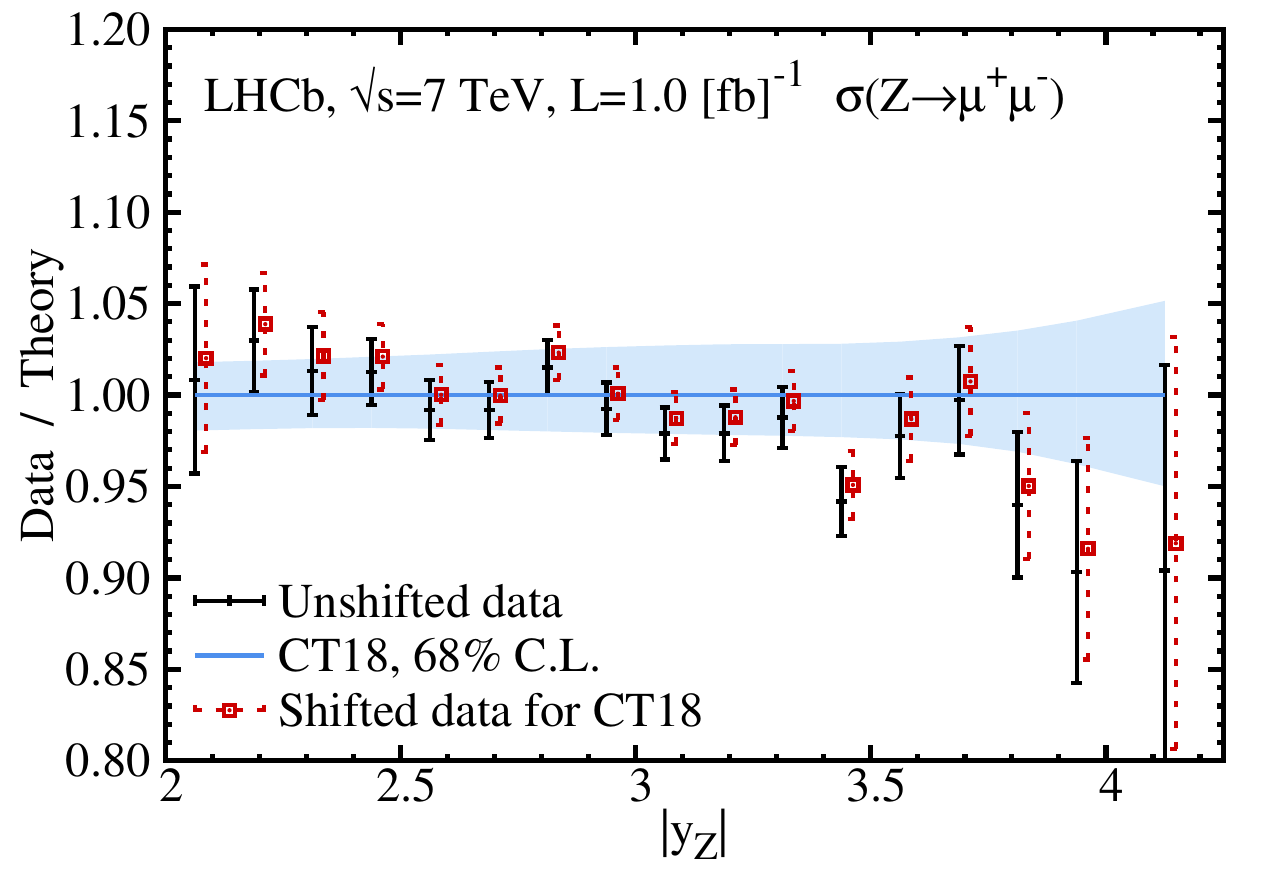}
	\caption{Same as Fig.~\ref{fig:id250}, for the LHCb 7 TeV (Exp.~ID=245).}
	\label{fig:id245}
\end{figure}

\begin{figure}[p]
\includegraphics[width=0.49\textwidth]{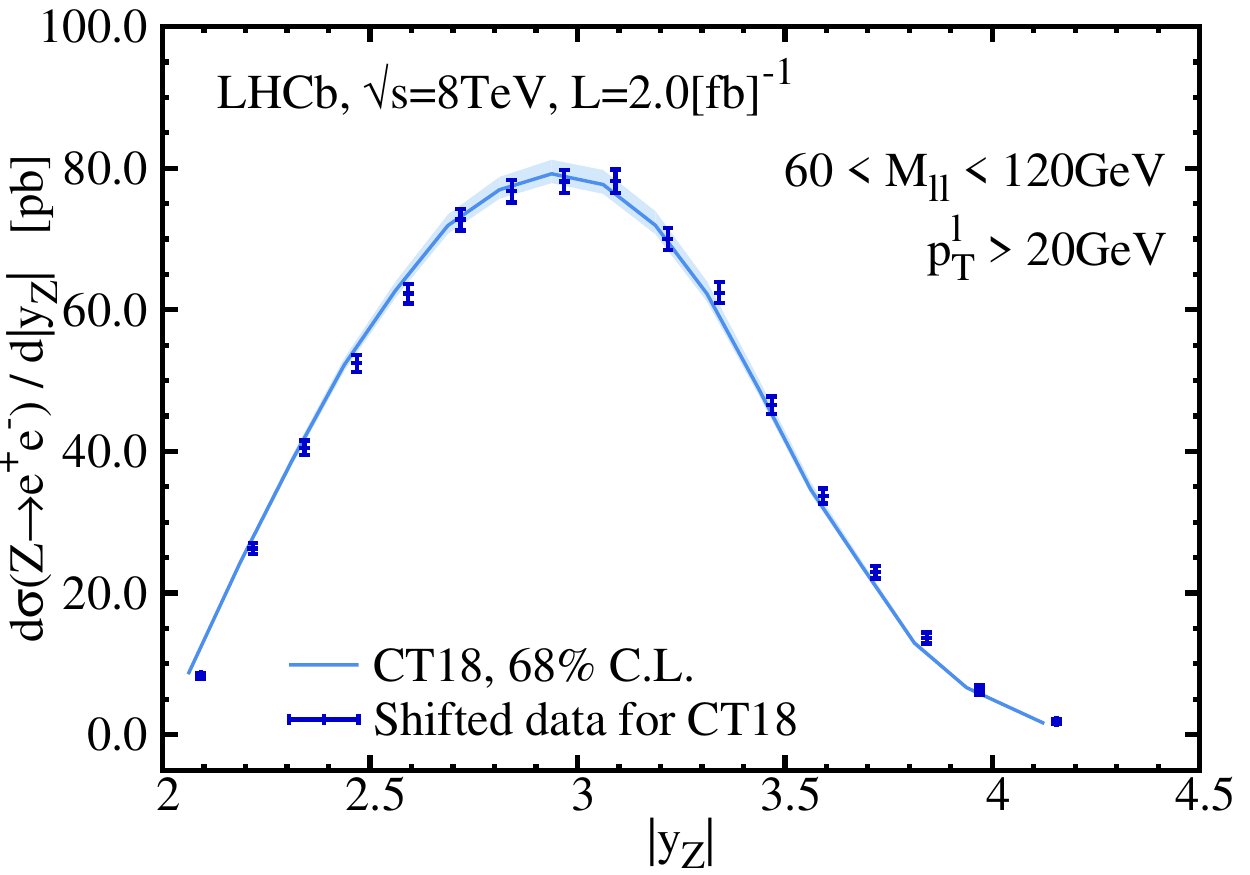}
\includegraphics[width=0.49\textwidth]{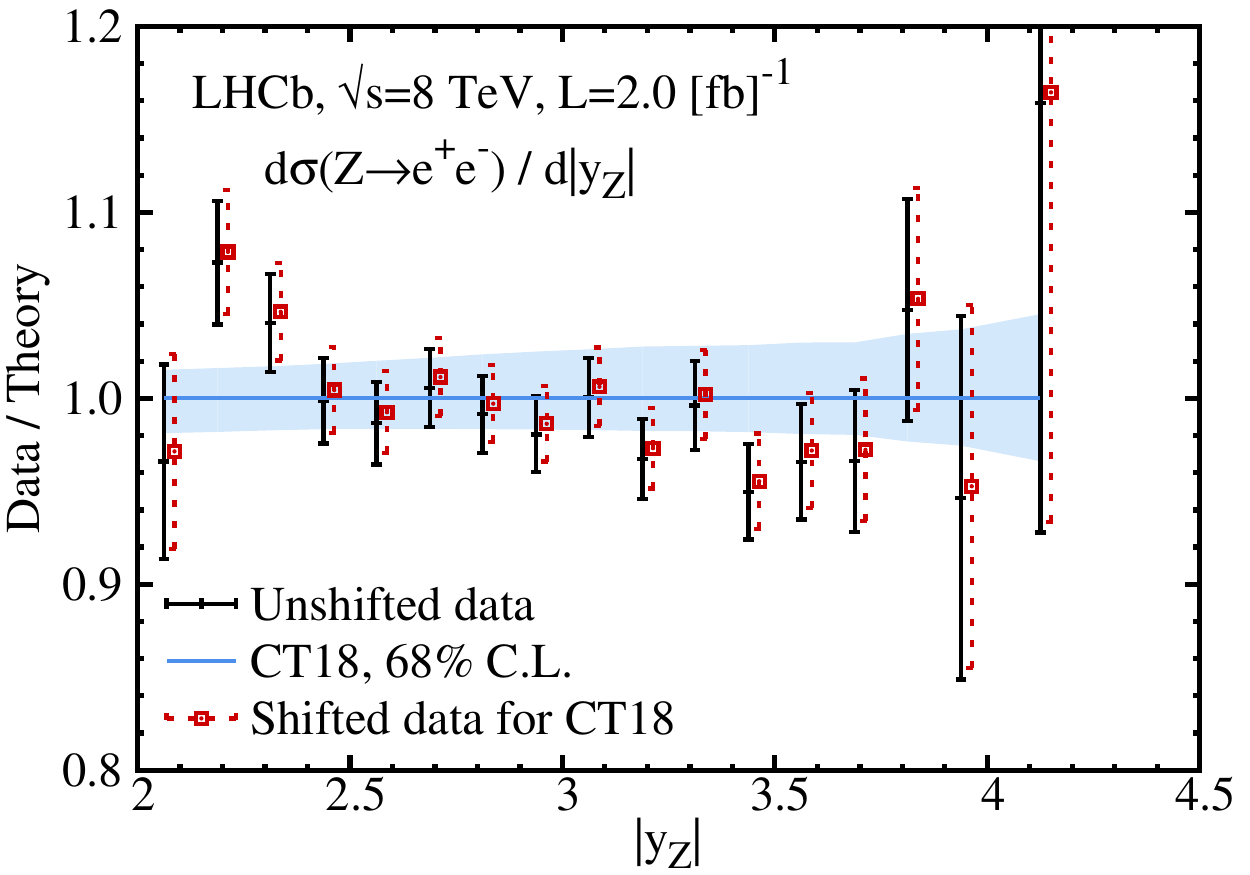}
\caption{A comparison of the CT18 theoretical predictions to the $Z$ rapidity distribution in $Z\to e^+e^-$ production by LHCb at 8 TeV (Exp.~ID=246) .
}
\label{fig:LHCb8Zee}
 \end{figure}

{\bf LHC data: LHCb.}
As discussed previously, Drell-Yan cross-section measurements from the LHCb collaboration (Exp.~IDs=250, 245 and 246, 
in that order of importance) produce the strongest impact on CT18 PDFs among the newly introduced LHC Drell-Yan data sets. Similarly to the D0 electron charge asymmetry, in Fig.~\ref{fig:id250}, 
the CT18 NNLO theory prediction is compared to both the shifted and the unshifted data of
$W/Z$ production  in the muon channel (Exp.~ID=250) at 8 TeV. 
The analogous comparisons to $W/Z$ production in the $\mu$ channel (Exp.~ID=245) at 7 TeV, and to $Z$ production in the $e$ channel (Exp.~ID=246) at 8 TeV are respectively shown in Fig.~\ref{fig:id245} and Fig.~\ref{fig:LHCb8Zee}.  
The NNLO theory is obtained using \texttt{APPLgrid} files generated with NLO \texttt{MCFM}, and multiplied by point-by-point $K$-factors computed 
with \texttt{FEWZ} and \texttt{MCFM-8.0}. 

In the case of $Z/W$ boson production at 8 TeV in Fig.~\ref{fig:id250}, 
theory and data agree well except for the data points near rapidity of 2. 
In $Z$ boson production in all three data sets (bottom rows), some disagreement between theory and data in shape at $y_Z <2.5$ and $y_Z=3-4$ remains in spite of systematic shifts. It leads to the  elevated $\chi^2_E$ for experiments 245 and 250 quoted in Table \ref{tab:EXP_2}, the discrepancy that is partially alleviated in the CT18Z fit after including the ATLAS 7 TeV $W/Z$ production data set (Expt.~ID=248).
At low rapidity in $Z$ production, there is a large modeling uncertainty for the kinematic acceptance of the observed leptons. 
On the other hand, the discrepancy at $y_Z\approx 4$ shows tension with pulls from other data sets included in the global fit, 
such as the CMS and D0 $W$ lepton-charge asymmetry data. This tension has been investigated using the \texttt{ePump} program. In particular, we compared updated fits in which we either had included, or had not included, the first (low rapidity) bin of the $Z$-boson distribution for Experiment 246. This choice had little impact on the resulting PDFs, though the $\chi^2_E/N_{pt,E}$ of the LHCb data had noticeably improved after dropping the first rapidity bin. 

\begin{figure}[t]
	\includegraphics[width=0.32\textwidth]{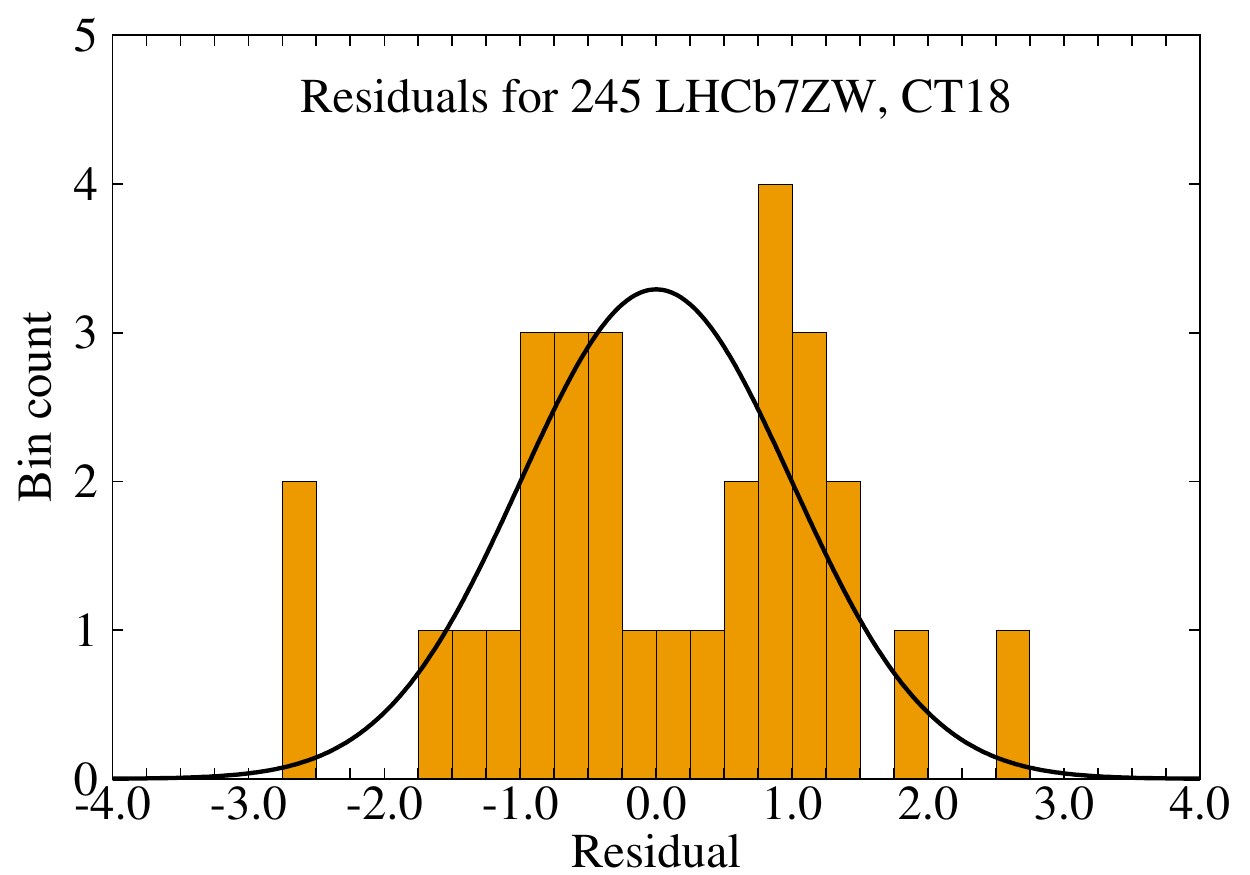}
	\includegraphics[width=0.32\textwidth]{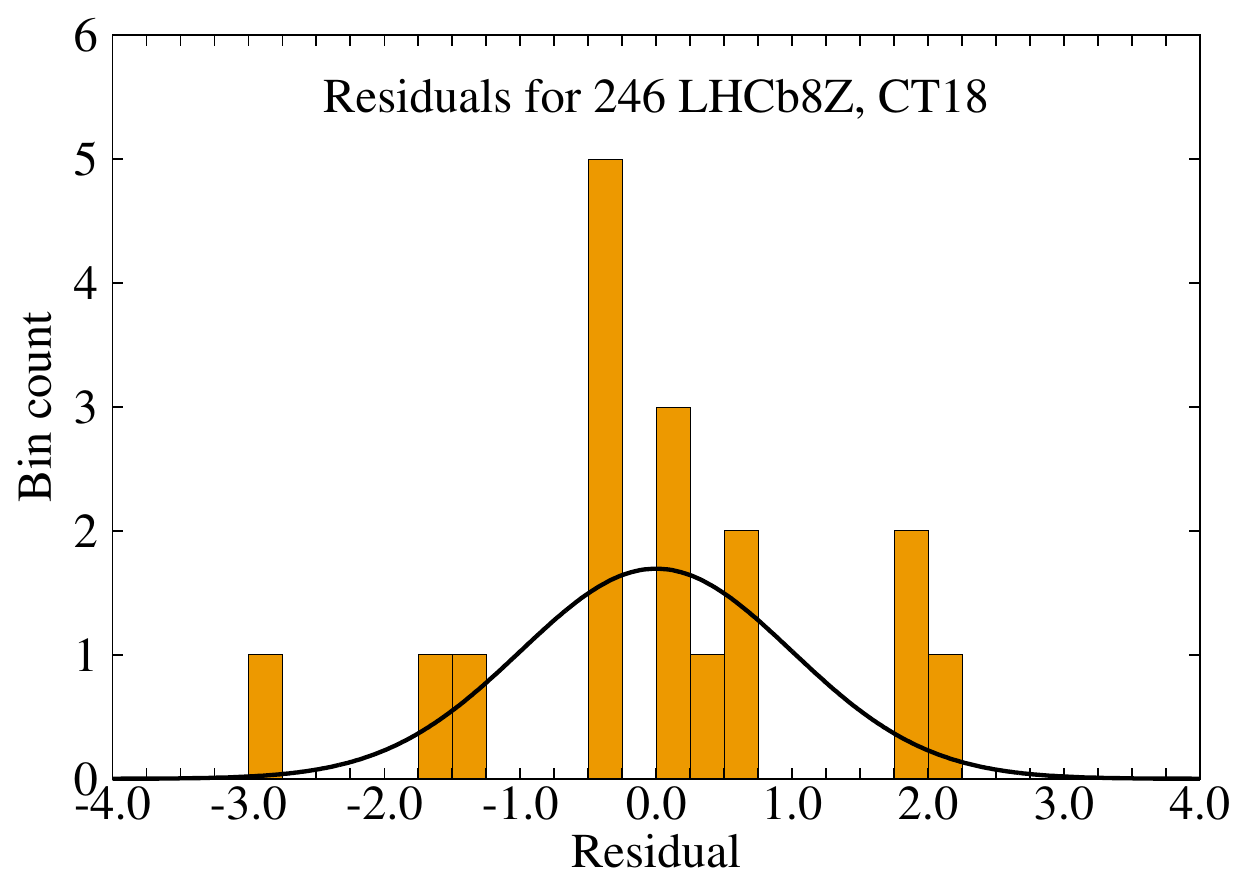}
	\includegraphics[width=0.32\textwidth]{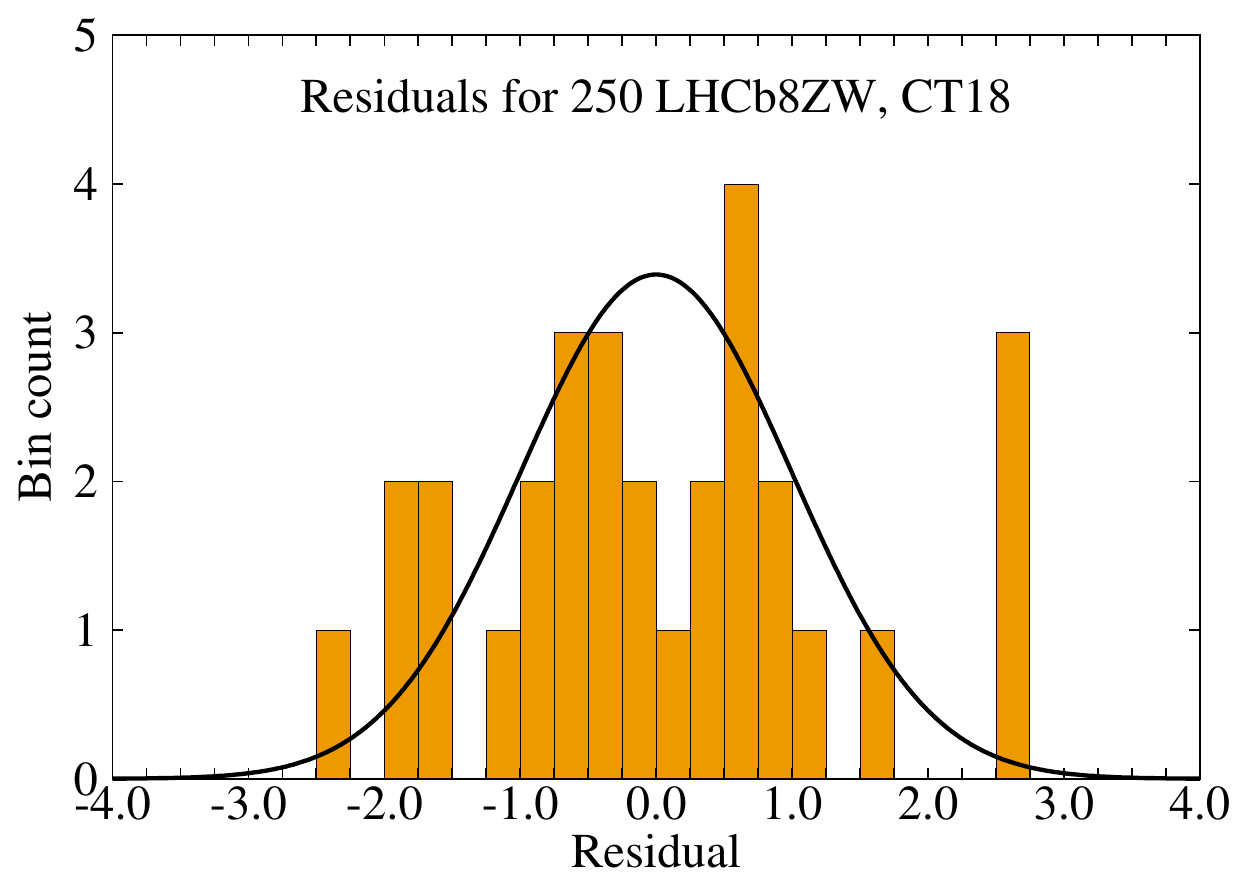}
	\caption{Distributions of the residuals for the LHCb $W/Z$ production cross sections at 7 and 8 TeV: Exp.~ID=245 (left), 246 (center), and 250 (right).
		\label{fig:res_rk_1}}
\end{figure}

\begin{figure}[t]
	\includegraphics[width=0.45\textwidth]{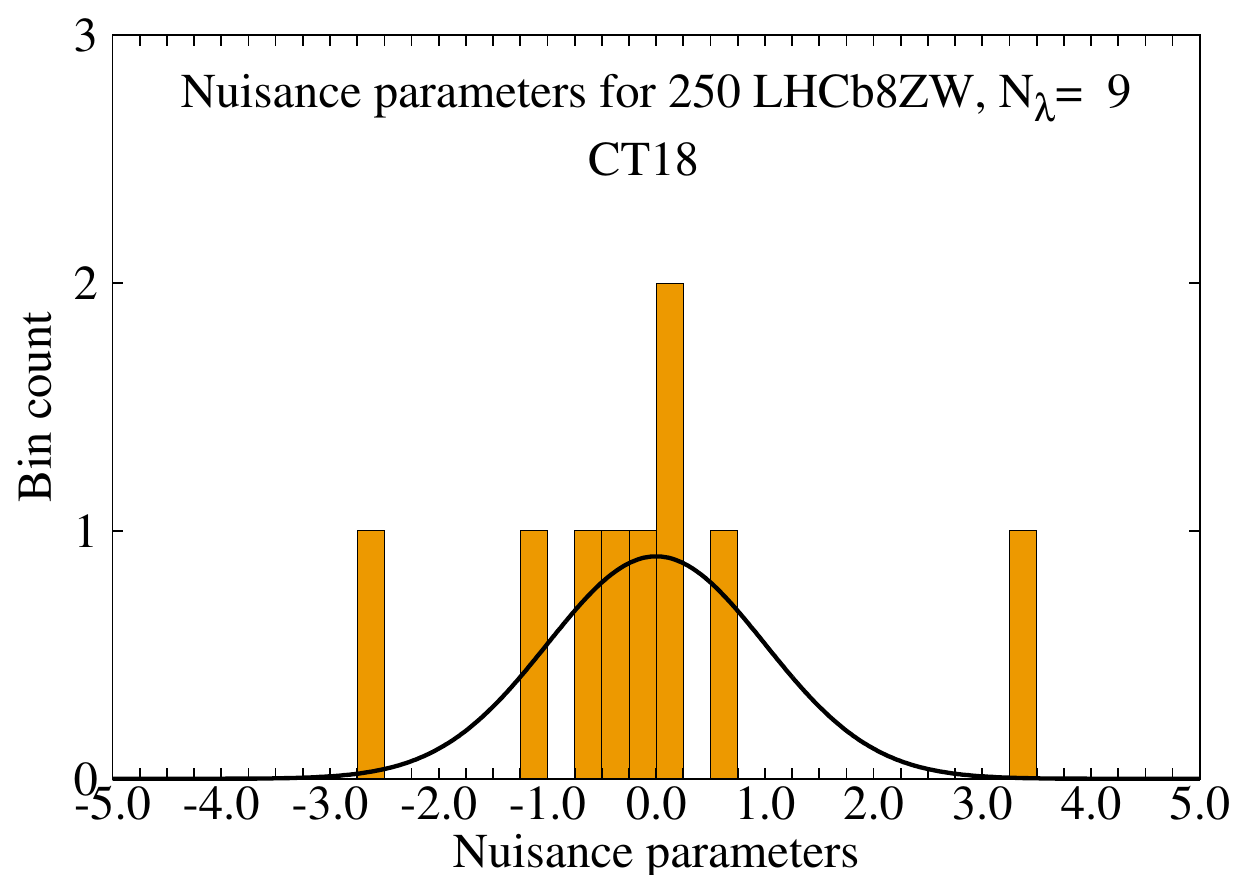}\quad\includegraphics[width=0.53\textwidth]{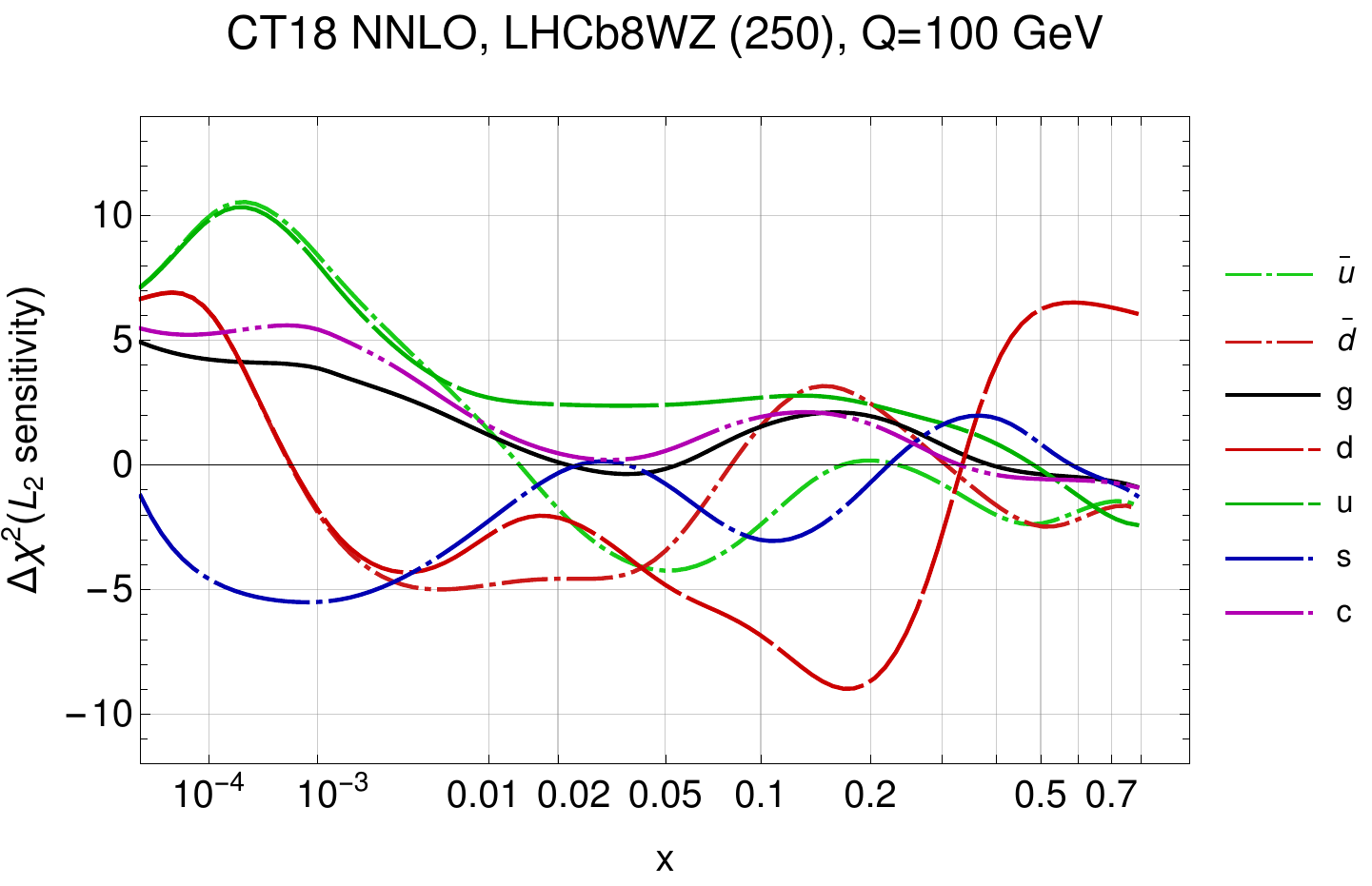}
	\caption{Left: distribution of nuisance parameters for the LHCb 8 TeV $W/Z$ cross sections (Exp.~ID=250).\\ Right: the pulls of these data on the CT18 NNLO PDFs at $Q=100$ GeV, computed in terms
of the $L_2$ sensitivity of Eq.~(\ref{eq:L2}) \cite{CT18L2Sensitivity}.
		\label{fig:res_rk_3}}
\end{figure}

The quality of the CT18 fit to the individual data points can be quantified by the histograms of the shifted residuals
shown in Fig.~\ref{fig:res_rk_1}. When the fit to experiment $E$ is good, the histograms of its shifted residuals $r_i\! =\! (D^\mathit{sh}_{i}-T_{i})/s_{i}$  and optimized nuisance parameters $\bar\lambda_\alpha$ are consistent with the standard normal distribution. For example, the third panel  illustrates the distribution of $r_i$ for
the LHCb 8 TeV $W^\pm$ and $Z$ data (Exp.~ID=250). It indicates that there are 
a few data points with large values in the Exp.~ID=250 data set. As expected, the large residuals result from the first 
few rapidity bins near $y=2$ in the $W^\pm$ and $Z$ data, and from $3.5\lesssim y_Z\lesssim 4$ between 3.5 and 4 in the $Z$ data.
Another useful criterion is the examination of the distribution of nuisance parameters needed 
to fit the Exp.~ID=250 data, which is shown in the left Fig.~\ref{fig:res_rk_3}. The distribution of nuisance parameters deviates from the normal distribution, with two nuisance parameters having particularly large values ($-3$ and $+3.7$).
The right panel of Fig.~\ref{fig:res_rk_3} represents 
the $L_2$ sensitivity of these data to various PDF flavors at $Q=100$ GeV. We see that the LHCb data prefers lower $u$, $\bar u$ PDFs at $x<10^{-2}$, as compared to the full global data, somewhat higher $s$ at $x < 10^{-2}$, and a higher $\bar d$ at $x\approx 0.2$. The plots of $L_2$ sensitivities for the other experiments and PDF combinations can be viewed at Ref.~\cite{CT18L2Sensitivity}.

\begin{figure}[t]
	\includegraphics[width=0.49\textwidth]{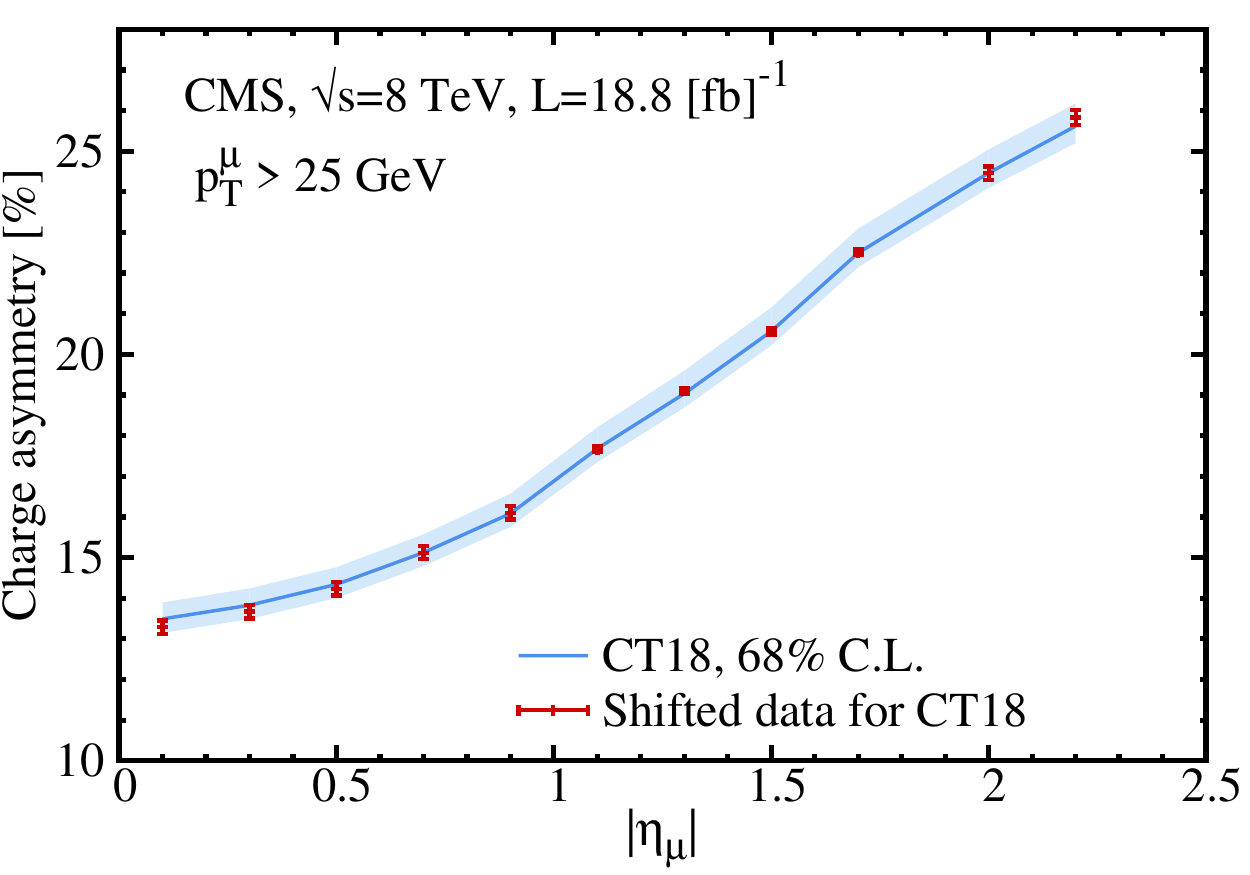}
	\includegraphics[width=0.49\textwidth]{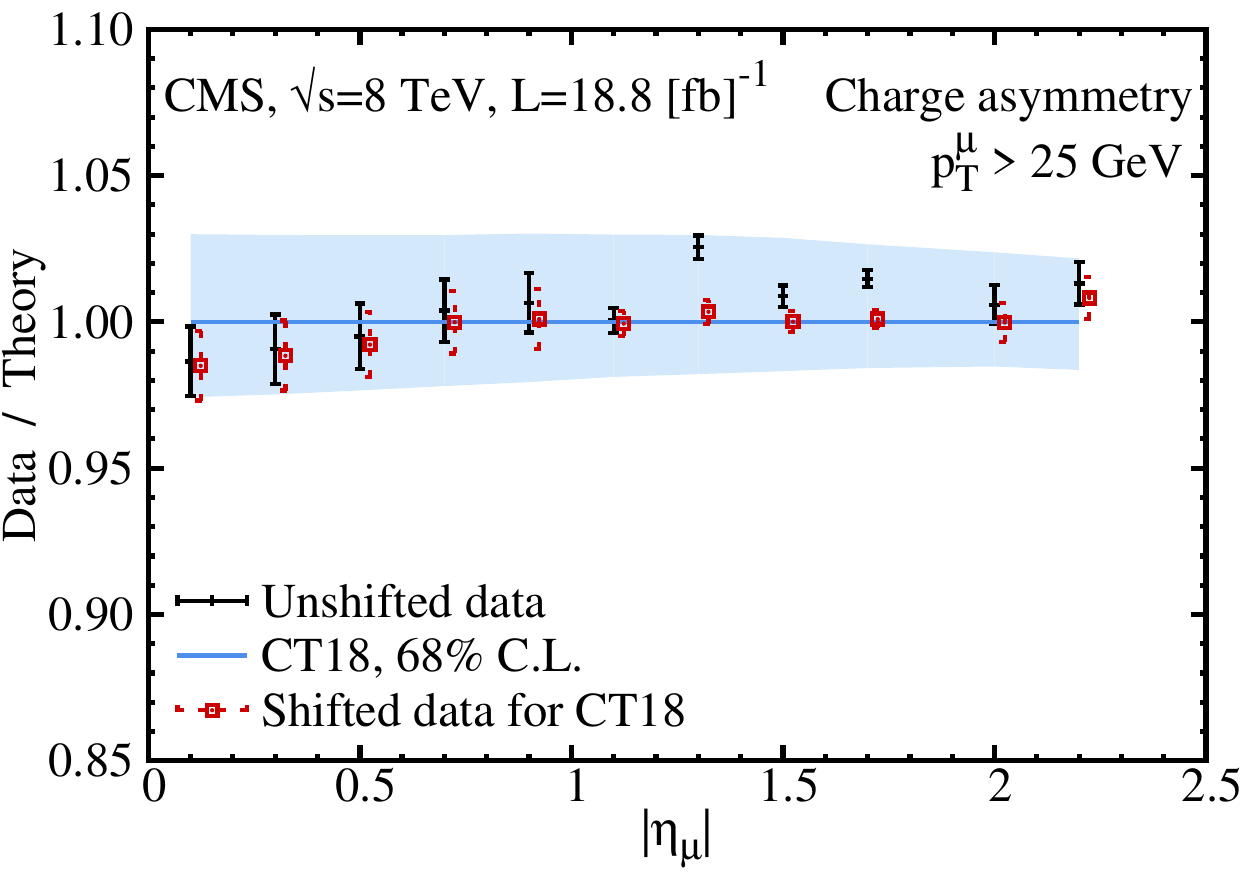}
	\caption{A comparison of the CT18 theoretical predictions to the CMS 8 TeV charge asymmetry data (Exp.~ID=249).}
\label{fig:id249}
\end{figure}

{\bf LHC data: CMS and ATLAS.}
Measurements of lepton charge asymmetry at 8 TeV (Exp.~ID=249) from the CMS collaboration are included  
in all the CT18 global fits. The theoretical predictions, compared with the shifted 
and unshifted data, are shown in Fig.~\ref{fig:id249}. We see that all the 
experimental data are fitted well within the 68\% C.L. PDF uncertainty.

\begin{figure}[p]
\begin{center}
\includegraphics[width=0.42\textwidth]{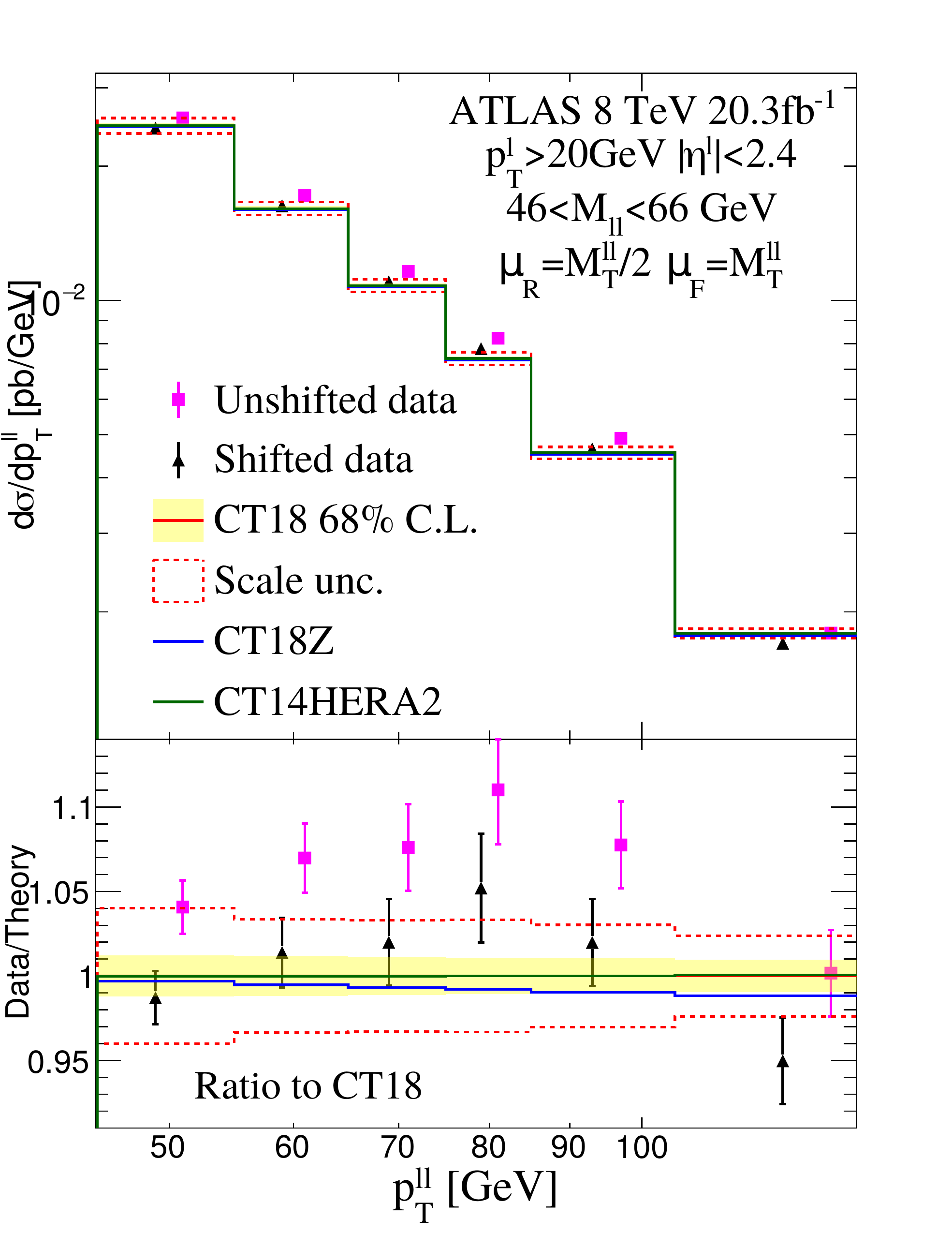}
\includegraphics[width=0.42\textwidth]{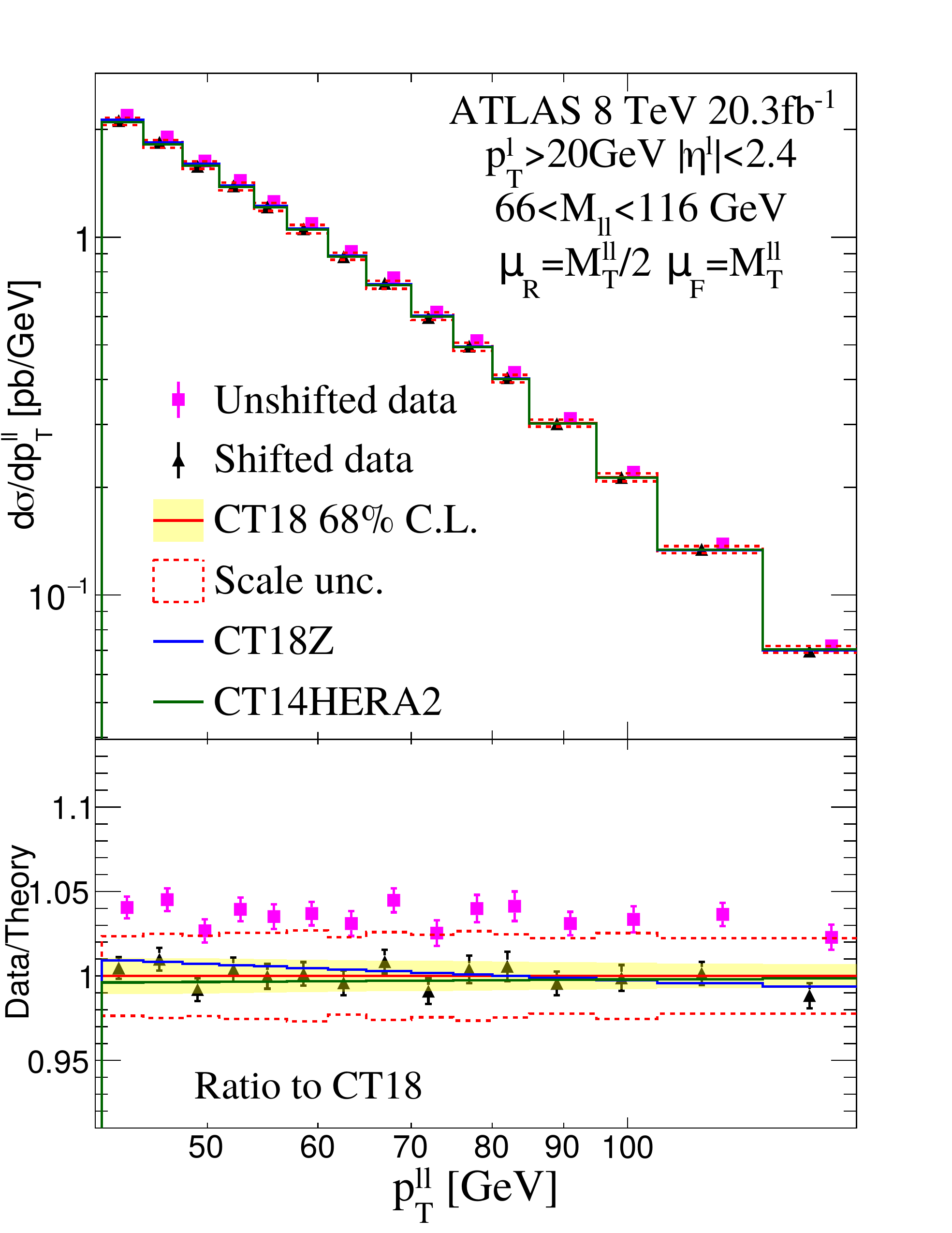}
\includegraphics[width=0.42\textwidth]{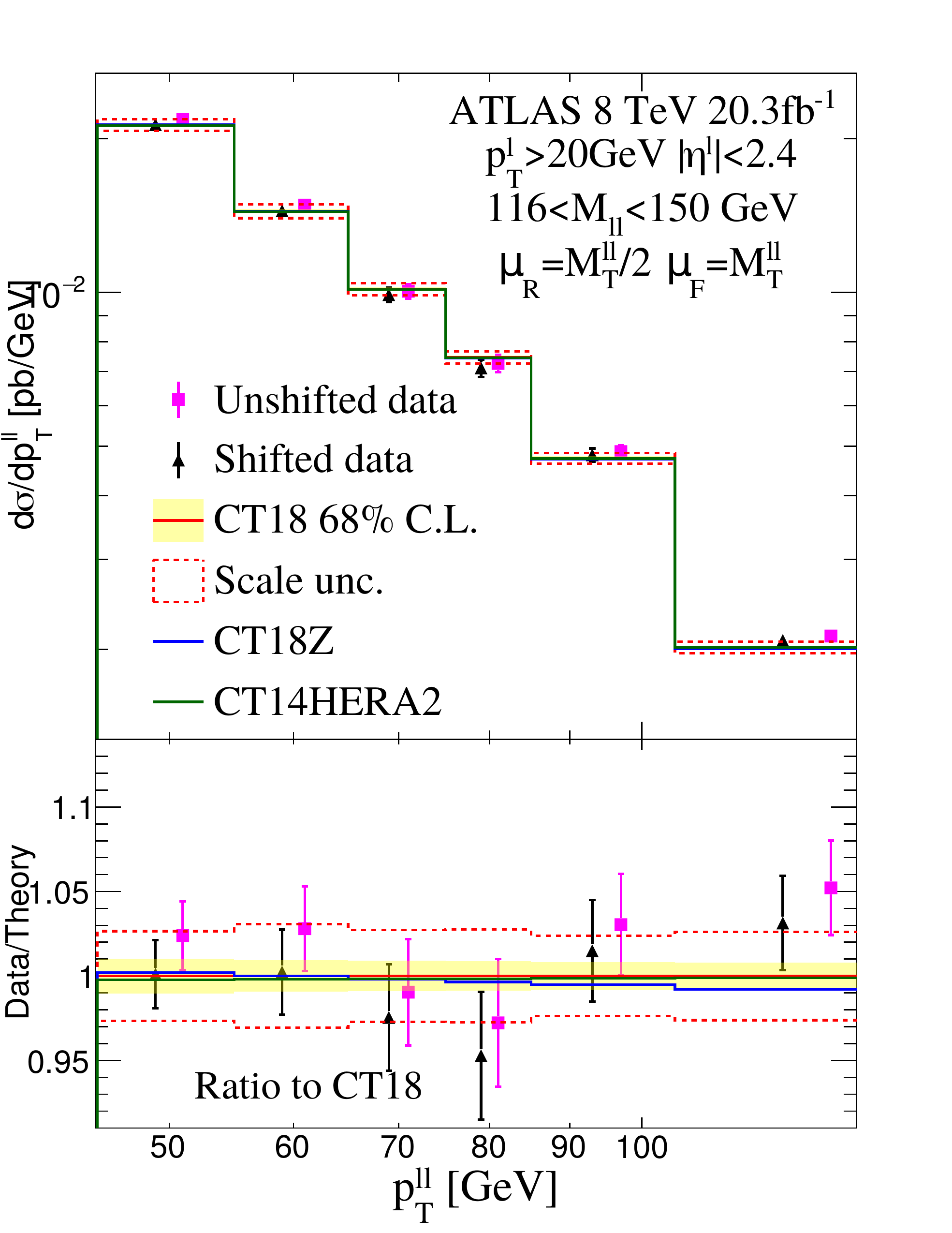}
\caption{Theoretical predictions for lepton pair transverse momentum distribution, $p_{T, \ell \bar \ell}$ based on the CT14$_\mathrm{HERAII}$, CT18, and CT18Z NNLO PDFs, using
 QCD scales $\mu_{R}=M_{T, \ell \bar \ell}/2$, $\mu_{F}=M_{T, \ell \bar \ell}$ and compared with the ATLAS 8 TeV measurements. The yellow band represents the PDF uncertainty calculated with the symmetric Hessian method at the 68\% C.L. The dashed band represents the scale uncertainty. }
\label{fig:ATL8ZpT}
\end{center}
\end{figure}

\begin{figure}[t]
	\includegraphics[width=0.55\textwidth]{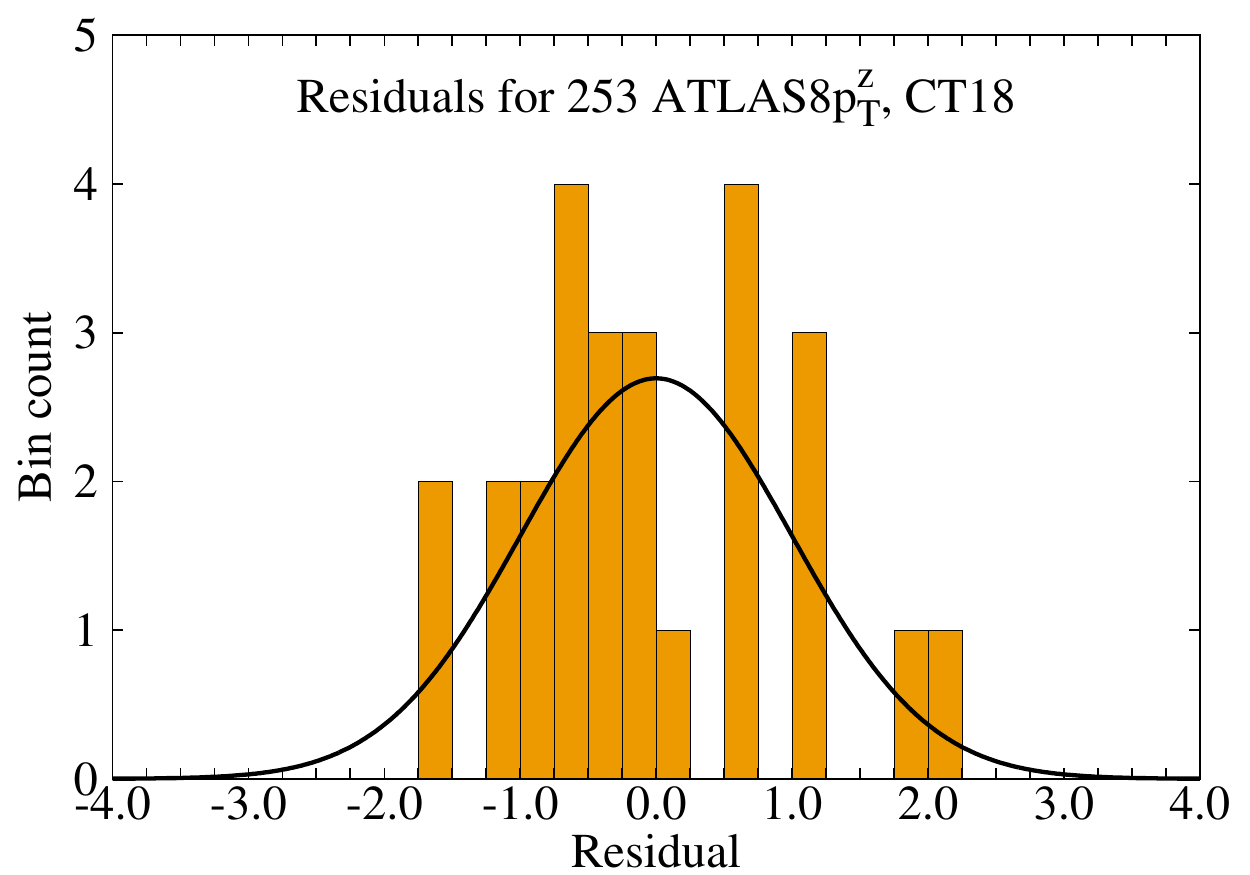}
	\includegraphics[width=0.44\textwidth]{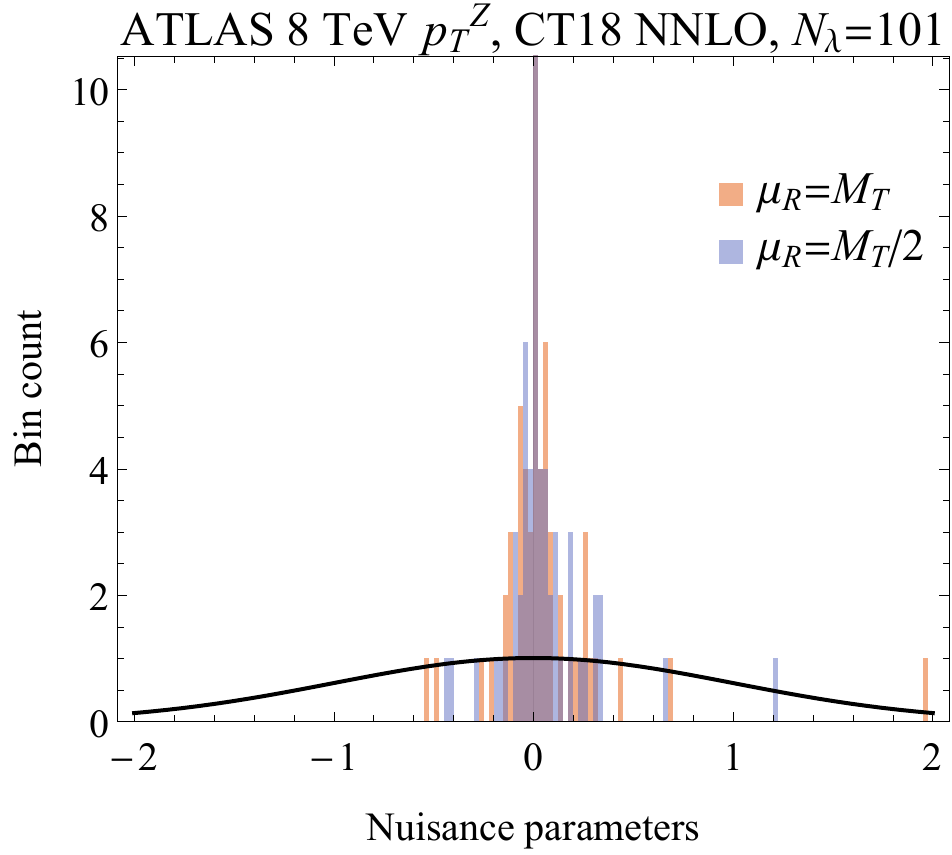}	
	\caption{Left: Distribution of the $\chi^2$ residuals for the ATLAS 8 TeV $Z$ $p_T$ data (Exp.~ID=253) obtained using the nominal QCD scales in Eq.~\ref{scales253_1}. Right: distributions of the respective nuisance parameters obtained for $\mu_R=M_{T,\ell\bar\ell}$ (default) and $M_{T,\ell\bar\ell}/2$, with $\mu_F=M_{T,\ell\bar\ell}$ in both cases. 
		\label{fig:res_rk_6}}
\end{figure}

In the CT18(Z) analysis, we have also included the transverse momentum ($p_{T}$) 
distributions of lepton pairs produced in $Z$ decays at ATLAS at 
$\sqrt{s}=8$ TeV (Exp.~ID=253).
The theoretical predictions for these data are obtained based on the NNLO fixed-order 
calculations for $Z+$jet production. 
We stress that we have imposed a kinematic cut $45<p_T^Z<150$ 
GeV to remove the low- and high-$p_T$ regions where this fixed-order calculation lacks the necessary accuracy.  The low-$p_T$ data are dropped 
because of the missing resummation effects in our fixed-order calculation. 
The high-$p_T$ data are dropped because (1) the constraining power of the data is small given the relatively large statistical errors, and (2) the EW corrections are non-negligible, as will be discussed in Sec.~\ref{sec:EW}.

As a practical implementation, we generated in-house NLO \texttt{APPLgrid} files with \texttt{MCFM} and multiplied them by the NNLO/NLO $K$-factors computed as the ratios of the NNLO and NLO cross sections published in
Refs.~\cite{Ridder:2015dxa,Gehrmann-DeRidder:2017mvr,Gehrmann-DeRidder:2016jns,Gehrmann-DeRidder:2016zml,Ridder:2016rzm,Ridder:2016nkl}.
To account for non-negligible fluctuations in the NNLO theoretical prediction, we have included an additional 0.5\% theoretical Monte-Carlo uncertainty, estimated by the standard deviation for a smooth curve fitted to discrete $K$-factors.
The nominal renormalization and factorization scales are chosen as

\begin{equation}
\mu_{R}=\mu_{F}=M_{T, \ell \bar \ell}=\sqrt{(p_{T, \ell \bar \ell})^{2}+M_{\ell \bar \ell}^{2}}\; ,
\label{scales253_1}
\end{equation}
assuming the unity prefactor in the scales. 
We have also investigated the QCD scale dependence by multiplying the renormalization and 
factorization scales independently by the scaling factors of 2 and 1/2.
Specifically, the scale uncertainty is estimated using the envelope of the 7-point scale variation:
\begin{equation}
(\mu_{R},\mu_{F}) = \big[(1/2, 1/2),(1, 1/2),(1/2, 1),(1, 1),(1, 2),(2, 1),(2, 2)\big] \times M_{T, \ell \bar \ell}
\;.
\end{equation}

All these combinations of QCD scales describe the shape of the ATLAS $Z p_T$ data fairly, however, the data prefer  higher-than-nominal normalizations, which can be accommodated either by increasing the overall normalization of theory by 1-2 standard deviations of the luminosity uncertainty, reducing $\mu_R$, or by increasing $\alpha_s$ to $0.120-0.124$ \cite{Forte:2020pyp}. As a result, we get a marginally better $\chi^2$ for ID=253 using the scales
\begin{equation}
\mu_{R}=M_{T, \ell \bar \ell}/2, ~\mu_{F}=M_{T, \ell \bar \ell}\;,
\label{scales253_2}
\end{equation}
with the negligible difference in the PDFs compared to the other scales. 
The CT18 NNLO theoretical predictions with the scales as in Eq.~(\ref{scales253_2}) are compared to the ATLAS 8 TeV data in Fig.~\ref{fig:ATL8ZpT}. We obtain $\chi^2_E/N_{pt,E}\sim 1$ and fairly describe all 3 invariant-mass bins after allowing for an upward shift by $1.2\sigma$ in the overall normalization. 
The complementary figure with the scales as in Eq.~(\ref{scales253_1}), also having $\chi^2_E/N_{pt,E} \approx 1$, cf.~Table~\ref{tab:EXP_2}, and requiring a shift in the overall normalization by $2\sigma$, is included in the supplementary material.

In Fig.~\ref{fig:res_rk_6}, the distributions of the residuals(in the left subfigure) and nuisance parameters (in the right subfigure) of these data are shown.  
We see excellent agreement of theory and data in the distribution of residuals. In the distribution of nuisance parameters, out of 101 nuisance parameters in this process, only one parameter, associated with the overall normalization, is increased by more than $1\sigma$ ($2\sigma$) for $\mu_R=M_{T, \ell \bar \ell}/2$ ($M_{T, \ell \bar \ell}$). More than seventy nuisance parameters are too 
close to zero in these fits, perhaps indicating that the experiment has tabulated too many evanescent systematic effects. [Such excess of very small nuisance parameters is not uncommon for the LHC experiments, as discussed at the beginning of Sec.~\ref{sec:Quality} and in Sec.~IV.E of \cite{Kovarik:2019xvh}.]

The alternative scales $\mu_{F,R}=M_{\ell \bar \ell}$ have been also tried and resulted in a worse description of the shape of the $p_T$ distribution (not only the normalization) and elevated $\chi^2$. 

The remaining difference cannot be explained by the EW corrections, 
since the EW corrections are small and negative (see Sec.~\ref{sec:EW}), 
pulling the theory further away from the data. Instead, the systematic shift in the normalization can possibly be ascribed to the missing higher-order (N$^{3}$LO) corrections, implied by two observations. First, the NNLO corrections to the $Z$ $p_{T}$ are generally as large as 10\%, which indicates 
slow convergence of the perturbative expansion. Second, the large 
scale uncertainty (about 3-4\%) is also an indication that the missing 
higher-order effects may be significant.

\subsubsection{Jet data}
\label{sec:Jet_fit}
Historically, inclusive jet production has played an important role in constraining the gluon density, $g(x,\Q)$, as evidenced by the impact that the older jet data from the Tevatron Run-II had on the CT10 and CT14 global analyses.  CT18 now also implements inclusive jet production data at even higher collider energies and  luminosities, measured by the ATLAS and CMS collaborations at the LHC, as described in Sec.~\ref{sec:DataJets}.

\begin{figure}[tb]
	\includegraphics[width=1.0\textwidth]{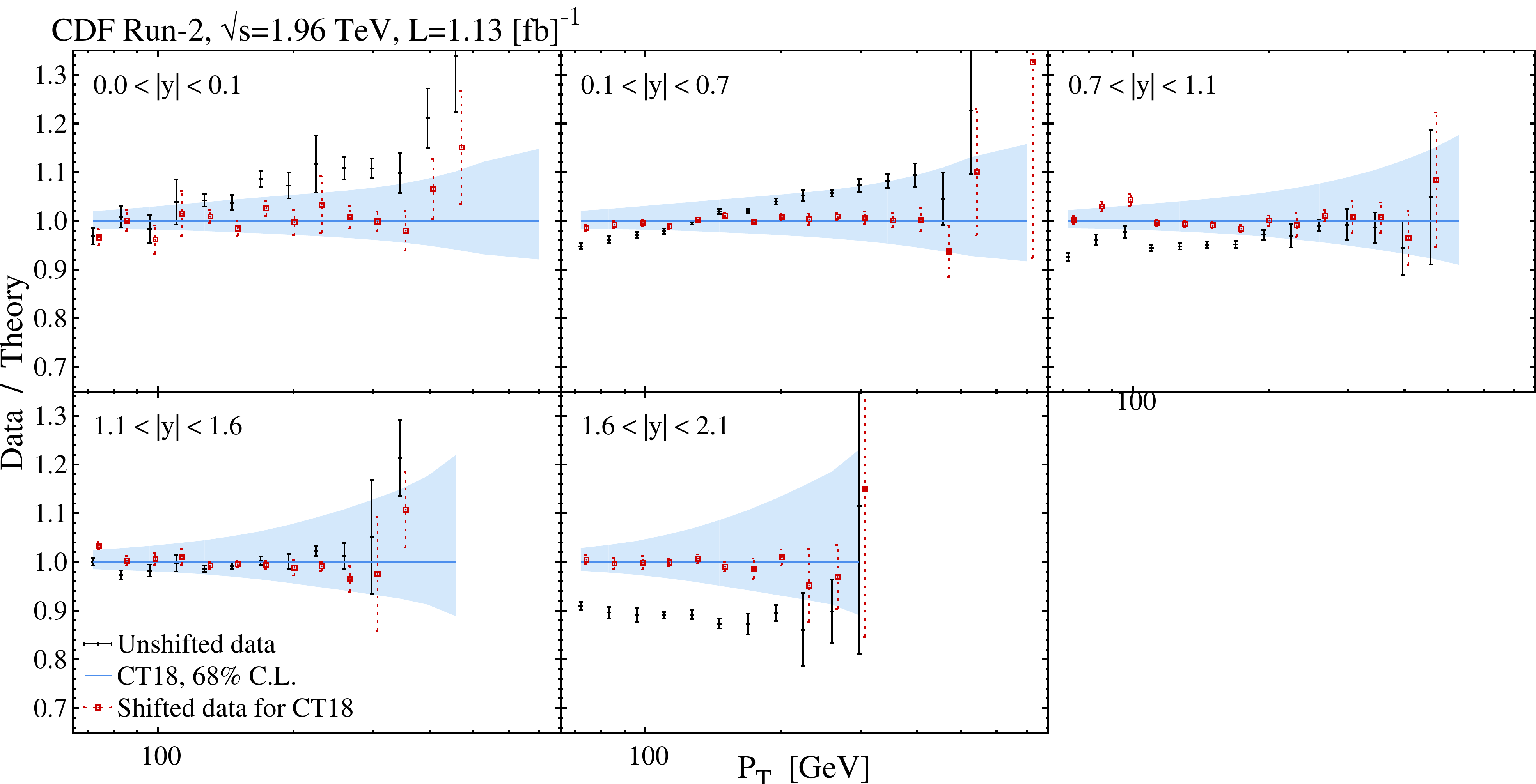}
	\caption{$\mathrm{Data}/\mathrm{Theory}$ values for CT18 NNLO and CDF Run 2 jet data (Exp.~ID=504).
		 \label{fig:id504}}
\end{figure}

\begin{figure}[tb]
	\includegraphics[width=1.0\textwidth]{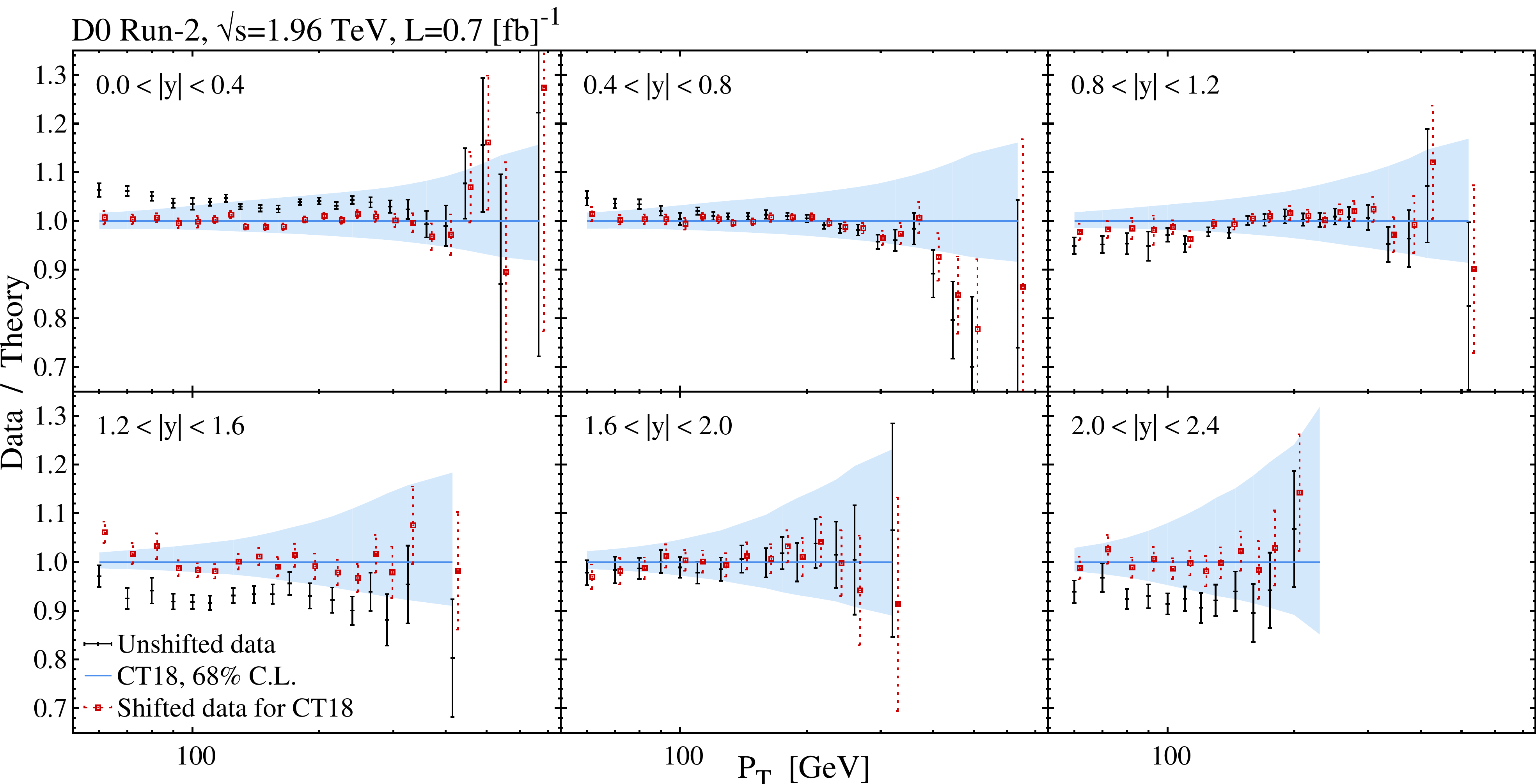}
	\caption{$\mathrm{Data}/\mathrm{Theory}$ values for CT18 NNLO and D0 Run-2 inclusive jet production (Exp.~ID=514).
	     \label{fig:id514}}
	
\end{figure}

{\bf Tevatron Run-II data.}
First, we examine the fits to the Tevatron Run-II jet data. 
The CDF Run-II 
data shown in Fig.~\ref{fig:id504} is not perfectly described by NNLO theory (has an elevated $\chi^2_E/N_{pt,E}\approx 1.7$ according to Table~\ref{tab:EXP_1}) and prefers a somewhat different shape of the gluon PDF $g(x,Q)$, compared to the average of all experiments, according to the $L_2$ sensitivity plot for $g(x,Q)$ in Fig.~\ref{fig:L2glu}. The D0 Run-II jet data, depicted in Fig.~\ref{fig:id514}, show better agreement with the rest of the data sets.

\begin{figure}[p]
\includegraphics[width=1.0\textwidth]{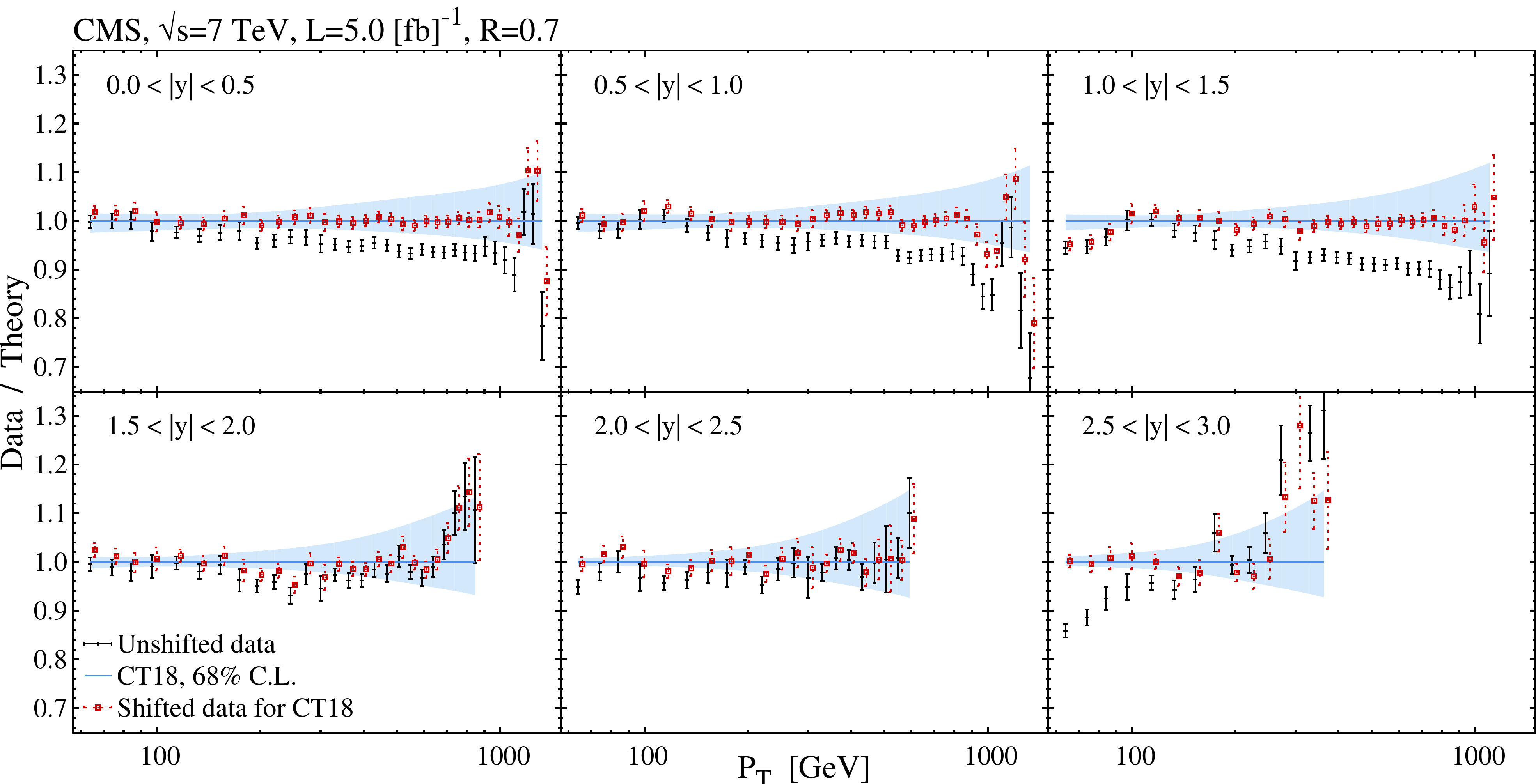}
\caption{$\mathrm{Data}/\mathrm{Theory}$ values for CT18 NNLO and  CMS 7 TeV inclusive jet production (Exp.~ID=542).
\label{fig:DoT542}
}
\end{figure}

\begin{figure}[p]
\includegraphics[width=1.0\textwidth]{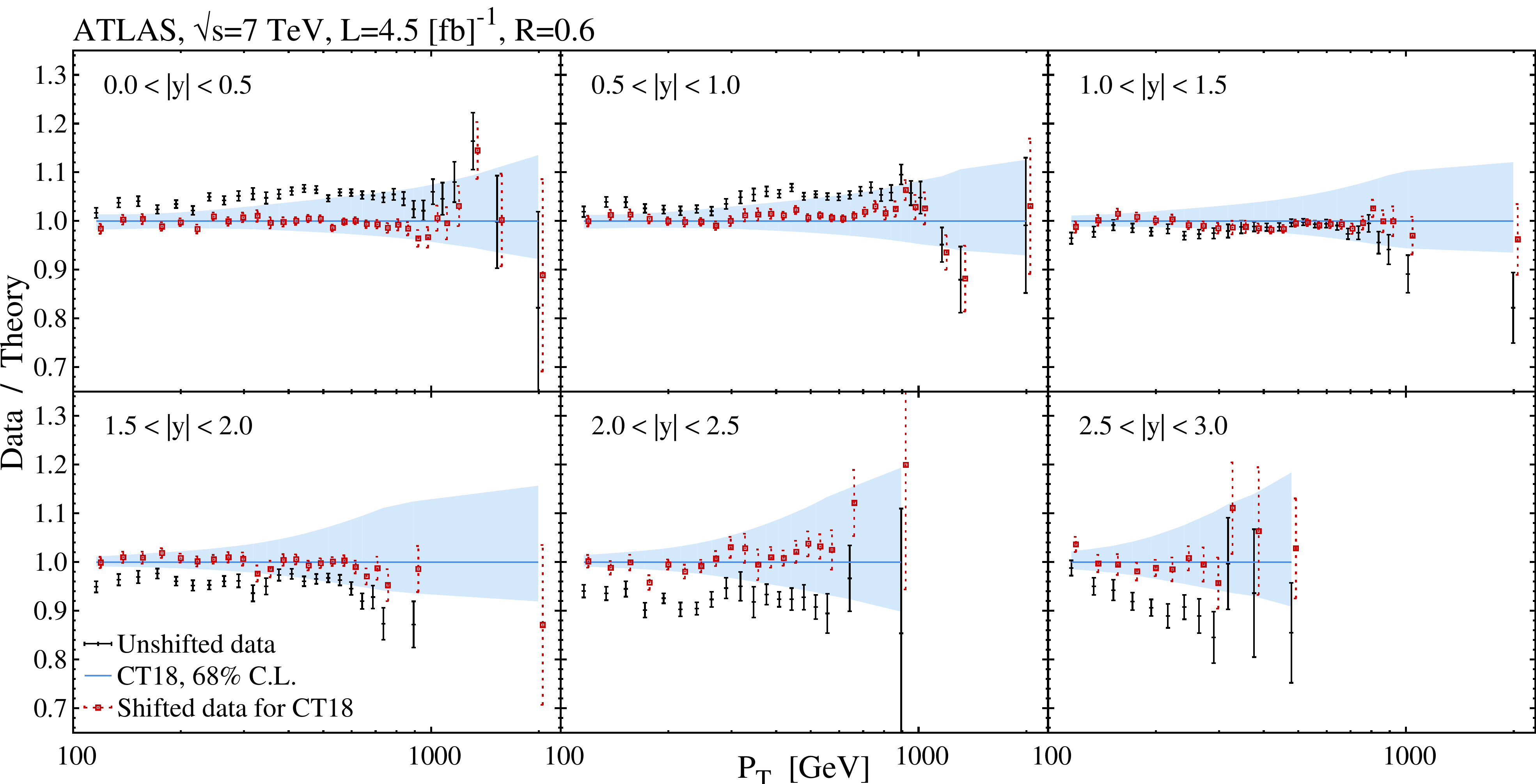}
\caption{$\mathrm{Data}/\mathrm{Theory}$ values for CT18 NNLO and ATLAS 7 TeV inclusive jet production (Exp.~ID=544).
\label{fig:DoT544}
}
\end{figure}

\begin{figure}[tb]
\includegraphics[width=1.0\textwidth]{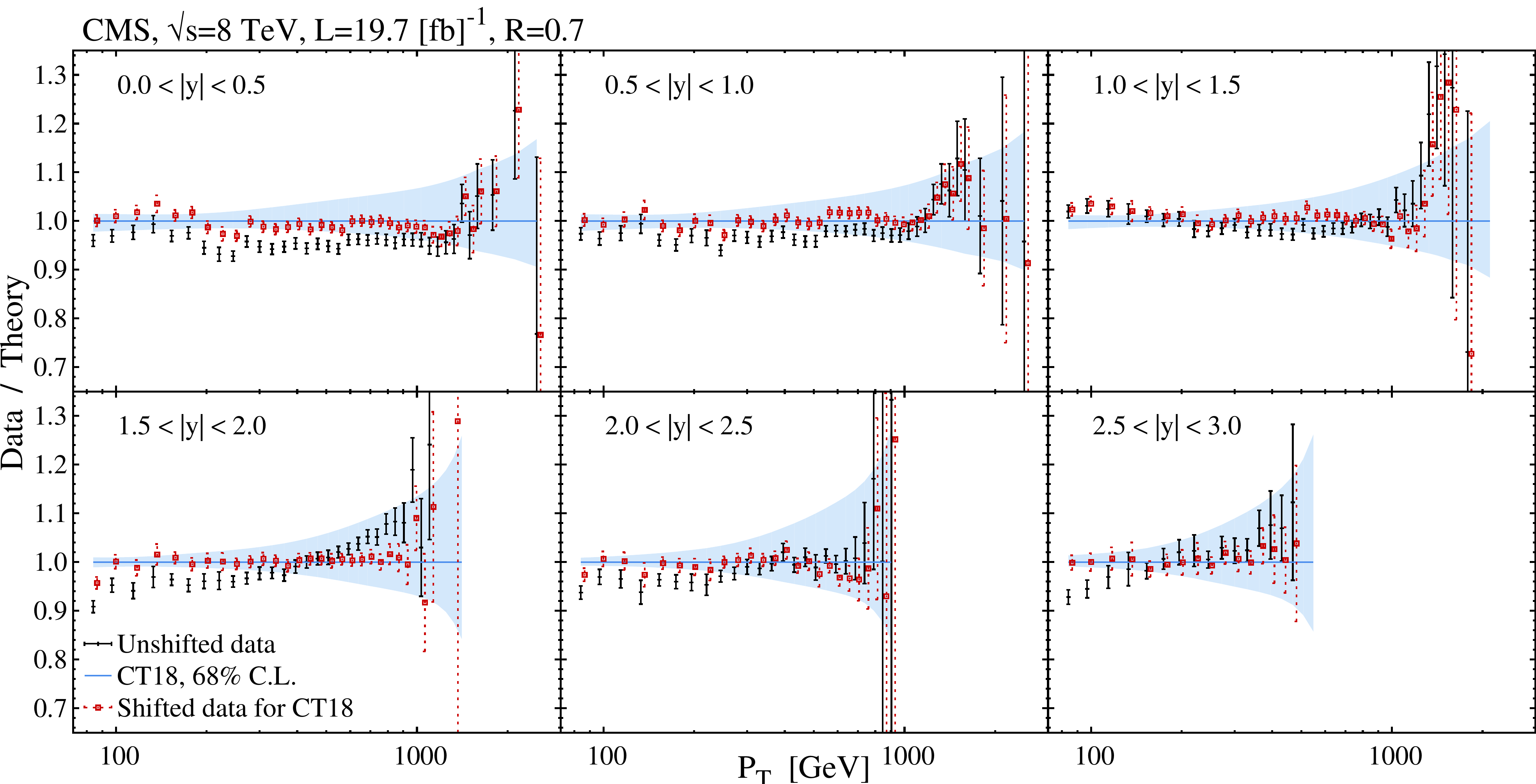}
\caption{$\mathrm{Data}/\mathrm{Theory}$ values for CT18 NNLO and CMS 8 TeV inclusive jet production (Exp.~ID=545).
\label{fig:DoT545}
}
\end{figure}

{\bf Run-1 LHC data.}
The CT18 fit can describe the LHC CMS and ATLAS jet data, depicted in Figs.~\ref{fig:DoT542}-\ref{fig:DoT545},
after the decorrelation of some correlated systematic errors, as 
laid out in Sec.~\ref{sec:DataJets} and App.~\ref{sec:ATLASjetdecorrel}, 
as well as the inclusion of a $0.5\%$ overall uncorrelated 
systematic error for all the LHC jet data,
as discussed in Sec.~\ref{sec:DataJets}.

Although the agreement with theory in the CT18 analysis is reasonable, 
we note some tensions among the LHC inclusive jet data sets themselves, especially the CMS 
results at 7 (Exp.~ID=542) and 8 TeV (Exp.~ID=545).
These tensions are particularly pronounced for some parton flavors in the specific kinematic regions --- most evidently, for the gluon PDF, as
quantified by the LM scans and $L_2$ sensitivity profiles plotted 
in Figs.~\ref{fig:LMg18} and~\ref{fig:L2glu}, respectively.
For the ATLAS inclusive jet data at 7 TeV, the best fit requires
the correlated errors to shift the raw data downward
in the smaller rapidity regions, but to shift the raw data upward at high rapidities. The majority of optimal nuisance 
parameters $\lambda_{\alpha}$ for the CT18 NNLO PDF set, shown in the histograms included in the supplementary material, are distributed 
narrowly about $|\lambda_{\alpha}|\! \sim\! 0$. For the CMS 8 TeV data set, four nuisance parameters out of 28 require absolute correlated shifts larger than two -- a larger count than is expected based on the assumed normal statistics.


\subsubsection{Top-quark pair production data
\label{sec:QualityTopData}
}

\begin{figure}[tb]
	\includegraphics[width=0.7\textwidth]{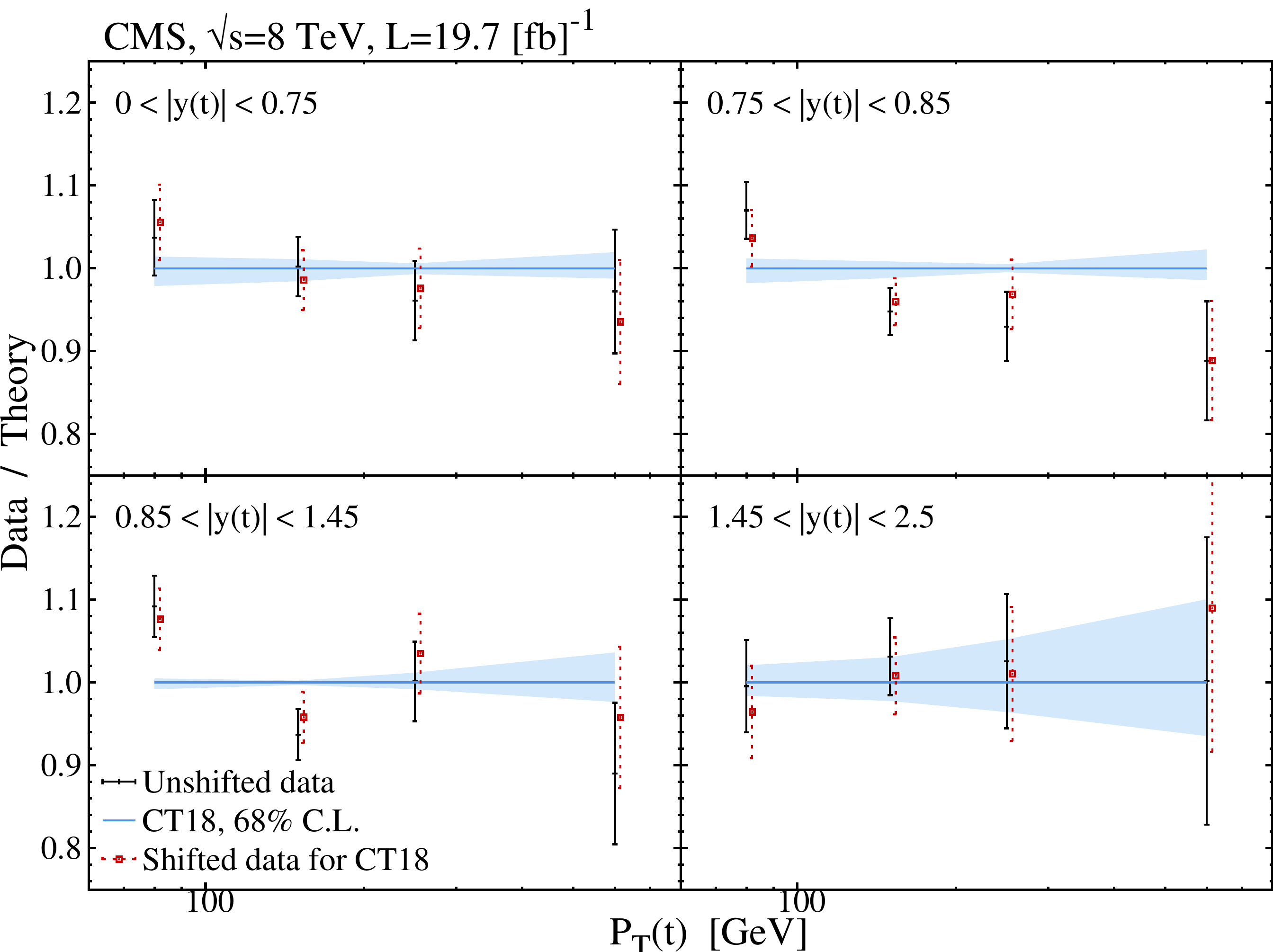}
	\caption{$(\mathrm{Data})/(\mathrm{Theory})$ comparison for the
	CMS 8 TeV $t\bar{t}$ production data (Exp.~ID=573) as a function of the transverse momentum of the top (anti-)quark.}
\label{fig:573}
\end{figure}

\begin{figure}[tb]
	\includegraphics[width=0.49\textwidth]{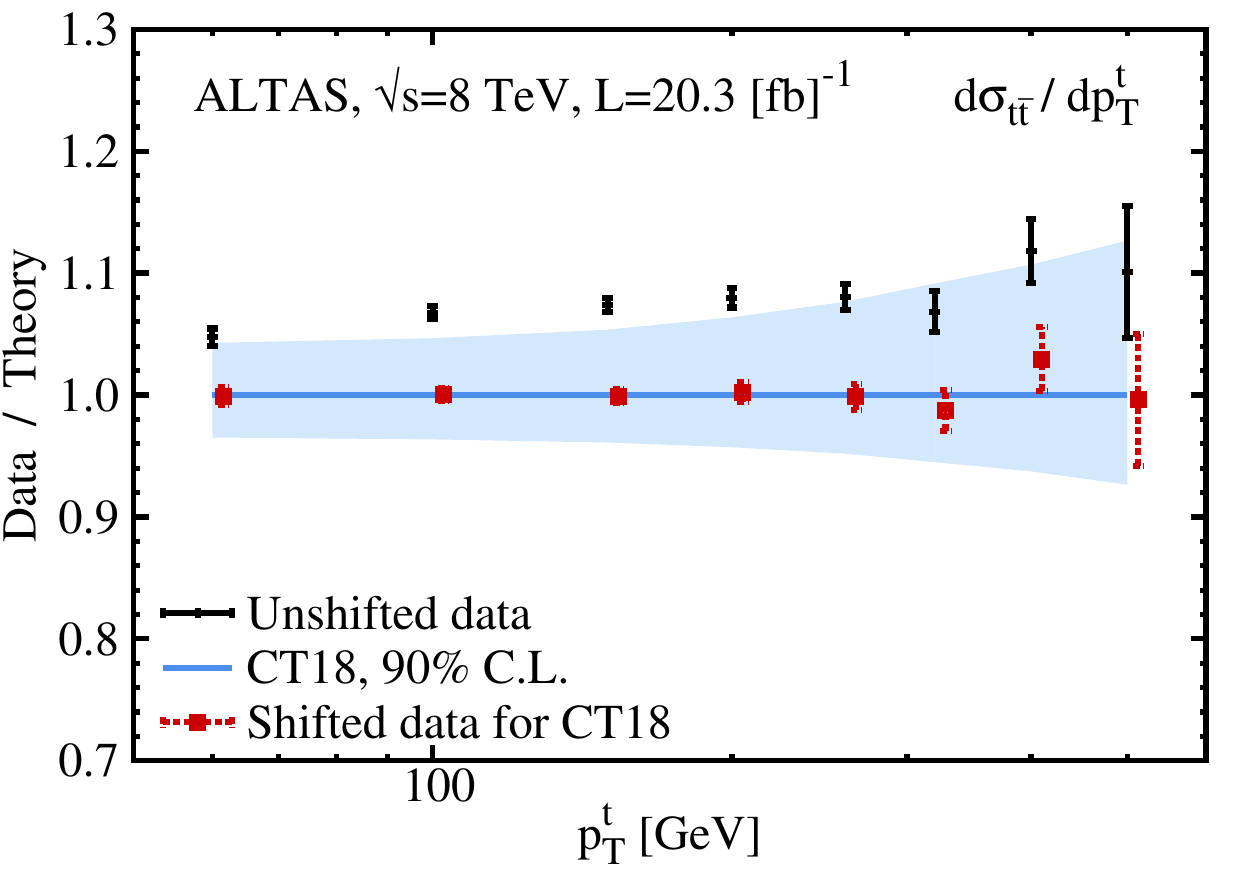}
	\includegraphics[width=0.49\textwidth]{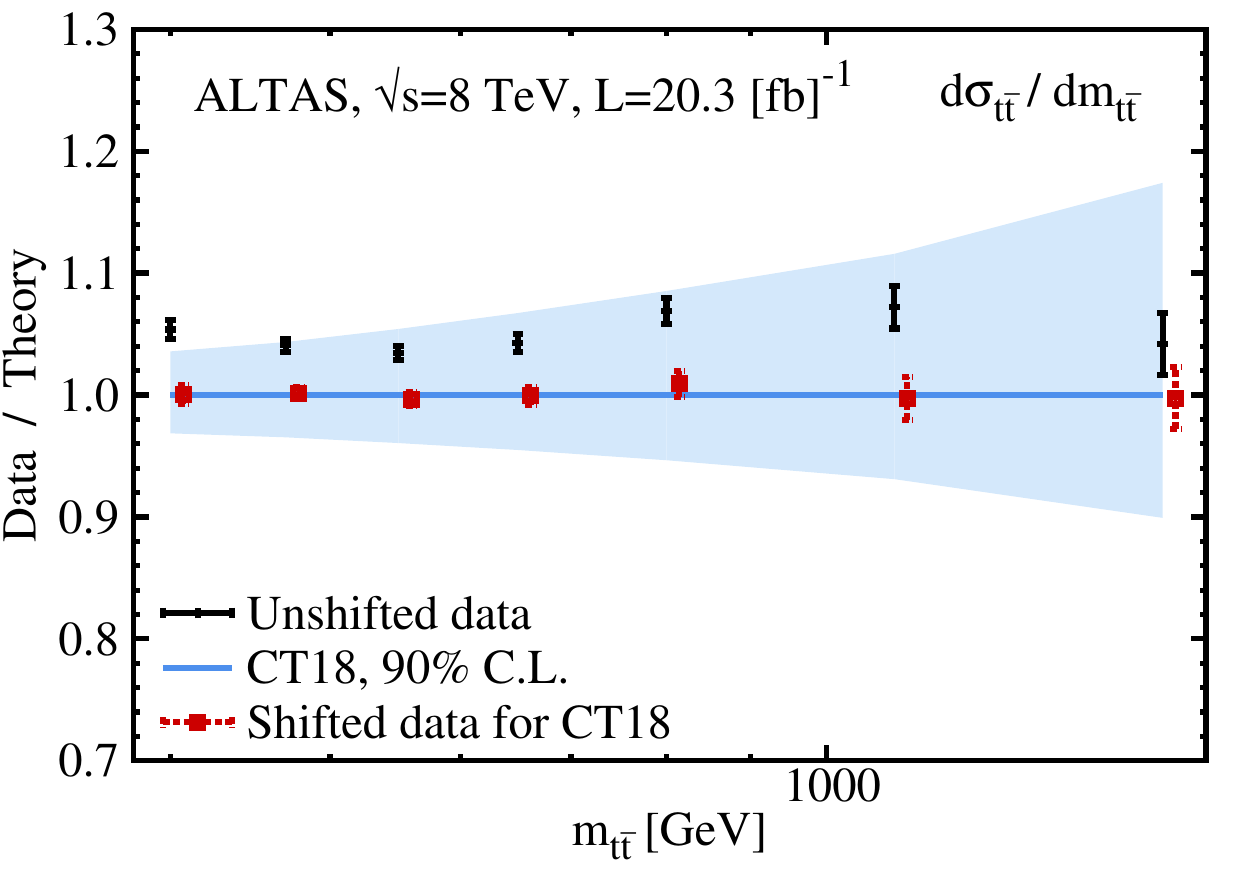}
	\caption{$(\mathrm{Data})/(\mathrm{Theory})$ comparison for the
	ATLAS 8 TeV $t\bar{t}$ production data (Exp.~ID=580) as a function of the $t\bar t$ transverse momentum (left) and  invariant mass (right). }
\label{fig:580}
\end{figure}

The two $t\bar{t}$ data sets included in the CT18 global analysis are well-described, as shown by the values of $\chi^2$ and effective Gaussian variable $S_E$ 
given in the latter two rows of Table~\ref{tab:EXP_2}. 
In particular, for the CMS (Exp.~ID=573) and ATLAS (Exp.~ID=580) data sets included in the fit,
we obtain $S_E\! =\! 0.6$ and $S_E\! =\! -1.1$, respectively.
For a detailed point-by-point description of the $t\bar{t}$ agreement with the theory,
in Figs.~\ref{fig:573} and~\ref{fig:580} we show plots of 
the $(\mathrm{Data})/(\mathrm{Theory})$ ratio
for both data sets.
For simplicity, the error bars for the CMS data in Fig.~\ref{fig:573} were calculated by including both statistical and correlated systematic errors listed in Tables 5 and 7 of Ref.~\cite{Sirunyan:2017azo}. 

The statistical correlations for these measurements (Table 6 in Ref.~\cite{Sirunyan:2017azo}) are not included in the CT18 analysis because of technical difficulties in the realization of the nuisance parameter representation for this statistical correlation information. The nuisance parameter representation is the one and only default representation utilized in all CT analyses. The statistical correlations released by the CMS collaboration are given in terms of the covariance matrix representation. 
Despite this, we independently cross-checked the impact of the statistical correlations by using the \texttt{ePumP} software which is able to process the correlation information when given in terms of covariance matrix. The conclusion is that the inclusion of statistical correlations has negligible impact on the resulting \texttt{ePumP} updated PDFs.
The CT18 baseline $\chi^2$ value obtained for the CMS (Exp.~ID=573) data with the inclusion of the statistical correlation covariance matrix increases by about 5 units, as compared to the value (18.9) given in Table II. (See Refs.~\cite{Czakon:2019yrx} and \cite{Bailey:2019yze} for related discussions on the inclusion of these measurements in global PDF analyses.)

In Fig.~\ref{fig:573}, the top-quark $p_T$ distribution at CMS  
is fitted reasonably well across the four rapidity bins examined here. We find
modest deviations between theoretical predictions and the (un)shifted data
for some points in the intermediate rapidity bins, $0.75\! <\! |y_t| \! <\! 0.85$
and $0.85\! <\! |y_t| \! <\! 1.45$, contributing to the somewhat broader
distribution of residuals. Notably, the effect of correlated errors
in fitting the CMS data is relatively minimal, given the fact that the shifted (red) and unshifted (black) data are very similar, as observed in  Fig.~\ref{fig:573}. 
Correlated systematics are nonetheless important for some cross section values, allowing the data values to shift enough to be within 
$1\sigma$ distance from the CT18 prediction.

In contrast, achieving a very good description of the analogous ATLAS 
$p_{T,t}$ and $m_{t\bar{t}}$ distributions shown in Fig.~\ref{fig:580}
critically depends on the use of nuisance parameters to compensate for
correlated systematics, as seen in Fig.~\ref{fig:580}. 
The uncorrelated errors are small (less than 1-2 percent) in most bins of this data set. On the other hand, the systematic errors are sizable and the systematic shifts lead to a very good agreement between theory and data, with $\chi^2_E/N_{pt,E}=9.4/15$ for CT18 NNLO.

\subsubsection{Dimuon production
\label{sec:Qualitydimuon}}

Charm-quark production cross sections in neutrino deep-inelastic scattering provide key low-$Q$ constraints on the strangeness PDF at $x > 10^{-2}$.
In the CT14 NNLO analyses, the charm-quark production cross section were calculated at NLO in
QCD~\cite{Gottschalk:1980rv,Gluck:1997sj,Blumlein:2011zu} in the S-ACOT-$\chi$ variable-flavor-number (VFN) scheme ~\cite{Aivazis:1993kh,Collins:1998rz,Kramer:2000hn,Tung:2001mv}.
Recently, charged-current coefficient functions in DIS have been calculated
to NNLO in QCD, including quark mass dependence ~\cite{Berger:2016inr,Gao:2017kkx}.
This calculation, in a fixed-flavor-number (FFN) scheme with 3 light-quark flavors,  is published in the form of fast interpolation tables for the kinematics of the CCFR and NuTeV dimuon experiments~\cite{Goncharov:2001qe,Mason:2006qa}.

The CT18 analysis still uses an NLO theory prediction in the S-ACOT-$\chi$ VFN scheme because it matches the precision of the CCFR and NuTeV experimental data sets. 
Implementation of charm-quark mass effects at NNLO
demands not only the NNLO charged-current cross section in an ACOT-like VFN scheme, which is not yet available, but also consistency in implementation of QCD radiative effects in the CCFR and NuTeV studies of their systematics. In the case of NuTeV \cite{Mason:2006qa}, unfolding of events, acceptance estimations,\footnote{CCFR and NuTeV collaborations apply significant acceptance corrections for extracting the charm-quark production cross sections from dimuon cross sections. These corrections were estimated at NLO precision only.} and studies of charm fragmentation were done using LO and NLO programs, with systematic uncertainties that exceed the magnitude of the NNLO radiative contribution, as concluded in Refs.~\cite{Berger:2016inr,Gao:2017kkx}, and  depend on the charm quark mass and an (arguably small \cite{Ball:2018twp}) nuclear correction. In the CT analyses, the CCFR and NuTeV dimuon cross sections are implemented by assuming the $c\to \mu$ branching ratio of 0.099, as in Section 5.2.1 of \cite{Mason:2006qa}.  The normalization uncertainty of 10\% is treated as fully correlated over the $\nu$ channel and similarly over the $\bar \nu$ channel. The rest of systematic uncertainties are added in quadrature.  
Overall, the discussion in Ref.~\cite{Gao:2017kkx} indicates that, for the kinematics of CCFR and NuTeV, the differences between the NNLO results from the FFN scheme and any VFN scheme are expected to be significantly smaller than the precision of experimental data.

	\begin{figure}[b]
		\begin{center}
			\includegraphics[width=0.47\textwidth]{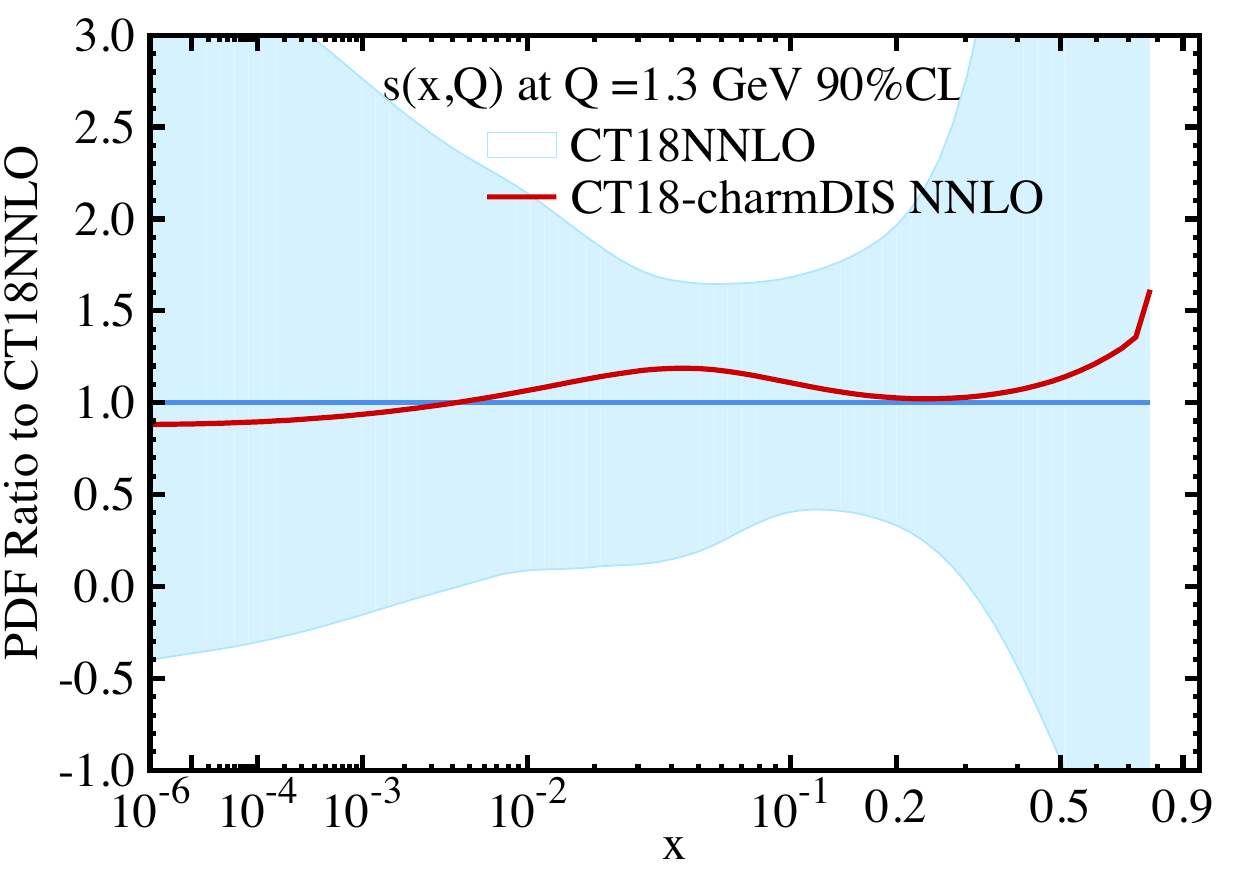}\hspace{0.1in}
			\includegraphics[width=0.47\textwidth]{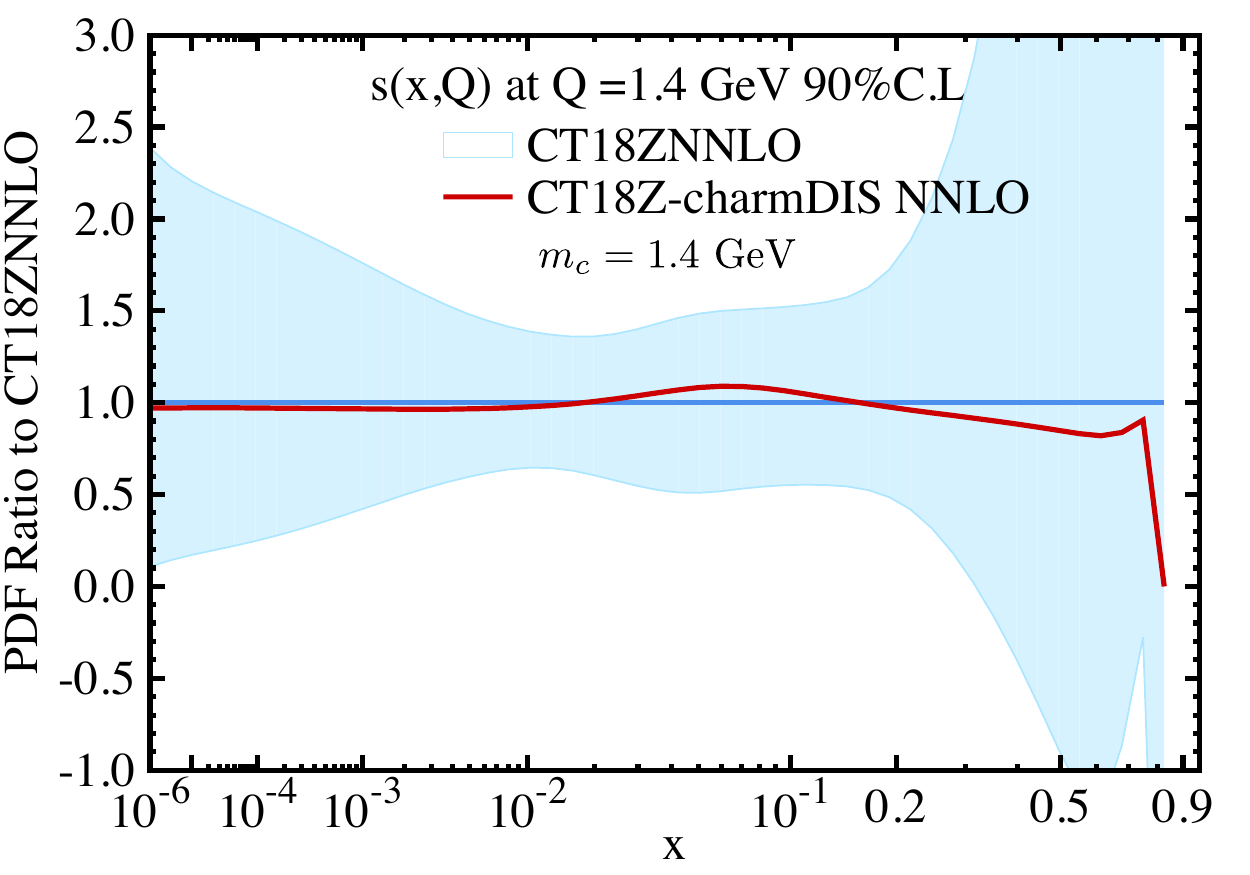}
			\includegraphics[width=0.47\textwidth]{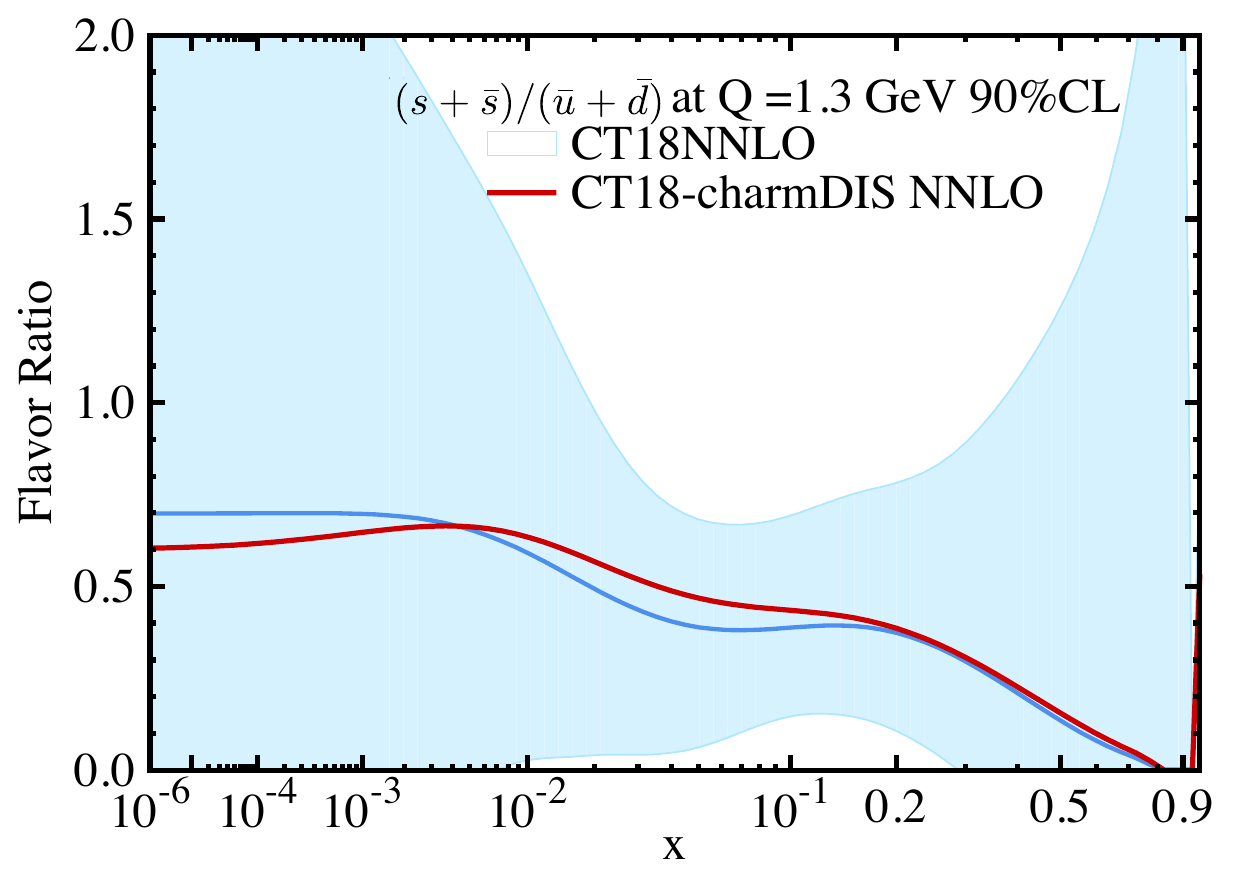}\hspace{0.1in}
			\includegraphics[width=0.47\textwidth]{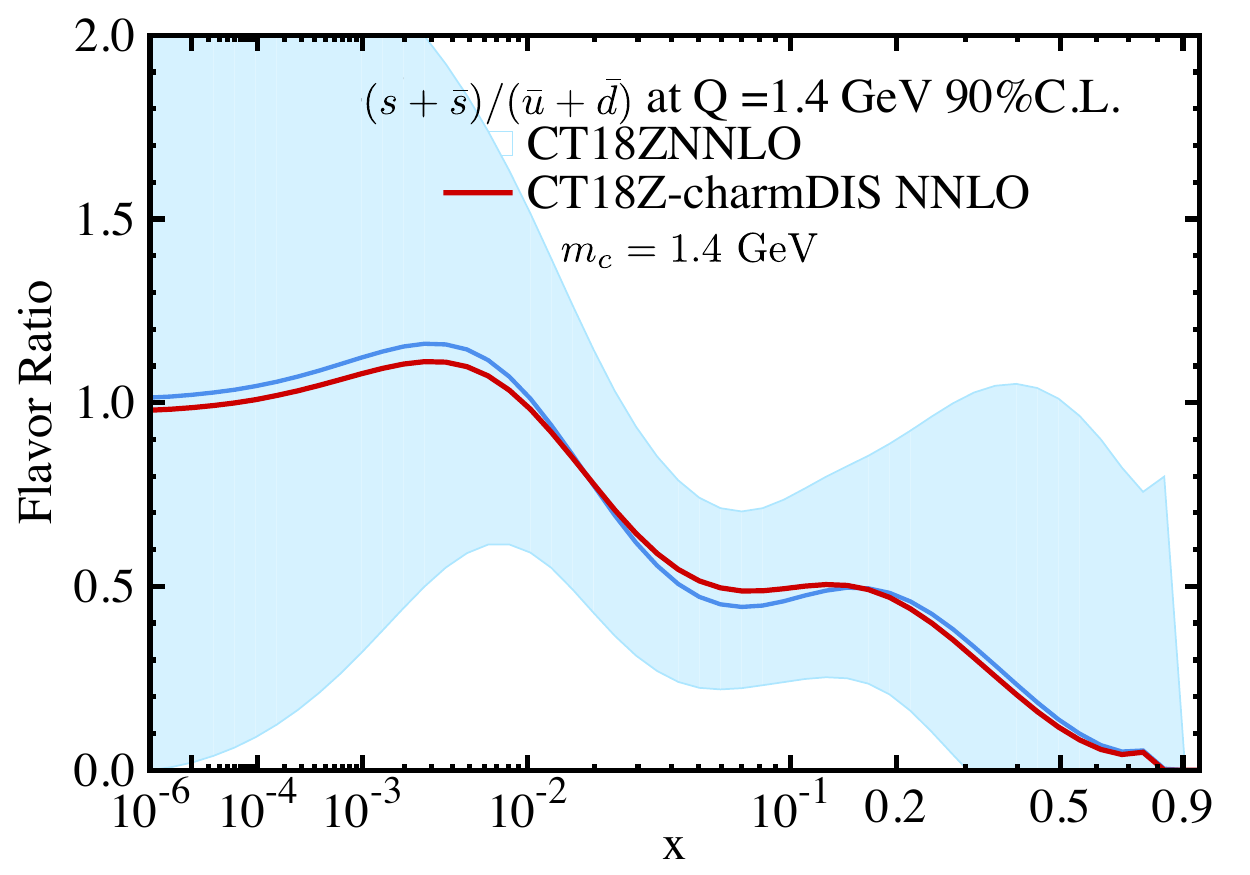}
		\end{center}
		\vspace{-2ex}
		\caption{\label{fig:dimuonalt}
			Strange-quark distribution $s(x,Q)$  and ratio $R_s(x,Q)$ in CT18 (left) and
			CT18Z (right) NNLO fits, compared with alternative fits using QCD NNLO cross sections for CCFR and NuTeV measurements.
		}
	\end{figure}

As a cross check, we have carried out alternative fits, labeled CT18(Z)-charmDIS NNLO, using the NNLO FFN calculations for dimuon production cross sections. 
In the case of CT18-charmDIS NNLO, the global $\chi^2$ is reduced by 6 units (compared to CT18), with the reduction in the $\chi^2_E$ for the dimuon data of the order of 1-2 units.
For CT18Z-charmDIS NNLO, the global $\chi^2$ and the $\chi^2_E$ for the dimuon data  are reduced by 11 and 8 units, respectively, compared to CT18Z.
In both cases the NNLO predictions  provide a marginally better agreement with the data. 

The impact of these choices  on the strange-quark PDF has also been cross checked. The strange-quark PDF $s(x,Q)$ and the ratio $R_s(x,Q)$ defined in Eq.~(\ref{eq:Rs}) are compared in Fig.~\ref{fig:dimuonalt} for the nominal CT18(Z) fits and their ``charmDIS NNLO'' alternatives. 
In the CT18-charmDIS fit, we observe a slight increase of the strange-quark PDF at $x\! \approx\! 0.1$. This outcome is consistent with the PDF profiling results in Ref.~\cite{Gao:2017kkx} and reflects negative NNLO QCD corrections in the same $x$ region. In the CT18Z-charmDIS fit, with the ATLAS 7 TeV
$W/Z$ data included, the PDFs change less as compared to the CT18-charmDIS fit.
Furthermore, the changes due to the NNLO contribution to dimuon production are small compared to the size of the PDF uncertainties, as one
can also infer from the relative stability of the $\chi^2$ values for the nominal and alternate fits.

\begin{figure}[htbp]
	\includegraphics[width=0.6\textwidth]{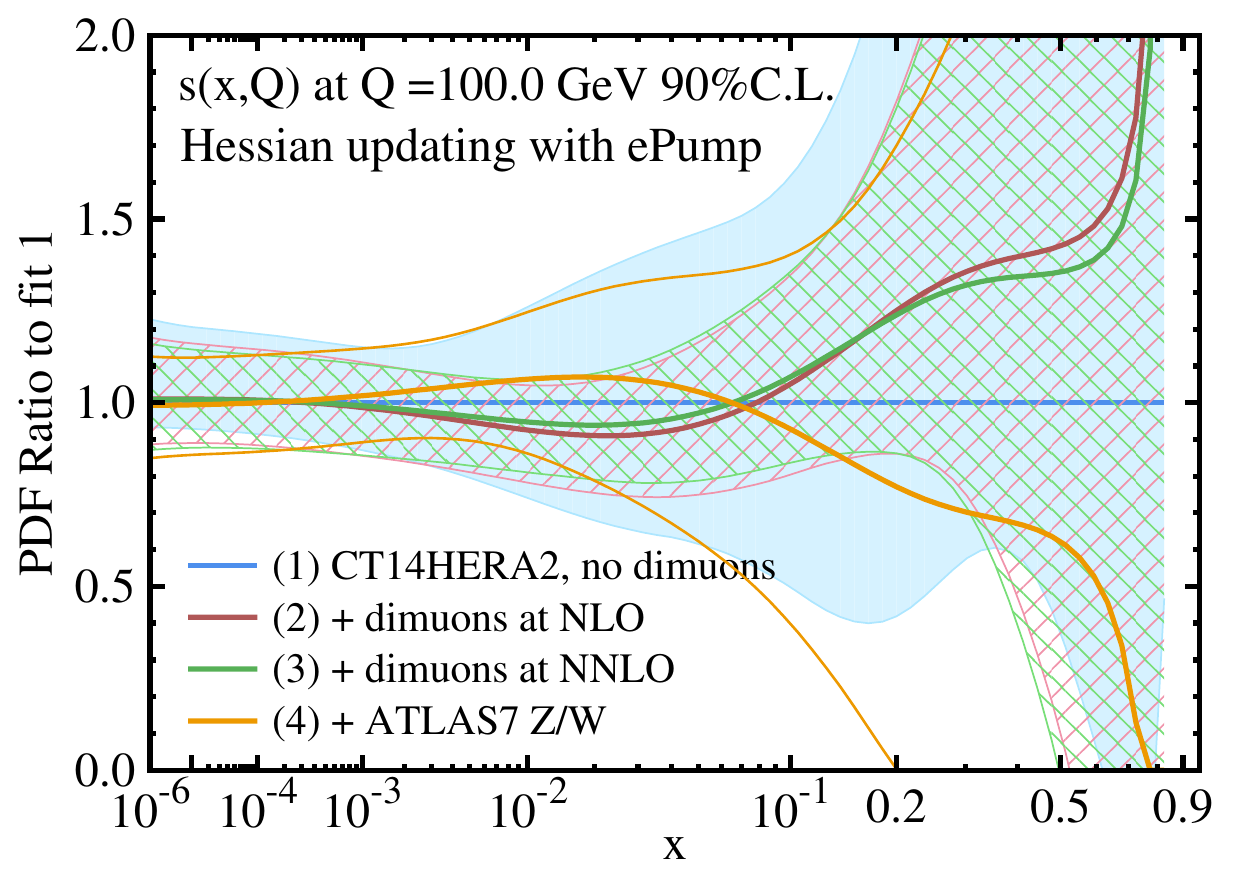}
	\caption{\label{fig:epump-dimuon}
		Comparison of $s$ PDF at $Q=100$ GeV for various fits. See the main text for its detail. }
\end{figure}

The tendency of the NNLO corrections to the dimuon cross sections to slightly increase the strangeness to higher values at $x\! \approx\! 0.1$ has  independently been confirmed by using the fast Hessian updating technique with \texttt{ePump}~\cite{Hou:2019gfw}, as well as by the MMHT group~\cite{Thorne:2019mpt}, cf.~Appendix~\ref{sec:AppendixCT18Z}. Increasing the $c\to \mu$ branching ratio from 0.099 to 0.092 adopted by MMHT \cite{Harland-Lang:2015nxa} only marginally increases $s(x,Q)$ in CT18 at $x>0.1$, while also slightly increasing the CCFR+NuTeV $\chi^2$ values. 

Finally, to estimate the impact of the NNLO corrections to the charm-quark production cross section on the simultaneous inclusion of the ATL7ZW and dimuon data sets,
we performed a series of NNLO fits illustrated in Fig.~\ref{fig:epump-dimuon}. There, we compare the strange-quark PDF obtained 
from four different fits that have been updated with \texttt{ePump}. PDF set (1)
is the base fit obtained from the \CTHERAII~data set by removing
the NuTeV and CCFR dimuon data. 
Adding back those four dimuon data sets, with NLO and NNLO predictions, yields the sets (2) and (3), respectively. PDF set (4) is obtained 
by adding the ATL7ZW data set, without the dimuon data sets. 
While PDF set (3) [found using the NNLO dimuon cross sections] yields an $s$ PDF that is marginally closer to
that constrained by the  ATL7ZW data for $10^{-3}\! \lesssim\! x\! \lesssim\! 10^{-1}$,  the improvement is still too weak to resolve the tension between the 
ATL7ZW and dimuon data sets.

\subsection{Electroweak corrections}
\label{sec:EW}

\begin{table}
\caption{A summary of electroweak corrections to the LHC precision data considered for CT18(Z).  For each process, we indicate the primary observable, an approximate upper bound
for the EW correction, references for computing the EW corrections, and whether the data were adopted in CT18(Z) with or without EW corrections.}
\label{tab:EWcorrections}
\hspace*{-0.75cm}\begin{tabular}{c|c|c|c|c|c}
\hline
\multirow{2}{*}{Data} & \multirow{2}{*}{Observables}  & Size of EW (and PI)  & Ref. & Data included & EW corrections  \\
 &    & corrections & & in the CT18(Z)? & included in the fits\\
\hline
	Inclusive jet & $p_{T}\sim1.4$ TeV, central  & 8\%  & \cite{Dittmaier:2012kx} & Yes & Yes \\
\hline
	$t\bar{t}$ &$p_{T}^{t}\sim500$ GeV   &  -5\% & \cite{Czakon:2017wor} & Yes & No \\
\hline
$W^{+}(W^{-})$ &     & -0.4(0.3)\%  & \multirow{4}{*}{\cite{Aaboud:2016btc}} & \multirow{4}{*}{CT18Z} & \multirow{4}{*}{Yes}  \\
DY low-mass & $46<M_{\ell \bar \ell}<66$ GeV central  & +1.5\%(PI) +6\%(EW) &  &     \\
DY $Z$-peak  & $66<M_{\ell \bar \ell}<116$ GeV central (forward) & $<$0.1\%(PI)-0.3(-0.4)\%(EW)  & & \\
DY high-mass &$116<M_{\ell \bar \ell}<150$ GeV central (forward) & +1.5\%(PI)-0.5(-1.2)\% (EW) & & \\
\hline
high-mass Drell-Yan & $M_{\ell \bar \ell}\sim1$ TeV  & +5\%(PI)-3\%(EW)  & \texttt{FEWZ} & No  & -- \\
\hline
\multirow{2}{*}{$Z$ $p_{T}$} & $p_{T}\sim m_Z$ & about -5\%   & \cite{Kallweit:2015fta} & Yes & No \\
 & $p_{T}\sim1$ TeV & about -30\%   & \cite{Kallweit:2015fta} & No & --  \\
\hline
\end{tabular}
\end{table}

In this subsection, we present a  summary of the electroweak (EW) corrections for the LHC data which were considered, and,
in some cases, applied, to the CT18(Z) fits. In general, we have not used data for which EW corrections are large, especially
if the data do not provide significant constraints to the PDFs. EW corrections tend to be larger in those kinematical regions
for which the statistical errors of the data are also sizable, such that those measurements which are most impacted by EW corrections
are often less sensitive to the PDFs. We note that photon-induced (PI) contributions are also important in kinematical regions afflicted
by large EW corrections, but these are of opposite sign, thus leading to partial cancellation. As we do not include an explicit photon PDF 
in the CT18(Z) PDFs, there is a potential to over-estimate the impact of EW corrections in the kinematical regions where they are greatest. 
For those EW corrections described below which were applied to the CT18(Z) fits, the implementation
was via multiplicative $K$-factors.

In Table~\ref{tab:EWcorrections}, we summarize the upper bounds upon the EW 
corrections to data considered for CT18(Z), indicating whether these data were
fitted and whether EW corrections were applied. Of these, the largest EW
corrections are for the inclusive jet cross section, being as large as 8\% in
the highest $p_{T}$ bins of the central rapidity region. 
The EW corrections for $t\bar{t}$ production have already been noted in 
Sec.~\ref{sec:TheoryTop}, with the largest EW corrections for the $p_{T}(t)$ 
distribution. At high-$p_{T}(t)$ values approaching 500 GeV, the EW correction is 
-5\% before decreasing rapidly at softer values of $p_{T}(t)$. 
For $t\bar{t}$ observables other than the $p_{T}(t)$ spectra, EW corrections are
negligible compared to the experimental uncertainty. Given the experimental precision of
the 8 TeV $t\bar{t}$ information over $p_T\! <\! 500$ GeV, we do not include EW corrections
when fitting these data, but such corrections will likely be required to describe future
measurements at higher $p_T$.

The EW corrections to the inclusive $W^+$, $W^-$ and $Z/\gamma^*$ 
production data have been investigated in Ref.~\cite{Aaboud:2016btc} 
using the \texttt{MCSANC} framework \cite{Arbuzov:2015yja}. 
For $W^{+}$ and $W^{-}$ production, the
EW corrections were found to be $-0.4$\% and $-0.3$\%, respectively. 
In the $Z$-peak region ($66\!<\!M_{\ell \bar \ell}\!<\!116$ GeV) for neutral-current (NC)
Drell-Yan (DY) with central (forward) selections\footnote{The central selection requires both leptons in the central region, $|\eta_l|<2.5$, while the forward one requires one central and one forward ($2.5<|\eta_l|<4.9$) leptons.}, the EW 
corrections are about $-0.3(-0.4)\%$, with only a weak kinematical dependence 
on the observables $M_{\ell \bar \ell}$ and $y_{\ell \bar \ell}$. We estimate that photon-induced dilepton
production ($\gamma\gamma\to l^{+}l^{-}$) contributes to $Z$-peak NC DY 
by less than 0.1\%. For the low-mass ($46\!<\!M_{\ell \bar \ell}\!<\!66$ GeV) region, the EW corrections 
are +6\% independent of rapidity selection criteria, and, for high-mass ($116\!<\!M_{\ell \bar \ell}\!<\!150$ GeV) NC DY production, the EW corrections are -0.5\%(-1.2\%) for the central (forward) selection, with a very weak dependence on the $\eta_l$ and $y_{\ell \bar \ell}$ bins. The PI contributions are 1.5\% for both $M_{\ell \bar \ell}$ bins.
Given the small impact of the low- and high-mass DY data on the PDF fits, we decided not to 
include the low- and high-mass and forward $Z$-peak DY data in the CT18A(Z) fits. 
For the $Z$-peak and $W^\pm$ data, the EW corrections are included in the multiplicative $K$-factors, while the PI contribution is ignored.
Finally, we note that, as discussed in Sec. 6.1.2 of Ref.~\cite{Aaboud:2016btc}, the background from the PI dilepton production has been subtracted from the ATLAS 7 TeV $W, Z$ data. 

\begin{figure}
 \includegraphics[width=0.45\textwidth]{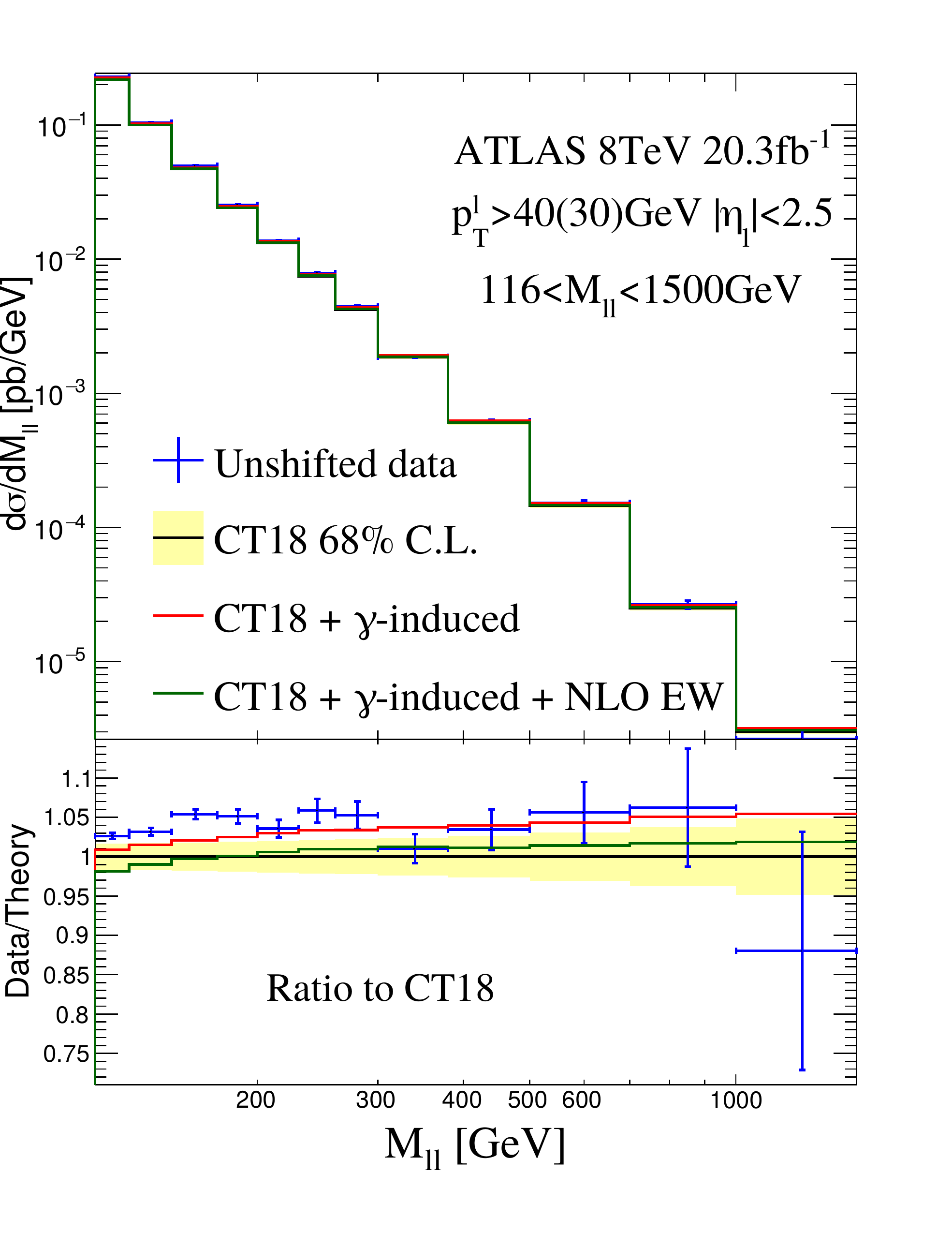}
\caption{The photon-induced contributions (photon PDF are taken from \texttt{LUXqed17\_plus\_PDF4LHC15}) and NLO EW corrections to ATLAS 8 TeV high-mass Drell-Yan production.}
\label{fig:EW4ATL8DY}
 \end{figure}

We also did not include the ATLAS 8 TeV very high-mass ($116\!<\!M_{\ell \bar \ell}\!<\!1500$ GeV) Drell-Yan data~\cite{ATLAS8DY} in our CT18(Z) fitting, due to non-negligible 
EW corrections and PI contributions. We find that, for very high invariant masses ($M_{\ell \bar \ell}\!\sim\!1$ TeV), the PI contribution can be as large as 5\% as computed 
with \texttt{LUXqed17\_plus\_PDF4LHC15} \cite{Manohar:2017eqh}. 
In comparison, the EW corrections can be calculated using the \texttt{FEWZ}
program as shown in Fig.~\ref{fig:EW4ATL8DY}, and are approximately -3\% in
this case. The partial cancellation of the PI contribution and EW correction
yields an increase in the cross section by less than 2\%. With the \texttt{ePump} 
program, we have also checked that the impact of these data on the CT18 fits 
is very small.

The only $Z$ $p_{T}$ distribution included in the CT18(Z) fits 
comes from the ATLAS 8 TeV measurements. We have dropped the high-$p_{T}$ 
data by imposing a kinematic cut of $p_{T}^{Z}\!<\!150$ GeV because the missing 
EW corrections to the high-$p_{T}$ data are significant. 
In general, we note that these corrections are negative. In terms of 
Refs.~\cite{Hollik:2015pja,Kallweit:2015fta}, the NLO EW corrections 
can be as large as several tens-of-percent when 
$p_{T}^{Z}\gg M_{Z}$, due to electroweak Sudakov logarithms. 
In the fitted region of $p_{T}^{Z}$, between 45 GeV and 150 GeV, 
the EW corrections are found to reduce the cross sections by several 
percent, thereby pulling the theory predictions further away from the ATLAS 8 TeV data.

\section{Standard candle cross sections
\label{sec:StandardCandles}
}

Measurements of total cross sections for inclusive hadroproduction at
colliders provide cornerstone tests of the Standard Model. These relatively simple
observables can both be measured with high precision and
predicted in NNLO QCD theory with small uncertainties.
In Sec.~\ref{sec:ellipse}, we collect NNLO theory predictions, based upon the CT14, \CTHERAII, and CT18(A/X/Z) 
NNLO PDFs, for the inclusive production cross sections of $W$ and $Z$ bosons, top-quark pairs, and Higgs bosons (through gluon-gluon fusion), at the
LHC with center-of-mass energies of $\sqrt{s} = 7$, 8, 13 and 14 TeV.
These theoretical predictions supersede similar comparisons made with the previous generations of CT10/CT14 PDFs \cite{Gao:2013xoa,Dulat:2015mca}
and can be compared to the corresponding experimental measurements. In addition, we also present theoretical predictions
for vector boson production at LHCb based on fixed-order and resummed calculations in Sec.~\ref{sec:Res}; explore
predictions for $W\!+\!c$ production at ATLAS in Sec.~\ref{sec:Wcharm} and 13 TeV $t\bar{t}$ production at CMS in Sec.~\ref{sec:tt13};
and show predictions for high-$x$ fixed-target Drell-Yan cross sections in Sec.~\ref{sec:SeaQuest}, in anticipation
of the forthcoming results of the SeaQuest experiment \cite{Aidala:2017ofy} at Fermilab.

\subsection{Inclusive total cross sections at the LHC}
\label{sec:ellipse}

\begin{figure}[p]
	\begin{center}
  \includegraphics[width=0.44\textwidth]{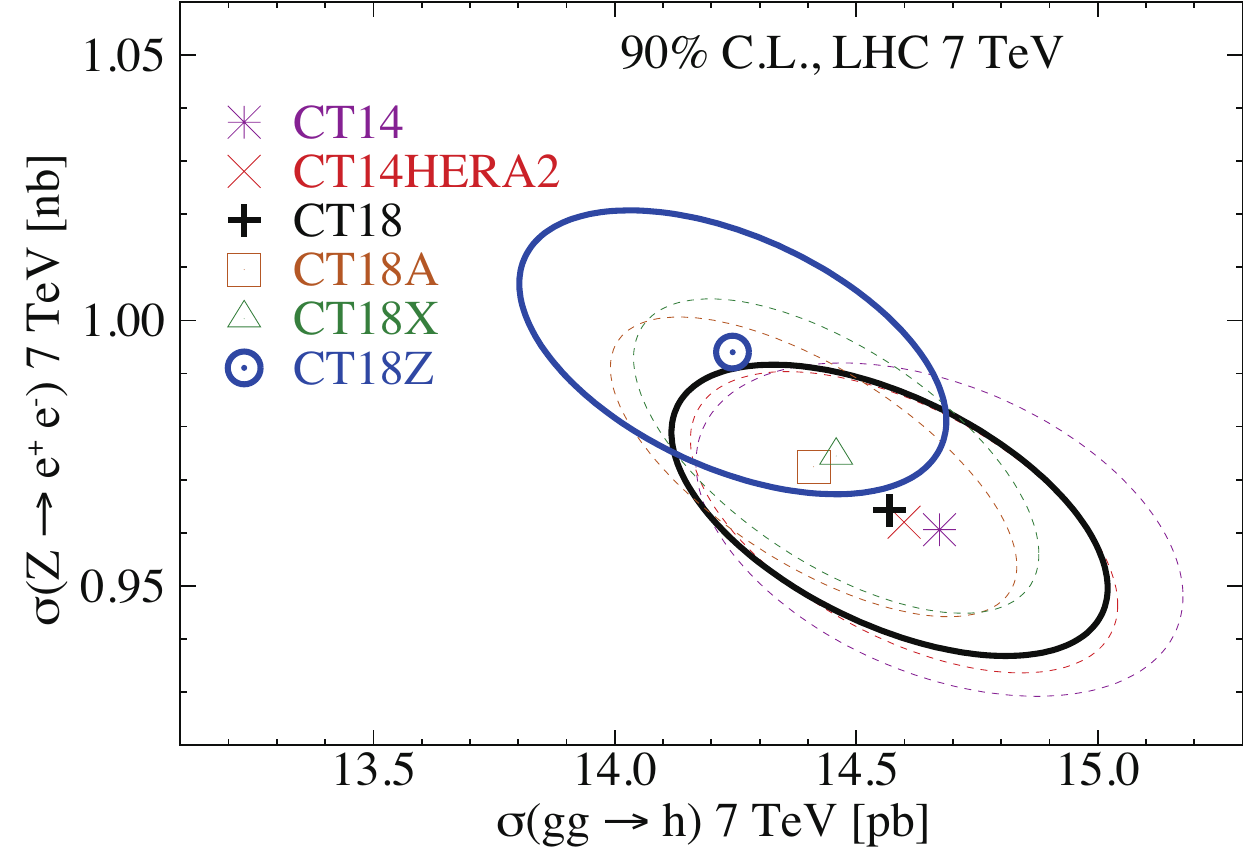}
  \includegraphics[width=0.44\textwidth]{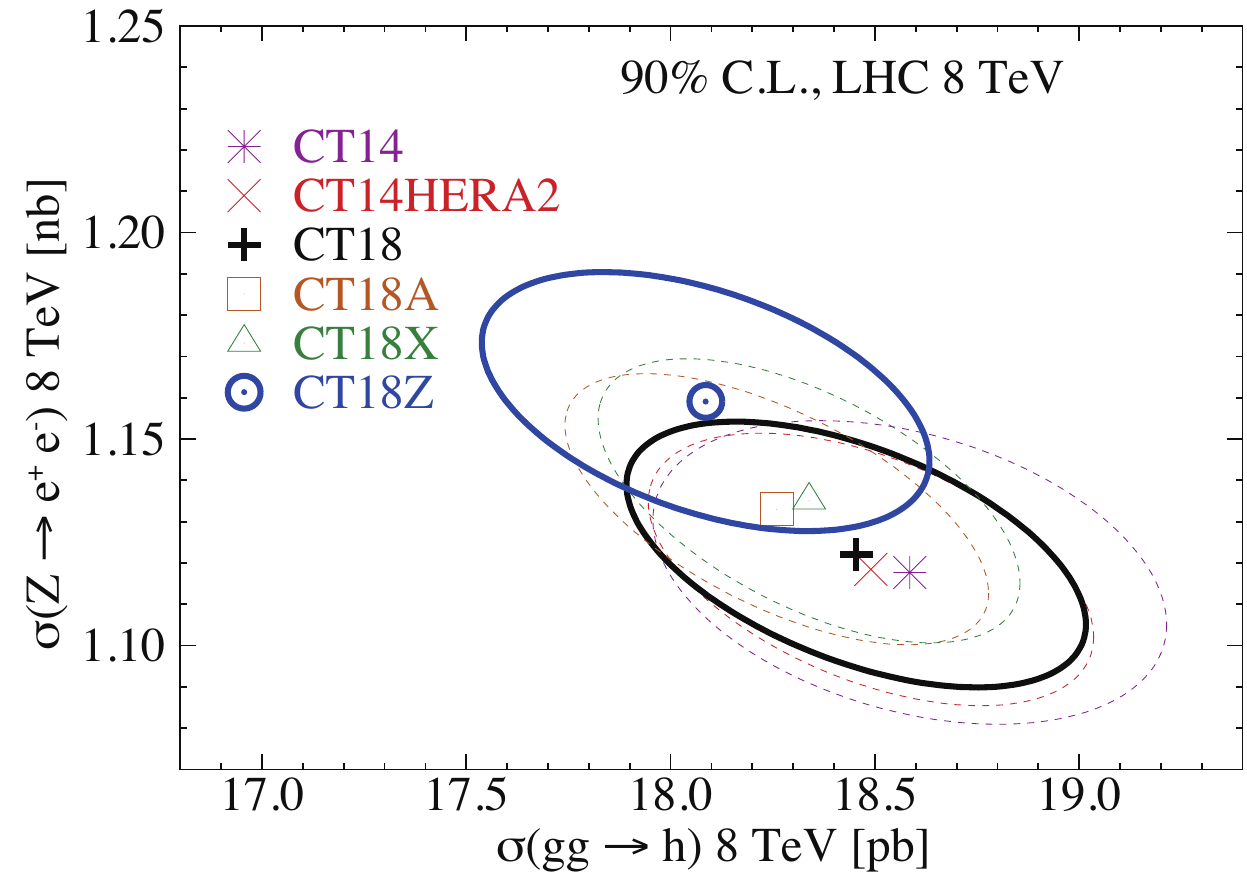} \\
  \includegraphics[width=0.44\textwidth]{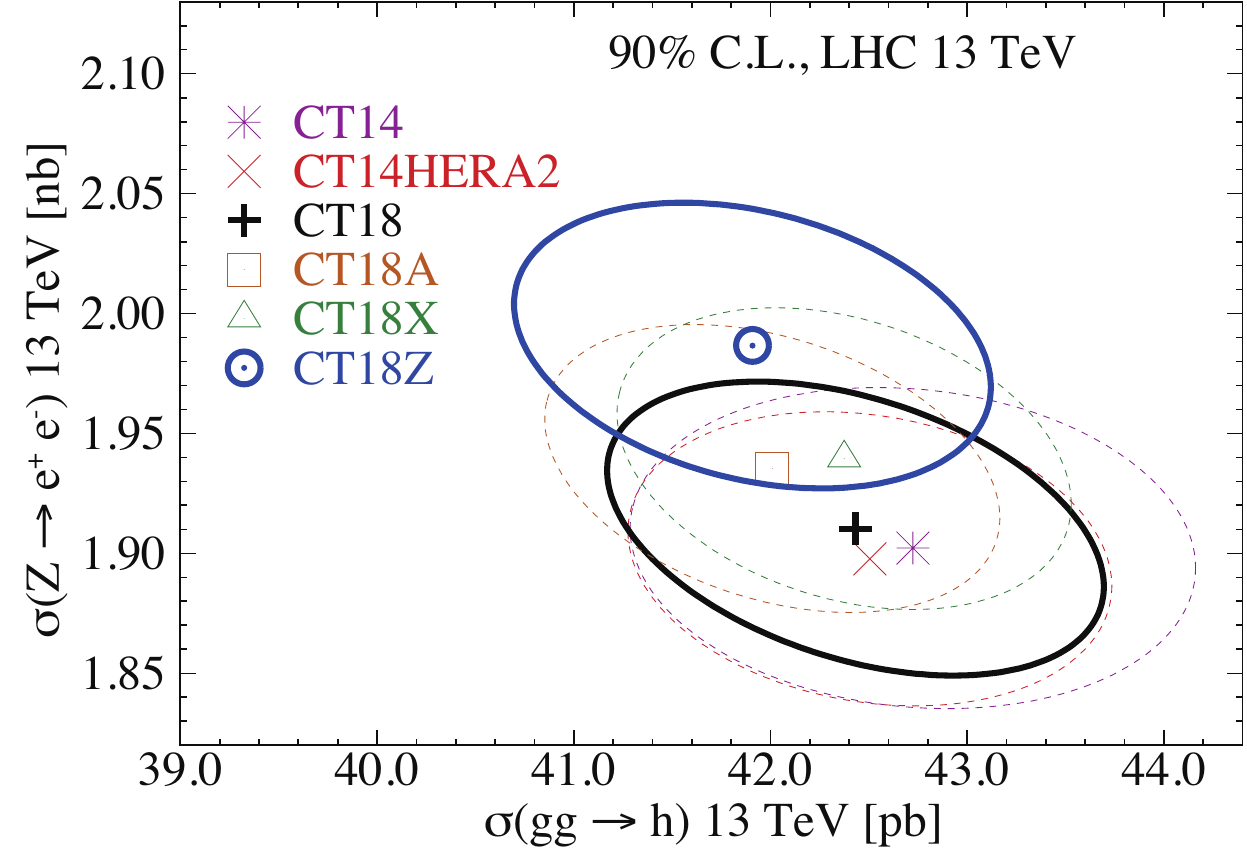}
  \includegraphics[width=0.44\textwidth]{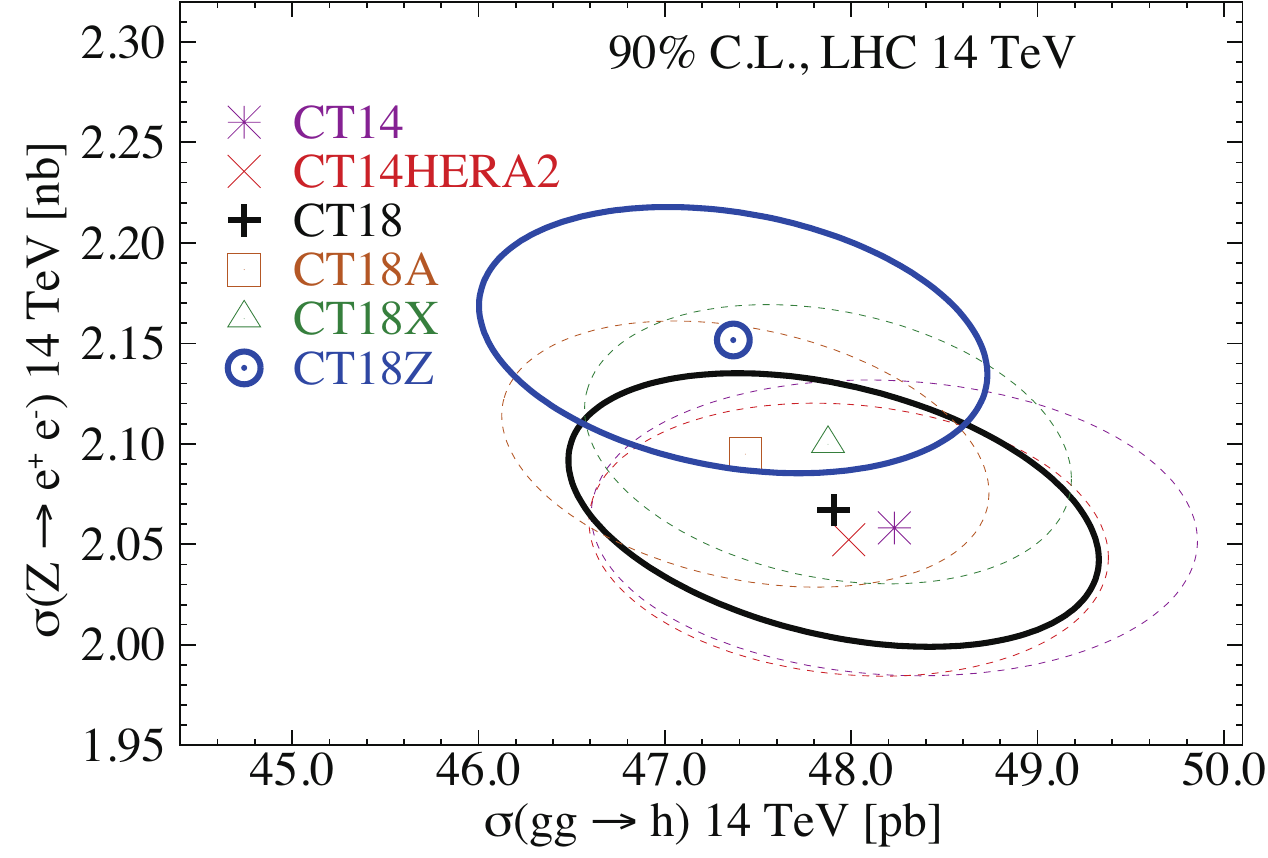} \\
  \includegraphics[width=0.44\textwidth]{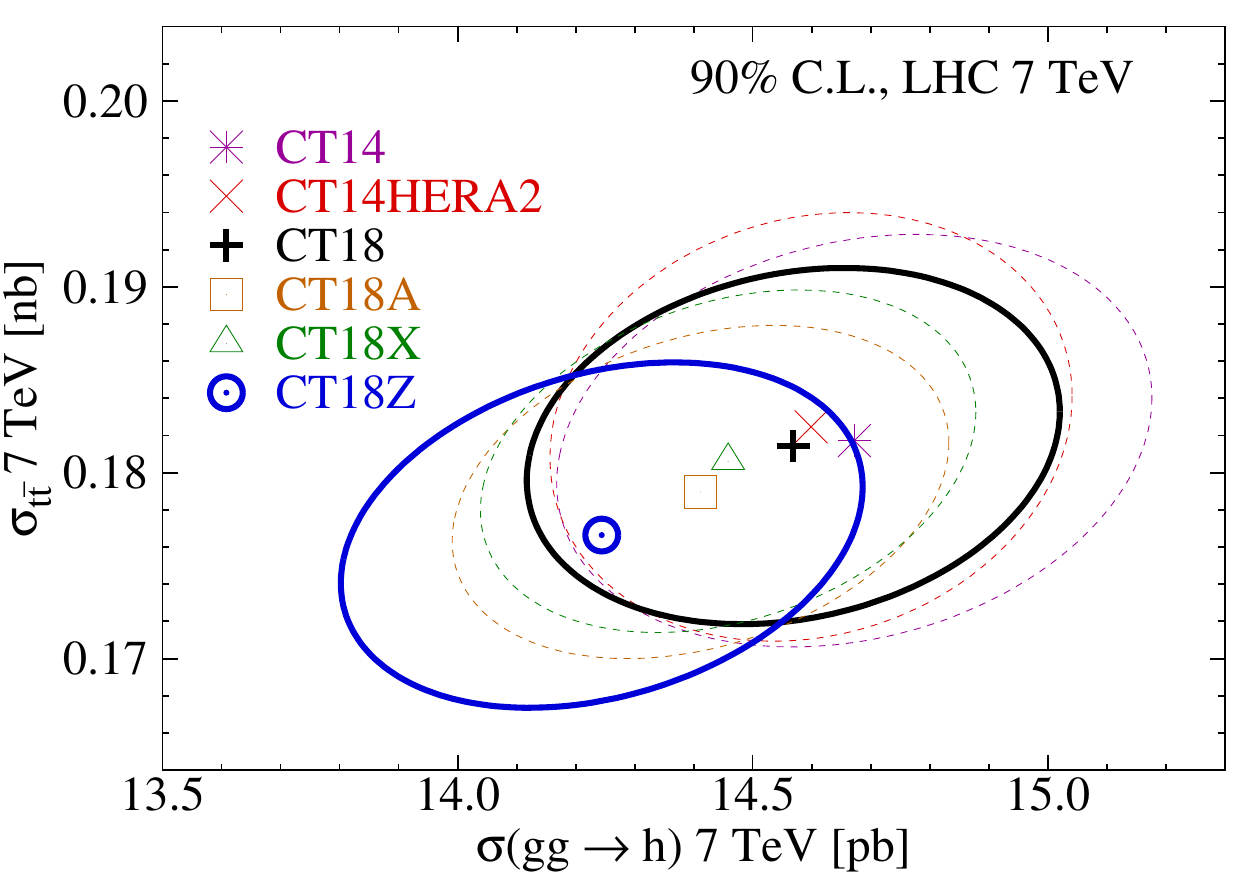}
  \includegraphics[width=0.44\textwidth]{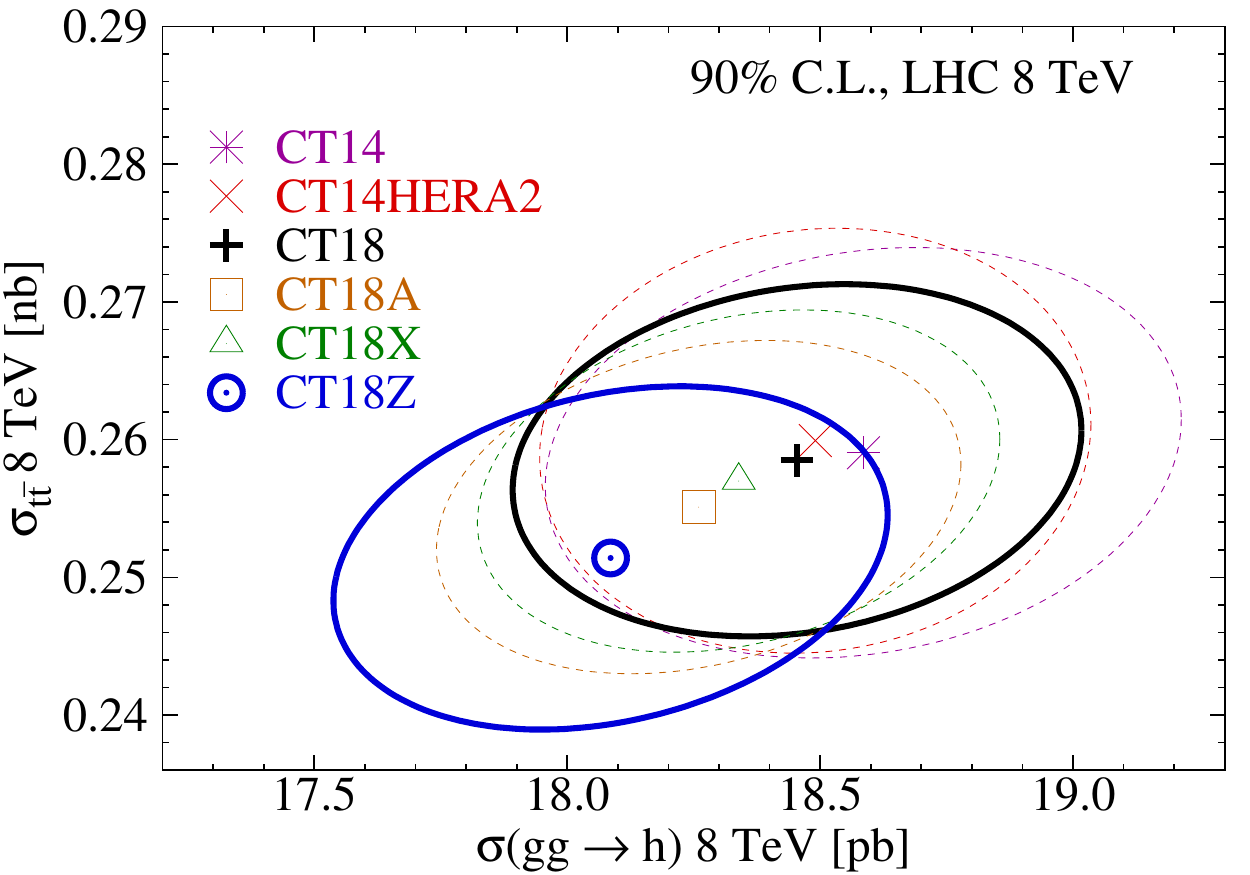} \\
  \includegraphics[width=0.44\textwidth]{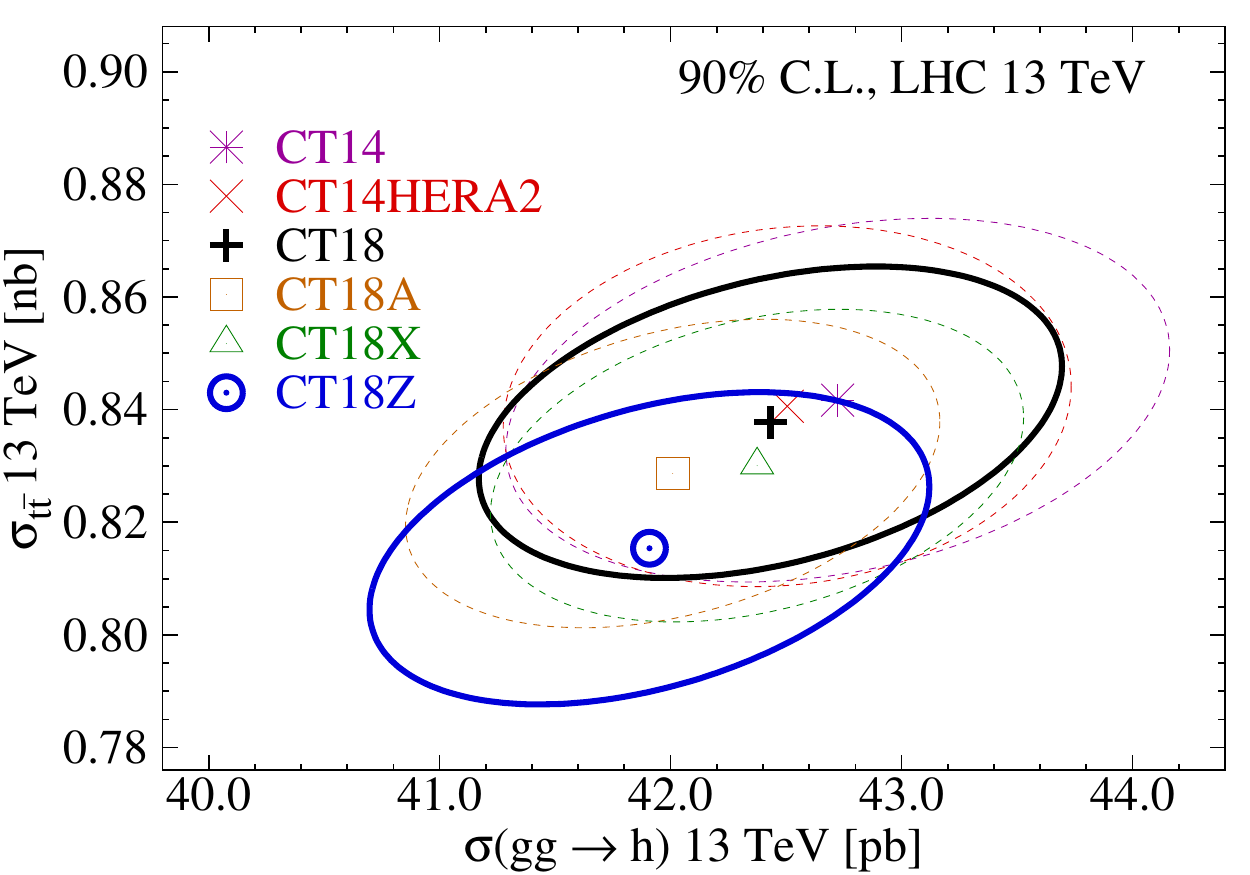}
  \includegraphics[width=0.44\textwidth]{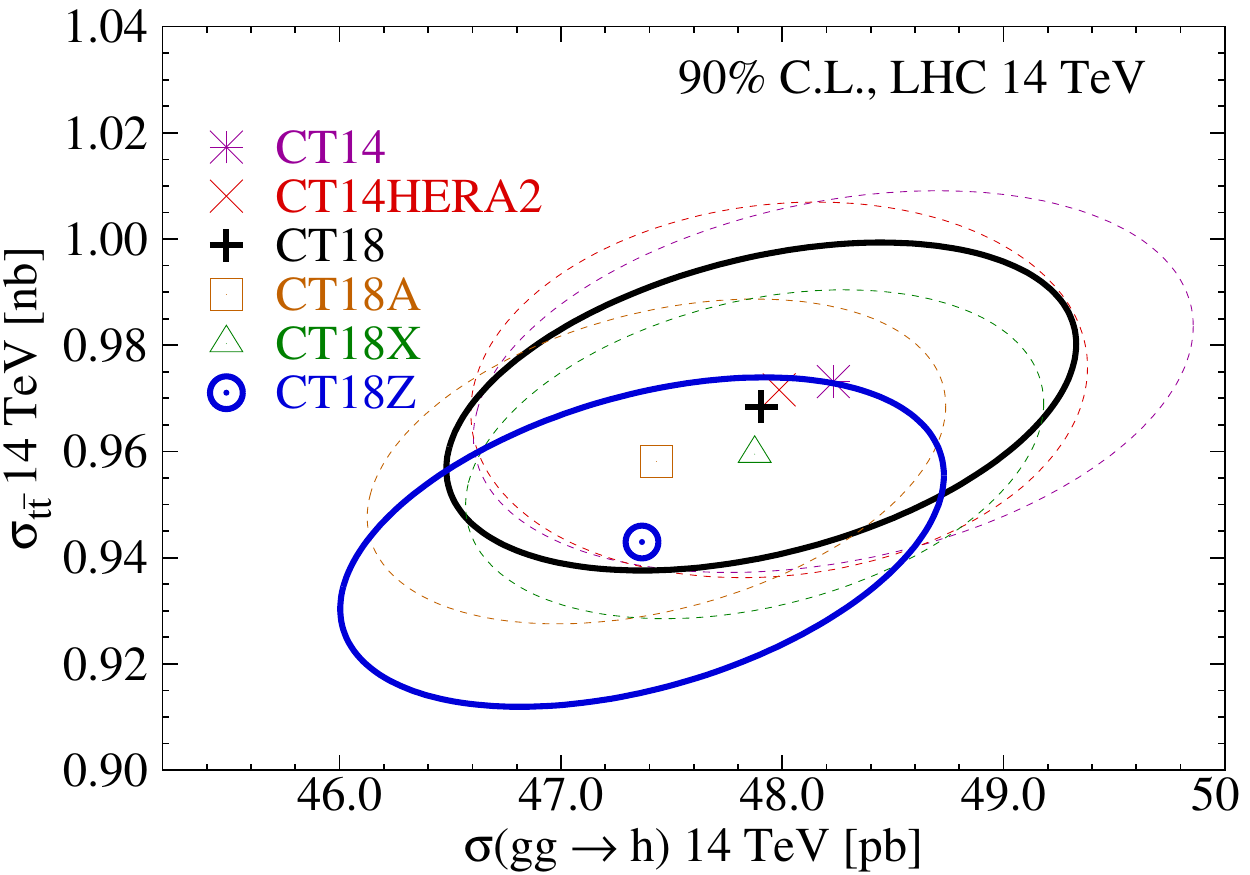} \\
	\end{center}
    \vspace{-2ex}
	\caption{The 90\% C.L. error ellipses for the $ggH$, $t \bar t$, and $Z^0$ inclusive cross sections computed with the CT18 NNLO family of PDFs at the LHC 7, 8, 13 and 14 TeV. 
		\label{fig:corr_ellipse1}}
\end{figure}

\begin{widetext}
\begin{figure}[p]
	\begin{center}
  \includegraphics[width=0.44\textwidth]{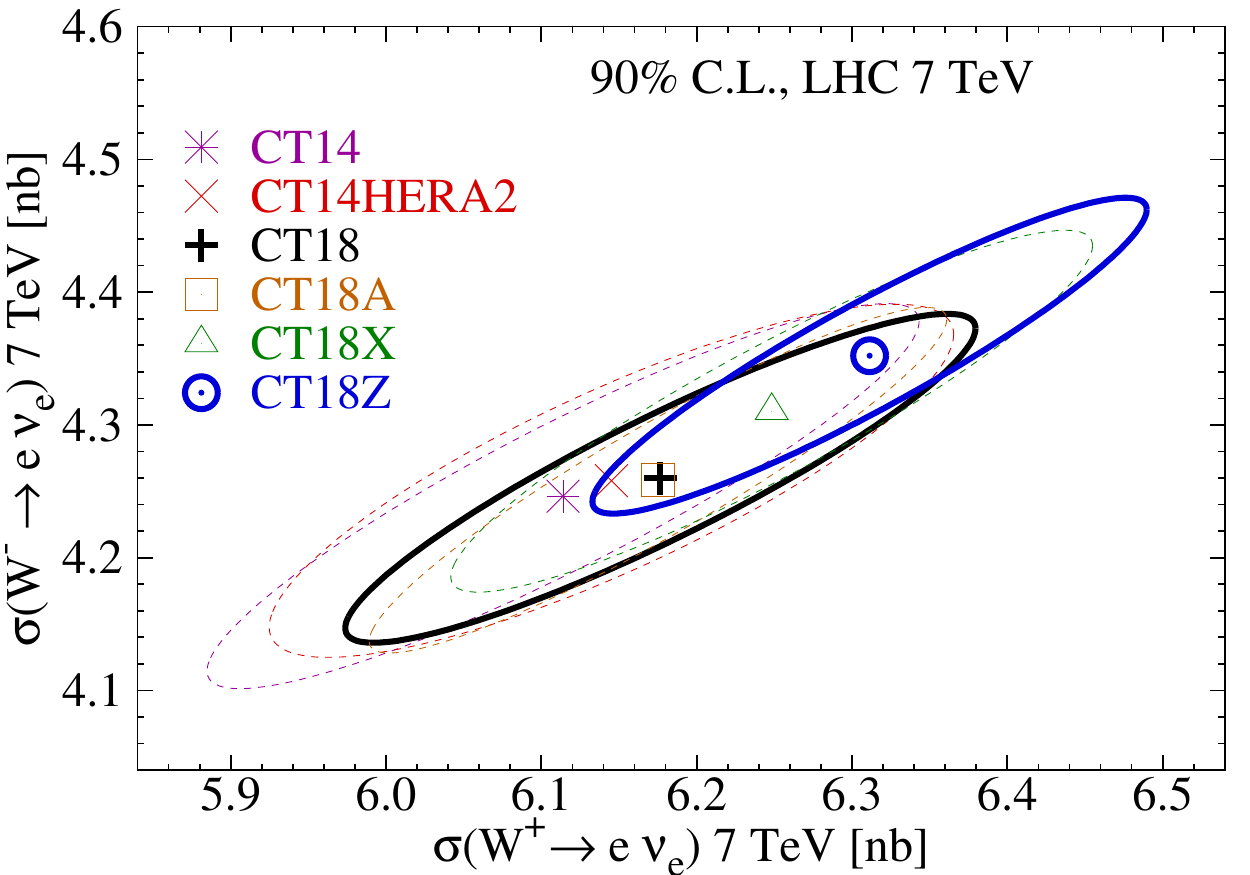} 
  \includegraphics[width=0.44\textwidth]{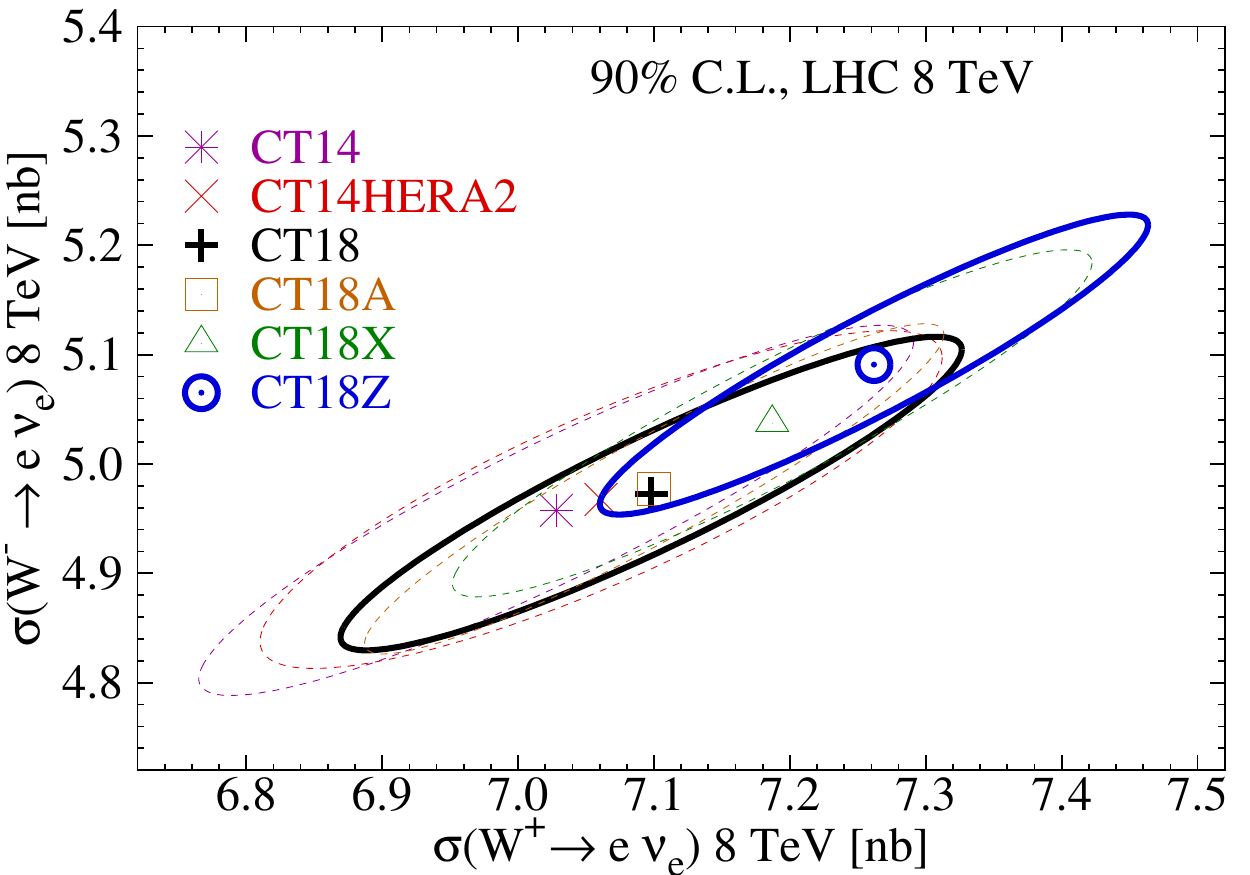} \\
  \includegraphics[width=0.44\textwidth]{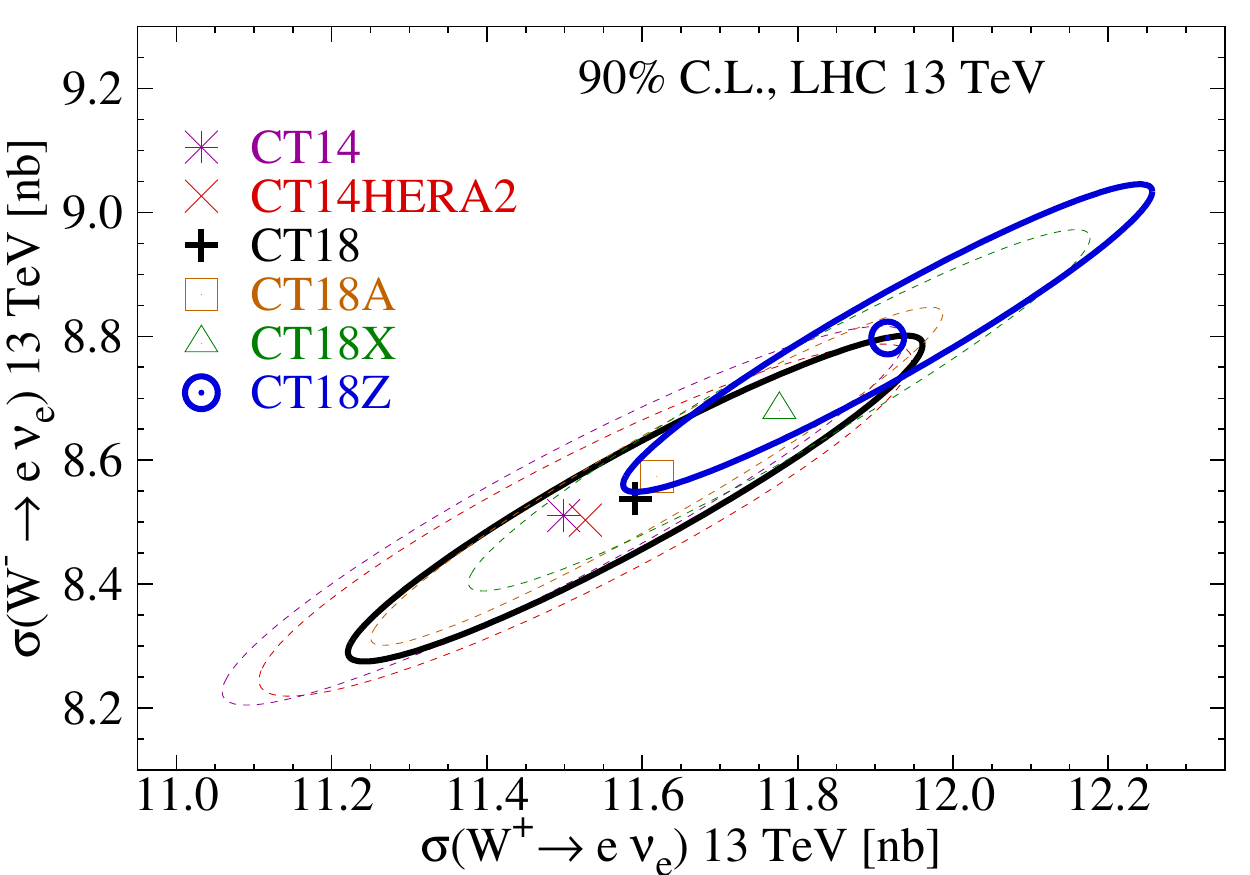} 
  \includegraphics[width=0.44\textwidth]{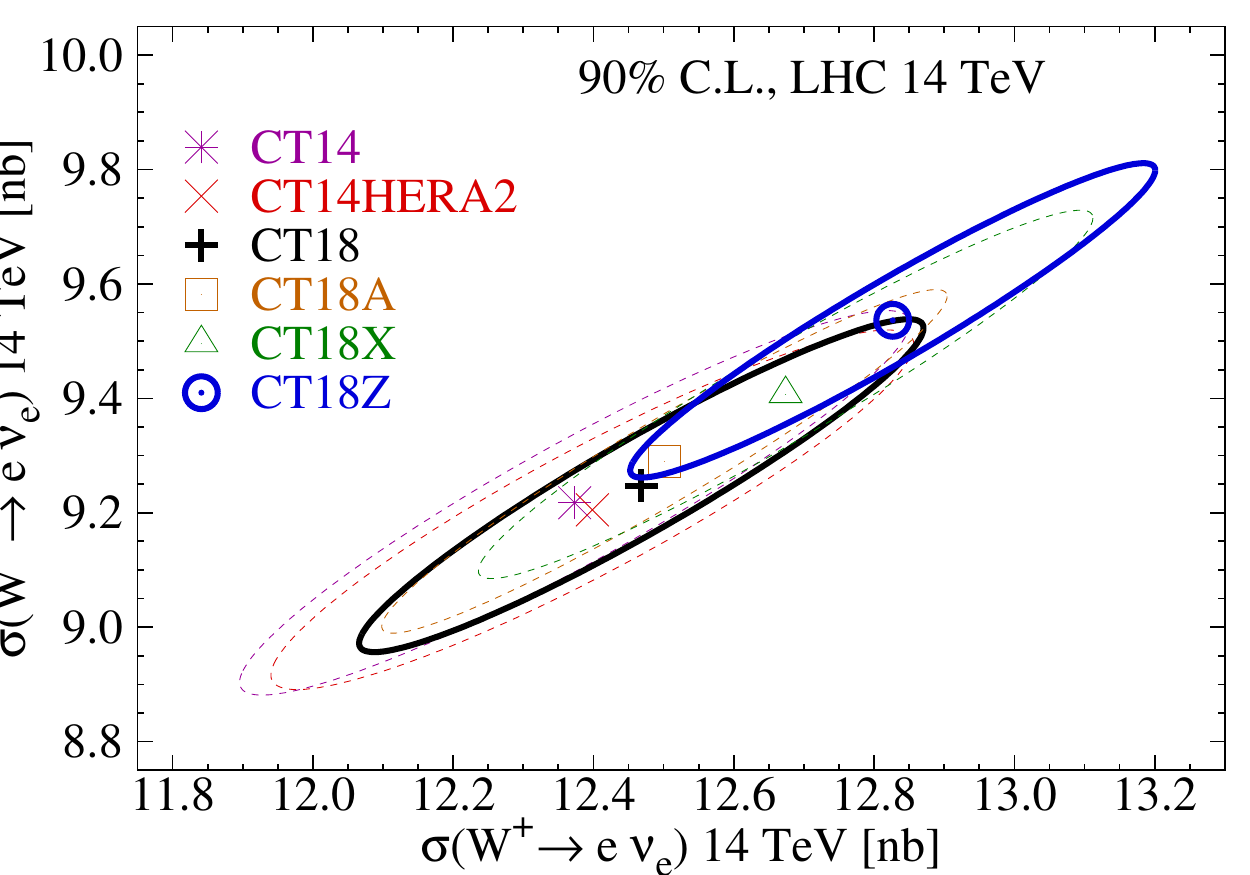} \\
  \includegraphics[width=0.44\textwidth]{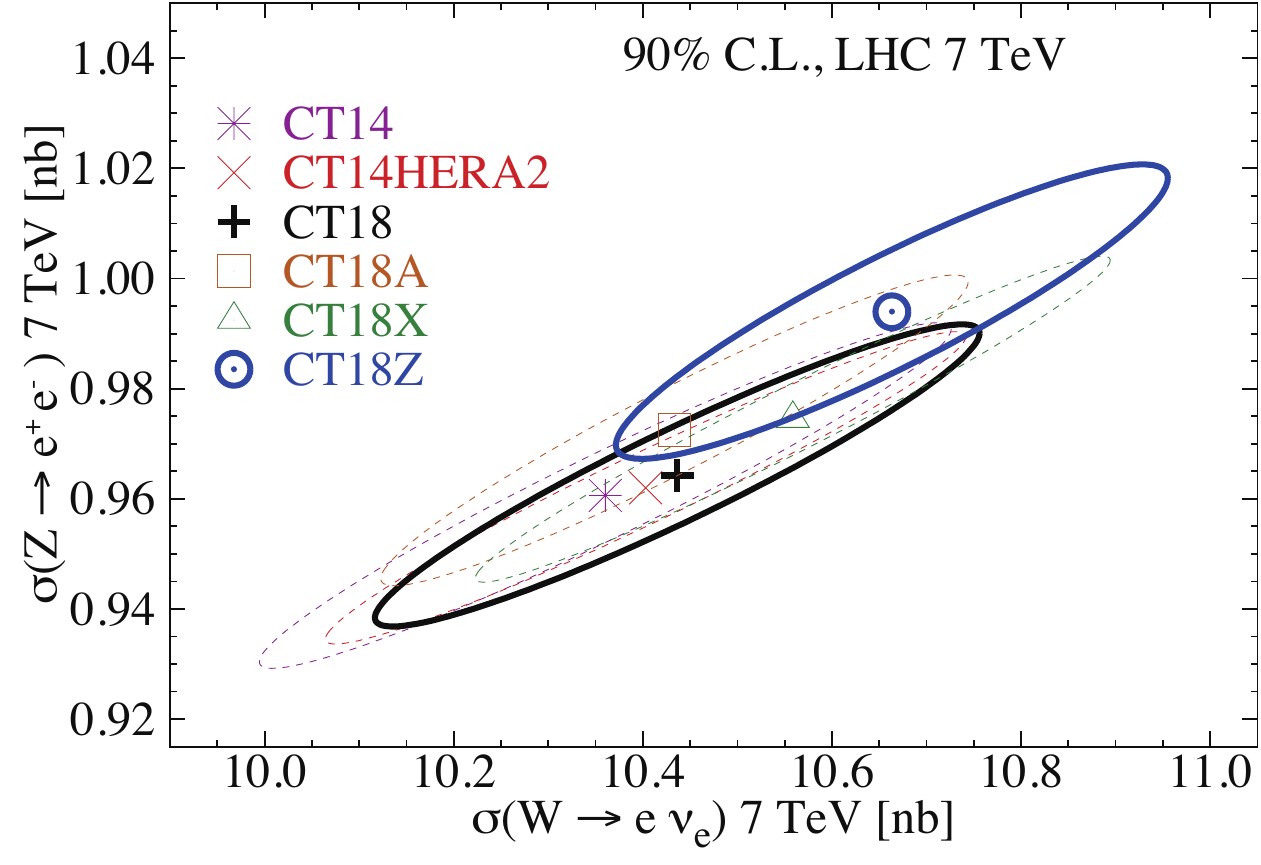} 
  \includegraphics[width=0.44\textwidth]{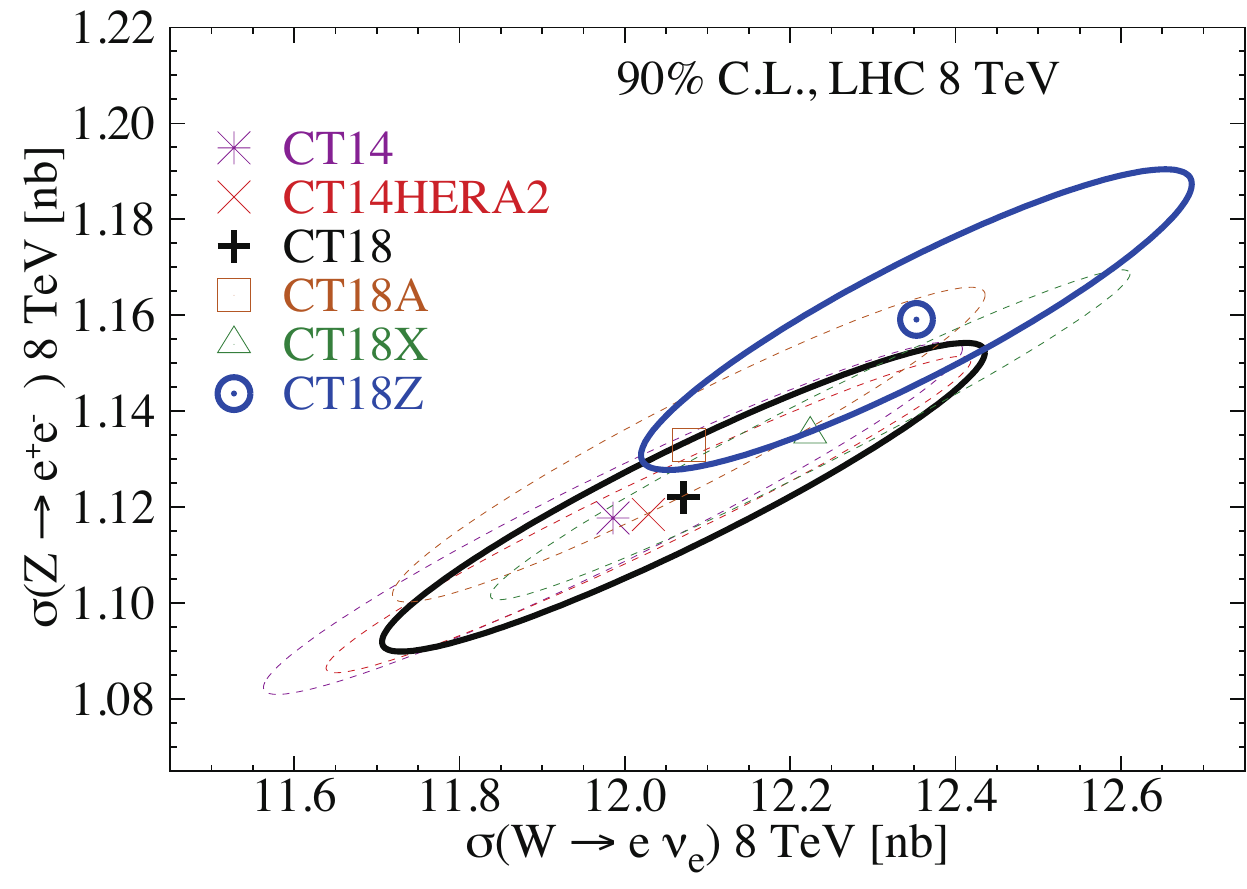} \\
  \includegraphics[width=0.44\textwidth]{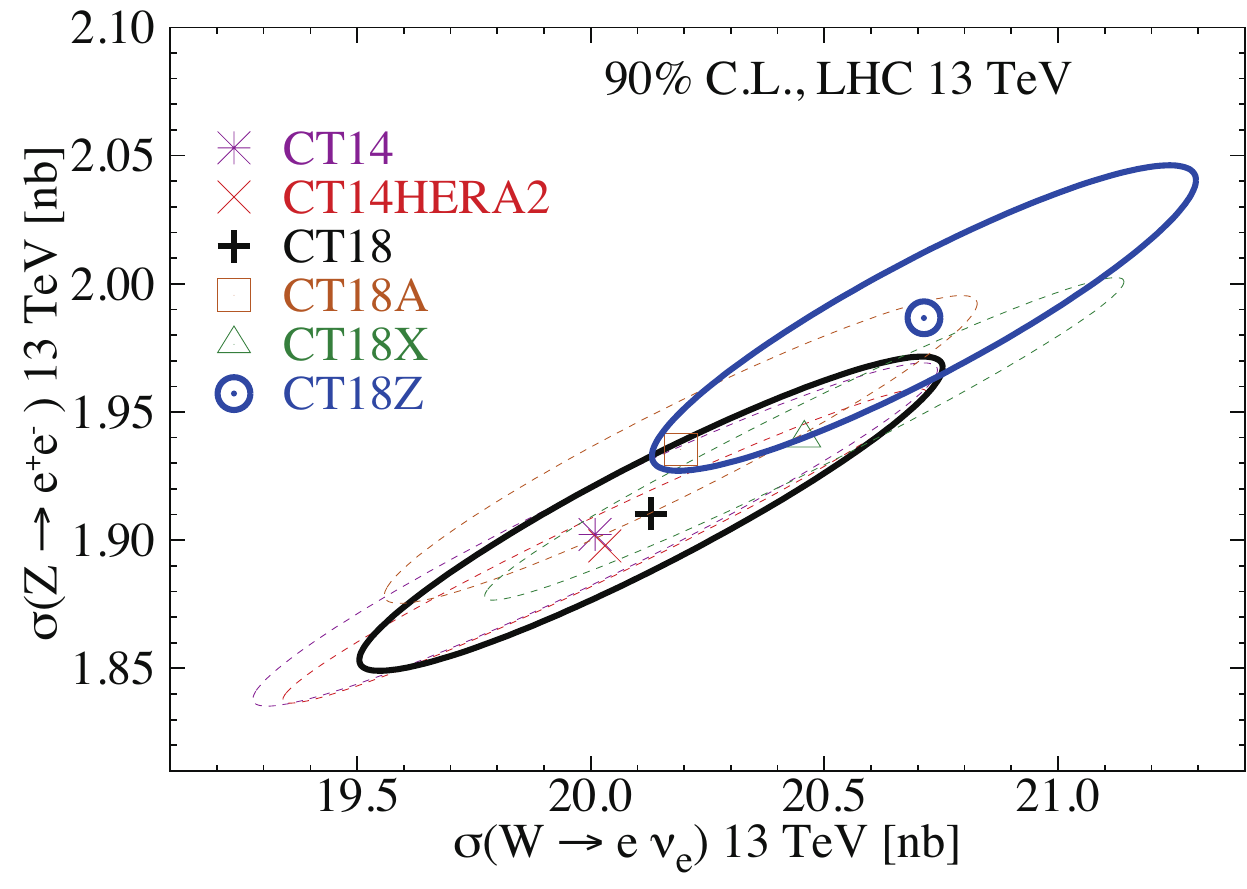} 
  \includegraphics[width=0.44\textwidth]{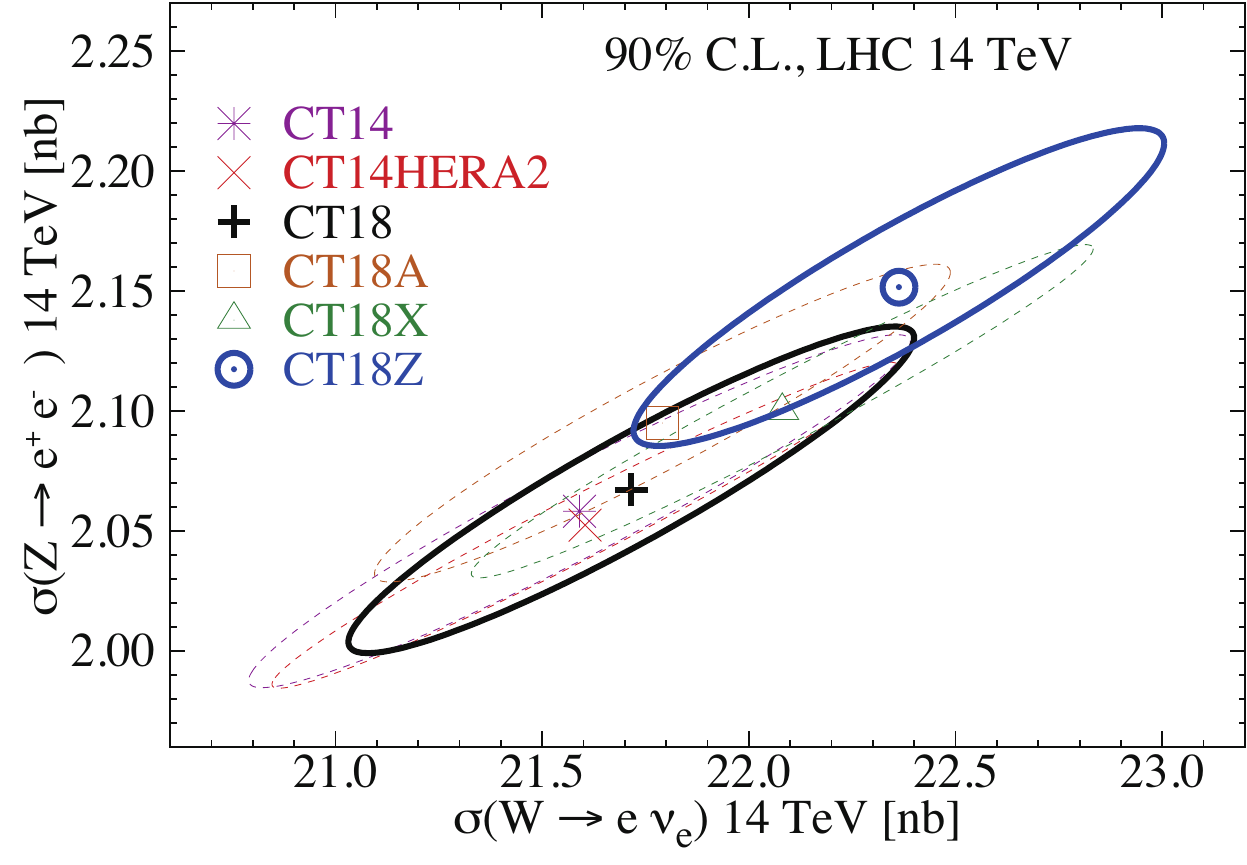} \\
	\end{center}
	\caption{Same as Fig.~\ref{fig:corr_ellipse1}, but for 
the $W^+$, $W^-$ and $Z^0$ inclusive cross sections.
		\label{fig:corr_ellipse2}}
\end{figure}
\end{widetext}

In this work, the masses of the top quark and Higgs boson
are set to $m_t^\mathit{pole}=173.3$ GeV and $m_H=125$ GeV, respectively.
The $W$ and $Z$ inclusive cross sections
(multiplied by branching ratios for the decay into one charged lepton flavor),
are calculated by using the \texttt{Vrap} v0.9 program~\cite{Anastasiou:2003ds,Anastasiou:2003yy} at NNLO in QCD, with the renormalization
and factorization scales ($\mu_R$ and $\mu_F$) set equal
to the invariant mass of the vector boson.
The total inclusive top-quark pair cross sections are calculated with
the help of the program \texttt{Top++}
v2.0~\cite{Czakon:2013goa,Top++} at NNLO+NNLL accuracy, with QCD
scales set to the mass of the top quark \cite{Czakon:2017wor} as is
the default in the \texttt{Top++} framework.
The Higgs boson cross sections via gluon-gluon fusion
are calculated at NNLO in QCD by using  the \texttt{iHixs} v1.3
program~\cite{Anastasiou:2011pi},
in the heavy-quark effective theory (HQET) with finite top quark mass correction,
and with the QCD scales set equal
to the Higgs boson mass.

\begin{figure}[tb]
	\begin{center}
		\includegraphics[width=0.65\textwidth]{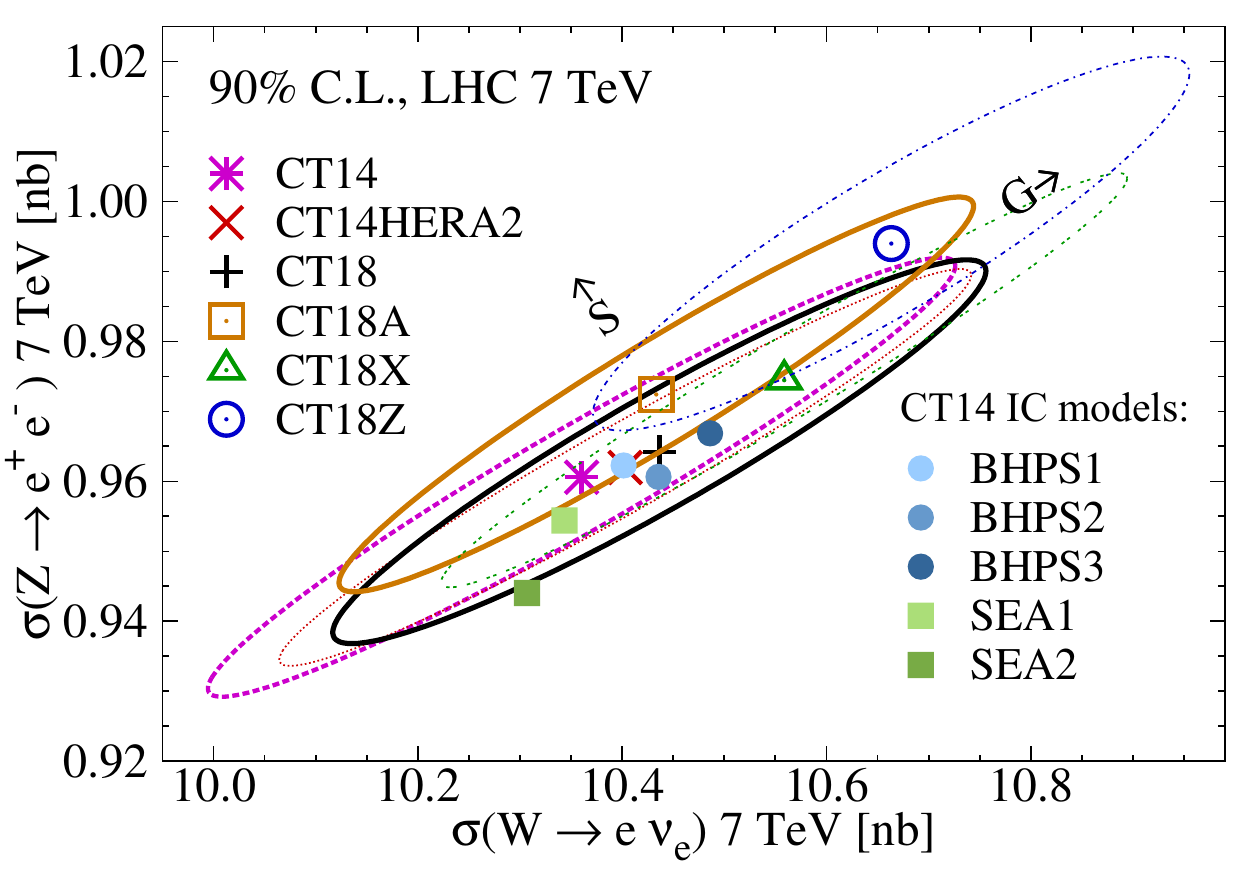}
	\end{center}
	\vspace{-2ex}
	\caption{Theoretical predictions of the total production cross sections of $W$ and $Z$ bosons
	at $\sqrt{s}=7$ TeV as relevant for the ATLAS 7 TeV $W/Z$ data (Exp.~ID=248).  Here, we also
	include several calculations which include an intrinsic charm (IC) PDF based upon either
	the BHPS valence-like model (with three different normalizations) or a sea-like model (with
	two different normalizations) in addition to CT14, as described in Ref.~\cite{Hou:2017khm}.
		\label{fig:IC}}
\end{figure}

Fig.~\ref{fig:corr_ellipse1} shows that the Higgs boson cross section through
gluon-gluon fusion ($ggH$) at the LHC does not have a pronounced correlation 
with the top-quark pair ($t \bar t$) cross section, because the two processes are dominated by the gluon
PDF in somewhat different $x$ regions. 
The degree of anti-correlation found in the $ggH$ and $Z$ boson cross sections decreases as the LHC energy increases. 
On the other hand, Fig.~\ref{fig:corr_ellipse2} shows that the electroweak gauge boson cross sections are highly correlated with each other at the LHC.
Generally speaking, the prediction of CT18 is closer to \CTHERAII, and the largest difference occurs between CT18Z and CT14. 
Furthermore, the CT18X prediction is closer to CT18Z for the electroweak gauge boson productions, cf. Fig.~\ref{fig:corr_ellipse2}, but not for the $ggH$ or $t \bar t$ inclusive cross sections, cf. Fig.~\ref{fig:corr_ellipse1}.

The mutual dispositions of the error ellipses for $W$ and $Z$ cross sections in the bottom half of Fig.~\ref{fig:corr_ellipse2} can be tied to the differences among the strangeness and other PDFs of the CT18, A, X, and Z ensembles discussed in Sec.~\ref{sec:CT18ZvsCT18}. In general, the orientations of all shown $W$-$Z$ ellipses are similar, with
the direction parallel to the semi-minor axes --- associated with the relative difference between the $W$ and
$Z$ production cross sections --- most closely identified with the strange PDF. 
The correlation between the $s$-PDF and the ratio of $W^\pm$ to $Z$ cross sections was first pointed out in the CTEQ6.6 analysis~\cite{Nadolsky:2008zw}.
The theory predictions based on
CT18A and CT18Z are both equally shifted in this direction. Meanwhile, CT18X, and especially, CT18Z, are
significantly offset along the semi-major axis [the ``$\sigma(Z)\!+\!\sigma(W)$ direction''], more related to the gluon at $x<10^{-2}$, as again was pointed out in \cite{Nadolsky:2008zw}. 
The close alignment of CT18Z and A in the perpendicular direction relates closely to similarity in the fitted strangeness
distributions obtained under these fits.

It is worthwhile to investigate whether the inclusion of nonperturbative charm may significantly alter these theoretical predictions, especially for electroweak boson production.  Ref.~\cite{Ball:2017nwa} suggested that tensions between the combined HERA data (Exp.~ID=160) and
ATLAS 7 TeV $W/Z$ data require that the charm PDF at the initial scale $Q_0$ be independently parameterized.  Such nonperturbative charm, of 
{\it indefinite sign} and shape, was thus implemented using the unique neural network approach of the NNPDF collaboration as a ``fitted charm'' contribution to the proton's structure, distinct from perturbative charm.  The question of intrinsic charm, including its dynamic origin in perturbative QCD, has also been studied by CT, most recently, in Ref.~\cite{Hou:2017khm}, which implemented positive nonperturbative charm as an explicitly twist-$2$ intrinsic PDF, informed by various models, at the scale $Q_0=m_c$ as a boundary condition for the perturbative evolution of charm.  

Following this work, we show in Fig.~\ref{fig:IC} theoretical predictions for the total $W$ and $Z$ production cross sections at 7 TeV, analogous to the third left-hand panel of Fig.~\ref{fig:corr_ellipse2}, but including several scenarios for IC. Correlation studies in the CTEQ6.6 analysis \cite{Nadolsky:2008zw} have shown that the central point $\{\sigma_W, \sigma_Z\}$ is shifted in the direction $G$ in Fig.~\ref{fig:corr_ellipse2} primarily by increasing the gluon PDF in the relevant region $x\approx 0.01$. It is shifted in the direction $S$ primarily by increasing the $s$ PDF in a similar $x$ region. 

By these rules of thumb, the upward shift of the ellipse for CT18A above CT18 is consistent with the increase of strangeness in CT18A upon the inclusion of the ATLAS 7 TeV $W/Z$ data. Inclusion of IC in CT14 shifts the theoretical prediction for the central CT14 against the direction $S$, which is consistent with some suppression of strangeness preferred by DIS experiments in CT14 after a {\it positive} IC PDF is included. The CT14 IC predictions are also shifted along direction $G$ in reflection of the different magnitude of the gluon PDF in CT14 IC models, as compared to the nominal CT14. 

The downward shift of the IC predictions in the figure, with respect to the purely extrinsic charm predictions of CT14, etc., thus appears to be a generic outcome of assuming a {\it non-negative charm} at the initial-scale $Q_0$, which would naturally arise from twist-4 contributions as discussed in~\cite{Hou:2017khm}. 
The reason is, again, some suppression of the strangeness PDF in such IC models, which only exacerbates the tension with the ATLAS 7 TeV $W/Z$ data. We conclude that the standard approach of including a non-negative intrinsic $c_\mathrm{IC}(x,Q=m_c)$ PDF is unlikely to resolve the tensions between ATLAS $W/Z$ 7 TeV data and HERA. The subtleties involved in the definition and dynamical origin of intrinsic/fitted charm are sufficiently complex that more forthcoming analyses will be required to disentangle them and understand their phenomenological implications.

\subsection{Vector boson differential cross sections at the LHC}
\label{sec:Res}

\begin{figure}[htb]
	\includegraphics[width=0.49\textwidth]{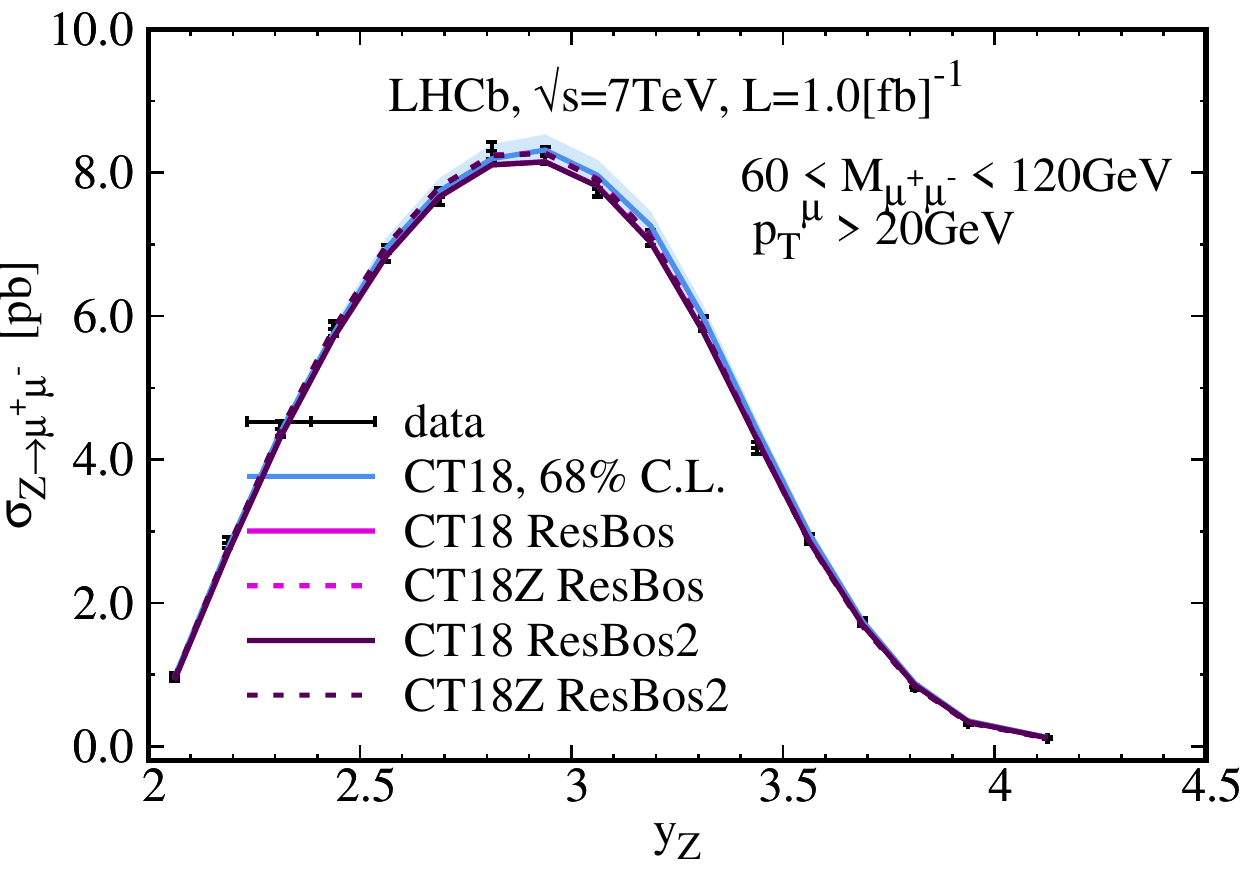}
	\includegraphics[width=0.49\textwidth]{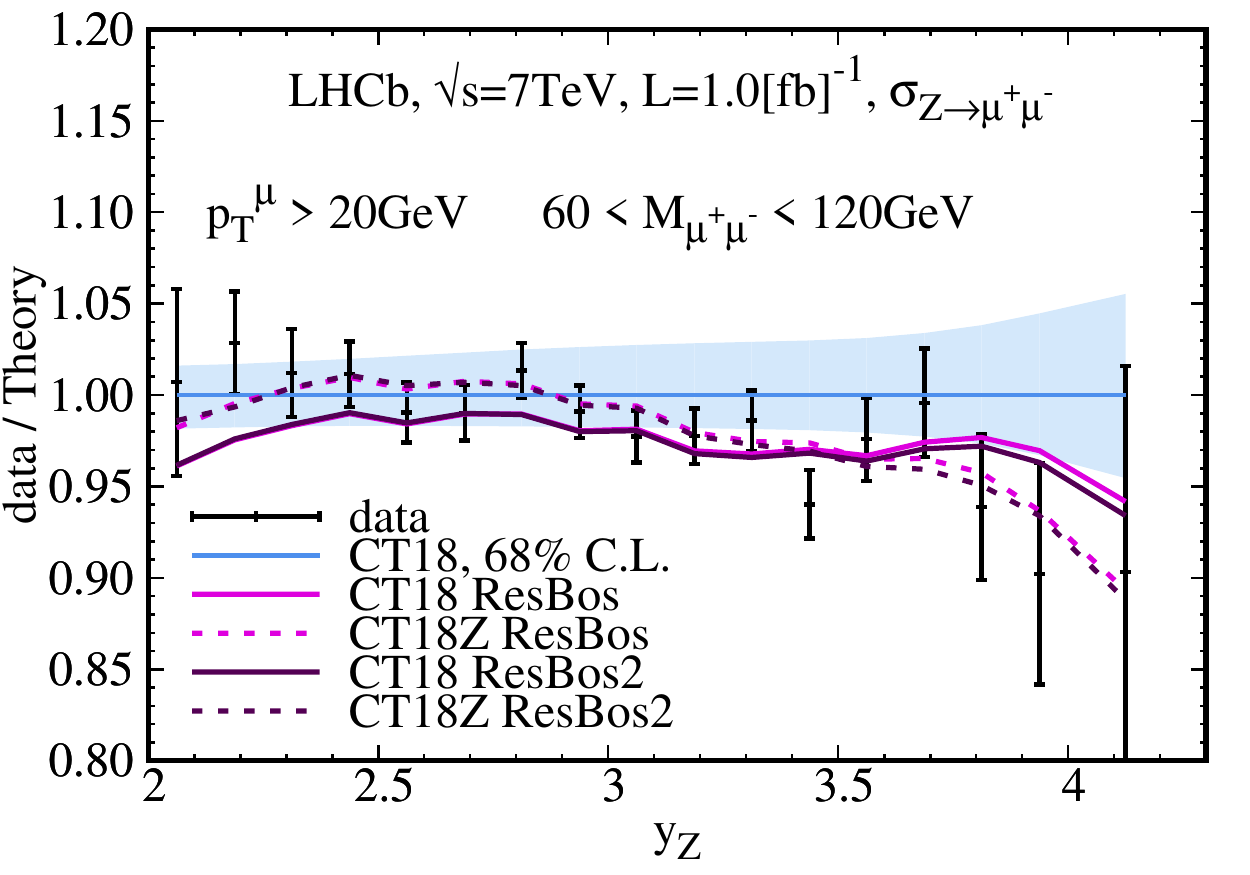}
	\includegraphics[width=0.49\textwidth]{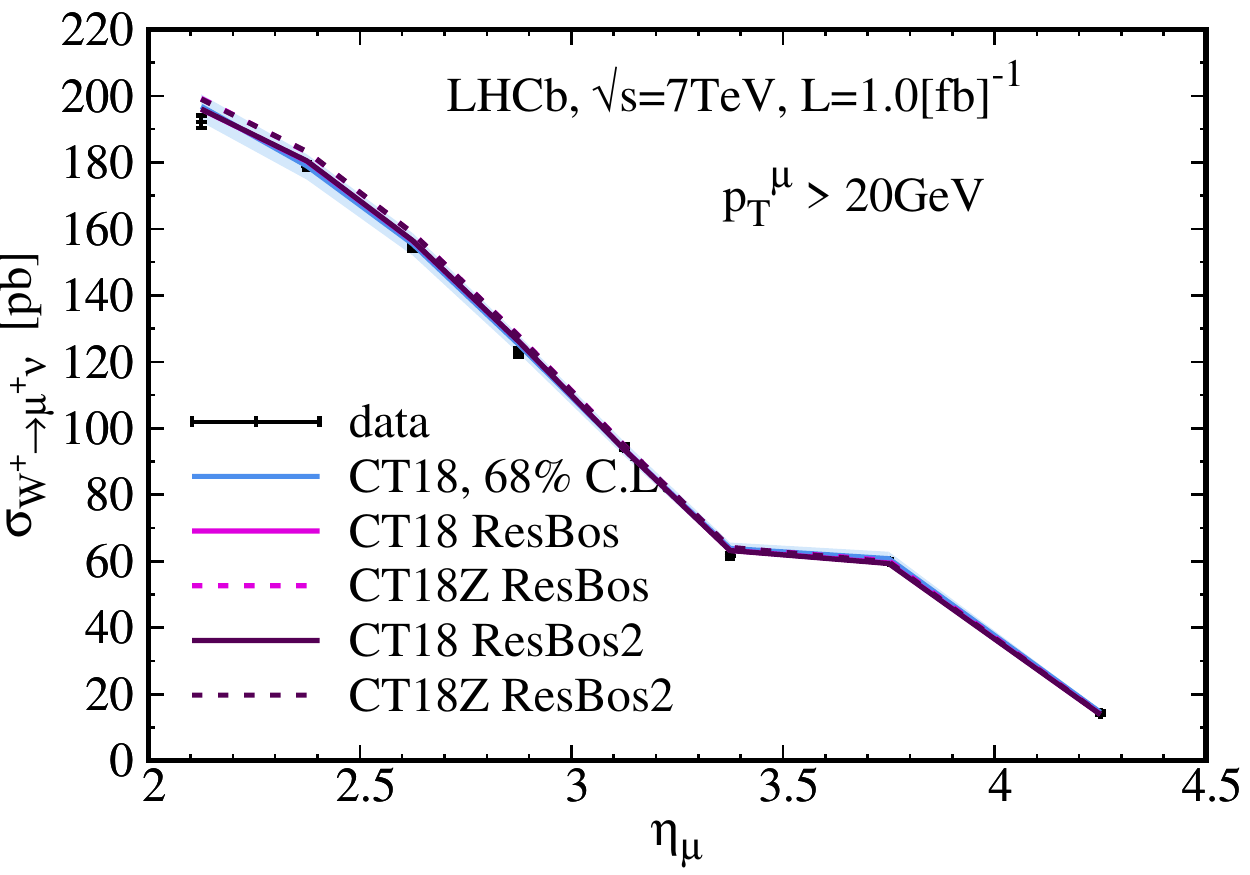}
	\includegraphics[width=0.49\textwidth]{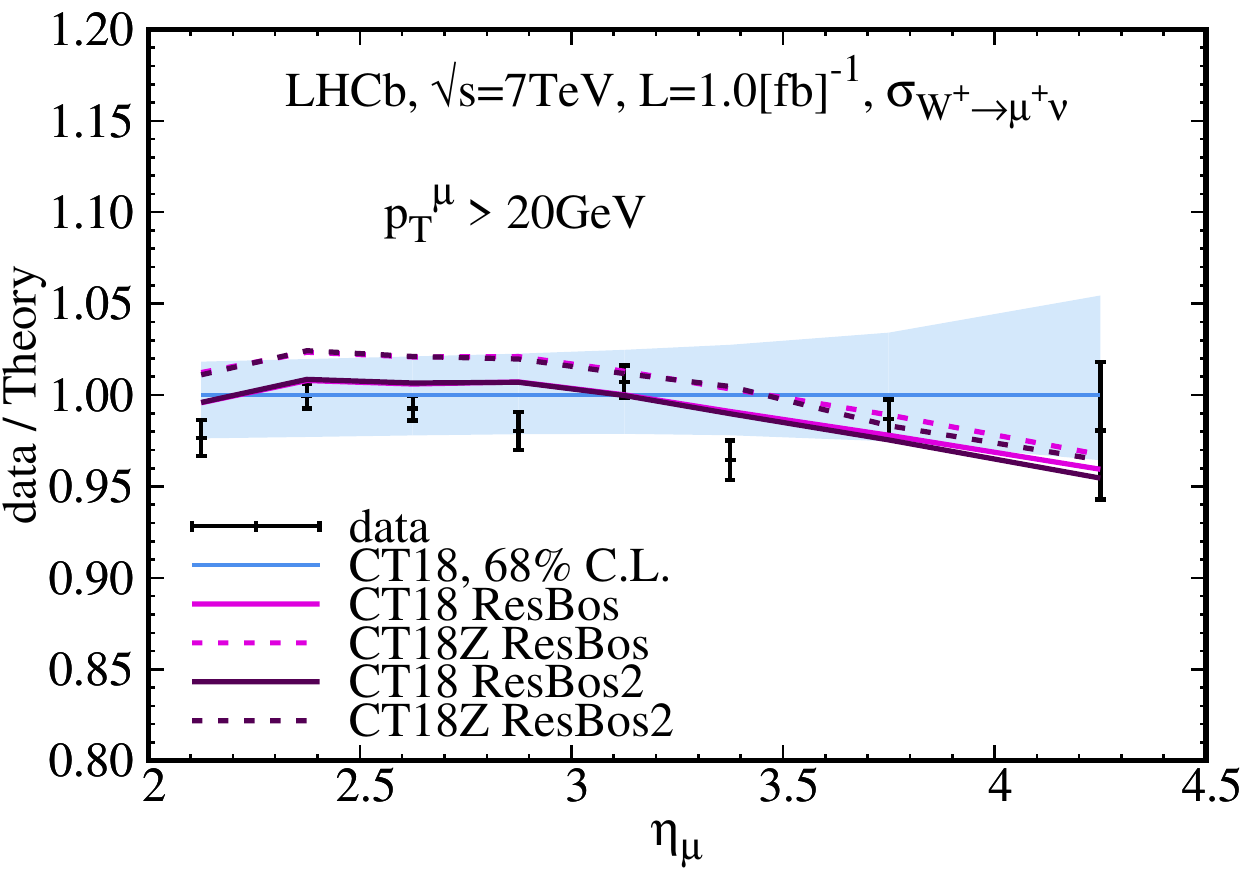}
	\includegraphics[width=0.49\textwidth]{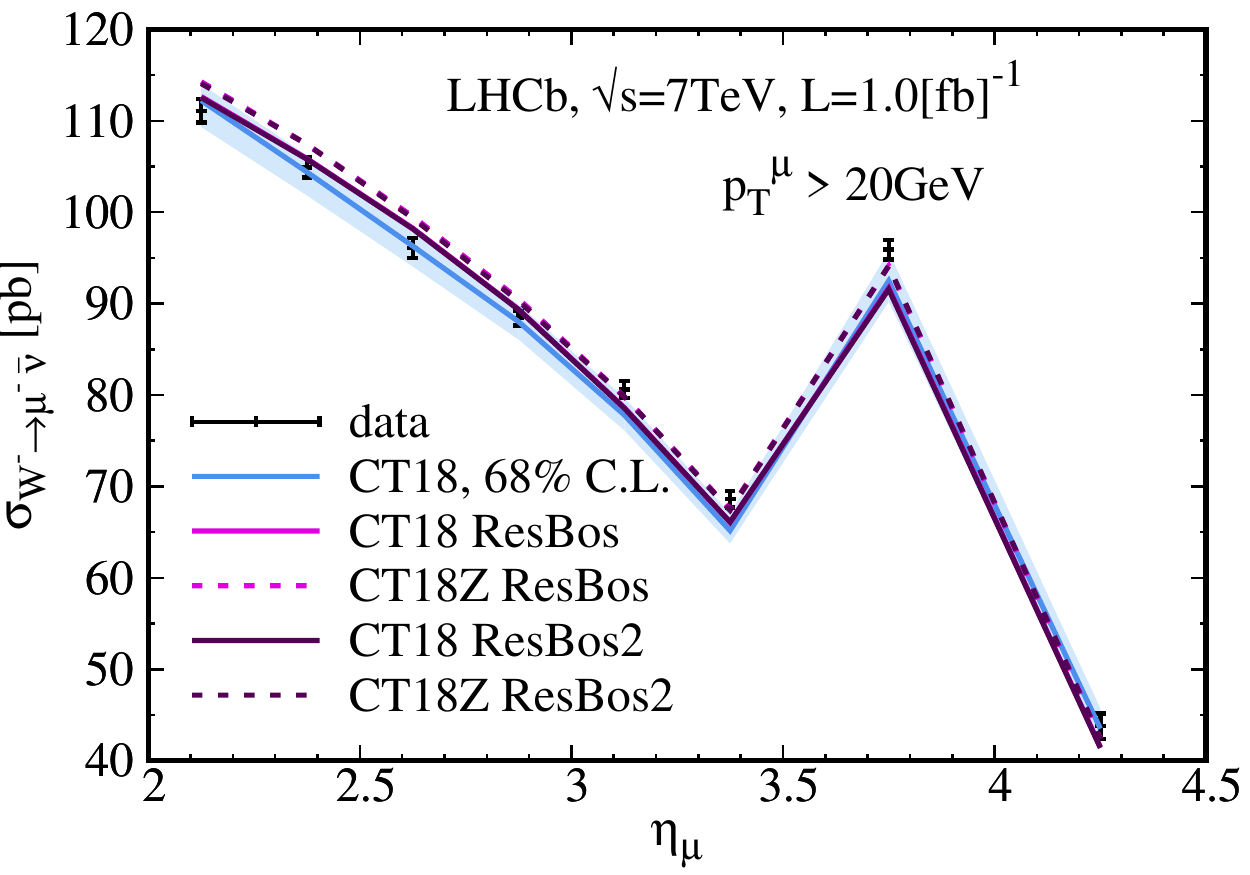}
	\includegraphics[width=0.49\textwidth]{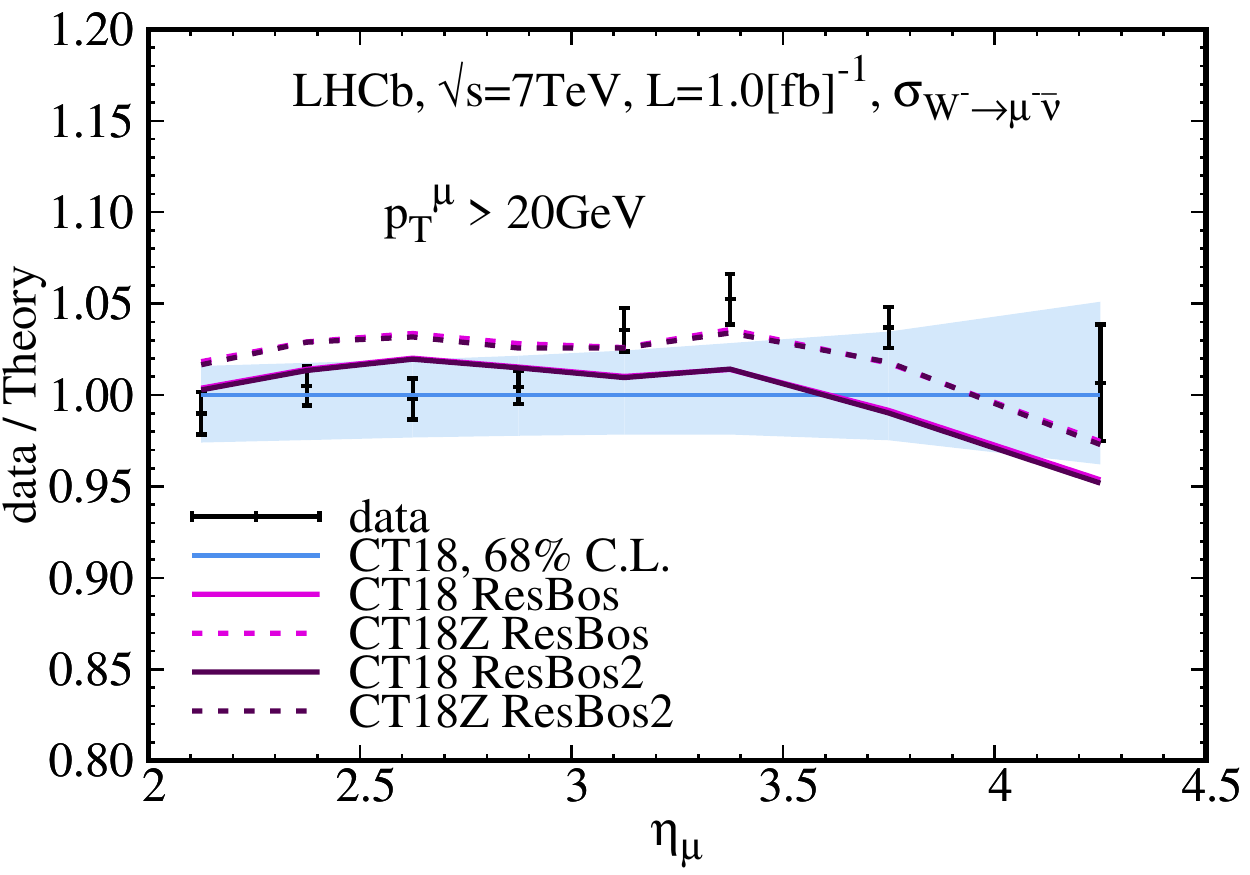}
	\caption{		
Comparison of the LHCb 7 TeV  $W$ and $Z$  data to CT18 predictions, with either NNLO (labeled as CT18), ResBos (labeled by CT18 ResBos) or ResBos2 (labeled by CT18 ResBos2) calculations. The prediction of CT18Z NNLO is also shown.
		\label{fig:245_res}}
\end{figure}

As described previously, NNLO calculations have been used formerly to predict vector boson production data at both the Tevatron and LHC. 
In the past, we have also compared this type of precision data to ResBos predictions, which include effects from multi-gluon emission \cite{Balazs:1997xd}, to produce the CTEQ6.6, CT10 and CT14 PDFs.  
For this reason, it is important to compare vector boson differential cross section measurements to predictions based on the CT18 (and CT18Z) PDFs with ResBos and NNLO calculations.
As an example, we compare the ResBos predictions to the LHCb 7 TeV $W$ and $Z$ boson differential distributions~\cite{Aaij:2015gna} in Fig.~\ref{fig:245_res}. 

For completeness, we have also included in the same figure the predictions from ResBos2, which is an updated version of the ResBos project to include full NNLO corrections, {\it i.e.}, the complete $\alpha_s^2$ contribution for Drell-Yan production of the dilepton pair has been included in this calculation \cite{resbos2-paper}. 
In contrast, the ResBos prediction only contains parts of the NNLO contribution. More specifically, it includes only the Wilson coefficient $C^{(1)}$, but not $C^{(2)}$, in the resummation calculation, cf.~Ref.~\cite{Balazs:1997xd}. 
As shown in Fig.~\ref{fig:245_res}, the predictions from ResBos and ResBos2 agree well for the LHCb kinematics, except in the very large rapidity region. 
The difference between the resummed and (NNLO) fixed-order predictions arises from multiple soft-gluon radiation, the effect of which tends to grow in the large-rapidity
region, where it becomes comparable in size to the LHCb 7 TeV experimental errors. Further detailed discussion about the difference between the resummation and fixed-order calculations will be presented elsewhere. 
In order to see how different PDFs might modify these comparisons between theory predictions and the LHCb 7 TeV data, we also present in the same plot predictions based upon the CT18Z PDFs, in which the gluon and sea quark distributions differ from those of CT18.

\subsection{$W$ plus charm-jet production at the LHC}
\label{sec:Wcharm}

\begin{figure}[tb]
	\begin{center}
		\includegraphics[width=0.45\textwidth]{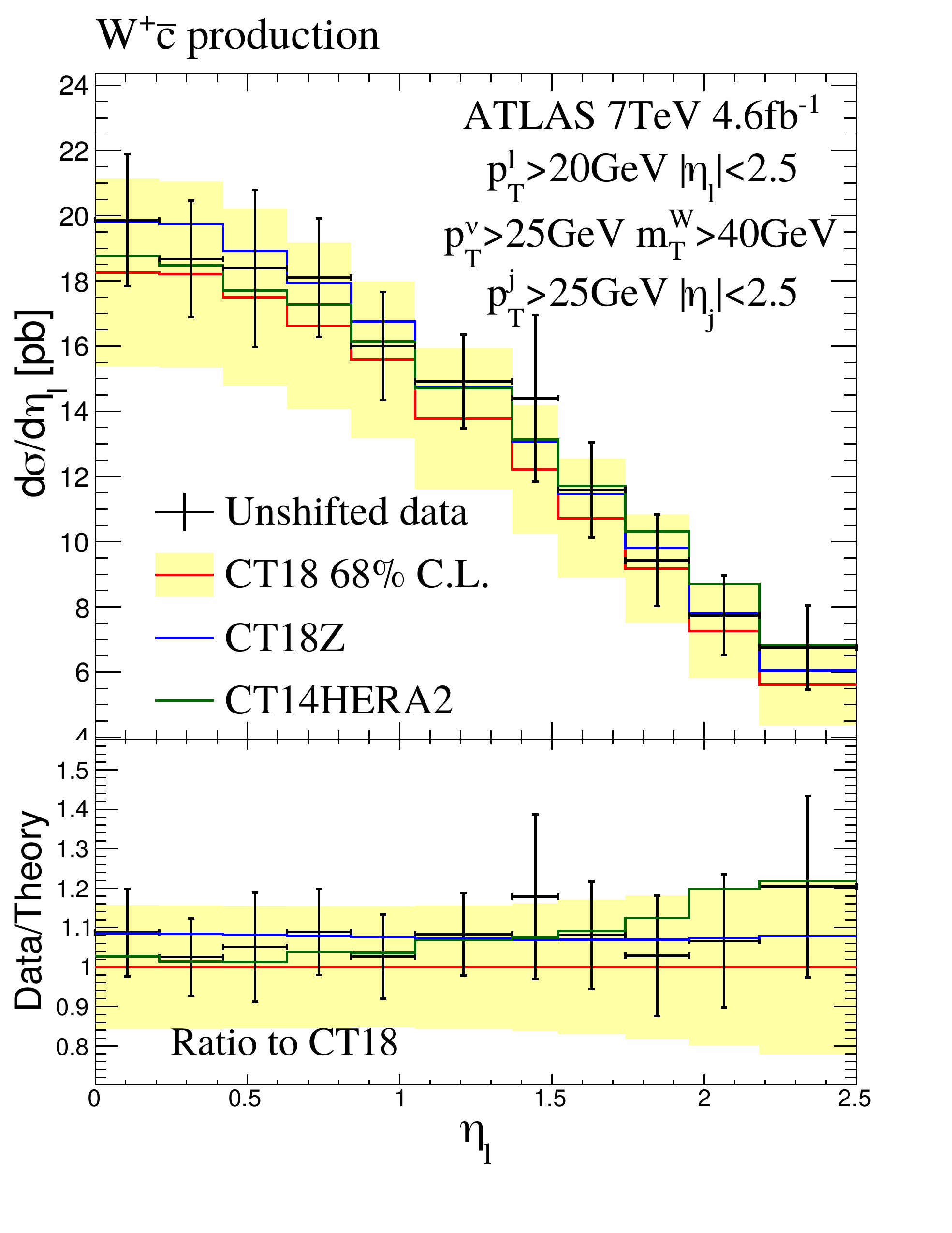}
		\includegraphics[width=0.45\textwidth]{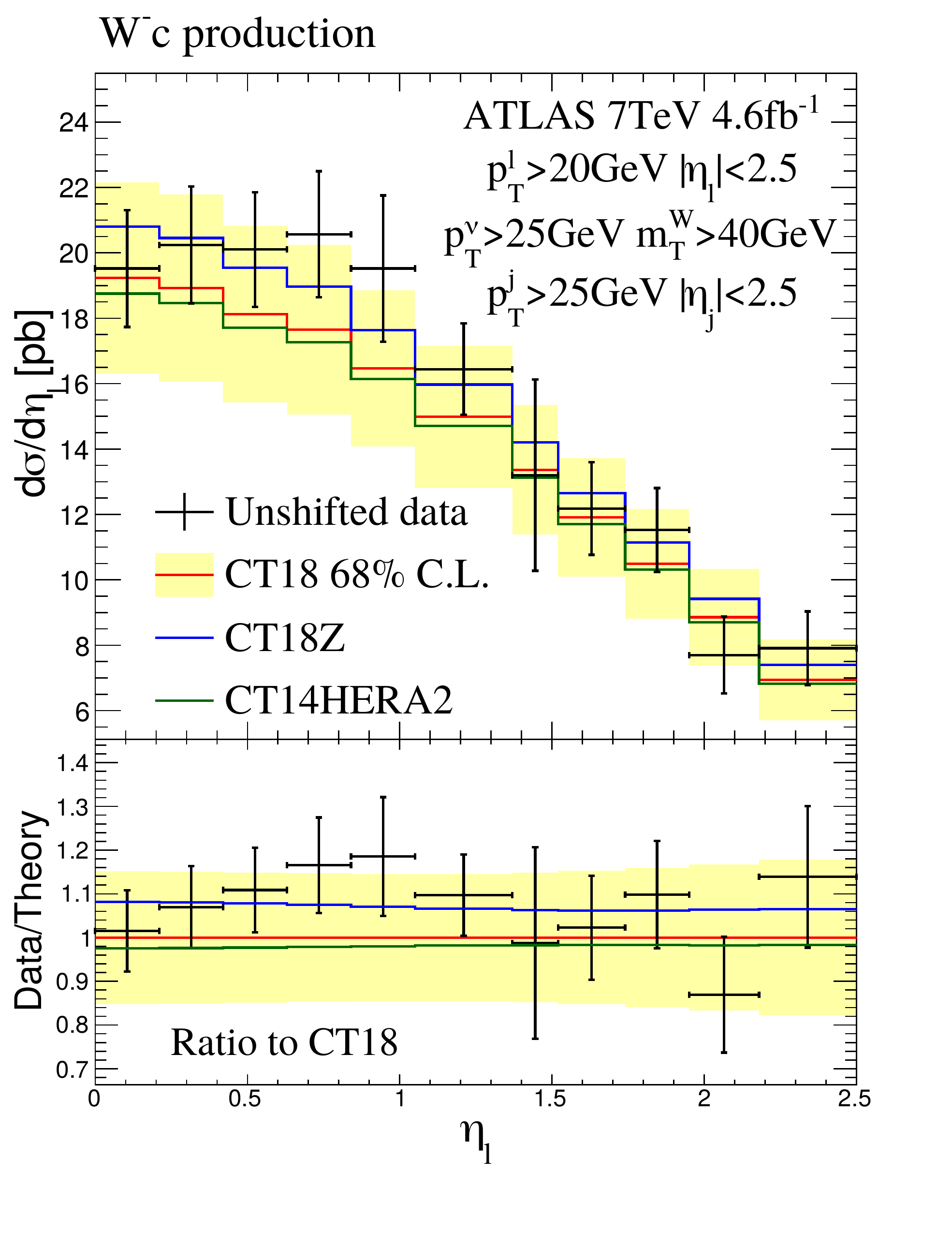}
	\end{center}
	\vspace{-2ex}
	\caption{Comparison of the CT18(Z) and \CTHERAII~NNLO predictions for ATLAS 7 TeV $W^+ + \bar{c}$-jet (left) 
		and $W^-\! +\! c$-jet (right) production,
	respectively, for the combined electron and muon decay channels.		
	The CT18 PDF uncertainty is evaluated at 68\% CL. 
		The scale choice is $\mu_R=\mu_F=M_W$.
		\label{fig:atl7wpcbar}}
\end{figure}

\begin{figure}[tb]
	\begin{center}
		\includegraphics[width=0.45\textwidth]{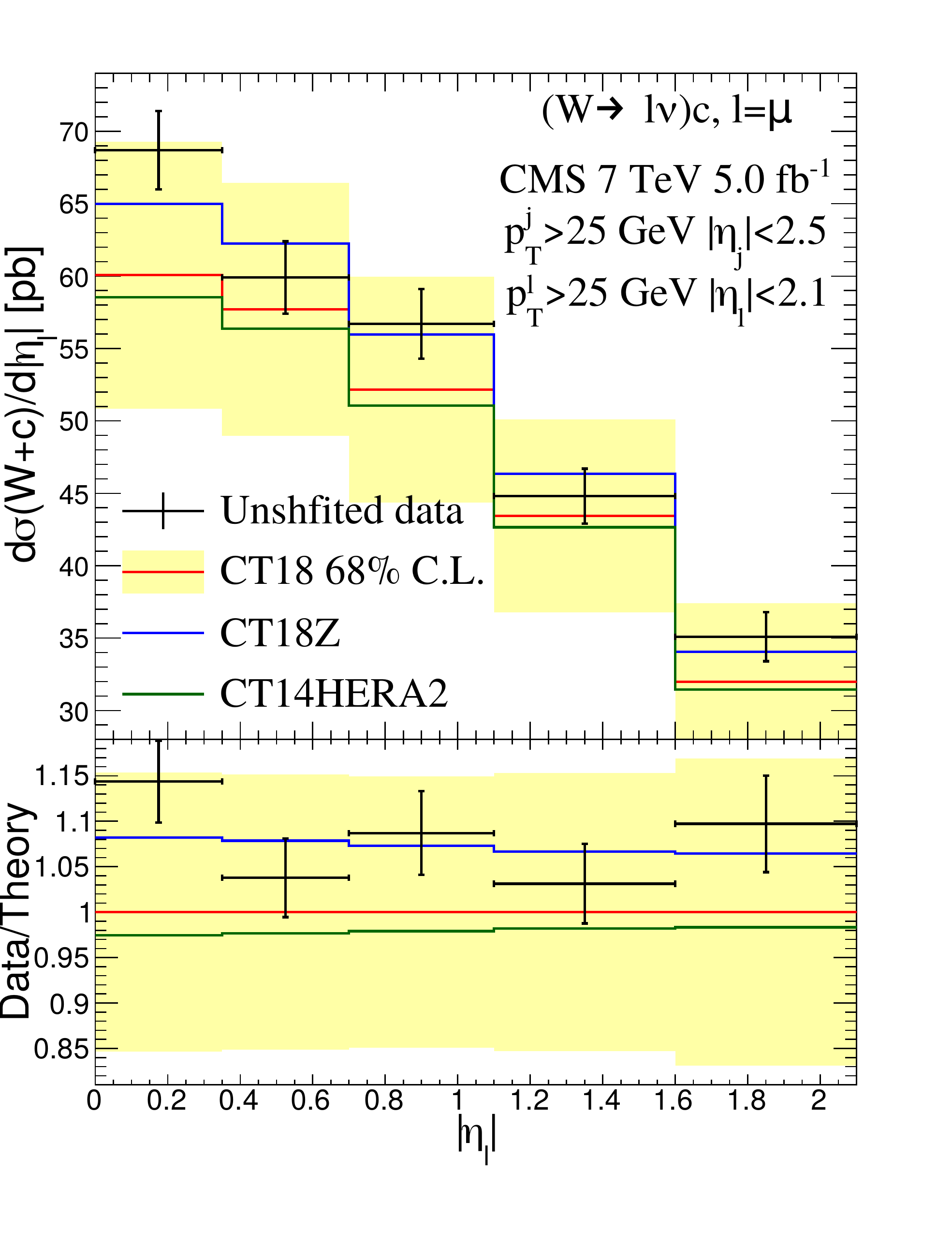}
		\includegraphics[width=0.45\textwidth]{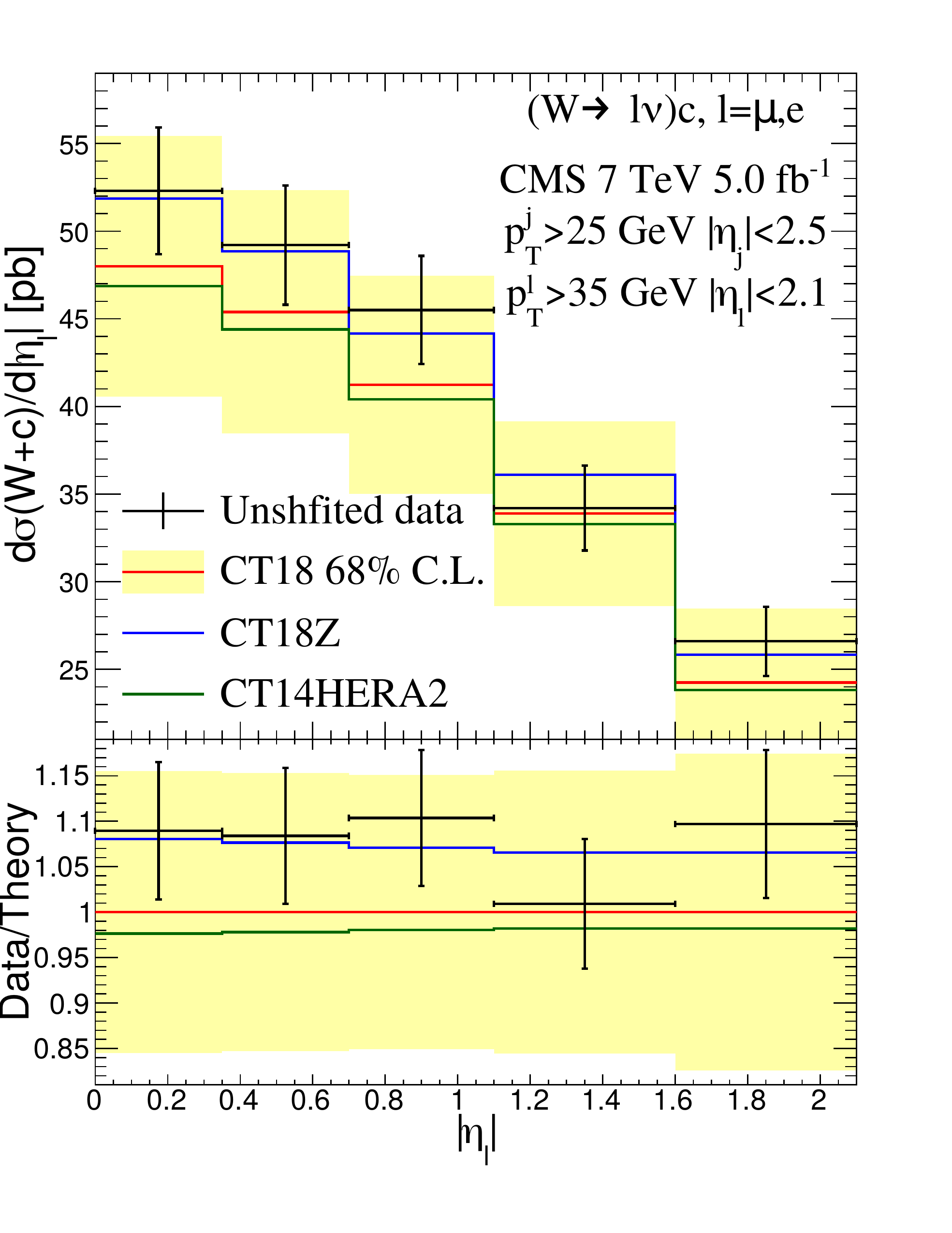}
	\end{center}
	\vspace{-2ex}
	\caption{Comparison of CT18(Z) and CT14$_{\textrm{HERAII}}$ predictions with CMS 7 TeV $W+c$ data, with lepton transverse momentum cut $p_T^l>25$ GeV (left), for the muon channel, and $p_T^l>35$ GeV (right), for the combined electron and muon channels, respectively. 
		The scale choice is $\mu_R=\mu_F=M_W$.
		\label{fig:cms7wpcbar}}
\end{figure}

The $s$-quark PDF of CT18 differs from CT14 at $x < 10^{-1}$ predominantly as a result of the inclusion of the new LHC vector boson production data from  LHCb and ATLAS 7 TeV. Independent constraints on 
the strange quark are provided by the cross sections with significant contributions of  processes initiated by $s$ quarks, such as $W$ plus charm-jet production~\cite{Yalkun:2019gah}. As this process has not yet been calculated to NNLO, the relevant data samples have not been included in the CT18(Z) NNLO PDF fit, but it is still instructive to compare the NLO predictions (but with the NNLO PDFs) with the data. 

Fig.~\ref{fig:atl7wpcbar} compares the \CTHERAII, CT18 and CT18Z predictions with ATLAS 7 TeV $W^+ + \bar{c}$-jet and $W^-\! +\! c$-jet data, respectively. The PDF uncertainty is evaluated at 68\% CL and represented by the yellow bands. The theoretical calculations are performed by using \texttt{APPLgrid} tables generated with \texttt{MCFM}, cross checked against \texttt{MadGraph\_aMC@NLO}+\texttt{aMCfast}. 
The scale choice for this calculation is $\mu_R=\mu_F=M_W$ and the running of $\alpha_s$ is at NNLO as provided by the LHAPDF tables used together with \texttt{APPLgrid}.
The $\chi^2/N_{pt}$ values are 0.59, 0.52 and 0.41, respectively, with the \CTHERAII, CT18 and CT18Z PDFs, for the total of $N_{pt}=22$ data points.
We observe an upward shift in the predictions based upon CT18Z compared with CT18 in both panels of Fig.~\ref{fig:atl7wpcbar} for $W^+$ and $W^-$. 

Interestingly, for $W^+\bar{c}$ production, the 
\CTHERAII~predictions tend to be even larger than those of CT18(Z) in the large-rapidity region, while for $W^-c$ production in the
right panel of Fig.~\ref{fig:atl7wpcbar}, the \CTHERAII~predictions lie well below the CT18(Z) predictions over the full plotted range. This 
nuanced behavior of the large-rapidity $W^+c$ cross section reflects not only the increase of $s$-quark PDFs in CT18Z, but also some compensating changes in $g$, $d$ and other PDFs that occur at large $x$ and have been independently verified by updating the \CTHERAII\, PDFs using $W+c$ cross sections with the \texttt{ePump} program. For completeness, we also show the similar comparison to CMS 7 TeV $W+c$ data in Fig.~\ref{fig:cms7wpcbar}.

\begin{figure}[tb]
\begin{center}
\includegraphics[width=8cm]{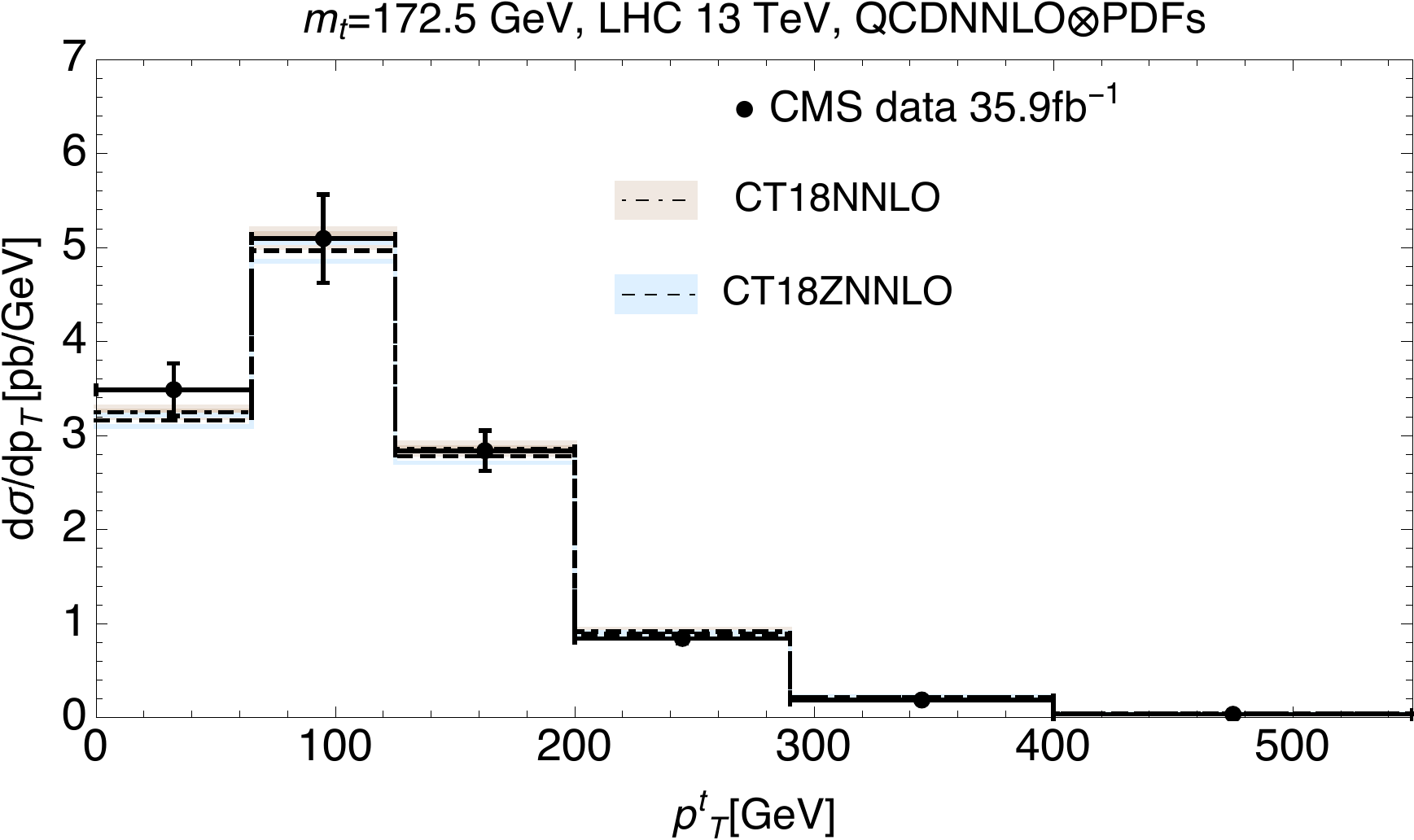}
\includegraphics[width=8cm]{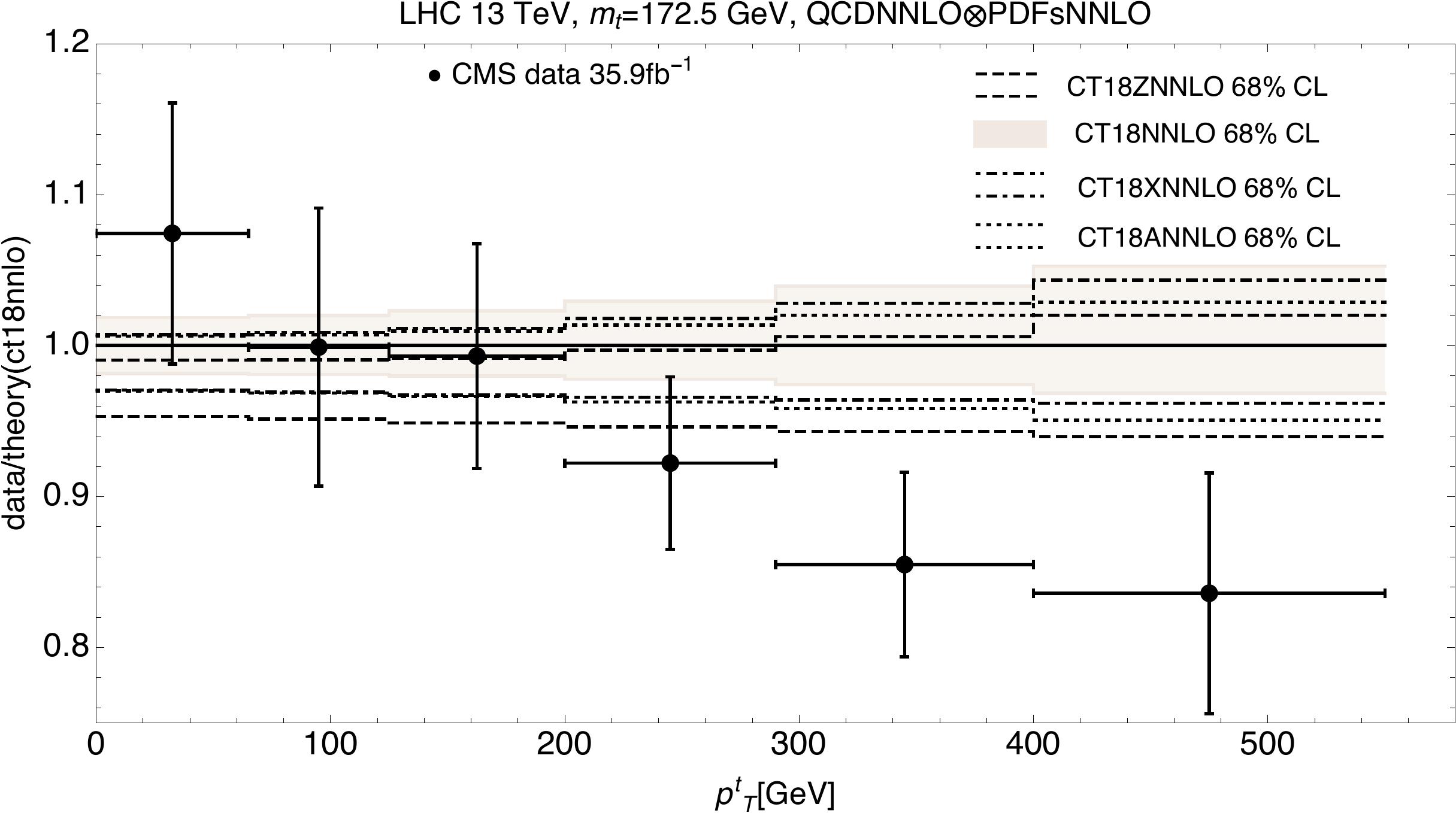}
\caption{Left: Top-quark transverse momentum $p_{T,t}$ distribution. 
Right:  Unshifted data vs. theory plot including CT18Z, CT18X, CT18A NNLO. In the right figure, data and theory predictions are normalized  to CT18NNLO theory.
The error bars indicate statistical and total systematic errors summed in quadrature.}
\label{pTt}
\end{center}
\end{figure}

\subsection{Top quark pair differential distributions at the LHC 13 TeV}
\label{sec:tt13}

\begin{figure}[tb]
\begin{center}
\includegraphics[width=8cm]{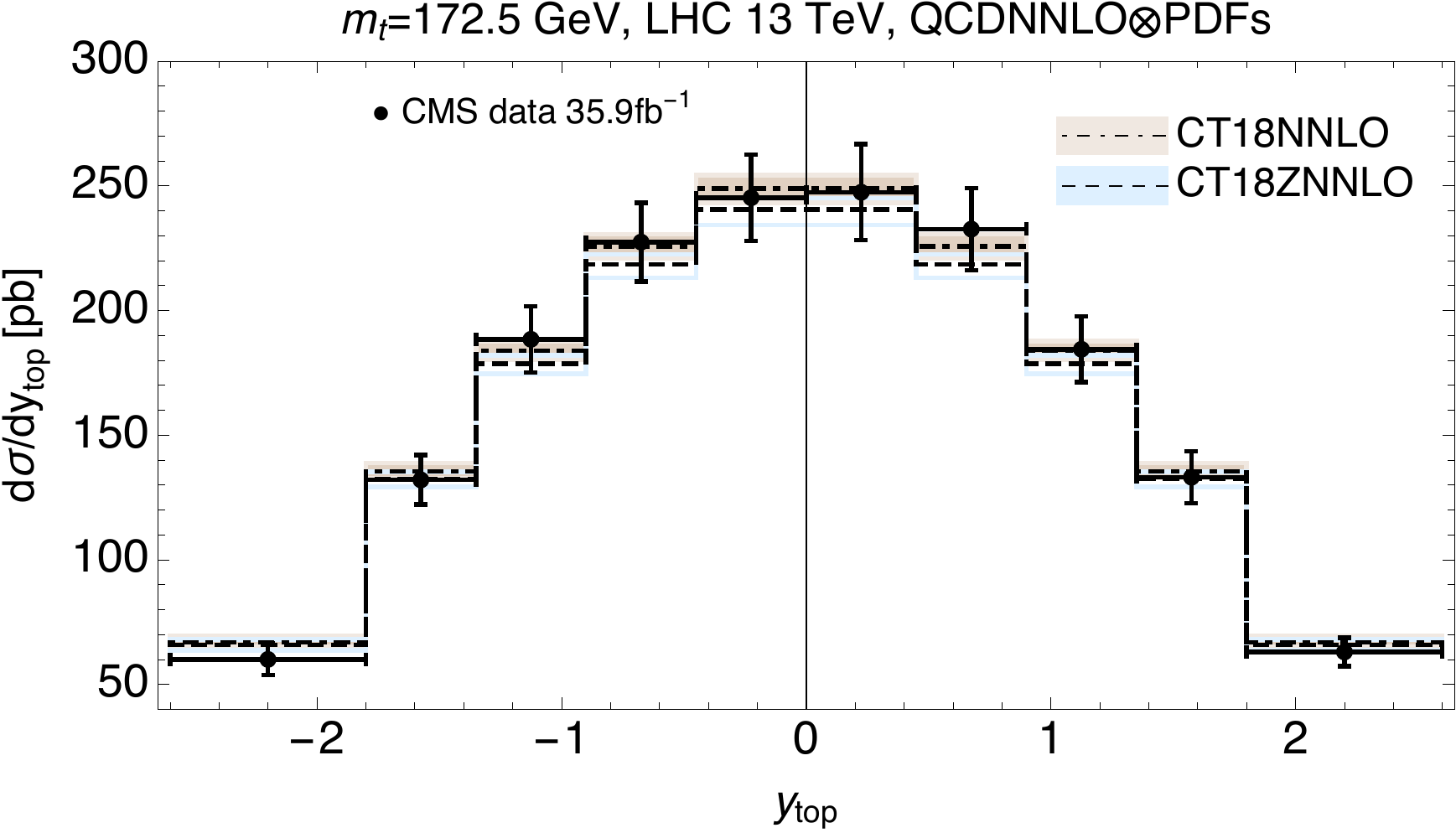}
\includegraphics[width=8cm]{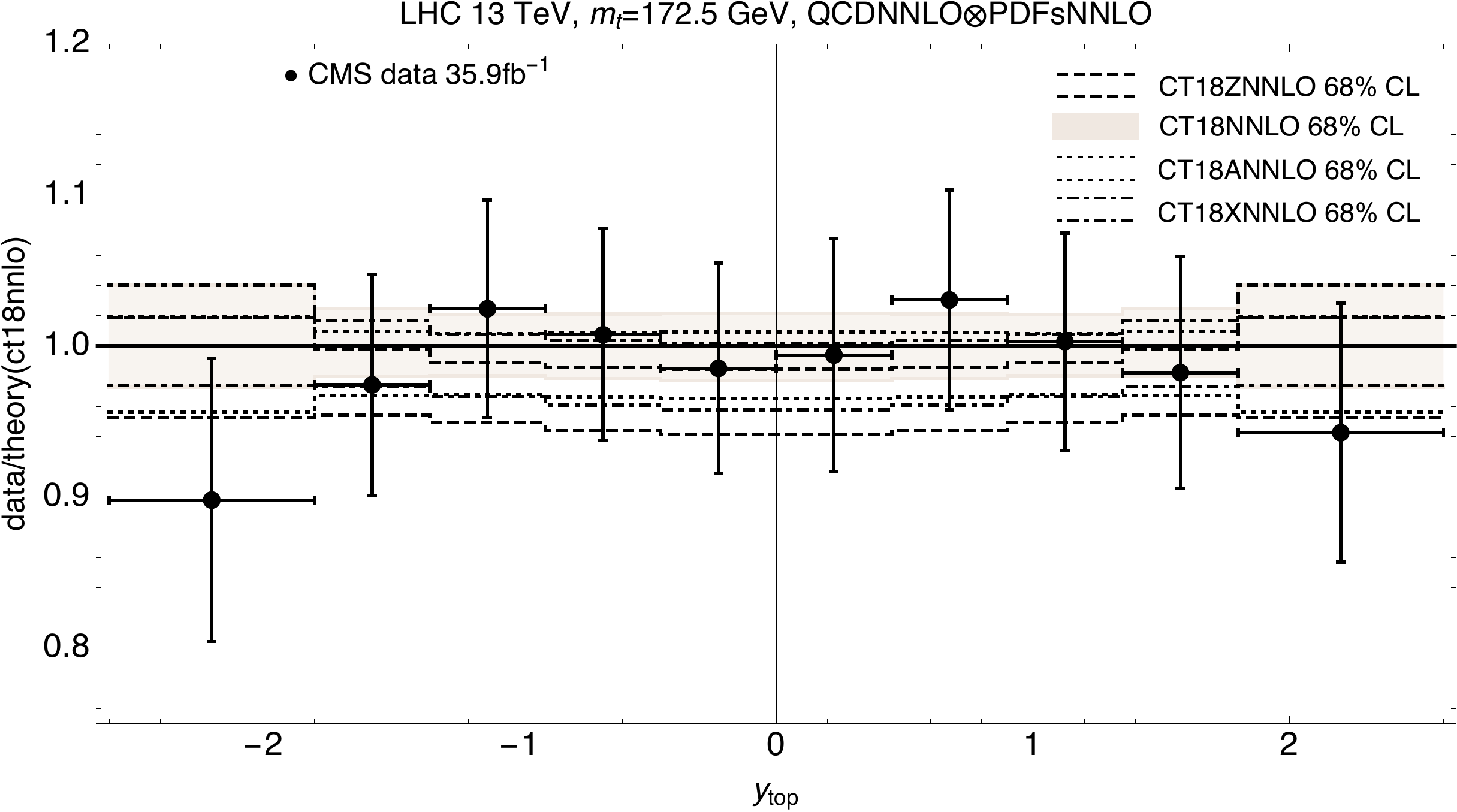}
\caption{Left: Top-quark rapidity $y_t$ distribution. 
Right: Unshifted data vs. theory plot including CT18, CT18Z, CT18X, CT18A NNLO. Data and theory predictions are normalized  to CT18NNLO theory.
The error bars indicate statistical and total systematic errors summed in quadrature.}
\label{yt}
\end{center}
\end{figure}

\begin{figure}[p]
\begin{center}
\includegraphics[width=8cm]{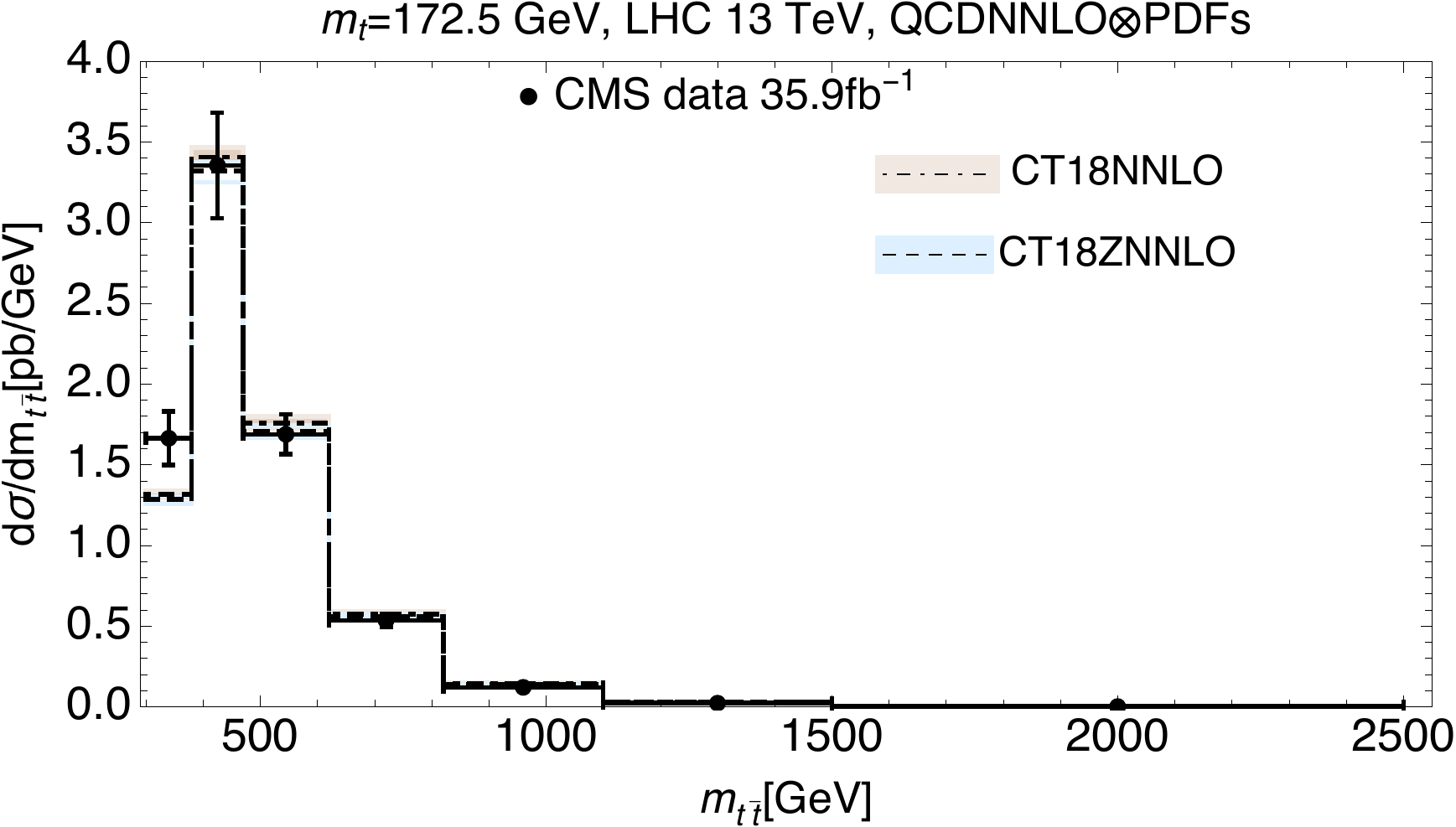}
\includegraphics[width=8cm]{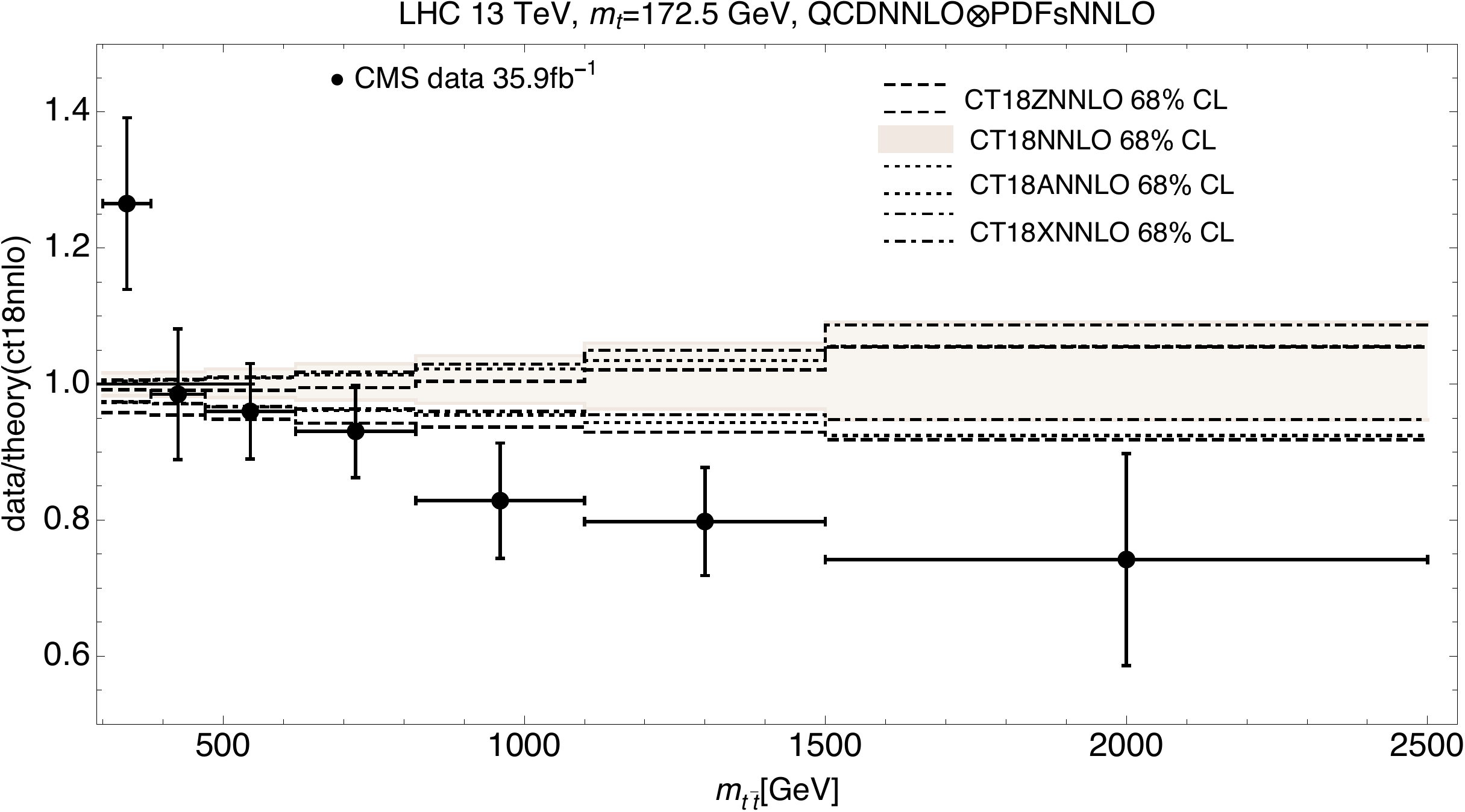}
\caption{Left: invariant mass distribution of the top-quark pair. 
Right: Unshifted data vs. theory plot including CT18, CT18Z, CT18X, CT18A NNLO. Data and theory predictions are normalized  to CT18NNLO theory.
The error bars are statistical and total systematic errors summed in quadrature.}
\label{Mtt}
\end{center}
\end{figure}

In Sec.~\ref{sec:QualityTopData}, we have shown data-to-theory comparisons to the ATLAS and CMS differential top-production data at 8 TeV, i.e., data which were included in the CT18(Z) fits. In this section we present analogous comparisons for the CMS 13 TeV measurement of $t\bar t$ differential cross 
sections in the dilepton channel \cite{Sirunyan:2018ucr}. These data have been released after the CT18(Z) data sets were frozen in the final form. The QCD theoretical predictions at NNLO in QCD are obtained by using \texttt{fastNNLO} tables~\cite{fastnnlo:grids} with CT18 NNLO PDFs. The value of the top-quark mass used to obtain the theory predictions in this case is $m_t^{\rm pole}=172.5$ GeV.
We also show the resultant theory predictions using CT18Z, CT18X, and CT18A NNLO PDFs,
with PDF uncertainties for the cross sections shown at the 68\% C.L.
Plots of the distributions and data-vs-theory comparisons are shown in Figs.~\ref{pTt}-\ref{Mtt}. 
In the data-vs-theory plots, all theory predictions are normalized to CT18NNLO theory. 
The error bars represent the quadrature sums of the statistical and total systematic errors.
We observe a clear difference in the slope between the theory and unshifted experimental data for both $d\sigma / dp^t_T$ and $d\sigma / dm_{t\bar{t}}$.
Those differences can be accommodated by systematic error shifts of the data, resulting in a good $\chi^2$ after all uncertainties are taken into account.
We notice that, in the case of the $p_T$ spectrum, the theory prediction obtained with CT18Z NNLO gives a slightly better description of the data at large $p_T$.

\begin{figure}
\begin{center}
\includegraphics[width=8cm]{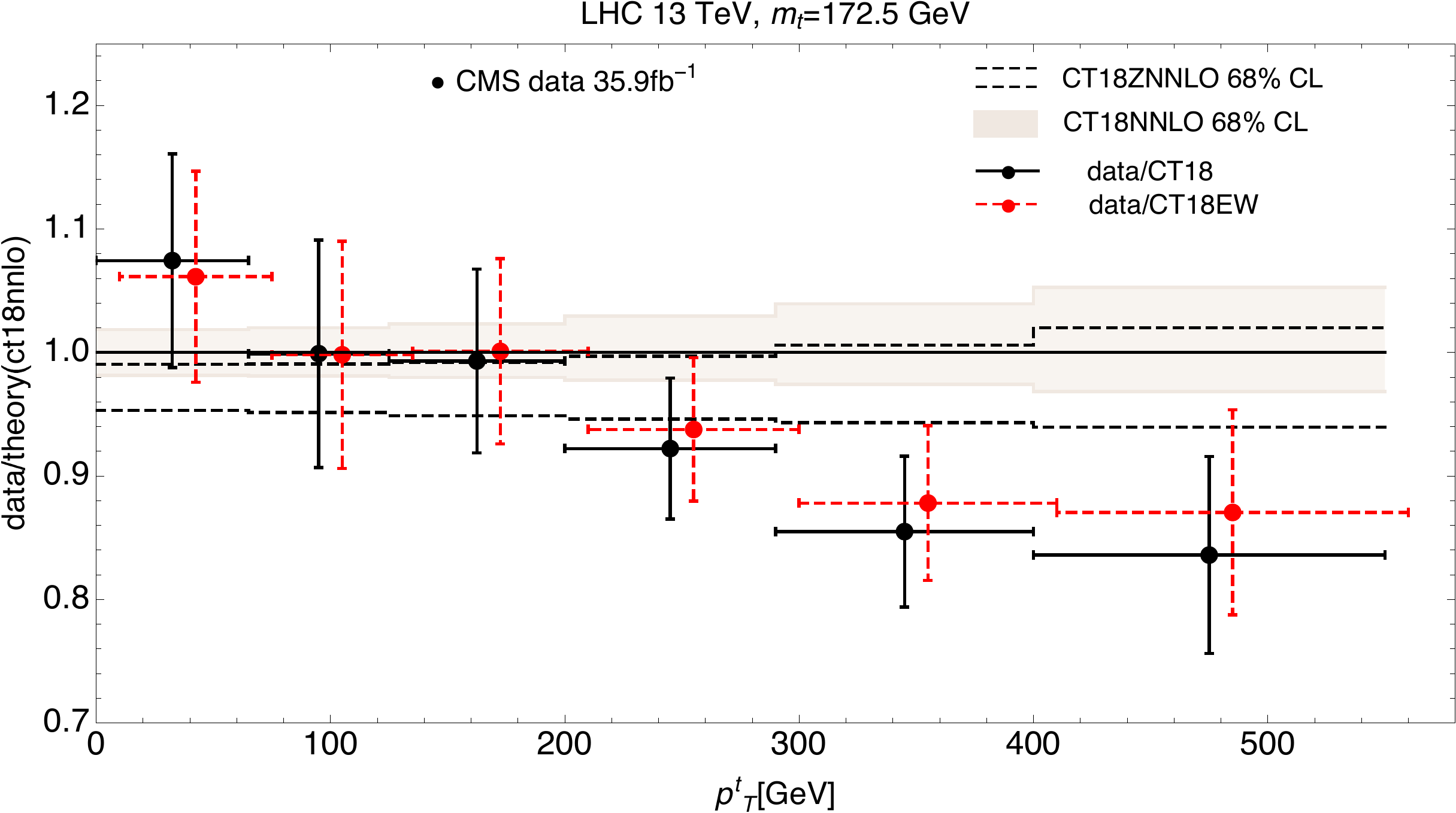}
\includegraphics[width=8cm]{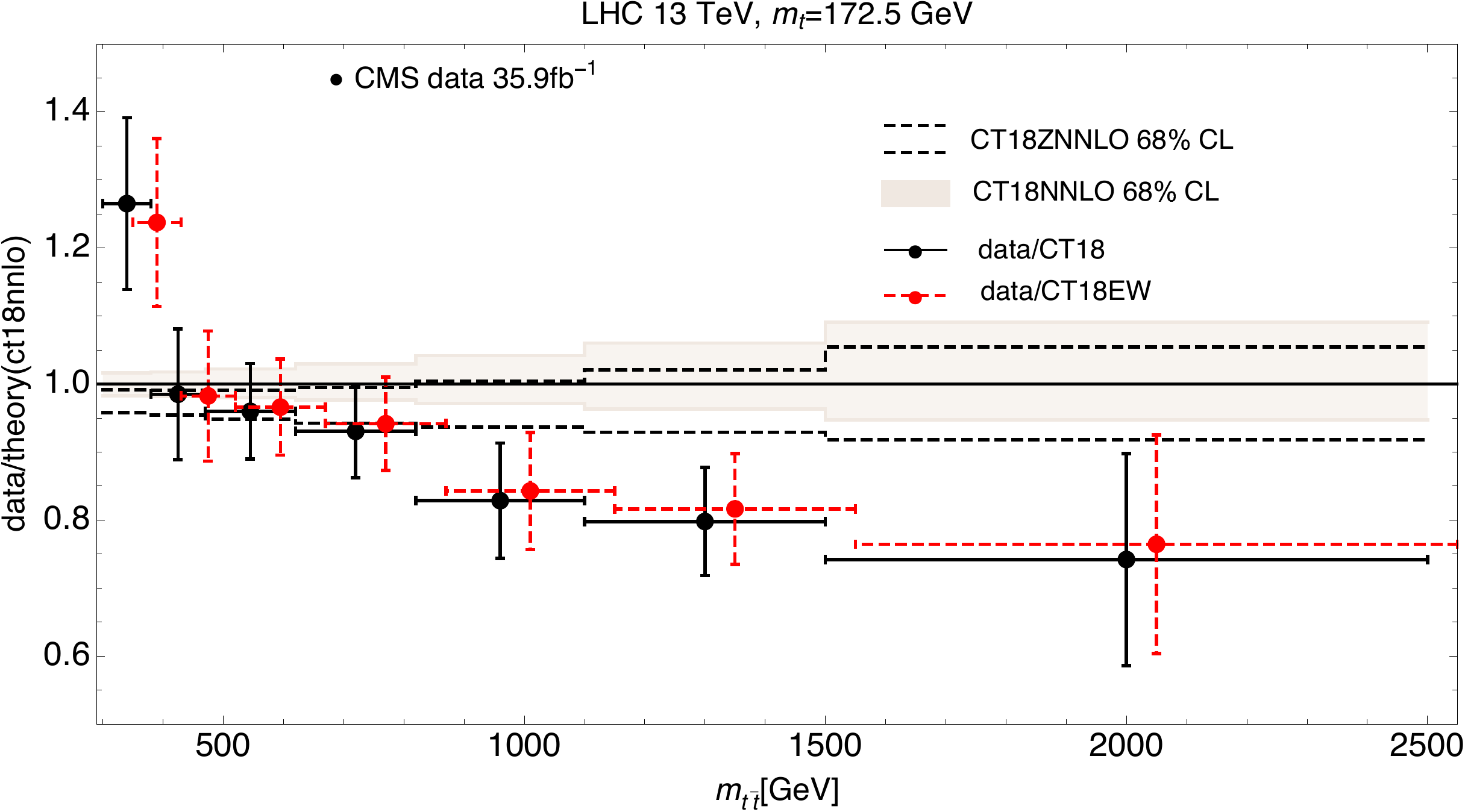}
\caption{Impact of NLO EW corrections. Unshifted data vs. theory plot for top-quark $p_T$ distribution and the invariant mass distribution of the $t\bar{t}$ pair. 
The error bars are statistical and total systematic errors summed in quadrature. The red data points with dashed error bars represent data divided by the CT18NNLO theory with NLO EW corrections. The black data points with solid error bars represent data normalized to the CT18NNLO theory. The CT18ZNNLO theory prediction (black dashed error band) is also normalized to CT18NNLO. The data vs (theory+EW) points are slightly shifted to the right in the same bin to improve visualization.}
\label{MttEW}
\end{center}
\end{figure}

The impact of the electroweak corrections on the CT18 theory is illustrated in Fig.~\ref{MttEW}. These corrections are included as $K$-factors using the multiplicative scheme according to Ref.~\cite{Czakon:2017wor}. They are available at~\cite{EW:kfac}. Large EW effects show up in the high $p_{T}^t$ tails. 
However, in the $p_T$ range 1 - 500 GeV shown in the figures, 
the EW corrections are not larger than 3-4\% in most cases. 
If one considers higher-$p_T$ regions, $K$-factors would be much larger there.
The EW corrections minimally improve the agreement of theory and data for the
top-quark $p_T$ and $m_{t\bar{t}}$ distributions. 
The $\chi^2/N_{pt}$ of the NNLO QCD $+$ NLO EW prediction using CT18 PDFs agrees well with the values presented in Table 49 of Ref.~\cite{Sirunyan:2018ucr}. 
For all other distributions, the EW corrections are negligible for the kinematic ranges studied.
The CT18 global analysis currently includes $\bar{t}t$ differential cross section measurements from ATLAS and CMS at 8 TeV only. The CT18 theory prediction for these distributions in the fit does not include EW corrections. If EW corrections were included in the fit their impact on the fitted PDFs would be negligible due to the size of the EW corrections in the kinematic range of the distributions currently considered.

\begin{figure}[h]
	\includegraphics[width=0.49\textwidth]{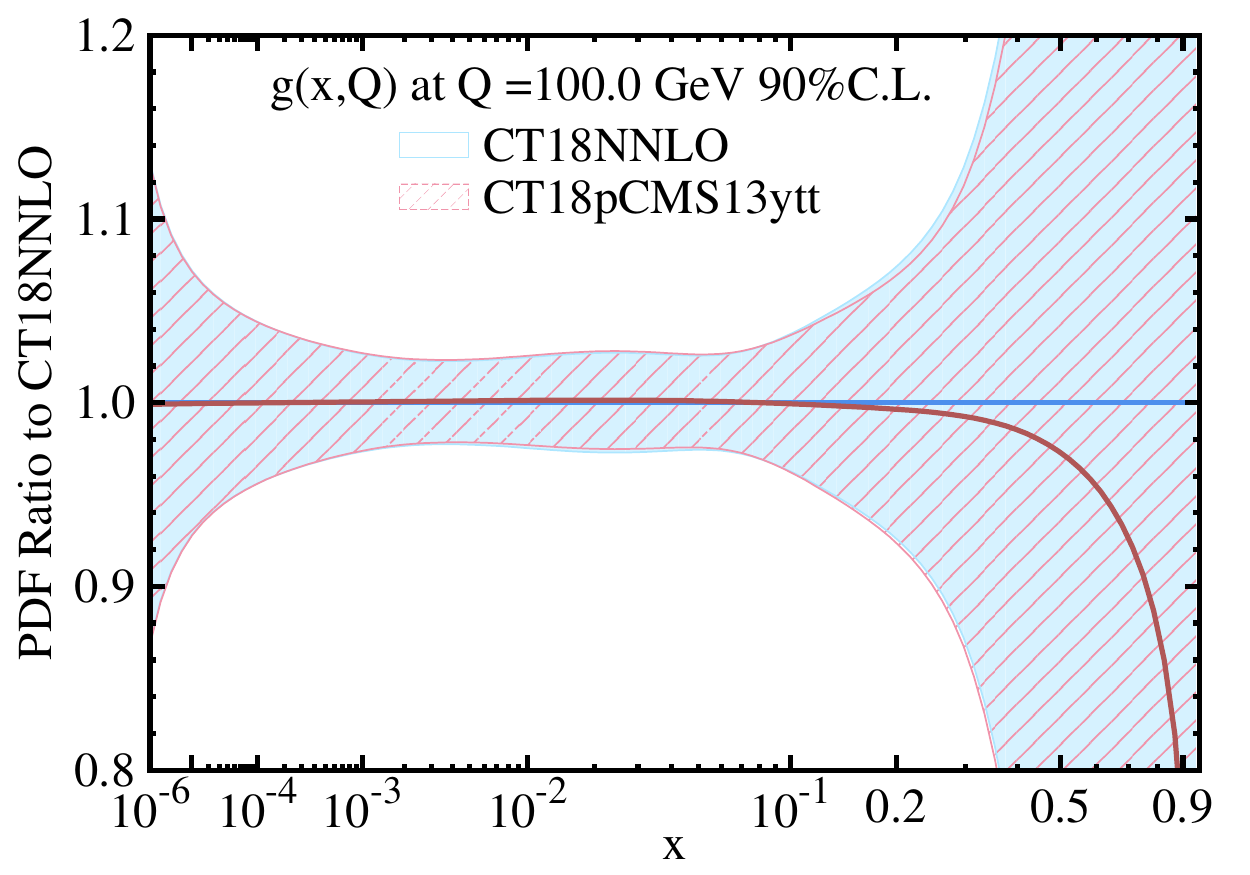}
	\includegraphics[width=0.49\textwidth]{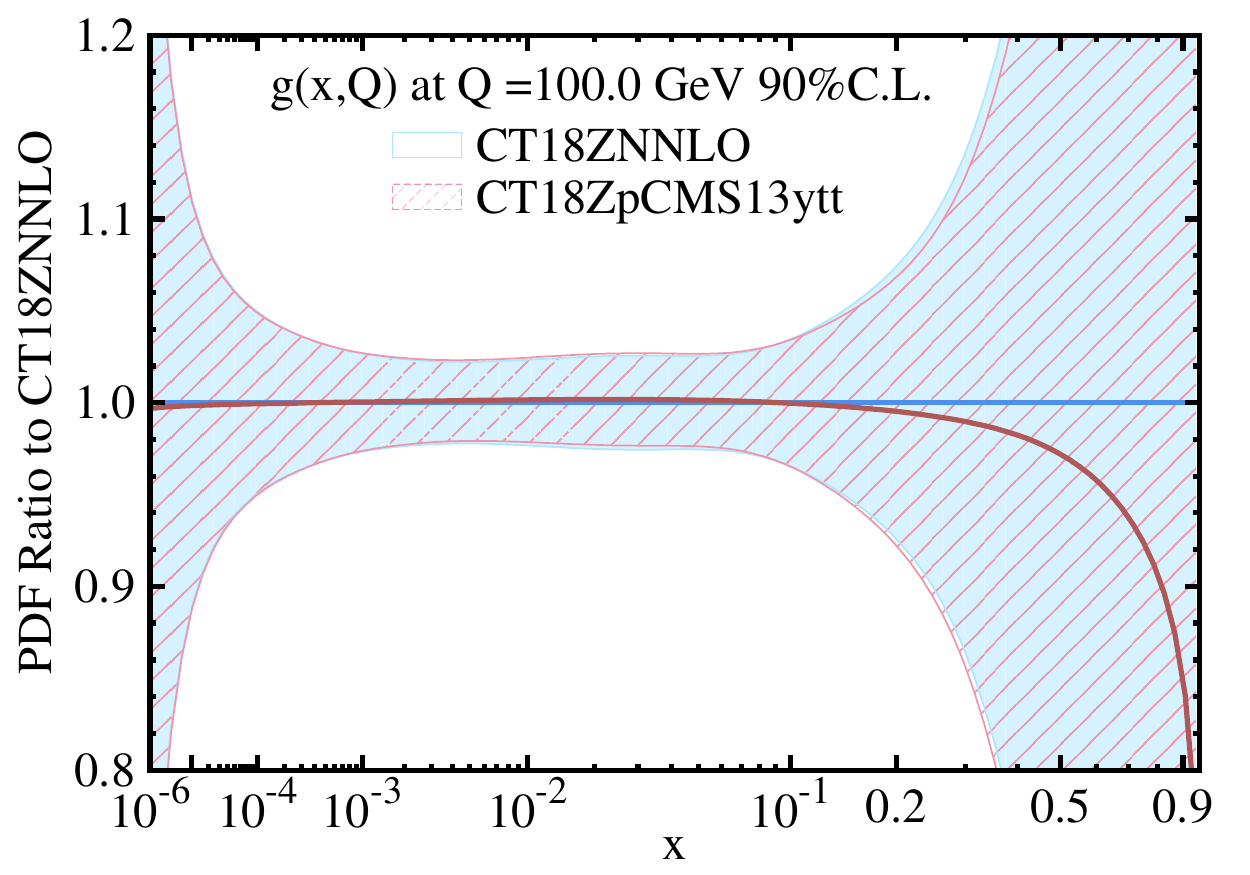}
	\caption{Impact of the $y_{t \bar t}$ differential cross section measurements of the CMS 13 TeV $t \bar t$ data on the CT18 (left) and CT18Z (right) gluon PDFs. 
	\label{fig:cms13ytt}}
\end{figure}

Among various one-dimensional $t \bar{t}$ differential distributions, the distribution of the top-quark pair rapidity, $y_{t \bar t}$, shows a good agreement between the CMS data and CT18 predictions. 
To examine how this data could modify the CT18(Z) gluon PDFs, we use the \texttt{ePump} program~\cite{Hou:2019gfw} to update the CT18(Z) PDFs, after including the CMS 13 TeV $y_{t \bar t}$ data in the fit. 
As shown in Fig.~\ref{fig:cms13ytt}, the updated gluon-PDF error band (labeled as CT18pCMS13ytt) is very slightly reduced for $x$ from 0.1 to 0.4 in both cases.
Further discussion about these data sets will be presented elsewhere.

\subsection{High-$x$ Drell-Yan predictions
\label{sec:SeaQuest}
}
Fixed-target Drell-Yan measurements provide an important probe of the $x$ dependence of the nucleon
(and nuclear) PDFs. This fact has motivated a number of experiments, including the Fermilab E866/NuSea
experiment \cite{Towell:2001nh}, which determined the normalized deuteron-to-proton cross section ratio
$\sigma_{pd} \big/ 2\sigma_{pp}$ out to relatively large $x_2$, the momentum fraction of the target. As can be seen
based upon a leading-order quark-parton model analysis, this ratio is expected to have
especially pronounced sensitivity to the $x$ dependence of the PDF ratio, $\bar{d}/\bar{u}$,
making it a favorable observable for investigations of flavor-symmetry breaking in the light-quark
sea. Breaking of $\mathrm{SU}(2)$ symmetry is understood to have a nonperturbative origin, as noted in
the discussion of the Gottfried Sum Rule in Sec.~\ref{sec:moments}.

\begin{figure}[h]
	\begin{center}
		\includegraphics[width=0.75\textwidth]{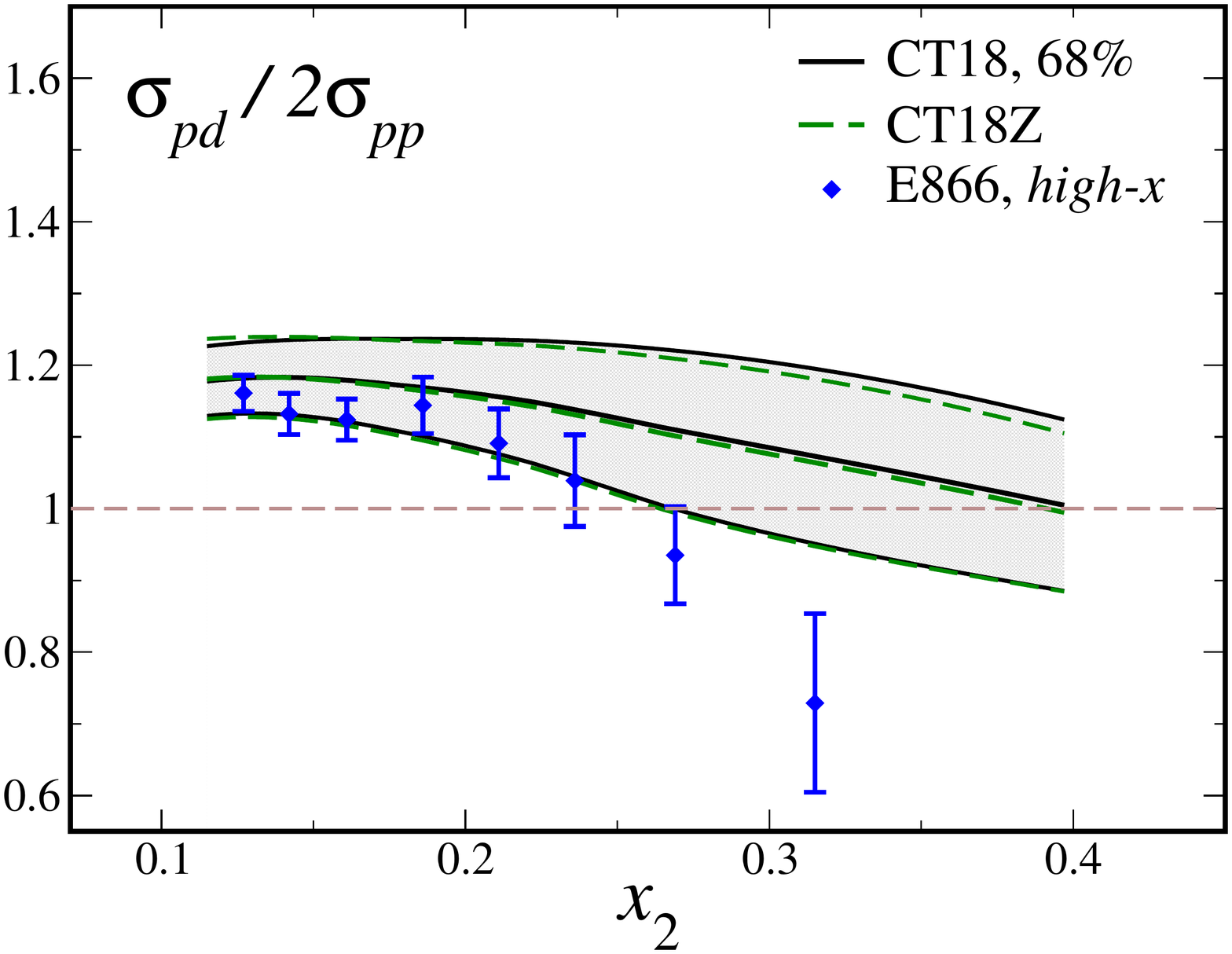}
	\end{center}
	\vspace{-2ex}
	\caption{ Theoretical predictions based on CT18 (black outer band) and CT18Z (Green inner band) for the fixed-target Drell-Yan
	cross section, $\sigma_{pd} \big/ 2\sigma_{pp}$, in the region of larger $x_2 \gtrsim 0.1$ to be probed by the SeaQuest experiment \cite{Aidala:2017ofy} at Fermilab. For comparison, we also plot the higher-$x_2$ portion of the older E866 data \cite{Towell:2001nh} (blue diamonds). }
\label{fig:hix_DY}
\end{figure}

Intriguingly, E866 \cite{Towell:2001nh} found evidence that the cross section ratio dropped below unity, $\sigma_{pd} \big/ 2\sigma_{pp} < 1$,
as $x_2$ approached and exceeded $x \gtrsim 0.25$, as seen by the higher $x_2$ portion of the E866 ratio points shown
in Fig.~\ref{fig:hix_DY}. This fact was surprising on the grounds of a number of theoretical models.
The E866 results therefore stimulated an interest in performing a similar measurement out to larger $x_2$ with higher
precision --- the main objective of the subsequent SeaQuest/E906 experiment at Fermilab \cite{Aidala:2017ofy}, from which
results are expected soon. For this reason, we illustrate in Fig.~\ref{fig:hix_DY} theoretical predictions based upon our
updated CT18 (black band) and CT18Z (green band) global analyses at the 68\% C.L.~to higher $x_2$ beyond that probed by E866.  While CT18
and CT18Z are constrained to the E866 ratio data, the theoretical prediction for
the deuteron-to-proton ratio remains above or consistent with unity out to
$x\! <\! 0.4$.  More precision data in the high-$x$ region will be instrumental in
resolving the behavior of the cross-section ratio and its implications for the
nucleon sea.
 
\clearpage

\section{Discussion and conclusions
\label{sec:Conclusions}
}

In this paper, we have presented the CT18 family of parton distribution functions (PDFs), including the CT18Z, CT18A, and CT18X alternative fits. CT18 is the next generation of NNLO (as well as NLO)
PDFs of the proton from a global analysis by the CTEQ-TEA group. It represents the next update following the release of the CT14 and \CTHERAII~NNLO distributions, the latter of which was prompted
by the release of precision HERA I and II combined data after the publication of CT14. 
CT18 is the nominal CTEQ-TEA PDF set, which we recommend for all general-use applications.  CT18A is the product of adding the ATLAS 7 TeV $W/Z$ data~\cite{Aaboud:2016btc})
into the CT18 fit; CT18X is a variation of CT18 with an $x$-dependent QCD scale for the low-$x$ DIS data (along with a slightly larger charm quark mass value of 1.4 GeV);
and CT18Z contains all the above variations and generally differs most significantly from CT18. The differences between  the CT18 and CT18Z data sets will be non-negligible only for a small range of applications, in which these differences will need to be folded into the total PDF uncertainty, for example, by taking the envelope of the CT18 and CT18Z uncertainties.  CT18A can be used for  more complete examination of the range of uncertainty for the strange quark distribution. Similarly, the possible impact of low $x$ resummation can be explored using CT18X.

Although some of the early 7 and 8 TeV LHC Run-1 data, including measurements of inclusive production of vector bosons
\cite{Aaij:2012vn, Chatrchyan:2013mza, Chatrchyan:2012xt, Aad:2011dm} and jets \cite{Aad:2011fc, Chatrchyan:2012bja}, were included as input for the CT14 fits, CT18 represents the first CT analysis that
substantially includes the most important experimental data from the full Run-1 of the LHC, including measurements of inclusive production of vector bosons, jets, and top quark pairs at 7 and 8 TeV.
Detailed information about the specific data sets included in the CT18 global analysis can be found in Tables~\ref{tab:EXP_1} and~\ref{tab:EXP_2}, with the newly included data in the latter table.
With rapid improvements in the precision of LHC measurements, the focus of the global analysis has shifted toward providing accurate predictions in the wide range of $x$ and $Q$ covered by the LHC data, by making use of the state-of-the-art theory calculations. To achieve this goal requires a long-term multi-pronged effort in
theoretical, experimental, and statistical areas.

The challenge from the side of the experimental information is to select and implement relevant and consistent data sets in the global analysis. Specifically, we have included processes that have a sensitivity for the PDFs of interest, and for which NNLO predictions are available. 
For example, we include as large a rapidity interval for the ATLAS jet data as we can, using the ATLAS decorrelation model, rather than using a single rapidity interval. We noted that using a single
rapidity interval may result in selection bias. The result may be a larger value of $\chi^2 / N_\mathit{pt}$, due to remaining tensions in the ATLAS jet data, as well as reduced PDF sensitivity compared
to the CMS jet data, cf. Sec.~\ref{sec:DataJets}.
Similarly, to incorporate the $t \bar t$ differential cross section measurements into the CT18 global analysis,  
we use two $t \bar t$ single differential observables from ATLAS (using statistical correlations) and doubly-differential measurements from CMS in order to include as much information as possible. Again, there is a risk of bias if we were to use only one differential distribution; however, some of the $t\bar t$ observables are in tension with each other, cf. Sec.~\ref{sec:DataTop}.
The CT18 global analysis shows that previous data sets, included in the CT14 global analyses, continue to have very strong pulls and tend to dilute the impact of new data. 
For example, low-energy DIS and Drell-Yan data, precision HERA data and precise measurements of the electron-charge asymmetry from D\O~ at $9.7\mbox{ fb}^{-1}$
\cite{D0:2014kma} remain important for probing combinations of quark flavors that cannot be resolved by the LHC Run 1 data alone. Furthermore, most experimental measurements contain
substantial correlated systematic uncertainties; we have taken account of these systematic errors in examining the PDF impacts of these measurements.
In addition, we have examined the PDF errors for important LHC processes and have tested the consistency of the Hessian and LM approaches.

The challenges from the theoretical side are threefold: to examine the dependence of theoretical predictions upon QCD-scale choices in comparison with experimental precision; to explore the
impact on the global analysis and
uncertainty in the chosen parametrization forms for the non-perturbative PDFs; and to be able to do fast and accurate theory calculations. 
In the nominal CT18 fits, we have used the canonical
choice of the QCD renormalization and factorization scales, which typically stabilize higher-order theoretical corrections. Fits with alternative scale choices were considered when studying the PDF uncertainty, cf. Sec.~\ref{sec:FinalPDFuncertainty}.

For the CT18 NNLO PDFs in particular, we have consistently applied NNLO calculations to precision DIS, Drell-Yan, jet and $t \bar t$ processes, cf.~Sec.~\ref{sec:Theory}. The specific QCD-scale choices
we take for various processes are listed in Tables~\ref{Theory-Calc-II} and~\ref{Theory-Calc-VB}.
For example, a non-negligible difference was found at low-jet transverse momentum between theory predictions at NNLO using as the momentum-scale choice either the inclusive jet or the
leading-jet transverse momentum~\cite{Currie:2018xkj}. The nominal choice adopted by the CTEQ-TEA group is to use the inclusive-jet $p_T$. 
We have observed that the fitted gluon PDF is not very sensitive to this choice even in kinematic regions where the difference in predictions between these two scale choices is important.
 This resilience in the global 
fit is due to the presence of other data constraining the gluon PDF in 
the relevant kinematic region and possibly due to the compensating effects from sizable systematic uncertainties.
To compare with the high precision data at the
LHC, electroweak corrections must also be included in theory predictions. Details can be found in Sec.~\ref{sec:calcs}, cf.~Table~\ref{tab:EWcorrections}.

To examine the dependence of the fits upon the non-perturbative functional forms chosen for the PDFs at the evolution starting scale $Q_0$ (around 1.3 GeV), we have sampled a large,
$\mathcal{O}(250)$, collection of candidate fitting forms, all having a comparable number of fitting parameters. (More flexible parametrizations are used to better capture variations in
the PDFs' $x$ dependence, cf.~Appendix \ref{sec:AppendixParam}.)
The result of this study can be seen in Fig.~\ref{fig:params}.
As we increase the number of fitting parameters in the global analysis,
we typically observe a steady improvement in $\chi^2$; this improvement generally
increases so long as $\lesssim\! 30$ parameters are fitted, beyond which fits tend
to destabilize as expanded parametrizations attempt to describe statistical noise.

In order to perform the CT18 global fits at NNLO for comparison with precision data whose {\it per datum} statistical error can be as small as 0.1\%, 
we require fast theory calculations with high numerical precision. 
Hence, the usage of various fast interfaces on the calculations of structure functions and cross sections becomes mandatory and conventional.
For that, we have internally developed fast ApplGrid and FastNLO calculations at NNLO accuracy in the QCD interaction, cf.~Sec.~\ref{sec:Theory}.
In addition, we have also parallelized our global-fitting algorithms to facilitate greatly accelerated convergence times, as discussed in Appendix~\ref{sec:AppendixCodeDevelopment}.

The experimental collaborations at the LHC have succeeded in taking copious high-precision data. To examine
the agreement with these precision data calls for advances in statistical methodology. 
Which of the eligible LHC experiments provide promising constraints on the CTEQ-TEA PDFs?
Do the LHC experiments agree among themselves and with other experiments?
A consistent answer emerges from a powerful combination of four methods:
1) \texttt{PDFSense} and $L_2$ sensitivity,
2) the \texttt{ePump} program, 
3) Effective Gaussian variables,
and 4) LM scans. 
While the last two methods had been introduced in the previous CTEQ-TEA global analysis, such as CTEQ6, CT10 and CT14, the first two techniques were invented in the process of CT18 global analysis. 

The \texttt{PDFSense} program~\cite{Wang:2018heo} provides an easy way to visualize the potential impact of data on PDFs in the $x$ and $Q$ plane. In addition, a simple $L_2$ sensitivity variable \cite{Hobbs:2019gob} is instructive for exploring agreement between different experiments similarly to the LM scans, but using a much faster Hessian formalism across the full range of $x$ or $Q$. See examples in Secs.~\ref{sec:L2} and \ref{sec:AppendixCT18Z}. 

The complementary \texttt{ePump} program~\cite{Schmidt:2018hvu} contains a fast and efficient method to estimate the effect of new data on a set of best-fit and Hessian error PDFs. Extensive validations against the previous CT14 global fits have also been performed~\cite{Hou:2019gfw}.
The application of the above four techniques in the CT18 analysis is illustrated throughout this paper. 

In the four CT18 fits, important impacts are found on PDFs from ATLAS and CMS inclusive jet production measurements, LHCb $W$ and $Z$ vector boson productions and ATLAS $\sqrt{s}=8$ TeV $Z$ boson transverse momentum data. We find contradictory preferences for the strange quark PDF between semi-inclusive (SI) DIS (e.g., NuTeV and CCFR dimuon production) experiments, on one hand, and some LHC experiments, especially ATLAS 7 TeV $W/Z$ production measurements and to some extent LHCb $W/Z$ measurements, on the other hand. Benchmarking of LHC measurements and theoretical predictions, as well as new (SI)DIS experiments can be highly effective for resolving these tensions. Going forward, to facilitate the discovery program of the HL-LHC, a sustained effort to navigate experimental tensions in collider data will be required to achieve the ultimate precision of these planned experiments.  We envision an interplay among theoretical and data-analytical methods (including those
used in this study to explore data compatibility), and additional high-precision experiments such as high-luminosity DIS colliders like the Electron-Ion Collider (EIC) \cite{Accardi:2012qut}, to be indispensable for making such progress.

The inclusion of new data and theoretical advances have resulted in the following changes in CT18, as compared to CT14: 
1) a smaller $g(x,Q)$ for $x \sim 0.3$ (mainly due to ATLAS and CMS 7 TeV jet data and ATLAS 8 TeV $Z$ $p_T$ data, with some tension found between CMS 7 and 8 TeV jet data); 
2) some changes in $u$, $d$, $\bar u$ and $\bar d$ at small $x$, such as a larger $d$ and $d/u$ and a smaller $\bar d / \bar u$ for $x \sim 0.2$ (mainly due to LHCb $W$ and $Z$ rapidity data and CMS 8 TeV $W$ lepton charge asymmetry data); and 
3) a larger $s$ and $(s+\bar s)/(\bar u + \bar d)$ at small $x$ (mainly due to LHCb $W$ and $Z$ rapidity data, and further enhanced by the ATLAS 7 TeV $W/Z$ data in the CT18A and CT18Z fits).
While the sensitivity of an individual $t \bar t$ data point can be similar to that of an individual jet data point at the LHC, the total sensitivity of the $t \bar t$ data is small due to the small number of  $t \bar t$ data points. Hence, we did not find noticeable impact from the double differential distributions of the $t \bar t$ data included in the CT18 analysis. A similar finding was also reported in Ref.~\cite{Hou:2019jxd}, in a CT14 analysis. 

Despite these changes in central predictions, the CT18 NNLO PDFs remain consistent with CT14 NNLO within the respective error bands.
More details about the comparison of CT18 and CT14 PDFs, as well as the quality of the fits to data can be found in Secs.~\ref{sec:OverviewCT18} and \ref{sec:Quality}. 

Some implications of CT18 predictions for phenomenological observables
were reviewed in Sec.~\ref{sec:StandardCandles}.
Compared to calculations with CT14 NNLO, both the $gg \rightarrow H$ and $t\bar t$  total
cross sections have decreased slightly in CT18. The $W$ and $Z$
cross sections, while still consistent with CT14, have slightly increased as a result of enhanced strangeness. Common ratios of
strange and non-strange PDFs for CT14 NNLO, shown in Sec.~\ref{sec:PDFratios}, are consistent with the independent ATLAS and CMS determinations within the PDF uncertainties.

We have also presented the implications of the CT18 global fits for the value of $\alpha_s$, as seen in Sec.~\ref{sec:AlphasDependence}.   
The full CT18 data set prefers, at NNLO, a value of $\alpha_s(M_Z)\! =\! 0.1164\! \pm\! 0.0026$, at 68\% C.L. The corresponding value for CT18Z is basically the same, $0.1169\pm0.0027$.
This is to be compared to the CT14 determination, which included very little LHC data, of $\alpha_s(M_Z)\! =\! 0.115^{+0.006}_{-0.004}$ at 90\% C.L.

The LM scans over the charm quark (pole) mass, $m_c$, as shown in Figs.~\ref{fig:lm_mc} and~\ref{fig:lm_mcZ}, support the usage of 1.3 GeV and 1.4 GeV in the CT18 and CT18Z fits,
respectively. Notably, the combined HERA charm data prefer a somewhat smaller $m_c$ value, while the ATLAS 7 TeV $W/Z$ data in the CT18Z fit prefer a larger $m_c$ value. Comments about the impact of fitted charm contributions on predictions for LHC $W/Z$ cross sections are made at the end of Sec.~\ref{sec:ellipse}.
Comparisons to the parton luminosities and predictions based on the PDFs from other groups can be found in Secs.~\ref{sec:PDFLuminosities}, \ref{sec:moments}, and \ref{sec:ATL7ZWchi2}.

To allow direct comparison to results obtained by the lattice QCD community, we have also presented the CT18 predictions for various PDF moments and sum rules in Sec.~\ref{sec:moments}.
In general, we find good agreement between CT18 and results from other phenomenological fitting efforts for most lattice observables.  At present, systematic
effects are such that many lattice calculations significantly overshoot the predictions of contemporary phenomenology, with the exception of the gluonic moment $\langle x \rangle_g$, which
is underpredicted by the lattice relative to PDF fits.  We expect complementary advances in lattice simulations and PDF phenomenology to improve this situation
in coming years and pave the way for a synergistic PDF-Lattice effort \cite{Hobbs:2019gob,Lin:2017snn} to determine the nucleon's longitudinal structure.

The final CT18 PDFs are presented in the form of 1 central and 58
Hessian eigenvector sets at NNLO and NLO. The 90\% C.L. PDF
uncertainties for physical observables can be estimated from these
sets using the symmetric \cite{Pumplin:2002vw} or asymmetric
\cite{Lai:2010vv,Nadolsky:2001yg} master formulas by adding
contributions from eigenvector pairs in quadrature. These PDFs are
determined for the central QCD coupling of $\alpha_s(M_Z)=0.118$,
consistent with the world-average $\alpha_s$ value. For estimation of
the combined PDF+$\alpha_s$ uncertainty, we provide two additional
best-fit sets for $\alpha_s(M_Z)=0.116$ and 0.120. The
90\% C.L. variation due to $\alpha_s(M_Z)$ can be estimated as a one-half of the
difference in predictions from the two $\alpha_s$ sets. The
PDF+$\alpha_s$ uncertainty, at 90\% C.L., and including correlations,
can also be determined by adding the PDF uncertainty and $\alpha_s$
uncertainty in quadrature~\cite{Lai:2010nw}.
Aside from these general-purpose PDF sets, we provide a series of (N)NLO
sets for $\alpha_s(M_Z)=0.111-0.123$
and additional sets using heavy-quark
schemes other than our standard 5-flavor method, with up to 3, 4, and 6 active flavors. 

Parametrizations for the CT18 PDF sets are distributed in a standalone
form via the CTEQ-TEA website \cite{CT18website}, or as a part of
the LHAPDF6 library \cite{LHAPDF6}. For backward compatibility with
version 5.9.X of LHAPDF, our website also provides CT18 grids in the
LHAPDF5 format, as well as an update for the CTEQ-TEA module
of the LHAPDF5 library, which must be included during compilation
to support calls of all eigenvector sets included with CT18 \cite{LHAPDF5}.

\begin{acknowledgments}
We are indebted to our friend and colleague Jon Pumplin for decades of fruitful collaboration and for his last major contribution to the CTEQ-TEA global analysis made in this article.
We thank Stefano Camarda, Amanda Cooper-Sarkar, Alexander Glazov, Lucian Harland-Lang, Jan Kretzschmar, Bogdan Malescu, Wally Melnitchouk, Dave Soper, and CTEQ colleagues for insightful discussions. This work is partially supported by the U.S.~Department of Energy under Grants No.~DE-SC0010129 (at SMU) and No.~DE-FG02-95ER40896 (at U. Pittsburgh); by the U.S.~National Science Foundation
under Grants No.~PHY-1719914 and No.~PHY-2013791 (at MSU) and No.~PHY-1820760 (at U.~Pittsburgh), and in part by the PITT PACC. T.~J.~Hobbs acknowledges support from a JLab EIC Center Fellowship.
The work of M.Guzzi is supported by the National Science Foundation under Grant No.~PHY-1820818. The work of J.G.~was supported by the National Natural Science Foundation (NNSF)
of China under Grants No.~11875189 and No.~11835005, and the work of S.D.~and I.S.~under NNSF Grants No.~11965020 and No.~11847160. C.-P.~Yuan is also grateful for the support from
the Wu-Ki Tung endowed chair in particle physics.

\end{acknowledgments}
\newpage
\appendix
\section{The alternative CT18Z global fit
\label{sec:AppendixCT18Z}
}

\begin{figure}[b]
	\includegraphics[width=0.49\textwidth]{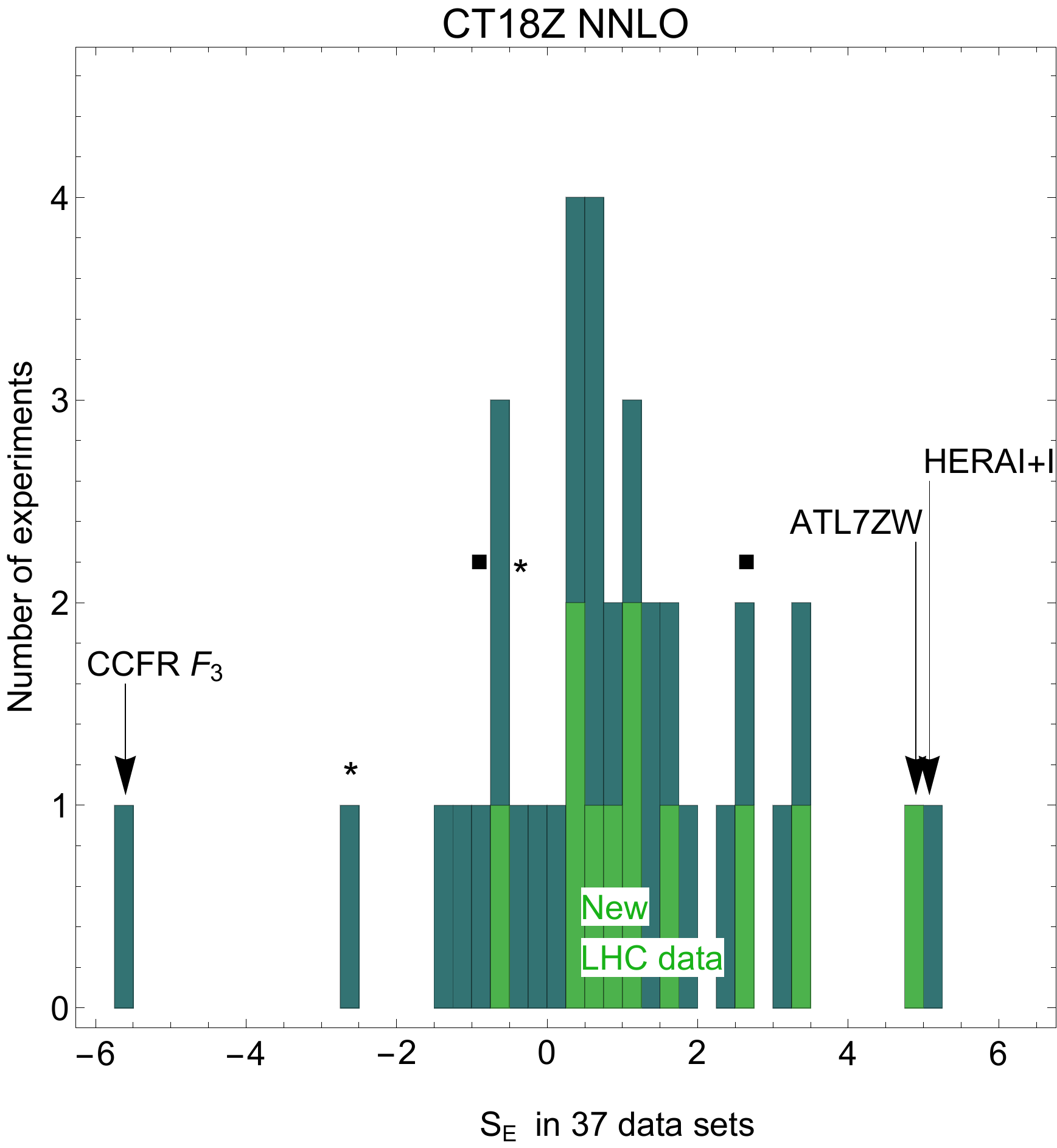}
	\caption{Analogous to Fig.~\ref{fig:sn_ct18}, the effective Gaussian variable ($S_E$) distribution of all CT18Z data sets. Two squares and two stars indicate the $S_E$ values
	for the NuTeV dimuon and CCFR dimuon experiments, respectively.
\label{fig:sn_ct18z}}
\end{figure}

In this appendix, we describe a series of fits leading to the CT18Z PDFs that provide a distinct alternative to the primary result of this analysis, CT18 NNLO. 
While CT18Z NNLO achieves a comparable level of success in describing the CTEQ-TEA data, producing $\chi^2/N_\mathit{pt}\! =\! 1.19$ as opposed to $\chi^2/N_\mathit{pt}\! =\! 1.17$
for the CT18 NNLO fit, the quality of the agreement for specific data sets undergoes a number of changes. This can be seen in part by comparing the distribution of $S_E$ values
obtained for CT18Z in Fig.~\ref{fig:sn_ct18z} with what we presented for CT18 in Fig.~\ref{fig:sn_ct18}.
The inclusion of the ATLAS 7 TeV $W/Z$ data (or, ATL7ZW data, with
CTEQ experimental ID=248) in CT18Z causes an upward shift of $S_E$ (or $\chi^2_E/N_{pt,E}$) for a number of experiments, notably for the dimuon data
from NuTeV (Exp.~IDs=124, 125) and CCFR (Exp.~IDs=126, 127), indicating that the ATL7ZW data is in some disagreement with these other data sets.

In Sec.~\ref{sec:alt}, we pointed out that we release two intermediate fits, CT18A (with addition of ATL7ZW data only) and CT18X (with a specially-chosen factorization scale in DIS cross sections).  
Compared to CT18A and X, the CT18Z PDFs produce maximal changes away from CT18 in the PDFs and their moments, the parton luminosities, and standard-candle predictions: those can be viewed in Figs.~\ref{fig:PDFbands1}, \ref{fig:DBandSBbands}, \ref{fig:DBandSBbands2}, \ref{fig:lumia}, \ref{fig:corr_ellipse1}, and \ref{fig:corr_ellipse2}.

We will now look into the accumulation of the modifications that led from CT18 to CT18Z NNLO in more detail. The role of the ATL7ZW and CDHSW experiments is reviewed in Sec.~\ref{sec:CT18Z_changes}. Sec.~\ref{sec:CT18ZvsCT18} summarizes the key differences between CT18, A, X, and Z PDFs, while the plots of error bands for CT18A and X NNLO and NLO PDFs are included in supplemental material. 
The agreement with the ATL7ZW data is explored in Sec.~\ref{sec:CT18Z_qual}. 
In Sec.~\ref{sec:LMCT18Z}, we examine $\chi^2$ scans to extract detailed information on the redistribution of constraints on the PDFs inside the CT18Z global fit, as well as on the CT18Z predictions for $\alpha_s(M_Z)$ and $m_c$.

The physics conclusions presented here have been verified using
several independent techniques. Initially, projections of the likely
impact of the data sets on the PDFs were obtained by applying the fast
Hessian techniques, \texttt{PDFSense}/$L_2$
sensitivity \cite{Wang:2018heo,Hobbs:2019gob}
and \texttt{ePump} \cite{Hou:2019gfw}, by starting from theoretical
predictions based on the previous \CTHERAII~ NNLO
PDFs \cite{Hou:2016nqm}. At this stage, we discovered that
the \texttt{ePump} and \texttt{xFitter} programs produce very
different results when profiling the ATL7ZW data. This discrepancy
is addressed in App.~\ref{sec:Appendix4xFitter}.
As a part of the fitting itself, we repeated
some fits multiple times while either constraining the PDFs at
specific values using Lagrange Multipliers (LM) or varying the
statistical weights of ATL7ZW and other data sets to explore their
mutual consistency within the approach by Collins and
Pumplin \cite{Collins:2001es}. All these methods render a coherent
physics picture that will be now summarized. 

\subsection{Alterations to data sets and theoretical settings}
\label{sec:CT18Z_changes}

\subsubsection{Modified data selection}
Let us first address some questions arising in the description
of two data sets: ({\it i}) the recent 7 TeV Drell-Yan data taken by ATLAS for the rapidity distributions for the inclusive production
of  $W$ and $Z$ bosons (ATL7ZW, Exp.~ID=248); and ({\it ii}) the $F^p_2$, $xF^p_3$ DIS structure function information
extracted by CDHSW from $\nu$-Fe data (Exp.~IDs=108, 109).

\begin{figure}[tb]
\includegraphics[width=0.7\textwidth]{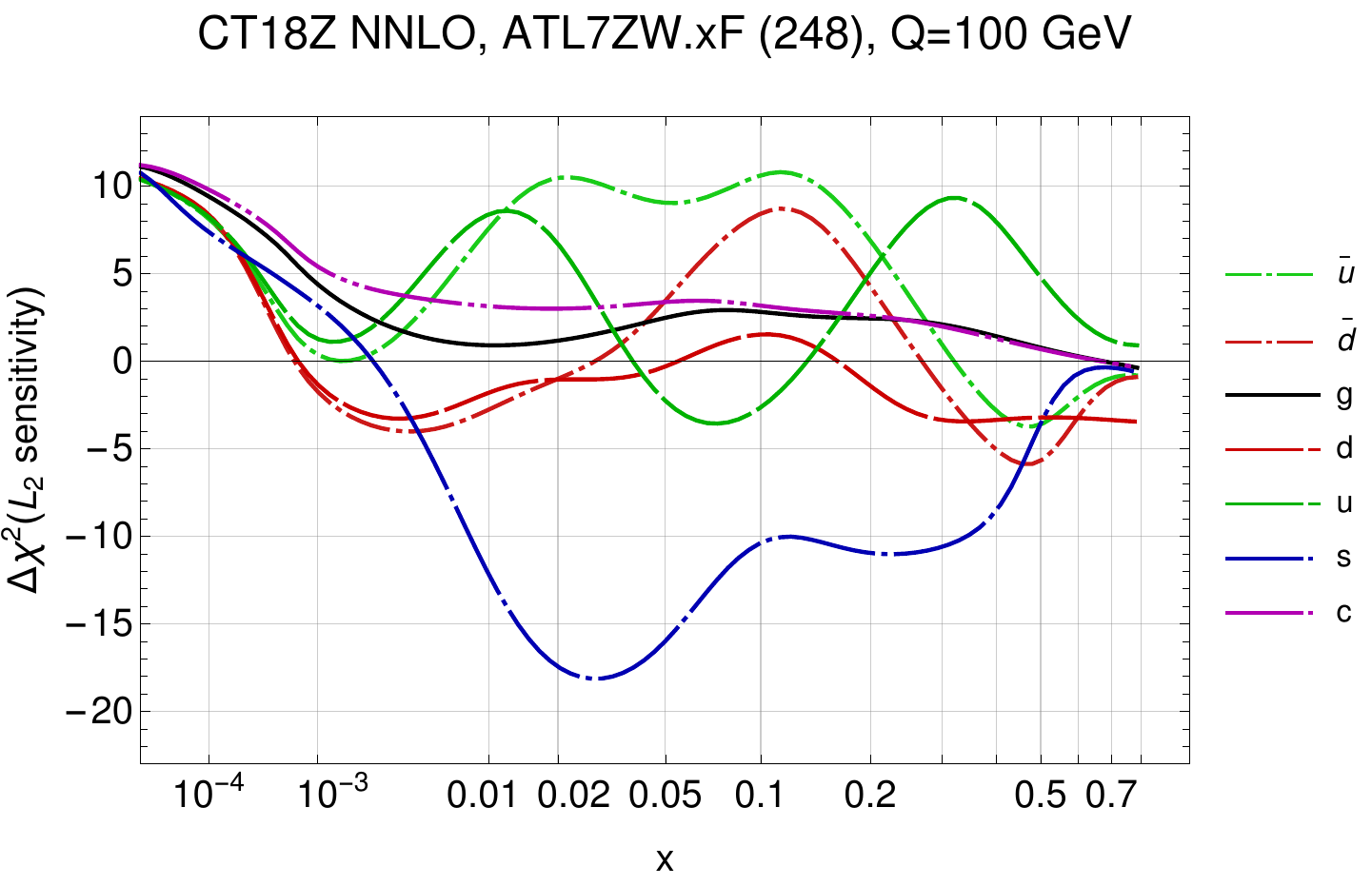}
\vspace{-2ex}
\caption{
	The $L_2$ sensitivity of the ATL7ZW data to the PDFs of
	several individual parton flavors. The pull on the strangeness distribution,
	$s(x,Q)$, is particularly large, peaking at $S_{s,\mathit{L2}}\! \sim\! -20$
	for $x\!\sim\! 0.02-0.05$; although opposing pulls on the $d$-, $\bar{u}$-,
	and $\bar{d}$-quark PDFs are also significant in a similar region of $x$.
}
\label{fig:L2_248}
\end{figure}

\begin{figure}[htbp]
        \begin{center}
                \includegraphics[width=0.95\textwidth]{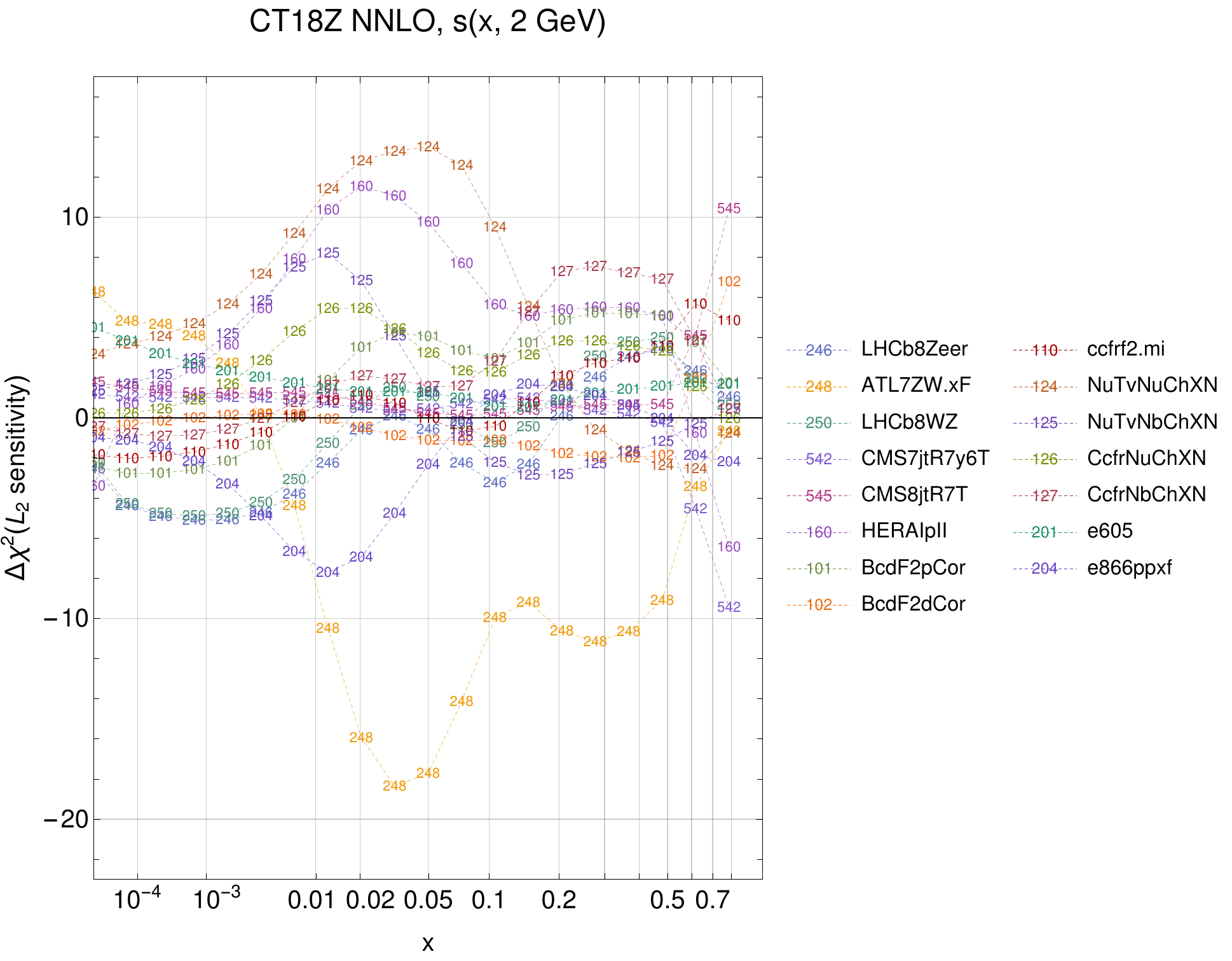}
        \end{center}
        \vspace{-2ex}
        \caption{
		The $L_2$ sensitivity of the data fitted in CT18Z to the $x$-dependent
		strange PDF at $Q\!=2$ GeV. Given that the fitted behavior for
		$R_s(x,Q)$ is substantially driven by $s(x,Q)$, the information
		shown here is complementary to the LM scans in Fig.~\ref{fig:lm_rsz}.
        }
\label{fig:L2sZ}
\end{figure}

{\bf ATLAS 7 TeV Inclusive $W/Z$-production data.}
Regarding case ({\it i}), the ATL7ZW data set is seen as providing
important information on the structure of the light-quark nucleon sea,
and, in particular, favoring an enhanced value for the strangeness
suppression factor, $R_s(x,Q)$ of Eq.~(\ref{eq:Rs})
compared to what has been found in the past global analyses
dominated by the DIS data.
The sensitivity analysis indicates that, while
the correlation cosines of the ATL7ZW measurements
with respect to various PDF flavors are quite modest
(typically, $|\cos \phi|\! <\! 0.6$), the sheer precision of the
ATL7ZW data creates a pronounced
pull on all light-antiquark flavors: $\bar u$, and $\bar d$, and
especially $\bar s$. For example, the pulls are revealed by the charts
showing $L_2$ sensitivities of ATL7ZW data to various CT18Z PDF flavors
in Fig.~\ref{fig:L2_248}. The strong positive pull on $s(x,Q)$
revealed by the corresponding negative $S_{L2,E}$ for $s(x,Q)$
at $x=0.01-0.1$ in Fig.~\ref{fig:L2_248} must be compensated in the
global fits by the
opposing pulls from other experiments. For example, we can compare
the Hessian estimates of the pulls on $s(x,Q=2\mbox{ GeV})$
by plotting the respective sensitivities for individual
experiments in Fig.~\ref{fig:L2sZ}. It is obvious that the prominent
negative pull on $\Delta \chi^2$ from ID=248 at $x=0.03$ is opposed by
the positive pulls from NuTeV (Exp.~ID=124, 125) and CCFR
(Exp.~ID=126, 127) dimuon SIDIS, and, also very prominently, the
inclusive HERAI+II data (Expt.~ID=160). On the other hand, the
fixed-target E866 Drell-Yan data on the $pp$ target (Expt.~ID=204)
weakly pulls in the same direction as ATL7ZW, although at smaller
$x \approx 0.01$.

While the $L_2$ sensitivity estimates contributions
to $\chi^2$, tensions with the ATL7ZW data are also revealed by other
statistical indicators, such as an effective Gaussian variable $S_E$
for experiment $E$ that quantifies how the change in $\chi^2_E$
compares to the respective statistical uncertainty. By this measure, 
the Hessian updating study \cite{Hou:2019gfw} based on \texttt{ePump} 
found that including the ATL7ZW data with
increasing statistical weights into the CT18 fit strongly increases
$S_E$ values for the NuTeV $\nu$ SIDIS (Exp.~ID=125),
the E866 $\sigma_{pd}/(2\sigma_{pp})$ Drell-Yan data (Exp.~ID=203),
and the CMS 7 TeV electron asymmetry data (Exp.~ID=267),
cf.~Fig.~25 of~\cite{Hou:2019gfw}. This change is accompanied by
modifications in the $s$-quark PDF and $\bar{d}/\bar{u}$ PDF ratio,
cf.~Fig.~26 of the same reference. Finally, we observe mild
suppression of $g(x,Q)$ at $x \gtrsim 10^{-2}$ after including the
ATL7ZW data set.

{\bf CDHSW data.}
Our LM scans, like the ones presented in Fig.~\ref{fig:LMg18} and the
$L_2$ sensitivity plot in Fig.~\ref{fig:L2glu}, reveal
that the CDHSW measurements of deep inelastic scattering
in charged-current neutrino interactions on iron
(Exp.~IDs=108 \cite{Berge:1989hr}, 109 \cite{Berge:1989hr})
are sensitive to the gluon distribution at $x > 0.2-0.5$ via $Q^2$
distributions of their cross sections. At $x <0.4$, the logarithmic
slopes of the structure functions $F_2$ and $xF_3$ measured by CDHSW
and CCFR are different, cf. Figs.~8.3 and 8.4 in
Ref.~\cite{Seligman:1997fe}, with CDHSW structure functions preferring
a harder $g(x,Q\!=\!100\,\mathrm{GeV})$ at $x\!\approx\!0.2$, but a softer
gluon at $x\!\gtrsim\!0.5$ --- similarly to the CDF Run-2 jet data
(Exp.~ID=504). It has been known that unresolved experimental issues may exist
in the CDHSW analysis~\cite{Barone:1999yv}, and nuclear corrections
may be non-negligible for the iron target. Thus,
one may wonder how the PDFs would change if
these two CDHSW data sets are excluded. This question was addressed
using  \texttt{ePump} in Ref.~\cite{Hou:2019gfw}, as well as by performing a special fit named
``CT18mCDHSW'' ({\it i.e.}, CT18 ``minus'' CDHSW), in which
the two CDHSW data sets were removed from the CT18 global data.

The plots of PDF error bands from the CT18mCDHSW NNLO fit, included in
the supplemental material, show that removing CDHSW data leads to a
slight reduction in the gluon PDF at $x=0.1-0.5$, combined with a
slight increase in the gluon PDF uncertainty, and compensating
increases in the $u$ and $d$ PDFs in the same $x$ region. 

\subsubsection{The $x_B$-dependent scale and modified global fits}
Another point of potential concern is the residual dependence on QCD
cross sections on the renormalization and factorization scales, which
we find to be non-negligible in some experiments, compared to the
latest experimental uncertainties, even when NNLO theoretical
expressions for QCD cross sections are used. In particular, by
evaluating the NNLO DIS cross sections at a carefully chosen
factorization scale $\mu_{F,x}$
dependent on Bjorken $x_B$ in Table~\ref{tab:AXZ},
we moderately improve agreement with HERA DIS cross sections:
the respective $\chi^2$ improves by 40 units
($\chi^2(\mbox{CT18})=1408$, $\chi^2(\mbox{CT18Z})=1378$ for
$N_{pt}=1120$ data points), and it can be improved by another 30 units
by increasing the statistical weight of the inclusive HERA data to 10 as in
the right Fig.~\ref{fig:saturation}. With this scale choice, we also obtain
larger PDFs for the gluon and strangeness PDFs at momentum fractions $x$
below 0.05, cf. left Fig.~\ref{fig:saturation}. At the same time,
$g(x,Q)$ and $s(x,Q)$ are reduced from $x=0.2$ to 0.4-0.5 to preserve
the sum rules.

Therefore, three modifications in the global fitting framework -- using
the $\mu_{F,x}$ scale in DIS, excluding the CDHSW data sets, and
adding the ATLAS7ZW data -- add up to suppress the gluon PDF at
$0.005 \lesssim x \lesssim 0.3$, across most of the interval
of $x$ relevant for the LHC
Higgs production via $gg$ fusion. Their combination also produces a
substantial increase of $s(x,Q)$ at all $x$. 

As discussed in Sec.~\ref{sec:alt}, this combination is adopted to
produce the CT18Z NNLO PDFs. The intermediate PDFs, CT18A and CT18X,
implement only the ATL7ZW data in the $66 < Q < 116$ GeV region (34
data points) or only the DIS scale $\mu_{F,x}$, as
indicated in Table~\ref{tab:AXZ}. [The low-luminosity
($35\, {\rm pb}^{-1}$) sample of the ATLAS 7 TeV  $W^\pm$ and $Z$
cross section data  (Exp.~ID=268) is removed from the CT18A and Z fits
to avoid double counting.]

\subsection{Comparisons between the four PDF ensembles}
\label{sec:CT18ZvsCT18}

The supplemental material includes a series of figures comparing the NNLO and NLO PDF uncertainty bands for CT18, A, X, and Z PDF flavors. The main characteristics of these comparisons can be distilled as follows.

\begin{enumerate}
\item {\bf $g(x,Q)$:} By comparing the PDF uncertainty bands for CT18,
CT18A, and CT18Z, on one hand, and CT18, CT18X, and CT18Z, on the
other hand, it is clear the bulk of the variation of CT18Z away from
CT18 is due to the modified DIS scale choice, $\mu_{F,x}$, with only
weaker changes resulting from the interplay of the removal of CDHSW
and inclusion of the ATL7ZW measurements. The deviations from CT18
are generally smaller at NLO, with the exception of the very low-$x$
region, where CT18Z and CT18X NLO ratios to CT18 are $\sim\! 50\%$ larger.
\item {\bf $d(x,Q)$:} In contrast to $g(x,Q)$, the sensitivity
of the Drell-Yan process to $d(x,Q)$ enhances the difference
between the CT18 and CT18A NNLO results for this flavor, realized as a
mild, $\sim\!1\%$ suppression of the central CT18A distribution for
$d(x,Q)$ relative to CT18 about $x\!\sim\!10^{-3}$, with a
few-percent reduction in the accompanying PDF uncertainty. The CT18Z
fit in this region is pulled in the opposing direction, being enhanced
by $\lesssim\!2\%$ with a comparable uncertainty for $x\!<\! 0.1$. In
the high-$x$ region,
$x \gtrsim 0.2$, the effect of fitting the ATL7ZW data
alone in CT18A NNLO boils down to a small shift in the $d(x,Q)$
uncertainty. CT18Z, in contrast, is suppressed relative to CT18 at higher values of $x$. 
The qualitative impact of fitting $d(x,Q)$ at NLO, as opposed
to NNLO, is fairly weak, being felt mostly at lower $x\!\lesssim\!
10^{-3}$.
\item {\bf $u(x,Q)$ and $\bar u(x,Q)$:} Here, the remarkable property
is the extent to which the deviations of CT18Z NNLO away from CT18
NNLO are driven by the modifications in CT18X, as attested by the very
close agreement between the central PDFs and uncertainties 
for CT18Z and X. The weak suppression of $\bar{u}(x,Q)$
in CT18A NNLO is almost entirely nulled once the DIS-scale
choice found in CT18X is implemented in CT18Z on top of the ATL7ZW measurements.
\item {\bf $s(x,Q)$ and $R_s(x,Q)$:} We have noted already
that the introduction of the ATL7ZW data in
CT18A/Z NNLO fits leads to a demonstrable enhancement of $s(x,Q)$ over
CT18, while the scale choice in CT18X mildly suppresses $s(x,Q)$ at
$0.01 \lesssim x \lesssim 0.5$ and enhances it at $x \lesssim 0.01$ and 
$x \gtrsim 0.5$. The strangeness suppression ratio, $R_s(x,Q)$,
is driven for the most part by the patterns observed for $s(x,Q)$ itself.
\end{enumerate}

\subsection{A closer look at the description of ATLAS 7 $W/Z$ data}
\label{sec:CT18Z_qual}

\subsubsection{Goodness of fit to ATL7ZW data in CT18A(Z)}
\label{sec:ATL7ZWchi2}

\begin{table}[tb]
\caption{The comparison of the $\chi^{2}$ values for ATLAS 7 TeV $W/Z$ data among the QCD analysis from different groups. For CT18A/Z PDFs, we show the 68\% C.L. uncertainties obtained by a Lagrange Multiplier scan on the $\chi^2$ weight of the ATLAS $W/Z$ data set, as illustrated in Fig.~\ref{fig:lm_wht} (left).}
\label{tab:CMN}
\begin{tabular}{c|c|c|c|c}
\hline\hline
PDF & $N_{\mathit{pt, E}}$ & $\chi^{2}_E$ & $\chi^{2}_E/N_{\mathit{pt}, E}$ & Ref. \\
\hline
CT18A &  34  & $88^{+68}_{-28}$  & $2.58^{+2.01}_{-0.84}$ & \\
CT18Z & 34 & $89^{+65}_{-31}$ & $2.61^{+1.90}_{-0.91}$  & \\
\hline
ATLAS-epWZ16 &  61 & 108    & 1.77 & \cite{Aaboud:2016btc}\\
MMHT (2019)  &  61 & 106.8  & 1.76 &  \cite{Thorne:2019mpt}\\
NNPDF3.1     &  34 & 73     & 2.14 & \cite{Ball:2017nwa} \\
\hline\hline
\end{tabular}
\end{table}

Turning now to the overall description of the ATL7ZW measurement, responsible for the strong modifications of the nucleon strangeness in CT18A/Z NNLO, we find
a significant improvement in $\chi^2$ for these data once they are actually fitted. Namely, CT18 NNLO produces a very large
$\chi^2_E/N_\mathit{pt,E}\!=\! 8.4$ ($S_E\! =\! 13.7$), which diminishes substantially to
$\chi^2_E/N_\mathit{pt,E}\!=\! 2.6$ ($S_E\! =\! 4.8$) in CT18Z NNLO, as we reported in
Table~\ref{tab:EXP_2}. We note that the corresponding values for CT18A NNLO, which deviates from
the settings of CT18 NNLO only in the implementation of the ATL7ZW data, are only a very
slight improvement over CT18Z, $\chi^2_E/N_\mathit{pt,E}\!=\! 2.6$ ($S_E\! =\! 4.7$). 
We also remind the reader that the CT18A/Z fits include only 34 ATL7ZW data points in the resonance region ($66 < Q < 116$ GeV), with the rest of 61 points of the published set being less precise and contributing mostly to the reduction of $\chi^2_E/N_{pt,E}$ without really improving the PDF constraints, while also enhancing the dependence of the PDFs on NLO EW corrections in the off-resonance regions. 

In comparison, the ATLAS group themselves obtained 
$\chi^2_E/N_{\mathit{pt,E}}=108/61$ in the ATLAS-epWZ16 fit \cite{Aaboud:2016btc} of the $W/Z$ and combined HERA I+II data \cite{Abramowicz:2015mha}. For CT14 PDFs, they obtained $\chi^2_E/N_{\mathit{pt,E}}=103/61=1.69$ after profiling the CT14 PDFs with the $W/Z$ data in \texttt{xFitter}. Their fitted strangeness fraction is
$R_s=(s+\bar{s})/(\bar{u}+\bar{d})=1.13\pm0.08$ at $x=0.023$, i.e., it
is significantly larger than what we obtained in the CT18A(Z) fits. 

We confirm these findings when including only the ATL7ZW and HERA I+II
data in the CT analysis. In fact, we get a better $\chi^2/N_{pt,E}\approx 1.4$ and even larger $R_s$, due to the more flexible CT parameterization, if the ATL7ZW data are included with an elevated statistical weight. 

\begin{figure}
    \centering
    \includegraphics[width=0.7\textwidth]{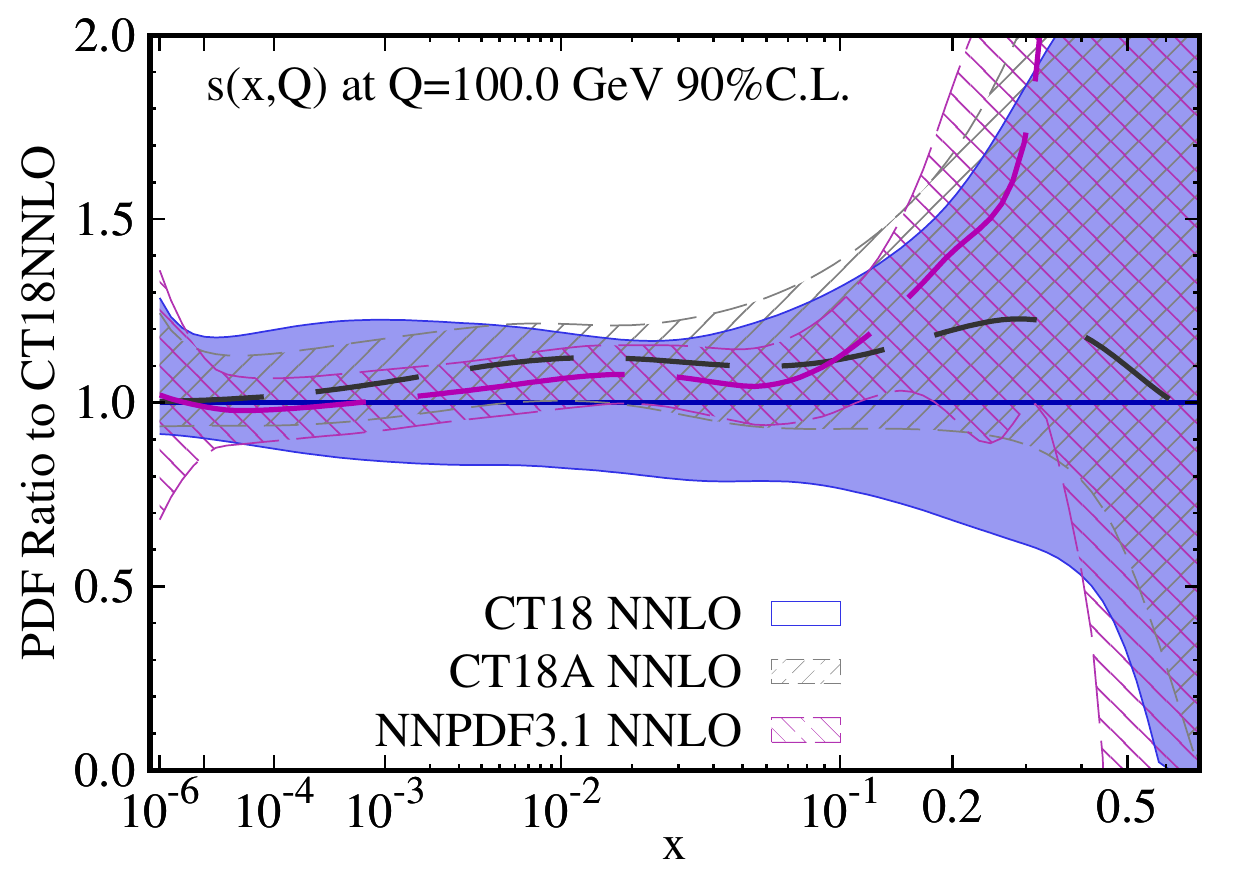}
    \caption{
        A comparison of the fitted strangeness PDF, $s(x,Q)$, at $Q\!=\!100$ GeV
		as obtained with CT18 (solid violet) as well as with CT18A (short-dashed gray)
		and NNPDF3.1 (long-dashed magenta).}
    \label{fig:CT18A2NNPDF31}
\end{figure}

However, we view this outcome as problematic on the grounds that
\begin{enumerate}
\item the above \texttt{xFitter} profiling analysis strongly deemphasizes the
experiments that show tension with the ATL7ZW data, as explained in
App.~\ref{sec:Appendix4xFitter};
\item our LM scans like the ones presented in Sec.~\ref{sec:LMCT18Z}
reveal that CT fits, with their increased
flexibility compared to HERAPDF fits, become unstable or have multiple
minima when $R_s(x,Q)$ is forced to be close to 1 at $x=0.01-0.1$, as
preferred by the ATL7ZW data set; the instability may reflect the still weak capability of data to discriminate between the $\bar s$, $\bar d$, and $\bar u$ contributions. 
\end{enumerate}

Other PDF fitting groups have also investigated the ATL7ZW data. Table~\ref{tab:CMN} summarizes the values of $\chi^{2}_E/N_{pt,E}$ for the ATL7ZW data obtained in CT18A, Z, ATLAS-epWZ16, MMHT (2019), and NNPDF3.1 NNLO fits. For the CT18A/Z fits, we quote the 68\% C.L. uncertainties on $\chi^2_{{\rm \small ATL7ZW}}$. We find them in Sec.~\ref{sec:ATL7ZWchi2wt}  from the LM scan on the weight of $\chi^2_{{\rm \small ATL7ZW}}$ shown in the left panel of  Fig.~\ref{fig:lm_wht}. The uncertainties are equal to the differences of $\chi^2_{{\rm \small ATL7ZW}}$ from the value of $\chi^2_{{\rm \small ATL7ZW}}$ at the best global fit, when the total $\chi^2$ (solid black line in the left panel of Fig.~\ref{fig:lm_wht}) is increased by one standard deviation (36 $\chi^2$ units). 

The respective $\chi^2_{{\rm \small ATL7ZW}}$ values quoted in Table~\ref{tab:CMN} 
are well within within the 68\% uncertainties of CT18A/Z, with the nominal CT18A/Z values 
for $\chi^2_{\rm \small ATL7ZW}$ being on the high side, but not significantly. The differences in $\chi^2_{\rm \small ATL7ZW}$ can be traced primarily to the magnitude of $s(x,Q)$ at moderate $x$, which is lower (within the uncertainties) in CT18A/Z NNLO than in some other fits. 

The ABM analysis \cite{Alekhin:2017olj} has emphasized tensions between ATL7ZW, NuTeV, and NOMAD data sets, as well as strong dependence of the preferred $R_s(x,Q)$ enhancement on the flexibility of the strangeness parametrization. The unpublished 2019 MMHT analysis \cite{Thorne:2019mpt}
obtains $\chi^2_E/N_\mathit{pt,E}\!=\! 1.76$ for 61 ATL7ZW data points by (a) 
including NNLO quark-mass corrections for inclusive charged-current (SI)DIS cross sections
~\cite{Berger:2016inr,Gao:2017kkx}, which slightly improve agreement between the DIS and ATL7ZW data sets, cf. Sec.~\ref{sec:Qualitydimuon}; and (b) using flexible 6-parameter parametrizations for $d$ and $s$ quarks. The 2019 MMHT analysis reports a significant enhancement in $s(x,Q)$ above that in MMHT'2014 for $x\! >\! 10^{-3}$, with a corresponding uncertainty reduction.

While NNPDF did not actively fit ATL7ZW in NNPDF3.0, which obtained $\chi_E^2/N_\mathit{pt,E}\!=\! 8.44$, these data were implemented in NNPDF3.1 \cite{Ball:2017nwa},
resulting in $\chi^2_E/N_\mathit{pt,E}\!=\! 2.14$ --- close to CT18A/Z NNLO. 
NNPDF3.1 observed an enhancement in the fitted strange PDF that is similar 
to that in CT18A NNLO for $x < 0.1$. Figure~\ref{fig:CT18A2NNPDF31} plots 
the ratios of the CT18A and NNPDF3.1 NNLO strangeness PDFs at $Q\!=\!100$ GeV 
to those in  CT18 (here, a baseline which did not include ATL7ZW).
In CT18A, we observe a $\sim\!10\%$ excess above CT18 
in the region $x\!\gtrsim\! 0.02$ where the ATL7ZW data
have the strongest pull. Especially in this region of $x$, the larger
$s$-PDF found in CT18A is closely reflected by the strangeness fitted in NNPDF3.1.

\begin{figure}[t]
	\begin{tabular}{cc}
		\subfloat[$W^-$ production]{\includegraphics[width=0.35\textwidth]{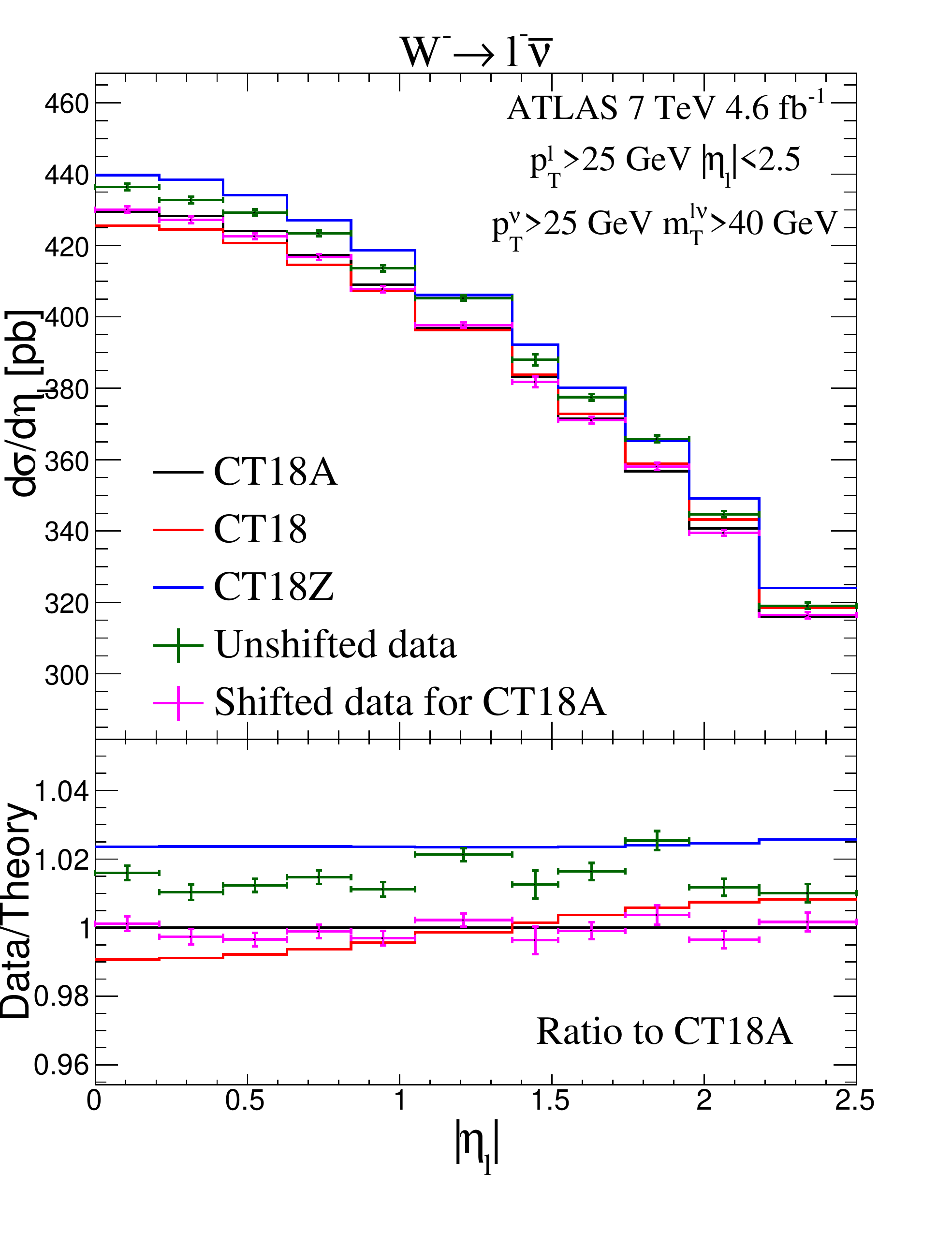}}
		\subfloat[$W^+$ production]{\includegraphics[width=0.35\textwidth]{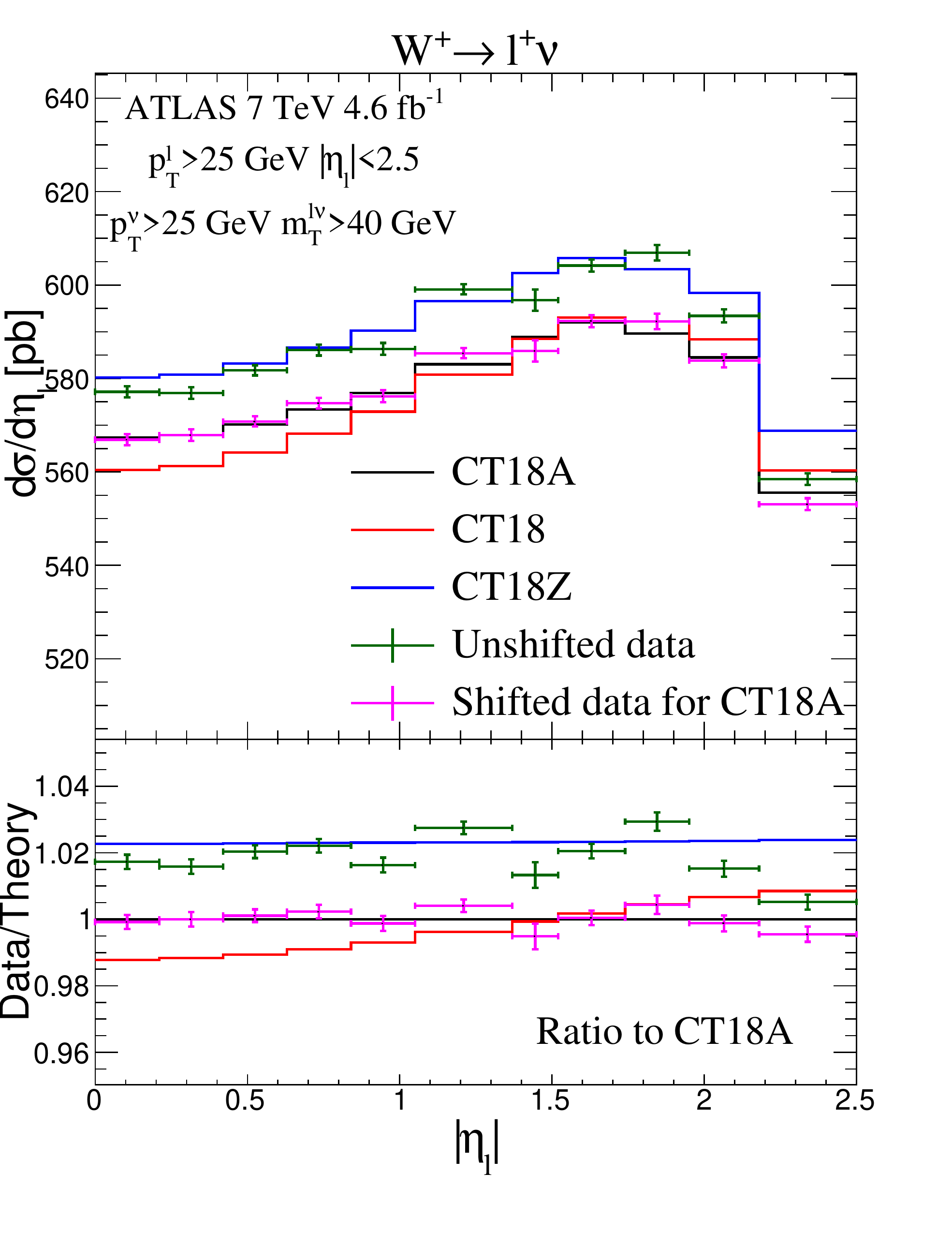}}
		\subfloat[$Z$ production]{\includegraphics[width=0.35\textwidth]{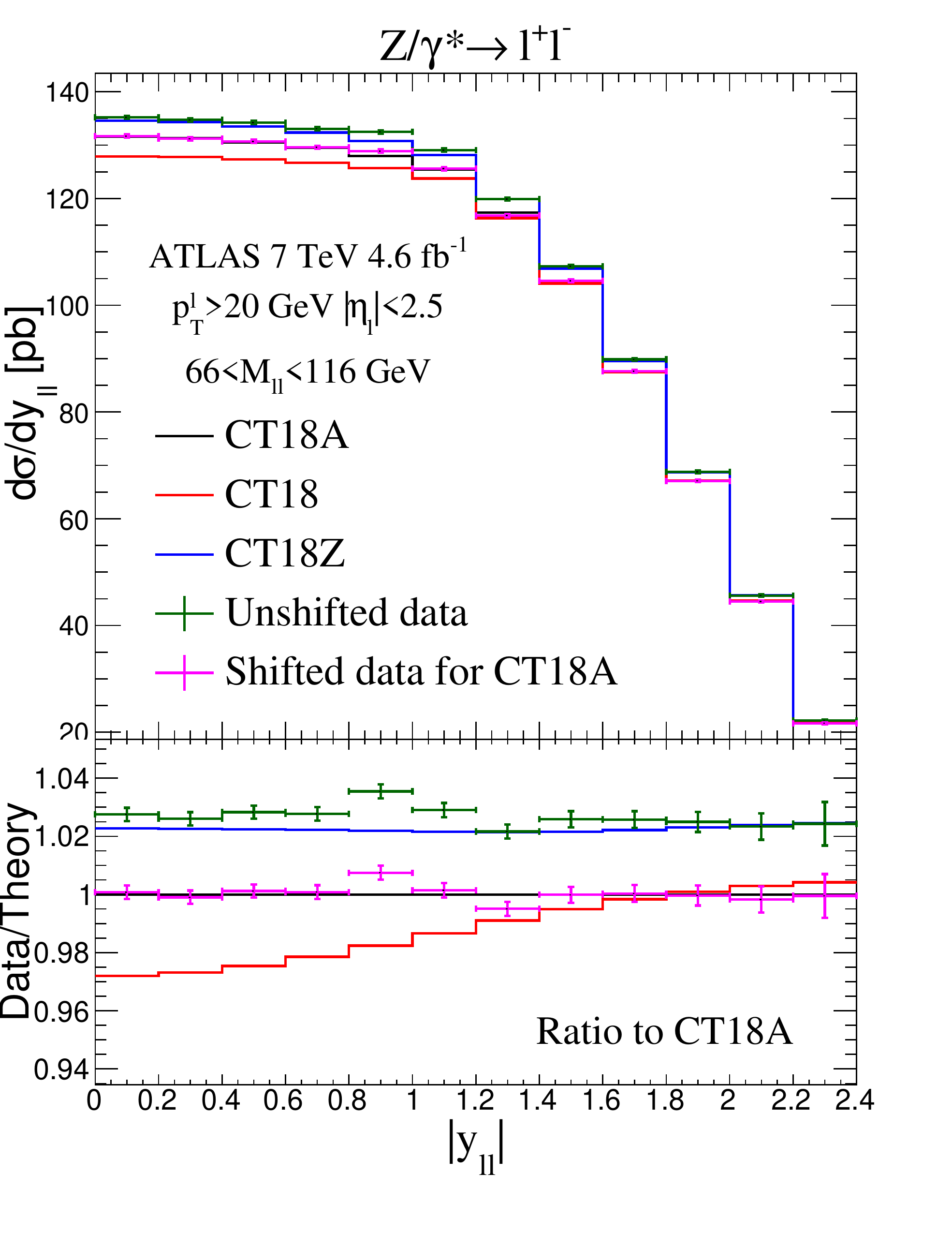}}
	\end{tabular}
	\caption{
		(a) A comparison of theoretical predictions for the ATL7ZW data based on CT18 and CT18A/Z NNLO.
		The shifted data (magenta crosses) are computed based on CT18A NNLO. The upper panels give rapidity distributions
		of the differential cross sections in $|\eta_l|$ (for the charge-current processes) or $|y_{\ell \bar \ell}|$ (for neutral-current), while the lower
		insets show the ratios of the data and theory, normalized to CT18A theory.
	}
\label{fig:248_DT}
\end{figure}

\begin{figure}[p]
	\begin{tabular}{cc}
\subfloat[$W^-$ production]{\includegraphics[width=0.49\textwidth]{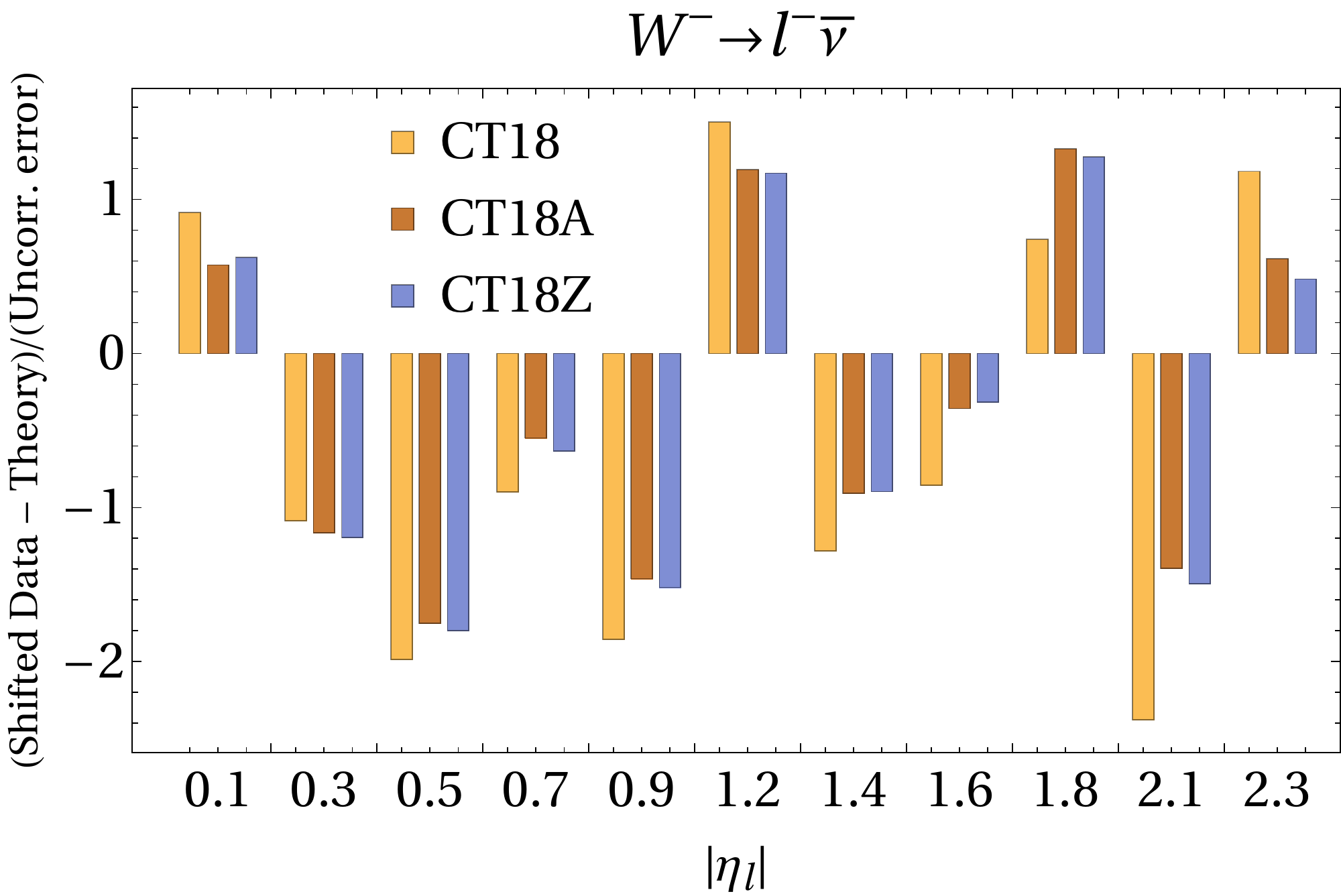}}
\subfloat[$W^+$ production]{\includegraphics[width=0.49\textwidth]{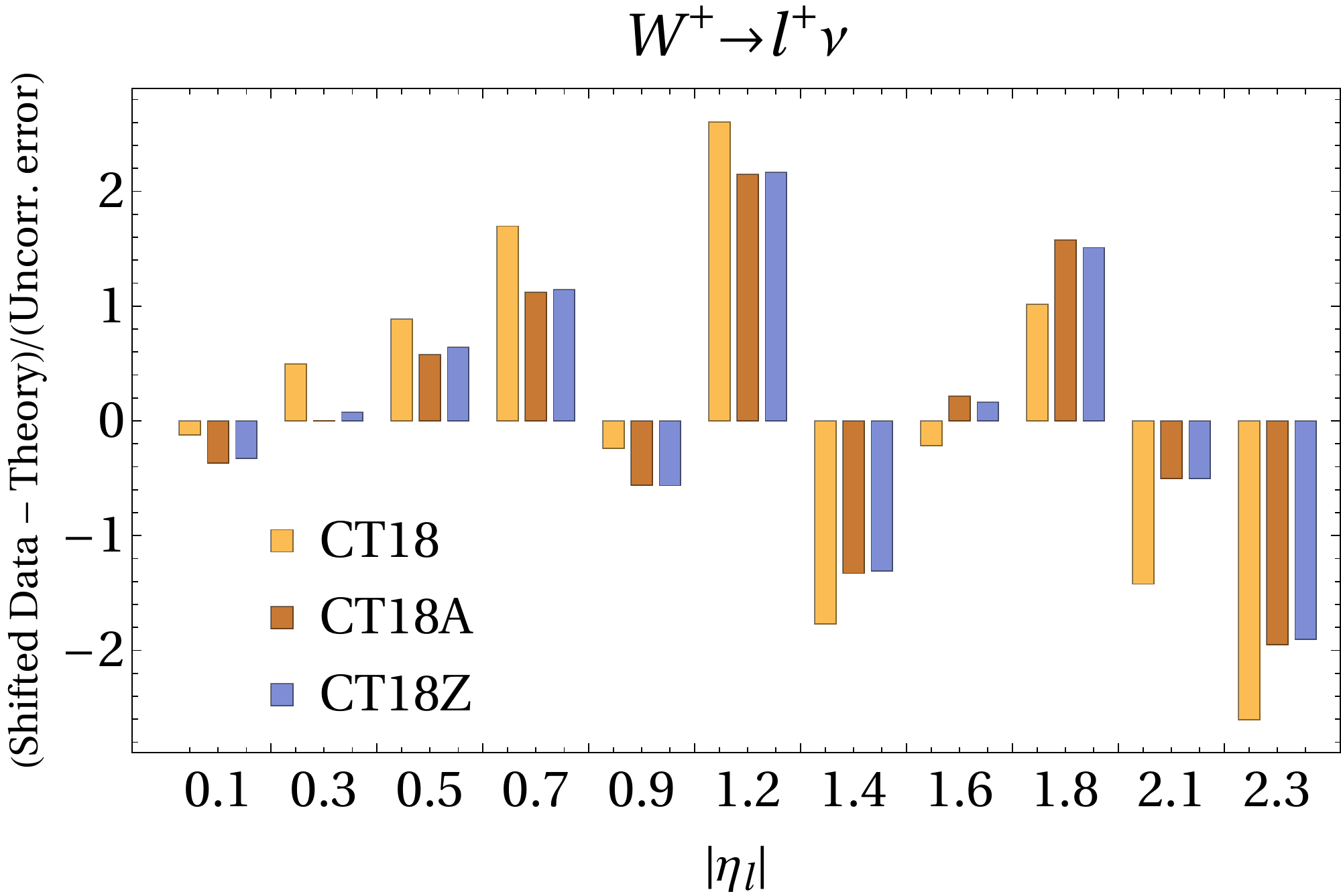}}\\
\subfloat[$Z$ production]{\includegraphics[width=0.49\textwidth]{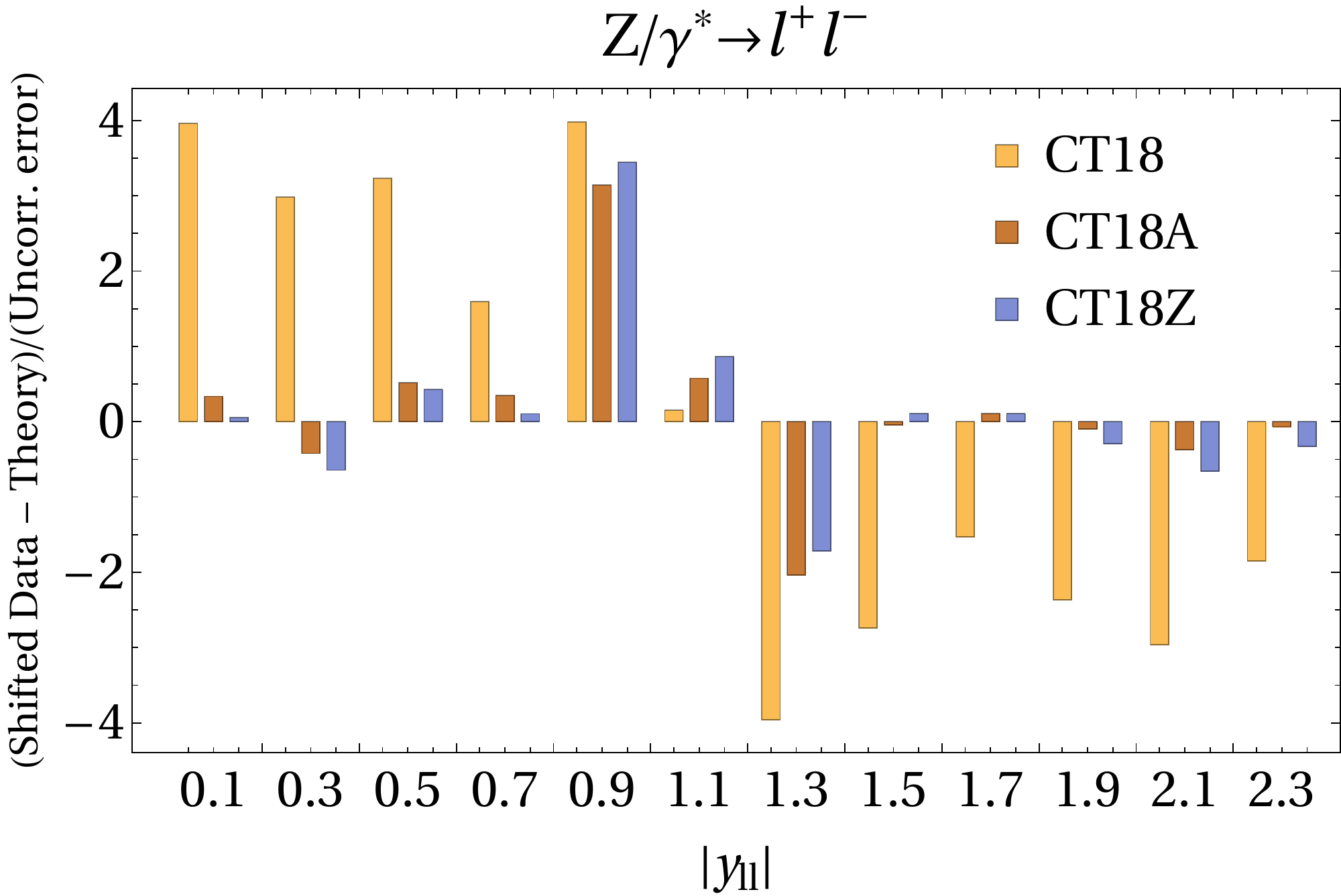}}
	\end{tabular}
	\caption{
		Bin-by-bin shifted $\mathrm{Theory}\!-\!\mathrm{Data}$ residuals computed based on CT18(A/Z) for each of the three ATL7ZW processes shown in Fig.~\ref{fig:248_DT}. 
	}
\label{fig:248_res}
\end{figure}

\begin{figure}[p]
\includegraphics[width=0.49\textwidth]{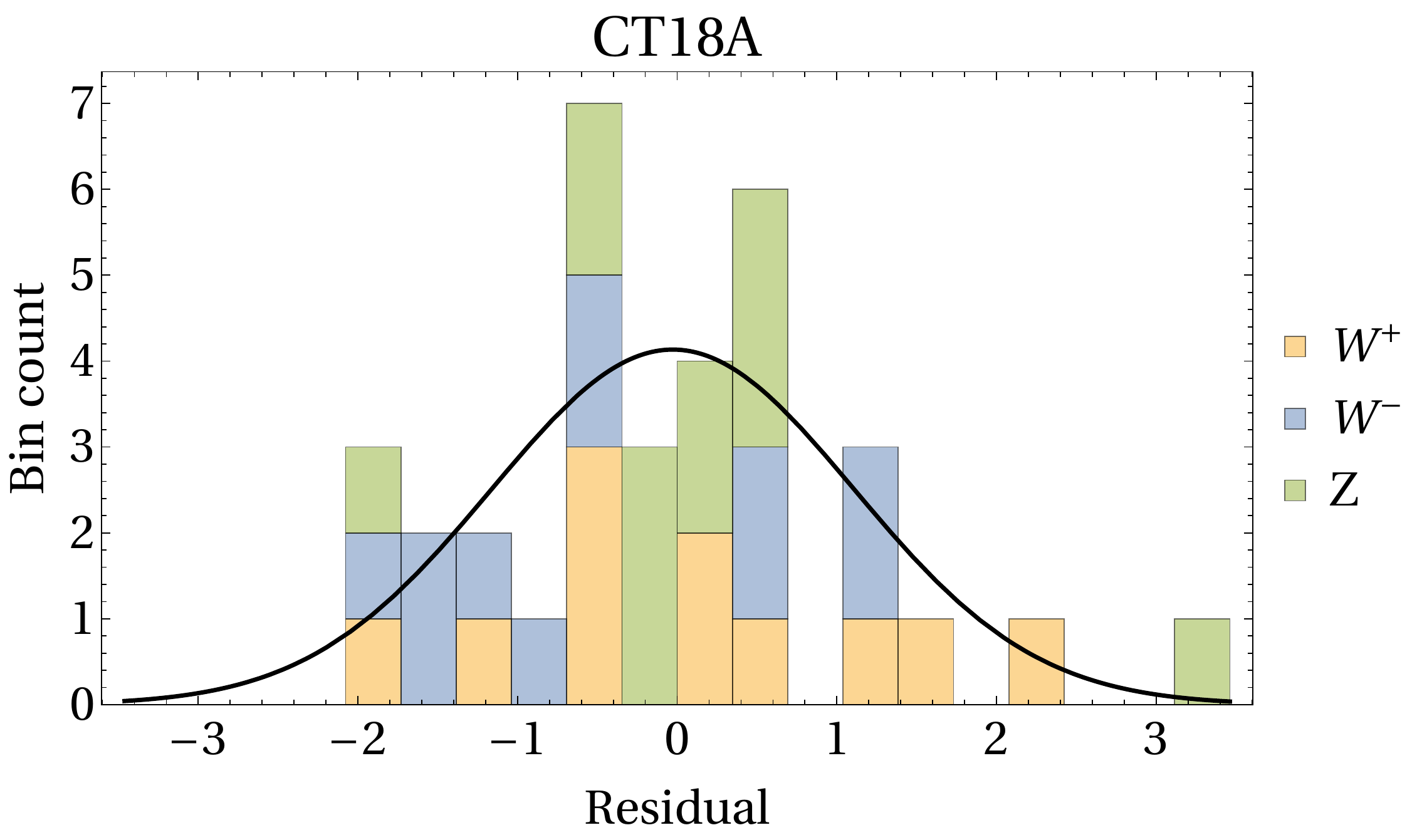}\quad \includegraphics[width=0.44\textwidth]{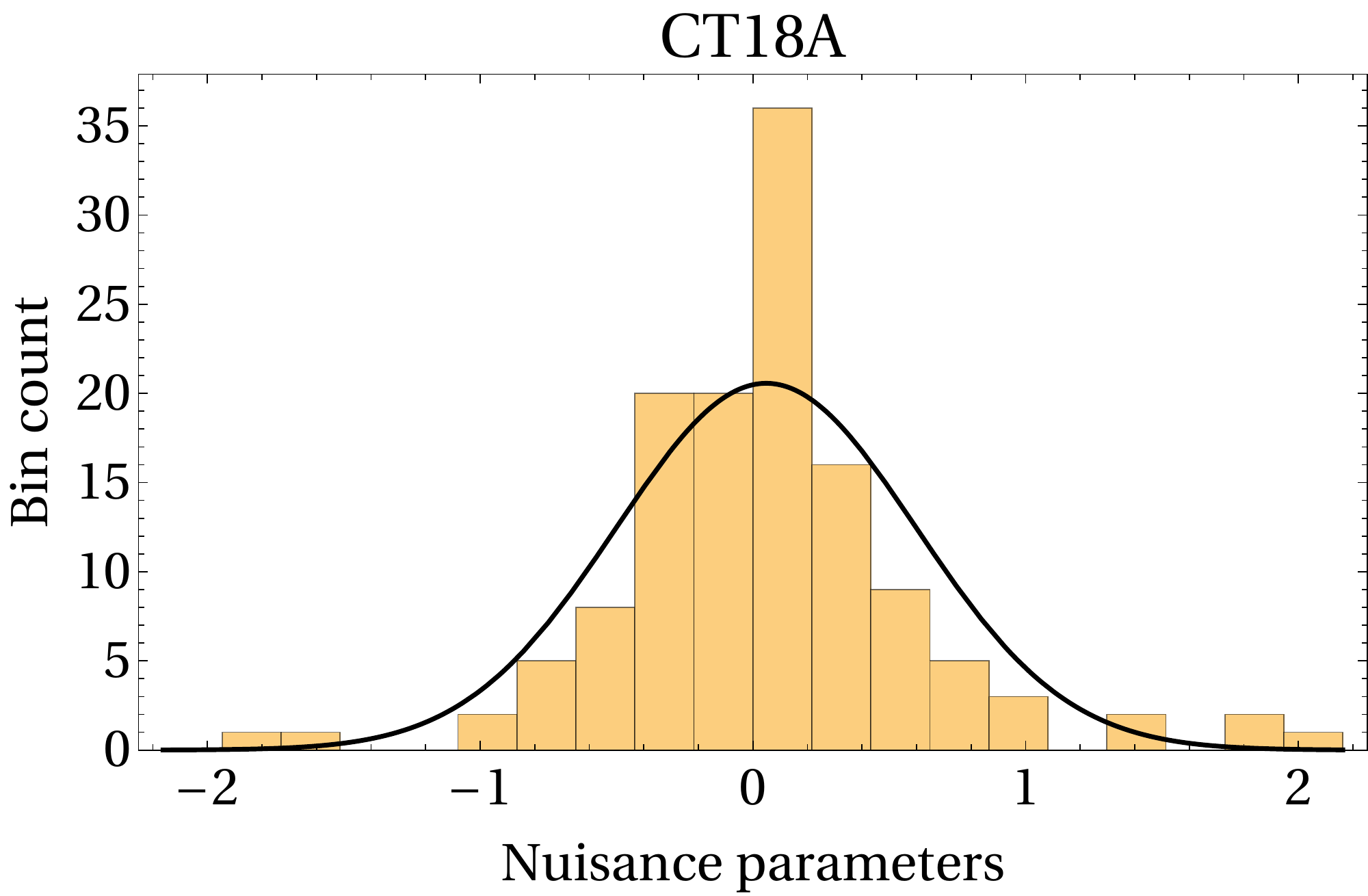}\\
(a)\hspace{2in} (b)
	\caption{
		Histogram of (a) the shifted residuals and (b) nuisance parameters for the ATL7ZW data obtained for CT18A NNLO.
	}
\label{fig:248_hist}
\end{figure}

The agreement with the ATL7ZW data set is impacted by the choice of the NNLO code for Drell-Yan pair production, as discussed at the end of App.~\ref{sec:Appendix4xFitter}, and also by the choice of the QCD scales, set equal to $\mu_{R,F}=Q$ in the CT18A/Z fits and to 
$\mu_{R,F}=Q/2$ in MMHT (2019), where $Q=M_W$  ($M_{\ell\bar \ell}$) in $W$ ($Z$) boson production.
Inclusion of associated $W$-boson and charm-jet ($W+c$)
production data in the fit, like the data set by CMS at 7 TeV \cite{Chatrchyan:2013uja} included by NNPDF3.1, tends to create an extra upward pull on $s(x,Q)$. We do not include $W+c$ measurements in CT18A(Z) yet because the full NNLO calculation is still unavailable. Neither are the NNLO massive heavy-quark contributions for differential cross sections of SIDIS dimuon production, needed to compute the detector acceptance when the fitted CCFR/NuTeV cross sections are reconstructed. 

\begin{figure}[tb]
\includegraphics[width=0.9\textwidth]{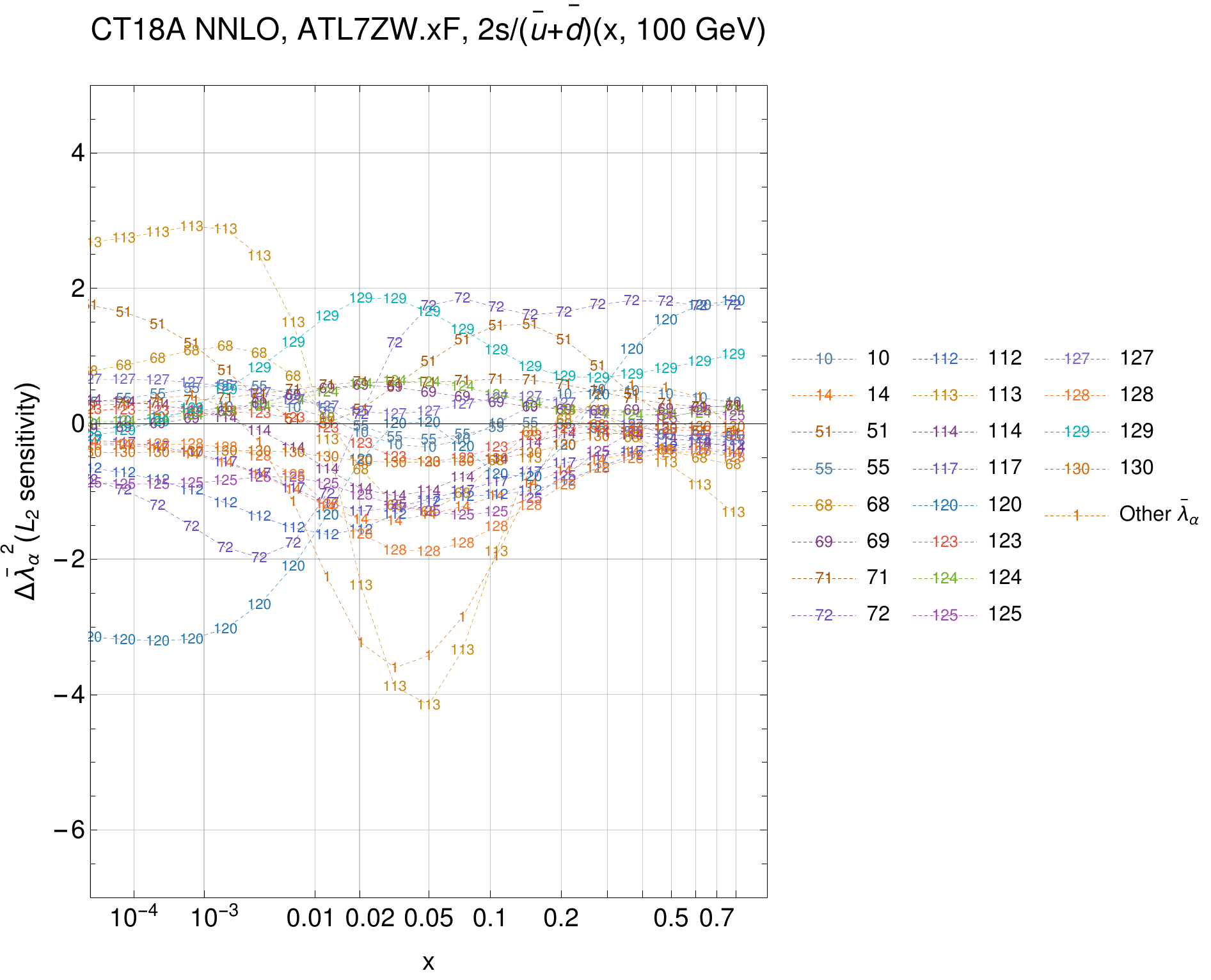}
	\caption{
		A plot, analogous to the $L_2$ sensitivity plots given above, showing the pulls of the correlated systematic uncertainties of the ATLAS $W/Z$ data
		on $R_s(x,Q=100\,\mathrm{GeV})$ in CT18A NNLO, as represented by shifts in the associated squared nuisance parameters, $\Delta \lambda^2_\alpha$.
	}
\label{fig:248_l2nui}
\end{figure}

\subsubsection{Comparisons to rapidity distributions}
\label{sec:ATL7ZWrap}

Figures~\ref{fig:248_DT}-\ref{fig:248_hist} show the theory-to-data
comparisons for CT18A/Z predictions against the ATL7ZW data set, the
distributions of shifted residuals in the final-state (pseudo)rapidity in
three individual channels, and the respective
cumulative histograms of the best-fit shifted
residuals $r_i$ and nuisance parameters $\bar\lambda_\alpha$.

In terms of the descriptions of the individual data points, the
figures indicate that CT18A/Z predictions describe $W^+$ production
well, while they show elevated differences with $W^-$ production
across the whole range of lepton pseudorapidities, as well with $Z$
production in the bins with the average $Z$-boson rapidity of 0.9 and 1.3.

Description of these data also require shifts of five nuisance
parameters $\bar \lambda_\alpha$,
labeled as 113, 129, 72, 125, and 128 in the ATLAS
data set, by $\approx \pm 2$ standard deviations. [These parameters
receive a mix of contributions from various systematic sources and
do not have a certain physics interpretation.]

A variation on the $L_2$ sensitivity technique allows us to demonstrate that
variations in these parameters are strongly linked to changes in the
strangeness ratio $R_s$ in the $x$ region probed by the ATL7ZW
measurement.
In the previous figures, the $L_2$ sensitivity approach explored the
connection between PDF variations and $\Delta \chi^2_E$
for individual experiments; but we can also compute
the $L_2$ correlation between the PDFs and contribution to $\chi^2$ from an individual nuisance parameter,  
$\Delta \bar \lambda^2_\alpha(a)$ 
for $\alpha=1,...,\ N_\lambda$, contributing to $\Delta \chi^2_E(a)$ according to Eq.~(\ref{Chi2a0l0}) for arbitrary $a\approx a_0$. In Fig.~\ref{fig:248_l2nui}, we demonstrate this for the ATL7ZW data, showing the
pulls of the ATL7WZ correlated systematic uncertainties on $R_s$ at $Q=100$ GeV in CT18A. While most nuisance parameter shifts
stay within $|\Delta \lambda^2_\alpha|\! \sim\! 1\!-\!2$ units, a
small collection of $\bar \lambda_\alpha$
are strongly sensitive to the fitted $R_s$ at various $x$,
notably parameters 113, 72, 120, and 129,
of which
some very nearly approach (parameters 72 and 129) or exceed (parameters 113 and 120)
the bound of $|\Delta \lambda^2_\alpha|\! <\!2$.

While we do not have information to reveal the specific causes driving
these systematic sources, it is clear that some have profound
effect on the preferred $R_s$ behavior in the intervals of $x$ that
can be read off Fig.~\ref{fig:248_l2nui}.

\subsubsection{Mini-summary}

The precision of the ATL7ZW data results in a strong impact on the strange quark distribution when included in the CT18 global PDF fits. The presence of the data in the fit (either CT18A or CT18Z) leads to a greatly improved $\chi^2$ for that data set, compared to a straight evaluation with the CT18 PDFs, but still significantly above one per degree-of-freedom. The effective Gaussian variable for this data set is also large, comparable only to that for HERA I+II. Overall, tensions 
with the other CT18 global data lead to a less consistent fit, as quantified by the strong goodness-of-fit criteria \cite{Kovarik:2019xvh}, and have resulted in the ATL7ZW data being included in the CT18A/Z PDFs, but not in the CT18 PDF. Additional precise LHC data may further resolve the strangeness issue. In the meantime, a full exploration of the strangeness uncertainty will require the use of the CT18A or CT18Z PDF sets.

\subsection{Scans on PDFs, $\alpha_s$, and $m_c$ in CT18Z
\label{sec:LMCT18Z}
}

We conclude this Appendix with a few results from Lagrange Multiplier
scans, the powerful technique applied at the end of the global
analysis cycle to obtain a close look at the exact probability
distributions that cannot be gleaned from the fast, but approximate,
Hessian studies. The results presented here for CT18Z NNLO
complement an analogous discussion for CT18
presented in Sec.~\ref{sec:QualityOverview}. By examining the
$\Delta \chi^2$ values that are returned from the full
fit rather than from the fast linearized approximations,
we discover subtle features such as the tensions
among experiments or instability/multiple solutions of the fit for
some PDF combinations. 

\begin{figure}[htbp]
\begin{center}
	\includegraphics[width=0.48\textwidth]{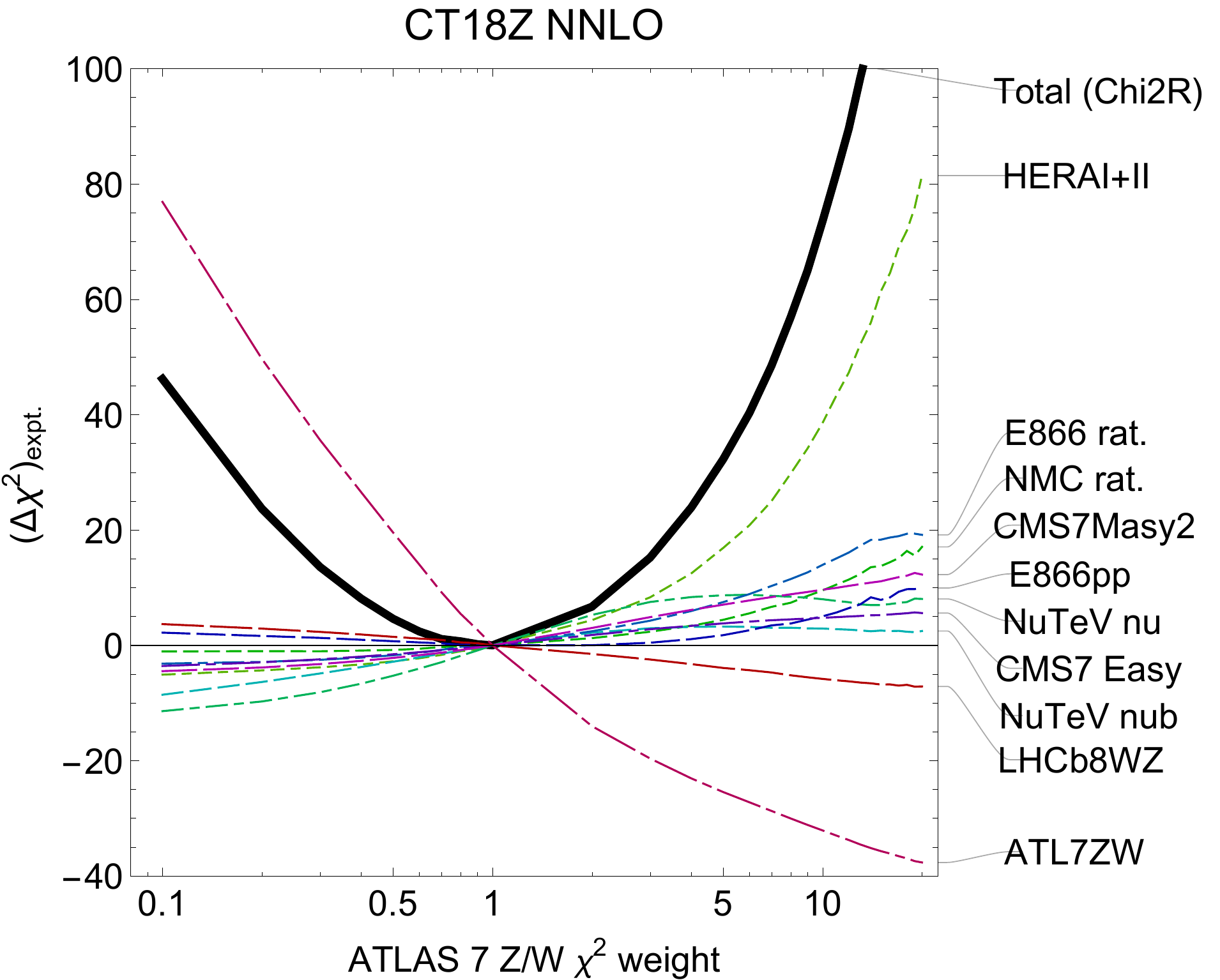}\quad
	\includegraphics[width=0.48\textwidth]{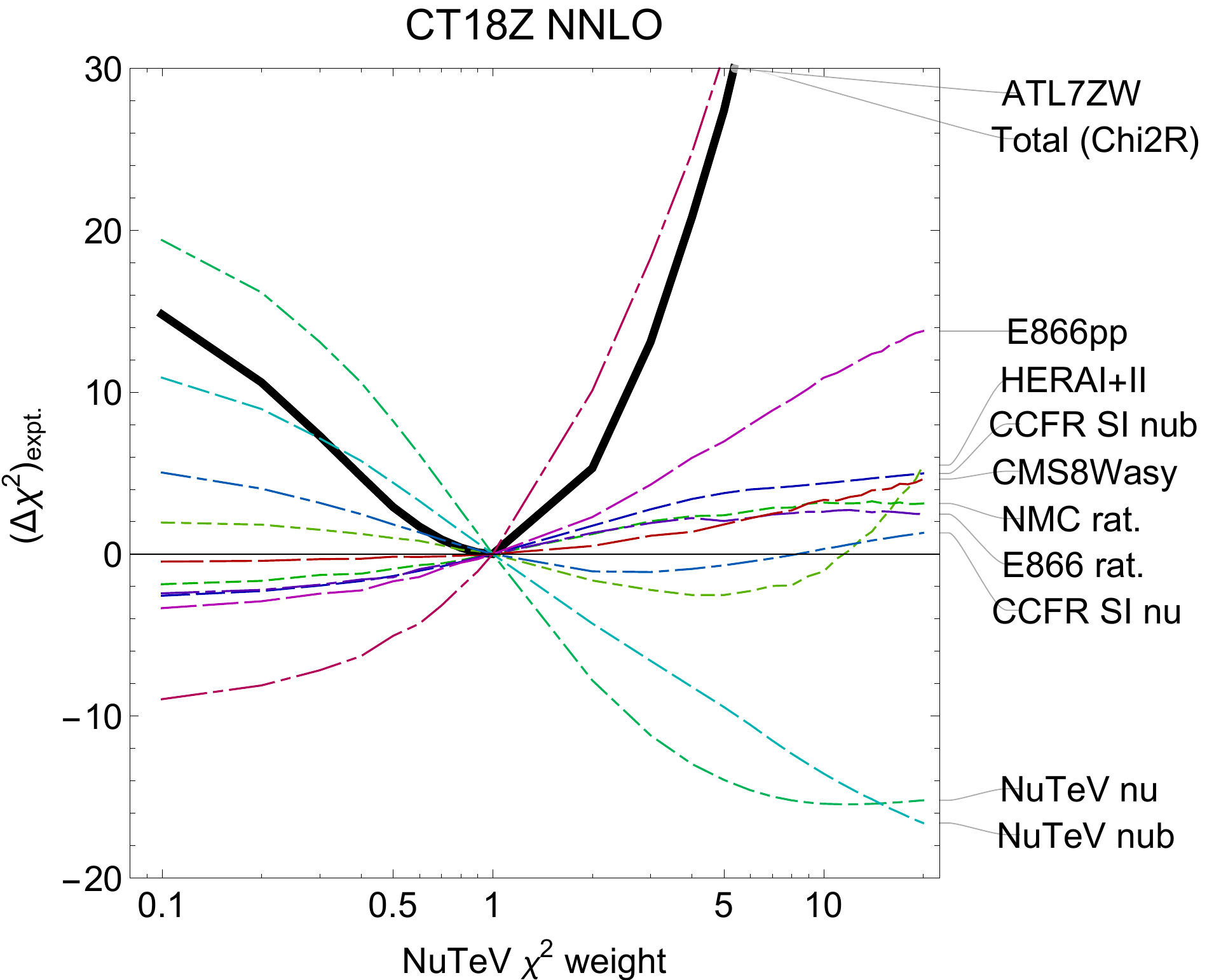}
	\caption{(Left panel) The change in total $\chi^2$, $\Delta \chi^2$, of the leading data sets included in the CT18Z NNLO fit
	when varying the weight of the ATL7ZW data away from weight $w\!=\!1$ as in the default fit. (Right panel)
	The analogous plot for the NuTeV $\nu,\, \bar{\nu}$ SIDIS dimuon data (Exp.~IDs=124, 125).
	}
\label{fig:lm_wht}
\end{center}
\end{figure}

\subsubsection{Scans with varied statistical weights of ATLAS 7 TeV $W/Z$ and NuTeV data}
\label{sec:ATL7ZWchi2wt}

The evident tension between the ATL7ZW data and other experiments can
be examined in terms of the variation of $\chi^2_E$ for a number of 
most sensitive experiments, when the weight of the ATL7ZW data is varied
within the CT18Z NNLO fit \cite{Collins:2001es}.  
The results of this are shown in the left panel of
Fig.~\ref{fig:lm_wht}, in which the weight of the ATL7ZW data is
continuously tuned from $w\!=\!0.1$ (sharply de-emphasizing this information, shifting $\chi^2_E$ by $\Delta \chi^2_E\! \sim\! +80$)
to $w\!=\!20$ (strongly over-weighting the ATLAS data, leading to a
$\Delta \chi^2_E\! \sim\! -40$ improvement in their description in CT18Z). The uncertainties in $\Delta \chi^2_E$ for the ATL7ZW data reported in Table~\ref{tab:CMN} correspond to an increase in the total $\chi^2$ of 36. 

\begin{figure}[tb]
\begin{center}
	\includegraphics[width=0.48\textwidth]{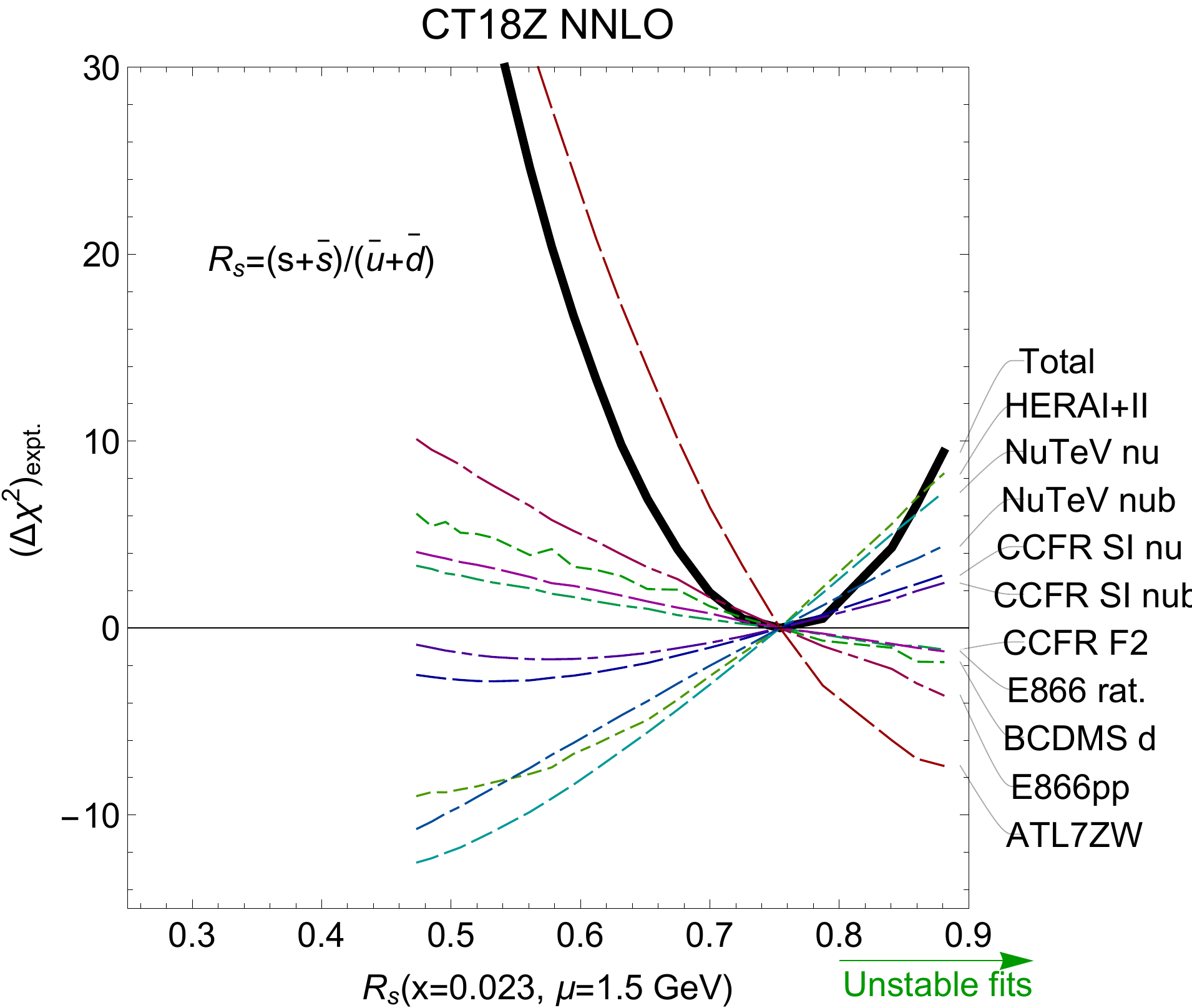}\quad
	\includegraphics[width=0.48\textwidth]{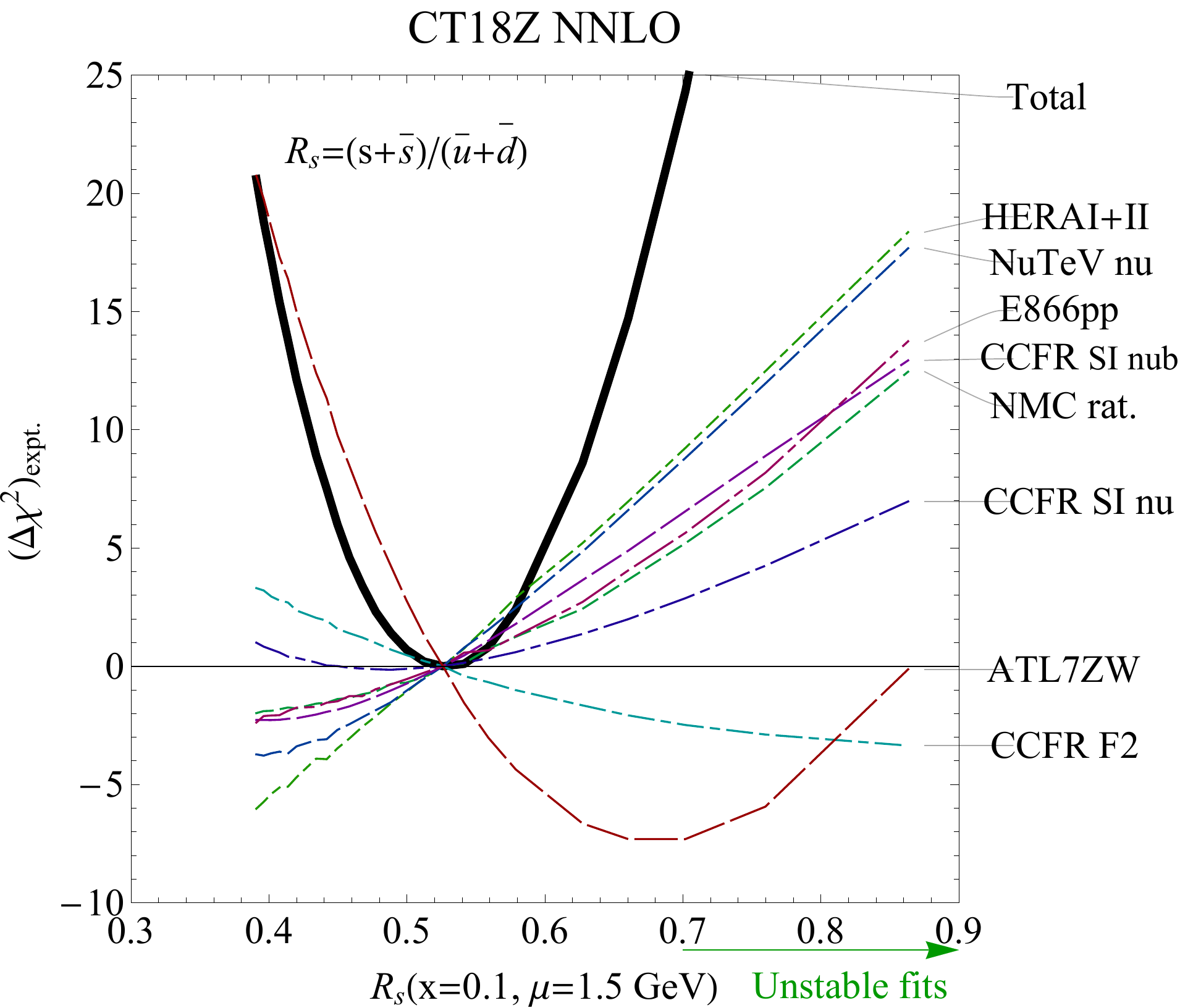}\\
(a)\hspace{2.6in}(b)\\
	\caption{The Lagrange Multiplier scan of $R_s$ at $Q=1.5$ GeV, $x=0.023$ and $x=0.1$ for the (a,b) CT18Z fit, analogous to Fig.~\ref{fig:LMRs} for CT18.
\label{fig:lm_rsz}}
\end{center}
\end{figure}

\begin{figure}[tb]
\begin{center}
	\includegraphics[width=0.6\textwidth]{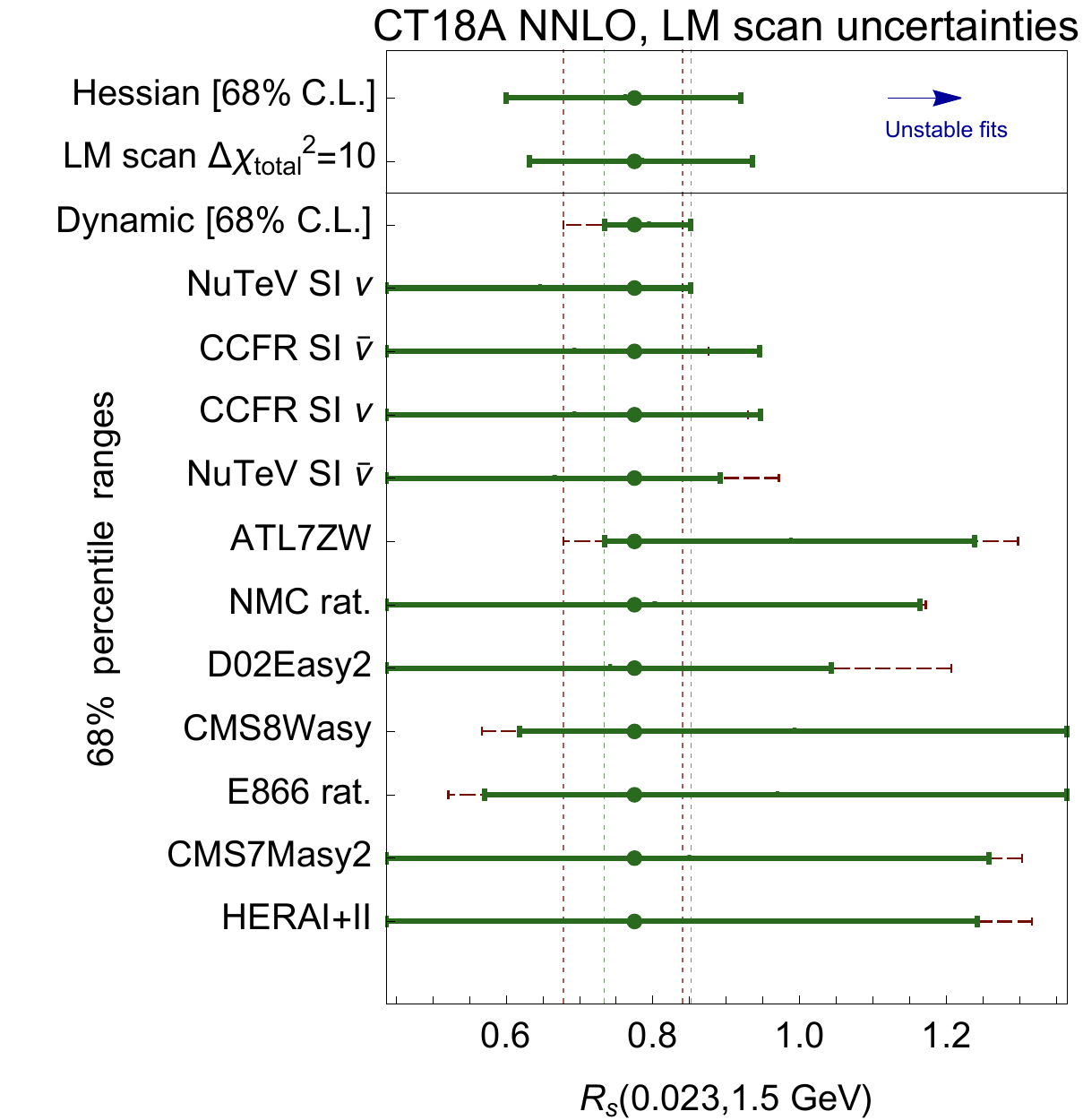}
	\caption{68\% C.\ L. uncertainty ranges on $R_s(0.023,1.5\mbox{ GeV})$ obtained based on $\chi^2$ variations and Hessian estimates. For the dynamic tolerance and tolerance ranges for individual experiments, the green solid (red dashed) bands correspond 
\label{fig:rserrors}}
\end{center}
\end{figure}

The figure also shows the curves for the experiments with the largest
variations in their $\Delta \chi^2_E$ when the $\chi^2$ weight of the
ATL7ZW data set is changed. 
A notable feature to observe in the left inset
is that it is the combined HERAI+II
inclusive DIS data set (Exp.~ID=160), rather than other experiments, that mostly
opposes including the ATL7ZW data set with a weight of 5-10.
With the exception of the LHCb 8 TeV $W/Z$
data (250), which is described better when the weight of ATL7ZW
is increased, the rest of the plotted experiments oppose such increase.
These experiments include E866 (203) and NMC (104) $p/d$ ratio,
NuTeV dimuon production (124, 125), and CMS 7 TeV $\mu$- and $e$-asymmetries
data (266 and 267).

As the ATLAS data are deemphasized ($w\!\sim\!0.1$),
the descriptions of the NuTeV SIDIS
dimuon-production data sets (Exp.~IDs=124, 125) are most improved,
further suggesting some tension.

These observations are consistent with the companion scan shown in
the right inset, which
similarly plots the $\Delta \chi^2$ variations of the experiments most
responsive to changing the weight of the NuTeV dimuon data.
Especially considering $N_{pt}$ for the plotted
experiments, the heavy over-weighing of the NuTeV data
leads to a very rapid deterioration of $\chi^2$ for the ATL7ZW points,
which in fact worsens more quickly than the full
CT18Z global analysis as the NuTeV weight is increased.
In fact, the $\chi^2_E$ for the inclusive HERA (160) and CCFR neutrino
dimuon production (126) mildly improves when the NuTeV weight is
increased to about 5. On the other hand, E866 $pp$ (204), CCFR
antineutrino dimuon (127), CMS 8 TeV charge asymmetry (249), and NMC $pd$
ratio (104) oppose such increase.

\subsubsection{LM scans on the strangeness ratio $R_s$ for CT18Z}
\label{sec:LMRsCT18Z}

Lagrange Multiplier scans on the strangeness ratio $R_s(x,Q)$, shown
for CT18Z NNLO in Fig.~\ref{fig:lm_rsz} at $Q\!=\!1.5$ GeV and two
representative momentum fractions, 
$x\!=\!0.023$ (in the left panel) and $x\!=\!0.1$ (in the right),
reveal several important features that are also observed in the other
LM scans we have performed. 

 Both panels of Fig.~\ref{fig:lm_rsz} reveal the opposing preferences
of, {\it e.g.}, the HERA DIS and NuTeV/CCFR SIDIS sets for a smaller value of
$R_s$, peaked more toward $R_s\!\sim\! 0.5$ at both values of $x$,
and the positive pull of the ATL7ZW data, supported by weaker pulls of
E866 $pd$ ratio (203), BCDMS $F_2^d$ (102), and especially
$F_2$ CCFR neutrino (110) data, which is very pronounced at $x=0.1$.

 When $R_s$ is forced to take values above 0.7-0.8, the $\chi^2$
starts to fluctuate irregularly in both scans and fails to converge in
many fits as $R_s$ is pushed to even higher values.

In reflection of the above strong trade-off between the pulls of the
DIS and ATL7ZW data sets, we observe that the Hessian uncertainty for $R_s$,
based either on the CT two-tier tolerance or dynamic tolerance used by MMHT, does not fully capture the true $\chi^2$ behavior revealed by the LM scans in Fig.~\ref{fig:lm_rsz}. We remind the reader that CTEQ-TEA Hessian eigenvector PDFs are constructed using a two-tier prescription \cite{Lai:2010vv,Gao:2013xoa} that prevents either too large increase in the global $\chi^2$ or too large increases in $\chi^2_E$ values of individual experiments. Figure~\ref{fig:rserrors} illustrates several estimates for 68\% C.L. uncertainty intervals for $R_s(0.023,1.5\mbox{ GeV})$ in the CT18A NNLO fit. [The findings are similar for the CT18Z NNLO fit.] The upper estimate is based on the CT18A Hessian eigenvector set, with the the 68\% C. L. estimate obtained by dividing the 90\% C.L. asymmetric uncertainty by 1.645. Right below the CT18A Hessian estimate, we show the interval corresponding to $\Delta \chi^2_{total}=10$ in the $R_s$ scan in the left Fig.~\ref{fig:lm_rsz}. We see that, while the CT Hessian uncertainty nominally corresponds to $\Delta \chi^2_{total}\approx 36$, in fact its uncertainty interval is comparable to that for the true $\Delta \chi^2_{total}\approx 10$ from the LM scan. The true uncertainty is wider than the Hessian estimate suggests. 

In the lower part of Fig.~\ref{fig:rserrors}, the solid green error bands show 68\% percentile ranges, $\xi_{68}$, plotted for $\chi^2_E$ distributions of individual experiments. The red dashed error bands are for these 68\% percentile ranges rescaled by $\chi^2_{E,0}/\xi_{50}$, where $\chi^2_{E,0}$ is the $\chi^2$ value achieved for experiment $E$ in the best global fit. From these, we construct an estimate, in the third line, for the 68\% C.L. uncertainty on $R_s(0.023,1.5\mbox{ GeV})$ according to the procedure in Sec. 6.2 of the MSTW'2008 analysis \cite{Martin:2009iq}. The resulting interval can be interpreted as the dynamic tolerance in the MSTW/MMHT approach for a particular eigenvector direction that is chosen to be along the gradient of $R_s(0.023,1.5\mbox{ GeV})$. [The other eigenvectors are perpendicular to the one shown and do not contribute to the uncertainty on $R_s(0.023,1.5\mbox{ GeV})$]. 

The dynamically estimated uncertainty in the third line is substantially narrower than the two uncertainty intervals above it. The dynamic uncertainty is equal to the small overlap between the rescaled 68\% percentiles for NuTeV neutrino and ATL7ZW data sets, which constrain the dynamic interval from above and from below. The dynamic uncertainty on $R_s(0.023,1.5\mbox{ GeV})$ computed this way will tend to be too small if some experiments strongly disagree with one another. 
Similarly, the tier-2 penalty in the CT analysis may result in too narrow uncertainties along some eigenvector directions if some experiments strongly disagree. 

\begin{figure}[tbp]
\centering
\includegraphics[width=0.49\textwidth]{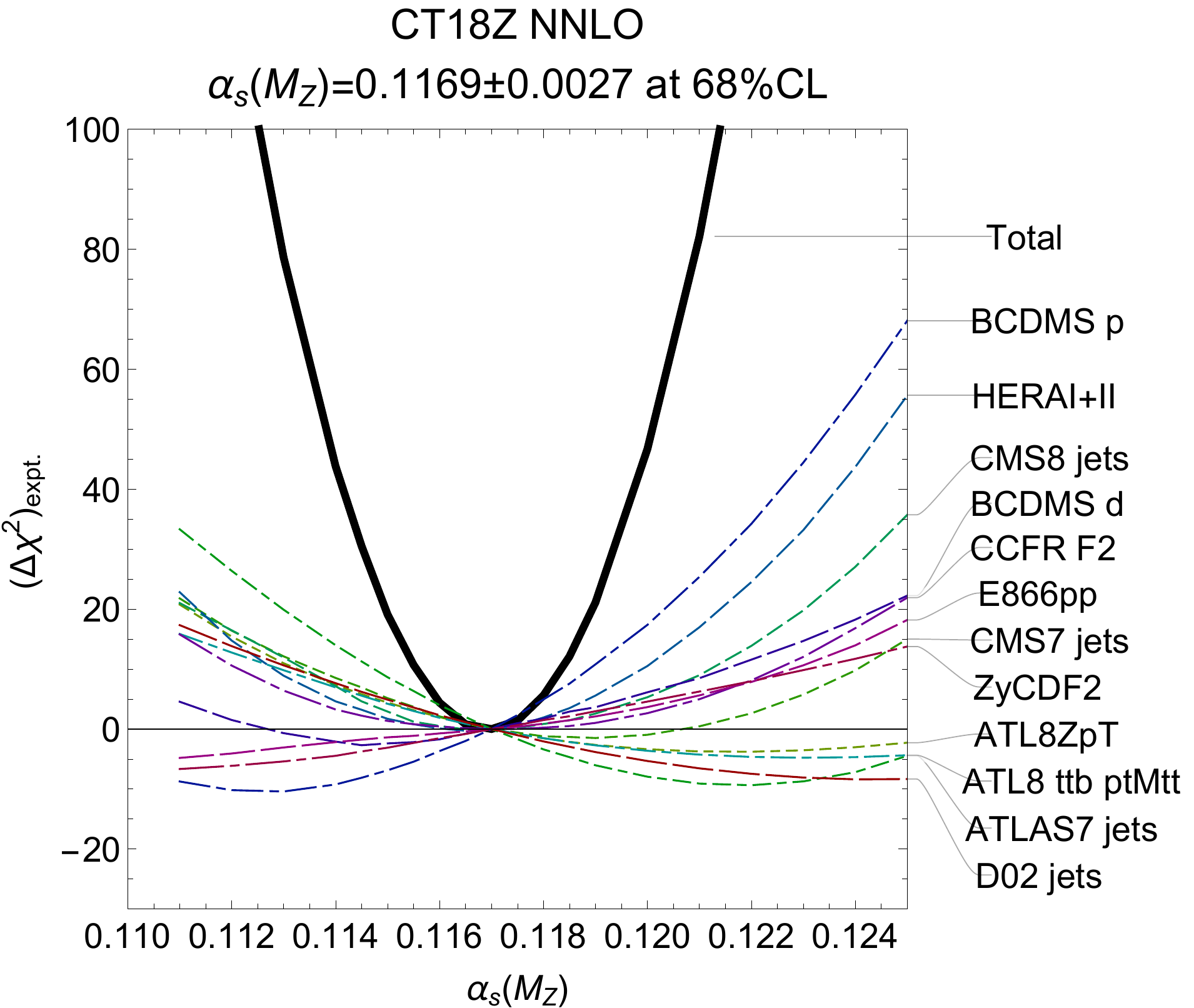}\quad
\includegraphics[width=0.46\textwidth]{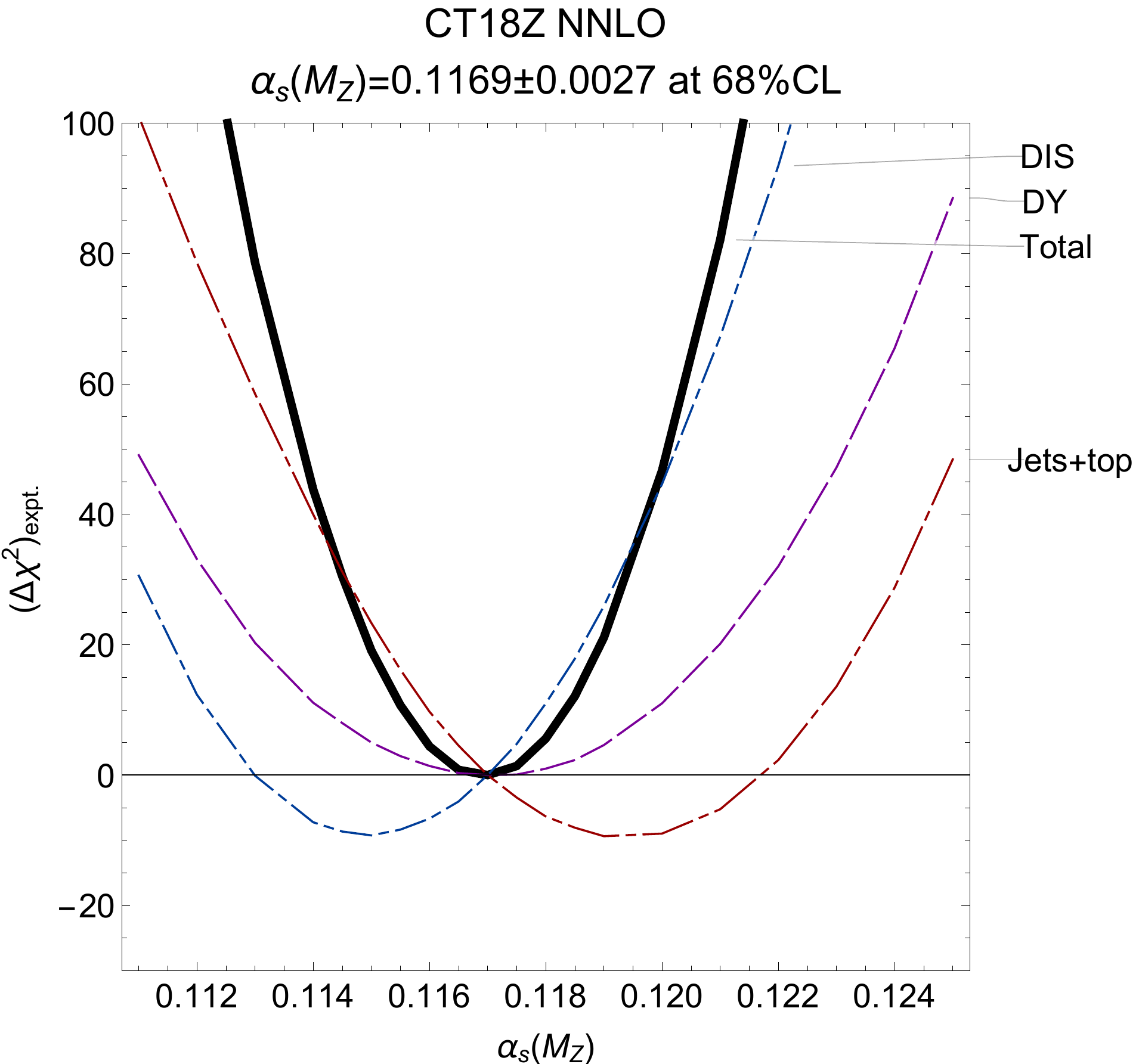}\quad
\caption{Scans over the strong coupling at the scale of $M_Z$ for CT18Z,
	analogous to Fig.~\ref{fig:lm_alphas} for CT18. As before, we show in the left panel
	the $\Delta \chi^2$ variations for a number of experiments with leading sensitivity
	to $\alpha_s(M_Z)$, while the right panel again shows the change in $\chi^2$ for all experiments
    fitted in CT18Z, separately collected into combined DIS, DY and top/jets data sets. Also as before,
    the ``Jets+top'' curve in the right panel is primarily influenced by the jet production data sets. Due to the
    intermediate impact of the $t\bar{t}$ data noted in Fig.~\ref{fig:lm_alphas}, variations in the
    CT18 fit leading to CT18Z are such that the ATLAS 8 TeV $t\bar{t}$ data (Exp.~ID=580) are now
    selected with the ensemble of sensitive experiments in the left panel.
\label{fig:lm_alphasz}}
\end{figure}

\subsubsection{CT18 LM scans on $\alpha_s(M_Z)$ and $m_c$}
\label{sec:CT18Zalphas}
The various modifications in CT18Z, especially the use of the
$x_B$-dependent scale $\mu_{F,x}$ in DIS experiments that provide the
largest coverage in the energy scale $Q$, modify 
the constraints on the strong coupling, $\alpha_s(M_Z)$,
for which we plot the scans in Fig.~\ref{fig:lm_alphasz}.

\begin{figure}[htbp]
\centering
\includegraphics[width=0.6\textwidth]{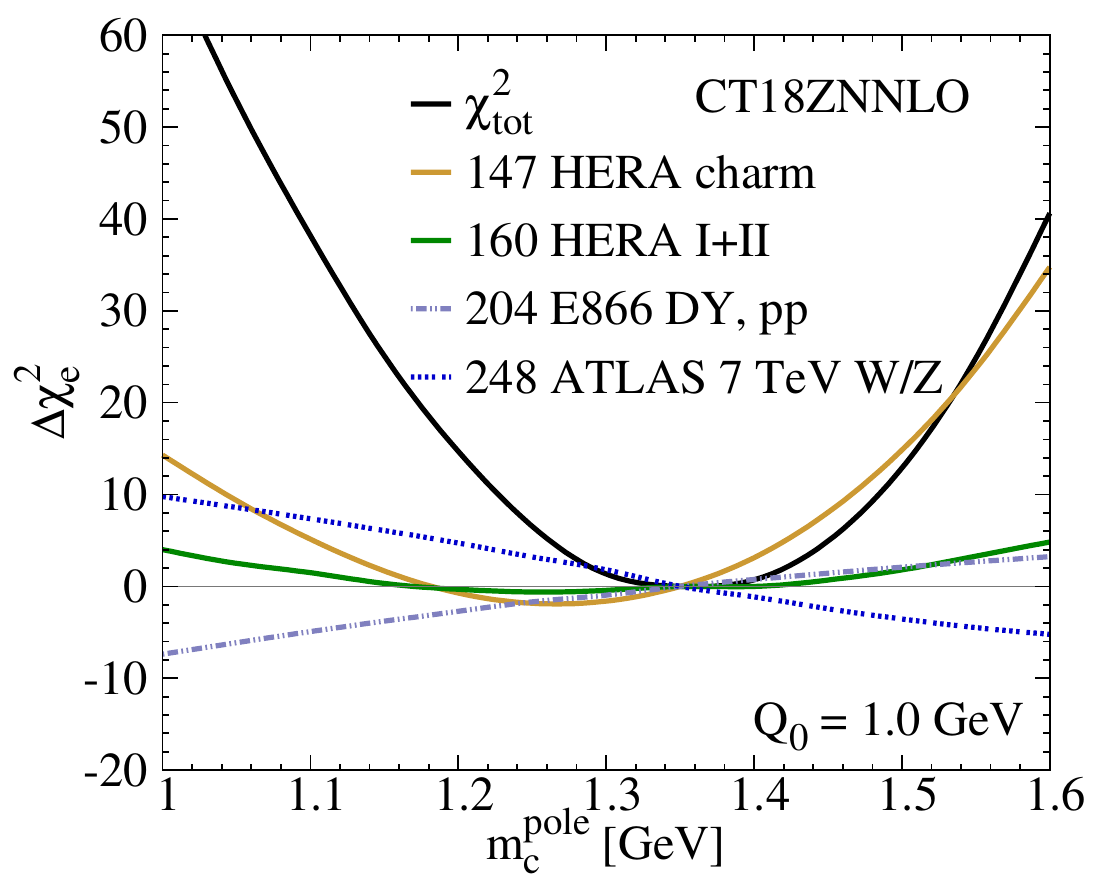}
\caption{
        Analogously to Fig.~\ref{fig:lm_mc}, the scan over values of the charm pole mass, $m_c$, computed here using CT18Z NNLO following the procedure described in Sec.~\ref{sec:AlphasDependence}. 
        }
\label{fig:lm_mcZ}
\end{figure}

Although the $68\%$ C.L.~determination of $\alpha_s(M_Z)\! =\!
0.1169\! \pm\! 0.0027$ agrees closely with the CT18-based
determination (with the latter being only slightly weaker and with an
identical uncertainty), the experimental pulls have notable
differences. Most strikingly, there is a separation in the preferences
of the combined inclusive jet-production data and Drell-Yan experiments,
which were aligned closely for CT18 in Fig.~\ref{fig:lm_mc}. The
combined Drell-Yan data (including ATL7ZW) now agree very closely with
the preferred value of the full CT18Z fit, but the DIS and
jet+$t\bar{t}$ data pull $\alpha_s(M_Z)$ in the opposite directions
more strongly than in CT18.

Such visible dependence of the preferred ranges for $\alpha_s(M_Z)$
from three categories of experiments on the DIS QCD scale may
indicate presence of important uncertainties beyond NNLO that are not
accounted in the nominal 68\% C.L. uncertainty of 0.0027.

Similarly, the choices made in the alternative CT18X/A/Z global
analyses can lead to different preferences in these fits for the
charm pole mass, $m_c^{pole}$.
While we described the $m_c$ scan in detail 
in the case of CT18 in Sec.~\ref{sec:AlphasDependence},
repeating this scan for CT18Z NNLO leads to a somewhat different
behavior shown in Fig.~\ref{fig:lm_mcZ}.
Although CT18Z ultimately arrives at a very similar central value of
$m_c$, the interplay of sensitive
experiments now is somewhat different than that
shown for CT18 in Fig.~\ref{fig:lm_mc}.
The pull of the combined HERA data (160) on $m_c$
decreases in CT18Z, the preferred $m_c$
of the HERA charm-production data (Exp.~ID=147)
increases slightly. In addition, the ATL7ZW data
also exhibit a modest preference for larger charm masses, and
these latter two experiments produce a small increase in the
central preferred value of $m_c$ in CT18Z relative to CT18,
although with a similar extent of uncertainty.

\clearpage

\section{CT18 goodness-of-fit function and treatment of correlated uncertainties}
\label{sec:chi2_app}
\newcommand{\betamat}{b}

In this appendix, we summarize the implementation of the
goodness-of-fit function $\chi^2$ and marginalization of nuisance
parameters in the CT18 family of fits. For the latter task, we
normally follow a procedure, adopted since the CTEQ6 analysis, to
estimate the correlated uncertainties using the correlation matrix
published by the experiment. For the few data sets that do not provide
the correlation matrix, we find it convenient to present the
covariance matrix in an approximate form that separates the
uncorrelated and correlated components. The algorithm for this
conversion is explained at the end of the appendix.

{\bf Definitions.}
In Eq.~(\ref{Chi2sys}) of Sec.~\ref{sec:chi2}, we introduced the standard
goodness-of-fit function $\chi^{2}$ used in the recent CT fits. Here
we review its treatment using a matrix notation.

We express $D_{k},$ $T_{k},$ and $\beta_{k\alpha}$
in units of $s_{k}$ for each $k$. That is, we introduce a vector
$d\equiv S^{-1}\left(D-T(a)\right)$ of length $N_{pt}$, where
\begin{equation}
S\equiv\mbox{diag}\left\{ s_{1},s_{2},...,s_{N_{pt}}\right\} .\label{S}
\end{equation}
Similarly, $\lambda\equiv\{\lambda_{\alpha}\}$ is a vector of length
$N_{\lambda}$; and $\betamat\equiv S^{-1}\beta$ is a rectangular matrix
of dimension $N_{pt}\times N_{\lambda}$.
In this matrix notation, Eq.~(\ref{Chi2sys}) takes the form
\begin{equation}
\chi_{E}^{2}(a,\lambda)=\left(d-\betamat\lambda\right)^{T}(d-
\betamat\lambda)+\lambda^{T}\lambda.\label{Chi2sys2}
\end{equation}

{\bf Minimization of $\chi^{2}$.}
The solution for the minimal value of $\chi^{2}$ (with respect to nuisance parameters) is found in terms
of the following matrices:
\begin{align}
\mathcal{A}\equiv\mathcal{I}+ {\betamat}^{T}\betamat;~~~ & C\equiv I+\betamat\betamat^{T};\label{ACm1}\\
\mathcal{A}^{-1}=\mathcal{I}-\betamat^{T}C^{-1}\betamat;~~~ & C^{-1}=I-\betamat\mathcal{A}^{-1}\betamat^{T}.\label{Am1C}
\end{align}
Uppercase roman and script letters denote matrices of dimensions $N_{pt}\times N_{pt}$
and $N_{\lambda}\times N_{\lambda}$, respectively. Therefore, $I$
is an $N_{pt}\times N_{pt}$ identity matrix, and $\mathcal{I}$ is
an $N_{\lambda}\times N_{\lambda}$ identity matrix. $C$ and $\mathcal{A}$
are covariance matrices (appropriately normalized by $s_{k}$) in
spaces of data values and nuisance parameters, respectively. The relations
between ${\cal A}^{-1}$ and $C^{-1}$ in Eq.~(\ref{Am1C}) can be
demonstrated by using $\betamat^{T}\betamat=\mathcal{A}-\mathcal{I}$ and
$\betamat\betamat^{T}=C-I$. The covariance matrices are symmetric: $\mathcal{A}^{T}=\mathcal{A},$
$C^{T}=C$.

For each theory input $a$, $\chi^{2}$ can be minimized with respect
to $\lambda$ analytically, by solving for $\partial\chi^{2}/\partial\lambda_{\alpha}=0$
\cite{Pumplin:2002vw}. The solution is
\begin{align}
 & \bar{\lambda}(a)=\mathcal{A}^{-1}\betamat^{T}d=\betamat^{T}C^{-1}d;\label{lambda0}\\
 & \bar{\lambda}^{T}(a)=d^{T}\betamat\mathcal{A}^{-1}=d^{T}C^{-1}\betamat.
\end{align}
 The global minimum $\chi^{2}(a_{0},\lambda_{0})$ for all experiments
$E$ can be found numerically as
\begin{equation}
\chi^{2}(a_{0},\lambda_{0})=\sum_{E}\ d_{0}^{T}C^{-1}d_{0},\label{Chi2CovMat}
\end{equation}
 with $d_{0}\equiv S^{-1}\left(D-T(a_{0})\right)$, $\lambda_{0}\equiv\bar{\lambda}(a_{0})$.

An equivalent form of this equation can be derived,
\begin{equation}
\chi^{2}(a_{0},\lambda_{0})=\sum_{E}\ \left(r_{0}^{T}r_{0}+\lambda_{0}^{T}\lambda_{0}\right),\label{Chi20rl}
\end{equation}
where $r_{0}$ are the best-fit shifted residuals:
\begin{equation}
r_{0}\equiv S^{-1}\left(d_{0}-\betamat\lambda_{0}\right)=\bar{r}(d_{0}),\mbox{~with~}\bar{r}(d)\equiv C^{-1}d.\label{r0}
\end{equation}
The representation (\ref{Chi20rl}) is particularly informative. We
anticipate that, in a good fit of theory to an experiment $E$, the
shifted residuals $r_{0}$, quantifying agreement with individual
data points, as well as the nuisance parameters $\bar{\lambda}_{0}$,
quantifying the systematic shifts, are distributed according to their
own standard normal distributions, ${\cal N}(0,1)$. Comparisons of
the two empirical distributions to the expected ${\cal N}(0,1)$ distributions
serve as the tests for the goodness of fit and for the implementation
of systematic errors \cite{Kovarik:2019xvh}.

{\bf Decomposition of the covariance matrix.}
The form of $\chi^2$ in Eq.~(\ref{Chi2CovMat1}) coincides with Eq.~(\ref{Chi2CovMat}),
obtained by optimizing the nuisance parameters $\lambda$ for a given $a$, in the prevalent
case when we separately know the uncorrelated and fully correlated
systematic errors. In this case, we identify
\begin{equation}
\mbox{cov}=SCS,~~~\mbox{cov}^{-1}=S^{-1}C^{-1}S^{-1}.\label{SCS}
\end{equation}
 The matrix elements, according to Eqs.~(\ref{ACm1}) and (\ref{Am1C}),
are
\begin{align}
\left(\mbox{cov}\right)_{ij} & =s_{i}^{2}\delta_{ij}+\sum_{\alpha=1}^{N_{\lambda}}\beta_{i\alpha}\beta_{j\alpha},\label{cov}\\
\left(\mbox{\mbox{cov}}^{-1}\right)_{ij} & =\frac{\delta_{ij}}{s_{i}^{2}}-\sum_{\alpha_1,\alpha_2=1}^{N_{\lambda}}\frac{\beta_{i\alpha_1}}{s_{i}^{2}}{\cal {\cal A}}_{\alpha_1\alpha_2}^{-1}\frac{\beta_{j\alpha_2}}{s_{j}^{2}}.\label{covm1}
\end{align}
In particular, a diagonal element $(\mbox{cov})_{ii}$ {[}no summation{]}
is the quadrature sum of the statistical, uncorrelated systematic,
and correlated systematic uncertainties:
\begin{equation}
(\mbox{cov})_{ii}=s_{i,\mbox{\scriptsize stat}}^{2}+s_{i,\mbox{\scriptsize uncor sys}}^{2}+s_{i,\mbox{\scriptsize cor sys}}^{2},\label{s2i}
\end{equation}
where $s_{i,\mbox{\scriptsize cor sys}}^{2}\equiv\sum_{\alpha}\beta_{i\alpha}^{2}$.
With the help of Eq.~(\ref{SCS}), the shifted residuals in Eq.~(\ref{r0})
then are written as
\begin{equation}
r_{i}\equiv\frac{D_{i}^{sh}(a)-T_{i}(a)}{s_{i}}=s_{i}\sum_{j=1}^{N_{pt}}\left(\mbox{\mbox{cov}}^{-1}\right)_{ij}\left(D_{i}-T_{i}(a)\right).\label{ricov}
\end{equation}
Computing them thus requires that we know the full uncorrelated error $s_{i}=\sqrt{s_{i,\mbox{\scriptsize stat}}^{2}+s_{i,\mbox{\scriptsize uncor sys}}^{2}}$.

{\bf Finding a correlation matrix from the covariance matrix.} In some experimental measurements, such as the LHCb 8 TeV W/Z production \cite{Aaij:2015zlq},
only the full covariance matrix is provided, making impossible the straightforward
reconstruction of the shifted residuals according to Eq.~(\ref{ricov}).
In several cases, we find it feasible to iteratively reconstruct the
approximate uncorrelated systematic and correlated systematic contributions,
$s_{k,\mbox{\scriptsize uncor sys}}$ and $\beta_{k\alpha}$, from the
provided covariance matrix $\mbox{cov}\equiv K_{0}$, by assuming
that the systematic shifts are dominated by a certain number $M_{\lambda}$,
with $M_{\lambda}\leq N_{\lambda}$, of fully correlated linear combinations.
The approximation makes use of the positive-definiteness of the covariance
matrix and its diagonal elements, cf.~Eq.~(\ref{s2i}). It represents
the original covariance matrix by a numerically close matrix
given by the sum of a diagonal matrix $\Sigma$, interpreted as consisting of total uncorrelated errors, and a non-diagonal square matrix $K$, interpreted as a product of the correlation matrix $\beta$ and its transpose.

In particular,
suppose we find the eigenvalues $x_{k}^{2}$ of $K_{0}$ and sort
them in the descending order:
\begin{align}
K_{0} & =O^{T}xO,\mbox{ with }x=\mbox{diag}\left\{ x_{1}^{2},x_{2}^{2},...,x_{N_{pt}}^{2}\right\} ,\label{K0}\\
 & x_{1}^{2}\geq x_{2}^{2}\geq...\geq x_{N_{pt}}^{2}>0.
\end{align}
Here $O$ is an orthogonal matrix.
We partition $x$ into a matrix $y$ containing the largest $M_{\lambda}$
eigenvalues and a matrix $z$ with the smallest ($N_{pt}-M_{\lambda})$
ones:
\begin{align}
y & \equiv\mbox{diag}\left\{ x_{1}^{2},x_{2}^{2},...,x_{M_{\lambda}}^{2},0,...,0\right\} ,\\
z & \equiv\mbox{diag}\left\{ 0,...,x_{M_{\lambda+1}}^{2},...,x_{N_{pt}}^{2}\right\} .\label{z}
\end{align}
Recalling that the diagonal elements of matrices $Y=O^{T}yO$ and
$Z=O^{T}zO$ are non-negative, we then construct a diagonal matrix
$\Sigma_{1}\equiv\mbox{diag}\left\{ Z_{11},...,Z_{N_{pt}N_{pt}}\right\} $
with $Z_{ii}>0,$ and another positive-definite matrix, $K_{1}\equiv Y+Z-\Sigma_{1}.$
We can iterate the steps in Eqs.~(\ref{K0})-(\ref{z}) by computing
$\Sigma_{a+1}$ and $K_{a+1}$ at step $a$ as
\begin{align}
\Sigma_{a+1} & \equiv\Sigma_{a}+\mbox{diag}\left\{ Z_{11},...,Z_{N_{pt}N_{pt}}\right\} ,\\
K_{a+1} & \equiv Y+Z-\mbox{diag}\left\{ Z_{11},...,Z_{N_{pt}N_{pt}}\right\} .
\end{align}
Here the matrices $Y$ and $Z$ are recomputed in each step using $K_{a}$ as the
input. After a sufficient number of steps $a$, the sum
\begin{equation}
C_{a}\equiv\Sigma_{a}+Y_{a}\label{Ca}
\end{equation}
approaches an asymptotic matrix that is close to the input matrix
$K_{0}$ in the sense of the $L_{p}$ norm $\sum_{i,j=1}^{N_{pt}}\left|(K_{0})_{ij}-(C_{a})_{ij}\right|^{p}$
with $p=2$ or $1$. If the extraction of the uncorrelated and fully
correlated components is feasible, the asymptotic $L_{p}$ distance
can be made small by choosing a large enough $M_{\lambda}$. For the
three experiments \cite{Aaij:2015gna,Aaij:2015zlq,Khachatryan:2016pev} in the CT18 data set that provide only the covariance
matrices, the $M_\lambda$ values giving good convergence
lie in the range between $N_{\lambda}/2$ and $N_{\lambda}$. 

By comparing
Eqs.~(\ref{cov}) and (\ref{Ca}), we identify
\begin{align}
(\Sigma_{a})_{ij} & \approx s_{i}^{2}\delta_{ij},\\
(Y_{a})_{ij} & \approx\sum_{\alpha=1}^{M_{\lambda}}\beta_{i\alpha}\beta_{j\alpha}.
\end{align}
Hence, $s_i$ is estimated from $(\Sigma_a)_{ij}$; and $\beta_{i\alpha}$ can be estimated from $(Y_{a})_{ij}$ by singular
values decomposition.

\section{Non-perturbative parametrization forms}
\label{sec:AppendixParam}

As noted in Sec.~\ref{sec:Paramstudies}, to obtain realistic estimates of the parametric PDF uncertainties, the CT global analyses explore a broad range of parametric forms for the PDFs at the starting scale, $Q\! =\! Q_0$. 
The goals of these investigations are ({\bf i}) to select a sufficiently flexible functional form capable of fitting an expansive high-energy data set without overfitting; and ({\bf ii}) to understand the uncertainties associated with the choice of parametrization. 

The appendix in the CT14 publication \cite{Dulat:2015mca} expounds our main rationales that guide the selection of the parametrization forms, including the ones adopted in the CT18 analysis. The general functional form in terms of the free parameters $a_k$ at the initial scale $Q_0$ is
\begin{equation}
    f_i(x,Q_0) = a_0 x^{a_1-1} (1-x)^{a_2} P_i(y; a_3, a_4,...).
    \label{eq:fiQ0}
\end{equation}
 The coefficients $a_1$ and $a_2$ control the asymptotic behavior of $f_i(x,Q_0)$ in the limits $x\rightarrow 0$ and $1$. $P_i(y;a_3,a_4, ...)$ is a sum of Bernstein polynomials (also called a B\'ezier curve) dependent on $y=f(x)$, such as $y\equiv \sqrt{x}$, that is very flexible across the whole interval $0<x<1$. While a variety of the functional forms $P_i(x; a_3, a_4,...)$ has been tried at the intermediate stages, in the nominal parametrization, we use the same parameters $a_1$ for valence PDFs and, separately, for sea PDFs. This choice guarantees that the ratios of the respective PDFs tend to finite values in the limit $x\to 0$. We also impose similar relations on the parameters $a_2$ for $x\to 1$. Finally, to reduce spurious correlations between the coefficients $a_2$ in $(1-x)^{a_2}$ with the rest of the  coefficients in $P_i(x;a_3,...)$ at $x\rightarrow 1$, we express some parameters in $P_i(x;a_3,...)$ in terms of the other parameters to eliminate the linear $(1-x)$ term in $P_i(x;a_3,...)$, {\it i.e.,} to have  
\begin{equation}
    \lim_{x\to 1} xf_i(x,Q_0) = a_0 (1-x)^{a_2} \left(1 + {\cal O}\left((1-x)^2\right)\right).
    \label{PixTo1}
\end{equation}
In the CT18 case, this procedure introduces relations in Eqs.~(\ref{eq:a6a1}) and (\ref{eq:a5a3}). For valence quarks, it allows us to achieve good $\chi^2$ values using four, rather than five or more, free parameters. The full procedure is explained around Eq.~(A.16) in \cite{Dulat:2015mca}.

\begin{table}[tb]
\begin{tabular*}{\textwidth}{c| @{\extracolsep{\fill}} cccccc}
\hline
best-fit parameters, &&&&&&\\              
  {\bf CT18} &  $u_v$          &   $d_v$          &    $g$            &   $u_\mathrm{sea}=\bar{u}$       &      $d_\mathrm{sea}=\bar{d}$       &    $s=\bar{s}$         \tabularnewline
\hline                                                                              
\% mom. fraction     & 32.5    & 13.4   & 38.5 & $2.8$ & $3.6$ & $1.3$ \tabularnewline
$a_0$                & $3.385^\mathrm{SR}$    & $0.490^\mathrm{SR}$   & $2.690$ & 0.414 & 0.414 & $0.288$ \tabularnewline                
$a_1$                                  &  $0.763$     &    $0.763$      &    $0.531$     &    $-0.022$      &      $-0.022$      &    $-0.022$     \tabularnewline
$a_2$                                  &  $3.036$     &    $3.036$      &    $3.148$     &    $7.737 $      &      $7.737$       &    $10.31$     \tabularnewline
$a_3$                                  &  $1.502$     &    $2.615$      &    $3.032$     &    $(4)$          &      $(4)$          &    $(4)$         \tabularnewline
$a_4$                                  &  $-0.147$    &    $1.828$      &    $-1.705$    &    $0.618 $      &      $0.292$       &    $0.466$      \tabularnewline
$a_5$                                  &  $1.671$     &    $2.721$      &      ---        &    $0.195 $      &      $0.647$       &    $0.466$      \tabularnewline
$a_6$                                  &     ---       &       ---        &      ---        &    $0.871 $      &      $0.474$       &    $0.225$      \tabularnewline
$a_7$                                  &     ---       &       ---        &      ---        &    $0.267 $      &      $0.741$       &    $0.225$      \tabularnewline
$a_8$                                  &     ---       &       ---        &      ---        &    $0.733 $      &        (1)          &      (1)         \tabularnewline
\hline                                                                                                              
\hline   \\                                                                                                          
{\bf CT18Z}          &  $u_v$          &   $d_v$          &    $g$            &   $u_\mathrm{sea}=\bar{u}$       &      $d_\mathrm{sea}=\bar{d}$       &    $s=\bar{s}$            \tabularnewline
\hline                                                                              \% mom. fraction     & 31.8    & 13.1   & 38.2 & $2.9$ & $3.7$ & $1.9$ \tabularnewline
$a_0$                & $3.631^\mathrm{SR}$    & $0.254^\mathrm{SR}$   & $0.668$ & 0.519 & 0.519 & 0.524 \tabularnewline         
                              
$a_1$                                  &  $0.787 $    &    $0.787$      &    $0.289 $    &    $0.0096 $      &      $0.0096$       &    $0.0096 $     \tabularnewline
$a_2$                                  &  $3.148 $    &    $3.148$      &    $1.872 $    &    $8.27 $      &      $8.27$       &    $11.4$     \tabularnewline
$a_3$                                  &  $1.559 $    &    $3.502$      &    $3.538 $    &    $(4)$          &      $(4)$          &    $(4)$         \tabularnewline
$a_4$                                  &  $-0.075$    &    $1.865$      &    $-1.665$    &    $0.679 $      &      $0.300$       &    $0.653 $     \tabularnewline
$a_5$                                  &  $1.605 $    &    $3.599$      &      ---        &    $-0.0016$      &      $0.532$       &    $0.653 $     \tabularnewline
$a_6$                                  &     ---       &       ---        &      ---        &    $1.085 $      &      $0.753$       &    $0.054 $     \tabularnewline
$a_7$                                  &     ---       &       ---        &      ---        &    $0.045 $      &      $0.440$       &    $0.054 $     \tabularnewline
$a_8$                                  &     ---       &       ---        &      ---        &    $0.759 $      &        (1)          &      (1)         \tabularnewline     %
\hline                                                                                                              
\end{tabular*}\caption{
	Percent momentum fractions at $Q_0=1.3 $ GeV and best-fit parameter values for the PDFs of the CT18 and CT18Z NNLO fits. The percent momentum fractions are
	evaluated as $\langle x \rangle_{f}$ based on the listed parameters for each parton flavor, such that the total sea-quark percent momenta entering the momentum
	sum rule are twice the values given above for $u_\mathrm{sea}$, $d_\mathrm{sea}$, and $s=\bar{s}$, in each case. The functional forms associated with each
	parametrization are defined explicitly in the text of this section, Eqs.~(\ref{eq:v_param})--(\ref{eq:ubar_exp}). Those entries
	corresponding to parameters that are not actively fitted for a given PDF flavor are indicated by a dash, ``---.'' Values in parentheses indicate fixed parameters. 
	As indicated above with the $\#^\mathrm{SR}$ annotation, normalizations $a_0$ for $u_v$, $d_v$, are derived using the valence sum rules (SR), while the rest of
	the normalizations are derived from $\langle x\rangle_g$ and $\langle x\rangle_{\bar s+\bar s}/\langle x\rangle_{\bar u+\bar d}$ fitted as free parameters, and
	$\langle x\rangle_{\bar u+\bar d+\bar s}$ computed using the momentum sum rule.  
}
\label{tab:parameters}
\end{table}

In CT18, the nominal nonperturbative parametrization is  generally similar to the one that served as the basis for \CTHERAII, but with a slightly more flexible parametrization for the sea-quark distributions. This enhanced flexibility is necessitated by the inclusion of LHC Run-1 data with direct sensitivity to the sea-quark content of the nucleon.
The functional form that parametrizes the starting-scale valence distributions, $u_v$ and $d_v$, is a polynomial in the
parameter $y\! \equiv\! \sqrt{x}$,
\begin{eqnarray}
\label{eq:v_param}
q_v(x,Q=Q_0) &=& a_0\, x^{a_1-1}(1-x)^{a_2}\,P^v_a(y), \\
P^v_a(y) &=& \sinh{[a_3]}         (1-y)^4 + 
\sinh{[a_4]}   4 y   (1-y)^3 + 
\sinh{[a_5]}   6 y^2 (1-y)^2 + 
a_6   4 y^3 (1-y)   + 
y^4,             \nonumber 
\end{eqnarray}
where 
\begin{eqnarray}
a_6 &=& 1 + \frac{1}{2} a_1\  \label{eq:a6a1}
\end{eqnarray}
 to satisfy Eq.~(\ref{PixTo1}). 

We emphasize that, while the nonperturbative forms for the
$u_v$ and $d_v$ distributions are the same, the parameters of $P^v_a(y)$ 
for these flavors are separately fitted; although we impose
constraints on the prefactor exponents, $a^{u_v}_1\! =\! a^{d_v}_1$ and
$a^{u_v}_2\! =\! a^{d_v}_2$, to ensure that flavor ratios are well-behaved
and consistent with Regge expectations and quark counting rules 
in the limits $x\! \to\! 0,1$.

For the gluon PDF, we also fit a polynomial in $y\! =\! \sqrt{x}$,
but with the form
\begin{eqnarray}
g(x,Q=Q_0) &=& a_0\, x^{a_1-1}(1-x)^{a_2}\,P^g_a(y), \\
P^g_a(y) &=& \sinh{[a_3]}         (1-y)^3 + 
\sinh{[a_4]}   3 y   (1-y)^2 + 
a_5   3 y^2 (1-y)   + 
y^3, \nonumber
\end{eqnarray}
in which the $a_5$ parameter is fixed to $a_1$ as
\begin{eqnarray}
a_5 &=& ( 3 + 2 a_1 ) \big/ 3. \label{eq:a5a3}
\end{eqnarray}

For the distributions of the light-quark sea, we fit somewhat
more flexible distributions relative to CT14. In these cases, we use
polynomials in $y\! \equiv\! 1\! -\! (1\! -\! \sqrt{x})^{a_3}$ for
$\bar{u}$, $\bar{d}$, and $\bar{s}=s$, where we fix $a_3\! =\! 4$ for
all three sea distributions.
We parametrize the sea-quark PDFs, $\bar{q}(x,Q_0)$ as
\begin{eqnarray}
\label{eq:s_param}
\bar{q}(x,Q=Q_0) &=& a_0\, x^{a_1-1}(1-x)^{a_2}\,P^{\bar{q}}_a(y), \\
P^{\bar{q}}_a(y) &=&             (1-y)^5 + 
a_4   5 y   (1-y)^4 + 
a_5  10 y^2 (1-y)^3 + 
a_6  10 y^3 (1-y)^2   \nonumber \\
&+& a_7   5 y^4 (1-y)   + 
a_8     y^5\ . \nonumber 
\end{eqnarray}
While the full parametric form of Eq.~(\ref{eq:s_param}) [with $a_3 = 4$ fixed] is used
for the $\bar{u}$-PDF, for $\bar{d}$ we fix $a_8=1$. Owing to the comparative lack of empirical constraints to nucleon strangeness, not all the parameters above are permitted to float freely for $s(x,Q)$ in CT18(Z); rather, we constrain $a_4\! =\! a_5$, $a_6\! =\! a_7$, and $a_8=1$ for the strangeness parameters.

As with the valence distributions, we constrain the prefactor exponents as
\begin{equation}
	a^{\bar{u}}_1 = a^{\bar{d}}_1 = a^s_1\ ,
\label{eq:ubar_exp}
\end{equation}
to ensure a finite value for the strangeness suppression ratio, $R_s\! =\! [s+\bar{s}] \big/ [\bar{u}+\bar{d}]$,
in the limit $x\!\to\!0$, and the convergence of $\int^1_0 dx\, [\bar{d}-\bar{u}]$. We bind the high-$x$ exponents of the $\bar{u}$- and $\bar{d}$-distributions, $a^{\bar{u}}_2\! =\! a^{\bar{d}}_2$, to stabilize $\bar{d}/\bar{u}$ for $x\!\to\!1$. Normalizations for individual sea quark PDFs are computed using the valence quark and momentum sum rules, and the first moments $\langle x \rangle_{g}$ and the ratio $\langle x\rangle_{\bar s+\bar s}/\langle x\rangle_{\bar u+\bar d}$ fitted as free parameters. Since the parametrizations do not determine the ratio of the strange-to-nonstrange PDFs, we restrict the ratio $\left(s(x,Q_0) + \bar s(x,Q_0)\right)/\left(\bar u(x,Q_0) + \bar d(x,Q_0)\right)$ to be in the intervals $[0.2, 2.0]$ at $x=10^{-8}$, and $[0.4, 1.8]$ at $x=10^{-5}$ by imposing appropriate Lagrange Multiplier constraints.

In Table~\ref{tab:parameters}, we summarize the central fitted values of the parameters noted above for CT18 (upper rows) and CT18Z (lower rows). Those parameters entries marked with ``$-$'' do not participate as degrees of freedom for the associated flavor.

\section{Fitting code developments}
\label{sec:AppendixCodeDevelopment}
The inclusion of more than 10 new LHC data sets into the CT18 analysis has necessitated substantial upgrades in the CTEQ global analysis software.
The usage of various fast interfaces for calculations of hard-scattering cross sections became mandatory and conventional.
Even after implementing the fast interfaces such as \texttt{ApplGrid}, the precision global fit is a time-consuming task due to the large size
of experimental data and the need to explore the multi-parametric probability distribution to estimate the PDF uncertainties. When transiting from CT14 to CT18, the CTEQ-TEA code was revised to parallelize various operations that were done sequentially in the past. 

\begin{figure}[tb]
	\begin{center}
		\includegraphics[width=0.99\textwidth]{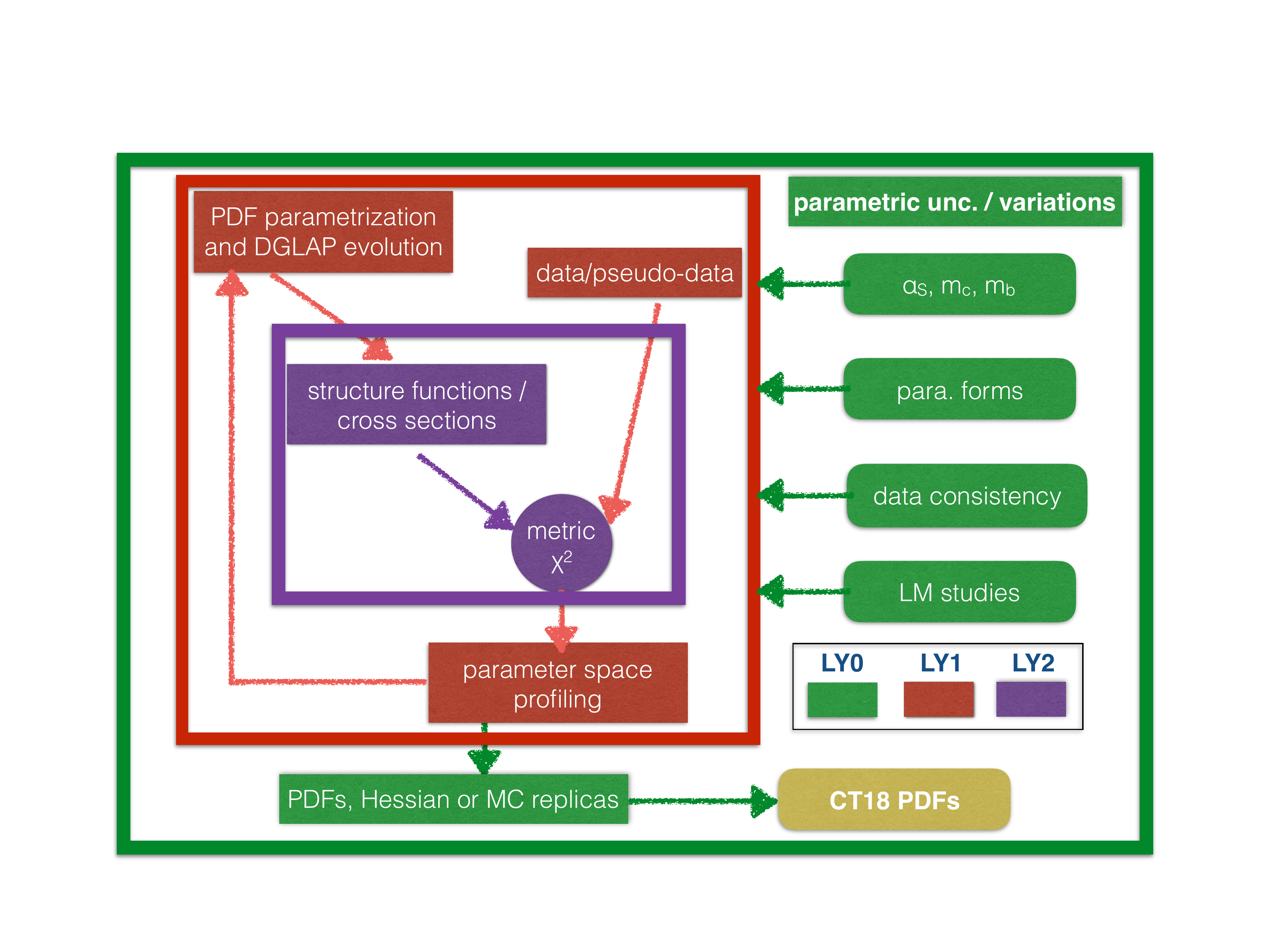}
	\end{center}
	\vspace{-2ex}
	\caption{\label{fig:gfit}
		A flowchart of the CT18 global analysis consists of the outer layer (LY0),
		middle layer (LY1), and core process (LY2).
	}
\end{figure}

The computational process of a typical CTEQ-TEA global analysis can be divided into three stages, or "layers", as visualized in Fig.~\ref{fig:gfit}.

The outer layer, LY0, corresponds to repeating the global fit with varied inputs or
constraints, e.g., for different values of the QCD coupling, heavy-quark masses, nonperturbative functional forms, or Lagrange Multiplier constraints.

The middle layer, LY1, corresponds to a single such fit, in which the program scans the PDF parameter space and constructs the probability ($\chi^2$) distribution for a fixed combination of inputs.
This step provides the best-fit and error PDF sets that quantify the uncertainty in the parameter space.

Within the LY1 layer, the calculation of global $\chi^2$, or "layer LY2", is the core part of the fit that will
be repeated for every combination of the PDF parameters.
LY2 calculates cross sections for thousands of data points and computes individual $\chi^2$ for each data set.

Most computational efforts shown in Fig.~\ref{fig:gfit} can be parallelized.
For instance, at the LY0 stage, one can simply submit thousands of simultaneous fits with varied inputs to a large computing cluster, since those fits are
independent. To obtain a diverse battery of results presented in the CT18(Z) analyses, including the computationally expensive Lagrange Multiplier scans, the high-performance computing clusters at MSU and SMU were used. 

At LY1, the major task is to find a global minimum of the $\chi^2$ in a parameter space with large dimensions ($N_{par}\sim 30$).
The choice of the suitable parallelization technique depends strongly on the minimization algorithm.
The CT18 analysis uses the variable-metric gradient descent method implemented in the \texttt{MINUIT} package~\cite{James:2004xla}.
It involves numerical calculations of the first- and 
second-order derivatives of $\chi^2$, combined with sequential minimum searches along fixed directions
in the PDF parameter space.
The calculations of derivatives are highly parallelizable, with the CPU-time expenditures scaling with the number $N_{par}$ of parameters approximately as
$1/N_\mathit{par}$ and $2/(N^2_\mathit{par}\!+\!N_\mathit{par})$, respectively.

In the core part, again the calculation of individual $\chi^2$ for different data sets, including their cross sections, is now done simultaneously.
While either \texttt{MPI} or \texttt{OpenMP} parallelization protocols are suitable at this stage, the latter is restricted to platforms with shared memory, but required fewer revisions in our fitting code.
Specifically, we used an approach similar to \texttt{OpenMP} to reduce the scope of changes inside our fitting code.
When computing $\chi^2_E$ values for the experiments, the \verb|fork| Linux command splits the main program into multiple threads,
each with a copy of the master memory and carrying out the calculations independently.
Later the \verb|join| command collects the  results from the individual threads and returns them to the main program.
The implementation of this \verb|fork-join| algorithm is borrowed from the widely-used
CUBA library~\cite{Hahn:2004fe} for multi-thread Monte-Carlo integration.   

\section{Decorrelation of ATLAS jet-production cross-section data}
\label{sec:ATLASjetdecorrel}
It has been observed in our global analysis, as well as others \cite{Harland-Lang:2017ytb}, that achieving a robust theoretical description of the Run-1 LHC jet production data generally requires numerical prescriptions to decorrelate select correlated systematic uncertainties. In our work, we follow the recommendations of experimental collaborations in applying such decorrelation procedures.

For example, while the ATLAS 7 TeV inclusive jet production data \cite{Aad:2014vwa} (Exp.~ID=544) produced an unacceptably high $\chi^2$ when fitted out of the box, the agreement of data and NNLO theory was improved by applying some decorrelation options
proposed by the ATLAS collaboration and summarized in the Appendix of Ref.~\cite{Aaboud:2017dvo}. 
The CT18(Z) fits implement ATLAS inclusive jet cross sections defined with an $R=0.6$ anti-$k_t$ jet algorithm. With this data set, we tested
the decorrelation procedures summarized in Table 6 of Ref.~\cite{Aaboud:2017dvo} and 
found some to have a substantial impact on the $\chi^2_E/N_\mathit{pt,E}$ of the ATLAS 7 TeV data. In the end, we followed the specific recommendations
of the ATLAS experimentalists themselves, who advocate \cite{Bogdan} the decorrelation of two jet energy scale (JES) uncertainties, associated with the JES MJB fragmentation (JES16) and the JES flavor response
(JES62). We obtained the greatest $\chi^2$ reduction using the upper portion of Table~6 in Ref.~\cite{Aaboud:2017dvo},
by decorrelating JES16 and JES62 according to Options 17 and 14 of Table~4, respectively. 

Ref.~\cite{Aaboud:2017dvo}
also details decorrelation options for select uncertainties in experimental simulations; 
of these, we specifically considered the effect of
decorrelating the error associated with the nonperturbative correction (`eNPC'), but found it to have negligible impact
on the quality of the \CTHERAII~fit according to $\chi^2$. 

The JES decorrelation that we have implemented proceeds by breaking a given correlated uncertainty into 2-3 subsidiary errors such that their sum in quadrature recovers the original correlated uncertainty. This procedure is bin-dependent; the resulting decorrelated
errors depend on the jet's rapidity and transverse momentum. For example, we decorrelate the JES MJB fragmentation error, $\delta_{16}$, into
three components given by
\begin{align}
	\delta^a_{16} &= \delta_{16}\, \Big\{ 1 - L^2\big( \ln(p_T[\mathrm{TeV}]),\ln(0.1),\ln(2.5) \big) \Big\}^{1/2}\, \left( 1 - L^2\big(|y|,0,1\big) \right)^{1/2}, \nonumber \\
	\delta^b_{16} &= \delta_{16}\, \Big\{ 1 - L^2\big( \ln(p_T[\mathrm{TeV}]),\ln(0.1),\ln(2.5) \big) \Big\}^{1/2}\, L\big(|y|,1,3\big), \nonumber \\
	\delta^c_{16} &= \big( \delta^2_{16} - (\delta^a_{16})^2 - (\delta^b_{16})^2 \big)^{1/2}\ ,
\label{eq:dec16}
\end{align}
where
\begin{equation}
L(x,x_\mathit{min},x_\mathit{max}) \equiv \frac{x - x_\mathit{min}}{x_\mathit{max} - x_\mathit{min}}\ \iff x \in [x_\mathit{min},x_\mathit{max}]\ ,
\end{equation}
and otherwise $L = 0$ for $x < x_\mathit{min}$ or $L = 1$ for $x > x_\mathit{max}$. We note that Eq.~(\ref{eq:dec16}) governs the magnitudes of the decorrelated uncertainties, which otherwise
inherit the sign of the original uncertainty. A similar algorithm is applied to decorrelate the JES flavor response JES62.

Analogous considerations were applied to the inclusive jet production data sets from CMS 7 and 8 TeV (experiments 542 and 545).  Namely, for the 7 TeV CMS jet data, the decorrelation methods of
Ref.~\cite{Khachatryan:2014waa} were applied to JEC2 with an additional
decorrelation\footnote{Private communication, M.~Voutilainen.} for the $2.5 < |y| < 3.0$ bin to obtain 6 subsidiary uncertainties. For the 8 TeV CMS jet data, systermatic uncertainties
were obtained using the xFitter framework according to the treatment in Ref.~\cite{Khachatryan:2016kdb}.
On the top of that, we found that the residual fluctuations
from the Monte-Carlo integration of the available NNLO predictions from \texttt{NNLOJet} \cite{Ridder:2015dxa,Gehrmann-DeRidder:2017mvr,Currie:2016bfm,Currie:2017ctp} lead to elevated $\chi^2$ for the inclusive jet production experiments, or equivalently, to some uncertainty in the tabulated NNLO/NLO corrections if they are fitted by a smooth function.   
In the CT18 analysis, the Monte-Carlo (MC) theoretical uncertainty for the three jet experiments is estimated by adding an overall {\it uncorrelated} uncertainty (`MC unc.') of 0.5\%, the typical magnitude of the intrinsic statistical noise associated with Monte-Carlo generation of NNLO/NLO $K$-factors. 
 
Table~\ref{tab-chi} summarizes the reduction in $\chi^2_E/N_{pt,E}$ for the three LHC jet data sets after performing the decorrelation and adding the Monte-Carlo uncertainties, on the example of the \CTHERAII~ PDFs. For example, for the ATLAS 7 TeV jet data, the decorrelation reduces $\chi^2_E$ by $\gtrsim\!\! 90$ units, yet adding the MC uncertainty is still necessary to reduce the $\chi^2_E/N_{pt,E}$ to a statistically plausible level (from 1.68 to 1.31 for $N_{pt,E}=140$). 

\begin{table}[tb]
	\begin{tabular}{c|ccc}
		\hline
		\hline
       	&	\multicolumn{3}{c}{evaluated, {\bf CT14$_\mathrm{HERAII}$ NNLO}} 	\ \ \ \\
		\ \ \  $\chi^2_E/N_{pt,E}$ \ \ \	& \ \ \ original data \ \ \  &  \ \ \  $+$\,decorr. \ \ \    &  \ \ \  $+$\,0.5\% MC unc. \ \ \   \\
		\hline
		\ \ \ ATLAS, 7 TeV \ \ \   &  \ \ \  2.34 \ \ \   &  \ \ \  1.68 \ \ \   &  \ \ \  1.31 \ \ \   \\
		\ \ \ CMS, 7 TeV   \ \ \   &  \ \ \  1.58 \ \ \   &  \ \ \  1.45 \ \ \   &  \ \ \  1.35 \ \ \   \\
		\ \ \ CMS, 8 TeV   \ \ \   &  \ \ \  1.90 \ \ \   &  \ \ \  1.34 \ \ \   &  \ \ \  1.23 \ \ \   \\
		\hline
		\hline
	\end{tabular}
	\caption{Values of $\chi^2_E/N_{pt,E}$ for the inclusive jet production data implemented in CT18,
		computed here using the \CTHERAII~PDFs \cite{Hou:2016nqm} under several error treatment scenarios. $\chi^2_E/N_{pt,E}$ is first given without
		implementing any decorrelation scheme (``original data''); using the decorrelation scheme described
		above (``$+$\, decorr.''); and, finally, adding an overall uncorrelated Monte Carlo uncertainty of 0.5\% on the top of decorrelations ("$+$\,0.5\% MC unc.").
	}
	\label{tab-chi}
\end{table}

\section{Hessian profiling of the ATLAS 7 TeV $W/Z$ Data}
\label{sec:Appendix4xFitter}

{\bf The $\chi^{2}$ definitions in \texttt{xFitter} and \texttt{ePump}.}
As an alternative to directly including new data inside a full QCD global fit, Hessian PDF-profiling techniques provide a fast and flexible approach to explore the
impact of new data on a given PDF set; these profiling methods are available within both the \texttt{xFitter} and \texttt{ePump} frameworks.
In \texttt{xFitter}, the $\chi^{2}$ function includes both experimental and theoretical uncertainties \cite{Paukkunen:2014zia,Camarda:2015zba,xFmanual}:	
\begin{equation}\label{eq:chi2xfitter}
\chi^{2}(\vec{\lambda}_{\textrm{exp}},\vec{\lambda}_{\textrm{th}})=
\sum_{i=1}^{N_{pt}}\frac{\left[D_{i}+\sum_{\alpha}\beta^{\textrm{exp}}_{i,\alpha}\lambda_{\alpha,\textrm{exp}}-
T_i-\sum_{\alpha}\beta_{i,\alpha}^{\textrm{th}}\lambda_{\alpha,\textrm{th}}
\right]^{2}}{s_{i}^{2}}+\sum_{\alpha}\lambda_{\alpha,\textrm{exp}}^{2}+\sum_{\alpha}T^2\lambda_{\alpha,\textrm{th}}^{2}.
\end{equation}
Here $D_{i}$ and $T_i$ are the $i$-th experimental datum and corresponding theoretical prediction, respectively, where the index $i$ runs over all $N_\mathit{pt}$ points in a given experiment. 
Meanwhile, $s_{i}$ denotes the total uncorrelated uncertainty, $\beta_{i,\alpha}^{\textrm{exp}}\, (\beta_{i,\alpha}^{\textrm{th}})$ represents the correlated experimental (theoretical) uncertainties,
and $\lambda_{\alpha,\textrm{exp}}\, (\lambda_{\alpha,\textrm{th}})$ are the corresponding nuisance parameters. In this case, the index $\alpha$ runs over the $N_\lambda$ correlated systematic
uncertainties in an experiment, or over the $N_\mathit{ev}$ Hessian eigenvector directions of the PDF error sets. Theoretical uncertainties are determined according to predictions
based upon the corresponding PDF error sets, $f_\alpha^\pm$, as
\begin{equation}
\beta_{i,\alpha}^{\textrm{th}}=\frac{T_i(f_{\alpha}^{+})-T_i(f_{\alpha}^{-})}{2},
\end{equation}
in which $f_{\alpha}^{\pm}$ are the PDF error sets corresponding to positive and negative variations along eigenvector $\alpha$. The tolerance parameter $T$ is set to 1 or 1.645 if the PDF error sets are defined at the 68\% or 90\% C.L., respectively. Instead of scaling down the theoretical uncertainties $\beta^{\textrm{th}}_{i,\alpha}$ as in Ref.~\cite{xFmanual}, we equivalently scale up the corresponding nuisance parameters $\lambda_{\alpha,\textrm{th}}$ in order to compare with \texttt{ePump} \cite{Schmidt:2018hvu,Hou:2019gfw} more transparently.

The $\chi^{2}$ function in Eq.~(\ref{eq:chi2xfitter}) can be converted into the form defined in \texttt{ePump} \cite{Schmidt:2018hvu,Hou:2019gfw}:
\begin{equation}\label{eq:chi2ePump}
\Delta\chi^{2}(\vec{\lambda}_{\textrm{th}})
=\sum_{i,j=1}^{N_\mathit{pt}}
[D_i-T_{i}(\vec{\lambda}_{\textrm{th}})]
\textrm{cov}_{ij}^{-1}[D_j-T_{j}(\vec{\lambda}_{\textrm{th}})]
+\sum_{\alpha}T^2\lambda_{\alpha,\textrm{th}}^{2},
\end{equation}
where $\textrm{cov}_{ij}^{-1}$ is the inverse experimental covariance matrix constructed from $s_i$ and
$\beta_{i,\alpha}^{\textrm{exp}}$ \cite{Gao:2013xoa}.
More generally, \texttt{ePump} also contains the option of the so-called ``dynamical tolerance'' $T^{(\alpha)}$ \cite{Schmidt:2018hvu,Hou:2019gfw}, whose specific value depends on the corresponding PDF eigenvector direction $\alpha$, for the purpose of taking into account additional constraints on the PDF error sets.
For example, in the CTEQ-TEA approach, the displacement along a given eigenvector direction is constrained either by the increase in the global $\chi^2$ by 100 units (at 90\% probability level) or by the tier-2 penalty for a too large increase in $\chi^2_E$ for one of the experiments \cite{Lai:2010vv,Gao:2013xoa}, which results in a $T^{(\alpha)}<100$. Dynamical tolerance is applied to MMHT PDFs as well, where separate constraints come from individual experiments \cite{Harland-Lang:2014zoa}.

The minimum of the $\chi^{2}$
function quantifies the compatibility between theory and data, and the minimization of $\chi^{2}$ optimizes the PDFs to describe the data \cite{Paukkunen:2014zia,Schmidt:2018hvu}.
In the linear approximation, the ``updated'' (or, in the language of \texttt{xFitter}, the ``profiled'') central PDF set, $f_{0}^{'}$, can be given in terms of nuisance parameters,
$\lambda_{\alpha,\textrm{th}}^{\textrm{min}}$, at the $\chi^{2}$ minimum:
\begin{equation}\label{eq:updatePDF}
f_{0}'=f_{0}+\sum_{\alpha} \lambda^{\textrm{min}}_{\alpha,\textrm{th}}\frac{f_{\alpha}^{+}-f_{\alpha}^{-}}{2}.
\end{equation}
The \texttt{xFitter}-profiled PDFs also include second-order diagonal terms, \linebreak
$\sim\! \frac{1}{2}(\lambda^{\textrm{min}}_{\alpha,\textrm{th}})^{2}(f_\alpha^{+}+f_\alpha^{-}-2f_0)$.
We note that the off-diagonal terms, $\sim\! \lambda^{\textrm{min}}_{\alpha,\textrm{th}}\lambda^{\textrm{min}}_{\alpha^\prime,\textrm{th}}$ for $\alpha\neq \alpha^\prime$, are not presently included,
since the off-diagonal, second-order partial derivatives cannot be constructed solely in terms of Hessian PDF error sets \cite{Hou:2016sho}.

 \begin{table}
\centering
\caption{The $\chi^2$ values for the ATLAS 7 TeV $W/Z$ data before and after \texttt{xFitter} profiling and \texttt{ePump} updating, using the CT14 and CT18 PDFs. 
}
\label{tab:chi2}
\begin{tabular}{c||c|c||c|c|c}
\hline\hline
Program & \multicolumn{2}{c||}{\texttt{xFitter}} & \multicolumn{3}{c}{\texttt{ePump}} \\
\hline
 PDFs & input & profiled & input & 
 $\begin{array}{c}
 \mbox{updated with}\\
 T^2=1.645^2,\\ 
 \mbox{as in \texttt{xFitter}}
 \end{array}
 $& 
 $\begin{array}{c}
 \mbox{updated with}\\
 \mbox{dynamical tolerance,}\\ 
 \mbox{as in CT18 fit}
 \end{array}
 $
 \\
\hline
\multicolumn{6}{c}{All the 7 measurements, $N_{\textrm{pt}}=61$}\\
\hline
CT14     & 290 & 106  & 285  & 104  &197  \\
CT18     & 362 & 104  & 356  & 103 &  199   \\
\hline
\multicolumn{6}{c}{$W^{+},W^{-}$, $Z$-peak DY (central), $N_{\textrm{pt}}=34$}\\
\hline
CT14     & 224 & 66  & 220  &  66  & 140\\
CT18     & 294 & 63 & 289  &  63 &  144\\
\hline
\hline
\end{tabular}
\end{table}

\begin{figure}
\includegraphics[width=0.45\textwidth]{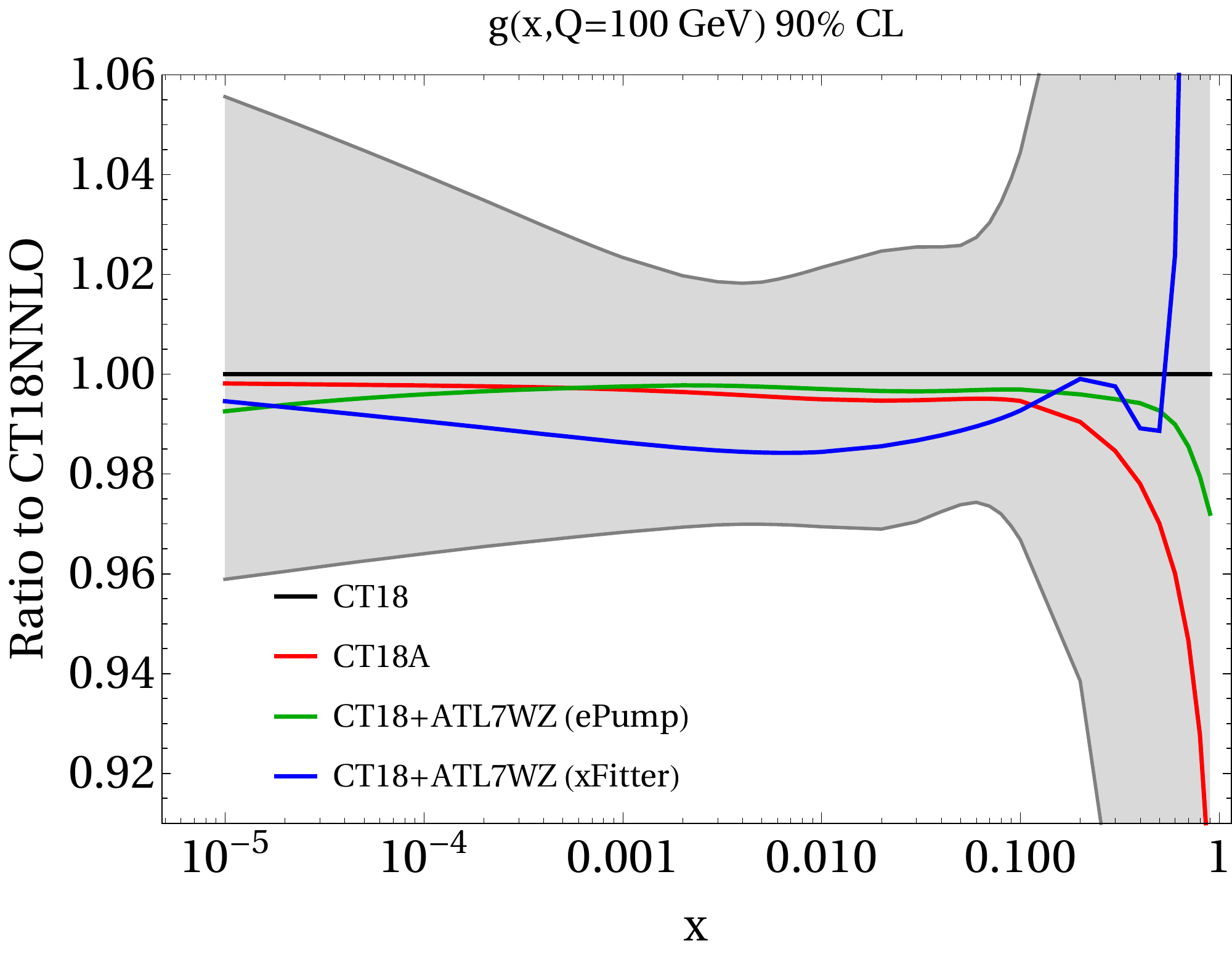}
\includegraphics[width=0.45\textwidth]{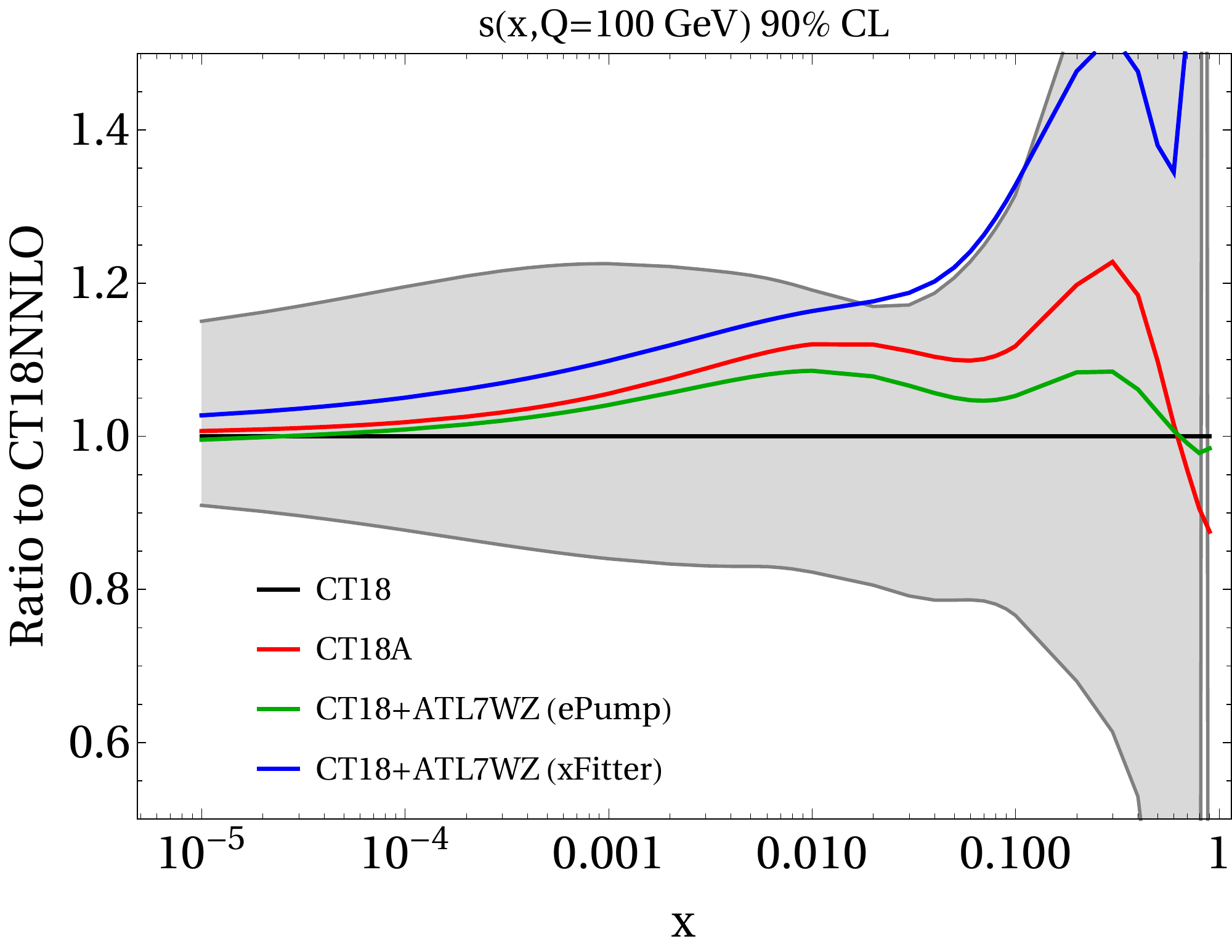}
\caption{Gluon and strangeness PDFs at $Q=100$ GeV  for the CT18 (at 90\% C.L.) and CT18A global fits, compared with the respective central PDFs obtained by \texttt{ePump} dynamical-tolerance updating and the \texttt{xFitter} profiling of the CT18 PDFs by the ATLAS 7 TeV $W/Z$-production data. 
The \texttt{xFitter}-profiled PDFs are obtained with $T=1.645$ and include the diagonal second-order terms.
    }\label{fig:epxf-strange}
\end{figure}

{\bf Impact of the ATLAS $W/Z$ data.}
Here, we use the Hessian-profiling method of \texttt{xFitter}, as well as \texttt{ePump} updating, to explore the impact of the ATLAS 7 TeV inclusive $W/Z$-production data (Expt.~ID=248, \cite{Aaboud:2016btc})
on several PDF sets. The change in the total $\chi^{2}$ values before and after profiling/updating with the ATLAS 7 TeV $Z/W$-production data
is presented in Table~\ref{tab:chi2} for each PDF set. 
First, we explore all 7 measurements, having a total of $N_\textit{pt}=61$ data points: $W^{+}$, $W^{-}$, neutral current DY in the low-mass, $Z$-peak, and high-mass regions
for the central and forward selections. In this case, we can directly compare with the ATLAS \cite{Aaboud:2016btc} and MMHT \cite{Thorne:2019mpt} analyses. 
As a second case, we take only the 3 most precise measurements, {\it i.e.}, the $W^{+},\, W^{-}$, and $Z$-peak DY data for the central selection, with $N_{\textit{pt}}=34$ data points in total,
which are included in the NNPDF3.1 \cite{Ball:2017nwa} and CT18A(Z) global analyses. The comparison of the fitted $\chi^2/N_{\textit{pt}}$ values for these data in CT18A(Z), NNPDF3.1 and
MMHT can be found in Sec.~\ref{sec:CT18Z_qual}.

Table~\ref{tab:chi2} shows that the $\chi^{2}$ values of CT14, before and after \texttt{xFitter} profiling, agree well with the results presented in Ref.~\cite{Aaboud:2016btc}.
Here, we should apply the tolerance $T=1.645$ in \texttt{xFitter} as the CT PDFs are defined according to a 90\%
C.L.~\cite{Dulat:2015mca,Hou:2016nqm}. 
As shown in Ref.~\cite{Hou:2019gfw}, the same results can be reproduced by the \texttt{ePump}
code when the tolerance is set to $T\! =\! 1.645$. However, as clearly discussed in~\cite{Hou:2019gfw}, setting $T=1.645$ in the \texttt{ePump} calculation for the CT PDFs is equivalent to
assigning a very large weight (about $100/1.645^2$) to the new data set included in the fit.
\texttt{xFitter} profiling therefore generally overestimates the impact of new data sets when using the CT PDFs.
Meanwhile, a universal tolerance value ($T=1.645$) is not able to capture the constraints of Tier-2 penalty to determine the CT PDF error sets.
An appropriate way to update an existing CT PDF set with the
inclusion of any new experimental data is to adopt a dynamical tolerance.
As such, \texttt{xFitter} profiling yields a smaller $\chi^2$ value (63) than
does~\texttt{ePump} updating with a dynamical tolerance (144), as can be seen by comparing the rightmost entries in the last row of Table~\ref{tab:chi2}.
This conclusion also holds when using~\texttt{xFitter} profiling with the MMHT~\cite{Harland-Lang:2014zoa} and PDF4LHC15~\cite{Butterworth:2015oua} PDFs.

The $\chi^2$ value of the 34 highest-precision ATLAS 7 TeV $W/Z$-production data points is found to be $\chi^2\!=\!87.6$ in the CT18A global fit, cf.~Table~\ref{tab:CMN}.  
To this we compare the corresponding value found using \texttt{ePump} updating with dynamical tolerance, for which we obtain $\chi^2\!=\!144$, as reported in Table~\ref{tab:chi2}.
This value includes two distinct contributions, $\chi^2_1=104$ from the difference between theory and data of the ATLAS 7 TeV $W/Z$-production itself, as well as from the quadrature sum $\chi^2_2=\sum_\alpha \lambda_{\alpha,\rm{th}}^2=40$ of theoretical nuisance parameters, which can be interpreted as the increase in $\chi^2$ of the other (``prior'') data sets included in the CT18 fit. 
The large increase in $\chi^2_2$, after \texttt{ePump} updating,  indicates the presence of some tensions between the ATLAS 7 TeV $W/Z$-production
data and the prior data sets, such as the CCFR/NuTeV SIDIS dimuon data~\cite{Hou:2019gfw}, cf.~Fig.~\ref{fig:lm_wht}.

The differences between the $\chi^2$ values in CT18A (87.6) and from \texttt{ePump} updating with a dynamical tolerance (104) indicate the breakdown of the linear approximation used
in the Hessian-updating method, when applied to this case. The breakdown can be confirmed by examining the updated central PDF set. In Fig.~\ref{fig:epxf-strange}, we show the gluon and strangeness
PDFs at $Q=100$ GeV for the CT18 (with 90\% C.L.~error) and CT18A global fits, compared with \texttt{ePump} dynamical-tolerance updating and \texttt{xFitter} profiling of CT18 with the ATLAS 7 TeV $W/Z$-production data.
As compared to the CT18 PDFs, we see that the updated $g$ and $s$ PDFs from the \texttt{ePump} program are similar to the CT18A PDFs, although with somewhat smaller shifts in the data-sensitive range, $10^{-3}\! <\! x\! <\! 10^{-1}$. In contrast, the default \texttt{xFitter} profiling produces a much larger shift in both $g$ and $s$ PDFs, than the
CT18A global fit.  As a result, \texttt{xFitter} profiling produces a too large change in the $s$-PDF, so that its central prediction touches the upper error band of CT18 at $x\! >\! 0.02$. 
We have found similar features in the comparison of other flavor PDFs. Namely, the default \texttt{xFitter} program generally overestimates the impact of new data sets in updating the
existing PDFs~\cite{Hou:2019gfw}. 

\begin{figure}[bhtp]
\begin{center}
\includegraphics[width=0.8\textwidth]{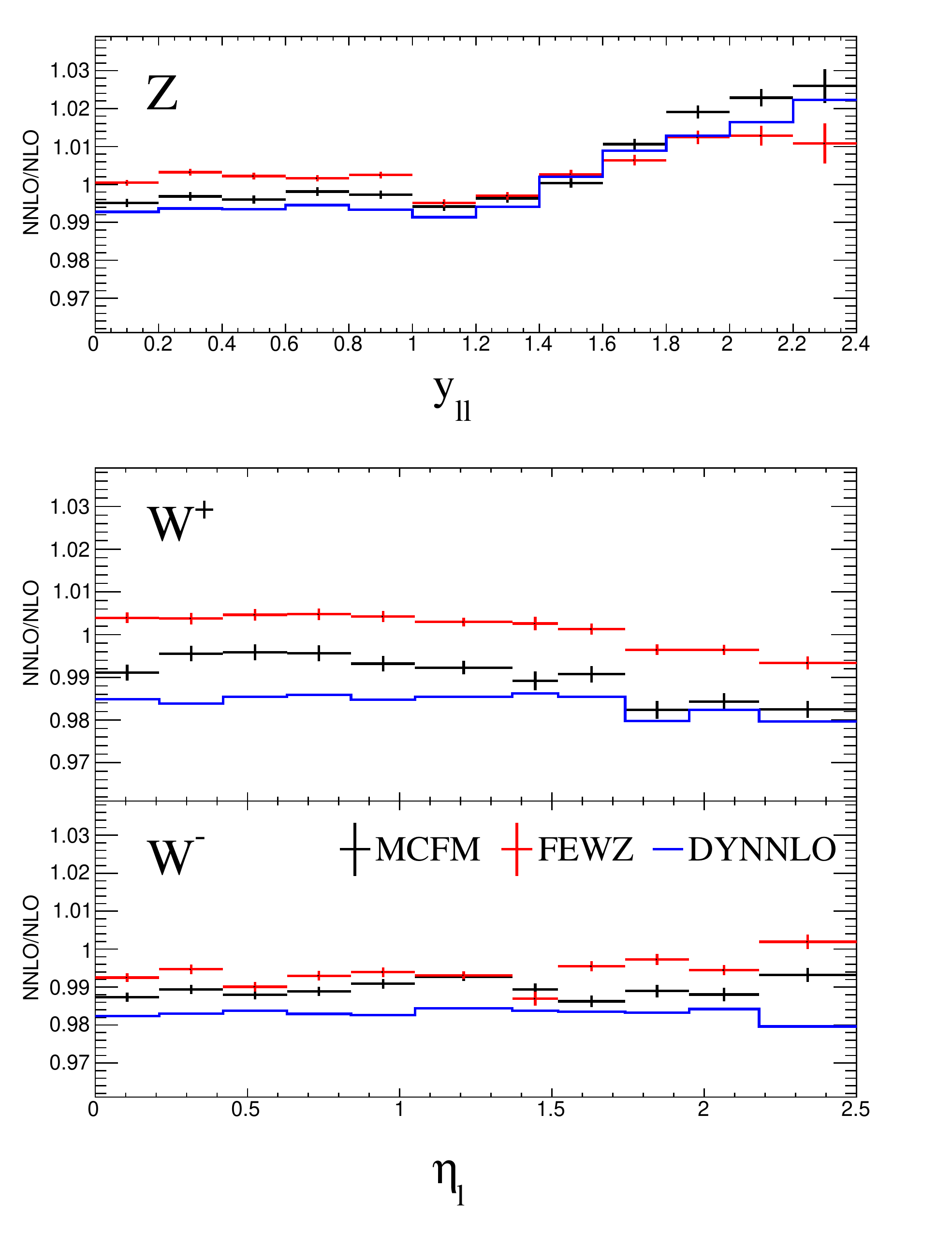}
\caption{The comparison of $K$-factors for the ATLAS 7 TeV $W/Z$ data calculated with \texttt{DYNNLO}, \texttt{FEWZ} and \texttt{MCFM}. The error bars indicate the theoretical Monte-Carlo uncertainties. The \texttt{DYNNLO} curves are extracted from \texttt{xFitter} and include NLO EW corrections.} 
\label{fig:Kfactors}
\end{center}
\end{figure}

\begin{table}
\caption{
The $\chi^2$ values for the ATLAS 7 TeV $W/Z$-production data (with 34 data points in total), before and after the \texttt{xFitter} profiling and the \texttt{ePump} dynamical-tolerance updating, with various NNLO predictions. (See the text for details.)
We do not update  CT18A(Z) fits to avoid 
double-counting the impact of the ATLAS 7 TeV $W/Z$-production data.
}
\label{tab:chi2Kfactors}
\begin{tabular}{c|c|c|c|c|c|c}
\hline
 &  \multicolumn{2}{c|}{\texttt{DYNNLO}} &    \multicolumn{2}{c|}{\texttt{MCFM}} &  \multicolumn{2}{c}{\texttt{FEWZ}}  \\
\hline
\multicolumn{7}{c}{\texttt{xFitter} profiling}\\
\hline
PDF & before & after   & before & after   & before & after \\
\hline
CT18 & 294 & 63    &  277 & 65  & 225  &  62   \\
CT18A & 87  & --   & 92   & --  & 109  & --    \\
CT18Z & 88 &   --  &  94  & --  & 109  &  --  \\
\hline
\multicolumn{7}{c}{\texttt{ePump} dynamic-tolerance updating}\\
\hline
CT18 & 289 & 144   &  273  & 144    & 223   & 135    \\
CT18A & 87 & --   &  91  & --  &  109 &  --  \\
CT18Z & 88 &   --   &  94  & --  &  109 & --  \\
\hline
\end{tabular}
\end{table}

{\bf Comparison of different NNLO predictions.}
In addition to the studies described above, we have also used the \texttt{xFitter} and \texttt{ePump} frameworks
to examine aspects of the theory calculations for the ATLAS 7 TeV $W/Z$ data.
In the CT18A(Z) global fits, the NNLO predictions for these measurements were calculated using NNLO/NLO $K$-factors combined with NLO  \texttt{APPLGrid} predictions.  Specifically, the $K$-factors used in our CT18A(Z) fits were 
directly extracted from \texttt{xFitter}, where they were calculated with the \texttt{DYNNLO} code~\cite{Catani:2007vq,Catani:2009sm}. 
It was noted by the ATLAS Collaboration in 
Ref.~\cite{Aaboud:2016btc} that the integrated fiducial  $Z$ , $W^{+}$ and $W^{-}$ cross sections predicted by the NNLO codes 
\texttt{FEWZ} \cite{Gavin:2010az,Gavin:2012sy,Li:2012wna} and \texttt{DYNNLO} \cite{Catani:2007vq,Catani:2009sm} differ 
by about 0.2\%, 1.2\% and 0.7\%, respectively.
Fig.~\ref{fig:Kfactors} shows a slightly larger difference, since the \texttt{DYNNLO} curves include the NLO EW corrections, while other two do not.

Next, we wish to compare various NNLO predictions for differential cross section measurements of the ATLAS 7 TeV $W/Z$-production data.  
First, we verified that the NLO predictions agree, within 0.2\%, among  \texttt{DYNNLO},  \texttt{MCFM}~\cite{Boughezal:2016wmq,MCFM8} 
and \texttt{FEWZ}, which reflects the fact that all three codes adopt the dipole formalism \cite{Catani:1996jh,Catani:1996vz} in the NLO calculations. 
Since the NNLO predictions can be expressed as the NLO predictions multiplied by their $K$-factors, we compare in Fig. \ref{fig:Kfactors} the $K$-factors obtained from each code,
for every ATLAS 7 TeV $W/Z$-production data point (with 34 data points in total).
It can be seen that the differences among the three $K$-factors vary as a function of the $Z$-boson rapidity, or the rapidity of the charged lepton from the $W$-boson decay. 
These differences can be sizable, $\gtrsim\!1\%$, as compared to the typically sub-percent statistical uncertainty found in the ATLAS 7 TeV $W/Z$ data.  
We also find that the predictions of \texttt{MCFM} generally lie between those of \texttt{DYNNLO} and \texttt{FEWZ}. This difference can be understood as a consequence
of the the different NNLO techniques: \texttt{FEWZ} adopts sector decomposition \cite{Binoth:2000ps,Anastasiou:2003gr,Anastasiou:2005qj}, while \texttt{DYNNLO} and \texttt{MCFM}
are based on transverse-momentum ($q_T$) \cite{Catani:2007vq,Catani:2009sm} and $N$-jettiness ($\mathcal{T}_N$) \cite{Boughezal:2016wmq} subtractions, respectively. 

It is also useful to investigate the dependence of the ATLAS 7 TeV $W/Z$ fit quality on the specific choice of NNLO calculation scheme. 
Table~\ref{tab:chi2Kfactors} summarizes the findings of such a study. 
After CT18 PDFs are updated using \texttt{ePump} with the ATLAS 7 TeV $W/Z$ data, we find that the final $\chi^2$ value for the ATLAS $W/Z$ data set
is equal to 144, 144, and 135 units when theory is predicted by the \texttt{DYNNLO}, \texttt{MCFM} and \texttt{FEWZ} NNLO calculations, respectively. 
Hence, we conclude that these three NNLO calculations lead to fits of similar quality for the ATLAS 7 TeV $W/Z$ data.
A similar conclusion also holds when using the \texttt{xFitter} framework, with the final $\chi^2$ values being 63, 65 and 62 units, respectively.
To recap, the default \texttt{xFitter} profiling (with tolerance $T\! =\! 1.645$ ) overestimates the impact of the
ATLAS 7 TeV $W/Z$ data when updating the CT18 PDFs, explaining why it yields a smaller $\chi^2$ value than \texttt{ePump} updating. We also confirm that the updated PDFs by the three $K$-factors differ slightly, but the difference is negligible compared with the same systematical shifts due to the experimental uncertainties.  
In conclusion, the NNLO theoretical predictions for the ATLAS 7 TeV $W/Z$ data by \texttt{DYNNLO}, \texttt{MCFM} and \texttt{FEWZ}  show perceptible differences when using the same PDF set. On the other hand, after the PDFs are updated by either \texttt{ePump} and \texttt{xFitter}, the final $\chi^2$ and PDFs show minor differences among the three codes.

\clearpage\newpage

\bibliographystyle{apsrev4-1}
\bibliography{ct18bibtex}

\clearpage\newpage

\section{Supplemental Material}
\label{sec:Supp}

\subsection{Additional comparisons between PDF sets}
\label{sec:Supp-PDFbands}

\begin{figure}[b]
\hspace*{-1.5cm}
\includegraphics[width=0.49\textwidth]{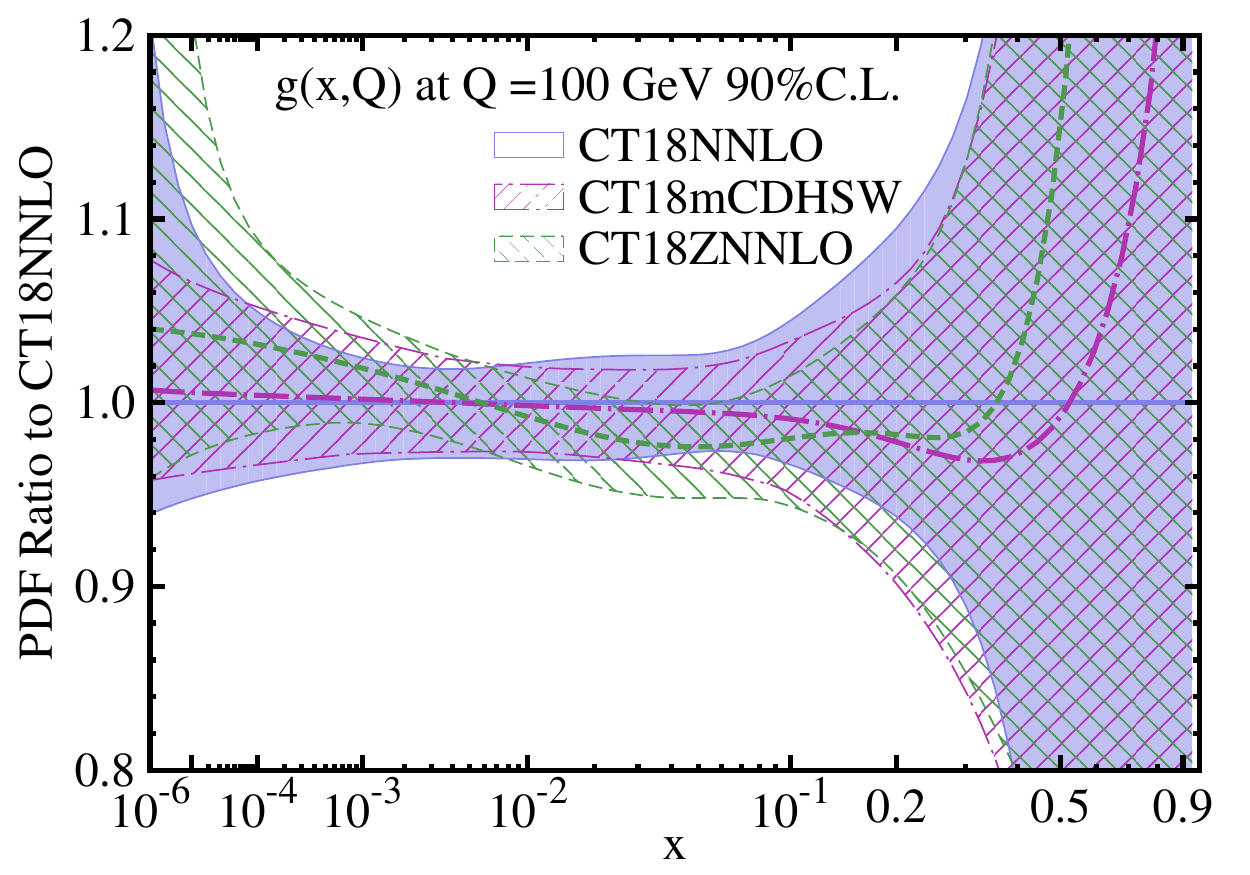}
\includegraphics[width=0.49\textwidth]{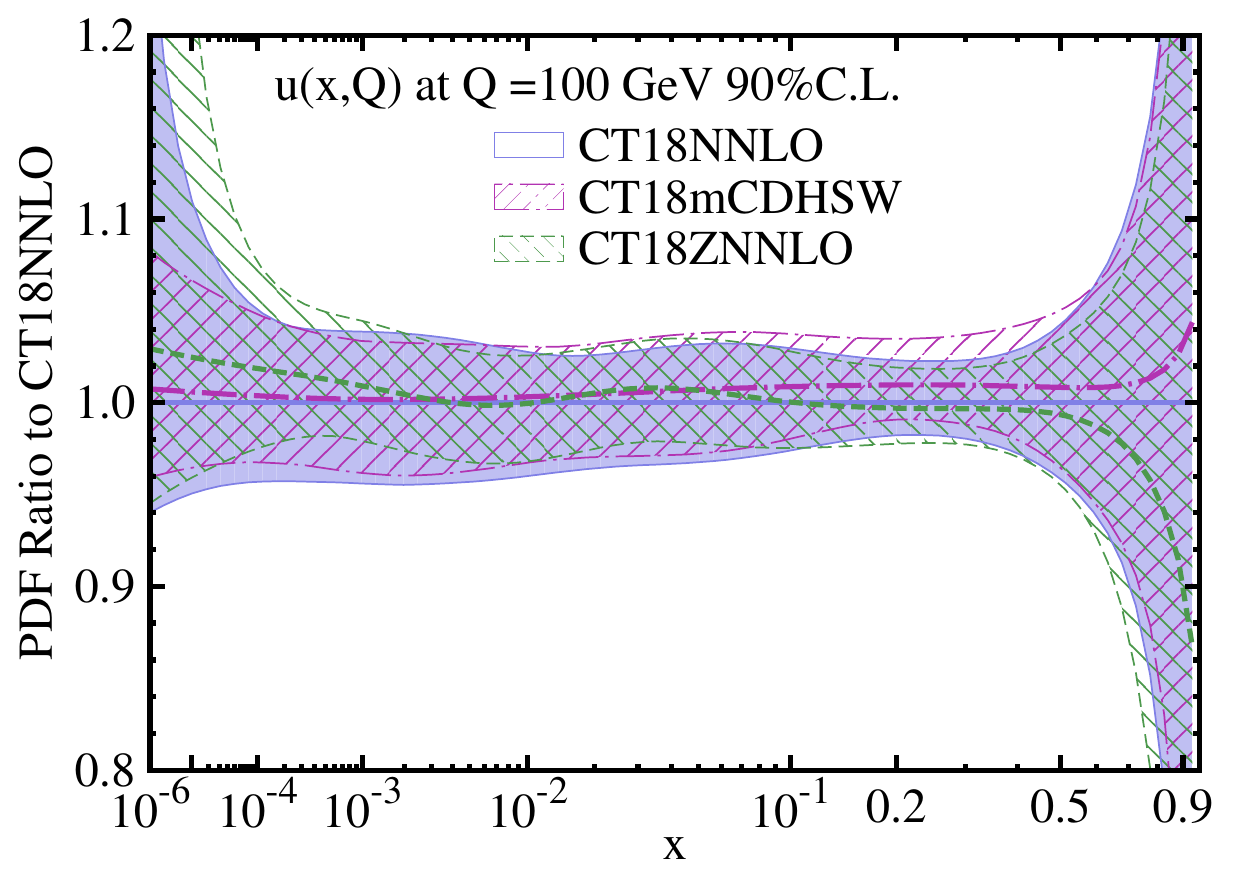}\\
\includegraphics[width=0.49\textwidth]{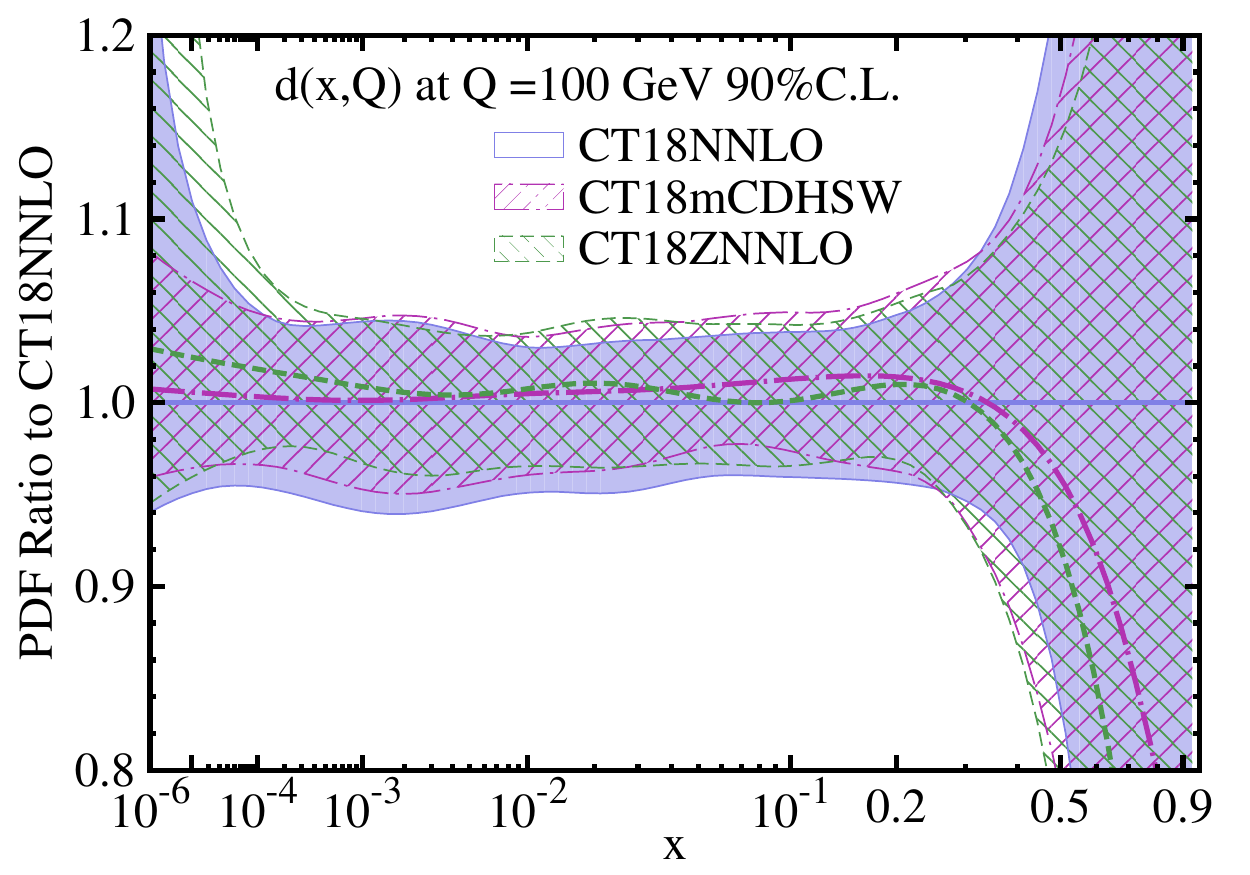}
\caption{Effect of eliminating CDHSW data (Exp.~IDs=108, 109) from the CT18 fit. CT18mCDHSW denotes the fit after removing the CDHSW data sets. The result of CT18Z fit is also shown for comparison. 
\label{fig:cdhsw}}
\end{figure}

\begin{figure}
	\begin{tabular}{cc}
		\subfloat[NNLO $g(x,Q = 100\,\mathrm{GeV})$, CT18(Z) vs.~CT18A]{\includegraphics[width=0.49\textwidth]{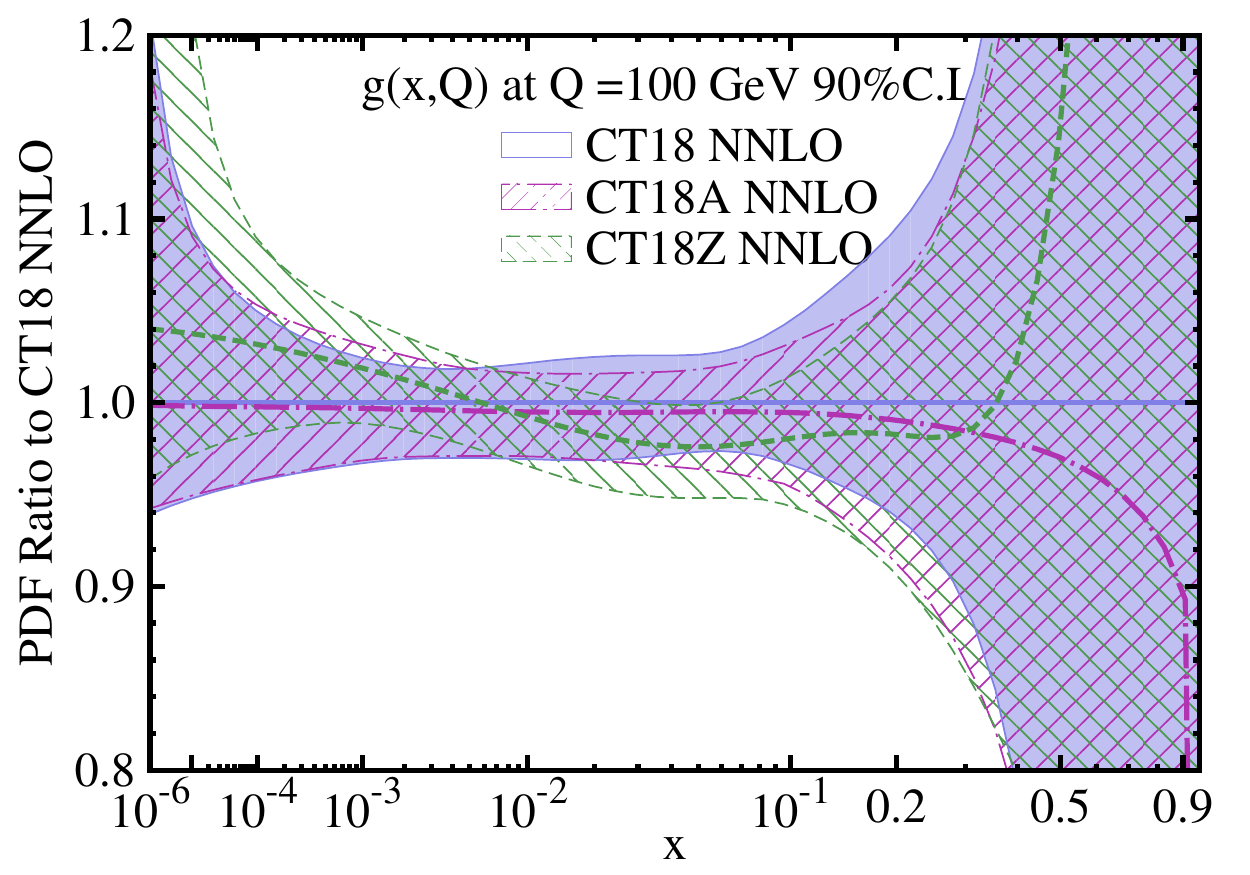}} &
		\subfloat[NNLO $g(x,Q = 100\,\mathrm{GeV})$, CT18(Z) vs.~CT18X]{\includegraphics[width=0.49\textwidth]{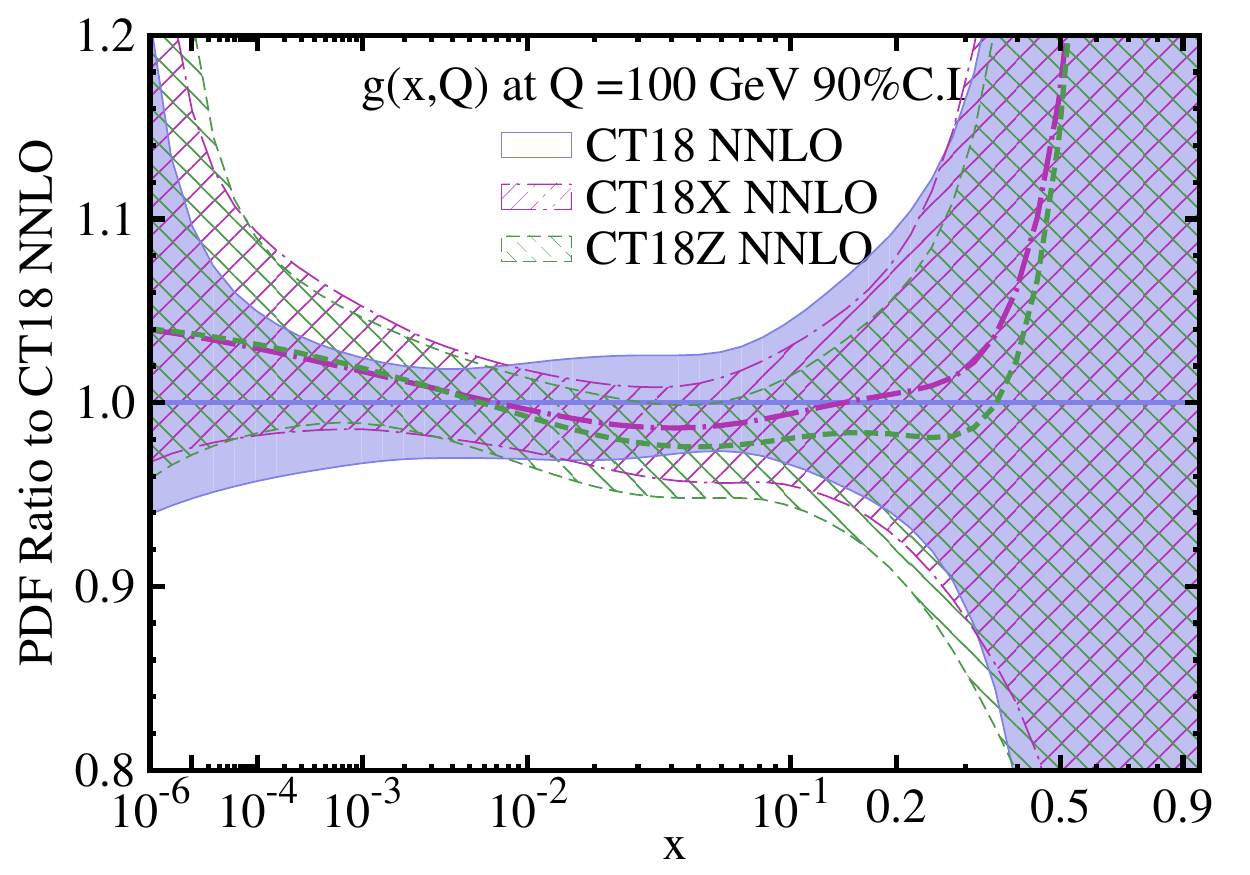}}\\
		\subfloat[NLO  $g(x,Q = 100\,\mathrm{GeV})$, CT18(Z) vs.~CT18A]{\includegraphics[width=0.49\textwidth]{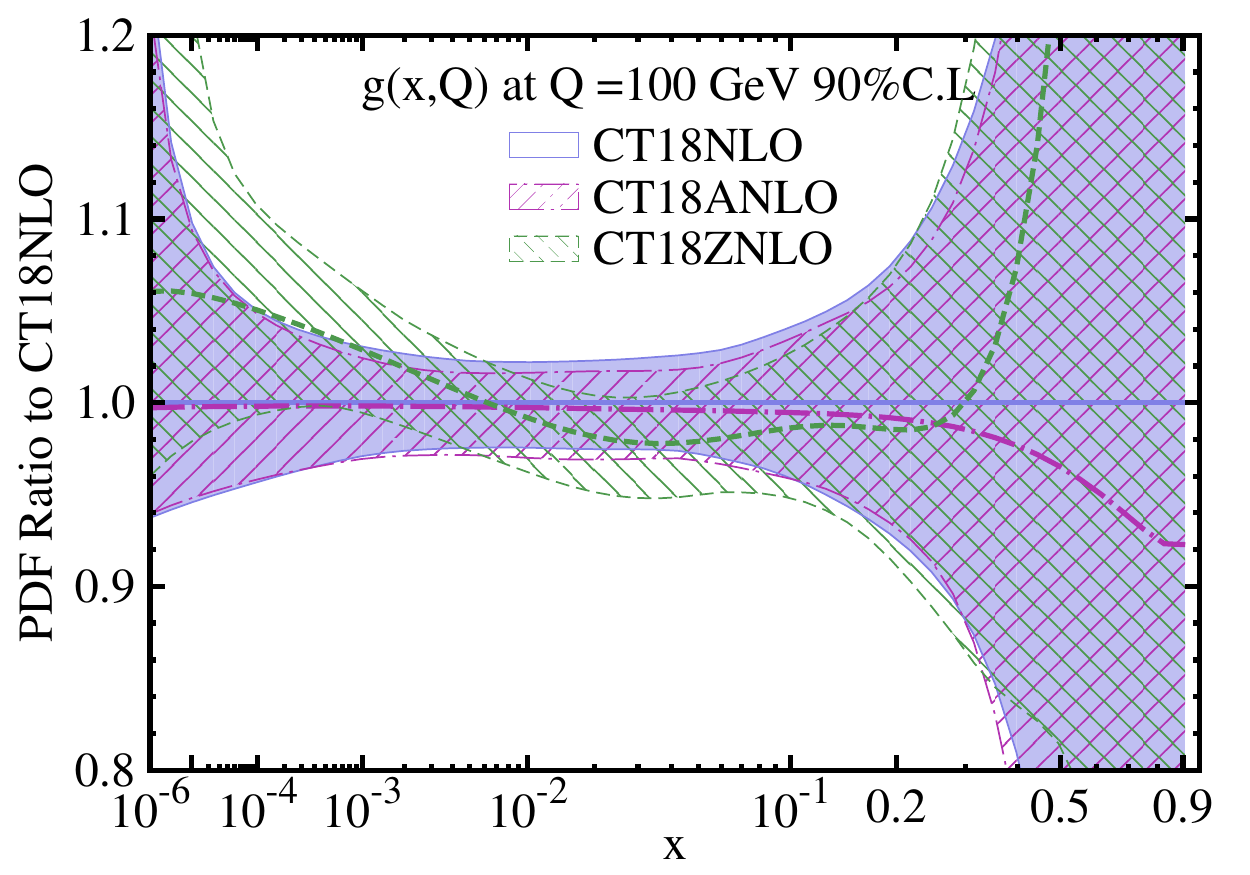}} &
		\subfloat[NLO  $g(x,Q = 100\,\mathrm{GeV})$, CT18(Z) vs.~CT18X]{\includegraphics[width=0.49\textwidth]{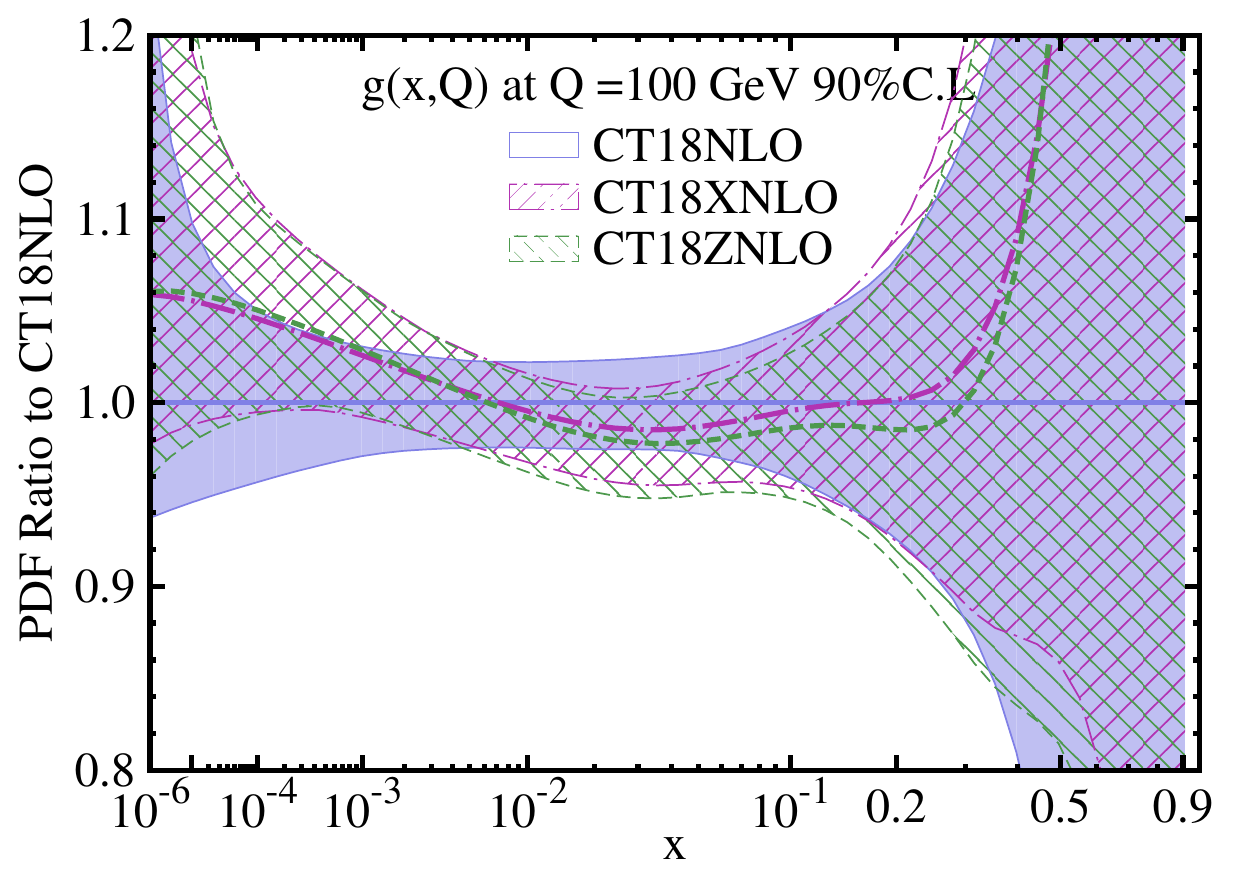}}
	\end{tabular}
	\caption{
		Gluon PDF ratios for the CT18Z and CT18A/X alternative fits, evaluated with respect to the primary, CT18 result. In
		the left panels [(a) and (c)], we compare CT18(Z) against CT18A, whereas the right panels [(b) and (d)] overlay CT18(Z) with CT18X. In addition,
		we examine differences for NNLO and NLO fits; the upper panels [(a) and (b)] are NNLO, and
		the lower panels [(c) and (d)] are NLO.
	}
\label{fig:glu_AXZ}
\end{figure}

\begin{figure}
	\begin{tabular}{cc}
		\subfloat[NNLO $d(x,Q = 100\,\mathrm{GeV})$, CT18(Z) vs.~CT18A]{\includegraphics[width=0.49\textwidth]{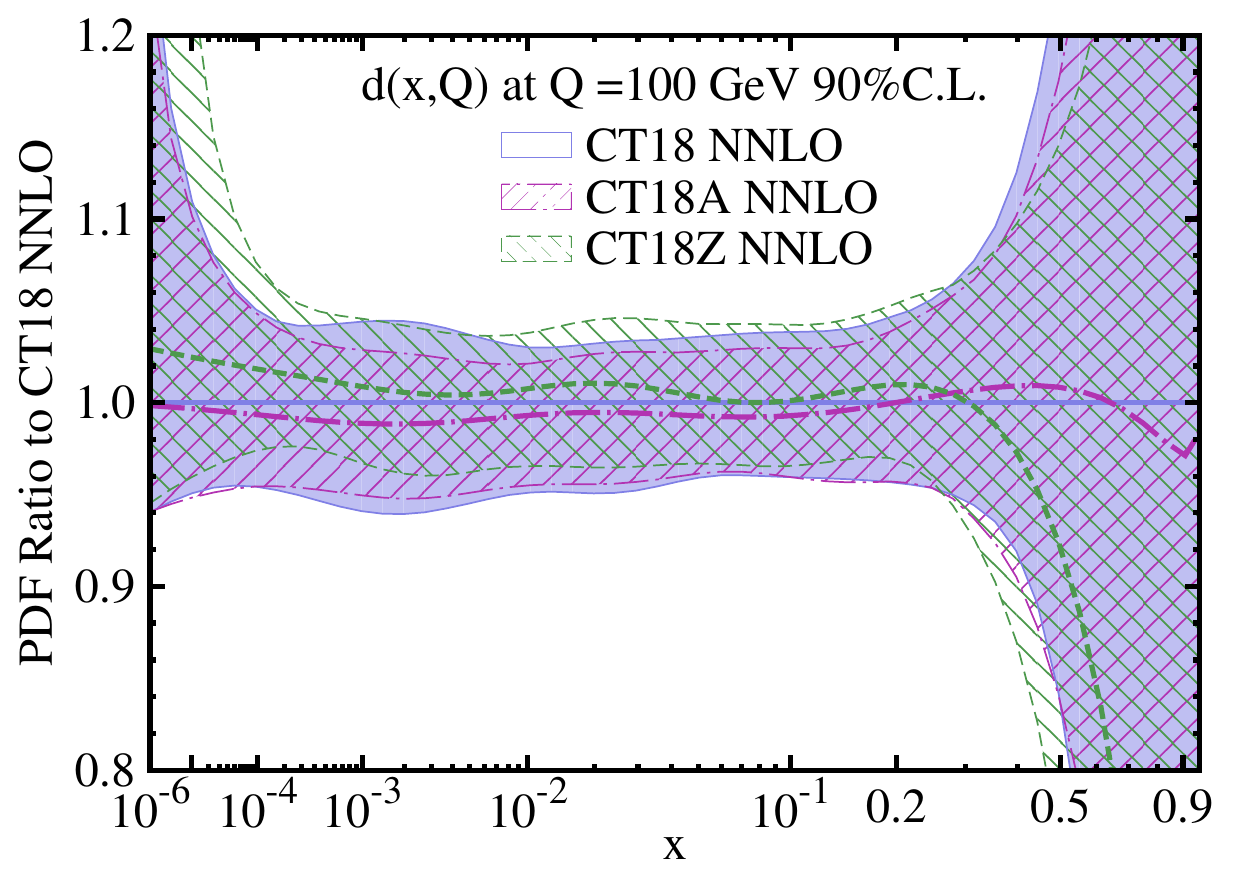}} &
		\subfloat[NNLO $d(x,Q = 100\,\mathrm{GeV})$, CT18(Z) vs.~CT18X]{\includegraphics[width=0.49\textwidth]{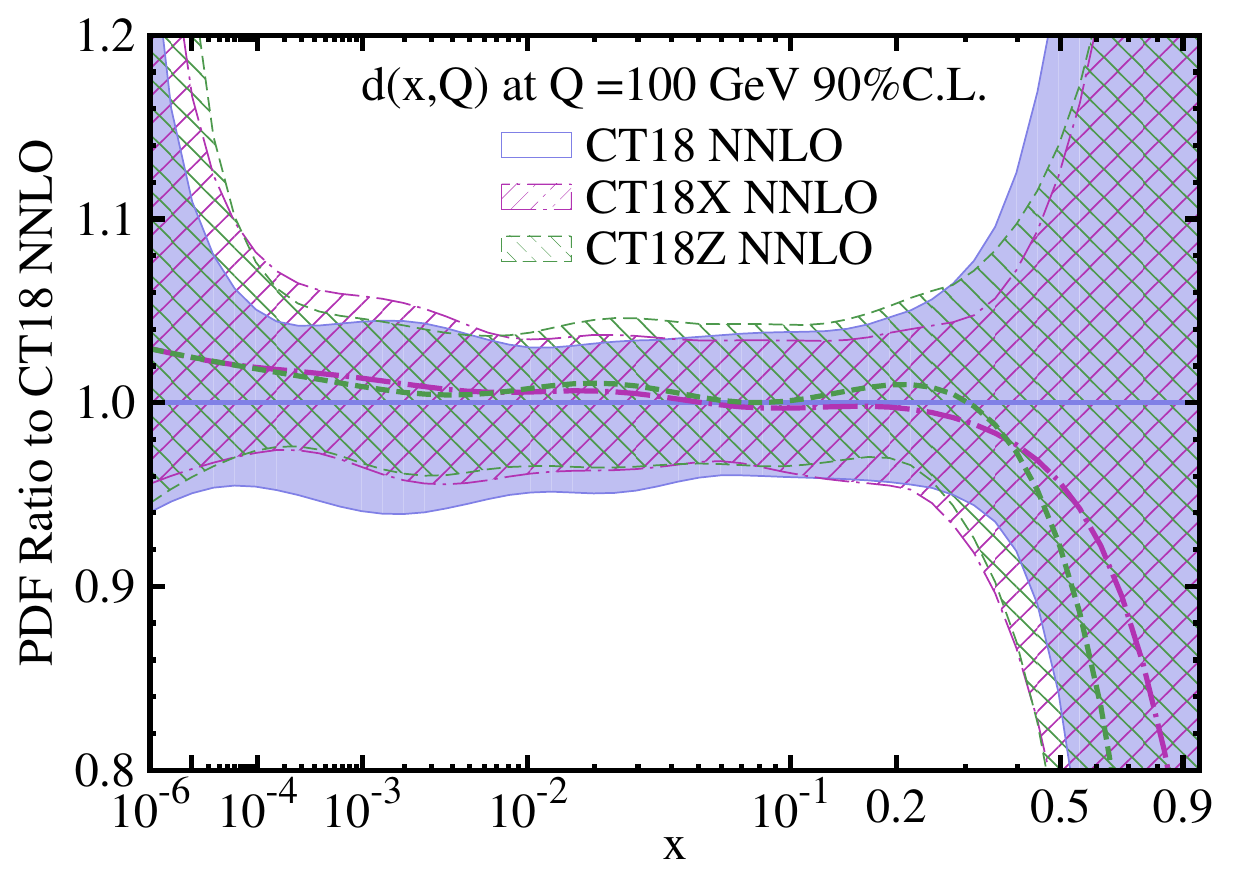}}\\
		\subfloat[NLO  $d(x,Q = 100\,\mathrm{GeV})$, CT18(Z) vs.~CT18A]{\includegraphics[width=0.49\textwidth]{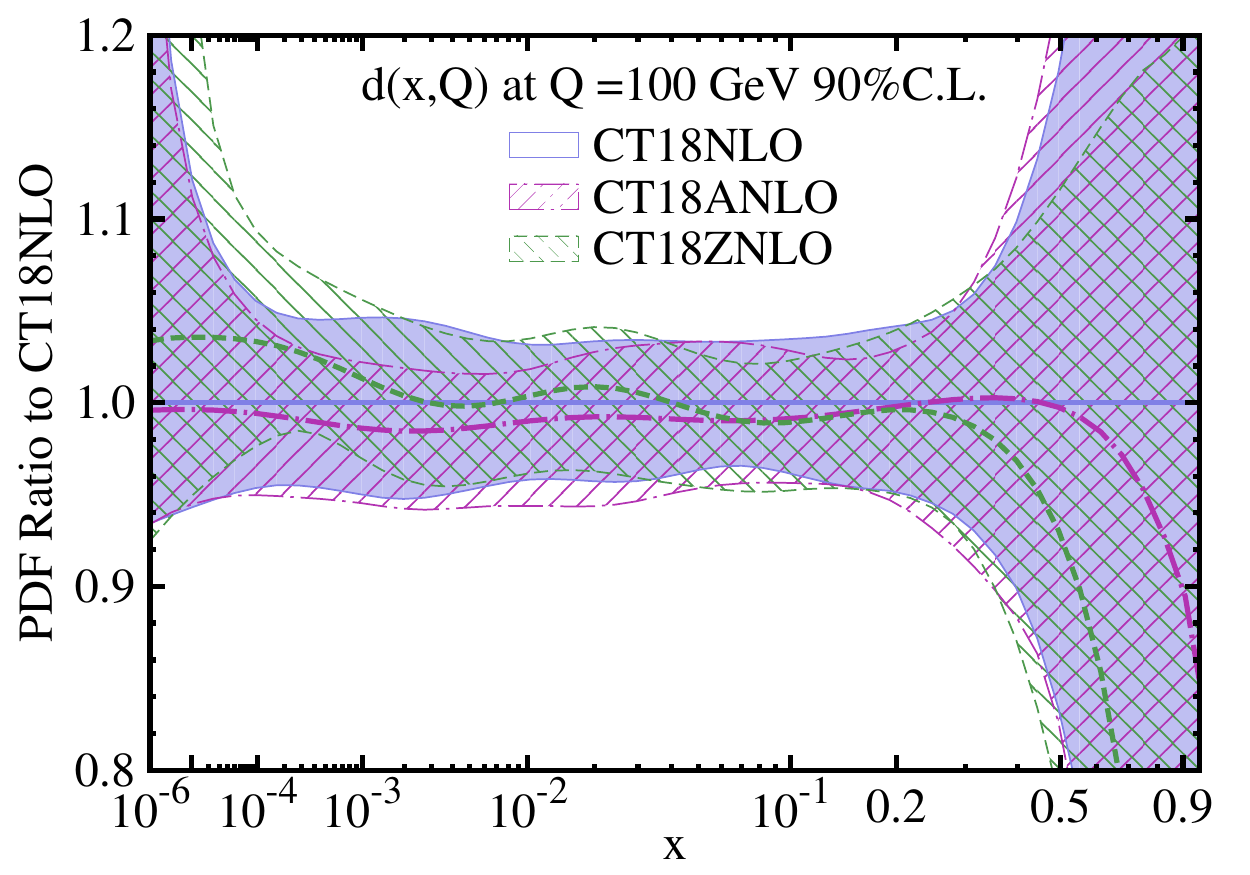}} &
		\subfloat[NLO  $d(x,Q = 100\,\mathrm{GeV})$, CT18(Z) vs.~CT18X]{\includegraphics[width=0.49\textwidth]{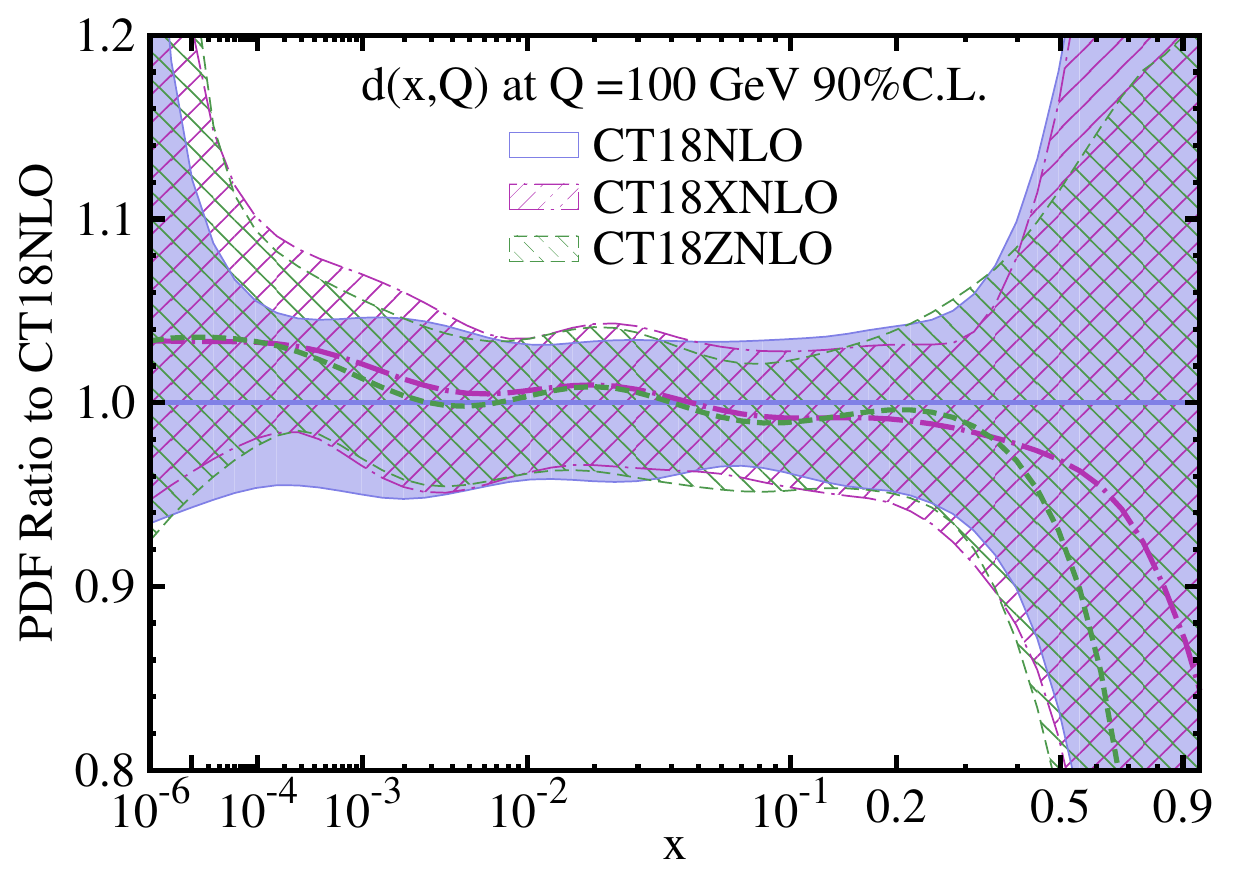}}
	\end{tabular}
	\caption{
		A comparison of the $d$-quark PDF ratios with respect to CT18 for the CT18(Z) PDFs vs.~the CT18A/X alternative fits.
		The plots here are analogous to those shown for the gluon in Fig.~\ref{fig:glu_AXZ}.
	}
\label{fig:d_AXZ}
\end{figure}

\begin{figure}
	\begin{tabular}{cc}
		\subfloat[NNLO $\bar{u}(x,Q = 100\,\mathrm{GeV})$, CT18(Z) vs.~CT18A]{\includegraphics[width=0.49\textwidth]{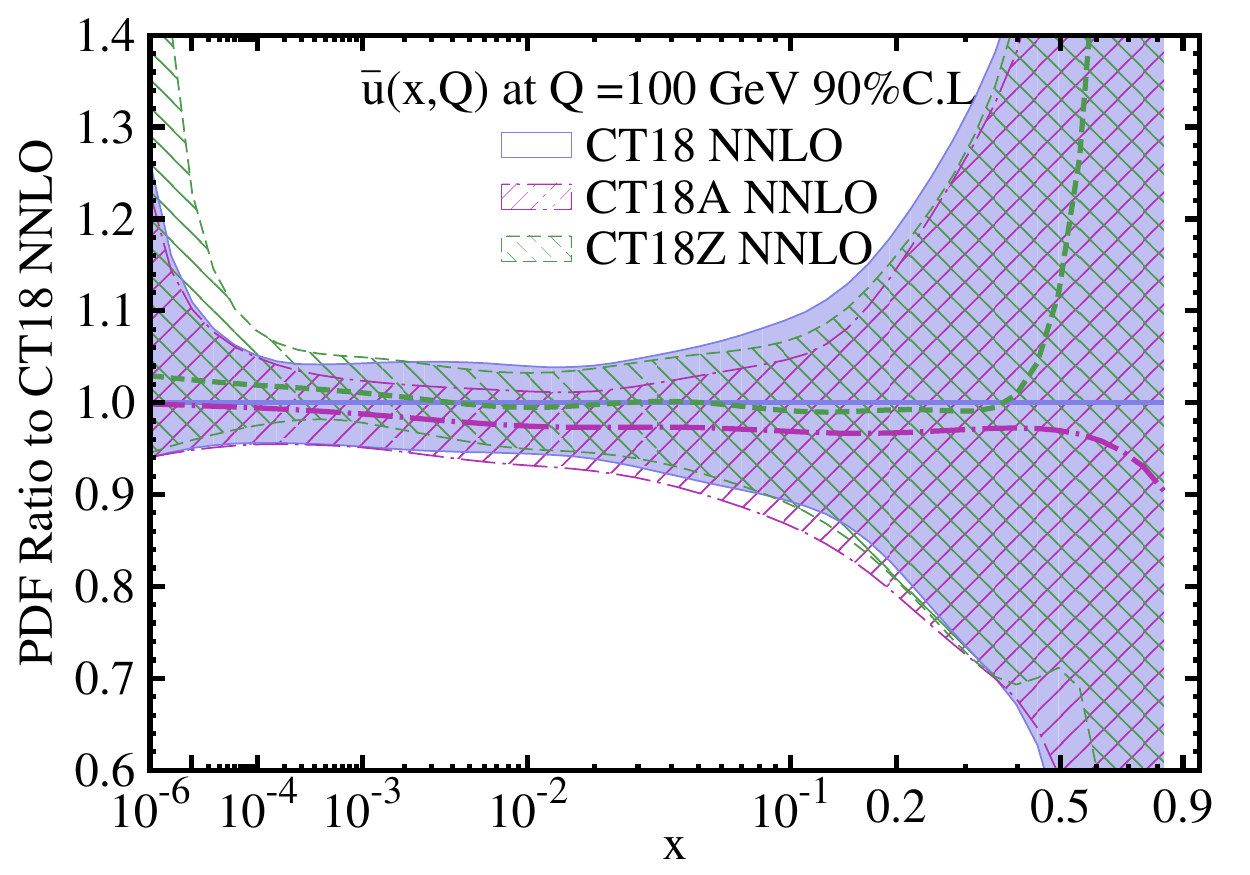}} &
		\subfloat[NNLO $\bar{u}(x,Q = 100\,\mathrm{GeV})$, CT18(Z) vs.~CT18X]{\includegraphics[width=0.49\textwidth]{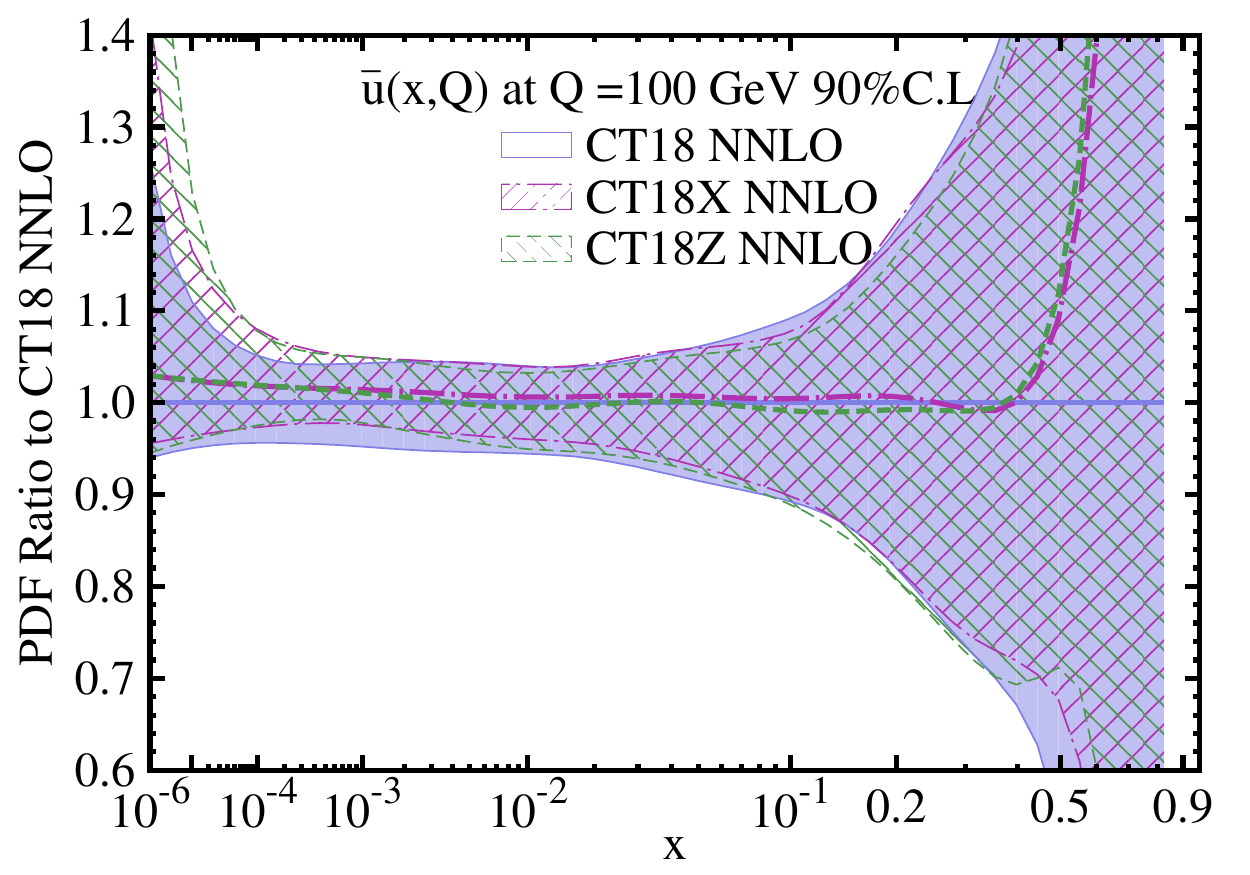}}\\
	\end{tabular}
	\caption{
		As with Fig.~\ref{fig:d_AXZ}, we explore the $\bar{u}$-antiquark PDF ratios at NNLO,
		comparing CT18(Z) with CT18A NNLO in panel (a), and with CT18X in panel (b).
	}
\label{fig:ubar_AXZ}
\end{figure}

\begin{figure}[tb]
	\begin{tabular}{cc}
		\subfloat[NNLO $s(x,Q = 100\,\mathrm{GeV})$, CT18(Z) vs.~CT18A]{\includegraphics[width=0.49\textwidth]{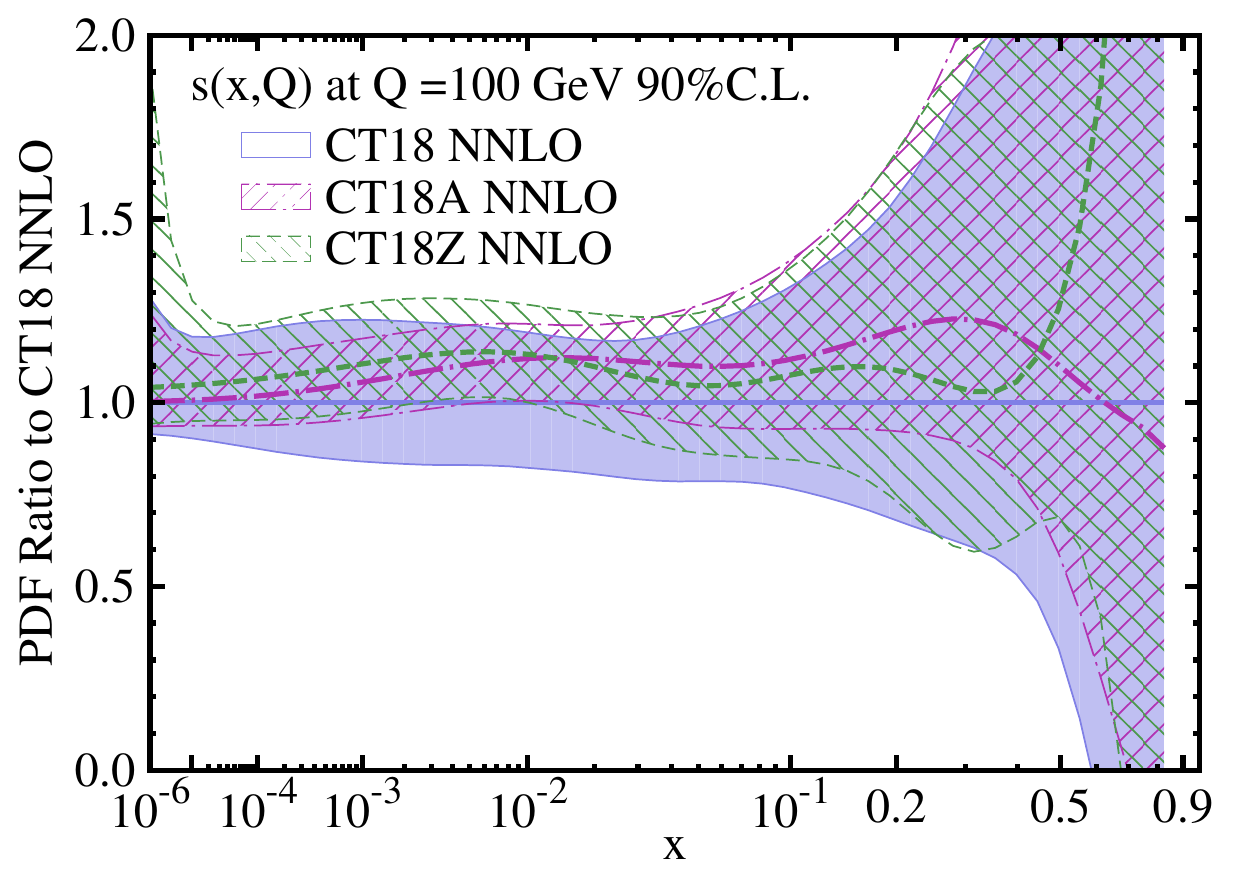}} &
		\subfloat[NNLO $s(x,Q = 100\,\mathrm{GeV})$, CT18(Z) vs.~CT18X]{\includegraphics[width=0.49\textwidth]{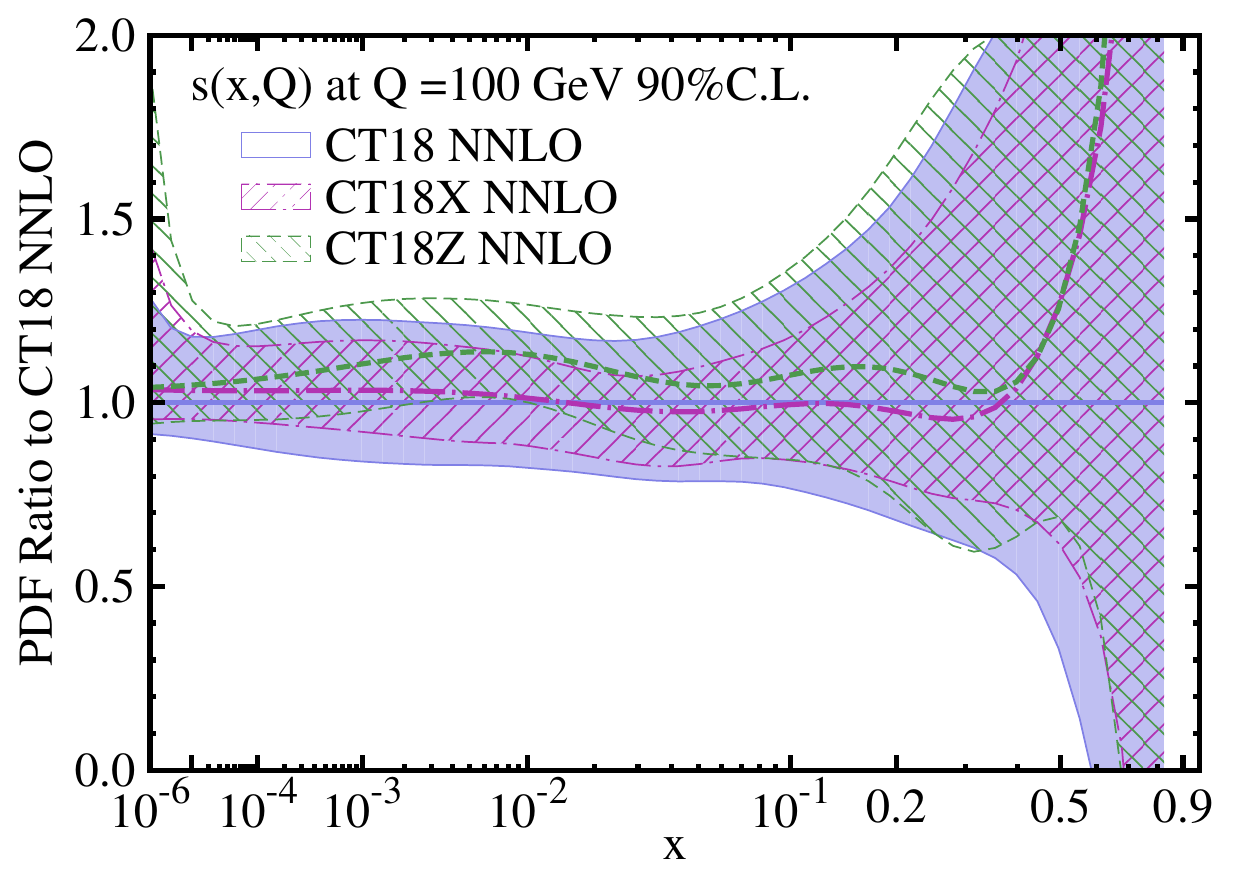}}\\
	\end{tabular}
	\caption{
		Like Fig.~\ref{fig:ubar_AXZ}, but now comparing alternative distributions for the nucleon strangeness, $s(x,Q)$.
	}
\label{fig:s_AXZ}
\end{figure}

\begin{figure}[tb]
	\begin{tabular}{cc}
		\subfloat[NNLO $R_s(x,Q = 100\,\mathrm{GeV})$, CT18(Z) vs.~CT18A]{\includegraphics[width=0.49\textwidth]{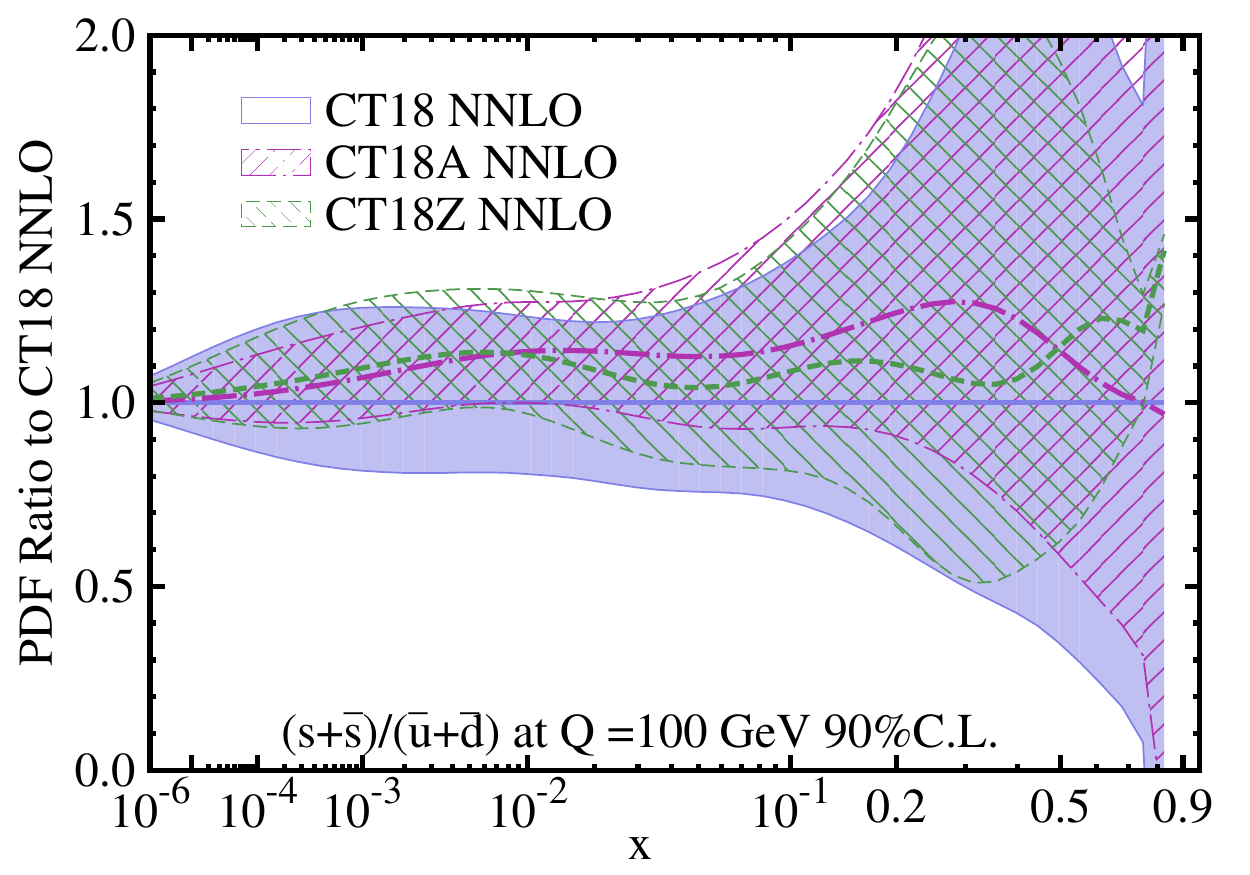}} &
		\subfloat[NNLO $R_s(x,Q = 100\,\mathrm{GeV})$, CT18(Z) vs.~CT18X]{\includegraphics[width=0.49\textwidth]{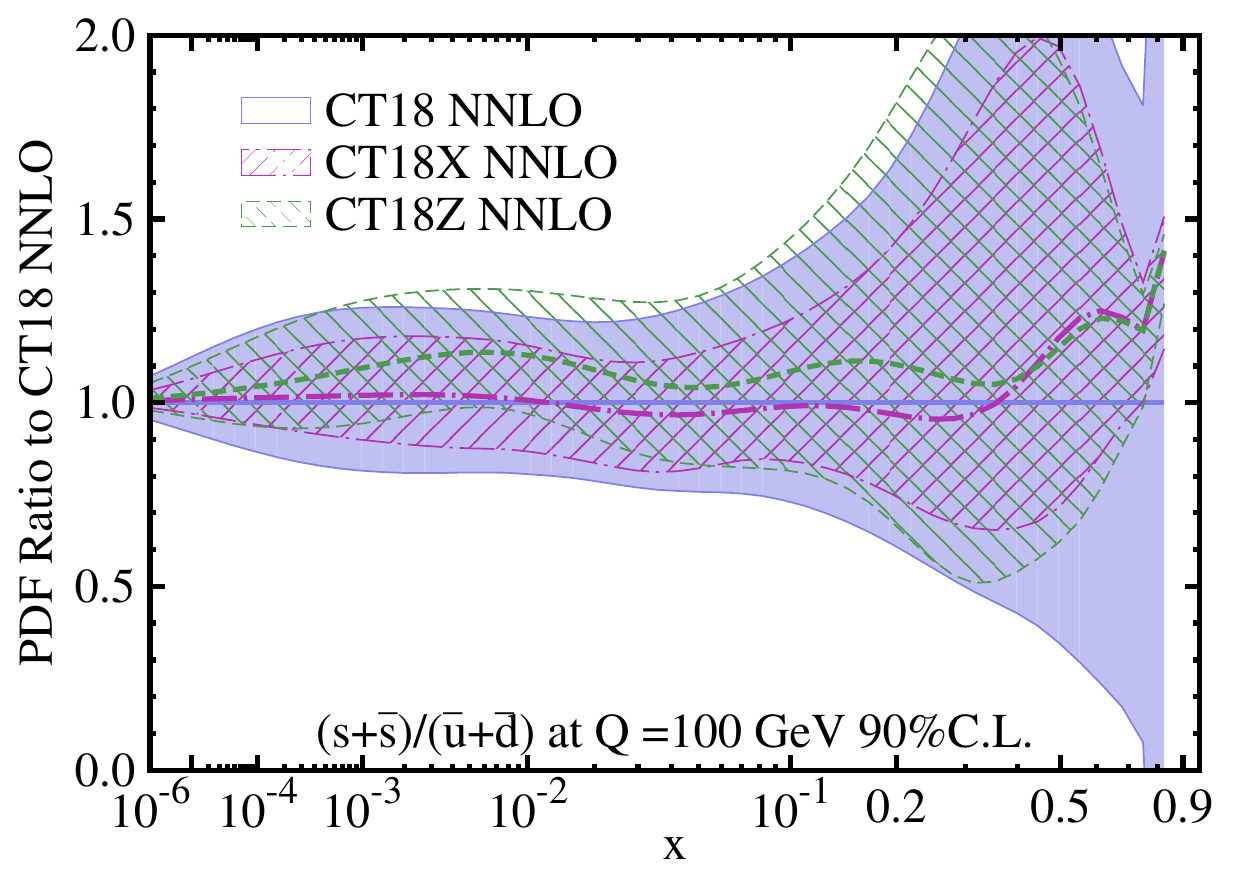}}\\
	\end{tabular}
	\caption{
		The strange suppression factor, $R_s\! \equiv\! (s+\bar s) / (\bar u + \bar d)$, for the CT18(Z) and CT18A/X NNLO alternative fits, evaluated with respect to the primary,
		CT18 NNLO result.
	}
\label{fig:Rs_AXZ}
\end{figure}

\begin{figure}[tb]
	\begin{tabular}{cc}
		\subfloat[$L_{gg}$ at $\sqrt{s}=14$ TeV, CT18(A/Z)]{\includegraphics[width=0.49\textwidth]{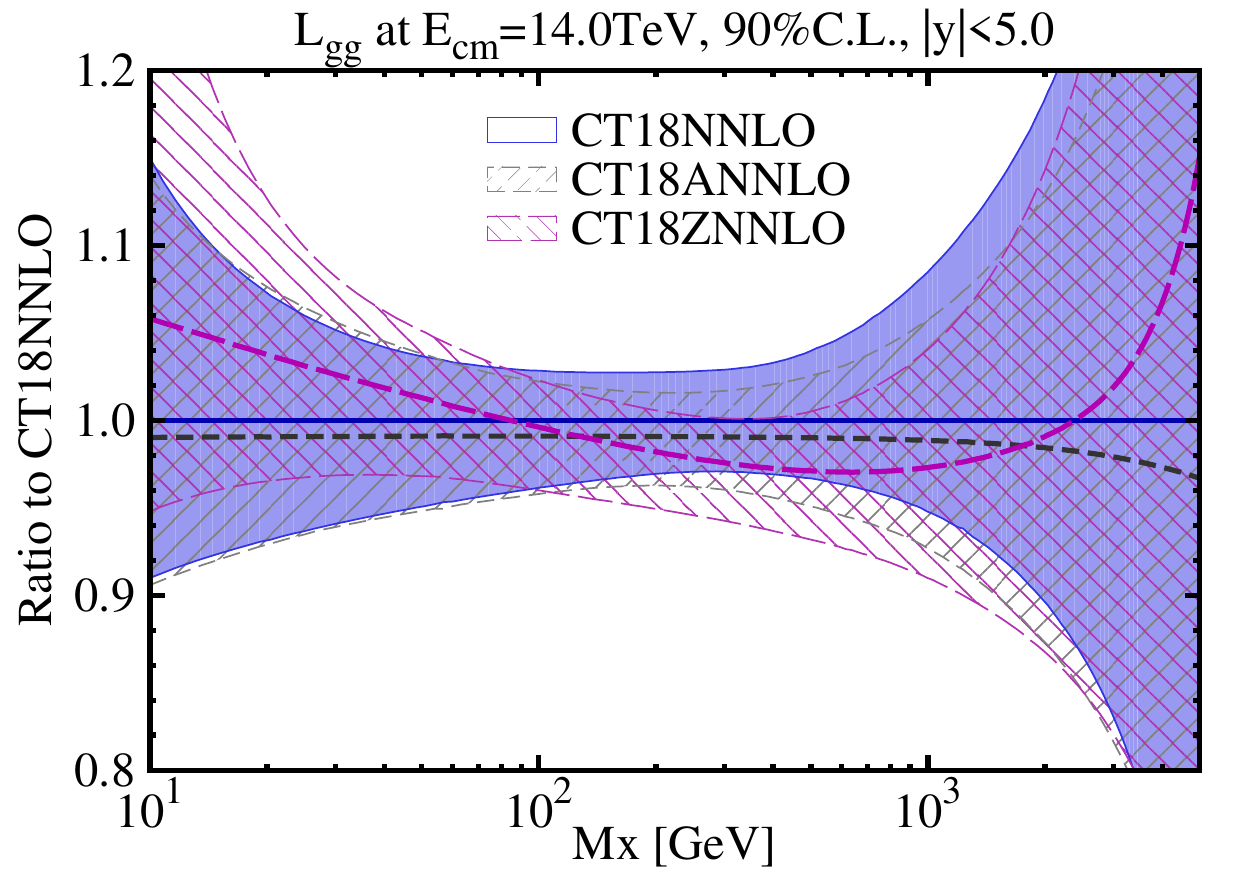}} &
		\subfloat[$L_{gg}$ at $\sqrt{s}=14$ TeV, CT18(X/Z)]{\includegraphics[width=0.49\textwidth]{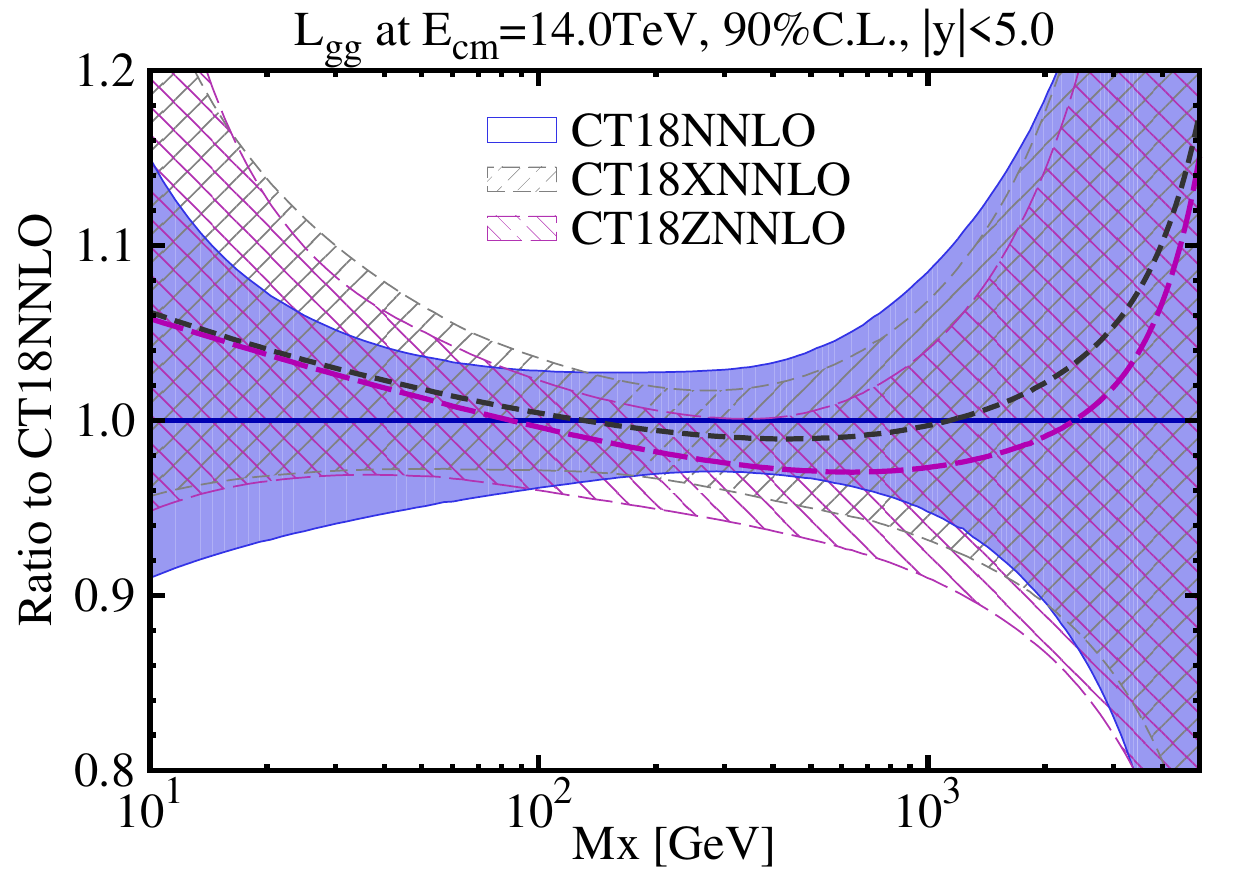}}\\
	\end{tabular}
	\caption{
		A comparison of the glue-glue parton luminosities, $L_{gg}$, at $\sqrt{s}=14$ TeV computed using CT18 NNLO as well as CT18A/Z (left panel) and
		CT18X/Z (right). All results are normalized to the central CT18 NNLO calculation.
	}
\label{fig:gg_lumi_AXZ}
\end{figure}

\begin{figure}[tb]
	\begin{tabular}{cc}
		\subfloat[$L_{gg}$ at $\sqrt{s}=14$ TeV, CT18(A/Z)]{\includegraphics[width=0.49\textwidth]{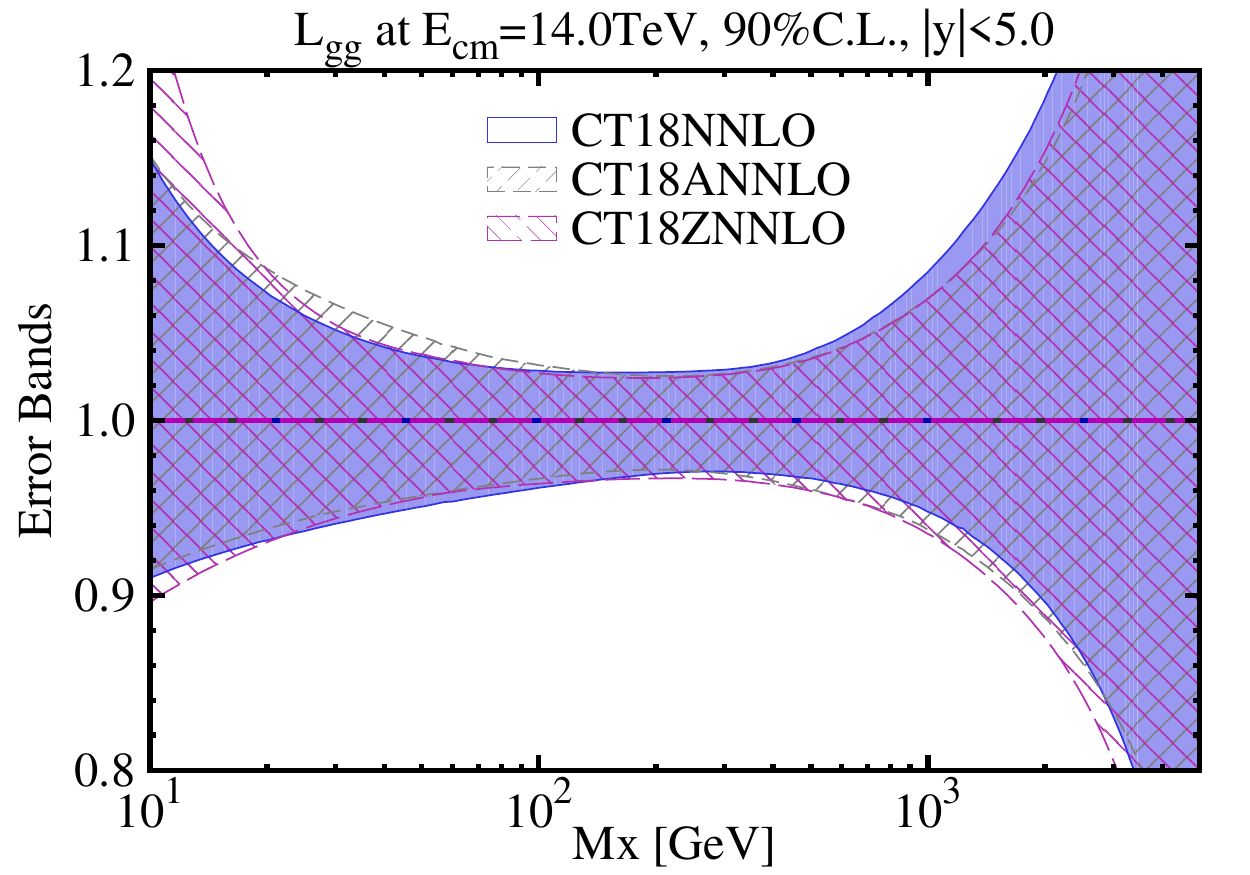}} &
		\subfloat[$L_{gg}$ at $\sqrt{s}=14$ TeV, CT18(X/Z)]{\includegraphics[width=0.49\textwidth]{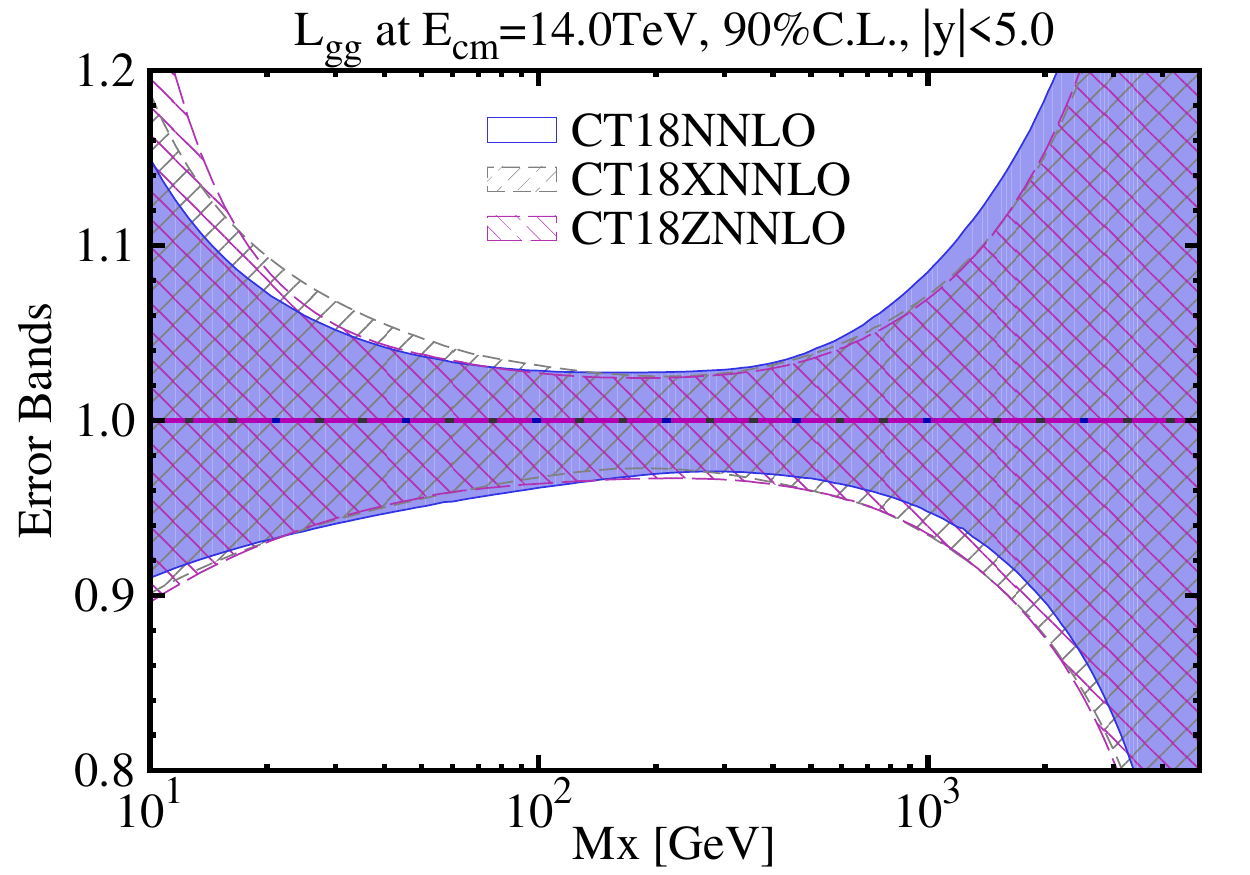}}\\
	\end{tabular}
	\caption{
		Like Fig.~\ref{fig:gg_lumi_AXZ}, but normalizing each result for $L_{gg}$ to its respective central-fit calculation to directly compare
		relative uncertainties.
	}
\label{fig:gg_lumi_AXZ_band}
\end{figure}

\begin{figure}[p]
\center
  \includegraphics[width=0.59 \textwidth]{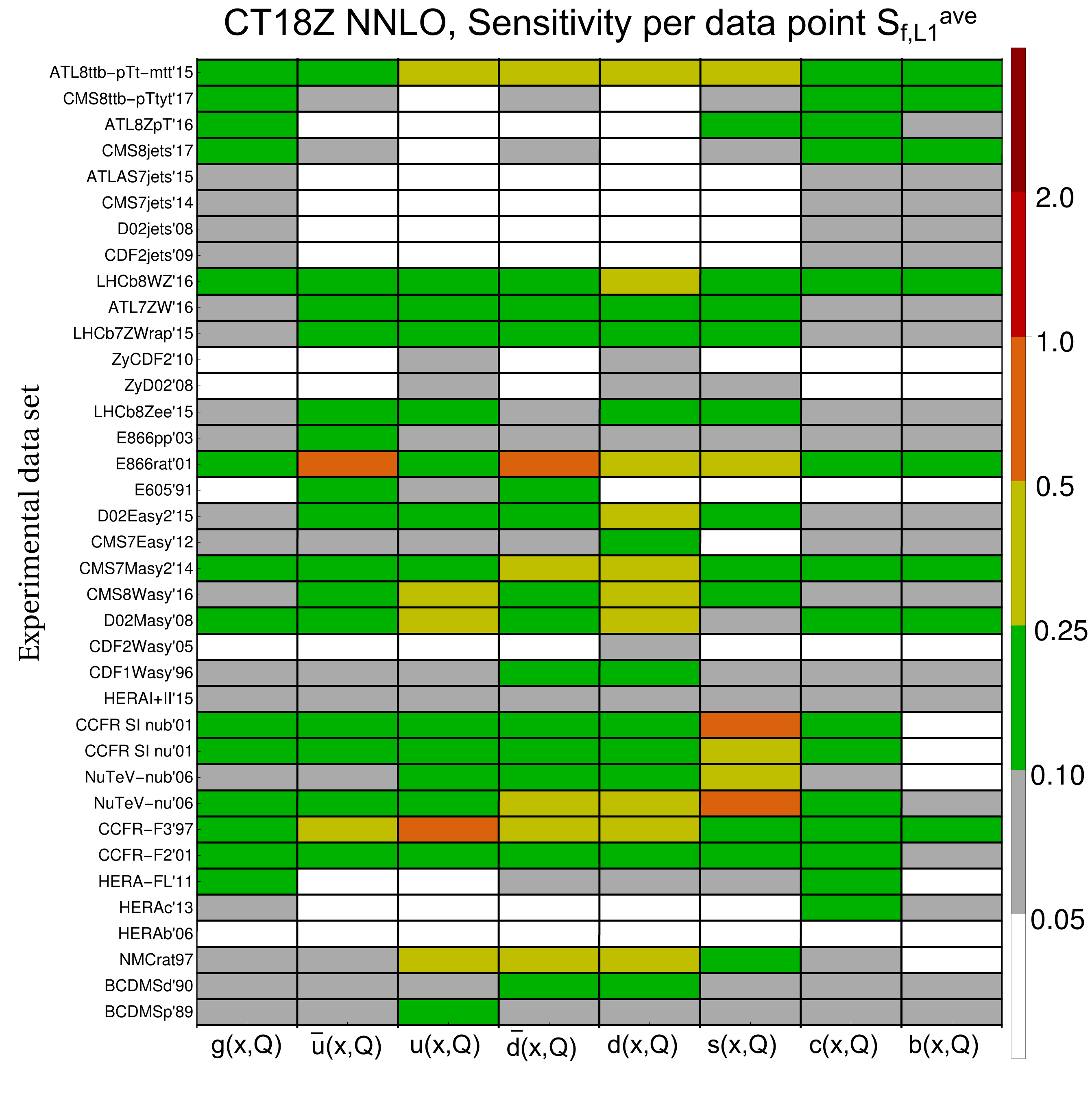}\\ 
  \includegraphics[width=0.59\textwidth]{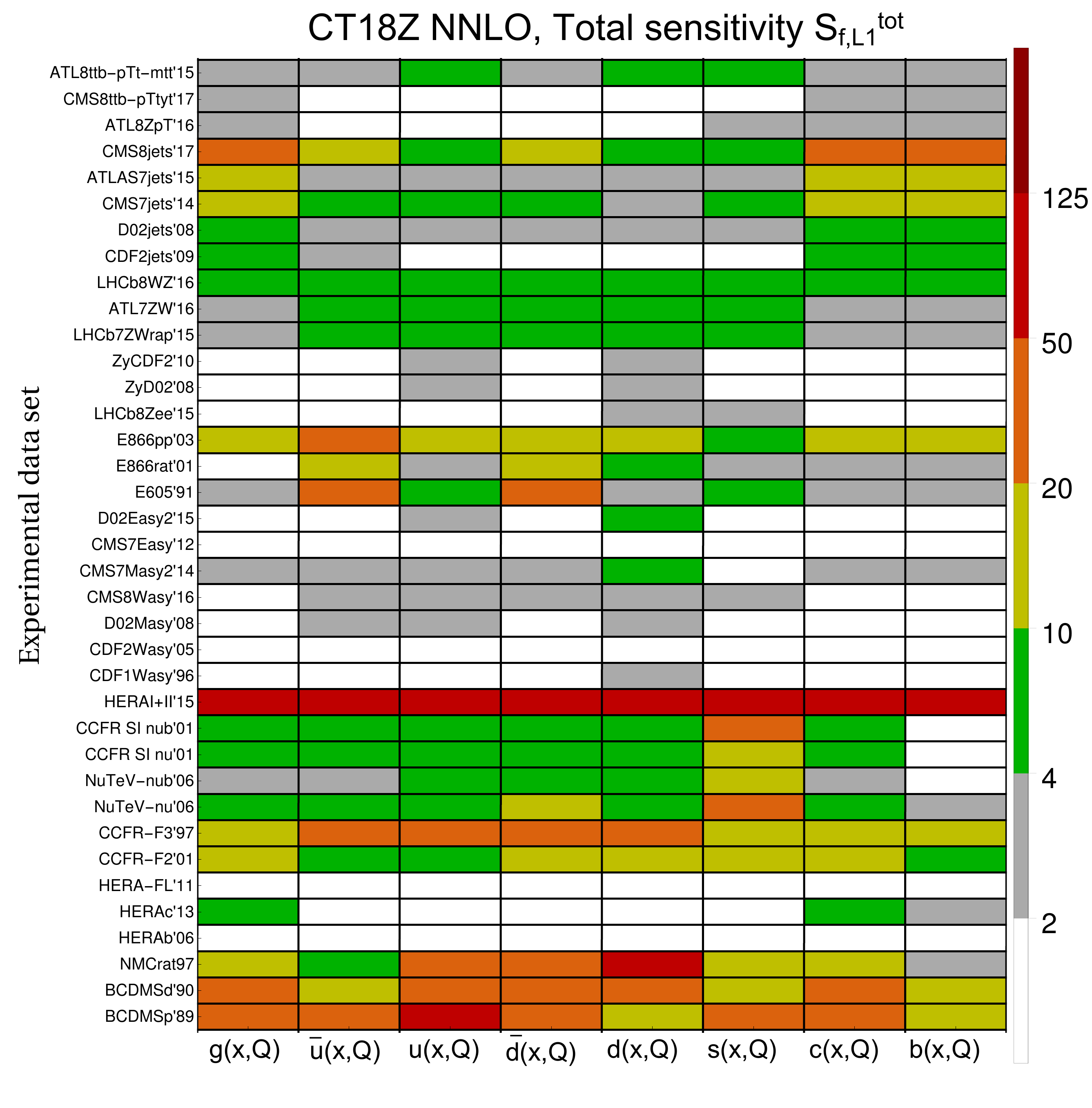}
        	\caption{$L_1$ sensitivities of experimental data
                  sets to PDF flavors in the CT18Z NNLO analysis,
                  computed according to the methodology in
               Ref.~\cite{Wang:2018heo}. The color of the cells
                  in the upper (lower) inset, 
                  chosen according to the palettes on the right,
                  indicates the point-average
                  (cumulative) sensitivity of the experimental set
                  on the vertical axis to the PDF flavor
                  on the horizontal axis. 
		\label{fig:CT18Zquilts}}
\end{figure}

\clearpage

\subsection{Additional histograms and comparisons to data}
\label{sec:Supp-hist}
\begin{figure}[h]
\includegraphics[width=0.46\textwidth]{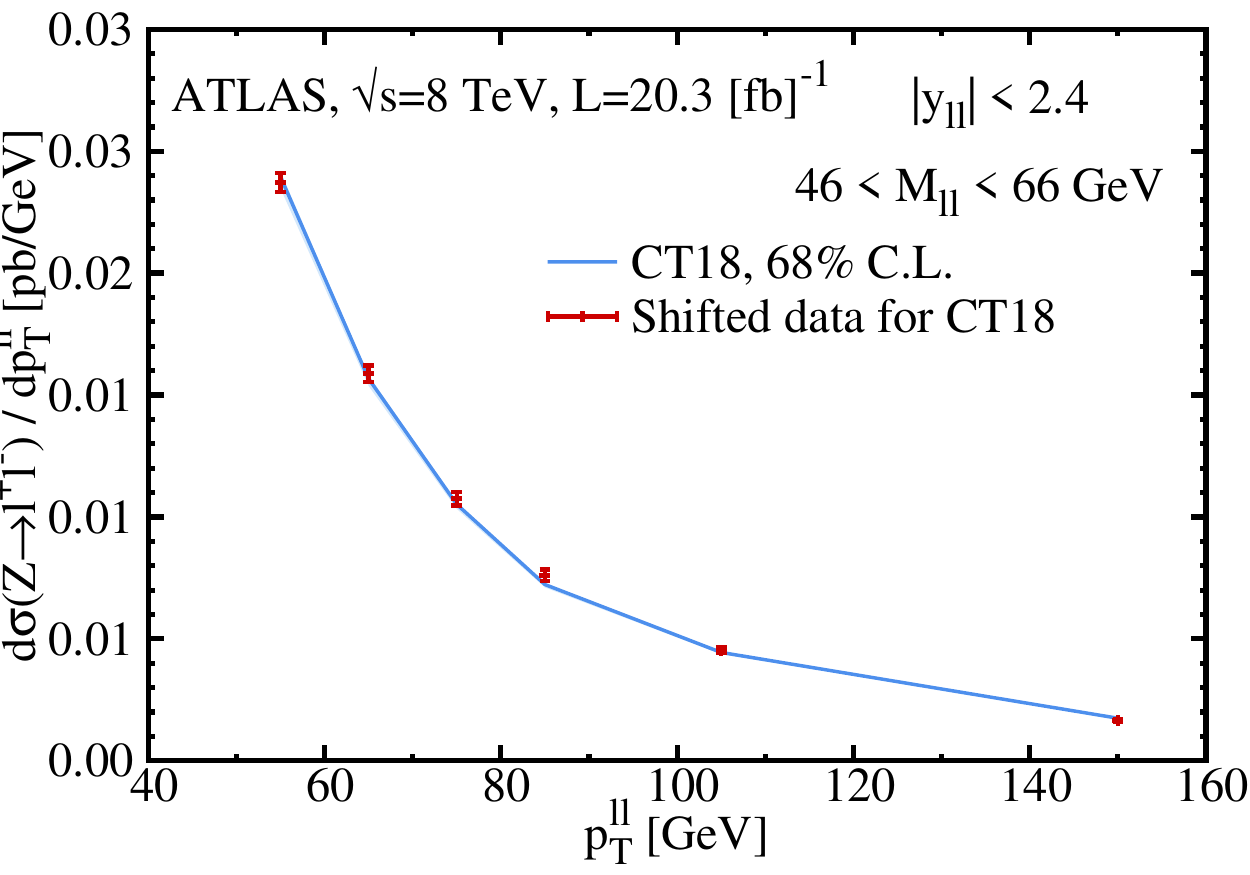}
\includegraphics[width=0.46\textwidth]{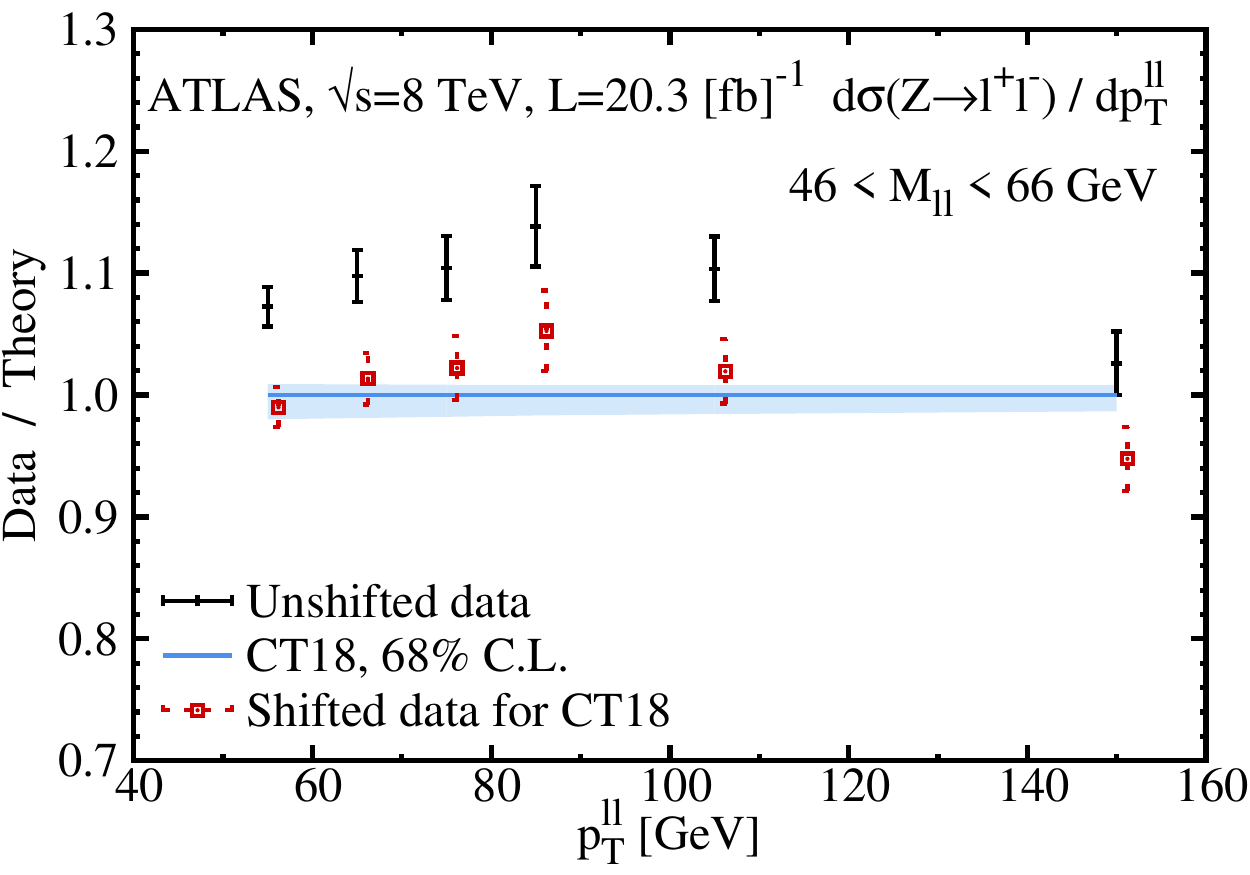}
\includegraphics[width=0.46\textwidth]{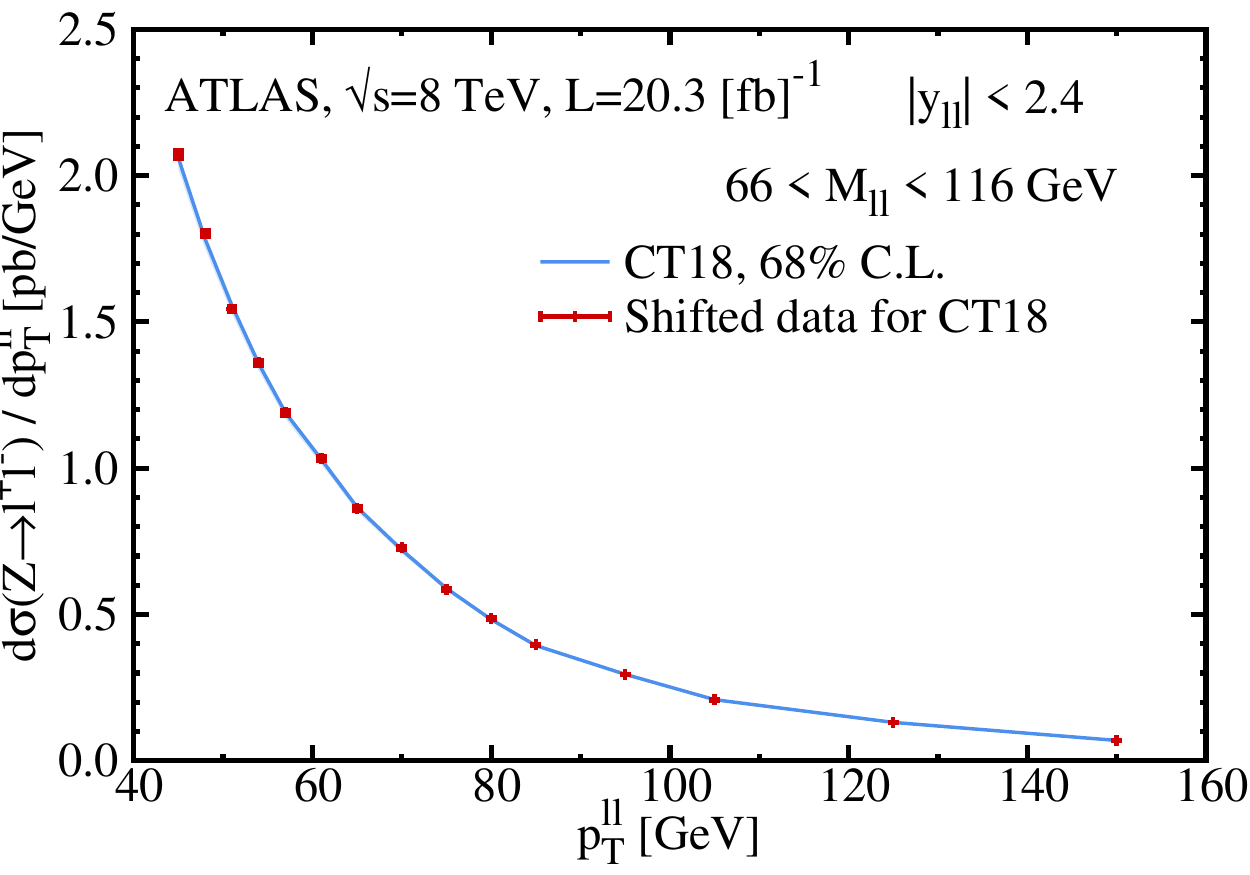}
\includegraphics[width=0.46\textwidth]{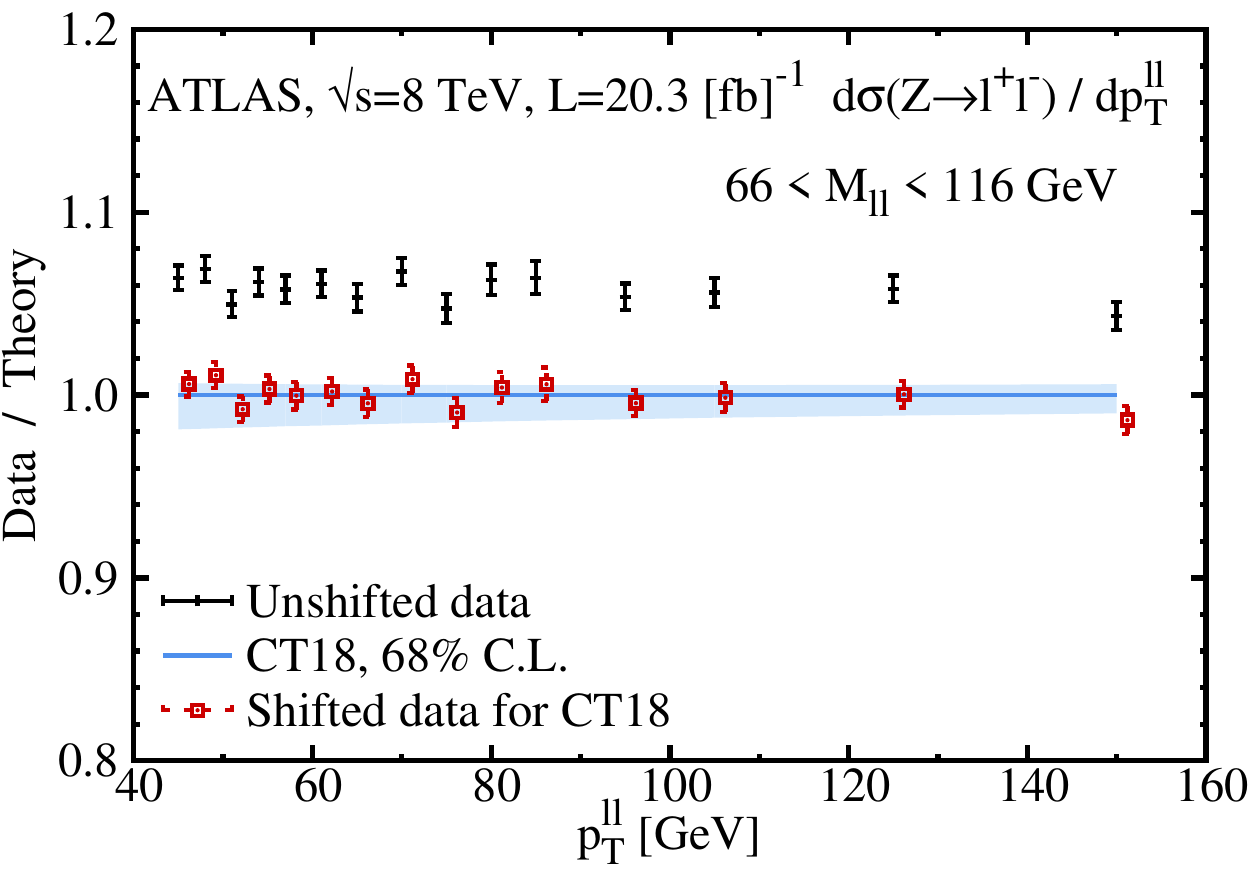}
\includegraphics[width=0.46\textwidth]{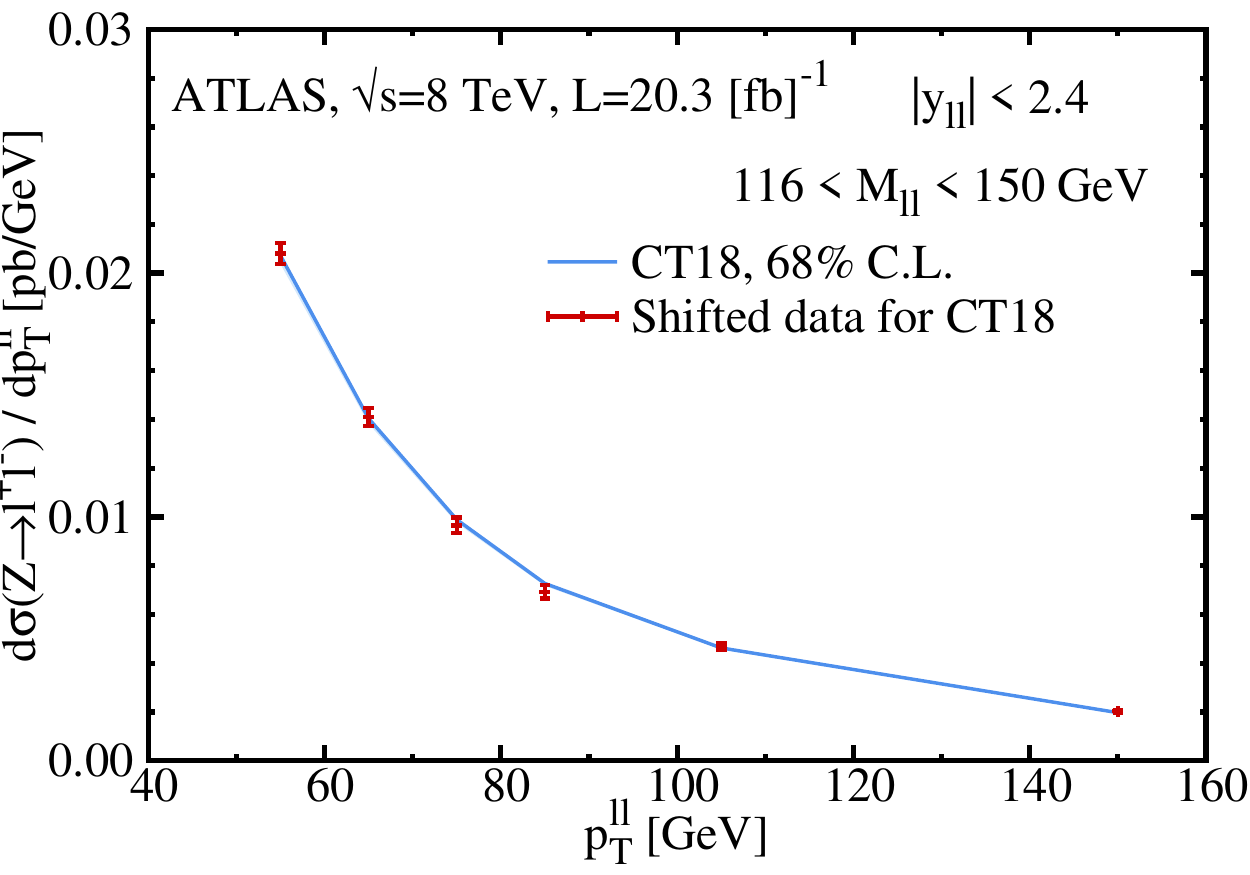}
\includegraphics[width=0.46\textwidth]{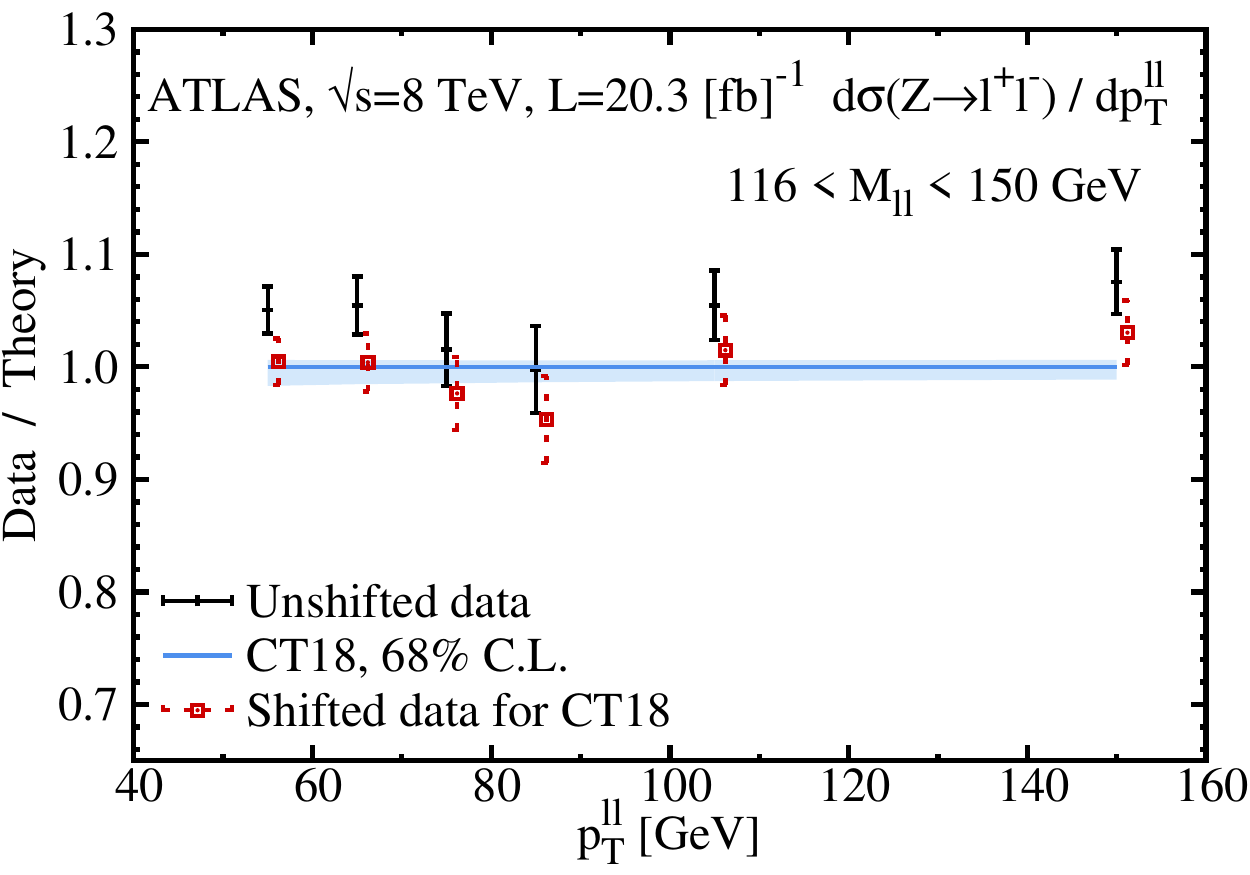}
\caption{A comparison of the CT18 theoretical predictions to the ATLAS 8 TeV $Z\, p_{T}$ data (Exp.~ID=253), using QCD scales $\mu_{R}=M_{T, \ell \bar \ell}$, $\mu_{F}=M_{T, \ell \bar \ell}$ . Predictions for the $p_T$ spectra measured by the ATLAS in
3 bins of the dilepton invariant mass,
$46\! <\! M_{\ell \bar \ell}\! <\! 66$ GeV,
$66\! <\! M_{\ell \bar \ell}\! <\! 116$ GeV, and
$116\! <\! M_{\ell \bar \ell}\! <\! 150$ GeV, are shown in the upper, center, and lower rows, respectively.
The right panels give the corresponding $\mathrm{Data}/\mathrm{Theory}$ profiles
for these data. The blue band represents the PDF uncertainty at the 68\% C.L. The renormalization and factorization scales are chosen as $\mu_{R}=\mu_{F}=M_{T}^{\ell \bar \ell}$.
\label{fig:id253}}
\end{figure}
\newpage

\begin{figure}[h]
	\includegraphics[width=0.49\textwidth]{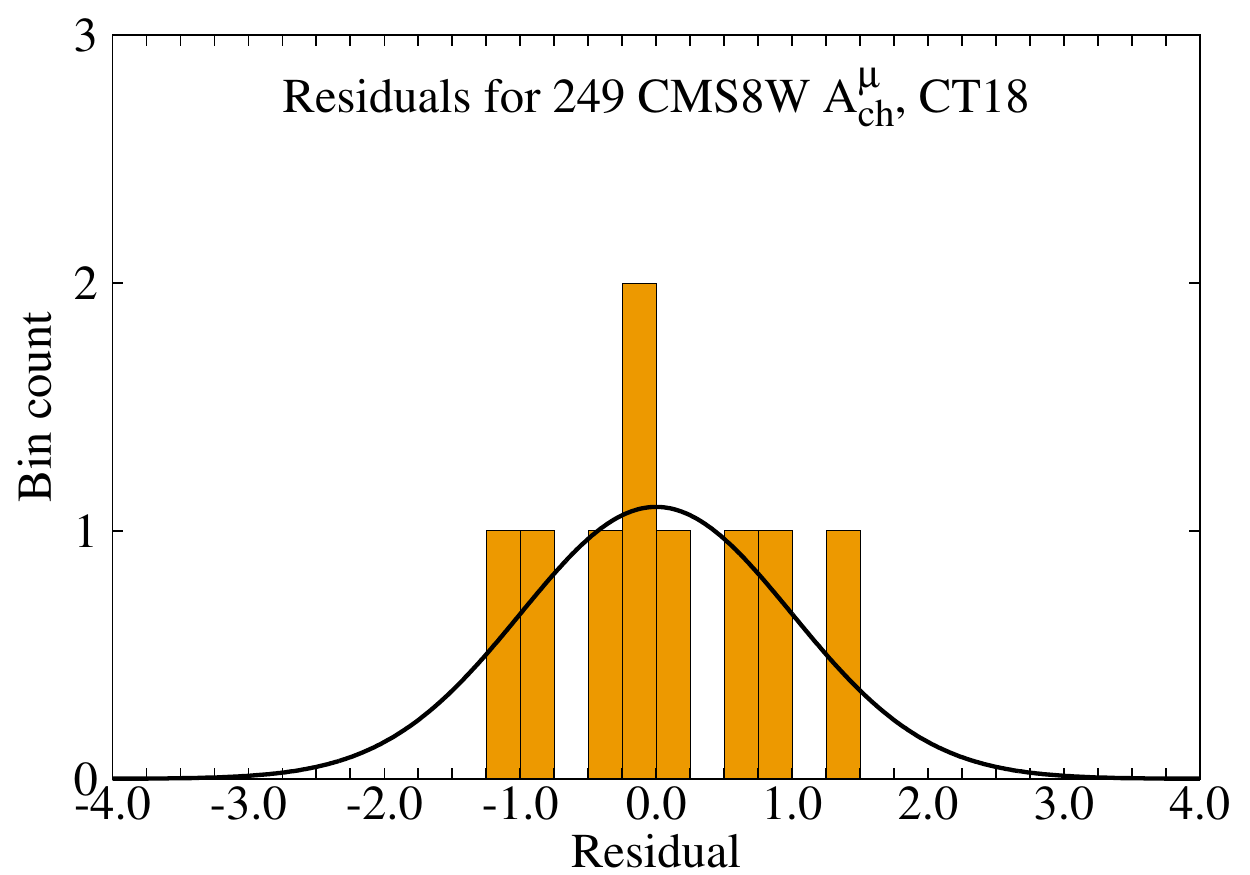}
	\includegraphics[width=0.49\textwidth]{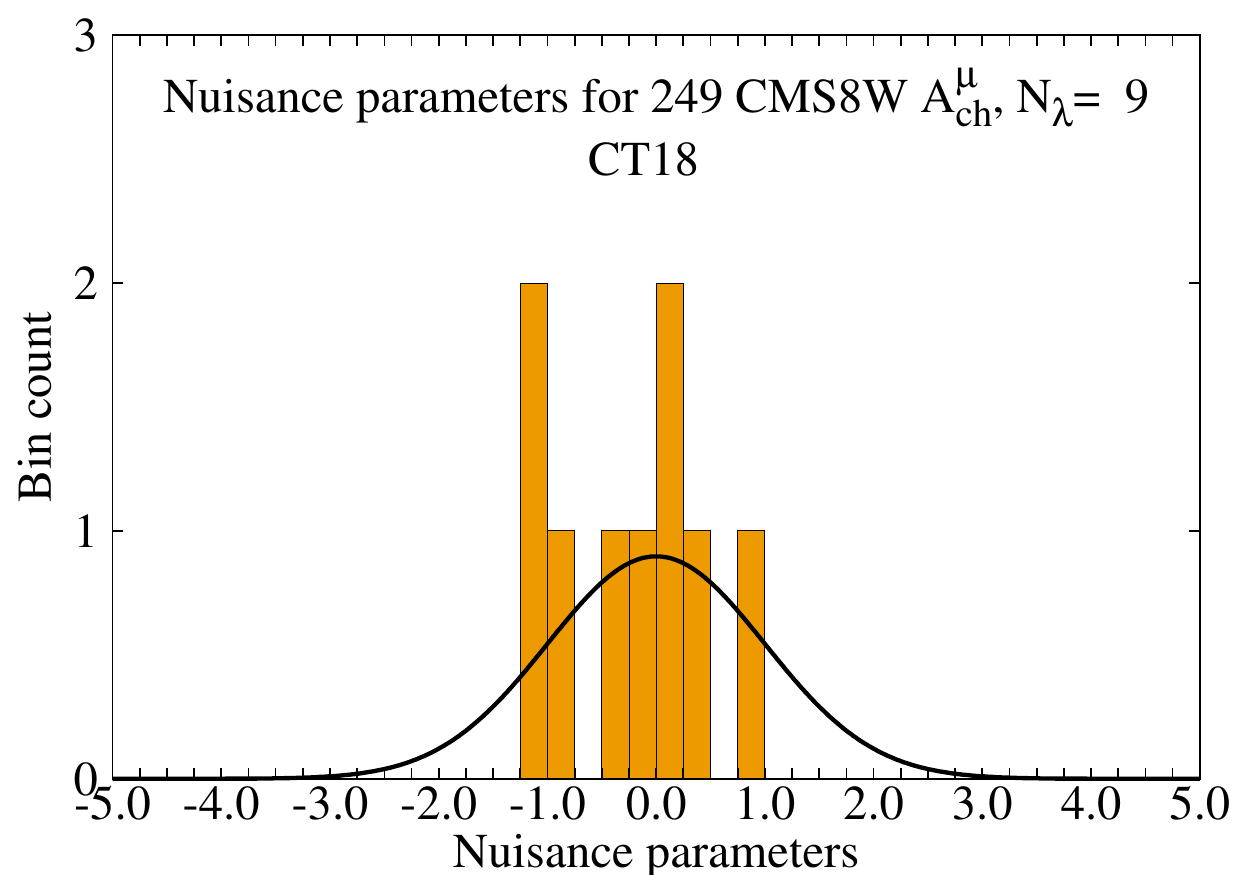}
	\caption{Distribution of residuals (left) and nuisance parameters (right) for the CMS 8 TeV $W$-lepton charge asymmetry data (Exp.~ID=249).
	}
\label{fig:resnui249}
\end{figure}

The residuals and nuisance parameters for Exp.~ID=249 are shown in
Fig.~\ref{fig:resnui249}. Both are reasonably compatible with the
normal distribution with mean 0 and standard deviation 1. 

\begin{figure}[htbp]
	\includegraphics[width=0.32\textwidth]{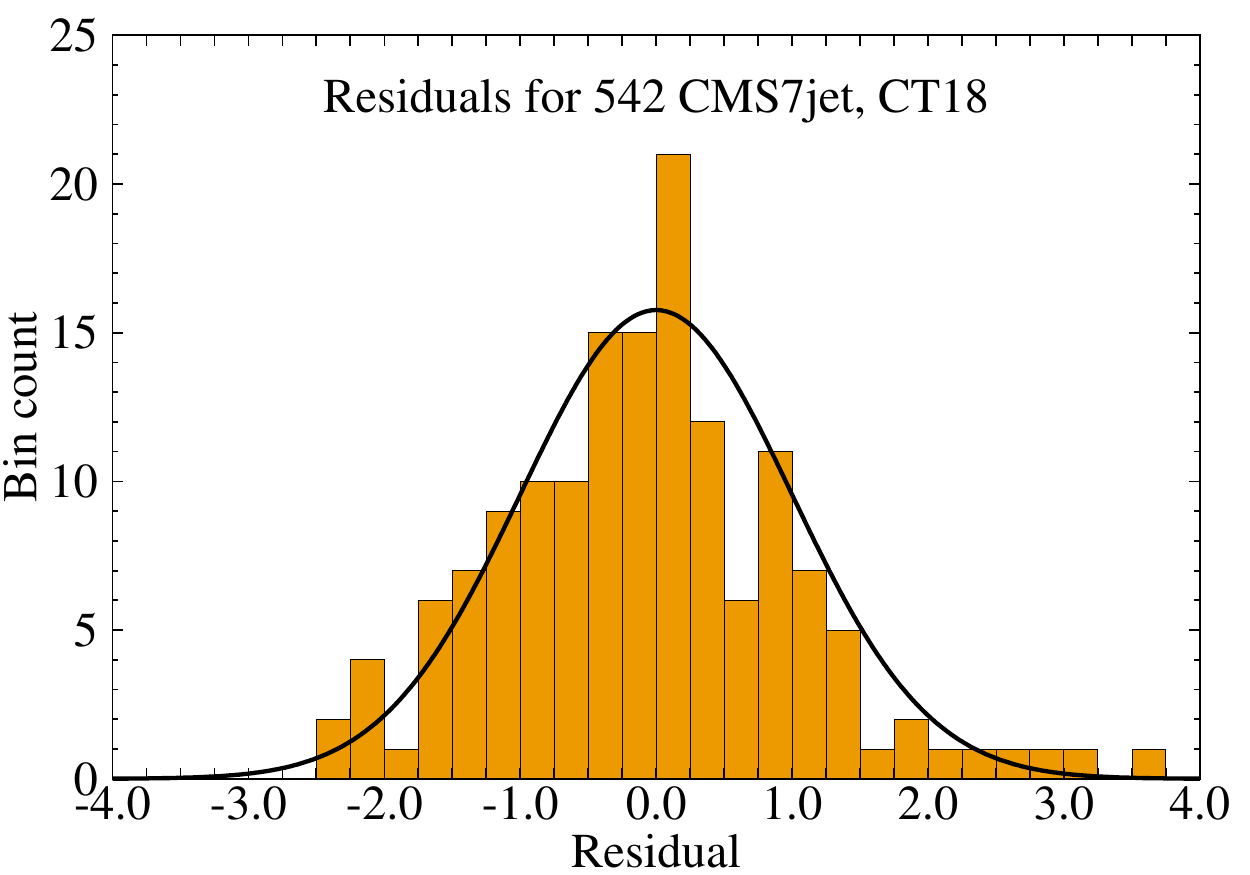}
	\includegraphics[width=0.32\textwidth]{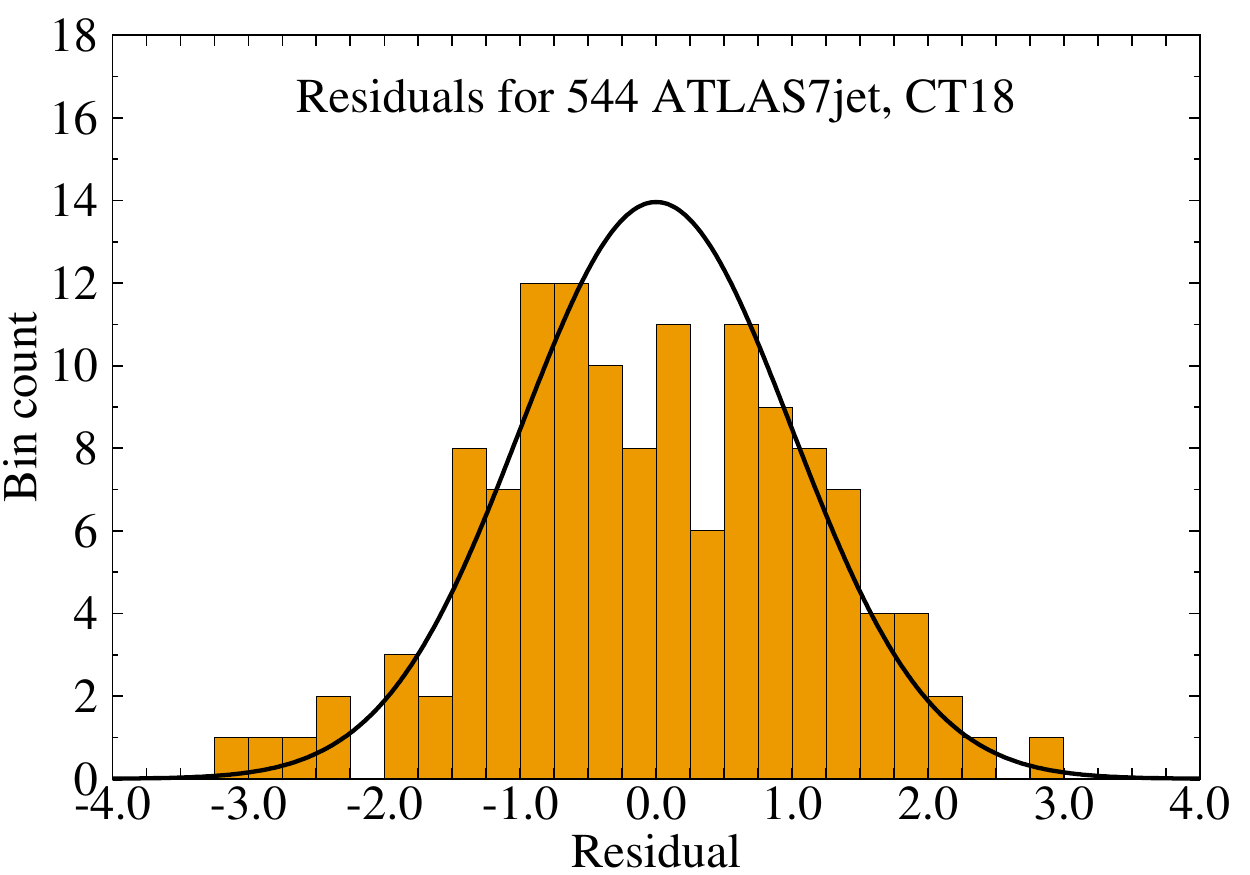}
	\includegraphics[width=0.32\textwidth]{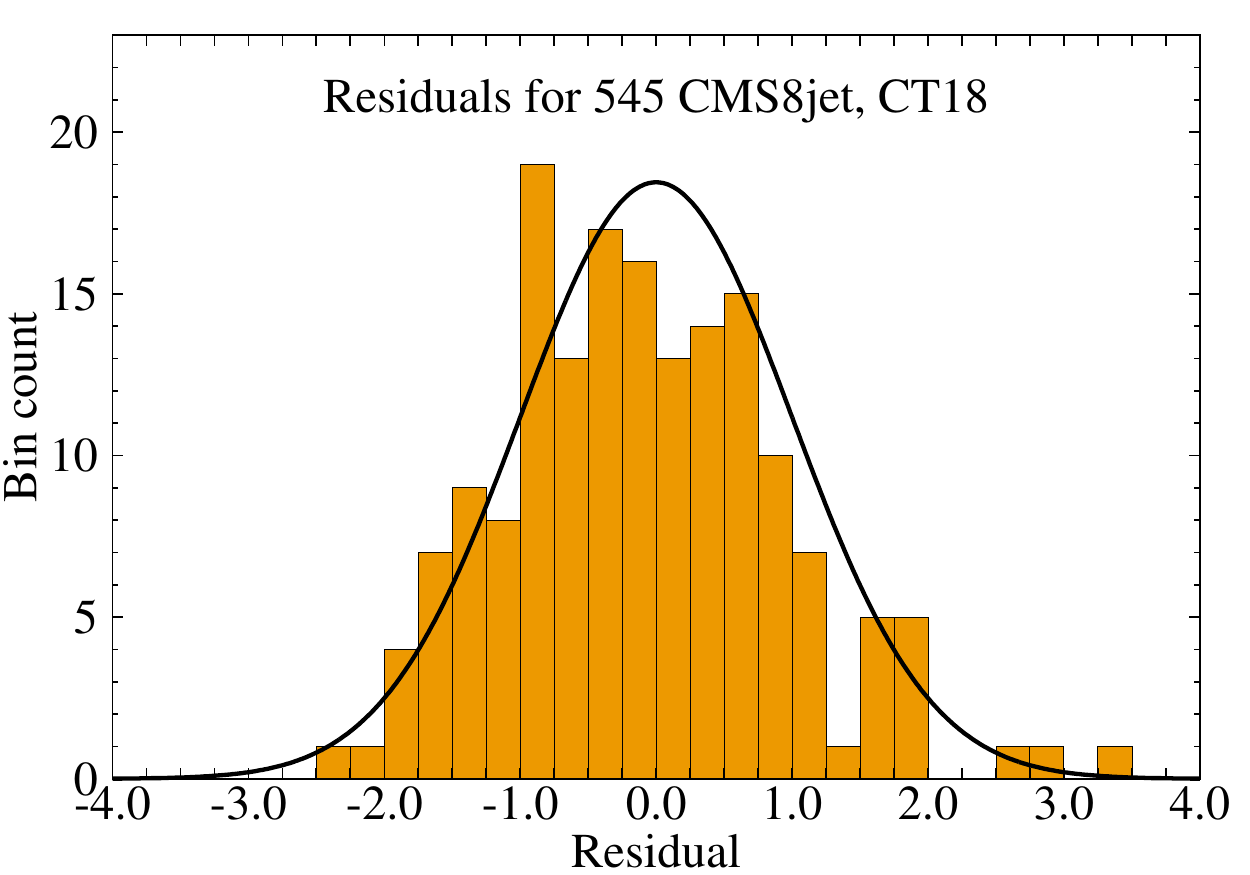}
	\caption{We plot histograms giving the distribution of shifted residuals, $r_i$ of Eq.~(\ref{eq:res-cov}), for each of the newly-included
	LHC jet experiments: the CMS 7 TeV data (Exp.~ID=542, left), ATLAS 7 TeV (Exp.~ID=544, center), and the CMS 8 TeV jet data (Exp.~ID=545, right).
		\label{fig:res_rk_2}}
\end{figure}

\begin{figure}[htbp]
\includegraphics[width=0.32\textwidth]{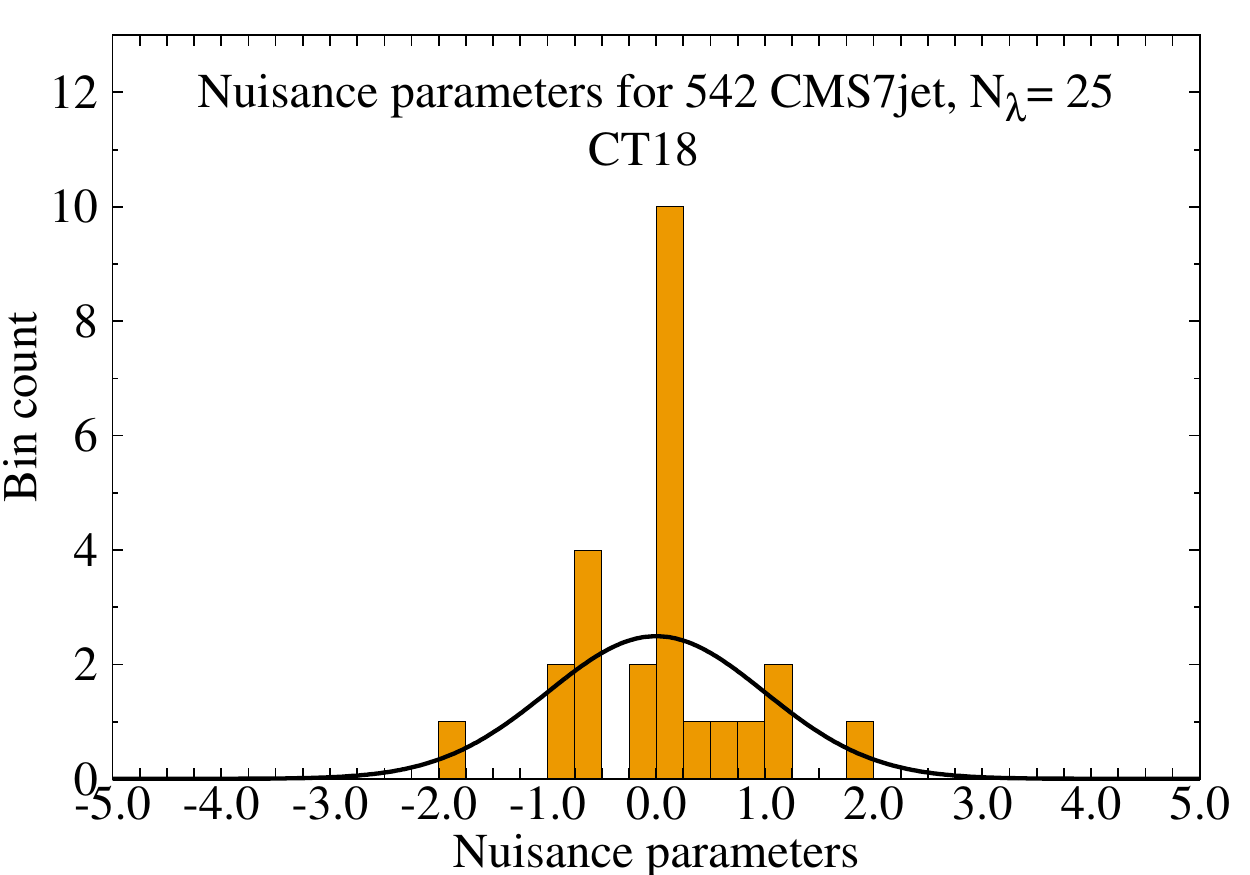}
\includegraphics[width=0.32\textwidth]{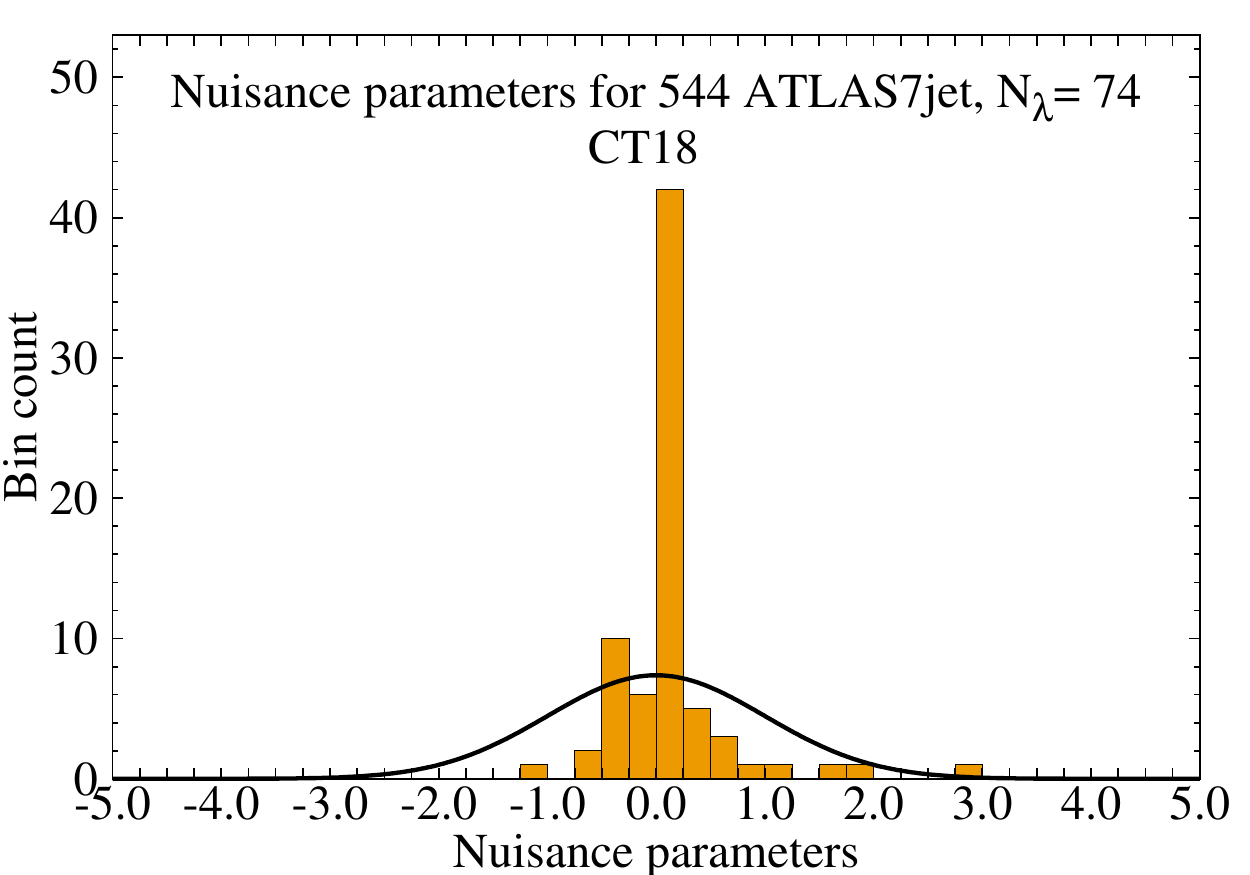}
\includegraphics[width=0.32\textwidth]{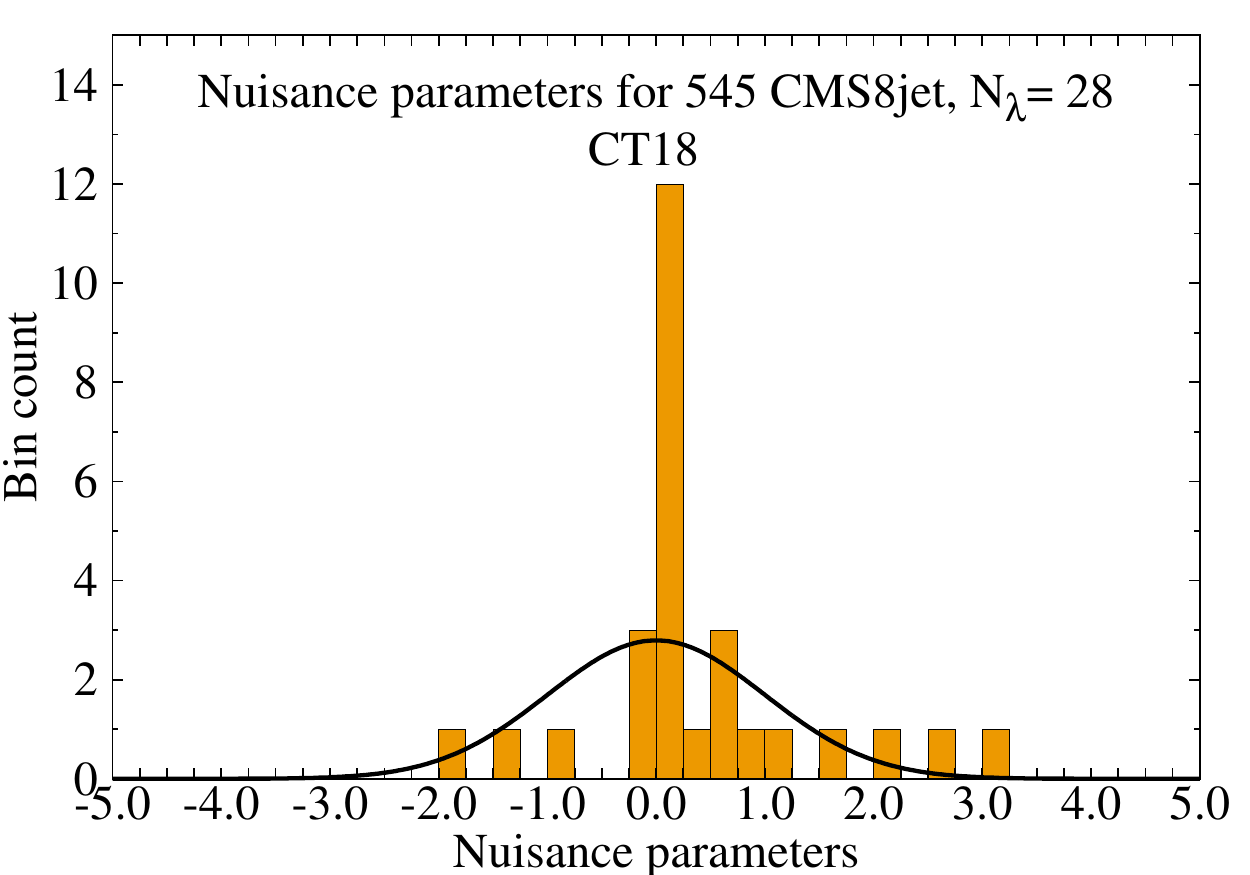}
\caption{Like Fig.~\ref{fig:res_rk_2}, but now for the distribution of nuisance parameters obtained for the CMS 7 TeV data (Exp.~ID=542, left), ATLAS 7 TeV
	(Exp.~ID=544, center), and the CMS 8 TeV jet data (Exp.~ID=545, right).
}
\label{fig:res_rk_4}
\end{figure}

The overall quality of the fit to the combined LHC jet data is 
demonstrated by the distributions of residuals and the fitted 
values of nuisance parameters, shown in
Figs.~\ref{fig:res_rk_2} and \ref{fig:res_rk_4}.

\begin{figure}[htbp]
	\includegraphics[width=0.4\textwidth]{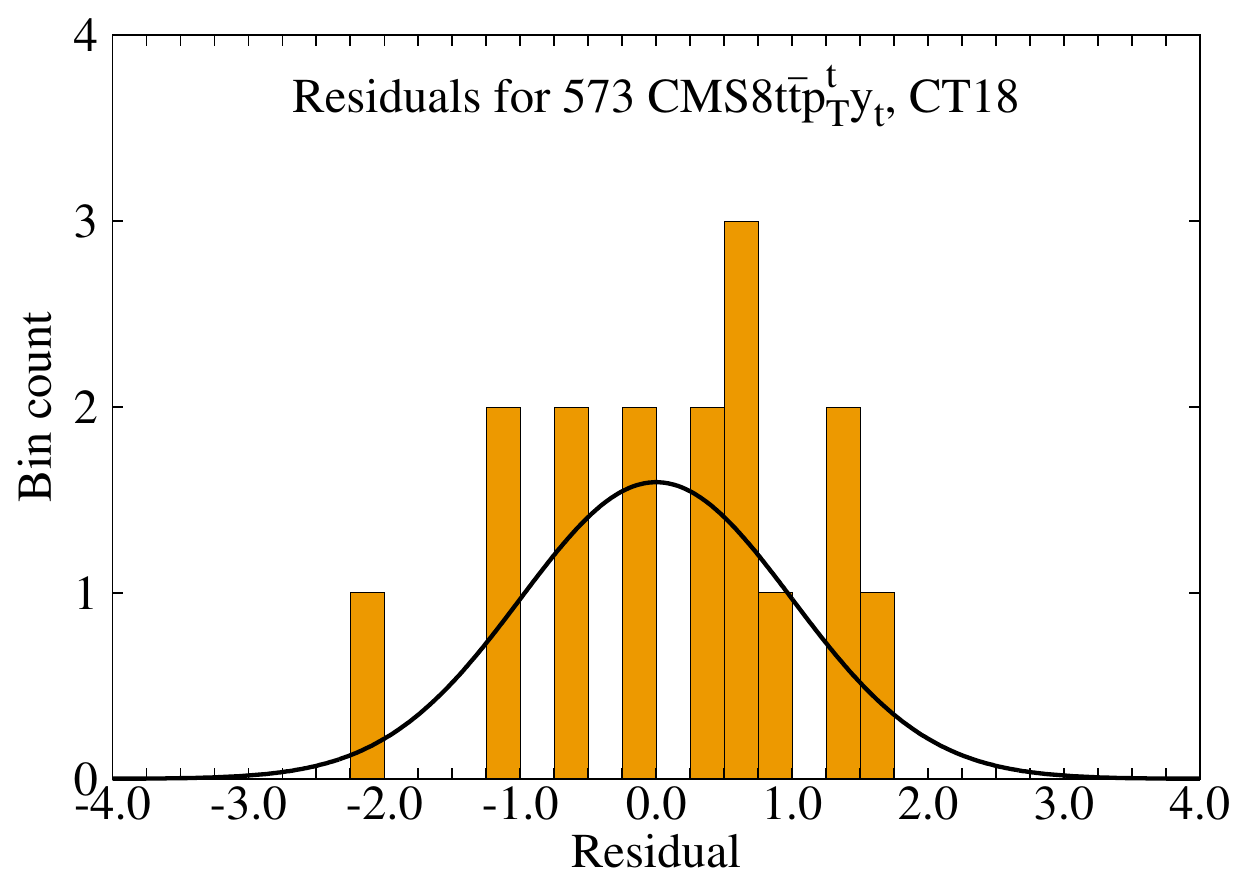}
	\includegraphics[width=0.4\textwidth]{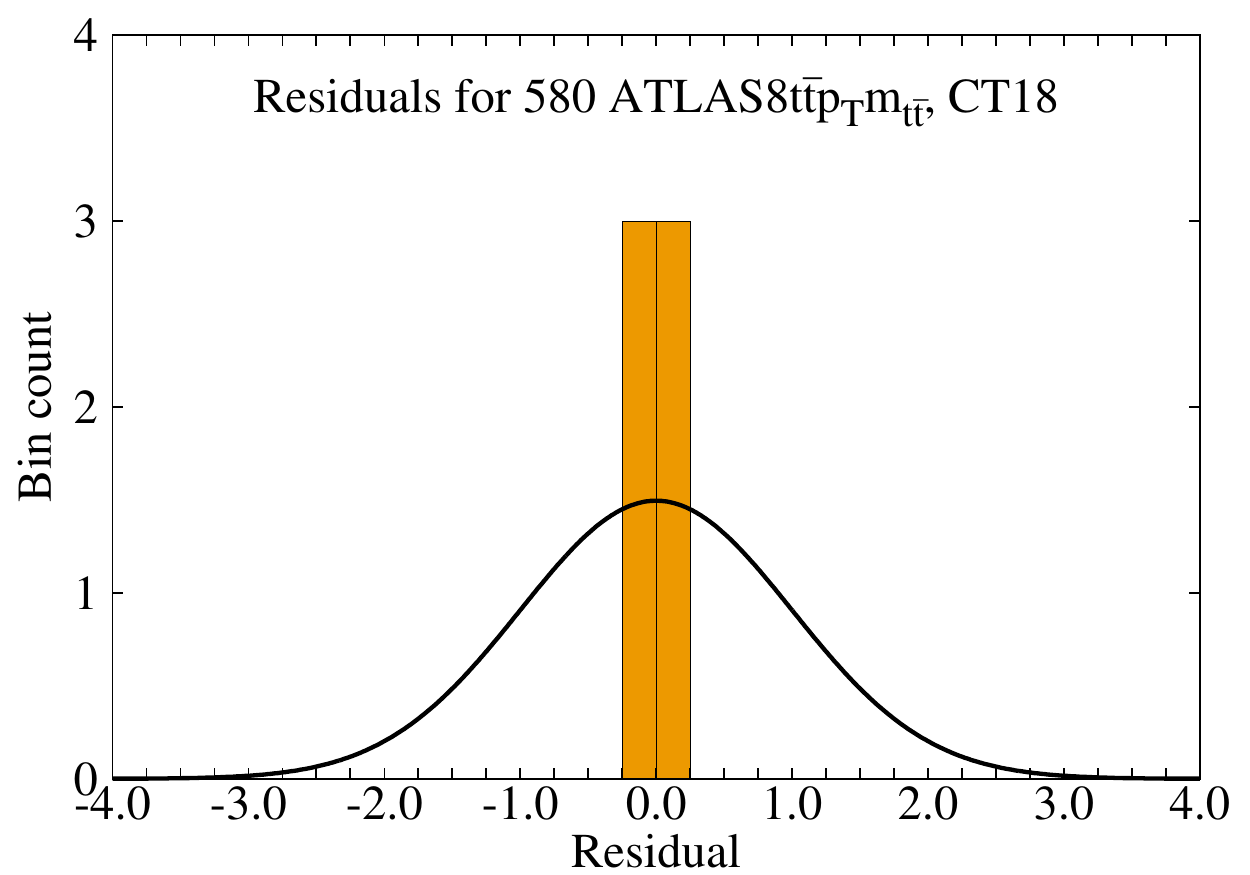}
	\includegraphics[width=0.4\textwidth]{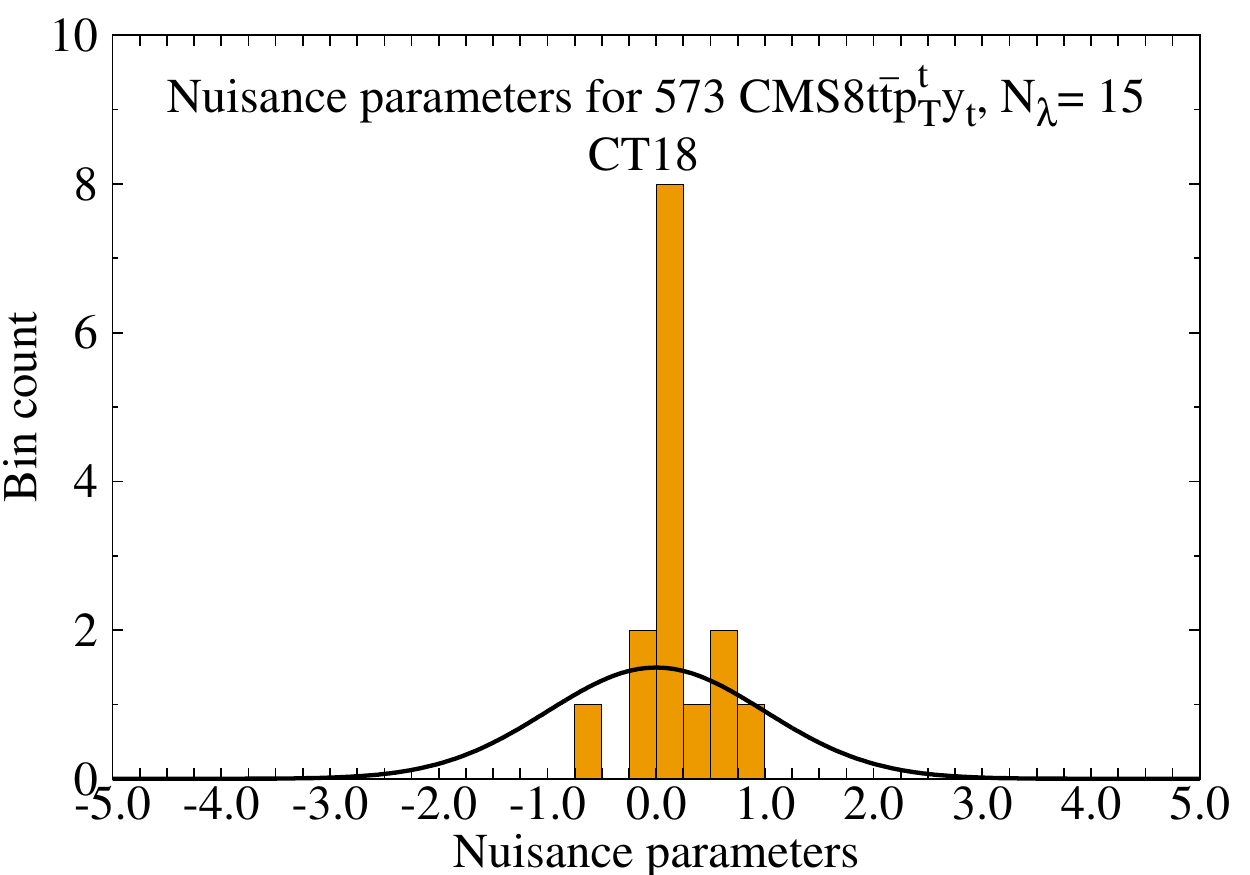}
	\includegraphics[width=0.4\textwidth]{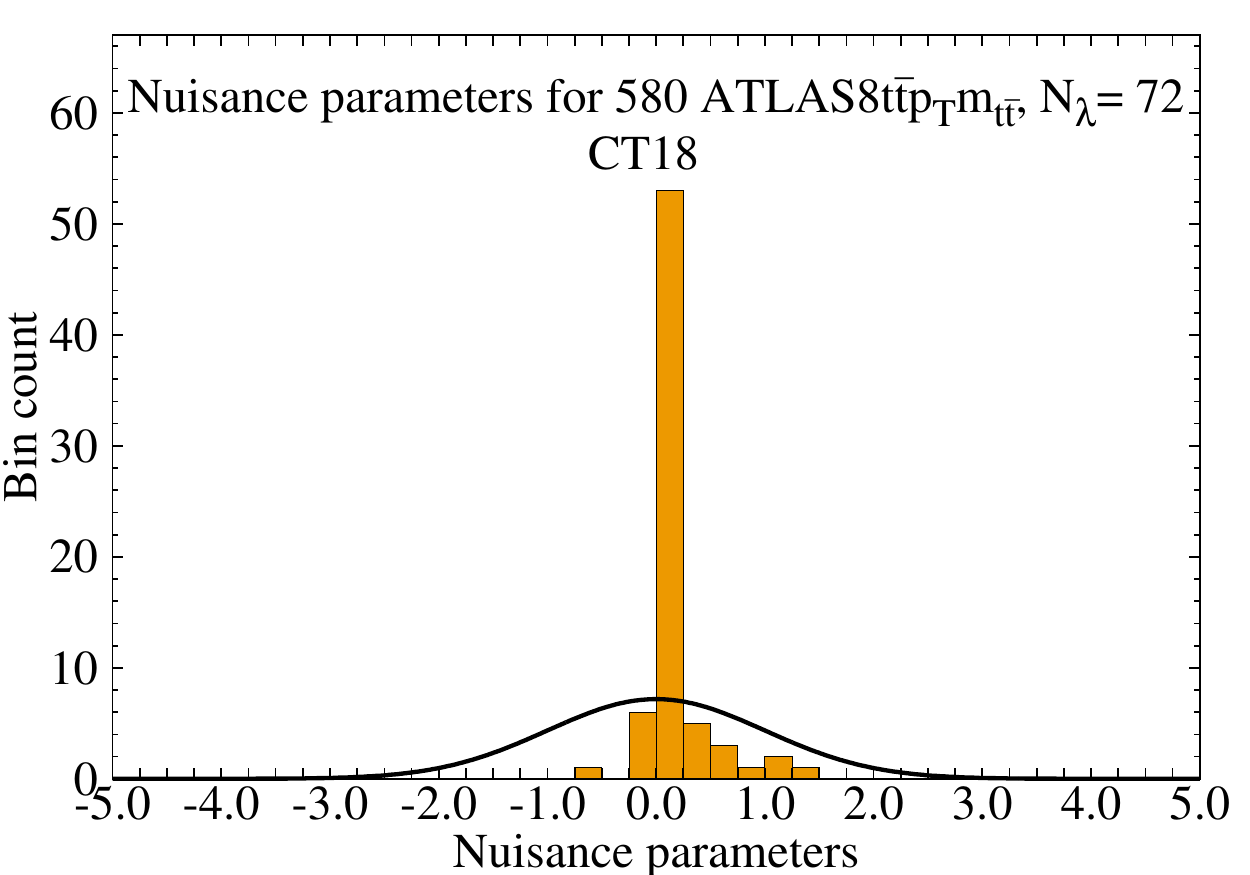}
	\caption{Distribution of residuals (upper panels) and nuisance parameters (lower panels)
	for the CMS (left panels, Exp.~ID=573) and ATLAS (right panels, Exp.~ID=580) 2D top
	quark pair data.
	\label{fig:res_rk_7}}
\end{figure}

\end{document}